\PassOptionsToPackage{unicode}{hyperref}
\PassOptionsToPackage{hyphens}{url}
\documentclass[11pt, ]{article}
\usepackage{mathtools}
\usepackage{amsmath}
\usepackage{amsthm}
\usepackage{amssymb}

\usepackage{iftex}
\ifPDFTeX
  \usepackage[OT1,T1]{fontenc}
  \usepackage[utf8]{inputenc}
  \usepackage{textcomp} 
    \usepackage[p,osf,swashQ]{cochineal}
  \usepackage[cochineal,vvarbb]{newtxmath}
      \usepackage[scale=0.95]{biolinum}
    \usepackage[scale=0.95,varl]{inconsolata}
\else 
  \usepackage[scale=0.95,varl]{inconsolata}
  \usepackage{newpxtext}
  \usepackage{mathpazo}
    \usepackage[scale=0.95]{biolinum}
  \fi
\ifLuaTeX
  \usepackage{selnolig}  
\fi
\IfFileExists{microtype.sty}{
  \usepackage[]{microtype}
  \UseMicrotypeSet[protrusion]{basicmath} 
}{}

\allowdisplaybreaks
\usepackage[dvipsnames,svgnames,x11names]{xcolor}
\usepackage[lmargin=1in,rmargin=1in,tmargin=1in,bmargin=1in]{geometry}
\usepackage[format=plain,
  labelfont={bf,sf,small,singlespacing},
  textfont={sf,small,singlespacing},
  justification=justified,
  margin=0.25in]{caption}

\usepackage{sectsty}
\usepackage[compact]{titlesec}
\makeatletter
\newcommand\@shorttitle{}
\newcommand\shorttitle[1]{\renewcommand\@shorttitle{#1}}
\usepackage{fancyhdr}
\fancyheadoffset{0pt}
\makeatother
\renewenvironment{abstract}{
  \centerline
  {\large\sffamily\bfseries Abstract}\vspace{-1em}
  \begin{quote}\small
}{
  \end{quote}
}

\makeatletter
\newcommand{\assumplabel}[2]{%
   \protected@write \@auxout {}{\string\newlabel{#1}{{#2}{\thepage}{#2}{#1}{}}}%
   \hypertarget{#1}{#2}%
}
\makeatother

\providecommand{\tightlist}{%
  \setlength{\itemsep}{0pt}\setlength{\parskip}{0pt}}\usepackage{longtable,booktabs,array}
\usepackage{calc} 
\usepackage{etoolbox}
\makeatletter
\patchcmd\longtable{\par}{\if@noskipsec\mbox{}\fi\par}{}{}
\makeatother
\IfFileExists{footnotehyper.sty}{\usepackage{footnotehyper}}{\usepackage{footnote}}
\makesavenoteenv{longtable}
\usepackage{graphicx}
\makeatletter
\newsavebox\pandoc@box
\newcommand*\pandocbounded[1]{
  \sbox\pandoc@box{#1}%
  \Gscale@div\@tempa{\textheight}{\dimexpr\ht\pandoc@box+\dp\pandoc@box\relax}%
  \Gscale@div\@tempb{\linewidth}{\wd\pandoc@box}%
  \ifdim\@tempb\p@<\@tempa\p@\let\@tempa\@tempb\fi
  \ifdim\@tempa\p@<\p@\scalebox{\@tempa}{\usebox\pandoc@box}%
  \else\usebox{\pandoc@box}%
  \fi%
}
\def\fps@figure{htbp}
\makeatother

\usepackage[italicdiff]{physics}

\usepackage{amsfonts}
\usepackage{amssymb}
\usepackage{bm} 
\usepackage{bbm}
\usepackage{xurl} 
\usepackage{algorithm,algorithmic}
\usepackage{pgfplots}
\usepgfplotslibrary{dateplot}

\usepackage{graphicx}
\usepackage{tikz}
\usetikzlibrary{arrows.meta,calc}

\usepackage[section]{placeins}

\DeclarePairedDelimiterX{\ainner}[2]{\langle}{\rangle}{#1, #2}
\DeclarePairedDelimiter{\aabs}{\lvert}{\rvert}
\DeclarePairedDelimiter{\anorm}{\lVert}{\rVert}
\DeclarePairedDelimiter{\acbrackets}{\{}{\}}
\DeclarePairedDelimiter{\aparens}{(}{)}
\DeclarePairedDelimiter{\asbrackets}{[}{]}

\newcommand{\bbN}{\mathbb{N}}

\newcommand{\bbR}{\mathbb{R}}

\let\abs\relax
\newcommand{\abs}[1]{\aabs*{#1}}
\newcommand{\inner}[2]{\ainner*{#1}{#2}}

\newcommand{\prob}[2][]{\mathbb{P}_{#1}\aparens*{#2}}
\newcommand{\set}[1]{\acbrackets*{#1}}
\newcommand{\E}[2][]{\mathbb{E}_{#1}\asbrackets*{#2}}

\let\exp\relax
\newcommand{\exp}[1]{\text{exp}\acbrackets*{#1}}

\newcommand{\I}[1]{\mathbbm{1}_{\{#1\}}} 
\newcommand{\Iset}[1]{\mathbbm{1}_{#1}} 
\newcommand{\iid}{\stackrel{iid}{\sim}}

\newcommand{\TV}[1]{\anorm*{#1}_{\text{TV}}}
\newcommand{\dTV}[2]{d_{\mathrm{TV}}\aparens*{#1,#2}}

\newcommand{\toD}[1][]{\xrightarrow[#1]{d}}
\newcommand{\toP}[1][]{\xrightarrow[#1]{P}}

\newcommand{\toAS}{\xrightarrow{a.s.}}

\newcommand{\toLnum}[1]{\xrightarrow{\mathcal{L}^{#1}}}

\newcommand{\gauss}[2]{\mathcal{N}\aparens*{#1,#2}}

\newcommand{\bern}[1]{\mathrm{Bern}\aparens*{#1}}

\newcommand{\dev}{\mathrm{Dev}}

\newcommand{\TreeCut}[1]{\acbrackets*{(T_{k}^{#1}, s_k), (T_{k'}^{#1}, s_{k'}), #1}}

\newcommand{\IGraph}[1]{\mathrm{G}\aparens*{#1}}
\newcommand{\IVertex}[1]{\mathrm{V}\aparens*{#1}}
\newcommand{\IEdge}[1]{\mathrm{E}\aparens*{#1}}
\newcommand{\AllTreeCuts}[2]{\mathcal{TC}\aparens*{#1, #2}}
\newcommand{\ITree}[2][]{\mathrm{T}_{#1}\aparens*{#2}}
\newcommand{\ITreeInv}[2][]{\mathrm{T}_{#1}^{-1}\aparens*{#2}}
\newcommand{\IRegion}[2][]{\mathrm{R}_{#1}\aparens*{#2}}
\newcommand{\IRegionInv}[2][]{\mathrm{R}_{#1}^{-1}\aparens*{#2}}
\newcommand{\Wilson}[2][]{\mathrm{W}_{#1}\aparens*{#2}}
\newcommand{\Split}[2][]{\mathrm{spl}_{#1}\aparens*{#2}}
\newcommand{\Components}[2][]{\mathrm{comp}_{#1}\aparens*{#2}}
\newcommand{\EffB}[2][]{\mathrm{Eff}_{#1}\aparens*{#2}}

\usepackage{pdflscape}
\usepackage{float}

\newcommand{\NaiveKernel}[1]{\mathrm{TopK}\aparens*{#1}}
\newcommand{\ArbKernel}[1]{p_{\text{cut}}\aparens*{#1}}

\usepackage{xcolor}
\definecolor{mycolor}{HTML}{008000}

\usepackage{inconsolata}

\usepackage{chngcntr}

\makeatletter
\@ifpackageloaded{caption}{}{\usepackage{caption}}
\AtBeginDocument{%
\ifdefined\contentsname
  \renewcommand*\contentsname{Table of contents}
\else
  \newcommand\contentsname{Table of contents}
\fi
\ifdefined\listfigurename
  \renewcommand*\listfigurename{List of Figures}
\else
  \newcommand\listfigurename{List of Figures}
\fi
\ifdefined\listtablename
  \renewcommand*\listtablename{List of Tables}
\else
  \newcommand\listtablename{List of Tables}
\fi
\ifdefined\figurename
  \renewcommand*\figurename{Figure}
\else
  \newcommand\figurename{Figure}
\fi
\ifdefined\tablename
  \renewcommand*\tablename{Table}
\else
  \newcommand\tablename{Table}
\fi
}
\@ifpackageloaded{float}{}{\usepackage{float}}
\floatstyle{ruled}
\@ifundefined{c@chapter}{\newfloat{codelisting}{h}{lop}}{\newfloat{codelisting}{h}{lop}[chapter]}
\floatname{codelisting}{Listing}

\usepackage{amsthm}
\theoremstyle{plain}
\newtheorem{corollary}{Corollary}[section]
\theoremstyle{plain}
\newtheorem{proposition}{Proposition}[section]
\theoremstyle{plain}
\newtheorem{lemma}{Lemma}[section]
\theoremstyle{definition}
\newtheorem{definition}{Definition}[section]
\theoremstyle{plain}
\newtheorem{theorem}{Theorem}[section]
\theoremstyle{remark}
\AtBeginDocument{}

\newtheorem{refremark}{Remark}[section]

\makeatother
\makeatletter
\@ifpackageloaded{caption}{}{\usepackage{caption}}
\@ifpackageloaded{subcaption}{}{\usepackage{subcaption}}
\makeatother

\usepackage[]{natbib}
\newenvironment{CSLReferences}[2]{
\bibliography{references.bib}
\clearpage
}{}

\usepackage{hyperref}
\usepackage{url}
\hypersetup{
  pdftitle={Generalized Sequential Monte Carlo Sampling for Redistricting Simulation},
  pdfauthor={Philip O'Sullivan; Kosuke Imai; Cory McCartan},
  pdfkeywords={redistricting simulation, Sequential Monte
Carlo, multi-member districts, gerrymandering},
  colorlinks=true,
  linkcolor={black},
  filecolor={Maroon},
  citecolor={VioletRed4},
  urlcolor={DodgerBlue4},
  pdfcreator={LaTeX via pandoc}}

\title{\sffamily\bfseries\huge\parfillskip=0pt
\rightskip=0pt plus .5\textwidth
\leftskip=0pt plus .5\textwidth
\emergencystretch=.3\textwidth Generalized Sequential Monte Carlo
Sampling for Redistricting Simulation}
\shorttitle{Generalized Sequential Monte Carlo Sampling for Redistricting Simulation}
\author{\textbf{Philip O'Sullivan}
\\Department of Statistics%
\\Harvard University%
\vspace{2pt}
 \and \textbf{Kosuke Imai}\footnote{
To whom correspondence should be addressed.
Email: \texttt{\href{mailto:imai@harvard.edu}{imai@harvard.edu}}.
Address:
1737 Cambridge Street, Cambridge, MA 02138.
We acknowledge partial support from the Sinnott-Wagner Geospatial
Statistics Fund.}
\\Department of Government and Department of Statistics%
\\Harvard University%
\vspace{2pt}
 \and \textbf{Cory McCartan}
\\Department of Statistics%
\\Pennsylvania State University%
\vspace{2pt}
 }
\date{September 8, 2026}

\begin{document}
\allsectionsfont{\sffamily}

\maketitle

\begin{abstract}
Simulation methods have become important tools for quantifying partisan
and racial bias in redistricting plans. We generalize the Sequential
Monte Carlo (SMC) algorithm of \citet{mccartan2023}, one of the commonly
used approaches. First, our generalized SMC (gSMC) algorithm can split
off regions of arbitrary size, rather than a single district as in the
original SMC framework, enabling the sampling of multi-member districts
with a varying number of representatives. Second, the gSMC algorithm can
operate over various sampling spaces, providing additional computational
flexibility. Third, we derive optimal-variance incremental weights and
show how to compute them efficiently for each sampling space, leading to
more efficient sampling. Finally, we propose a hybrid gSMC-MCMC
algorithm by incorporating Markov chain Monte Carlo (MCMC) steps to
handle large-scale redistricting applications without changing the
target distribution. We demonstrate the effectiveness of the proposed
methodology through analyses of the Irish Parliament, which uses
multi-member districts of varying sizes, and the Pennsylvania House of
Representatives, which has more than 200 single-member districts.
\end{abstract}

\textbf{\textit{Keywords}}\quad redistricting
simulation~\textbullet~Sequential Monte Carlo~\textbullet~multi-member
districts~\textbullet~gerrymandering


\section{Introduction}\label{sec-intro}

Evaluation of districting plans for racial and partisan bias
increasingly relies on simulation-based methods. Analysts use a
simulation algorithm to sample large ensembles of alternative
redistricting plans from a target distribution that encodes relevant
state and federal legal requirements, along with other constraints of
interest. By comparing an enacted plan to this ensemble, researchers can
conduct statistical hypothesis tests for detecting gerrymandering and
quantify partisan and racial bias in the enacted plan. Such
simulation-based approaches can account for a state's physical and
political geography, as well as its jurisdiction-specific redistricting
rules. As a result, these methods have gained broad acceptance among
scholars and courts as a principled and empirically grounded framework
for evaluating redistricting plans
\citep[e.g.,][]{chen2013, Magleby_Mosesson_2018, gerrymandering2022, Becker2021-ws, Herschlag2020-vj, chen_stephanopoulos_2021_race_blind_future, DeFord_Eubank_Rodden_2022, Kaufman2025, 2018league, 2019rucho, 2024utah}.

\begin{figure}[t]

\centering{

\includegraphics[width=0.8\linewidth,height=\textheight,keepaspectratio]{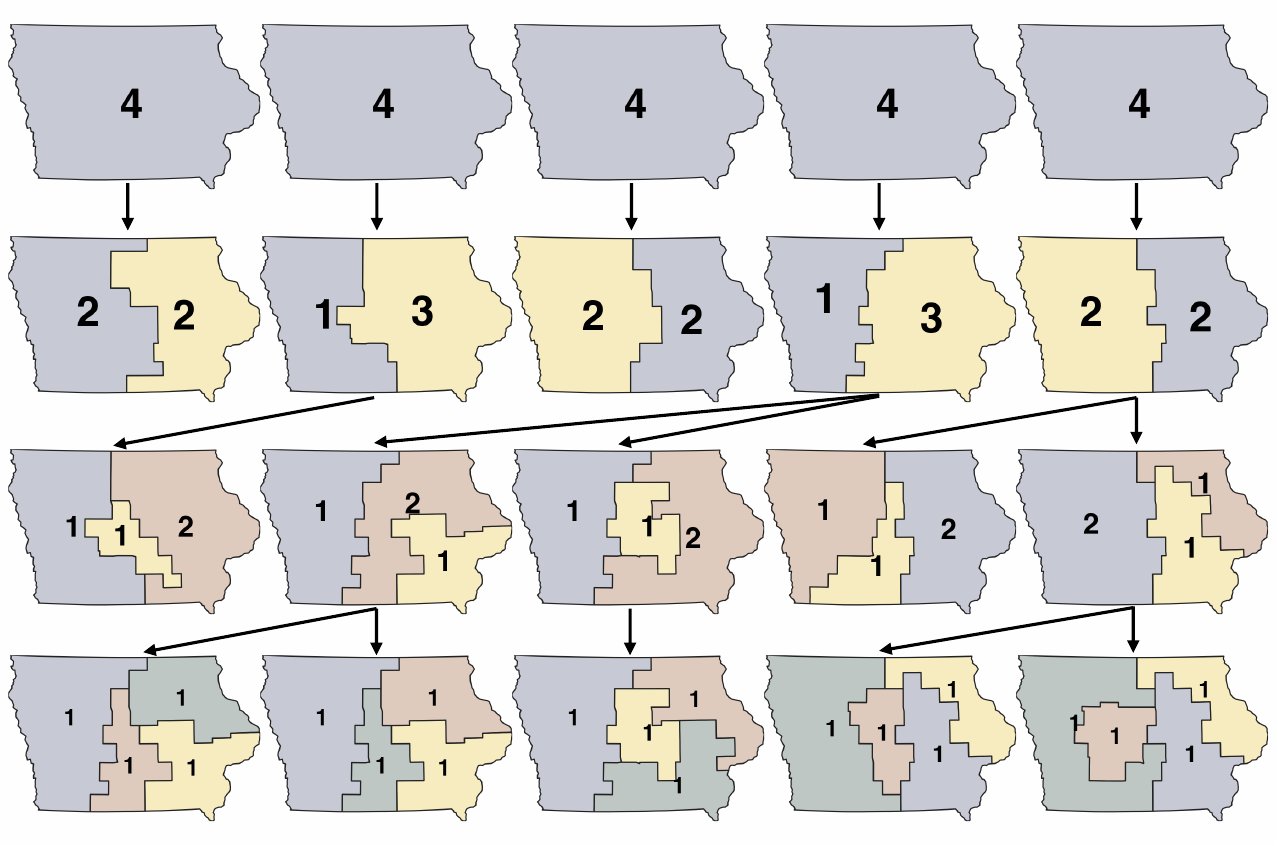}

}

\caption{\label{fig-iowa-smc-example}gSMC schematic for congressional
districting in Iowa with \(N = 5\) plans. At each step an ancestor
partial plan is selected with probability proportional to its importance
weights and split. Newly split plans with positive probability under the
target distribution are kept. Arrows indicate which previous stage's
plan was split to form the current one. Numbers on the region indicate
its size, i.e., how many seats it contains.}

\end{figure}

Markov chain Monte Carlo (MCMC) and Sequential Monte Carlo (SMC) methods
are two widely used approaches. MCMC algorithms begin with an existing
plan and produce new samples by iteratively modifying it according to
specified probabilistic transition rules
\citep[e.g.,][]{mattingly2014, wu2015, chikina2017, deford2019, carter2019, fifield2020mcmc, cannon2022spanning}.
In contrast, SMC algorithms construct plans from scratch. Starting with
an unassigned map, they iteratively and probabilistically carve off
districts in parallel, weight the intermediate plans, resample, and
repeat, until complete districting plans are formed
\citep{mccartan2023}. Figure~\ref{fig-iowa-smc-example} visualizes this
process for our proposed algorithm for congressional redistricting in
Iowa.

In this paper, we introduce a new sampling framework that generalizes
the SMC algorithm while maintaining the same class of target
distributions. The proposed generalized SMC (gSMC) algorithm offers
several improvements. First, we introduce a new splitting procedure that
enables the sampling of multi-member districts with a varying number of
representatives, which are used in some U.S. local elections and many
other countries. The gSMC framework can sample a \emph{region} of
varying size, which may comprise one or more districts. This allows for
the efficient sampling of multi-member district plans from a class of
target distributions with asymptotic convergence guarantees.\footnote{One
  could use existing sampling algorithms to generate single-member
  districts and then merge them to form multi-member districts. However,
  this approach faces computational challenges when multi-member
  districts vary in size or when the population of some geographic units
  exceeds the mandated population limit per seat, as in our Irish
  example. Moreover, even when all multi-member districts are of the
  same size, the distribution over districting plans induced by this
  procedure remains unknown and the computational cost is greater.} In
addition to expanding modeling flexibility, this splitting procedure
often reduces computational cost. Empirical results demonstrate runtime
reductions of anywhere up to one-fifth to one-half. This reduction is
generally greater when the size of the map is larger and more MCMC moves
are added.

Second, gSMC can sample districting plans by first sampling more complex
mathematical objects, each of which induces a unique districting plan.
Working in these alternative \emph{sampling spaces} does not change the
final target distribution over districting plans, but it can improve
computational efficiency. Specifically, alternative spaces allow for
more efficient probabilistic splitting of districts, but require more
complex weight calculations, creating a computational trade-off.
Empirically, particularly for large maps with many districts, we find
these alternative sampling spaces reduce overall runtime without harming
performance.

Third, gSMC improves the calculation of intermediate resampling weights,
which are used to reweight sampled plans to approximate the target
distribution. The overall performance of SMC algorithms is heavily
influenced by the variance of these weights. We derive the
minimum-variance weights and provide efficient algorithms for computing
them within each sampling space, thereby improving the statistical and
computational efficiency of the gSMC algorithm.

Finally, we incorporate a merge--split MCMC step, similar to those of
\citep{deford2019, carter2019, McmcForests, McmcLinkingEdge}, into the
gSMC algorithm. The resulting hybrid gSMC--MCMC sampler targets the same
distribution as the original gSMC sampler and retains all of its
theoretical guarantees without requiring additional assumptions about
the irreducibility or aperiodicity of the Markov kernel. At each stage
of gSMC, we randomly select two adjacent regions, merge them, and then
probabilistically re-split them. While SMC samplers constructs plans
from scratch in parallel---often producing less dependent samples than
MCMC samplers---they can become inefficient when there are many
resampling steps, which reduces plan diversity, a problem known as
ancestor extinction. Interspersing merge--split MCMC steps within
splitting steps mitigates this issue by promoting greater diversity
among partial plans throughout the gSMC's iterative procedure.

We show empirically that the combination of these improvements
drastically boost the performance and scalability of the original SMC
algorithm. Section~\ref{sec-applications} applies gSMC to two real-world
scenarios to which the original SMC algorithm would not be applicable:
Ireland's Dáil Éireann, a multi-member map with 43 districts and 174
seats, and Pennsylvania's State House, a large districting map with 203
single-member districts. Appendix Section~\ref{sec-addl-empirical}
demonstrates the improved performance on a range of additional empirical
examples, including several U.S. congressional maps with a deviation of
\emph{two or fewer} people per district.

\section{The Problem Formulation}\label{the-problem-formulation}

This section first formalizes the problem of redistricting simulation
for single-member districts and then generalizes the framework to
multi-member district settings. Specifically, we characterize the target
distribution over districting plans from which we aim to draw a
representative sample.

\subsection{Setup}\label{setup}

We consider sampling districting plans with \(D\) single-member
districts. We represent the map of a state as a graph \(G = (V, E)\),
where \(V = \{v_1, \dots, v_m\}\) denotes the set of geographic units
partitioning the state (e.g., precincts), and \(E\) is the set of edges
denoting units that are considered legally adjacent. Legal adjacency
does not necessarily imply geographical adjacency. For example, some
jurisdictions consider units that are physically separated by bodies of
water to be legally adjacent when there is a ferry line connecting them.
Each geographic unit \(v_i\) has an associated population
\(\mathrm{pop}(v_i)\). For any subset of vertices \(V' \subseteq V\),
its population is defined as
\(\mathrm{pop}(V') = \sum_{v \in V'} \mathrm{pop}(v)\).

\begin{figure}[t]

\begin{minipage}{0.33\linewidth}

\centering{

\includegraphics[width=\linewidth,height=1.125in,keepaspectratio]{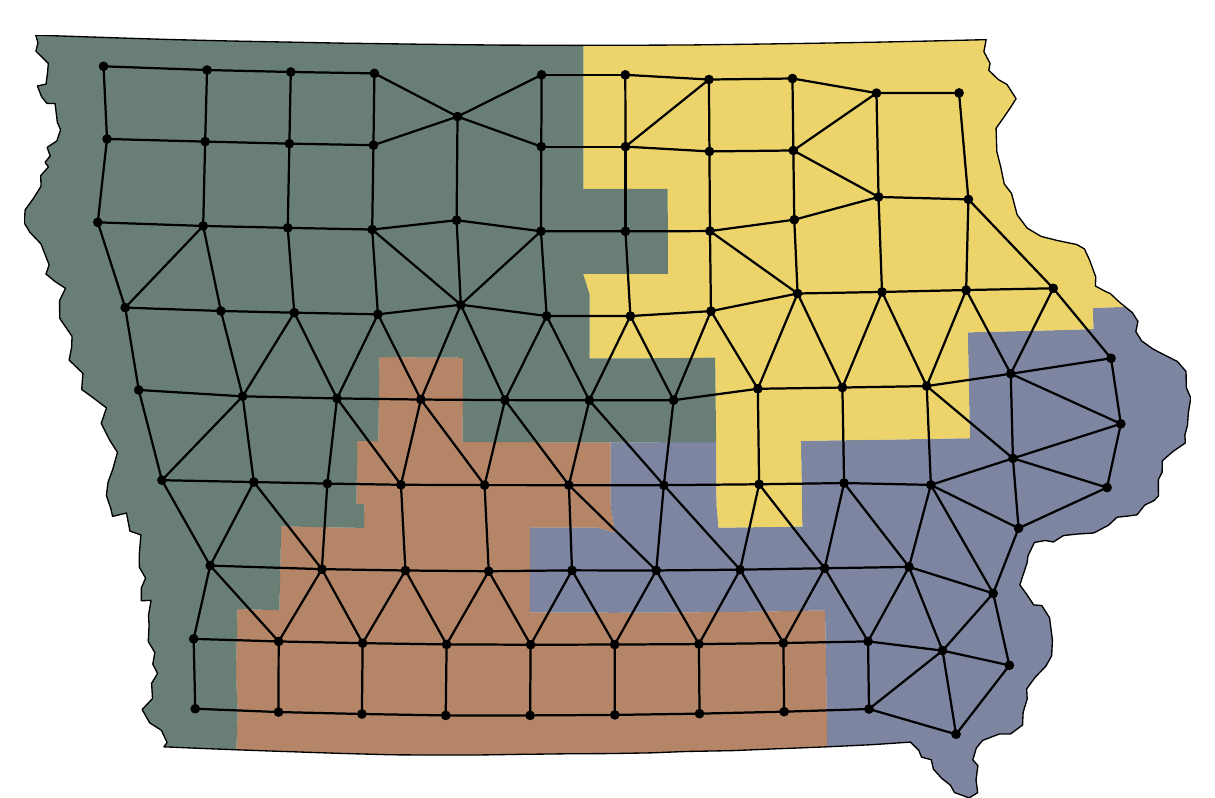}

}

\subcaption{\label{fig-iowa-plan}Iowa as a graph}

\end{minipage}%
\begin{minipage}{0.33\linewidth}

\centering{

\includegraphics[width=\linewidth,height=1.125in,keepaspectratio]{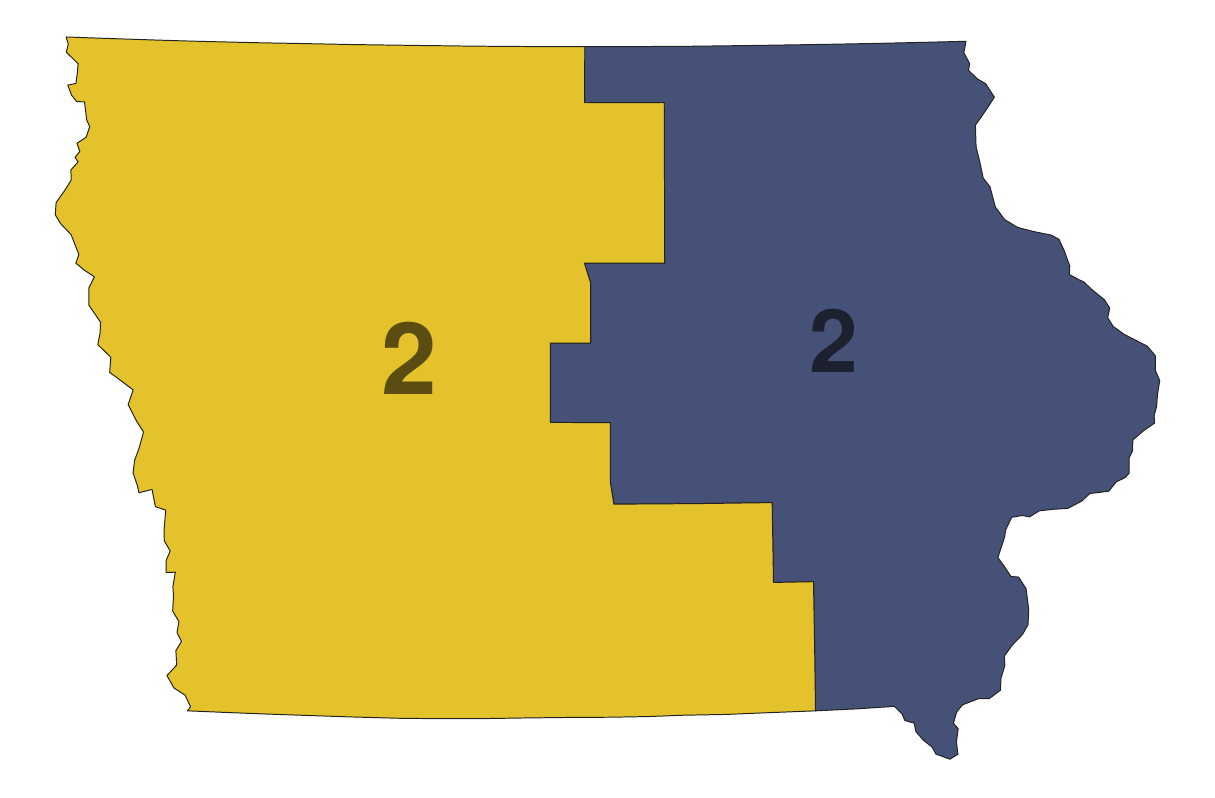}

}

\subcaption{\label{fig-iowa-partial1}Two regions of size two}

\end{minipage}%
\begin{minipage}{0.33\linewidth}

\centering{

\includegraphics[width=\linewidth,height=1.125in,keepaspectratio]{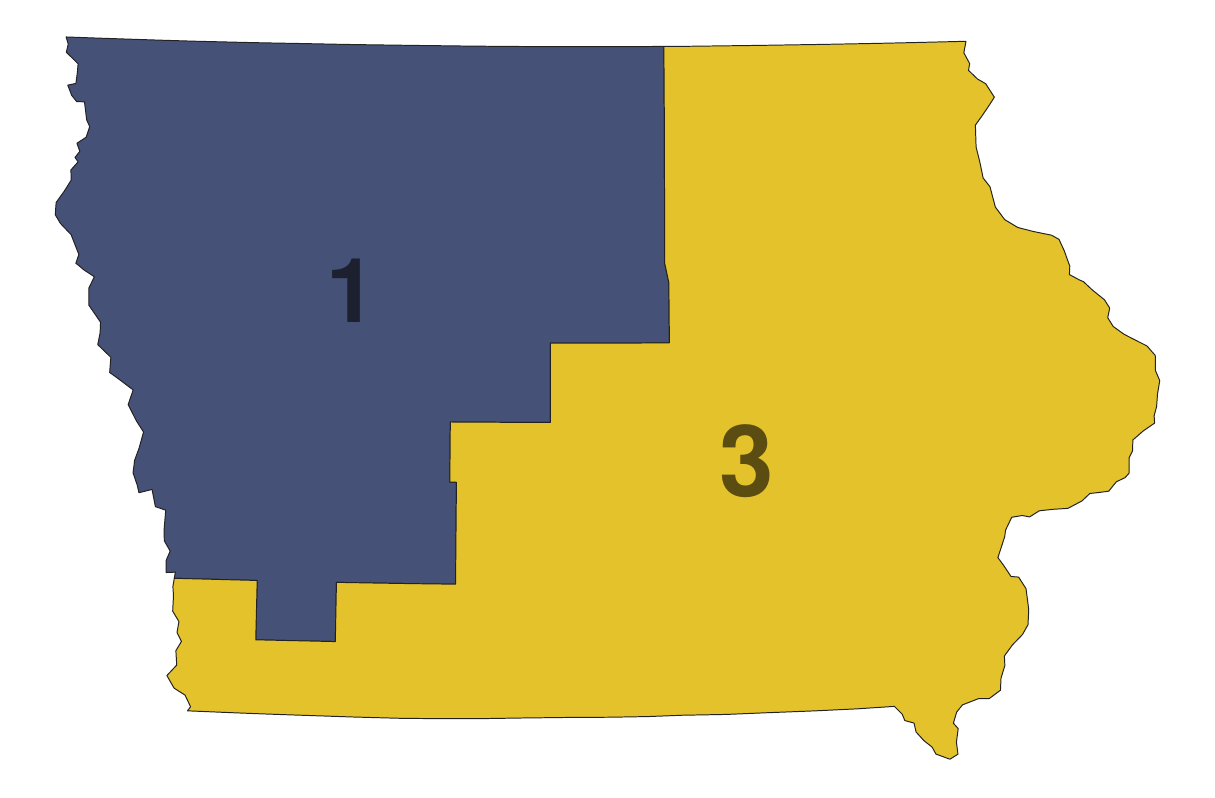}

}

\subcaption{\label{fig-iowa-partial2}One district and one region of size
three}

\end{minipage}%

\caption{\label{fig-iowa-examples}Panel (a) shows the 2020 enacted plan
for Iowa as a graph where counties are geographical units. Panels (b)
and (c) present examples of regions. Regions that are not districts are
called multidistricts.}

\end{figure}

A single-member district
\(\widetilde{G} = (\widetilde{V}, \widetilde{E})\) is defined as the
subgraph of \(G\) induced by a connected vertex subset
\(\widetilde{V} \subset V\). A redistricting plan with \(D\)
single-member districts is represented by the collection of districts
\(\xi = \{G_k\}_{k=1}^D\), where the vertex sets \(V_k\) corresponding
to districts \(G_k\) form a partition of \(V\), i.e.,
\(V = \bigcup_{k=1}^D V_k\) and
\(V_k \cap V_\ell = \emptyset \ \text{for } k \neq \ell\). For
notational convenience, we index the districts in a plan as
\(G_1, \dots, G_D\). Formally, however, districts are \emph{unlabeled}
subgraphs of \(G\), and any given plan admits \(D!\) equivalent
labelings corresponding to permutations of the district
indices.\footnote{Unlike in \citet{mccartan2023}, no labeling correction
  is needed in the weights.} Neither the sampling algorithm nor the
target distribution introduced below depends on a particular labeling of
districts. As an illustration, Figure~\ref{fig-iowa-plan} displays the
2020 enacted Congressional plan for the U.S. state of Iowa, comprising
four single-member districts with counties serving as geographic
units.\footnote{Throughout the paper, we use Iowa as a running example
  as its size and adjacency make visualization interpretable.}

We now extend the preceding notation to the multi-member district
setting. We define a \emph{region} as a tuple \((G_k, s_k)\) consisting
of a connected subgraph \(G_k\) of \(G\), induced by a vertex set
\(V_k \subset V\), together with an associated size \(s_k\), a positive
integer representing the number of representatives elected from that
region. For notational simplicity, we will often refer to a region
simply by \(G_k\), suppressing the explicit notation for its size
\(s_k\). Any region with \(s_k\) greater than the number of
representatives a district is allowed to elect is referred to as a
\emph{multidistrict}. Figures \ref{fig-iowa-partial1} and
\ref{fig-iowa-partial2} provide illustrative examples of regions and
multidistricts.

Finally, we define a \emph{redistricting scheme}, which specifies the
fundamental parameters governing the redistricting simulation, as a
tuple \((G, D, S, [d^-, d^+])\), where \(G\) denotes the state map,
\(S\) is the total number of seats to be allocated, and \(D\) is the
total number of districts, with each district's size constrained to lie
within the interval \([d^-, d^+]\). A \emph{redistricting plan} is then
given by \(\xi = \{(G_k, s_k)\}_{k=1}^D\), a partition of \(G\) into
\(D\) districts whose associated sizes sum to the total number of seats,
i.e., \(S = \sum_{k=1}^D s_k\). The single-member district setting is
recovered as the special case where \(D = S\) and \(d^- = d^+ = 1\).

\subsection{Target Distribution}\label{sec-target}

Our goal is to sample redistricting plans from a flexible class of
target distributions defined under a redistricting scheme. In most
applications, we seek plans in which district populations are comparable
across districts. We formalize this requirement through the notion of a
\emph{balanced plan}. Let \([P^-, P^+]\) denote \emph{per-seat}
population bounds satisfying \(P^- \leq \mathrm{pop}(V)/S \leq P^+\). A
plan \(\xi\) is said to be balanced if, for every district, the
population per assigned seat lies within these bounds, i.e., \[
\mathrm{pop}(G_k) \in [s_k \cdot P^-,\; s_k \cdot P^+], \quad \text{for all } k = 1, \ldots, D.
\]

We wish to sample from a class of target distributions over balanced
plans of the form \begin{equation}\phantomsection\label{eq-target}{
    \pi(\xi) \propto \exp{-J(\xi)} \prod_{k=1}^D \tau(G_k)^\rho,
}\end{equation} where \(J(\xi)\) is a scoring function described below
that encodes constraints on the plans and \(\rho \in [0, \infty)\) is a
compactness parameter, with larger values favoring more compact
districts. The function \(\tau(G_k)\) is defined as the number of
spanning trees that can be drawn on the subgraph \(G_k\). Since more
compact districts have more internal edges and thus more spanning trees,
\(\tau(G_k)\) serves as a measure of the compactness of district
\(G_k\).\footnote{The idea is more ``blobby'' districts will have lots
  of internal edges leading to a higher spanning tree count, in contrast
  thin ``spindly'' districts where few trees can be drawn. The most
  extreme case would be a district consisting of a single line where
  only one spanning tree can be drawn. The total number of spanning
  trees which can be drawn on a plan \(\xi\) is also sometimes written
  as \(\tau(\xi)^\rho\) in the literature where
  \(\tau(\xi)^\rho = \prod_{k=1}^D \tau(G_k)^\rho\), i.e., the product
  of the number of spanning trees that can be drawn on each district
  \citep[see][]{mccartan2023}.} It is closely related to the
\emph{edge-cut compactness} measure, which computes the proportion of
edges that must be removed from the original graph to obtain a given
plan \citep{dube2016, deford2019}, \[
  \mathrm{rem}(\xi) \ \coloneqq \ 1 - \frac{\sum_{k=1}^D |E(G_k)|}{|E(G)|}.
\]

Empirically, edge-cut compactness is highly correlated with
\(\log \tau(G) - \log \tau(\xi)\)
\citep[see][]{mccartan2023, clelland2021compactness}. In principle, one
can therefore adjust \(\rho\) to control compactness. In practice,
however, values of \(\rho\) far from 1 can substantially reduce sampling
efficiency.\footnote{Characterizing precisely how far away from 1
  \(\rho\) must be for efficient sampling to become computationally
  infeasible is challenging but as a general rule of thumb performance
  will degrade heavily for \(\rho \not\in [.7, 1.3]\). See Appendix
  Section~\ref{sec-rho-tests} for more discussion on the issue.}
Moreover, although \(\rho=1\) is a computationally convenient default,
it has no normative significance.\footnote{When \(\rho \neq 1\), for
  both SMC and MCMC steps, it is necessary to compute the number of
  spanning trees, which can be drawn on the involved region subgraphs.
  This involves calculating the determinant of any minor of the graph
  Laplacian, thereby drastically slowing down computation.} We therefore
recommend that researchers evaluate the compactness of the resulting
simulated plans using alternative measures, such as the Polsby-Popper
score, when choosing a value of \(\rho\). If desired, these alternative
measures can also be directly incorporated into the \(J\) function.

The scoring function \(J(\cdot) \in (-\infty, \infty]\) encodes
additional preferences or constraints on the types of plans considered,
where \(J(\xi) = \infty\) assigns zero probability to plan \(\xi\). In
practice, it is convenient to specify \(J\) so that it decomposes into a
sum of district-level terms and global plan-level terms. This
formulation accommodates both soft and hard constraints at the district
and plan levels. In particular, \(J\) should always be understood as
implicitly encoding population and contiguity constraints, i.e., \[
    \exp{-J(\xi)} = \exp{-J_\text{other}(\xi)} \cdot \prod_{k=1}^D {\bm 1}\{G_k \text{ is connected}\} {\bm 1}\{\mathrm{pop}(G_k)\in [s_k \cdot P^-, s_k \cdot P^+]\},
\] where \(J_\text{other}(\xi)\) encodes other desired constraints. See
\citet{mccartan2023} for practical and theoretical justifications of
this target distribution.

We extend this target distribution to \emph{partial plans} \(\xi_r\),
which are partially complete plans with \(r < D\) regions. Some regions
in a partial plan can be districts, but they need not be. For each
partial plan with \(r\) regions, we can define an associated target
distribution \(\pi_r\) in an analogous manner,
\begin{equation}\phantomsection\label{eq-target-r}{
    \pi_r(\xi_r) \propto \exp{-J(\xi_r)} \prod_{k=1}^r \tau(G_{k})^\rho,
}\end{equation} where the same contiguity and population constraints are
part of the \(J(\xi_r)\) term. As shown below, the proposed gSMC
algorithm samples partial plans from \(\pi_r\) at each step of the
algorithm.\footnote{A preference for certain types of multidistricts can
  be encoded into the intermediate target distributions via the \(J\)
  function.}

\section{The Proposed Algorithm}\label{sec-GraphAlgSection}

In this section, we introduce our generalized Sequential Monte Carlo
(gSMC) algorithm, which extends and improves upon the original SMC
algorithm of \citet{mccartan2023}. At a high level, gSMC generates
districting plans by initializing a collection of blank maps and then
sequentially partitioning regions, including those created in earlier
splitting steps. At each splitting step \(r\), the current collection of
plans is reweighted and resampled to target the intermediate
distribution \(\pi_r\). This ``split-and-resample'' procedure is
repeated for \(D - 1\) rounds, progressively increasing the number of
regions, until a final collection of plans is obtained.

The precise mechanics of how plans are split and which plans are
selected for splitting depend on the choice of sampling space, the
splitting schedule, and the corresponding importance weights. Although
these choices do not change the final target distribution, they can
substantially affect runtime and performance. We begin by presenting an
overview of the gSMC algorithm in Section~\ref{sec-SMCS}, followed by a
more detailed description of the splitting procedure and optimal weights
derivation in Sections \ref{sec-GraphSplitting} and
\ref{sec-graph-optimal-weights}, respectively. Finally,
Section~\ref{sec-diagnostics} discusses diagnostics and practical
considerations for implementing the gSMC algorithm.

\subsection{The Overview of the gSMC Algorithm}\label{sec-SMCS}

The gSMC algorithm belongs to the broader class of Monte Carlo methods
known as Sequential Monte Carlo (SMC) samplers \citep{DelMoralSMCS}.
These methods are designed to generate samples from a sequence of target
distributions (\(\pi_1, \dots, \pi_{D-1}, \pi_D = \pi\) in our case)
when direct sampling from the final distribution is infeasible. In the
literature, these individual samples generated at each step are referred
to as \emph{particles}. SMC samplers operate by constructing samples
sequentially, using particles (partial plans in our case) obtained at
earlier stages to guide and inform sampling at subsequent stages.

Algorithm \ref{alg-gsmcs} formally defines the gSMC algorithm, which is
parameterized by a choice of redistricting scheme and target
distribution \(\pi\), together with user-specified \emph{forward} and
\emph{backward} Markov kernels, \(M_r(\xi_r \mid \xi_{r-1})\) and
\(L_{r-1}(\xi_{r-1} \mid \xi_r)\), respectively. We let
\(\gamma_r(\cdot)\) and \(Z_r\) denote the unnormalized density and
normalizing constants, respectively, such that
\(\pi_r(\xi_r)=\gamma_r(\xi_r)/Z_r\) holds. The forward kernels encode
the ``splitting procedure,'' which specifies how a plan \(\xi_{r-1}\)
with \(r-1\) regions generates a plan \(\xi_r\) with \(r\) regions by
partitioning one of its regions. While one must be able to sample from
each forward kernel \(M_r\), the backward kernels are only used for the
computation of weights.

To generate a sample of \(N\) plans, the gSMC algorithm initializes with
a collection of blank (i.e., single-region) plans
\(\{\xi_1^{(i)}\}_{i=1}^N\) and uniform weights
\(\{W_1^{(i)}\}_{i=1}^N\), where \(W_1^{(i)} = 1/N\) for all \(i\). A
split-and-resample procedure is then applied for \(D - 1\) stages. At
stage \(r\), to construct the next collection of partial plans
\(\{\xi_r^{(i)}\}_{i=1}^N\), we first select an ancestor plan
\(\xi_{r-1}^{(j)}\) at random with probability \(W_{r-1}^{(j)}\) (When
\(r = 2\), all ancestor plans are identical blank maps). Conditional on
the selected ancestor, we propose a new partial plan by sampling
\(\xi_r^\ast \sim M_r(\cdot \mid \xi_{r-1}^{(j)})\). If the proposed
plan \(\xi_r^\ast\) has zero target density, e.g., because it violates
population constraints, it is rejected, and a new ancestor plan is drawn
according to the weights, after which the proposal step is repeated.

Once a proposed partial plan satisfies the constraint
\(\pi_r(\xi_r^\ast) \neq 0\), we accept it by setting
\(\xi_r^{(i)} = \xi_r^\ast\) and compute the associated incremental
weight \(w_r^{(i)}(\xi_{r-1}^{(j)}, \xi_r^{(i)})\) according to the
general formula in Equation \ref{eq-weights}. This procedure is repeated
for each particle \(i\), until \(N\) new partial plans have been
generated. The weights are then normalized so that
\(W_r^{(i)} \propto w_r^{(i)}\), and the algorithm proceeds to the next
stage \(r+1\). Figure~\ref{fig-iowa-smc-example} provides an
illustration of this process for Iowa with \(N=5\) samples.

\begin{algorithm}[t]
\begin{algorithmic}[1]
\REQUIRE{target distributions $\set{\pi_r}_{r=1}^D$, forward kernels $\{M_r\}^D_{r=2}$, backward kernels $\{L_r\}^{D-1}_{r=1}$, redistricting scheme $(G, D, S, d^-, d^+)$ }
\STATE{Set initial weights $w^{(i)}_1 = \gamma_1(\xi_1)$ for $i = 1, \dots, N$}
\FOR{$r=1$ to $D-1$}
\STATE{Set $T_{r+1} = 0$}
\FOR{$i=1$ to $N$}
\STATE{set $p=0$}
\WHILE{$p = 0$}
\STATE{sample parent index $i^\prime \in \set{1,\dots,N}$ with $\Pr(i^\prime = j) \propto w_r^{(j)}$ for each $j=1,\ldots,N$}
\STATE{sample $\xi_{r+1}^\ast$ from the forward kernel $M_{r+1}(\cdot \mid \xi^{(i^\prime)}_{r})$}
\STATE{set $p=\pi_{r+1}(\xi^\ast_{r+1})$}
\STATE{Set $T_{r+1} = T_{r+1} + 1$}
\ENDWHILE
\STATE{set $\xi^{(i)}_{r+1}=\xi^\ast_{r+1}$}
\STATE{compute the incremental weight as
\begin{equation}\label{eq-weights}
w_{r+1}^{(i)}(\xi^{(i')}_{r}, \xi_{r+1}^{(i)})
= \frac{\gamma_{r+1}(\xi^{(i)}_{r+1}) L_{r}( \xi^{(i^\prime)}_{r} \mid \xi_{r+1}^{(i)}) }{ \gamma_{r}(\xi^{(i')}_{r}) M_{r+1}( \xi_{r+1}^{(i)}\mid \xi^{(i^\prime)}_{r})}
\end{equation}
}
\ENDFOR
\ENDFOR
\end{algorithmic}
\caption{The gSMC algorithm}
\label{alg-gsmcs}
\end{algorithm}

This sequential construction induces a joint distribution on
\((\xi_1, \dots, \xi_D)\) under which each \(\xi_r\) has a marginal
distribution \(\pi_r\). After stage \(r\), the algorithm yields a
collection of weighted samples \(\{(\xi_r^{(i)}, W_r^{(i)})\}_{i=1}^N\),
whose (random) weighted empirical measure
\(\hat{\pi}^N_r(\cdot) = \sum_{i=1}^N W_r^{(i)} \, \delta_{\xi_r^{(i)}}(\cdot)\)
converges to the target distribution \(\pi_r\) almost surely in the
total variation distance metric as \(N \to \infty\), as shown in
Theorem~\ref{thm-smc-convergence} below.

In particular, Algorithm \ref{alg-gsmcs} can be viewed as a type of
particle filter algorithm designed to keep the particle system alive in
a manner similar to \citet{legland2005sequential} and \citet{AliveSMC}.
We relax the assumptions to only require that for any
\(\xi_r \in \pi_r\), there exists some \(\xi_{r-1} \in \pi_{r-1}\) where
\(M_r(\xi_r\mid \xi_{r-1}) > 0\) and \(\pi_{r-1}(\xi_{r-1})> 0\). Unless
additional hard constraints imposed through the \(J\) function violate
this, Theorem~\ref{thm-gsmc-convergence} shows that these conditions are
always satisfied when using the optimal weights derived in
Section~\ref{sec-graph-optimal-weights}. The fact that the sampling
space is finite allows us to prove strong convergence results despite
these minimal assumptions. \newpage

\begin{theorem}[Strong Convergence of the gSMC
Algorithm]\protect\hypertarget{thm-smc-convergence}{}\label{thm-smc-convergence}

For \(r = 1, \dots, D\), let
\(\hat{\pi}^N_r(\cdot)=\sum_{i=1}^N W^{(i)} \delta_{\xi_r^{(i)}}(\cdot)\)
be the weighted particle approximation generated by Algorithm
\ref{alg-gsmcs}. Let \(\TV{\cdot}\) represent the total variation
distance between probability measures. Then, as \(N \to\infty\), the
weighted empirical measure \(\hat{\pi}^N_r(\cdot)\) converges almost
surely in total variation distance, i.e., \[
\TV{\hat{\pi}^N_r - \pi_r} \toAS 0.
\] Moreover, for all functions \(h\) on unlabeled plans, as
\(N \to\infty\), we have, \[
    \E[\hat{\pi}^N_r]{h(\xi)} \toAS \E[\pi_r]{h(\xi)},
\] and \[
    \sqrt{N} (\E[\hat{\pi}^N_r]{h(\xi)} - \E[\pi_r]{h(\xi)})
    \toD \mathcal{N}(0, V_\text{SMC}(h)),
\] for some asymptotic variance \(V_\text{SMC}(h)\).

\end{theorem}

The proof is in Appendix
Section~\ref{sec-gsmc-specific-convergence-results}. Under these
assumptions, the algorithm may yield a set of partial plans that are
impossible to split. However, we prove that for a sufficiently large
\(N\) the probability of such event tends to \(0\). In practice, for
maps with approximately \(30\) districts or less, we have not
experienced any issues even for relatively small sample sizes (e.g.,
\(N=1\,000\)). For larger maps, the addition of even a few MCMC steps
appears to eliminate this problem completely. See Appendix
Section~\ref{sec-addl-empirical} for more discussion.

Another benefit of the SMC sampler in general is that the unnormalized
weights and the number of tries \(T_r\) can be used to construct a
biased but strongly consistent estimator of the normalizing constants
\(Z_r\). Corollary~\ref{cor-gsmc-norm-estimate} provides information
about how to construct these estimators of \(\set{Z_r}\) while
Remark~\ref{rem-unbiased-gsmc-norm-estimator} proposes how Algorithm
\ref{alg-gsmcs} can be slightly modified to obtain an unbiased
estimator. These estimators can be useful when comparing different
target distributions or combining samples across different target
distributions.

\subsection{The Splitting Procedure}\label{sec-GraphSplitting}

We now describe the splitting procedure at a high level, with details
provided in Appendix Section~\ref{sec-graph-space-splitting}. To split a
partial plan \(\xi_{r-1}\), we first randomly select a multidistrict
\(H_\ell\) of size \(s_\ell\), according to a pre-specified probability
distribution \(\varphi(\cdot \mid \xi_{r-1})\) defined over all
multidistricts in \(\xi_{r-1}\). We then sample a spanning tree
\(T^\ast\) uniformly at random on the selected subgraph \(H_\ell\) using
Wilson's algorithm \citep{wilson1996}. We consider all possible splits
formed by removing an edge \(e^\ast\) from the sampled spanning tree
\(T^\ast\) and assigning two new sizes \(s_k^\ast\) and \(s_{k'}^\ast\)
(where \(s_\ell = s_k^\ast + s_{k'}^\ast\)) to the split subgraphs, and
randomly choose one of the splits to make. This split creates two
distinct regions, \(G_k^\ast\) and \(G_{k'}^\ast\), with associated
sizes \(s_k^\ast\) and \(s_{k'}^\ast\), respectively. By construction,
these satisfy \(H_\ell = G_k^\ast \cup G_{k'}^\ast\) with
\(G_k^\ast \cap G_{k'}^\ast=\emptyset\).

The splitting algorithm has three inputs. First, we define a probability
function \(\varphi\) that selects a particular multidistrict \(H_\ell\)
from the partial plan \(\xi_{r-1}\). Second, we define a \emph{splitting
schedule}, which specifies the allowable sizes of the two regions
produced by splitting a multidistrict of size \(s\) at stage \(r\). For
example, when sampling single-member district plans and splitting off
one district at a time, the splitting schedule would begin by splitting
a multidistrict of size \(D\) into sizes \((D-1, 1)\) and then proceed
with splitting the resulting multidistrict into sizes \((D-2, 1)\) at
the second stage. Third, we define a \emph{splitting parameter}
\(\mathcal{K}_r\), which determines the random selection of splits.
Specifically, for any given spanning tree, we sort all possible splits
by population deviation and then choose one of the top \(\mathcal{K}_r\)
most balanced splits uniformly at random.

With these inputs, the splitting procedure yields a candidate partial
plan \(\xi_r^\ast\) that contains one additional region relative to
\(\xi_{r-1}\). If the candidate plan satisfies all hard constraints, we
accept it and set \(\xi_r = \xi_r^\ast\), with the two new regions
denoted by \((G_k, s_k)\) and \((G_{k'}, s_{k'})\). Otherwise, we repeat
the splitting procedure until a valid partial plan is obtained. This
iterative splitting is visible in Figure~\ref{fig-iowa-smc-example},
culminating in maps with four single-member districts.

When \(\xi_r\) and \(\xi_{r-1}\) are balanced and \(\mathcal{K}_r\) is
chosen properly, we show in Appendix
Section~\ref{sec-graph-space-splitting} that the forward kernel has the
following splitting probability,
\begin{equation}\phantomsection\label{eq-graph-forward-kernel-eq}{
M_r(\xi_r \mid \xi_{r-1}) =  \varphi(H_{\ell}\mid \xi_{r-1}) \frac{\tau(G_{k})\tau(G_{k'}) }{\mathcal{K}_r \cdot \tau(H_{\ell})} \abs{\mathcal{C}(G_{k}, G_{k'})}.
}\end{equation}

\subsection{Optimal Weights}\label{sec-graph-optimal-weights}

Given our sequence of target distributions \(\{\pi_r\}_{r=1}^D\) and
forward kernels \(\{M_r\}_{r=2}^D\), we are free to choose any backward
kernels \(\{L_r\}_{r=1}^{D-1}\) provided that they are valid Markov
kernels, i.e., each \(L_r\) is a valid probability distribution over
plans with \(r-1\) regions. It is known that the optimal backward
kernel, which yields the minimum-variance weights, is given by the
following general formula \citep{dai2022}, \[
L^\text{opt}_{r-1}(\xi_{r-1}\mid \xi_r) = \frac{\pi_{r-1}(\xi_{r-1})M_r( \xi_r \mid \xi_{r-1})}{f_r(\xi_r)},
\] where \(f_r\) denotes the marginal proposal density obtained by
integrating out the previous partial plan \(\xi_{r-1}\) according to its
target distribution,
\begin{equation}\phantomsection\label{eq-general-marginal-proposal-integral}{
f_r(\xi_r) = \int \pi_{r-1}(\xi_{r-1}) M_r(\xi_r \mid \xi_{r-1}) \; d\xi_{r-1}.
}\end{equation} This backward kernel leads to the following formula for
the optimal weights, which only depend on the current plan \(\xi_r\), \[
w_r(\xi_{r-1}, \xi_r) = \frac{1}{Z_{r-1}} \frac{\gamma_r(\xi_r)}{ f_r(\xi_r)}.
\]

For many problems, the marginal proposal density in
Equation~\ref{eq-general-marginal-proposal-integral} does not have a
closed form and cannot be efficiently computed. Fortunately, for the
gSMC algorithm, the marginal proposal density is readily available due
to the fact that the term \(\pi_{r-1}(\xi_{r-1})\) is non-zero only if
the partial plan \(\xi_{r-1}\) is balanced. In addition, given that
\(\xi_{r-1}\) is balanced, \(M_r(\xi_r \mid \xi_{r-1})\) can only be
non-zero if it is a plan with \(r-1\) regions that can be split into
\(\xi_r\). This means that \(\xi_r\) and \(\xi_{r-1}\) share all but two
regions in common, and thus \(\xi_{r-1}\) must be a plan formed by
merging two adjacent regions in \(\xi_r\) while leaving all the other
regions unchanged.

It is straightforward to show that the set of all such possible
\(\xi_{r-1}\) consists of plans made by taking a pair of adjacent
regions in \(\xi_{r}\) and merging them together. Therefore,
Equation~\ref{eq-general-marginal-proposal-integral} is equal to a sum
over all pairs of adjacent regions in \(\xi_r\) where we take
\(\xi_{r-1}\) to be the plan made by merging the adjacent region pair.

\begin{figure}[t]
\centering
\begin{tikzpicture}[every node/.style={inner sep=0}]

\node (a1) at (0,0)
  {\includegraphics[width=0.17\textwidth]{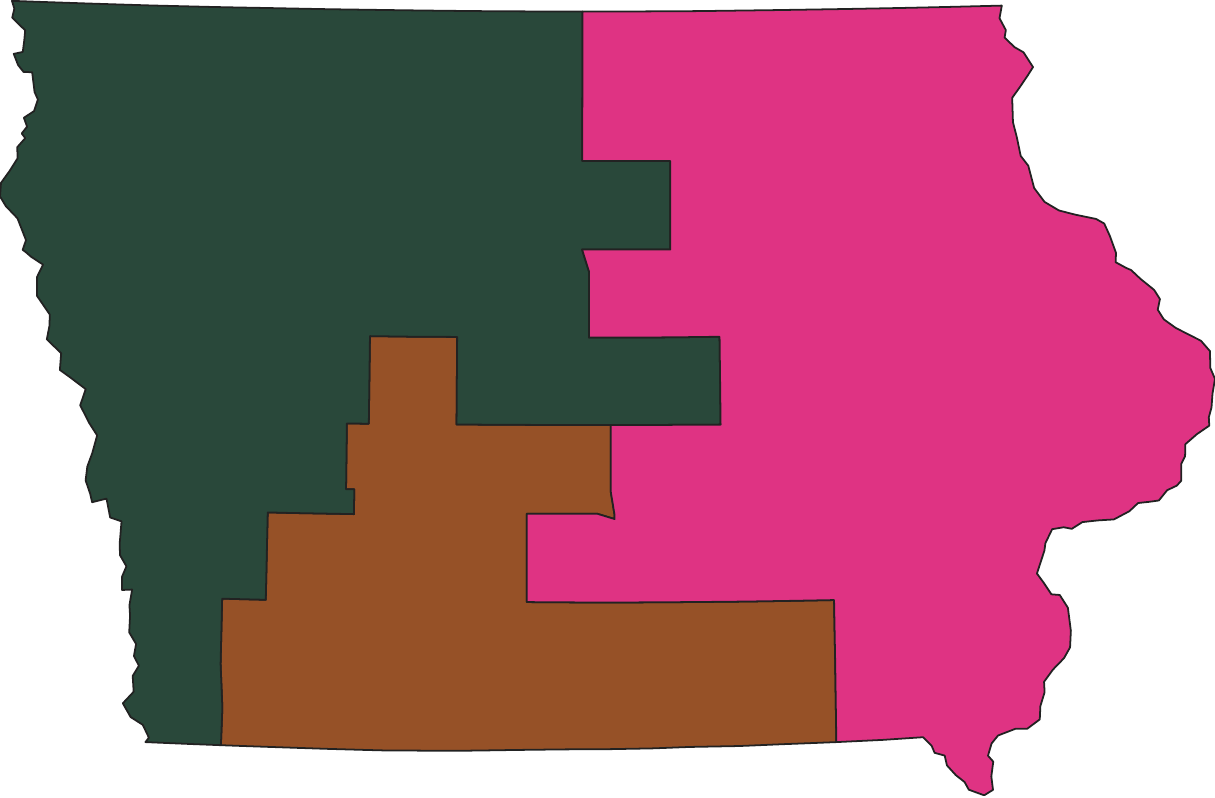}};
\node (a2) at (3.6,0)
  {\includegraphics[width=0.17\textwidth]{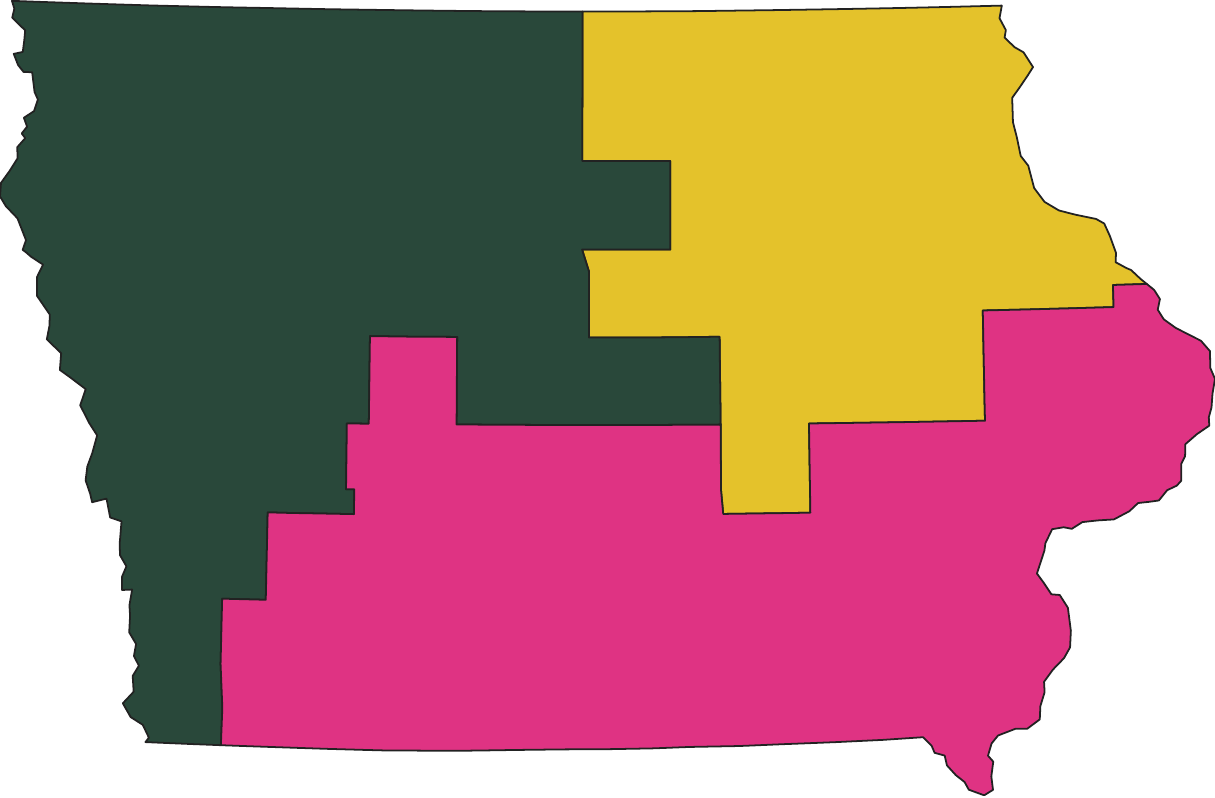}};
\node (a3) at (7.2,0)
  {\includegraphics[width=0.17\textwidth]{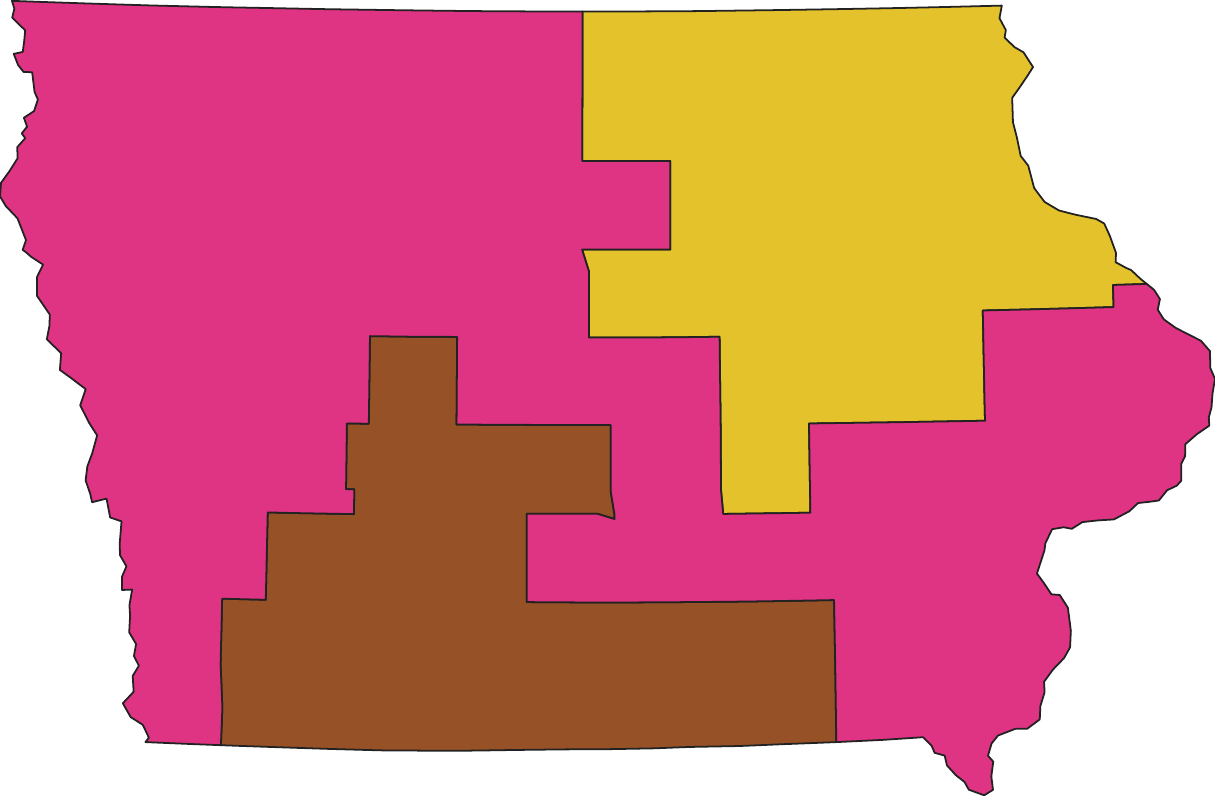}};
\node (a4) at (10.8,0)
  {\includegraphics[width=0.17\textwidth]{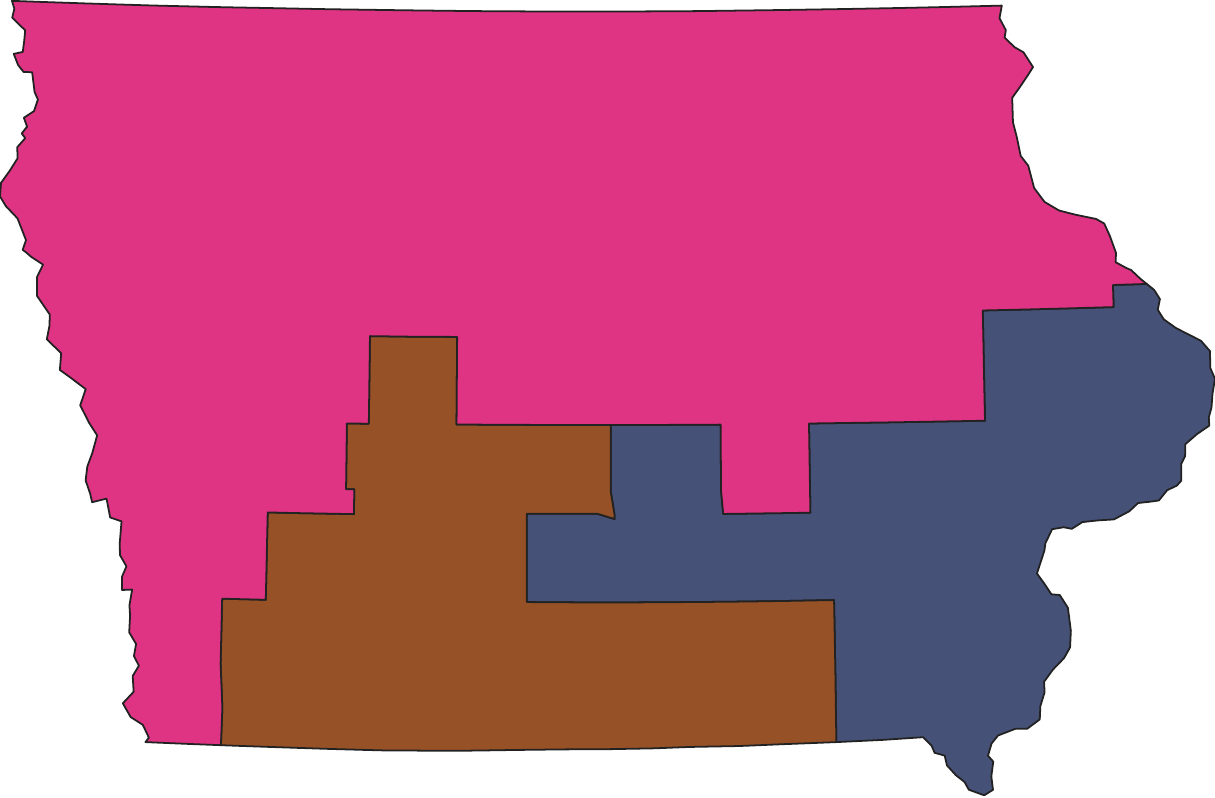}};
\node (a5) at (14.4,0)
  {\includegraphics[width=0.17\textwidth]{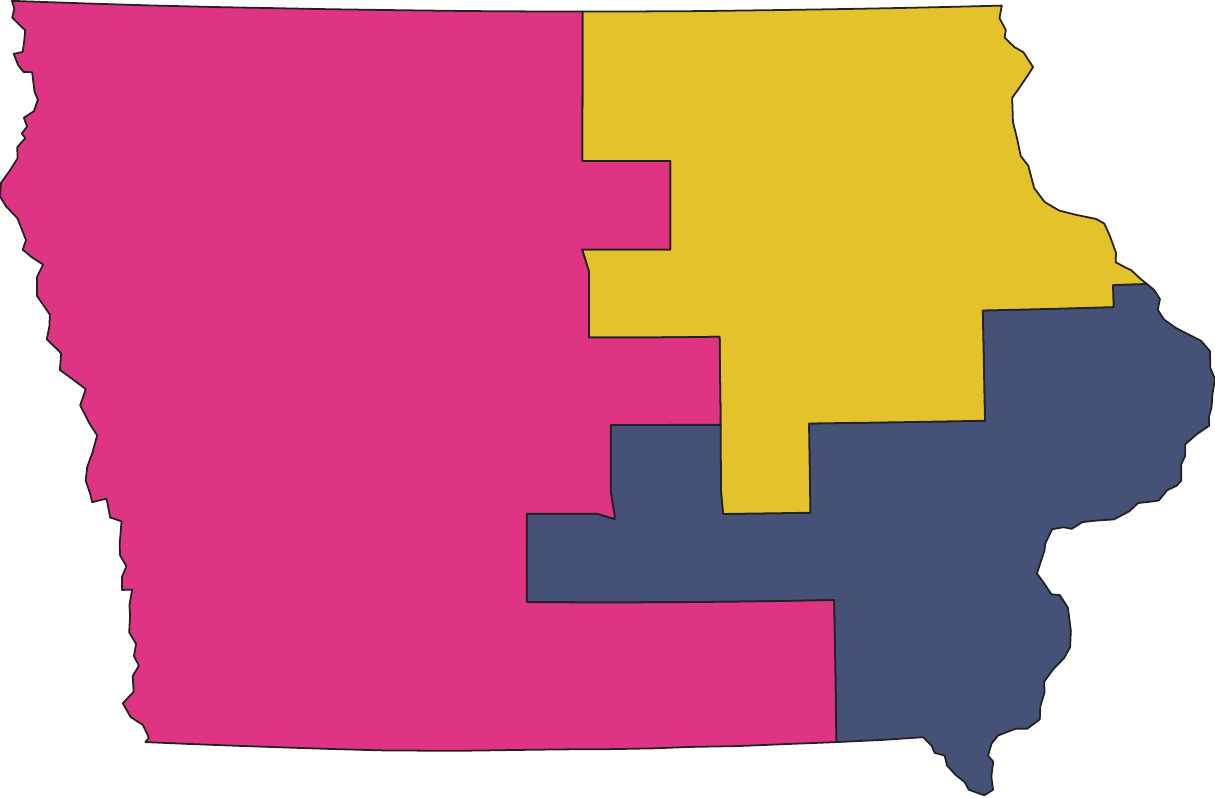}};

\node (main) at (7.2,-3.6)
  {\includegraphics[width=0.22\textwidth]{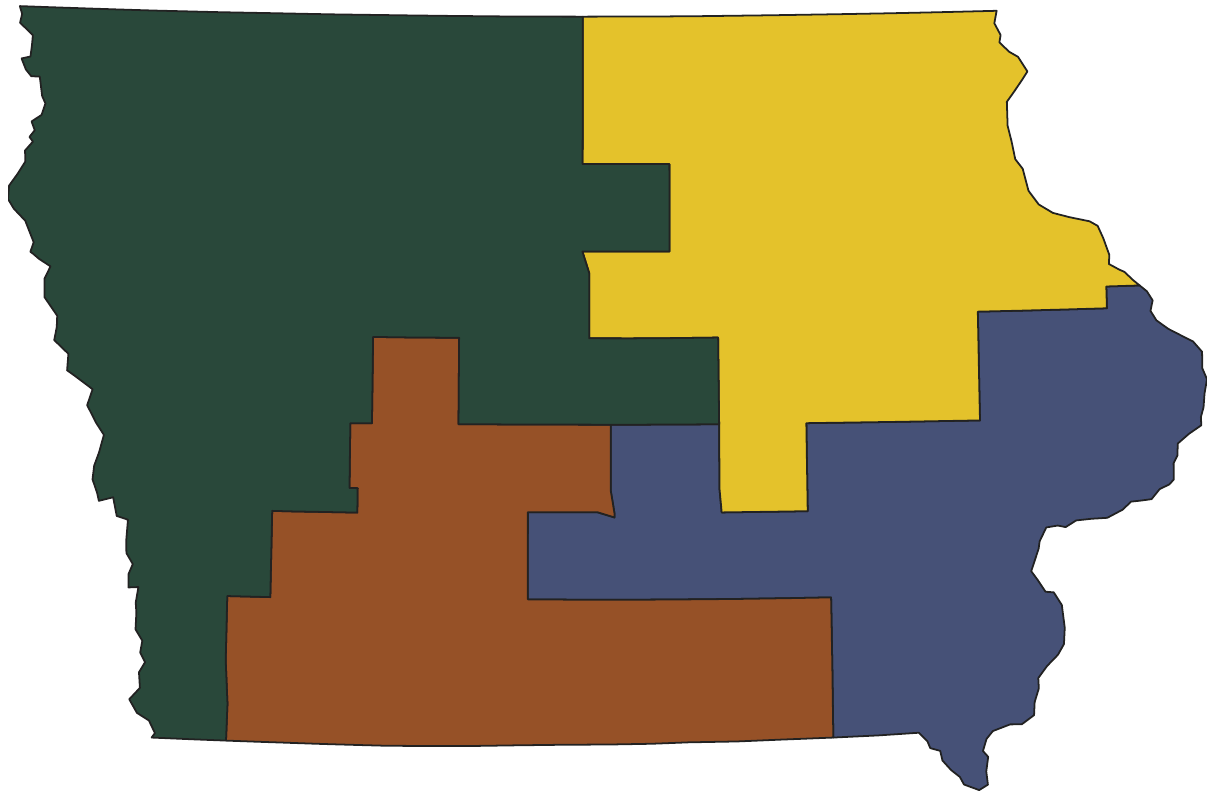}};

\coordinate (src) at ($(main.north)+(0,0.05)$);

\draw[-{Latex[length=3mm,width=2.0mm]}, line width=1.0pt]
  (src) -- ($(a1.south)+(0.15,-0.1)$);
\draw[-{Latex[length=3mm,width=2.0mm]}, line width=1.0pt]
  (src) -- ($(a2.south)+(0.05,-0.1)$);
\draw[-{Latex[length=3mm,width=2.0mm]}, line width=1.0pt]
  (src) -- ($(a3.south)+(0,-0.1)$);
\draw[-{Latex[length=3mm,width=2.0mm]}, line width=1.0pt]
  (src) -- ($(a4.south)+(-0.05,-0.1)$);
\draw[-{Latex[length=3mm,width=2.0mm]}, line width=1.0pt]
  (src) -- ($(a5.south)+(-0.15,-0.1)$);

\end{tikzpicture}

\caption{Iowa's 2020 congressional plan and all previous plans with 3 regions (top) that could have been split to create the 4-region plan (bottom)}

\label{fig-6SampleIowaN4SMCPaths}
\end{figure}

Figure \ref{fig-6SampleIowaN4SMCPaths} illustrates the idea, where there
are five pairs of adjacent districts in the plan \(\xi_4\) (bottom). For
each pair of adjacent districts, there is an associated plan \(\xi_3\)
(top) formed by replacing the two adjacent districts with the region
created by merging them. For an arbitrary plan \(\xi_r\), this yields
the following closed form expression of the optimal weights.

\begin{proposition}[Optimal
weights]\protect\hypertarget{prp-opt-wt-main}{}\label{prp-opt-wt-main}

For \(r=2,\dots,D\) the optimal minimal variance incremental weights are
\begin{equation}\phantomsection\label{eq-graph-optimal-weights-main}{
w_r(\xi_{r-1}, \xi_r) = \mathcal{K}_r \cdot \left( \sum_{\substack{ G_{k} \sim G_{k'} \in \xi_r}} \varphi(G_{k} \cup G_{k'} \mid \tilde{\xi}_{r-1})  \frac{\exp{-J(\tilde{\xi}_{r-1})}}{\exp{-J(\xi_{r})}} \left(\frac{\tau(G_{k} \cup G_{k'})}{\tau(G_k) \tau(G_{k'})}\right)^{\rho-1} \abs{\mathcal{C}(G_{k}, G_{k'})}  \right)^{-1},
}\end{equation} where \(G_{k} \sim G_{k'}\) denotes adjacent regions in
\(\xi_r\), \(\tilde{\xi}_{r-1}\) is the plan formed by merging \(G_{k}\)
and \(G_{k'}\), and \(\mathcal{C}(G_{k}, G_{k'})\) is the set of edges
in \(G\) with one vertex in \(G_k\) and one in \(G_{k'}\).

\end{proposition}

The number of adjacent regions in a plan is on the order of \(O(r)\),
and the pairs of adjacent regions and boundary lengths can all be
efficiently computed via a single pass through the graph.

\subsection{Diagnostics and Practical Implementation
Details}\label{sec-diagnostics}

While Theorem~\ref{thm-gsmc-convergence} guarantees asymptotic
convergence of the algorithm's output, in practice, the quality of
samples depends on the choice of a splitting schedule, the multidistrict
selection probability \(\varphi(\cdot \mid \cdot)\), and any other
relevant splitting parameters. Appendix Section~\ref{sec-appendixC}
provides detailed discussion of this point. Diagnostics also play an
essential role in implementing the gSMC algorithm. While diagnostics can
never prove the convergence of the algorithm in any particular
application, they help researchers detect convergence failure and other
potential issues. We follow the recommendation of \citet{mccartan2023},
in particular computing the Gelman-Rubin \(\hat R\) statistic for
various summary statistics based on multiple independent runs of the
gSMC algorithm \citep{gelman1992inference, vehtari2019rank}.

\section{Extensions}\label{extensions}

The gSMC algorithm presented in Section~\ref{sec-GraphAlgSection} can be
extended in several ways: (1) we can accommodate different sampling
spaces and splitting procedures; (2) we can limit administrative
boundary splits; and (3) MCMC steps can be added to boost performance.
We briefly discuss these extensions while leaving the details to the
appendix.

\subsection{Different Sampling Spaces}\label{sec-OtherSpaces}

Drawing on previous work \citep{McmcLinkingEdge, McmcForests}, we can
modify the procedure in Sections \ref{sec-GraphSplitting} and
\ref{sec-graph-optimal-weights}, which operates on the space of graph
partitions, by using two different sampling spaces: spanning forest and
linking edge spaces. While these sampling spaces ultimately lead to the
same target distribution given in Equation~\ref{eq-target}, they can
flexibly address the computational trade-offs between splitting plans
and computing weights. These two alternative sampling spaces can be
viewed as intermediaries in the sampling process for graph partitions.
Spanning forest space represents ``stopping'' the splitting process in
Algorithm \ref{alg-naive-k-split} after we remove the edge but before we
remove the trees. Linking edge space represents an even earlier step
where we select a tree cut and then save both the trees and edges. For
both of these spaces, there is a many-to-one function of the graph
partition they induce.

\begin{figure}[t]

\begin{minipage}{0.33\linewidth}

\centering{

\pandocbounded{\includegraphics[keepaspectratio]{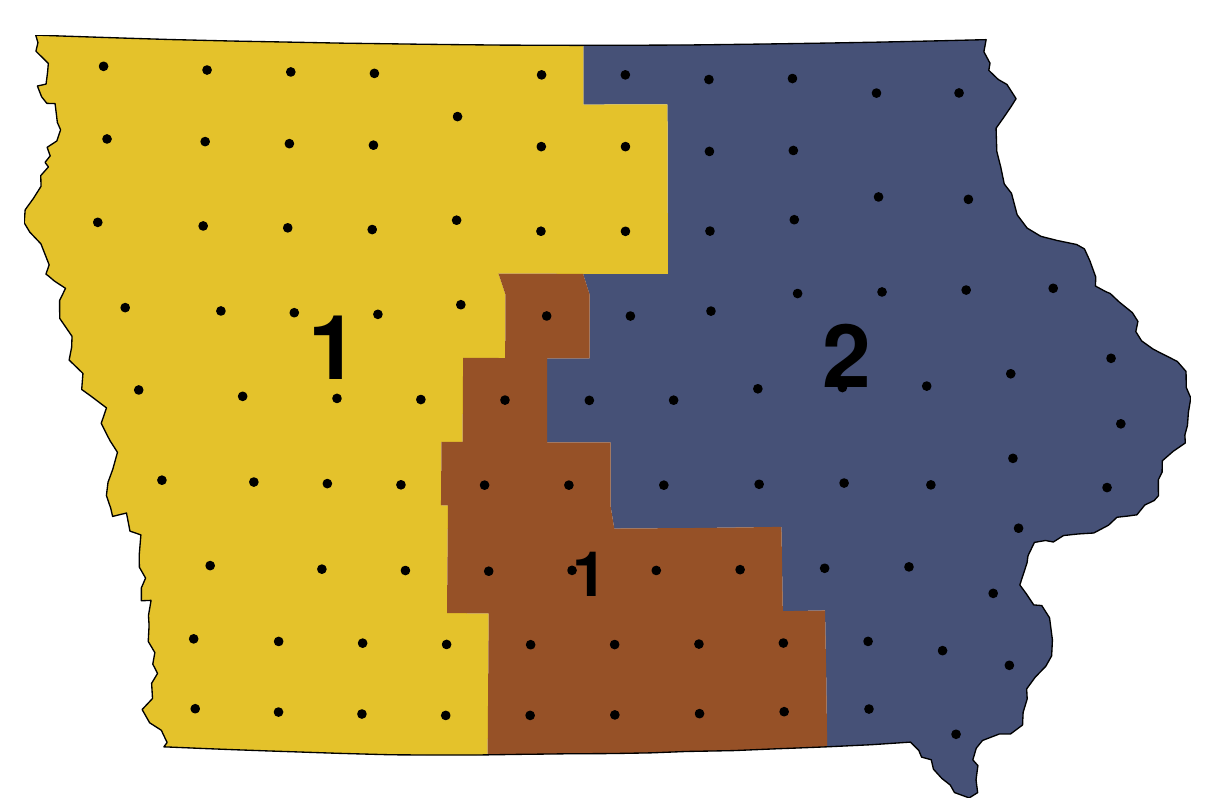}}

}

\subcaption{\label{fig-graph-example}Graph partition partial plan}

\end{minipage}%
\begin{minipage}{0.33\linewidth}

\centering{

\pandocbounded{\includegraphics[keepaspectratio]{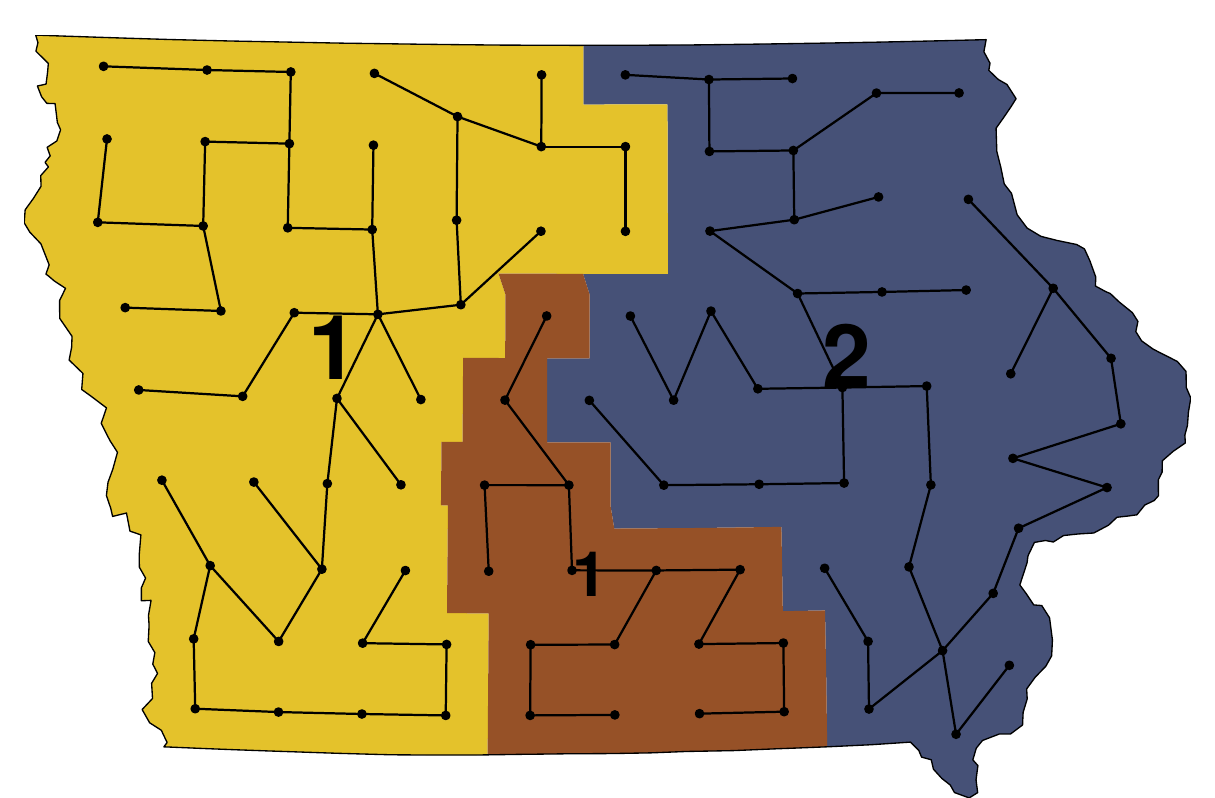}}

}

\subcaption{\label{fig-forest-example}Spanning forest partial plan}

\end{minipage}%
\begin{minipage}{0.33\linewidth}

\centering{

\pandocbounded{\includegraphics[keepaspectratio]{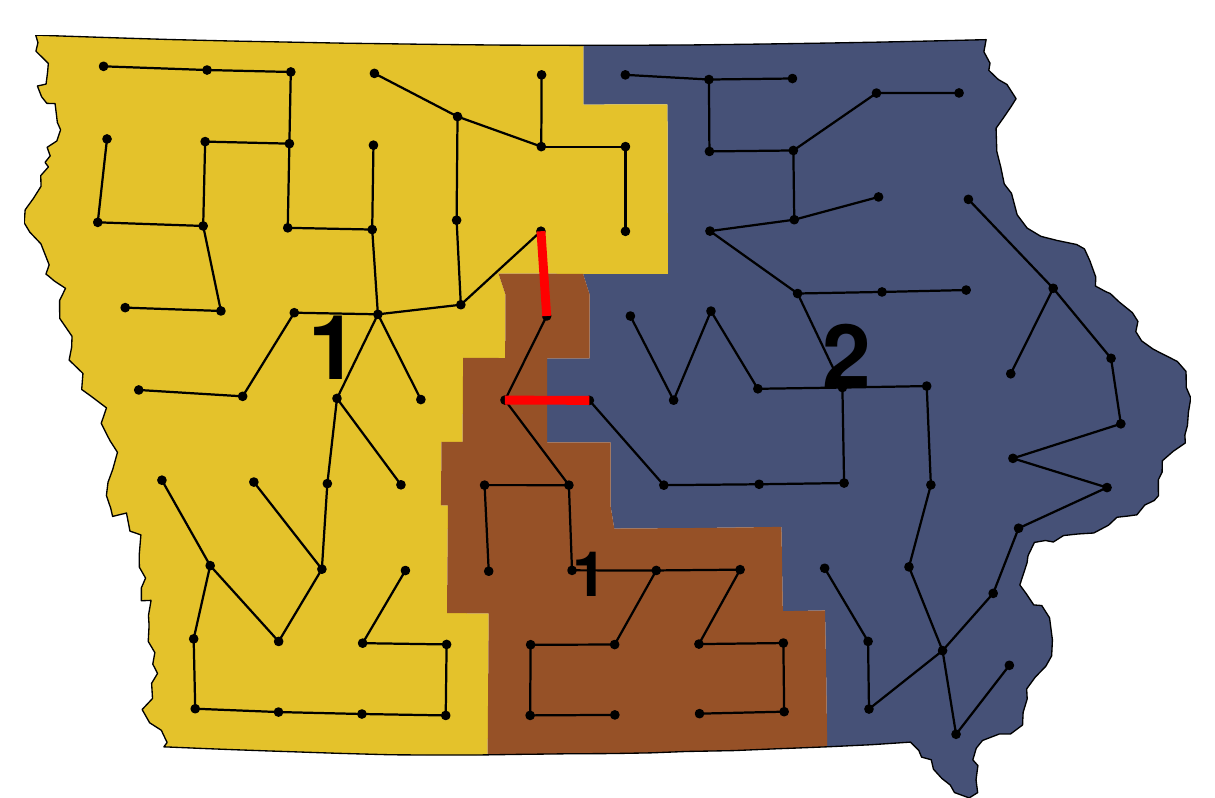}}

}

\subcaption{\label{fig-linking-edge-example}Linking edge partial plan}

\end{minipage}%

\caption{\label{fig-new-sample-spaces-iowa}Example of a partial plan on
each of the three sampling spaces. The graph partition plan in
Figure~\ref{fig-graph-example} stores only the graph partition of \(G\)
into three regions. The spanning forest partial plan in
Figure~\ref{fig-forest-example} stores the partition pieces from
Figure~\ref{fig-graph-example} and augments the space with spanning tree
on each region, forming a spanning forest on the entire graph \(G\). The
linking edge partial plan in Figure~\ref{fig-linking-edge-example}
retains the trees on each region and also stores two additional
``linking edges'', marked in red, which, when combined with the spanning
forest, form a spanning tree on the entire graph.}

\end{figure}

Figure~\ref{fig-new-sample-spaces-iowa} displays an example of a forest
and linking edge space plan associated with an underlying graph space
plan. In contrast with Figure~\ref{fig-graph-example}, where we only
store the graph partitions, as Figure~\ref{fig-forest-example}
demonstrates, for spanning forest space we save the trees drawn in the
splitting process to create each region.
Figure~\ref{fig-linking-edge-example} shows how for linking edge space
we save not just the trees drawn to create regions but also the removed
edges themselves (highlighted in red).

The use of alternative sampling spaces allows one to select tree cuts
according to an arbitrary distribution over splits rather than selecting
one of the top \(\mathcal{K}\) splits uniformly at random. This makes
the splitting stage much less computationally costly because we can
ensure whenever there is at least one balanced split it will be chosen.
This contrasts with the original splitting procedure where there is a
greater chance of choosing a non-balanced split even when balanced
splits are present, requiring the entire procedure to be repeated. The
trade off is that the computational complexity of the optimal weights is
now greater than \(O(V)\). Empirically, across a variety of maps and
schemes, we find that these two new sampling spaces outperform graph
space sampling in terms of pure wall time (see Appendix
Section~\ref{sec-addl-empirical}). In particular, since linking edge
space performs the best, we recommend using it in practice. Appendix
Section~\ref{sec-new-sampling-space-proofs} provides a detailed
discussion on these new sampling spaces and the proofs of their optimal
weights.

\subsection{Limiting the Splits of Administrative
Boundaries}\label{sec-boundary}

In many applications, analysts may also wish to discourage districts
from crossing county, municipal, or other administrative boundaries
without just using those entities as the geographic units of the map.
Rather than encoding this preference solely through a constraint in
\(J\), we can efficiently incorporate it directly into the sampling
mechanism by drawing plans hierarchically with respect to the
administrative boundaries.

We do this in the same manner as in \citet{mccartan2023}: spanning trees
are drawn within units first, and then linked across units by drawing
another spanning tree on the unit-level multigraph. These hierarchically
sampled plans are guaranteed to have no more than \(D-1\) administrative
unit splits (and a preference for even fewer splits can be incorporated
through the \(J\) function). However, the inclusion of hierarchical
sampling requires modifications to the backwards kernel, weights, and
MCMC moves (discussed next). See Appendix
Section~\ref{sec-hierarchical-sampling-proofs} for a formal
characterization of hierarchical plans and details of how the algorithm
changes.

\subsection{Incorporating MCMC Moves for Improving
Convergence}\label{sec-addl-mcmc}

Lastly, we improve the convergence of gSMC by incorporating Markov chain
Monte Carlo (MCMC) moves after each step of the algorithm. In the
original gSMC algorithm, the forward Markov kernel \(M_r\) always splits
off a new region. Here, we allow for some Markov kernel
\(\widetilde{M}_r(\cdot\mid\cdot)\) operating on the space of partial
plans \(\set{\xi_r}\) to conditionally sample a new partial plan while
keeping the number of regions the same. So long as this Markov kernel
does not alter the target distribution \(\pi_r\), i.e., it is
\(\pi_r\)-invariant, the convergence properties of the algorithm are not
affected. In fact, we can carry forward the SMC weights from the
previous iteration without modification. Unlike Markov chains using the
same kernel, we do not need any assumptions about irreducibility or
aperiodicity of the state space for convergence results to hold. See
Appendix Section~\ref{sec-gsmc-specific-convergence-results} for
details.

We leverage this ability to add MCMC moves by incorporating a
Metropolis-Hastings merge-split kernel inspired by the existing work
\citep{deford2019, carter2019, cannon2025, McmcForests, McmcLinkingEdge}.
The merge-split kernels, which can operate on all three of the different
sampling spaces, work by probabilistically selecting two adjacent
regions in a given partial plan \(\xi_r\), merging and then splitting
them as if they were a multidistrict (i.e., drawing a tree using
Wilson's algorithm and then selecting a tree cut according to the rules
of the sampling space), and lastly implementing the Metropolis-Hastings
rejection step. The formulas for transition probabilities are presented
in Appendix Section~\ref{sec-mcmc-appendix}. Critically, the merge-split
kernel is \(\pi_r\)-invariant for all \(r\).

Empirically, we find that the merge-split MCMC steps substantially
improves convergence. With typical sample sizes (\(N<20\,000\)), the SMC
algorithm often struggles to handle more than about 25 districts at a
time. In contrast, using the gSMC-MCMC algorithm, we have been able to
effectively analyze much more difficult problems such as a large-scale
map with more than 200 districts (Section~\ref{sec-pa-house-main}), and
districts with extremely tight population bounds up to a 2-3 person
difference (Appendix Section~\ref{sec-extreme-balance-sims}).\footnote{While
  MCMC steps can somewhat improve performance, convergence performance
  still degrades for target distributions when \(\rho\) is far away from
  one and when extreme constraints are imposed through \(J\). See
  Appendix Section~\ref{sec-rho-tests} for more discussion.} Appendix
Section~\ref{sec-choose-mcmc} discusses how to choose the number of MCMC
steps in the gSMC+MCMC algorithm.

\section{Validation}\label{sec-paper-validation}

In Appendix Section~\ref{sec-validation}, we assess the empirical
performance of the gSMC algorithm by comparing the sampled plans with
the ground truth in small-scale problems where it is possible to
enumerate all plans \citep{fifield2020enum}. We examine both a
single-member and multi-member map example. These two examples
demonstrate our open-source implementation \texttt{redist} correctly
samples from the given target distributions and the diagnostics in
Section~\ref{sec-diagnostics} perform well in practice.

\section{Empirical Applications}\label{sec-applications}

We apply the proposed gSMC algorithm to two real-world redistricting
problems: drawing multi-member districts for the lower house of the
Irish parliament, i.e., Dáil Éireann, and drawing many single-member
districts for the Pennsylvania House of Representatives. These
applications highlight the advantages of gSMC over the original SMC
algorithm.

\subsection{Multi-member redistricting: Ireland's Dáil
Éireann}\label{multi-member-redistricting-irelands-duxe1il-uxe9ireann}

The Dáil Éireann consists of 174 representatives elected from 43
multi-member constituencies. Each constituency elects either three,
four, or five representatives. One policy question of interest is how
districting plans would change under an alternative redistricting
scheme, in which each constituency instead elects four, five, or six
members \citep{oireachtas_lrs_2023_electoral_amendment_bill_digest}. We
conduct a simulation study to investigate the redistricting plans under
such a scheme.

\begin{figure}[t]

\centering{

\pandocbounded{\includegraphics[keepaspectratio]{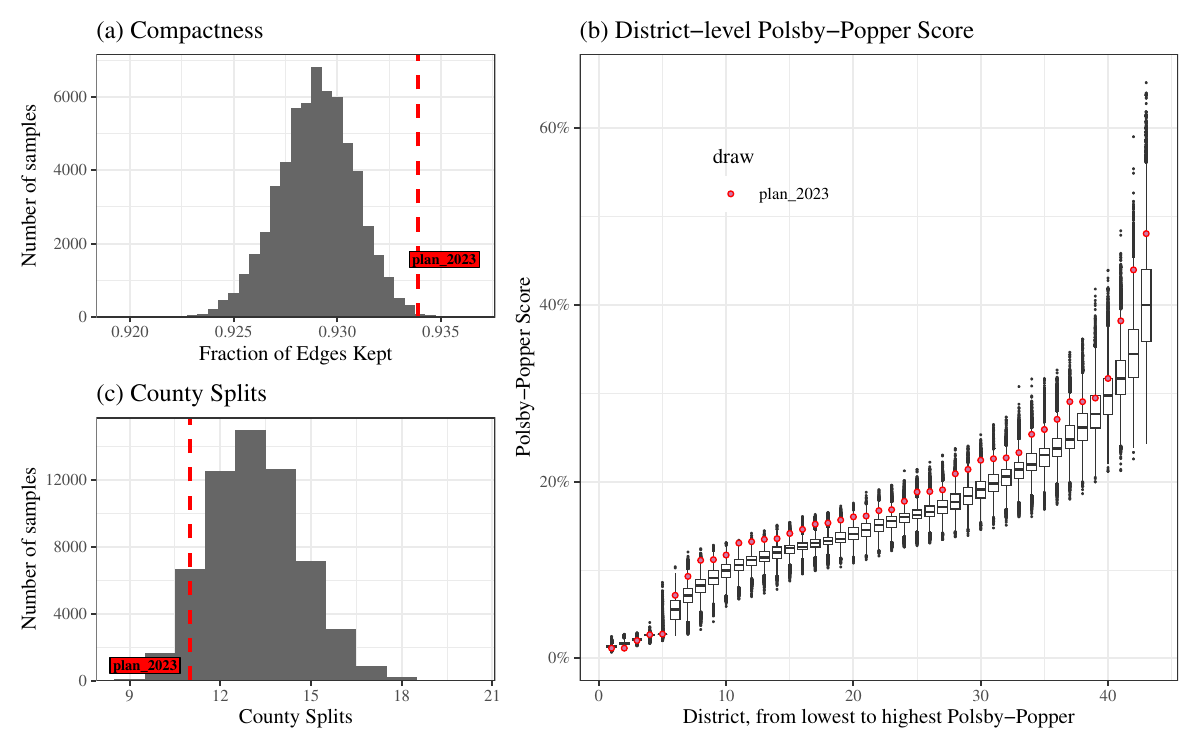}}

}

\caption{\label{fig-ireland-sims}Summary statistics for the sampled and
enacted plans for Irish redistricting: (a) compactness (larger means
more compact) and (c) county splits. Panel (b) shows the order
statistics of the Polsby-Popper score by district (i.e., within each
plan, districts are ordered by Polsby-Popper score).}

\end{figure}%

\subsubsection{Setup}\label{setup-1}

Ireland uses electoral divisions (EDs) as the smallest geographic units
for redistricting. Although 3,441 EDs are legally defined, some have
been amalgamated in publicly released data for privacy protection,
resulting in 3,420 published EDs. We use these published EDs as the
units of analysis. When originally drawn in the nineteenth century, EDs
were designed to have roughly equal populations. Over time, however,
population sizes have diverged substantially. In the 2022 Census, the
smallest ED had a population of 71, whereas the largest had 43,905
residents.

To account for Dublin's distinct administrative and demographic
characteristics, we partition the map into two district
clusters---Dublin and the remainder of Ireland---as was done in the most
recent redistricting in Ireland. We then sample plans separately within
each district cluster and combine them for the final analysis. In both
clusters, we draw plans from the target distribution specified in
Equation~\ref{eq-target} with \(\rho = 1\). We also impose the same
maximum population deviation based on the current plan, i.e.,
\(\mathrm{dev}(\xi) \leq 0.0812\). For the non-Dublin cluster, we
incorporate the administrative boundary constraint described in
Section~\ref{sec-boundary} to limit the number of county splits For
Dublin, we instead apply this constraint at the municipality level and
impose an additional soft constraint, where \(J(\xi)\) is proportional
to the number of splits, to further discourage municipality splits.

Under the alternative seat-size redistricting scheme (4--6
representatives per constituency), we reduce the number of
constituencies from 43 to 35 so that the average constituency size is
approximately five. We generate six independent runs of 10,000 samples
each under both redistricting schemes. Sampling is conducted in the
linking-edge space described in Appendix Section~\ref{sec-linking-edge},
using district-only splits and MCMC updates after each SMC step. Under
the enacted scheme, we used 10 and 100 expected successful MCMC moves
after each SMC step for Dublin and the remainder of Ireland,
respectively. For the alternative scheme, we used 10 and 20 MCMC moves,
respectively. The diagnostics discussed in Section~\ref{sec-diagnostics}
show no issues; all \(\hat R\) values for split and compactness
statistic are well under the threshold of \(1.05\).

\subsubsection{Results}\label{results}

\begin{figure}[t]

\begin{minipage}{0.50\linewidth}

\centering{

\pandocbounded{\includegraphics[keepaspectratio]{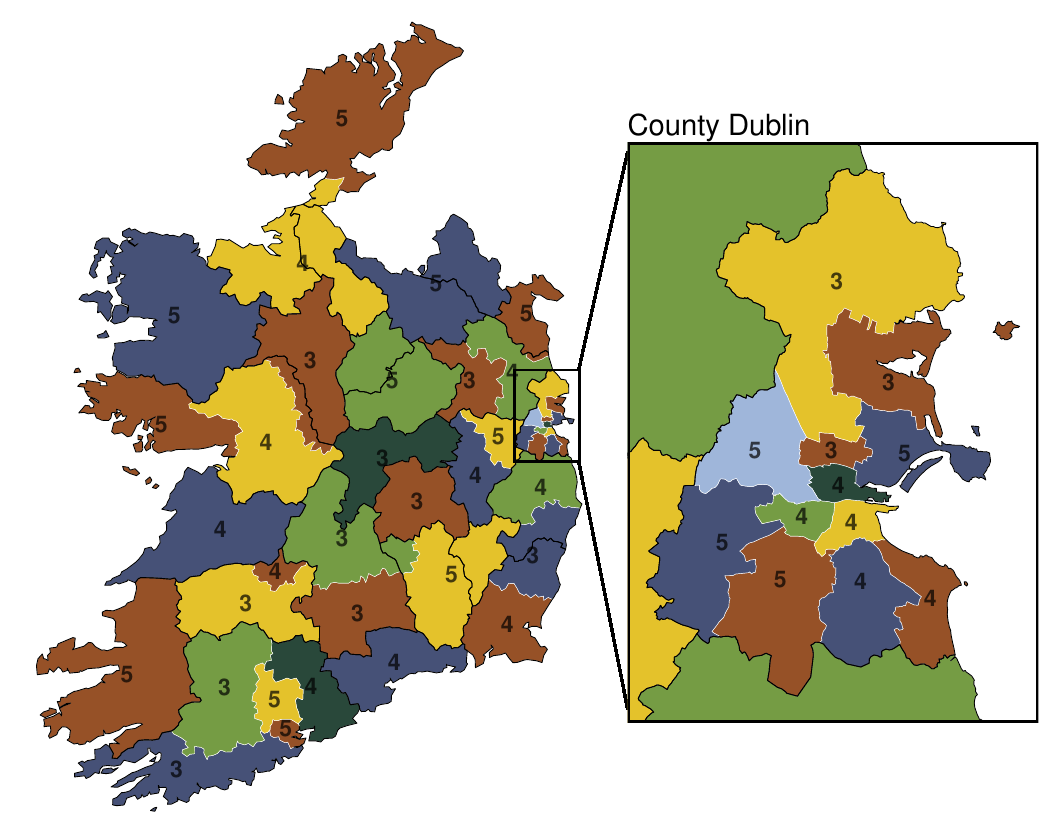}}

}

\subcaption{\label{fig-ireland-enacted}Ireland's 2023 enacted plan}

\end{minipage}%
\begin{minipage}{0.50\linewidth}

\centering{

\pandocbounded{\includegraphics[keepaspectratio]{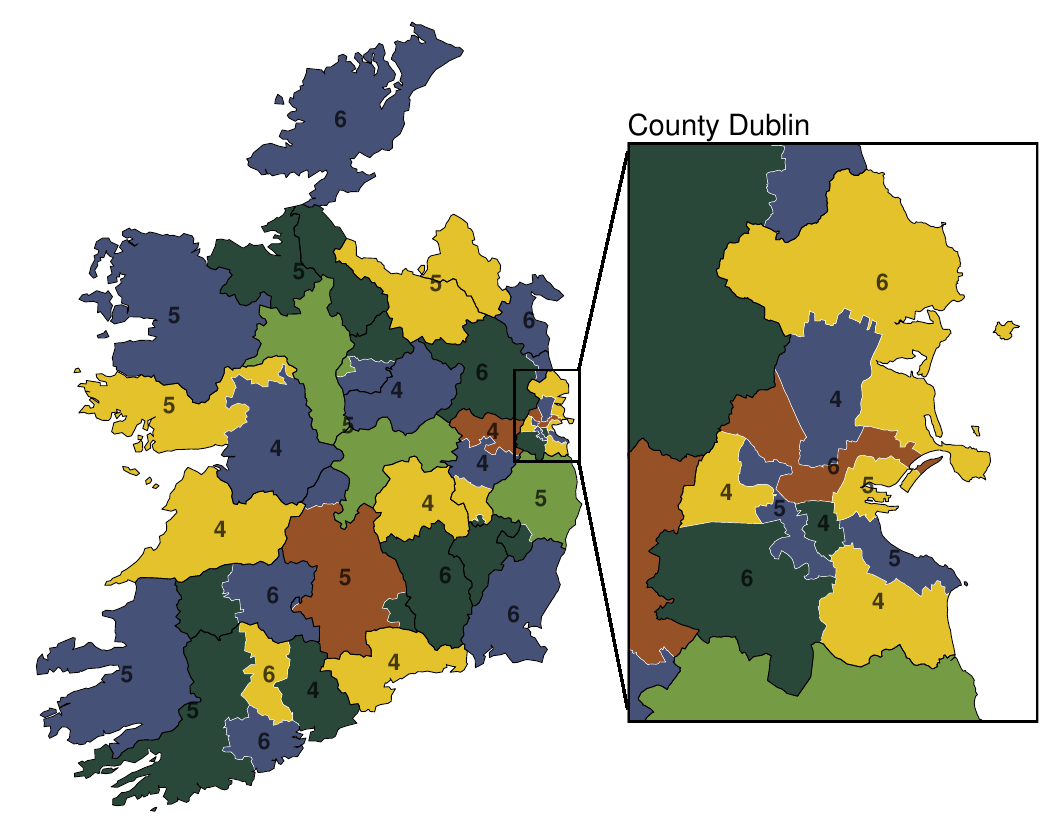}}

}

\subcaption{\label{fig-ireland-alt-example}A simulated plan under the
4-5-6 constituency scheme}

\end{minipage}%

\caption{\label{fig-ireland-examples}Ireland's enacted 2023 plan (panel
(a)) and a simulated plan under the alternative 4-5-6 seat scheme that
splits 9 counties. Constituencies labeled by number of seats.}

\end{figure}%

Figure~\ref{fig-ireland-sims}(a) displays the fraction of edges kept for
the enacted plan (red dashed line) and for the simulated plans
(histogram), where larger values indicate greater compactness. According
to this measure, the enacted plan is more compact than 99.8\% of
simulated plans.\footnote{This extreme difference between the enacted
  and simulated plans is at least in part due to how Ireland has
  historically drawn its maps. In the past two enacted plans, certain
  counties like Dublin, Cork, and Waterford have never had districts
  breaching their boundaries (i.e., being contained partially in
  multiple counties). Restricting which counties are split partially
  explains the difference in compactness between the enacted and
  simulated plans. One could enforce the same county splits in simulated
  plans to try to improve compactness though we do not take this
  approach in our analysis presented here.} Panel (b) reports the order
statistics of the district-level Polsby--Popper scores, with the
simulated order statistics shown as box plots and the enacted plan shown
as red solid circles. Thus, the leftmost box plot and circle on the
\(x\)-axis corresponds to the least compact district in each plan.
Consistent with panel (a), the enacted plan appears generally more
compact than the simulated plans at the district level, though the
simulations are as or more compact for the 5 least-compact districts.
Overall, there is substantial variation in compactness across districts.
Lastly, panel (c) shows the number of county splits for the enacted plan
and the simulated plans. The enacted plan has fewer county splits than
86.0\% of the simulated plans.

Finally, Appendix Figure~\ref{fig-ireland-props} presents the
distribution of constituency sizes under both schemes. The red
horizontal line shows the counts of each constituency size under the
enacted plan whereas a boxplot presents the corresponding counts under
the simulated plans. Interestingly, under the current 3-4-5 scheme, the
simulated plans tend to have a greater number of constituencies with
three or five seats as opposed to the enacted plans. Under the
alternative 4-5-6 scheme, constituencies with five seats are the most
common.

These simulations demonstrate that it is possible to draw plans which
achieve similar scores on traditional redistricting metrics such as
county splits and compactness while increasing the average district
magnitude (seats per district) and thus potentially increasing
proportionality. Figure~\ref{fig-ireland-alt-example} demonstrates one
such example under the alternative 4-5-6 scheme. That plan is more
compact and splits two fewer counties than the enacted plan.

\subsection{Large-scale redistricting: Pennsylvania State
House}\label{sec-pa-house-main}

Pennsylvania's state House has 203 single-member districts, making it
the second largest state house in the U.S, and thus an excellent
opportunity to study the empirical performance of the proposed
algorithms. We examine the Pennsylvania state House map enacted in 2022
by the Legislative Reapportionment Commission and how its partisan
balance compares to simulated alternatives. Our simulation analysis is
illustrative, as it does not systematically model partisan swings across
elections nor incorporate other important considerations, such as the
Voting Rights Act.\footnote{The Supreme Court's recent decision in
  \emph{Louisiana v. Callais} has significantly weakened the
  requirements of the Voting Rights Act (VRA)
  \citep[see][]{kenny2026callais}.}

\subsubsection{Setup}\label{setup-2}

We sample maps from the target distribution given in
Equation~\ref{eq-target} with \(\rho=1\), while limiting splits of
Pennsylvania's 1,083 municipalities and all 67 counties using the
hierarchical splitting procedure described in
Section~\ref{sec-boundary}.\footnote{Specifically, we construct
  pseudo-counties by replacing counties whose populations exceed 4.3
  times the target district population (64,052.71 people) with their
  constituent municipalities, when available; otherwise, we retain the
  counties themselves. Creating pseudo-counties in this manner allows
  one to trade off between county and municipality splits. A larger
  cutoff value tends to lead to more municipality and fewer county
  splits. In this case 4.3 was chosen in conjunction with the additional
  \(J\) constraint to balance the two. We leave the development of a
  more efficient multi-level hierarchical sampling strategy for future
  work.} We also impose an additional constraint through the \(J\) term
to favor plans with fewer total county splits. We choose the strength of
this constraint so that a majority of the sampled plans have no more
county splits than the enacted plan. Finally, we impose a population
deviation constraint of \(\mathrm{dev}(\xi) \leq .05\), which is
slightly stricter than the deviation of the enacted plan (5.2\%). This
population constraint translates to a tolerance of approximately
\(\pm 3\,200\) people per district. We use precincts as geographical
units. Pennsylvania has 9,178 precincts with a median population of
1,176.

We generate 16 independent runs of 10,000 samples each. Plans were
sampled using the linking edge sampling space described in Appendix
Section~\ref{sec-linking-edge} and any-valid splits. A number of MCMC
steps were taken after each SMC step, chosen so that the expected number
of successful MCMC steps would be 100. The overall sample size and
number of MCMC steps were chosen to ensure that all the diagnostics
discussed in Section~\ref{sec-diagnostics} indicated convergence. These
include \(\hat R\) values for 9,623 summary statistics of potential
interest, all of which are below our recommended threshold of 1.05;
96.6\% are below the stronger threshold of 1.01.

\subsubsection{Results}\label{results-1}

\begin{figure}

\centering{

\includegraphics[width=0.95\linewidth,height=\textheight,keepaspectratio]{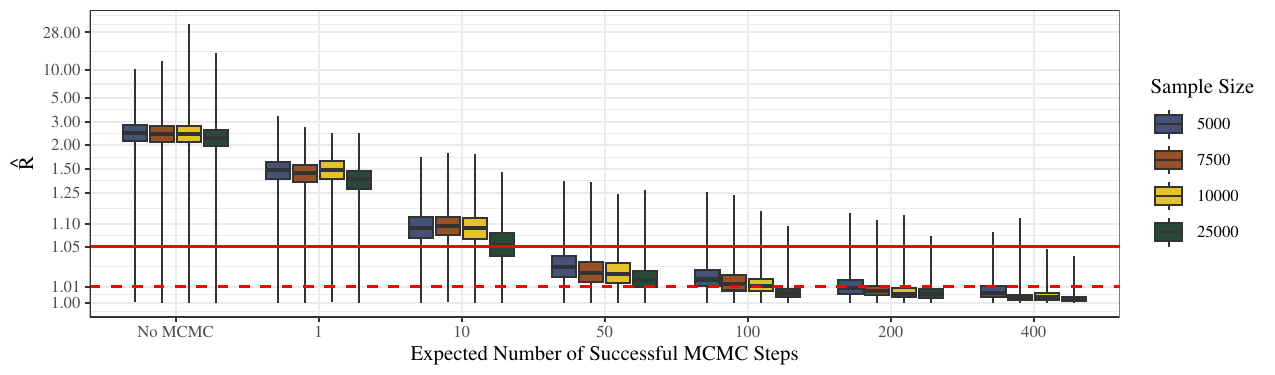}

}

\caption{\label{fig-PA_mcmc_comparison}Boxplots of
pseudo-logarithmically spaced \(\hat R\) values for the 9,623 summary
statistics of interest for a range of expected successful MCMC steps.
Nine \(\hat R\) values less than 1 were truncated to 1. Whiskers are
drawn such that they span the full observed range.}

\end{figure}%

We begin by examining how the performance of the gSMC algorithm improves
by the inclusion of MCMC steps. Figure~\ref{fig-PA_mcmc_comparison}
plots the \(\hat R\) values of the 9,623 statistics of interest against
the number of expected successful MCMC moves for several fixed sample
sizes (\(N=2\,500\), \(5\,000\), and \(10\,000\)) using 16 independent
runs. A greater number of expected successful MCMC steps corresponds to
performing more MCMC moves in between each SMC step. The figure
demonstrates that for a fixed sample size, increasing the number of MCMC
moves greatly boosts convergence as measured by \(\hat{R}\) statistics.
We find that somewhere between 200 and 400 MCMC steps are sufficient for
obtaining satisfactory convergence results for these sample sizes.

Appendix Figure~\ref{fig-PA-House-Sims-Compactness} shows the fraction
of edges kept for each plan. According to this measure, the enacted plan
is more compact than 98.6\% of simulated plans. Unfortunately, in this
case, increasing the value of \(\rho\) is not feasible because it
dramatically reduces sampling efficiency.\footnote{For example,
  increasing the value of \(\rho\) to 1.065 and adjusting the strength
  of the total county splits parameter produced samples where
  approximately 59\% had higher edges kept compactness values. However,
  these runs took over triple the wall time to generate compared to the
  \(\rho=1\) samples and while the \(\hat{R}\) values for fraction of
  edges kept and expected democratic seats were well under the 1.05
  threshold, about 15\% of the \(\hat{R}\)s for the 9,623 statistics
  were between 1.05 and 1.32.} In Appendix Section~\ref{sec-local-sens},
we present a local sensitivity analysis to examine the relationship
between compactness and partisanship. This analysis suggests that the
partisan conclusions are not sensitive to small changes in the value of
\(\rho\).

Appendix Figure~\ref{fig-PA_split_figs} shows the number of splits and
total splits of both counties and municipalities. 80.1\% of the sampled
plans have the same or fewer county splits than the enacted. 66.1\% of
them have the same or fewer total county splits (the total number of
contiguous county-intersect-district pieces) than the enacted. For
municipality splits and total municipality splits, those numbers are
35.7\% and 62.0\% respectively.

\begin{figure}

\begin{minipage}{0.50\linewidth}

\centering{

\pandocbounded{\includegraphics[keepaspectratio]{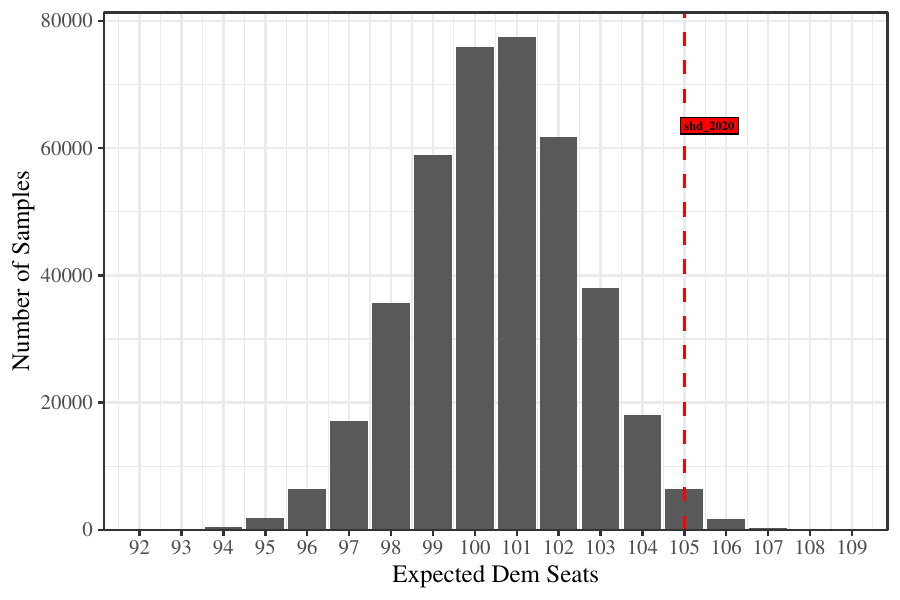}}

}

\subcaption{\label{fig-edem-counts}Expected Democratic Seats}

\end{minipage}%
\begin{minipage}{0.50\linewidth}

\centering{

\pandocbounded{\includegraphics[keepaspectratio]{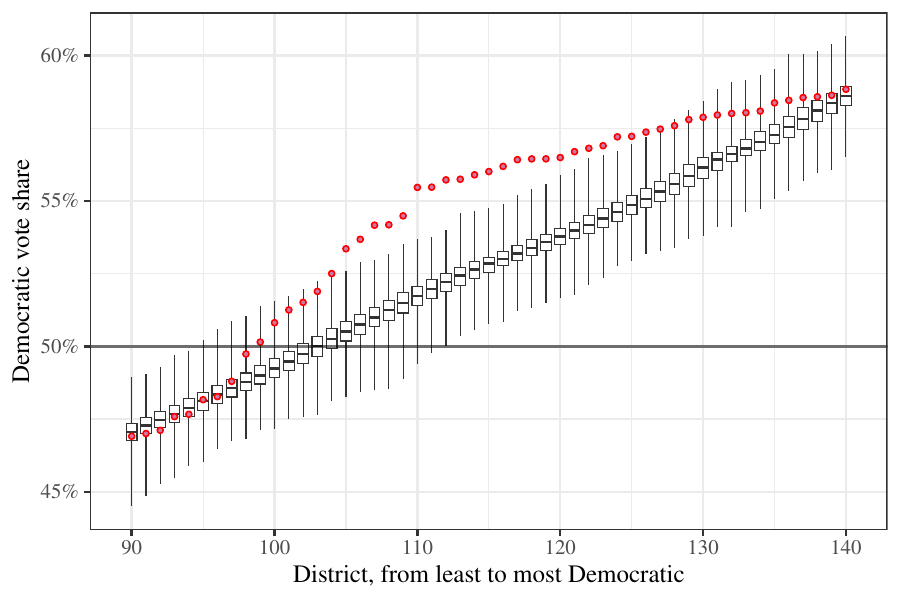}}

}

\subcaption{\label{fig-paper-dem-share}District-level Democratic Vote
Shares}

\end{minipage}%

\caption{\label{fig-PA-partisan-plots}Partisan statistics for the
sampled and enacted plans. Panel (a) shows the expected number of
democratic seats for each plan. Panel (b) shows the order statistics of
Democratic two-party vote share (i.e., within each plan districts are
ordered by Democratic vote share) for ordered districts 90 through 140.
Whiskers are drawn to span the observed range of each statistic.}

\end{figure}%

To examine the partisan bias of the enacted plan, we use an aggregation
of precinct-level voting patterns. The baseline partisan estimates are
calculated by averaging the vote totals for several statewide elections
based on the data from the \texttt{alarmdata} package
\citep{alarmdata}.\footnote{These elections are the 2016 and 2020
  presidential elections, the 2016 and 2018 US Senate elections, the
  2016 and 2020 state attorney general elections, and the 2018
  gubernatorial election.} Panel (a) of
Figure~\ref{fig-PA-partisan-plots} presents the expected number of
Democratic seats, which are calculated by counting the number of
districts in each (enacted or simulated) plan where the estimated
Democratic vote share is greater than 50\%. On average, the simulated
plans yield about 101 Democratic seats compared to an estimated 105
Democratic seats for the enacted plan. The 95\% interval of the expected
Democratic seats ranges from 97 to 104 seats, and the enacted plan falls
outside of that range (\(p=0.0431\)). One may be concerned that the
simulated plans significantly differ from the enacted plan in terms of
compactness. However, our local sensitivity analysis in Appendix
Section~\ref{sec-local-sens} suggests that the partisan conclusions
based on the expected Democratic seats in the simulated plans are
relatively robust to a small change in compactness, and thus the results
likely would not change if more compact plans were simulated.

To investigate partisan bias further, we plot in panel (b) the
district-level order statistics of Democratic vote share for ordered
districts 90 through 140, analogously to the compactness order
statistics in Figure~\ref{fig-ireland-sims}(b) (Appendix
Figure~\ref{fig-full-dem-share} shows the full plot for all districts).
When compared to the simulated maps, the enacted plan delivers more
Democratic vote share in some of the competitive districts. In
particular, the enacted plan has 25 seats with a Democratic vote share
between 45\% and 55\%, compared to an average of approximately 43 in the
simulated plans (\(p<0.0001\)). These results imply that the enacted
plan favors Democratic party when compared to simulated
plans.\footnote{ While we only examine two statistics here, multiple
  testing correction becomes important if many statistics are used for
  hypothesis testing \citep{bhq}.}

\section{Conclusion}\label{conclusion}

We have introduced a new generalized Sequential Monte Carlo (gSMC)
algorithm that substantially widens the applicability of
simulation-based redistricting analysis. By allowing for new splitting
schedules, alternative sampling spaces, and the incorporation of MCMC
moves, gSMC makes it possible to analyze redistricting problems with
multi-member districts with a variable number of representatives and
larger-scale districting maps. We mention some specific possible
applications next.

First, simulation-based analyses of partisan bias have been largely
confined to U.S. congressional redistricting, where the number of
districts is relatively small and all districts are single-member
\citep{chen2013, 2018league, 2019rucho, chen_stephanopoulos_2021_race_blind_future, gerrymandering2022, 2024utah, jasny2025}.
Our algorithms enable researchers to extend these methods to U.S. state
legislatures, which often have many more districts, as well as to
redistricting settings outside the United States, where multi-member
districts are common.

Second, the improved convergence properties of gSMC make the proposed
algorithms suitable for settings with relatively stringent constraints,
under which the original SMC algorithm can struggle. For example, we
have shown that the gSMC and gSMC--MCMC algorithms can accommodate
extremely tight population tolerances, as are often required in
practice. In some of the congressional simulations in Appendix
Section~\ref{sec-addl-empirical} the algorithms were performant for maps
with a maximum deviation of \emph{two or fewer people}. The proposed
algorithms can also better accommodate constraints arising from the
Voting Rights Act \citep{kenny2026callais}.

The proposed algorithms still leave several methodological challenges
unresolved. For example, although spanning-tree-based methods, including
the algorithms proposed here, are widely used because of their
computational convenience
\citep{deford2019, carter2019, fifield2020mcmc, cannon2022spanning, cannon2025},
alternative parameterizations of compactness may be desirable. For
example, \citet{cyclewalk} discusses research on how boundaries between
rural and urban areas can be treated differently from rural--rural or
urban--urban boundaries under spanning-tree-based and
Polsby--Popper-based parameterizations of compactness. Recently proposed
splitting methods offer a promising avenue for accommodating such
alternative notions of compactness \citep{cyclewalk, BUD}. Building on
our approach, future work could incorporate these splitting methods into
new forward kernels for the gSMC algorithm.

\counterwithin*{figure}{section}
\counterwithin*{table}{section}

\renewcommand{\thetable}{\thesection\arabic{table}}

\hypertarget{refs}{}

\begin{CSLReferences}{0}{0}\end{CSLReferences}

\appendix

\renewcommand\thefigure{\thesection\arabic{figure}}

\setcounter{figure}{0}

\section{Proofs}\label{sec-proofs}

\setcounter{algorithm}{0}
\renewcommand{\thealgorithm}{A\arabic{algorithm}}

\subsection{General Redistricting Definitions and
Results}\label{general-redistricting-definitions-and-results}

We begin by formally defining some general terms, special functions, and
results used in the proceeding sections. Below, we provide a list of
important notation with references to their formal definitions when
applicable.

\textbf{Subscript Notation}

\begin{itemize}
\tightlist
\item
  \(D\) denotes the total number of districts.
\item
  \(S\) denotes the total number of seats in the entire state (in a
  single-member districting scheme this is equal to \(D\), but for
  multi-member districting \(S > D\)).
\item
  \(N\) denotes the number of Monte Carlo samples.
\item
  \(r\) indexes the number of regions (which will be equivalent to
  time).
\item
  \(k\) indexes specific regions or districts.
\end{itemize}

\textbf{Redistricting Notation}

\begin{itemize}
\tightlist
\item
  \(G = (V, E)\) is used to the map (Definition~\ref{def-map}).
\item
  \(\mathrm{pop}(V')\) denotes the population of a vertex set
  \(V' \subset V\)
\item
  \((G_k, s_k)\) is used to denote a region
  (Definition~\ref{def-Region}).
\item
  \(\xi_r\) is used to denote a (partial) plan
  (Definition~\ref{def-rRegionPartialPlan}).
\item
  \(\tau(\cdot)\) is used to denote the number of spanning trees that
  can be drawn on a graph or multigraph (Definition~\ref{def-ST-set}).
\item
  \(G_k \sim G_{k'}\) denotes two adjacent regions
  (Definition~\ref{def-AdjacentRegions}).
\item
  \(\mathcal{S}(r, s, \xi_r)\) denotes a splitting schedule
  (Definition~\ref{def-split-schedule}).
\item
  \(F_r\) is used to denote a (partial) forest plan
  (Definition~\ref{def-forest-plan}).
\item
  \(L_r\) is used to denote a (partial) linking edge plan
  (Definition~\ref{def-linking-plan}).
\item
  \(G/\xi_r\) is used to denote the plan quotient multigraph
  (Definition~\ref{def-plan-quotient-graph}).
\end{itemize}

\textbf{SMC Notation}

\begin{itemize}
\tightlist
\item
  \(M_r\) denotes the forward kernel at time \(r\).
\item
  \(L_r\) denotes the backward kernel at time \(r\).
\item
  \(w_r\) denotes the incremental weight function at time \(r\).
\end{itemize}

\begin{definition}[Map]\protect\hypertarget{def-map}{}\label{def-map}

We define the map of a state to be the graph \(G = (V,E)\). This graph
represents the state on some level of geographic precision. The vertex
set \(V = \set{v_1,\dots,v_m}\) is composed of the base geographic units
(precincts, counties, etc) and each \(v_i\) has an associated population
\(\mathrm{pop}(v_i)\). The population of a set of vertices
\(U \subseteq V\) is defined as
\[\mathrm{pop}(U)=\sum_{v \in U} \mathrm{pop}(v).\]

The edge set \(E\) represents units that are considered legally
adjacent.

\end{definition}

\begin{definition}[Region]\protect\hypertarget{def-Region}{}\label{def-Region}

Given a map \(G\), we define a region \((G_{k}, s_{k})\) as the pair
consisting of

\begin{itemize}
\tightlist
\item
  a connected subgraph \(G_k \subset G\), with the property that \(G_k\)
  is the induced (connected) subgraph in \(G\) of its vertex set, i.e.,
  \[\IGraph{\mathrm{V}(G_{k})} = G_{k}\]
\item
  a size \(s_{k} \in \bbN\). When \(G_k\) is a district the size \(s_k\)
  is equal to the number of seats in that district.
\end{itemize}

Note we often refer to the region \((G_{k}, s_{k})\) solely using
\(G_{k}\) but this is just notational shorthand, the size is always
implicitly included.

\end{definition}

\begin{definition}[Plan]\protect\hypertarget{def-rRegionPartialPlan}{}\label{def-rRegionPartialPlan}

Given a districting scheme \((G, D, S, [d^-, d^+])\) and \(r \leq D\),
we define a \(r\)-region (graph) plan \(\xi_r\) as an unordered
collection of \(r\) regions \[\xi_r = \set{(G_k, s_k)}_{k=1}^r\] such
that

\begin{enumerate}
\def\labelenumi{\arabic{enumi}.}
\tightlist
\item
  The region sizes form an integer partition of \(S\),
  i.e.~\(\sum_{k=1}^r s_k = S\).\\
\item
  The regions' vertex sets form a partition of the vertex set of the map
  \(G\), i.e.~\(\cup_{k=1}^r V_k = V\)
\end{enumerate}

It is crucial to emphasize that plans are an \textit{unordered}
collection of regions and number of final districts. The use of \(k\)
for indexing above is due to notational convenience, and any arbitrary
permutation of the labeling of the regions is valid.

\end{definition}

\begin{definition}[Graph Plan
Space]\protect\hypertarget{def-GraphPlanSpace}{}\label{def-GraphPlanSpace}

Given a districting scheme \((G, D, S, [d^-, d^+])\) and \(r \leq D\),
we define \(\mathcal{P}_r(G, D, S, [d^-, d^+])\) to be the set of all
\(r\)-region plans under that districting scheme. For notational
convenience, we will write it simply as \(\mathcal{P}_r\). Note that the
normalizing constants \(Z_r\) are merely the sum, \[
Z_r = \sum_{\xi \in \mathcal{P}_r} \gamma_r(\xi_r).
\]

\end{definition}

\begin{definition}[Seat Population
Bounds]\protect\hypertarget{def-seat-pop-bounds}{}\label{def-seat-pop-bounds}

Given a map \(G = (V,E)\) and a total number of seats \(S\), we define
valid seat population bounds as any closed interval \([P^-, P^+]\) such
that \(\mathrm{pop}(V) /S \in [P^-, P^+]\).

\end{definition}

\begin{definition}[Balanced Regions and
Plans]\protect\hypertarget{def-balance}{}\label{def-balance}

Given population bounds \([P^-,P^+]\), we say that a region
\((G_k, s_k)\) is \textit{balanced} or has a
\textit{balanced population} if its population is within \([P^+,P^-]\)
scaled by \(s_k\), i.e., \[
\mathrm{pop}(G_k) \in [P^- \cdot s_{k}, P^+ \cdot s_{k}]
\] Similarly, we say that a plan \(\xi_r\) is \textit{balanced} or has a
\textit{balanced population} if all of its regions are balanced.

\end{definition}

\begin{definition}[]\protect\hypertarget{def-boundary-set}{}\label{def-boundary-set}

Let \(G = (V,E)\) be a graph and let
\(H_1 = (V_1, E_1), H_2 = (V_2, E_2)\) be subgraphs. We then define
\(\mathcal{C}(H_1,H_2)\) as the set of all edges where one vertex is in
\(H_1\) and the other is in \(H_2\). In other words \begin{align*}
\mathcal{C}(H_1,H_2) = \set{(v_1,v_2) = e \in E \mid v_1 \in H_1, v_2 \in H_2}
\end{align*}

\end{definition}

In essence, \(\mathcal{C}(H_1,H_2)\) counts the number of boundary edges
between two subgraphs in the original graph \(G\). It is important to
note that the boundary edge set only depends on the vertices of the two
subgraphs. That is, if we have subgraphs \(H_1, \tilde{H}_1\) where
\(\mathrm{V}(H_1) = \mathrm{V}(\tilde{H}_1)\) but
\(\tilde{H}_1 \neq H_1\), then
\(\mathcal{C}(H_1, H_2) = \mathcal{C}(\tilde{H}_1, H_2)\) still holds.

\begin{definition}[Districting
Scheme]\protect\hypertarget{def-districting-scheme}{}\label{def-districting-scheme}

We define a districting scheme \((G, D, S, [d^-, d^+])\) as the tuple
consisting of

\begin{itemize}
\tightlist
\item
  a map \(G = (V,E)\)
\item
  \(D\) total number of districts to be drawn on the map
\item
  \(S\) total seats that the map will contain
\item
  \([d^-, d^+]\) is a range of integer values for how many seats a
  district can contain.
\end{itemize}

\end{definition}

\begin{definition}[Plan
Operations]\protect\hypertarget{def-plan-operations}{}\label{def-plan-operations}

Let \(\xi_r\) and \(\xi_m\) be \(r\) and \(m\) region plans respectively
(\(r\) does not have to equal \(m\)). Then, we define the following
operations

\begin{itemize}
    \item Intersection - $\xi_r \cap \xi_m$ \\
We define the intersection of the two partial plans $\xi_r \cap \xi_m$ to be the set of regions in both plans i.e.,
\begin{align*}
    \xi_r \cap \xi_m = \set{G_k \; | \; G_k \in \xi_r \text{ and } G_k \in \xi_m}
\end{align*}
    \item Vertex Set enumeration - $\mathrm{V}(\xi_r)$ \\
We define $\mathrm{V}(\xi_r)$ to be the union of vertex sets of each region in $\xi_r$ i.e.,
\begin{align*}
    \mathrm{V}(\xi_r) = \bigcup_{k=1}^r V_{r,k}
\end{align*}
Note it is necessarily true that $\mathrm{V}(\xi_r) = \mathrm{V}(\xi_m) = \mathrm{V}(G)$
\end{itemize}

\end{definition}

\begin{definition}[Adjacent
Regions]\protect\hypertarget{def-AdjacentRegions}{}\label{def-AdjacentRegions}

Let \(\xi_r\) be an \(r\)-region partial plan. We define
\(\mathrm{AR}(\xi_r)\) to be the set of pairs of regions in \(\xi_r\)
such that they share at least one edge in \(G\). (in other words they
are adjacent). We can write this as \begin{align*}
\mathrm{AR}(\xi_r) = \set{G_{k}, G_{k'} \in \xi_r  \; \big\vert \; 0 < \abs{\mathcal{C}(G_{k}, G_{k'})} , \; G_{k} \neq G_{k'}}
\end{align*} If \(G_{k} \in \xi_r\) is a region then we define
\(\mathrm{AR}(G_{k})\) to be the set of regions in \(\xi_r\) that are
adjacent in \(G\). In other words \begin{align*}
    \mathrm{AR}(G_{k}) = \set{G_{k'} \in \xi_r \; \big\vert \; 0 < \abs{\mathcal{C}(G_{k}, G_{k'})} , \; G_{k} \neq G_{k'}}
\end{align*}

We also denote regions being adjacent as \(G_k \sim G_{k'}\)

\end{definition}

\begin{definition}[Population Tolerance
Function]\protect\hypertarget{def-pop-tol-func}{}\label{def-pop-tol-func}

Given population bounds \([P^+,P^-]\), we define the population
tolerance function \(\Iset{\mathrm{PT}}(\cdot)\) that indicates whether
or not a given region or plan is balanced. For a region
\(G_{k} \subset G\) and associated \(s_{k}\), we define it as
\begin{align*}
    \Iset{\mathrm{PT}}(G_{k}, s_{k}) = \begin{cases}
        1 & \text{if } \mathrm{pop}(G_{k}) \in [P^- \cdot s_{k}, P^+ \cdot s_{k}] \\
        0 & \text{else}
    \end{cases}
\end{align*} If it is clear from context what \(s_{k}\) is, we will
write this as \(\Iset{\mathrm{PT}}(G_{k})\). For a plan \(\xi_r\), we
define it as \begin{align*}
    \Iset{\mathrm{PT}}(\xi_r) = \prod_{k=1}^r \Iset{\mathrm{PT}}(G_{k}, s_{k})
\end{align*}

\end{definition}

\begin{definition}[Splittable
Plan]\protect\hypertarget{def-splittable-plans}{}\label{def-splittable-plans}

Let \(\set{\pi_r}^D_{r=1}\) be a sequence of target distributions and
\(\set{M_r}_{r=2}^D\) be a sequence of forward kernels. We say an
\(r\)-region partial plan \(\xi_r\) is splittable if there exists a
sequence of partial plans \(\xi_1,\dots,\xi_{r-1}\) such that they all
have positive intermediate target distribution probability and positive
transition probability. In other words,
\(\pi_1(\xi_1), \dots, \pi_{r-1}(\xi_{r-1})\) and
\(M_{r}(\xi_r|\xi_{r-1}), \dots, M_2(\xi_2|\xi_1) > 0\).

\end{definition}

\begin{definition}[Splittable Score
Function]\protect\hypertarget{def-splittable-score-func}{}\label{def-splittable-score-func}

A score function \(J\) is splittable if for all \(r\), all partial plans
in the support of \(\pi_r\) are still splittable. In other words, for
all \(r = 1,\dots, D\) \[
\forall \xi_r \in \mathrm{Supp}(\pi_r), \; \; \exists \xi_1,\dots,\xi_{r-1} \text{ s.t. } \pi_1(\xi_1) M_1(\xi_2|\xi_1), \dots, \pi_{r-1}(\xi_{r-1})M_{r}(\xi_r|\xi_{r-1}) > 0
\]

\end{definition}

That is, any hard constraints set by the \(J\) function must make it
possible to split a plan. In what follows, we will assume that \(J\) is
splittable unless stated otherwise. If the only hard constraints encoded
by \(J\) are population balance and connectedness, \(J\) will be
splittable for all Wilson-based forward kernels discussed. However, if
additional hard constraints are introduced, then some care is needed.

\begin{definition}[Region
Deviance]\protect\hypertarget{def-Region-Deviance}{}\label{def-Region-Deviance}

Given a region \((G_k, s_k)\), we define its seat population deviance as
\begin{align*}
    \mathrm{Dev}(G_{k}, s_{k}) = \frac{\mathrm{pop}(G_{k}) - s_{k} \cdot \frac{\mathrm{pop}(V)}{S}}{s_{k} \cdot \frac{\mathrm{pop}(V)}{S}} = \frac{\mathrm{pop}(G_{k})}{s_{k} \cdot \frac{\mathrm{pop}(V)}{S}} - 1.
\end{align*} We define the absolute deviance as \begin{align*}
    \abs{\dev(G_{k}, s_{k})} = \frac{\abs{\mathrm{pop}(G_{k}) - s_{k} \cdot \frac{\mathrm{pop}(V)}{S}}}{s_{k} \cdot \frac{\mathrm{pop}(V)}{S}} = \abs{ \frac{\mathrm{pop}(G_{k})}{s_{k} \cdot \frac{\mathrm{pop}(V)}{S}} - 1}.
\end{align*}

\end{definition}

Notice that there is an equivalence between the population of a region
being within bounds and the deviance being within a certain range,
namely: \begin{align*}
\mathrm{pop}(G_k) \in [s_k \cdot P^-, s_k \cdot P^+] \iff \frac{P^-}{s_k \cdot \frac{\mathrm{pop}(V)}{S}} - 1 \leq  \dev(G_k, s_k) \leq \frac{P^+}{s_k \cdot \frac{\mathrm{pop}(V)}{S}} - 1.
\end{align*} If our population bounds are symmetric, i.e.,
\(P^+ - \frac{\mathrm{pop}(V)}{S} = \frac{\mathrm{pop}(V)}{S}- P^-\)
then it is enough to check the absolute deviance, i.e., \begin{align*}
\mathrm{pop}(G_k) \in [s_k \cdot P^-, s_k \cdot P^+]  \iff \abs{\dev(G_k, s_k)} \leq \frac{P^+}{s_k \cdot \frac{\mathrm{pop}(V)}{S}} - 1.
\end{align*}

\begin{definition}[Region
Tree]\protect\hypertarget{def-RegionTree}{}\label{def-RegionTree}

Given a map \(G\), we define a region tree \((T, s_k)\) as the pair
consisting of

\begin{itemize}
\tightlist
\item
  a spanning tree \(T \subset G\) defined on some subgraph
  \(H \subset G\)
\item
  a size \(s \in \bbN\)
\end{itemize}

\end{definition}

Note we define the notions of being balanced and population deviation
the same for a region tree as a region.

\begin{definition}[Spanning Tree
Set]\protect\hypertarget{def-ST-set}{}\label{def-ST-set}

Let \(H\) be a subgraph of \(G\). We define \(\mathcal{T}(H)\) as the
set of all spanning trees that can be drawn on \(H\) so \begin{align*}
    \mathcal{T}(H) = \set{T \mid \text{$T \subset G$ is spanning tree and } \mathrm{V}(T) = \mathrm{V}(H)},
\end{align*} We denote the size of this set as \(\tau(H)\) where
\[\abs{\mathcal{T}(H)} = \tau(H).\] For a plan \(\xi_r\) we define
\(\tau(\xi_r)\) to be the product of the number of spanning trees that
can be drawn on each of its regions, i.e., \[
\tau(\xi_r) = \prod_{k=1}^r \tau(G_k)
\]

\end{definition}

\begin{definition}[Induced Region
Function]\protect\hypertarget{def-RegionFunction}{}\label{def-RegionFunction}

We define the induced region function \(\IRegion{\cdot}\) as the
function which maps a region tree \((T_k, s_k)\) to a region
\((G_k, s_k)\) by setting \(G_k\) equal to the subgraph induced by the
vertices of \(T_k\). More formally, we define
\(\IRegion{T_k, s_k}  = (G_k, s_k)\) where
\(\IGraph{\IVertex{T_k}}  = G_k\).

\end{definition}

The region function maps region trees to a region via the subgraph
induced by the vertex set of the trees. Therefore, for a region
\(G_{k}\), the only trees that could produce it are trees drawn on the
vertex set of \(G_{k}\). We formally state this and provide its proof.

\begin{lemma}[Pre-Images of Induced Region
Function]\protect\hypertarget{lem-RegionFuncPreImg}{}\label{lem-RegionFuncPreImg}

Let \((G_{k}, s_{k})\) be a region on the map \(G\). The pre-image of
\((G_{k}, s_{k})\) under the induced region function \(\IRegion{\cdot}\)
from Definition~\ref{def-RegionFunction} is equal to \begin{align*}
    \IRegionInv{G_{k}, s_{k}} = \set{(T, s_{k}) \; \middle\vert \; T \in \mathcal{T}(G_{k})} = \mathcal{T}(G_k) \times \set{s_k},
\end{align*} which also implies \begin{align*}
    \abs{ \IRegionInv{G_k, s_k}} = \tau(G_{k}).
\end{align*}

\begin{proof}  
Let $(T, s_k) \in \IRegionInv{G_{k}, s_{k}}$. The definition of $\IRegion{\cdot}$ implies $\IGraph{\IVertex{T}} = G_k$, which in turn implies $\mathrm{V}(T) = \mathrm{V}(G_k)$. 
Since $T$ is a spanning tree with the same vertex set as $G_k$ it must be a spanning tree on $G_k$ and thus  $(T, s_k) \in \set{(T, s_{k}) \mid T \in \mathcal{T}(G_{k})}$.

Let $(T, s_k) \in \set{(T, s_{k}) \; \middle\vert \; T \in \mathcal{T}(G_{k})}$. By the definition of being a spanning tree on $G_k$, $\mathrm{V}(T) = \mathrm{V}(G_k)$. Now, consider by definition of being a region we know $G_k$ has the property that $\IGraph{\mathrm{V}(G_k)} = G_k$. Together, we have
$\IRegion{T, s_k} = (\IGraph{\mathrm{V}(T)}, s) = (G_k, s)$,
which implies $(T, s_k) \in \IRegionInv{G_{k}, s_{k}}$.
\end{proof}

\end{lemma}

\begin{definition}[Tree
Cut]\protect\hypertarget{def-tree-cut}{}\label{def-tree-cut}

~

Let \((T, s)\) be a region tree and let \(e \in \IEdge{T}\) be an edge
in \(T\). We define a tree cut \(\TreeCut{e}\) to be the tuple
consisting of

\begin{itemize}
    \item two disjoint region trees $(T^e_k, s_k), (T^e_{k'}, s_{k'})$ where the sizes satisfy the property that 
\begin{align*}
    1 \leq s_k, s_{k'} < s && s_k + s_{k'} = s
\end{align*}
    \item the trees and the edge satisfy the property that 
    \begin{align*}
        T^e_k \cup \set{e} \cup T^e_{k'} = T
    \end{align*}
\end{itemize}

So essentially a tree cut of a tree is an edge from a tree, the two
trees created by removing that edge and sizes we are assigning to those
new trees.

\end{definition}

For tree cuts, it is convenient to frame things in terms of a directed
tree. In other words, pick some arbitrary \(v \in T\) and fix it as the
root of the tree (viewing the tree here as directed), orienting all
edges relative to \(v\). Then, for an edge \(e\), we can consider the
two trees created by removing the edge as \(T_k^e\) corresponding to the
tree ``above'' the edge \(e\) and \(T_{k'}^e\) corresponds to the tree
``below'' the edge \(e\).

\begin{definition}[Tree Cut
Operations]\protect\hypertarget{def-tree-cut-ops}{}\label{def-tree-cut-ops}

~

Given a tree \(T\) and a tree cut \(\TreeCut{e}\) we define the
following functions

\begin{itemize}
    \item $\dev\left(\TreeCut{e}\right)$ returns the population deviation of the two regions induced by $(T^e_{k}, s_{k})$ and $ (T^e_{k'}, s_{k'})$
     \item $\mathrm{MaxAbsDev}\left(\TreeCut{e}\right)$ returns the larger of the two absolute deviations so
    \begin{align*}
        \mathrm{MaxAbsDev}\left(\TreeCut{e}\right) = \max\left(\abs{\dev\left(\TreeCut{e}\right)}\right)
    \end{align*}
\end{itemize}

\end{definition}

\begin{definition}[Splitting
Schedule]\protect\hypertarget{def-split-schedule}{}\label{def-split-schedule}

Given a fixed number of seats \(S\) and districts \(D\), a splitting
schedule \(\mathcal{S}(\cdot, \cdot, \cdot)\) is a function defined as
\[
\mathcal{S}(r, s, \xi_r) :[D] \times [S] \times [S]^r \to \{(s_1, s_2)\in [s - 1]^2 : s_1\le s_2, s_1 + s_2 = s\},
\] where \(r\) denotes the number of regions in the current partial
plan, \(s\) is the size of the multidistrict being split,
\([n] := \{1,\dots,n\}\) for any \(n\in\mathbb{Z}^+\), and \([n]^r\)
denotes the Cartesian product of \([n]\). Note that even though we write
\(\xi_r\) in \(\mathcal{S}(r, s, \xi_r)\) that term is really only a
function of the region sizes in \(\xi_r\), however we write \(\xi_r\)
for notational convenience. A key requirement for the splitting schedule
is that it must not yield sizes that would make it impossible to
continue sampling a full plan.

To see why the sizes of \(\xi_r\) term is necessary, consider the
following multi-member scheme
\[(G, D = 5, S=20, [d^-, d^+] = \set{3,4,5}).\] Suppose that we have
\(r=3\) and a plan \(\xi_3\) with region sizes \((5,5,10)\). For this
plan, we cannot split \(10\) into \((5,5)\) as we would then be left
with four districts of size \(5\), which is impossible to split further.
In contrast, suppose that we have \(\tilde{\xi}_3\) with region sizes
\((3,7,10)\). Under this scenario, we can split \(10\) into \((5,5)\) as
we would be left with sizes \((3,7,5,5)\) and that can be split into
\((3,3,4,5,5)\). In other words,
\((5,5) \notin \mathcal{S}(3, 10, \xi_3)\) but
\((5,5) \in \mathcal{S}(3, 10, \tilde{\xi}_3)\).

Luckily for single-member districting schemes and multi-member
district-only splitting schedules, the sizes of the existing regions in
\(\xi_r\) do not matter and the schedule is only a function of \(s\).
For notational convenience, we will suppress any notation of \(\xi_r\)
in the schedule going forward.

We also provide brief examples of splitting schedules. When sampling
single-member district plans and splitting off one district at a time,
the splitting schedule takes the form
\(\mathcal{S}(1, 10) = \{(1, 9)\}\), \(\mathcal{S}(2, 9) = \{(1, 8)\}\),
and so on. In contrast, if splits of any size are permitted, the
splitting schedule would be
\(\mathcal{S}(1, 10) = \set{(1,9), (2, 8), (3,7), \dots}\),
\(\mathcal{S}(2, 9) = \set{(1,8), (2, 7), (3,6), \dots}\), and so forth.

\end{definition}

\begin{refremark}[Splitting Schedules Effect on Intermediate Target
Distributions]
While the splitting schedule has no impact on the target distribution
\(\pi = \pi_D\), it has potentially significant effects on the space of
intermediate target distributions \(\pi_1,\dots, \pi_{D-1}\) by altering
their support. These effects manifest themselves in the intermediate
target distributions through the score function \(J\), assigning
probability zero to partial plans that are impossible to create under
that splitting schedule. Consider the district-only schedule, under
which the intermediate target distribution \(\pi_r\) assigns probability
zero to plans that do not have \(r-1\) districts. For example,
Figure~\ref{fig-iowa-partial1} would have probability zero under the
\(\pi_2\) of a district-only schedule because this splitting schedule
does not allow for having more than one multidistrict in a plan. In
contrast, the any-valid schedule imposes the minimum additional
restrictions on the support of the intermediate target distributions.

\label{rem-split-effect-itargets}

\end{refremark}

\begin{definition}[All Tree Cuts
Set]\protect\hypertarget{def-AllTreeCuts}{}\label{def-AllTreeCuts}

Let \((T, s)\) and \(\mathcal{S}(\cdot)\) be a region tree and a
splitting schedule, respectively. We define
\(\AllTreeCuts{T}{\mathcal{S}(r-1, s)}\) to be the set of all possible
tree cuts of \(T\) compatible with the splitting schedule value
\(\mathcal{S}(\cdot)\). More formally: \begin{align*}
    \AllTreeCuts{T}{\mathcal{S}(r-1, s)} = \set{\TreeCut{e} \mid (s_a, s_b) \in \mathcal{S}(r-1, s), e \in \mathrm{E}(T)}.
\end{align*}

\end{definition}

\begin{definition}[Balanced Tree
Cut]\protect\hypertarget{def-BalancedTreeCut}{}\label{def-BalancedTreeCut}

Given a region tree \((T,s)\) and population bounds \([P^-, P^+]\), we
say a tree cut \(\TreeCut{e}\) is balanced if the two region trees
associated with it both have balanced populations. In other words:
\begin{align*}
    \TreeCut{e} \text{ is balanced } \iff   \Iset{\mathrm{PT}}(T^{e}_{k}, s_k) = \Iset{\mathrm{PT}}(T^{e}_{k'}, s_{k'}) = 1.
\end{align*}

\end{definition}

Note that a tree cut is balanced if, and only if, the two region trees
associated with it are balanced.

\begin{definition}[Balanced Tree Cut
Set]\protect\hypertarget{def-BalancedTreeCutSet}{}\label{def-BalancedTreeCutSet}

Let \((T, s)\) be a region tree with \(T\) drawn on the region
\((G_k, s)\) and let \(\mathcal{S}(\cdot)\) be a splitting schedule. We
define the set \(\mathrm{ok}(T, \mathcal{S}(r-1, s))\) to be the subset
of \(\AllTreeCuts{T}{\mathcal{S}(r-1, s)}\) which are balanced. In other
words \begin{align*}
    \mathrm{ok}(T, \mathcal{S}(r-1, s)) = \set{\TreeCut{e} \in \AllTreeCuts{T}{\mathcal{S}(r-1, s)}  \mid \Iset{\mathrm{PT}}(T^{e}_{k}, s_k) = \Iset{\mathrm{PT}}(T^{e}_{k'}, s_{k'}) = 1}.
\end{align*}

\end{definition}

\begin{definition}[Induced Region Tree
Function]\protect\hypertarget{def-ITree-cut-func}{}\label{def-ITree-cut-func}

Let \(\TreeCut{e}\) be a tree cut of the region tree \((T, s)\). We
define the induced region tree function \(\ITree{\cdot}\) as the
function which maps a tree cut to the two region trees associated with
the tree cut. In other words \begin{align*}
    &\mathrm{T}: \AllTreeCuts{T}{\mathcal{S}(r-1, s)} \longrightarrow \mathcal{T}\set{G_k} \times \set{s_k} \times  \mathcal{T}\set{G_{k'}} \times \set{s_{k'}} \\
    &\ITree{\TreeCut{e}} = \set{(T_e^k, s_k), (T_e^{k'}, s_{k'})}.
\end{align*}

\end{definition}

The induced region tree cut function essentially just ``erases'' the
edge removed and leaves one with the two new region trees. Given two
adjacent trees \(T_k, T_{k'}\), the only tree cuts that could have
produced are the ones from the tree made by taking
\(e \in \mathcal{C}(T_k, T_{k'})\) and combining
\(T_k \cup \set{e} \cup T_{k'}\). We will formally state this claim and
provide a proof.

\begin{lemma}[Induced Region Tree Function
Pre-Image]\protect\hypertarget{lem-TreeFuncPreImg}{}\label{lem-TreeFuncPreImg}

~

Let \(\set{(T_k, s_k), (T_{k'},s_{k'})}\) be two adjacent region trees.
Then the preimage of \(\set{(T_k, s_k), (T_{k'},s_{k'})}\) under the
induced region tree function \(\ITree{\cdot}\) is \begin{align*}
     \ITreeInv{\set{(T_k, s_k), (T_{k'},s_{k'})}} = \bigcup_{e \in \mathcal{C}(T_k, T_{k'})} \TreeCut{e}
\end{align*}

\begin{proof}

Let $\TreeCut{e} \in \ITreeInv{\set{(T_k, s_k), (T_{k'},s_{k'})}}$. By definition of being a tree cut, this means that $T_k \cup \set{e} \cup T_{k'}$ must be a tree on the region $G_L = \IRegion{\IGraph{T_k \cup T_{k'}}, s_k+s_{k'}}$. By definition of being a region, this means that $e \in G_L \subset G$. Further, by definition of being a tree cut, $e$ cannot be an edge in either $T_k$ or $T_{k'}$ so it is an edge in $G$ where one vertex is in $T_k$ and one is in $T_{k'}$. Therefore, by definition  $e \in \mathcal{C}(T_k, T_{k'})$, we have proved the first part.

Let $\TreeCut{e} \in \bigcup_{e \in \mathcal{C}(T_k, T_{k'})}$. By definition of $\mathcal{C}(T_k, T_{k'})$, this means that $e \in G$. Furthermore, we know that $T = T_k \cup \set{e} \cup T_{k'}$ is a valid spanning tree and thus $\TreeCut{e}$ is an edge cut where the induced region trees are exactly $(T_k,s_k), (T_{k'},s_{k'})$
\end{proof}

\end{lemma}

\begin{definition}[Wilson's Algorithm
Function]\protect\hypertarget{def-WilsonFunction}{}\label{def-WilsonFunction}

Given a subgraph \(H\), we define the Wilson's algorithm function
\(\Wilson{H} = T\) as the random function which returns a spanning tree
on \(H\) with uniform probability. In other words \begin{align*}
    \prob{\Wilson{H} = T | H} = \frac{1}{\tau(H)}.
\end{align*}

\end{definition}

\subsection{Graph Space}\label{graph-space}

We now present the results for both the forward kernel expression from
Section~\ref{sec-GraphSplitting} and the derivation of the optimal
weights for graph space plans from
Section~\ref{sec-graph-optimal-weights}.

\subsubsection{Graph Space Splitting}\label{sec-graph-space-splitting}

We describe the splitting procedure in more detail before presenting the
necessary definitions, lemmas, and the final propositions. We first
specify a splitting schedule (see Definition~\ref{def-split-schedule}).
Intuitively, the splitting schedule specifies the allowable sizes of the
two regions produced by splitting a multidistrict of size \(s\) at stage
\(r\). For a given SMC step, we choose a \emph{splitting parameter}
\(\mathcal{K}_r \in \bbN\) and apply Algorithm \ref{alg-naive-k-split}
to divide the selected multidistrict \(H_\ell\) into two new regions.

Given a plan \(\xi_{r-1}\), we first select a multidistrict
\(H_\ell \in \xi_{r-1}\) according to some distribution
\(\varphi(\cdot \mid \xi_{r-1})\), which can be any probability
distribution over the multidistricts in \(\xi_{r-1}\). We then use
Wilson's algorithm to sample a spanning tree \(T^\ast\) on \(H_\ell\)
uniformly at random. We then iterate over all possible tree cuts that
can be formed on \(T^\ast\) according to the splitting schedule value
\(\mathcal{S}(r-1, s_\ell)\) and sort them by maximum absolute pairwise
deviation (Definition~\ref{def-tree-cut-ops}). We choose one of the
\(\mathcal{K}_r\) cuts with the smallest values uniformly at random and
set our new regions to be the two regions associated with the selected
tree cut. The pseudo-code of this is given in Algorithm
\ref{alg-naive-k-split}. We denote the sampling probability of this
splitting procedure throughout as \(q(\cdot)\).

\begin{algorithm}[t]
\begin{algorithmic}[1]
\REQUIRE{Splitting schedule $\mathcal{S}_{r-1}(\cdot)$, multidistrict $(H_{\ell}, s_\ell) \in \xi_{r-1}$, and a splitting parameter $\mathcal{K} \in \bbN$}
\STATE{Draw spanning tree $T^\ast$ on $H_{\ell}$ using Wilson's algorithm}
\STATE{Let \texttt{TreeCuts} be a list of tree cuts}
\FOR{each edge $e_i \in E(T^\ast)$}
\STATE{Let $T^{e_i}_{k}$,$T^{e_i}_{k'}$ be the two trees formed by removing $e_i$ from $T^\ast$}
\FOR{$(s_{k}, s_{k'}) \in \mathcal{S}_{r-1}(s)$}
\STATE{\texttt{TreeCuts.append($\{(T^{e_i}_{k}, s_{k}), (T^{e_i}_{k'}, s_{k'}), e_i\}$)}}
\ENDFOR
\ENDFOR
\STATE{Order \texttt{TreeCuts} by their maximum absolute deviation and choose one of the smallest $\mathcal{K}$ valued tree cuts uniformly at random}
\end{algorithmic}
\caption{Graph space splitting algorithm}
\label{alg-naive-k-split}
\end{algorithm}

Given this splitting procedure, we can express the forward transition
kernel \(M_r(\xi_r \mid \xi_{r-1})\) in closed form as the product of
two components: the probability \(\varphi\) of selecting a particular
multidistrict \(H_\ell\) from the partial plan \(\xi_{r-1}\), and the
conditional probability \(q(G_k, G_{k'} \mid H_\ell)\) of splitting that
multidistrict into two new regions \((G_k, s_k)\) and
\((G_{k'}, s_{k'})\), \[
M_r(\xi_r \mid \xi_{r-1}) = \varphi(H_\ell \mid \xi_{r-1}) \, q(G_k, G_{k'} \mid H_\ell)
\]

In Proposition~\ref{prp-split-prob-graph}, we prove the following
result. When \(G_k\) and \(G_{k'}\) are balanced and \(\mathcal{K}_r\)
is chosen such that it is greater than or equal to the number of
balanced tree cuts that can be made on \(H_\ell\), i.e., \[
\mathcal{K}_r
\ge
\max_{T \in \mathcal{T}(H_\ell)}
\left|
\mathrm{ok}\bigl(T, \mathcal{S}_{r}( s_\ell)\bigr)
\right|
\] Then, we can derive a closed form expression for
\(q(G_k, G_{k'} \mid H_\ell)\). This leads to the following closed form
expression for the forward kernel \[
     M_r(\xi_{r}|\xi_{r-1}) = \varphi(H_\ell \mid \xi_{r-1}) \frac{1}{\mathcal{K}_r} \cdot \frac{\tau(G_{k})\tau(G_{k'})}{\tau(H_{\ell})} \cdot \abs{\mathcal{C}(G_{k}, G_{k'})},
\]

If we choose \(\mathcal{K}_r\) to be greater than or equal to the
maximum number of balanced tree cuts over all regions and plans, i.e.,
\(\mathcal{K}_r
\ge
\max_{\xi_{r-1} \in \mathcal{P}_{r-1}(G)} \max_{H_\ell \in \xi_{r-1}} \max_{T \in \mathcal{T}(H_\ell)}
\left|
\mathrm{ok}\bigl(T, \mathcal{S}_{r}( s_\ell)\bigr)
\right|\), then, we are guaranteed to always be able to derive the
closed form of our forward kernel for balanced plans. See
Section~\ref{sec-split-k-estimation} for how to implement choosing
\(\mathcal{K}_r\) in practice.

Note that depending on the nature of \(G\) and the algorithm parameters,
this splitting procedure can be very inefficient. Whenever we draw a
tree \(T\) where the number of balanced tree cuts
\(b = \mathrm{ok}(T, \mathcal{S}(r-1, s_\ell))\) is smaller than
\(\mathcal{K}_r\), there is a
\(\frac{\mathcal{K}_r - b}{\mathcal{K}_r}\) chance we will select an
unbalanced split and thus be forced to run the procedure all over again.
This can be particularly inefficient when there is large variance in the
number of balanced tree cuts between sampled trees.

We now present all the necessary definitions, lemmas, and propositions
in full detail.

\begin{definition}[Naive Top K Tree Cutting
Kernel]\protect\hypertarget{def-NaiveTreeSplitterKernel}{}\label{def-NaiveTreeSplitterKernel}

Let \(\mathcal{K} \in \mathbb{N}\) and
\(R = \mathrm{Rank}(\TreeCut{e}, \AllTreeCuts{T}{\mathcal{S}(r-1, s)})\)
denote the rank of a tree cut's maximum absolute deviation among all
tree cuts in \(\AllTreeCuts{T}{\mathcal{S}(r-1, s)}\) (so a rank of 1
means its the smallest maximum absolute deviation). Then, the Naive Top
K Tree Cutting Kernel \(\NaiveKernel{T,\mathcal{S}(r-1, s), k}\) is the
random function which takes a spanning tree \(T\) as input and returns
one of the \(\mathcal{K}\) tree cuts in
\(\AllTreeCuts{T}{\mathcal{S}(r-1, s)}\) with rank
\(R \leq \mathcal{K}\) with equal probability \(\frac{1}{\mathcal{K}}\).
More formally, we can write this as \begin{align*}
\prob{\NaiveKernel{T, \mathcal{S}(r-1, s),  k} = \TreeCut{e}} = \begin{cases}
    0 & \text{ if } T \neq T_k \cup \set{e} \cup T_{k'} \\
    \frac{1}{\mathcal{K}} & \text{if } \mathrm{Rank}(\TreeCut{e}, \mathcal{S}(r-1, s)) \leq \mathcal{K} \\
    0 & \text{else}
\end{cases}
\end{align*}

\end{definition}

Now, we show that the naive top \(K\) tree splitting probability is
given by \(1/\mathcal{K}\)

\begin{lemma}[Naive Top K Tree Splitting
Probability]\protect\hypertarget{lem-NaiveTreeCutProb}{}\label{lem-NaiveTreeCutProb}

Let \((H_{L}, s_{L})\) be a region and let \(\mathcal{K}\) be an integer
such that \begin{align*}
    \mathcal{K} \geq \max_{T \in \mathcal{T}(G_{L})} \abs{\mathrm{ok}(T, \mathcal{S}(r-1, s))}
\end{align*} Now let \(T \in \mathcal{T}(H_{L})\) be a tree drawn on
\(H_{L}\) and let \(t = \TreeCut{e}\) be a tree cut of
\(\AllTreeCuts{T}{\mathcal{S}(r-1, s)}\). If this tree cut is a balanced
tree cut, then the probability that tree cut is selected by
Definition~\ref{def-NaiveTreeSplitterKernel} is
\(\frac{1}{\mathcal{K}}\). In other words: \begin{align*}
     \TreeCut{e} \in \mathrm{ok}(T, \mathcal{S}(r-1, s)) \implies q\left(\NaiveKernel{T, \mathcal{S}(r-1, s),  \mathcal{K}} = \TreeCut{e}| T \right) = \frac{1}{\mathcal{K}}.
\end{align*}

\begin{proof} Note that if the tree cut $t = \TreeCut{e}$ is balanced, then we know 
\begin{align*}
    &\mathrm{Rank}(t, \AllTreeCuts{T}{\mathcal{S}(r-1, s)}) \leq \abs{\mathrm{ok}(T, \mathcal{S}(r-1, s))} \leq \max_{\tilde{T} \in \mathcal{T}(G_{L})} \abs{\mathrm{ok}(\tilde{T}, \mathcal{S}(r-1, s))} \leq \mathcal{K} \implies \\
    &\mathrm{Rank}(t, \AllTreeCuts{T}{\mathcal{S}(r-1, s)})  \leq \mathcal{K}.
\end{align*}
Since the rank of $t$ is always guaranteed to be less than or equal to $\mathcal{K}$, by definition (\ref{def-NaiveTreeSplitterKernel}), it must have probability $\frac{1}{\mathcal{K}}$ of being selected.
\end{proof}

\end{lemma}

\begin{definition}[Naive Top K
Kernel]\protect\hypertarget{def-NaiveTopKKernel}{}\label{def-NaiveTopKKernel}

Given a multidistrict \((H_\ell, s_\ell)\), the graph space splitting
procedure outlined in \ref{alg-naive-k-split} can be formally written as
the composition of the Wilson function
(Definition~\ref{def-WilsonFunction}), the naive top K tree splitter
kernel (Definition~\ref{def-NaiveTreeSplitterKernel}), the induced
region tree function (Definition~\ref{def-ITree-cut-func}), and finally
the induced region function (Definition~\ref{def-RegionFunction}).
Altogether it can be written as \[
\IRegion{\ITree{\NaiveKernel{(\Wilson{H_{\ell}}, s_\ell), \mathcal{S}, \mathcal{K}}}} = \set{(G_k, s_k), (G_{k'}, s_{k'})}
\]

Note that all the randomness comes from the \(\Wilson{\cdot}\) and
\(\NaiveKernel{ \cdot, \mathcal{S}, k}\) terms.

\begin{proposition}[Splitting
probability]\protect\hypertarget{prp-split-prob-graph}{}\label{prp-split-prob-graph}

Let \((H_{\ell}, s_{\ell})\) be a multidistrict and let \((G_{k}, s_k)\)
and \((G_{k'}, s_{k'})\) be the two newly split balanced regions
resulting from Algorithm \ref{alg-naive-k-split}. If we choose
\(\mathcal{K}\) such that \[
    \mathcal{K} \geq \max_{T \in \mathcal{T}(H_{\ell})} \abs{\mathrm{ok}(T_\ell, \mathcal{S}_{r-1}(s_\ell))},
\] then the probability of splitting the two new regions given the old
one is \[
     q\left(G_{k}, G_{k'} \mid H_{\ell}\right)= \frac{1}{\mathcal{K}} \cdot \frac{\tau(G_{k})\tau(G_{k'})}{\tau(H_{\ell})} \cdot \abs{\mathcal{C}(G_{k}, G_{k'})},
\] where \(\mathcal{C}(G_{k}, G_{k'})\) is the set of edges in \(E(G)\)
that connect \(G_k\) and \(G_{k'}\).

\begin{proof}
Consider that we can formally write the event we sample the new regions given the old and $k$ in terms of our special functions as 
\begin{align*}
    q\left(G_{k}, G_{k'} | H_\ell\right) &= q \left( \IRegion{\ITree{\NaiveKernel{\Wilson{H_{\ell}}, \mathcal{S}(r-1, s_\ell),  \mathcal{K}}}} = \set{(G_k, s_k), (G_{k'}, s_{k'})} \mid H_\ell\right).
\end{align*}
The details are below but the high level derivation is 
\begin{align*}
q\left(G_{k}, G_{k'} \mid H_\ell\right) &= q \left( \IRegion{\ITree{\NaiveKernel{\Wilson{H_{\ell}}, \mathcal{S}(r-1, s_\ell),  \mathcal{K}}}} = \set{(G_k, s_k), (G_{k'}, s_{k'})} \mid H_\ell\right) \\
&= \E{q \left( \IRegion{\ITree{\NaiveKernel{\Wilson{H_{\ell}}, \mathcal{S}(r-1, s_\ell),  \mathcal{K}}}} = \set{(G_k, s_k), (G_{k'}, s_{k'})} \mid \Wilson{H_\ell} = T, H_\ell \right) \mid H_\ell } \\
&= \E{\sum_{T_k \in \mathcal{T}(G_k)} \sum_{T_{k'} \in \mathcal{T}(G_{k'})} \sum_{e \in \mathcal{C}(T_k, T_{k'})} \frac{1}{\mathcal{K}} \cdot \I{T = T_{k} \cup \set{e} \cup T_{k'}}|H_\ell} \\
&= \sum_{T_k \in \mathcal{T}(G_k)} \sum_{T_{k'} \in \mathcal{T}(G_{k'})} \sum_{e \in \mathcal{C}(T_k, T_{k'})} \frac{1}{\mathcal{K}} \cdot \E{\I{T = T_{k} \cup \set{e} \cup T_{k'}} \mid H_\ell} \\
&= \sum_{T_k \in \mathcal{T}(G_k)} \sum_{T_{k'} \in \mathcal{T}(G_{k'})} \sum_{e \in \mathcal{C}(T_k, T_{k'})} \frac{1}{\mathcal{K}} \cdot q\left(T = T_{k} \cup \set{e} \cup T_{k'} \mid H_\ell\right) \\
&= \sum_{T_k \in \mathcal{T}(G_k)} \sum_{T_{k'} \in \mathcal{T}(G_{k'})} \sum_{e \in \mathcal{C}(T_k, T_{k'})} \frac{1}{\mathcal{K}} \cdot \frac{1}{\tau(H_\ell)} \\
&= \frac{1}{\mathcal{K}} \cdot \frac{1}{\tau(H_\ell)} \sum_{T_k \in \mathcal{T}(G_k)} \sum_{T_{k'} \in \mathcal{T}(G_{k'})} \sum_{e \in \mathcal{C}(T_k, T_{k'})} \\
&= \frac{1}{\mathcal{K}} \cdot \frac{1}{\tau(H_\ell)} \sum_{T_k \in \mathcal{T}(G_k)} \sum_{T_{k'} \in \mathcal{T}(G_{k'})} \abs{\mathcal{C}(T_k, T_{k'})} \\
&= \frac{1}{\mathcal{K}} \cdot \frac{1}{\tau(H_\ell)} \sum_{T_k \in \mathcal{T}(G_k)} \sum_{T_{k'} \in \mathcal{T}(G_{k'})} \abs{\mathcal{C}(G_k, G_{k'})} \\
&= \frac{1}{\mathcal{K}} \cdot \frac{\tau(G_k)\tau(G_{k'})}{\tau(H_\ell)}  \abs{\mathcal{C}(G_k, G_{k'})}
\end{align*}
The more detailed derivation is as follows: for a given $T \in \mathcal{T}(H_\ell)$ consider 
\begin{align*}
&q \left( \IRegion{\ITree{\NaiveKernel{T, \mathcal{S}(r-1, s_\ell), \mathcal{K}} }} = \set{(G_k, s_k), (G_{k'}, s_{k'})} \mid H_\ell, \Wilson{H_\ell}=T \right) \\
&= q \left( \ITree{\NaiveKernel{T, \mathcal{S}(r-1, s_\ell), \mathcal{K}} } \in \IRegionInv{\set{(G_k, s_k), (G_{k'}, s_{k'})}} \mid H_\ell,  \Wilson{H_\ell}=T \right).
\end{align*}
Now, by Lemma \ref{lem-RegionFuncPreImg}, we have
\begin{align*}
    \IRegionInv{\set{(G_k, s_k), (G_{k'}, s_{k'})}} = \mathcal{T}(G_k) \times \set{s_k} \times \mathcal{T}(G_{k'}) \times \set{s_{k'}}
\end{align*}
So our expression becomes 
\begin{align*}
&= q \left( \ITree{\NaiveKernel{T, \mathcal{S}(r-1, s_\ell), \mathcal{K}} } \in \IRegionInv{\set{(G_k, s_k), (G_{k'}, s_{k'})}} \mid H_\ell,  \Wilson{H_\ell}=T \right) \\
&= q \left( \ITree{\NaiveKernel{T, \mathcal{S}(r-1, s_\ell), \mathcal{K}} } \in \mathcal{T}(G_k) \times \set{s_k} \times \mathcal{T}(G_{k'}) \times \set{s_{k'}} \mid H_\ell, \Wilson{H_\ell}=T \right) \\
&= \sum_{T_k \in \mathcal{T}(G_k)} \sum_{T_{k'} \in \mathcal{T}(G_{k'})} q \left( \ITree{\NaiveKernel{T, \mathcal{S}(r-1, s_\ell), \mathcal{K}} } = \set{(T_k, s_k), (T_{k'}, s_{k'})} \mid H_\ell, \Wilson{H_\ell}=T \right) \\
&= \sum_{T_k \in \mathcal{T}(G_k)} \sum_{T_{k'} \in \mathcal{T}(G_{k'})} q \left( \NaiveKernel{T, \mathcal{S}(r-1, s_\ell), \mathcal{K}}  \in \ITreeInv{\set{(T_k, s_k), (T_{k'}, s_{k'})}} \mid H_\ell, \Wilson{H_\ell}=T \right)
\end{align*}
Now, by Lemma \ref{lem-TreeFuncPreImg} we have
\begin{align*}
    \ITreeInv{\set{(T_k, s_k), (T_{k'},s_{k'})}} = \bigcup_{e \in \mathcal{C}(T_k, T_{k'})} \TreeCut{e}
\end{align*}
Thus, we see 
\begin{align*}
&= \sum_{T_k \in \mathcal{T}(G_k)} \sum_{T_{k'} \in \mathcal{T}(G_{k'})} q \left( \NaiveKernel{T, \mathcal{S}(r-1, s_\ell), \mathcal{K}}  \in \ITreeInv{\set{(T_k, s_k), (T_{k'}, s_{k'})}} \mid H_\ell,  \Wilson{H_\ell}=T \right) \\
&= \sum_{T_k \in \mathcal{T}(G_k)} \sum_{T_{k'} \in \mathcal{T}(G_{k'})} q \left( \NaiveKernel{T, \mathcal{S}(r-1, s_\ell), \mathcal{K}}  \in \bigcup_{e \in \mathcal{C}(T_k, T_{k'})} \TreeCut{e} \mid H_\ell,  \Wilson{H_\ell}=T \right) \\
&= \sum_{T_k \in \mathcal{T}(G_k)} \sum_{T_{k'} \in \mathcal{T}(G_{k'})} \sum_{e \in \mathcal{C}(T_k, T_{k'})} q \left( \NaiveKernel{T, \mathcal{S}(r-1, s_\ell), \mathcal{K}}  = \TreeCut{e} \mid H_\ell,  \Wilson{H_\ell}=T \right). \\
\end{align*}
Now, consider that since $G_k, G_{k'}$ are balanced regions then that means $\TreeCut{e}$ is a balanced tree cut so by Lemma \ref{lem-NaiveTreeCutProb} we know the selection probability is either 0 if $T$ does not equal the tree implied by the edge cut (i.e., $T \neq T_{k} \cup \set{e} T_{k'}$) or merely $\frac{1}{\mathcal{K}}$ if its the same tree so our expression becomes 
\begin{align*}
&= \sum_{T_k \in \mathcal{T}(G_k)} \sum_{T_{k'} \in \mathcal{T}(G_{k'})} \sum_{e \in \mathcal{C}(T_k, T_{k'})} q \left( \NaiveKernel{T, \mathcal{S}(r-1, s_\ell), \mathcal{K}}  = \TreeCut{e} | H_\ell,  \Wilson{H_\ell}=T \right) \\
&= \sum_{T_k \in \mathcal{T}(G_k)} \sum_{T_{k'} \in \mathcal{T}(G_{k'})} \sum_{e \in \mathcal{C}(T_k, T_{k'})} \frac{1}{\mathcal{K}} \cdot \I{T = T_{k} \cup \set{e} \cup T_{k'}} 
\end{align*}
Suppose that we take the expectation of this with respect to $\Wilson{H_\ell}$. 
Since we sample trees uniformly at random, the indicator becomes $\frac{1}{\tau(H_\ell)}$. 
\end{proof}

\end{proposition}

Now, we obtain the forward kernel for graph space sampling.

\begin{corollary}[Graph Space Forward
Kernel]\protect\hypertarget{cor-graph-forward-kernel}{}\label{cor-graph-forward-kernel}

Let \(\xi_{r-1}\) be a balanced plan. If \(\xi_r\) is a balanced plan
such that there exists some \(H_{\ell} \in \xi_{r-1}\),
\(G_{k}, G_{k'} \in \xi_r\) where \(H_{\ell} = G_{k} \cup G_{k'}\) and
if we choose \(\mathcal{K}\) such that \[
    \mathcal{K} \geq \max_{T \in \mathcal{T}(H_{\ell})} \abs{\mathrm{ok}(T_\ell, \mathcal{S}_{r-1}(s_\ell))},
\] then the forward kernel probability is \begin{align*}
    M_r(\xi_r \mid \xi_{r-1}) = \varphi(H_{\ell}\mid \xi_{r-1}) \cdot \frac{1}{\mathcal{K}_r} \frac{\tau(G_{k})\tau(G_{k'}) }{\tau(H_{\ell})} \abs{\mathcal{C}(G_{k}, G_{k'})} 
\end{align*}

\begin{proof} By the law of total probability, we have 
\begin{align*}
    M_r(\xi_r \mid \xi_{r-1}) &= \sum_{\tilde{H}_{\ell} \in \xi_{r-1}} \prob{\tilde{H}_{\ell} \mid \xi_{r-1}} \prob{\xi_r \mid \xi_{r-1}, \tilde{H}_{\ell}}  \\
    &= \sum_{\tilde{H}_{\ell} \in \xi_{r-1}} \varphi \left(\tilde{H}_{\ell} \mid \xi_{r-1} \right) q \left( G_{k}, G_{k'} \mid \tilde{H}_{\ell} \right)  
\end{align*}

For all $\tilde{H}_{\ell} \neq G_{k} \cup G_{k'}$, we know $q \left( G_{k}, G_{k'} \mid \tilde{H}_{\ell} \right) = 0$ so our sum just reduces to $\tilde{H}_\ell = H_\ell$. Now, if we apply Proposition \ref{prp-split-prob-graph}, we have
\begin{align*}
    M_r(\xi_r \mid \xi_{r-1}) &=  \varphi \left(H_{\ell} \mid \xi_{r-1} \right) q \left( G_k, G_{r,k'} \mid  H_\ell \right)  \\
    &= \varphi \left(H_{\ell} \mid \xi_{r-1} \right) \cdot \frac{1}{\mathcal{K}_r} \frac{\tau(G_{k})\tau(G_{k'}) }{\tau(H_{\ell})} \abs{\mathcal{C}(G_k, G_{r,k'})} 
\end{align*}

\end{proof}

\end{corollary}

We can also now easily prove when the only hard constraints imposed are
population balance and connectedness then under the Wilson based forward
kernel described above all plans are splittable.

\begin{corollary}[Splittability of Graph Space Forward
Kernel]\protect\hypertarget{cor-graph-plan-splittability}{}\label{cor-graph-plan-splittability}

If the only hard constraints are imposed through the score function
\(J\) are population balance and connectedness then under the naive top
K forward kernel all plans \(\xi_r\) in the support of \(\pi_r\) are
splittable for \(r=2,\dots, D\).

\begin{proof} Let $\xi_{r} \in \mathrm{Supp}(\pi_{r})$ for $r > 1$. Let $(G_k, s_k), (G_{k'}, s_{k'})$ be any two adjacent regions in $\xi_{r}$ (we know at least one such pair must exist if there are at least 2 regions). Consider the plan $\xi_{r-1}$ formed by merging $G_k, G_{k'}$ into the region $H_\ell = G_k \cup G_{k'}$. We know that since $G_k, G_{k'}$ are balanced then $H_\ell$ is also balanced. Since all $r-2$ other regions in $\xi_{r}$ are also balanced we thus know every region in $xi_{r-1}$ is balanced. Further, since $G_k, G_{k'}$ are adjacent in $\xi_{r}$ we know that $H_\ell$ must be connected, thus every region in $\xi_{r-1}$ is connected. Therefore, since every region is $\xi_{r-1}$ is balanced and connected we know it must have positive probability under $\pi_{r-1}$. Further, since $H_\ell = G_k \cup G_{k'}$ then by Corollary \ref{cor-graph-forward-kernel} we know that $M_r(\xi_r|\xi_{r-1}) > 0$. 

We can repeat this same merging construction for to find a $\xi_{r-2}$ such that $\pi_{r-2}(\xi_{r-2}) M_{r-1}(\xi_{r-1}|\xi_{r-2}) > 0$ and so on until we have $\xi_1,\dots,\xi_{r-1}$ such that they all have positive probability under their respective target distributions and positive transition probability under their respective forward kernels.  

\end{proof}

\end{corollary}

\subsubsection{Graph Space Optimal
Weights}\label{graph-space-optimal-weights}

We must first prove some supporting results that establish .

\begin{definition}[L-Region
Ancestor]\protect\hypertarget{def-LRegionAncestor}{}\label{def-LRegionAncestor}

~

Let \(\xi_r\) be an \(r\)-region plan. For a fixed \(0 < L < r\) we say
an \((r-L)\)-region plan \(\xi_{r-L}\) is an \(L\)-region ancestor of
\(\xi_r\) if

\begin{itemize}
    \item $\xi_{r-L}$ has $r-L$ regions or equivalently $\xi_r$ has $L$ more regions than $\xi_{r-L}$.
    \item All but $L$ of the regions in $\xi_{r-L}$ are also in $\xi_r$. In other words $r-L-1$ of the regions in $\xi_{r-L}$ are also in $\xi_r$. 
\end{itemize}

\end{definition}

This definition captures the notion of a ``path'' from \(1\)-region
plans to \(N\)-region plans where a forward kernel splits one region
into two at each step.

\begin{definition}[L-Region Ancestor
Set]\protect\hypertarget{def-SetOfLDistrictAncestors}{}\label{def-SetOfLDistrictAncestors}

Given an \(r\)-region plan \(\xi_r\) and \(0 < L < r\) we define
\(\mathcal{A}_L(\xi_r)\) to be the set of all \(L\)-region ancestors of
\(\xi_r\).

\end{definition}

\begin{definition}[New, Old, and Common
Regions]\protect\hypertarget{def-new-old-common-regions}{}\label{def-new-old-common-regions}

Let \(\xi_r\) be an \(r\)-region plan. Let \(\xi_{r-1}\) be a 1-region
ancestor so \(\xi_{r-1} \in \mathcal{A}_1(\xi_r)\). Then we define the
new regions of \(\xi_r\) relative to \(\xi_{r-1}\) as the set of regions
in \(\xi_r\) but not \(\xi_{r-1}\). We will denote this as
\(\mathrm{NR}(\xi_{r-1}, \xi_r)\) and it is defined as \begin{align*}
    \mathrm{NR}(\xi_{r-1}, \xi_r) = \xi_{r} \setminus \xi_{r-1}
\end{align*} We define the old region of \(\xi_{r-1}\) relative to
\(\xi_r\) as the region in \(\xi_{r-1}\) but not \(\xi_r\). We denote it
as \(\mathrm{OR}(\xi_{r-1}, \xi_r)\) and formally define it as
\begin{align*}
    \mathrm{OR}(\xi_{r-1}, \xi_r) = \xi_{r-1} \setminus \xi_{r}
\end{align*} Finally we define the common regions of \(\xi_{r-1}\)
relative to \(\xi_r\) as the regions in both \(\xi_{r-1}\) and
\(\xi_r\). We denote it as \(\mathrm{CR}(\xi_{r-1}, \xi_r)\) and
formally define it as \begin{align*}
    \mathrm{CR}(\xi_{r-1}, \xi_r) = \xi_{r-1} \cap \xi_{r}
\end{align*}

\end{definition}

\begin{refremark}[Region Ancestor Decomposition]
Notice by definition for \(\xi_{r-1} \in \mathcal{A}_1(\xi_r)\) we can
decompose \(\xi_r\) and \(\xi_{r-1}\) as follows \begin{align*}
    \xi_r = \mathrm{NR}(\xi_{r-1}, \xi_r) \sqcup \mathrm{CR}(\xi_{r-1}, \xi_r) && \xi_{r-1} = \mathrm{OR}(\xi_{r-1}, \xi_r) \sqcup \mathrm{CR}(\xi_{r-1}, \xi_r)
\end{align*} This will prove very useful later on.

\label{rem-RegionAncestorDecomp}

\end{refremark}

\begin{lemma}[Old Region
Characterization]\protect\hypertarget{lem-OldRegionChar}{}\label{lem-OldRegionChar}

~

Let \(r > 1\) and let \(\xi_r\) be an \(r\)-region plan. Let
\(\xi_{r-1} \in \mathcal{A}_1(\xi_r)\) be a 1-region ancestor. Further,
let \(H_{\ell}\) denote the old region and \(G_{k}, G_{k'}\) denote the
new regions, i.e.,: \begin{align*}
    \abs{\mathrm{OR}(\xi_{r-1}, \xi_r)} = \set{H_{\ell}} && \abs{\mathrm{NR}(\xi_{r-1}, \xi_r)} = \set{G_k, G_{k'}}
\end{align*} Then the old region \(H_{\ell}\) is equal to the subgraph
induced by the union of the vertices in the two new regions
\(G_{k}, G_{k'}\). In other words: \begin{align*}
    H_{\ell} = \mathrm{G}(V_{k} \sqcup V_{k'})
\end{align*}

\begin{proof} Recall that we can decompose $\xi_{r-1}$ and $\xi_r$ as follows:
\begin{align*}
    \xi_r = \mathrm{NR}(\xi_{r-1}, \xi_r) \sqcup \mathrm{CR}(\xi_{r-1}, \xi_r) && \xi_{r-1} = \mathrm{OR}(\xi_{r-1}, \xi_r) \sqcup \mathrm{CR}(\xi_{r-1}, \xi_r)
\end{align*}
Since we know $\mathrm{V}(\xi_r) = \mathrm{V}(\xi_{r-1})$ this implies that the set of vertices in $G_{k}$ and $G_{k'}$ is equal to the vertices in $H_{\ell}$. To formally show this consider 
\begin{align*}
    \mathrm{V}(\xi_r) &= \mathrm{V}(\xi_{r-1}) \\
    \mathrm{V}(\mathrm{NR}(\xi_{r-1}, \xi_r)) \sqcup \mathrm{V}(\mathrm{CR}(\xi_{r-1}, \xi_r)) &= \mathrm{V}(\mathrm{OR}(\xi_{r-1}, \xi_r)) \sqcup \mathrm{V}(\mathrm{CR}(\xi_{r-1}, \xi_r)) \implies \\
    \mathrm{V}(\mathrm{NR}(\xi_{r-1}, \xi_r)) &= \mathrm{V}(\mathrm{OR}(\xi_{r-1}, \xi_r)) \\
    V_{k} \sqcup V_{k'} &= V_{\ell} 
\end{align*}
Since we know that $H_{\ell} = \mathrm{G}(V_{\ell})$ by definition then that tells us that $H_{\ell} = \mathrm{G}(V_{k} \sqcup V_{k'})$
\end{proof}

\end{lemma}

This lemma formalizes the result that if a plan \(\xi_r\) only differs
from \(\xi_{r-1}\) by one region, then this region in \(\xi_{r-1}\) must
be formed by the union of the vertices in the two new regions in
\(\xi_r\).

\begin{corollary}[Connectedness of New District and
Remainder]\protect\hypertarget{cor-RemainderAndNewDistConnected}{}\label{cor-RemainderAndNewDistConnected}

Let \(r >1\). Let \(\xi_r\) be an \(r\)-region plan. Let
\(\xi_{r-1} \in \mathcal{A}_1(\xi_r)\) be a 1-region ancestor. Further,
let \(H_\ell\) denote the old region and \(G_{k}, G_{k'}\) denote the
new regions. Then \(G_{k}\) and \(G_{k'}\) share at least one edge in
\(G\), i.e., \begin{align*}
    \abs{\mathcal{C}(G_{k}, G_{k'})} > 0
\end{align*}

\begin{proof} Recall by Lemma \ref{lem-OldRegionChar} above we know $H_{\ell} = \mathrm{G}(V_{k} \sqcup V_{k'})$. Since $H_{\ell}$ is a valid region and thus connected that implies $G_{k}$ and $G_{k'}$ must share at least one edge in $G$, else $\mathrm{G}(V_{k} \sqcup V_{k'})$ would not be a connected subgraph. 
\end{proof}

\end{corollary}

\begin{proposition}[Ancestor Set
Equivalency]\protect\hypertarget{prp-AncestorSetCharacterization}{}\label{prp-AncestorSetCharacterization}

Fix \(r > 1\) and let \(\xi_r\) be a \(r\)-region plan. Then the set of
\(1\)-region ancestors \(\mathcal{A}_1(\xi_r)\) is in bijective
correspondence with the set of adjacent regions in \(\xi_r\) i.e.,
\(\mathrm{AR}(\xi_r)\).

Specifically, the bijection is the function that maps adjacent regions
\((G_{k}, G_{k'}) \in \mathrm{AR}(\xi_r)\) to the \(r-1\)-region partial
plan made by taking the other \(r-2\) regions and a region formed by
combining \((G_{k}, G_{k'})\). So
\(\xi_r \setminus \set{G_{k}, G_{k'}}\) and
\(\mathrm{G}(V_{k} \sqcup V_{k'})\).

\begin{proof}
Define the following functions 
\begin{align*}
f: \mathcal{A}_1(\xi_r) \longrightarrow \mathrm{AR}(\xi_r)  \\
f(\xi_{r-1}) = \mathrm{NR}(\xi_{r-1},\xi_r)
\end{align*}
So $f$ maps an $n-1$-region plan $\xi_{r-1} \in \mathcal{A}_1(\xi_r)$ to the two new regions in $\xi_r$ relative to $\xi_{r-1}$ which we proved in \ref{cor-RemainderAndNewDistConnected} are adjacent in $G$. We will show that this function is a bijection. 
\item Part 1. $f$ is injective \\
Let $\xi_{r-1}, \tilde{\xi}_{r-1} \in \mathcal{A}_1(\xi_r)$ such that $f(\xi_{r-1}) = f(\tilde{\xi}_{r-1}) = (G_{k}, G_{k'})$ (for some $G_{k}, G_{k'}\in \xi_r$). \\
Consider that since $\xi_{r-1}, \tilde{\xi}_{r-1} \in \mathcal{A}_1(\xi_r)$ we know by \ref{rem-RegionAncestorDecomp} that 
\begin{align*}
    \xi_r &= \mathrm{NR}(\xi_{r-1}, \xi_r) \sqcup \mathrm{CR}(\xi_{r-1}, \xi_r) = \mathrm{NR}(\tilde{\xi}_{r-1}, \xi_r) \sqcup \mathrm{CR}(\tilde{\xi}_{r-1}, \xi_r) \\
    \xi_{r-1} &= \mathrm{OR}(\xi_{r-1}, \xi_r) \sqcup \mathrm{CR}(\xi_{r-1}, \xi_r) \\
    \tilde{\xi}_{r-1} &= \mathrm{OR}(\tilde{\xi}_{r-1}, \xi_r) \sqcup \mathrm{CR}(\tilde{\xi}_{r-1}, \xi_r)
\end{align*}
By definition of $f$ we have $f(\xi_{r-1}) = \mathrm{NR}(\xi_{r-1}, \xi_r)$ and $f(\tilde{\xi}_{r-1}) = \mathrm{NR}(\tilde{\xi}_{r-1}, \xi_r)$ so we see $f(\xi_{r-1}) = f(\tilde{\xi}_{r-1})$ implies 
\begin{align*}
    \xi_r = \mathrm{NR}(\xi_{r-1}, \xi_r) \sqcup \mathrm{CR}(\xi_{r-1}, \xi_r) &= \mathrm{NR}(\tilde{\xi}_{r-1}, \xi_r) \sqcup \mathrm{CR}(\tilde{\xi}_{r-1}, \xi_r) \implies \\
    f(\xi_{r-1}) \sqcup \mathrm{CR}(\xi_{r-1}, \xi_r) &= f(\tilde{\xi}_{r-1}) \sqcup \mathrm{CR}(\tilde{\xi}_{r-1}, \xi_r) \implies \\
    \mathrm{CR}(\xi_{r-1}, \xi_r) &=  \mathrm{CR}(\tilde{\xi}_{r-1}, \xi_r) 
\end{align*}
Now consider that by Lemma \ref{lem-OldRegionChar} we know that 
\begin{align*}
\mathrm{OR}(\xi_{r-1}, \xi_r) = \mathrm{V}(\mathrm{NR}(\xi_{r-1}, \xi_r))  && \mathrm{OR}(\tilde{\xi}_{r-1}, \xi_r) = \mathrm{V}(\mathrm{NR}(\tilde{\xi}_{r-1}, \xi_r)) 
\end{align*}
So we see $f(\xi_{r-1}) = f(\tilde{\xi}_{r-1})$ implies $\mathrm{OR}(\xi_{r-1}, \xi_r) = \mathrm{OR}(\tilde{\xi}_{r-1}, \xi_r)$ as well. Since $\xi_{r-1}$ and $\tilde{\xi}_{r-1}$ can be decomposed into their new and common regions with respect to $\xi_r$ and we've just shown both of those are equal then we can conclude $\tilde{\xi}_{r-1} = \xi_{r-1}$.

\item Part 2. $f$ is surjective \\
Let $\set{G_{k}, G_{k'}} \in \mathrm{AR}(\xi_r)$. Define an $r-1$-region partial plan $\xi_{r-1}$ as the $r-2$ regions $\xi_r \setminus \set{G_{k}, G_{k'}}$ and a new region $\mathrm{G}(V_{k} \cup V_{k'})$. This is clearly an $r-1$ region partial plan. Moreover since all but 1 of $\xi_{r-1}$'s regions are shared with $\xi_r$ so we see it is actually a 1-region ancestor. So by construction we have found $\xi_{r-1} \in \mathcal{A}_1(\xi_r)$ such that $f(\xi_{r-1}) = \set{G_{k}, G_{k'}}$ and thus $f$ is surjective.
\end{proof}

\end{proposition}

\begin{lemma}[Ancestor Density
Factorization]\protect\hypertarget{lem-OneRegionAncestorDensityFactorization}{}\label{lem-OneRegionAncestorDensityFactorization}

Let \(\xi_r\) be an \(r\)-region plan and \(\xi_{r-1}\) be a 1-region
ancestor (\(\xi_{r-1} \in \mathcal{A}_1(\xi_r)\)). Let \(H_\ell\) be the
old region and \(G_{k}, G_{k'}\) be the new regions. Then we can factor
their unnormalized densities as \begin{align*}
    \gamma_r(\xi_r) &= e^{-J(\xi_{r})} \cdot \tau(G_{k})^\rho \tau(G_{k'})^\rho \cdot \tau(\mathrm{CR}(\xi_{r-1}, \xi_r))^\rho \\ 
    \gamma_{r-1}(\xi_r) &= e^{-J(\xi_{r-1})} \cdot \tau(H_{\ell})^\rho \cdot \tau(\mathrm{CR}(\xi_{r-1}, \xi_r))^\rho
\end{align*} Where \begin{align*}
\tau(\mathrm{CR}(\xi_{r-1}, \xi_r))^\rho = \prod_{\substack{G_{n} \in \\ \mathrm{CR}(\xi_{r-1}, \xi_r)}} \tau(G_{n})^\rho 
\end{align*}

\begin{proof} This proof is very straight forward. Recall as noted in remark \ref{rem-RegionAncestorDecomp} that we can express $\xi_r$ and $\xi_{r-1}$ as 
\begin{align*}
    \xi_r = \mathrm{NR}(\xi_{r-1}, \xi_r) \sqcup \mathrm{CR}(\xi_{r-1}, \xi_r) && \xi_{r-1} = \mathrm{OR}(\xi_{r-1}, \xi_r) \sqcup \mathrm{CR}(\xi_{r-1}, \xi_r)
\end{align*}
Further recall we know from Lemma \ref{lem-OldRegionChar} that $\mathrm{OR}(\xi_{r-1}, \xi_r) = H_{\ell}$ and  $\mathrm{NR}(\xi_{r-1}, \xi_r) = \set{G_{k}, G_{k'}}$. Since $\gamma_r$ and $\gamma_{r-1}$ are a product over the regions of a plan times the score of the entire plan it immediately follows that then that we can use the decomposition to see that 
\begin{align*}
    \gamma_r(\xi_r) &= e^{-J(\xi_r)} \prod_{n=1}^r \tau(G_{n})^\rho \\
    &= e^{-J(\xi_r)} \cdot \tau(\mathrm{NR}(\xi_{r-1}, \xi_r))^\rho \cdot \tau(\mathrm{CR}(\xi_{r-1}, \xi_r))^\rho \\
    &= e^{-J(\xi_r)} \cdot \tau(G_{k})^\rho \tau(G_{k'})^\rho \cdot \tau(\mathrm{CR}(\xi_{r-1}, \xi_r))^\rho
\end{align*}
Similarly for $\gamma_{r-1}$ we have 
\begin{align*}
    \gamma_{r-1}(\xi_{r-1}) &= e^{-J(\xi_{r-1})} \prod_{n=1}^{r-1} \tau(H_{n})^\rho \\
    &= e^{-J(\xi_{r-1})} \cdot \tau(\mathrm{OR}(\xi_{r-1}, \xi_r))^\rho \cdot \tau(\mathrm{CR}(\xi_{r-1}, \xi_r))^\rho \\
    &= e^{-J(\xi_{r-1})} \cdot \tau(H_{\ell})^\rho \cdot \tau(\mathrm{CR}(\xi_{r-1}, \xi_r))^\rho
\end{align*}
and we are done
\end{proof}

\end{lemma}

\begin{theorem}[Marginal proposal
density]\protect\hypertarget{thm-marg-dens}{}\label{thm-marg-dens}

Given a forward kernel \(M_r\) and a target distribution \(\pi_r\), the
marginal proposal density is given by, \[
f_r(\xi_r)  = \frac{1}{\mathcal{K}_r} \frac{\gamma_r(\xi_r)}{ Z_{r-1}} \sum_{\substack{ G_{k} \sim G_{k'} \in \xi_r}} \varphi(G_{k} \cup G_{k'}  \mid \tilde{\xi}_{r-1})  \frac{\exp{-J(\tilde{\xi}_{r-1})}}{\exp{-J(\xi_{r})}} \frac{\tau(H_\ell)^{\rho-1}}{\tau(G_k)^{\rho-1} \tau(G_{k'})^{\rho-1}} \abs{\mathcal{C}(G_{k}, G_{k'})},
\] where \(\tilde{\xi}_{r-1}\) is the plan formed by replacing the
adjacent regions \(G_{k}\) and \(G_{k'}\) in \(\xi_r\) with the merged
region \(H_{\ell} = G_k \cup G_{k'}\).

\begin{proof} 
Consider the integral needed to derive a closed form of the marginal proposal density 
$$
f_r(\xi_r) = \int \pi_{r-1}(\xi_{r-1}) M_r(\xi_r \mid \xi_{r-1}) \; d\xi_{r-1}.
$$

The general idea of the proof is that we know from \ref{cor-graph-forward-kernel} that $M_r(\xi_r \mid \xi_{r-1})$ is only non-zero when $\xi_r$ and $\xi_{r-1}$ share all but one region in common. Moreover, we know that for the non-shared region $H_\ell \in \xi_{r-1}$ it must be the case that $H_\ell = G_k \cup G_{k'}$ where  $G_k, G_{k'}$ are the two non-shared regions in $\xi_r$. Notice this is exactly the definition of $\xi_{r-1}$ being a 1-region ancestor of $\xi_r$. Thus we see that the integral actually reduces to a sum over all $\xi_{r-1} \in \mathcal{A}_1(\xi_r)$ so  

$$
f_r(\xi_r) = \sum_{\xi_{r-1} \in \mathcal{A}_1(\xi_r)} \pi_{r-1}(\xi_{r-1}) M_r(\xi_r \mid \xi_{r-1}) .
$$

Next we can simplify the terms even further by leveraging the fact that if $\pi_{r-1}(\xi_{r-1})$ is non-zero then $\xi_{r-1}$ must be balanced. We thus know by \ref{cor-graph-forward-kernel} the closed form expression of $M_r(\xi_r \mid \xi_{r-1})$ in the sum is

$$
f_r(\xi_r) = \sum_{\xi_{r-1} \in \mathcal{A}_1(\xi_r)} \pi_{r-1}(\xi_{r-1}) \varphi(H_{\ell}\mid \xi_{r-1}) \cdot \frac{1}{\mathcal{K}_r} \frac{\tau(G_{k})\tau(G_{k'}) }{\tau(H_{\ell})} \abs{\mathcal{C}(G_{k}, G_{k'})}  .
$$

We will now simplify each term of the sum. Consider: 
\begin{align*}
&\pi_{r-1}(\xi_{r-1}) \varphi(H_{\ell}\mid \xi_{r-1}) \cdot \frac{1}{\mathcal{K}_r} \frac{\tau(G_{k})\tau(G_{k'}) }{\tau(H_{\ell})} \abs{\mathcal{C}(G_{k}, G_{k'})} \\
&= \frac{1}{Z_{r-1}} \gamma_{r-1}(\xi_{r-1})  \cdot \varphi(H_\ell|\xi_{r-1}) \frac{\tau(G_{k})\tau(G_{k'}) }{\mathcal{K}_r \cdot \tau(H_\ell)} \abs{\mathcal{C}(G_{k}, G_{k'})} \\
&= \frac{1}{Z_{r-1}\mathcal{K}_r } \gamma_{r-1}(\xi_{r-1})  \cdot \varphi(H_\ell|\xi_{r-1}) \frac{\tau(G_{k})\tau(G_{k'}) }{\tau(H_\ell)} \abs{\mathcal{C}(G_{k}, G_{k'})} \\
&=  \frac{\gamma_{r-1}(\xi_{r-1}) \tau(G_{k})\tau(G_{k'}) }{\tau(H_\ell)} \frac{1}{Z_{r-1}\mathcal{K}_r}    \varphi(H_\ell|\xi_{r-1})  \cdot \abs{\mathcal{C}(G_{k}, G_{k'})}  
\end{align*}
Now consider for $\gamma_{r-1}(\xi_{r-1}) \tau(G_{k})\tau(G_{k'})$ we can use \ref{lem-OneRegionAncestorDensityFactorization} to simplify things 
\begin{align*}
&\; \; \; \; \gamma_{r-1}(\xi_{r-1}) \tau(G_{k})\tau(G_{k'}) \\
&= \exp{-J(\xi_{r-1})}\tau(H_\ell)^\rho \cdot \tau(\mathrm{CR}(\xi_{r-1}, \xi_r))^\rho \tau(G_{k})\tau(G_{k'}) \; \; \text{by }\ref{lem-OneRegionAncestorDensityFactorization} \\
&= \frac{\exp{-J(\xi_r)}}{\exp{-J(\xi_r)}} \frac{\tau(G_k)^{\rho-1} \tau(G_{k'})^{\rho-1} }{\tau(G_k)^{\rho-1} \tau(G_{k'})^{\rho-1} } \cdot \exp{-J(\xi_{r-1})} \tau(H_\ell)^\rho \cdot \tau(\mathrm{CR}(\xi_{r-1}, \xi_r))^\rho \tau(G_{k})\tau(G_{k'}) \\
&= \frac{\exp{-J(\xi_{r-1})}}{\exp{-J(\xi_r)}}\frac{ \tau(H_\ell)^\rho}{\tau(G_k)^{\rho-1} \tau(G_{k'})^{\rho-1}} \cdot \exp{-J(\xi_r)} \tau(\mathrm{CR}(\xi_{r-1}, \xi_r))^\rho \tau(G_{k})^\rho \tau(G_{k'})^\rho  \\
&= \frac{\exp{-J(\xi_{r-1})}}{\exp{-J(\xi_r)}}\frac{ \tau(H_\ell)^\rho}{\tau(G_k)^{\rho-1} \tau(G_{k'})^{\rho-1}} \cdot \gamma_{r}(\xi_r) \; \; \text{by }\ref{lem-OneRegionAncestorDensityFactorization} 
\end{align*}
Thus we have
\begin{align*}
\gamma_{r-1}(\xi_{r-1}) \tau(G_{k})\tau(G_{k'}) &= \frac{\exp{-J(\xi_{r-1})}}{\exp{-J(\xi_r)}}\frac{ \tau(H_\ell)^\rho}{\tau(G_k)^{\rho-1} \tau(G_{k'})^{\rho-1}} \cdot \gamma_{r}(\xi_r)  
\end{align*}
So plugging that back in above we get 
\begin{align*}
\pi_{r-1}(\xi_{r-1}) \cdot M_n(\xi_r \; | \; \xi_{r-1}) &= \frac{\gamma_{r-1}(\xi_{r-1}) \tau(G_{k})\tau(G_{k'}) }{\tau(H_\ell)} \frac{1}{Z_{r-1}\mathcal{K}_r}    \varphi(H_\ell|\xi_{r-1})  \cdot \abs{\mathcal{C}(G_{k}, G_{k'})}  \\
&=  \frac{\frac{\exp{-J(\xi_{r-1})}}{\exp{-J(\xi_r)}}\frac{ \tau(H_\ell)^\rho}{\tau(G_k)^{\rho-1} \tau(G_{k'})^{\rho-1}} \cdot \gamma_{r}(\xi_r) }{\tau(H_\ell)} \frac{1}{Z_{r-1}\mathcal{K}_r}    \varphi(H_\ell|\xi_{r-1})  \cdot \abs{\mathcal{C}(G_{k}, G_{k'})}  \\
&=  \frac{\exp{-J(\xi_{r-1})}}{\exp{-J(\xi_r)}}\frac{ \tau(H_\ell)^{\rho-1}}{\tau(G_k)^{\rho-1} \tau(G_{k'})^{\rho-1}} \cdot \gamma_{r}(\xi_r)  \frac{1}{Z_{r-1}\mathcal{K}_r}    \varphi(H_\ell|\xi_{r-1})  \cdot \abs{\mathcal{C}(G_{k}, G_{k'})}  \\
&=   \frac{\gamma_{r}(\xi_r)}{Z_{r-1}\mathcal{K}_r} \varphi(H_\ell|\xi_{r-1}) \frac{\exp{-J(\xi_{r-1})}}{\exp{-J(\xi_r)}}\frac{ \tau(H_\ell)^{\rho-1}}{\tau(G_k)^{\rho-1} \tau(G_{k'})^{\rho-1}}      \cdot \abs{\mathcal{C}(G_{k}, G_{k'})}  
\end{align*}

Now note in this expression above the term $\frac{\gamma_{r}(\xi_r)}{Z_{r-1}\mathcal{K}_r}$ in the product is constant with respect to $\xi_{r-1}$ so that can be pulled outside the sum. Now returning to our sum we have

\begin{align*}
f_r(\xi_r) &= \sum_{\xi_{r-1} \in \mathcal{A}_1(\xi_r)} \pi_{r-1}(\xi_{r-1}) \varphi(H_{\ell}\mid \xi_{r-1}) \cdot \frac{1}{\mathcal{K}_r} \frac{\tau(G_{k})\tau(G_{k'}) }{\tau(H_{\ell})} \abs{\mathcal{C}(G_{k}, G_{k'})} \\
&= \sum_{\xi_{r-1} \in \mathcal{A}_1(\xi_r)} \frac{\gamma_{r}(\xi_r)}{Z_{r-1}\mathcal{K}_r} \varphi(H_\ell|\xi_{r-1}) \frac{\exp{-J(\xi_{r-1})}}{\exp{-J(\xi_r)}}\frac{ \tau(H_\ell)^{\rho-1}}{\tau(G_k)^{\rho-1} \tau(G_{k'})^{\rho-1}}      \cdot \abs{\mathcal{C}(G_{k}, G_{k'})} \\
&= \frac{\gamma_{r}(\xi_r)}{Z_{r-1}\mathcal{K}_r} \sum_{\xi_{r-1} \in \mathcal{A}_1(\xi_r)}  \varphi(H_\ell|\xi_{r-1}) \frac{\exp{-J(\xi_{r-1})}}{\exp{-J(\xi_r)}}\frac{ \tau(H_\ell)^{\rho-1}}{\tau(G_k)^{\rho-1} \tau(G_{k'})^{\rho-1}}      \cdot \abs{\mathcal{C}(G_{k}, G_{k'})} 
\end{align*}

Now finally by Proposition \ref{prp-AncestorSetCharacterization} we know that that a sum over 1-region ancestor plans $\xi_{r-1} \in \mathcal{A}_1(\xi_r)$ is equivalent to summing over pairs of adjacent regions $G_k \sim G_{k'} \in \xi_r$ and taking $\tilde{\xi}_{r-1}$ to be the plan made by merging $G_k, G_{k'}$ in $\xi_r$ so we have 

\begin{align*}
f_r(\xi_r) &= \frac{\gamma_{r}(\xi_r)}{Z_{r-1}\mathcal{K}_r} \sum_{\xi_{r-1} \in \mathcal{A}_1(\xi_r)}  \varphi(H_\ell|\xi_{r-1}) \frac{\exp{-J(\xi_{r-1})}}{\exp{-J(\xi_r)}}\frac{ \tau(H_\ell)^{\rho-1}}{\tau(G_k)^{\rho-1} \tau(G_{k'})^{\rho-1}}      \cdot \abs{\mathcal{C}(G_{k}, G_{k'})}  \\
&= \frac{1}{\mathcal{K}_r} \frac{\gamma_r(\xi_r)}{ Z_{r-1}} \sum_{\substack{ G_{k} \sim G_{k'} \in \xi_r}} \varphi(G_{k} \cup G_{k'}  \mid \tilde{\xi}_{r-1})  \frac{\exp{-J(\tilde{\xi}_{r-1})}}{\exp{-J(\xi_{r})}} \frac{\tau(H_\ell)^{\rho-1}}{\tau(G_k)^{\rho-1} \tau(G_{k'})^{\rho-1}} \abs{\mathcal{C}(G_{k}, G_{k'})}
\end{align*}

\end{proof}

\end{theorem}

\begin{proposition}[Optimal
weights]\protect\hypertarget{prp-opt-wt}{}\label{prp-opt-wt}

Given a forward kernel \(M_r\) and target distribution \(\pi_r\), the
optimal minimal variance incremental weights are
\begin{equation}\phantomsection\label{eq-graph-optimal-weights}{
w_r(\xi_{r-1}, \xi_r) = \mathcal{K}_r \cdot \left( \sum_{\substack{ G_{k} \sim G_{k'} \in \xi_r}} \varphi(G_{k} \cup G_{k'} \mid \tilde{\xi}_{r-1})  \frac{\exp{-J(\tilde{\xi}_{r-1})}}{\exp{-J(\xi_{r})}} \left(\frac{\tau(G_{k} \cup G_{k'})}{\tau(G_k) \tau(G_{k'})}\right)^{\rho-1} \abs{\mathcal{C}(G_{k}, G_{k'})}  \right)^{-1},
}\end{equation} where \(G_{k} \sim G_{k'}\) denotes adjacent regions in
\(\xi_r\), \(\tilde{\xi}_{r-1}\) is the plan formed by merging \(G_{k}\)
and \(G_{k'}\), and \(\mathcal{C}(G_{k}, G_{k'})\) is the set of edges
in \(G\) with one vertex in \(G_k\) and one in \(G_{k'}\).

\begin{proof}
Recall from \citep{dai2022} the incremental weights for a SMC sampler are given by 
\begin{align*}
    w_r(\xi_{r-1}, \xi_r) = \frac{\gamma_r(\xi_r) L_{r-1}(\xi_{r-1} \;| \; \xi_r) }{\gamma_{r-1}(\xi_{r-1} ) M_{r}(\xi_r \;| \; \xi_{r-1}) }
\end{align*}
and the minimum variance weights are achieved by setting 
\begin{align*}
    L_{r-1}(\xi_{r-1} \;| \; \xi_r)  = \frac{\pi_{r-1}(\xi_{r-1} ) M_{r}(\xi_r \;| \; \xi_{r-1})}{f_r(\xi_r)}
\end{align*} 
which for us simplifies to 
\begin{align*}
     w_r(\xi_{r-1}, \xi_r) &= \frac{\gamma_r(\xi_r) L_{r-1}(\xi_{r-1} \;| \; \xi_r) }{\gamma_{r-1}(\xi_{r-1} ) M_{r}(\xi_r \;| \; \xi_{r-1}) } \\
&= \frac{\gamma_r(\xi_r) \pi_{r-1}(\xi_{r-1} ) M_{r}(\xi_r \;| \; \xi_{r-1}) }{\gamma_{r-1}(\xi_{r-1} ) M_{r}(\xi_r \;| \; \xi_{r-1}) f_r(\xi_r)}  \\
&= \frac{\gamma_r(\xi_r) \gamma_{r-1}(\xi_{r-1} ) M_{r}(\xi_r \;| \; \xi_{r-1}) }{Z_{r-1} \gamma_{r-1}(\xi_{r-1} ) M_{r}(\xi_r \;| \; \xi_{r-1}) f_r(\xi_r)} \\
&= \frac{1}{Z_{r-1}} \frac{\gamma_r(\xi_r)}{f_r(\xi_r)} 
\end{align*}
Now using by Theorem \ref{thm-marg-dens} we can simplify to 

\begin{align*}
     w_r(\xi_{r-1}, \xi_r) &=  \frac{1}{Z_{r-1}} \frac{\gamma_r(\xi_r)}{f_r(\xi_r)} \\
&= \frac{1}{Z_{r-1}} \frac{\gamma_r(\xi_r)}{\frac{1}{\mathcal{K}_r} \frac{\gamma_r(\xi_r)}{ Z_{r-1}} \sum_{\substack{ G_{k} \sim G_{k'} \in \xi_r}} \varphi(G_{k} \cup G_{k'}  \mid \tilde{\xi}_{r-1})  \frac{\exp{-J(\tilde{\xi}_{r-1})}}{\exp{-J(\xi_{r})}} \frac{\tau(H_\ell)^{\rho-1}}{\tau(G_k)^{\rho-1} \tau(G_{k'})^{\rho-1}} \abs{\mathcal{C}(G_{k}, G_{k'})}} \\
&= \mathcal{K}_r  \frac{1}{ \sum_{\substack{ G_{k} \sim G_{k'} \in \xi_r}} \varphi(G_{k} \cup G_{k'}  \mid \tilde{\xi}_{r-1})  \frac{\exp{-J(\tilde{\xi}_{r-1})}}{\exp{-J(\xi_{r})}} \frac{\tau(H_\ell)^{\rho-1}}{\tau(G_k)^{\rho-1} \tau(G_{k'})^{\rho-1}} \abs{\mathcal{C}(G_{k}, G_{k'})} } \\
&= \mathcal{K}_r  \left(  \sum_{\substack{ G_{k} \sim G_{k'} \in \xi_r}} \varphi(G_{k} \cup G_{k'}  \mid \tilde{\xi}_{r-1})  \frac{\exp{-J(\tilde{\xi}_{r-1})}}{\exp{-J(\xi_{r})}} \frac{\tau(H_\ell)^{\rho-1}}{\tau(G_k)^{\rho-1} \tau(G_{k'})^{\rho-1}} \abs{\mathcal{C}(G_{k}, G_{k'})} \right)^{-1} 
\end{align*}
\end{proof}

\end{proposition}

\begin{refremark}[Effect of Splitting Schedule on the Optimal Weights]
Recall from Remark~\ref{rem-split-effect-itargets} that the choice of
splitting schedule implicitly influences the space of intermediate
distributions. This is reflected in the weights as well meaning when the
schedule is not any-valid splits then some of the terms in the sum may
become zero because the merged plan \(\tilde{\xi}_{r-1}\) is not
possible under that splitting schedule. The easiest example to see that
is the district-only schedule where unless \(r=D\) two merged districts
will have probability zero in the sum as that would create a plan with
more than one multidistrict.

\label{rem-split-weight-interaction}

\end{refremark}

\subsection{New Sampling Spaces}\label{sec-new-sampling-space-proofs}

As mentioned in the paper, gSMC can be expanded to operate on new
sampling spaces. While the final output is still plans distributed
according to the same target distribution as before, these new sampling
spaces allow for trading off some of the computational cost of splitting
plans in exchange for increased computational complexity of computing
the weights.

\subsubsection{Spanning Forest Space}\label{sec-spanning-forest}

The first sampling space the algorithm can be lifted to is the space of
spanning forest plans which we will call spanning forest space. Instead
of a plan \(\xi_r\) consisting of a collection of \(r\) regions, we
instead have a forest plan \(F_r\) consisting of a collection of \(r\)
region trees \(F_r = \set{(T_k, s_k)}_{k=1}^r\). Formally, we can define
it as follows:

\begin{definition}[Forest Plan
Definition]\protect\hypertarget{def-forest-plan}{}\label{def-forest-plan}

Given a districting scheme \((G, D, S, [d^-, d^+])\) and \(r \leq D\),
we define a \(r\)-region forest plan \(F_r\) as an unordered collection
of \(r\) region trees \[\xi_r = \set{(T_k, s_k)}_{k=1}^r\] such that the
region sizes form an integer partition of \(S\),
i.e.~\(\sum_{k=1}^r s_k = S\) and the vertex sets of the trees form a
partition of the vertex set of \(G\).

\end{definition}

We now also define a new special function, the induced plan function.

\begin{definition}[Induced Plan
Function]\protect\hypertarget{def-induced-plan-func}{}\label{def-induced-plan-func}

We define the induced plan function \(\xi(\cdot)\) as the function that
takes a forest plan \(F_r\) maps it to the associated graph plan
\(\xi_r\) induced by the vertex set of its region trees. In other words
\[
\xi(F_r) = \xi_r = \set{(\IRegion{T_k, s_k})}_{k=1}^r.
\]

\end{definition}

\begin{definition}[Forest Plan Space
Definition]\protect\hypertarget{def-ForestPlanSpace}{}\label{def-ForestPlanSpace}

Given a districting scheme \((G, D, S, [d^-, d^+])\) and \(r \leq D\),
we define \(\mathcal{F}_r(G, D, S, [d^-, d^+])\) to be the set of all
\(r\)-region forest plans under that districting scheme. For notational
convenience we will write it simply as \(\mathcal{F}_r\).

Note that \(\mathcal{F}_r\) can be equivalently expressed as the
preimage of the space of graph plans, \(\mathcal{P}_r\), under the
induced plan function. In other words, \[
\mathcal{F}_r = \xi^{-1}(\mathcal{P}_r).
\]

\end{definition}

\begin{lemma}[Induced Plan Function
Preimage]\protect\hypertarget{lem-induced-plan-func-preim}{}\label{lem-induced-plan-func-preim}

Let \(\xi_r = \set{(G_k, s_k)}_{k=1}^r\) be an \(r\)-region graph space
plan. The preimage of \(\xi_r\) under the induced plan function
\(\xi(\cdot)\) is all forest plans that can be made from a combination
of the Cartesian product of all spanning trees which can be drawn on the
regions \(G_1,\dots,G_r \in \xi_r\). In other words

\[
\xi^{-1}(\xi_r) = \bigcup_{T_1 \in \mathcal{T}(G_1)} \dots \bigcup_{T_r \in \mathcal{T}(G_r)} \set{(T_k, s_k)}_{k=1}^r
\] where \(\set{(T_k, s_k)}_{k=1}^r\) is a forest plan.

\begin{proof}
This is a straightforward application of Lemma \ref{lem-RegionFuncPreImg}. Since $\xi(\cdot)$ is just the application of the induced region function \ref{def-RegionFunction} to each region tree in a forest plan then it follows by property of function pre-images

\begin{align*}
\xi^{-1}(\xi_r) &= \set{\IRegionInv{G_k, s_k}}_{k=1}^r \\
&= \bigcup_{T_1 \in \mathcal{T}(G_1)} \dots \bigcup_{T_r \in \mathcal{T}(G_r)} \set{(T_k, s_k)}_{k=1}^r
\end{align*}
\end{proof}

\end{lemma}

\begin{corollary}[Induced Plan Preimage
Size]\protect\hypertarget{cor-iplan-preim-size}{}\label{cor-iplan-preim-size}

Note that Lemma~\ref{lem-induced-plan-func-preim} immediately tells us
that the size of \(\xi^{-1}(\xi_r)\) is equal to the product of the
number of spanning trees that can be drawn on each region. In other
words

\[
\abs{\xi^{-1}(\xi_r)} = \prod_{k=1}^r \tau(G_k) = \tau(\xi_r)
\]

Moreover, the size of set of all \(r\)-region forest plans
\(\mathcal{F}_r\) is equal to sum of the number of spanning trees that
can be drawn on all \(r\)-region graph plans, i.e.,

\[
\abs{\mathcal{F}_r} = \sum_{\xi_r \in \mathcal{P}_r} \tau(\xi_r).
\]

\end{corollary}

We now define a new, modified sequence of target distributions
\(\tilde{\pi}_1, \dots, \tilde{\pi}_D\) on the space of forest plans. We
will define it in such a way as to ensure that the pushforward density
of the forest plans under \(\xi(\cdot)\) is exactly our original target
distribution \(\pi_r(\cdot)\).

\begin{definition}[Forest Space Target
Distributions]\protect\hypertarget{def-forest-target}{}\label{def-forest-target}

Given a sequence of target distributions \(\pi_1,\dots,\pi_D\) we define
our associated forest space unnormalized densities
\(\tilde{\gamma}_1, \dots, \tilde{\gamma}_D\) as the unnormalized
density of the induced graph plan divided by the number of spanning
trees which can be drawn on the induced graph plan, so

\[
\tilde{\gamma}_r(F_r) = \gamma_r(\xi(F_r)) \prod_{k=1}^r
\frac{1}{\tau(\mathrm{G}(T_{k}))} = \gamma_r(\xi(F_r)) \frac{1}{\tau(\xi(F_r))}.
\]

We thus define our associated target distributions
\(\tilde{\pi}_1, \dots, \tilde{\pi}_D\) as

\begin{equation}\phantomsection\label{eq-forest-target}{
\tilde{\pi}_r(F_r) = \frac{\tilde{\gamma}_r(F_r)}{\tilde{Z}_r}
}\end{equation}

Where \(\tilde{Z}_r\) is the normalizing constant. We will soon prove
that this normalizing constant \(\tilde{Z}_r\) is actually the same as
the graph space normalizing constant \(Z_r\).

\end{definition}

\begin{lemma}[Pushforward Forest Space Target
Distributions]\protect\hypertarget{lem-forest-pushforward-target}{}\label{lem-forest-pushforward-target}

If we have forest plans \(\set{F_r}\) distributed according to
\(\tilde{\pi}_r\) then the pushforward measure of the associated graph
space plans under \(\xi(\cdot)\) is distributed according to \(\pi_r\).
Moreover, the normalizing constant \(\tilde{Z}_r\) of \(\tilde{\pi}_r\)
is equivalent to the normalizing constant \(Z_r\) of \(\pi_r\).

\begin{proof} 

We begin by proving that $\tilde{Z}_r = Z_r$. Recall that 

$$
Z_r = \sum_{\xi_r \in \mathcal{P}_r} \gamma_r(\xi_r)
$$
and 
$$
\tilde{Z}_r = \sum_{F_r \in \mathcal{F}_r} \tilde{\gamma}_r(F_r)
$$

Now recall that by definition we know that $\mathcal{F}_r = \xi^{-1}(\mathcal{P}_r)$ so the sum can be equivalently expressed as a double sum over first, each graph plan $\xi_r \in \mathcal{P}_r$, and second, over each forest plan $F_r \in \xi^{-1}(\xi_r)$, so we can express the sum as 
$$
\begin{aligned}
\tilde{Z}_r
&= \sum_{\xi_r \in \mathcal{P}_r} \sum_{F_r \in \xi^{-1}(\xi_r)} \tilde{\gamma}_r(F_r) 
\end{aligned}
$$

Now, leveraging the fact that if $F_r \in \xi^{-1}(\xi_r)$ then $\xi(F_r) = \xi_r$ and $\tilde{\gamma}_r(F_r)$ can be expressed solely as a function of $\xi(F_r)$ we have
$$
\begin{aligned}
\tilde{Z}_r
&= \sum_{\xi_r \in \mathcal{P}_r} \sum_{F_r \in \xi^{-1}(\xi_r)} \tilde{\gamma}_r(F_r) \\
&= \sum_{\xi_r \in \mathcal{P}_r} \sum_{F_r \in \xi^{-1}(\xi_r)} \gamma_r(\xi(F_r)) 
\frac{1}{\tau(\xi(F_r))}  \\
&= \sum_{\xi_r \in \mathcal{P}_r} \sum_{F_r \in \xi^{-1}(\xi_r)} \gamma_r(\xi_r)  
\frac{1}{\tau(\xi_r)}  &&\text{ because } F_r \in \xi^{-1}(\xi_r) \\
&= \sum_{\xi_r \in \mathcal{P}_r} \gamma_r(\xi_r)  
\frac{1}{\tau(\xi_r)} \sum_{F_r \in \xi^{-1}(\xi_r)} 1  
\end{aligned}
$$
Now recall from Corollary \ref{cor-iplan-preim-size} that $\abs{\xi^{-1}(\xi_r)} = \tau(\xi_r)$ so we see $\frac{1}{\tau(\xi_r)}$ term cancels and we are left with 

$$
\begin{aligned}
\tilde{Z}_r
&= \sum_{\xi_r \in \mathcal{P}_r} \gamma_r(\xi_r)  
\frac{1}{\tau(\xi_r)} \sum_{F_r \in \xi^{-1}(\xi_r)} 1   \\
&= \sum_{\xi_r \in \mathcal{P}_r} \gamma_r(\xi_r)  
\frac{1}{\tau(\xi_r)} \tau(\xi_r) \\
&= \sum_{\xi_r \in \mathcal{P}_r} \gamma_r(\xi_r)  \\
&= Z_r
\end{aligned}
$$

We now prove the pushforward of $\tilde{\pi}_r$ under $\xi(\cdot)$ is exactly $\pi_r$. Assume $F_r \sim \tilde{\pi}_r$ and consider 
$$
\begin{aligned}
\mathrm{Pr}_{\tilde{\pi}_r }(\xi(F_r) = \xi_r) &= \sum_{\tilde{F}_r \in \xi^{-1}(\xi_r)} \tilde{\pi}_r(\tilde{F}_r) \\
&= \sum_{T_1 \in \mathcal{T}(G_1)} \dots \sum_{T_r \in \mathcal{T}(G_r)} \tilde{\pi}_r(\set{(T_k, s_k)}_{k=1}^r) \; \; \text{by \ref{lem-induced-plan-func-preim}} 
\end{aligned}
$$

Notice that $\tilde{\pi}_r$ only depends on $F_r$ through its induced graph plan $\xi(F_r)$
That means $\tilde{\pi}_r$ is the same for any two $F_r, \tilde{F}_r$ that map to the same plan (i.e., $\xi(F_r) = \xi(\tilde{F}_{r})$), so since we are summing over forest plans in the preimage of $\xi_r$ we see every sum in that term is equal and by Corollary \ref{cor-iplan-preim-size} we know that simplifies to 
$$
\begin{aligned}
\mathrm{Pr}_{\tilde{\pi}_r }(\xi(F_r) = \xi_r) &= \frac{1}{Z_r} \tilde{\gamma}_r(F_r)  \tau(\xi(F_r)) \\
&= \frac{1}{Z_r} \gamma_r(\xi(F_r)) 
\frac{1}{\tau(\xi(F_r))} \tau(\xi(F_r)) \\
&= \frac{1}{Z_r} \gamma_r(\xi(F_r))  \\
&= \frac{1}{Z_r} \gamma_r(\xi_r) \\
&= \pi_r(\xi_r)
\end{aligned}
$$
So we see the pushforward measure of the forest plans to the space of plans is exactly the target distribution in Equation \ref{eq-target}.
\end{proof}

\end{lemma}

To sample forest plans from \(\tilde{\pi}_r\) we use the same SMC
sampler algorithm as in Algorithm \ref{alg-naive-k-split} with some
slight modifications to the splitting procedure and weight calculations.
For splitting, we still have the same parameters \(\varphi\) and a
splitting schedule \(\mathcal{S}(\cdot, \cdot)\), except now instead of
selecting a splitting parameter \(\mathcal{K} \in \mathbb{N}\), we must
specify a tree cut selection probability
\(p_\text{cut}( \cdot | \cdot)\) which given a region tree
\((T, s_\ell)\) selects some tree cut \(\TreeCut{e}\) with probability
\(p_\text{cut}(\TreeCut{e}|(T, s_\ell), \mathcal{S})\). (Going forward
for notational simplicity we will suppress the reference to \(s_\ell\)
and \(\mathcal{S}\) but it is always implicitly there). We also require
that \(p_\text{cut}\) does not render any forest plans unsplittable but
this will always be satisfied it some positive probability is always put
on all balanced tree cuts. Further, we require for \(\varphi\) that the
multidistrict tree chosen must only be a function of the induced plan,
not just the forest plan itself. In other words,
\(\varphi(\cdot|\xi(F_{r-1}))\), not \(\varphi(\cdot|F_{r-1})\). For
notational convenience we denote this \(\varphi(G_\ell|\xi(F_{r-1}))\)
where \(G_\ell\) is the subgraph induced by the chosen multidistrict
tree's vertex set. For example, a balanced tree cut could be selected
uniform over the set of balanced tree cuts (if they exist), i.e.,
\(p_\text{cut}(\TreeCut{e}|T) \propto \abs{\mathrm{ok}(T_\ell,  \mathcal{S}(r-1, s_\ell))}\).

\begin{algorithm}[H]
\begin{algorithmic}[1]
\REQUIRE{Splitting schedule $\mathcal{S}(\cdot, \cdot)$, region tree $(T_{\ell}, s_\ell) \in F_{r-1}$, and a tree cut selection rule $p_\text{cut}( \cdot | \cdot)$}
\STATE{Draw spanning tree $T^\ast$ on $\IGraph{T_{\ell}}$ using Wilson's algorithm}
\STATE{Let \texttt{TreeCuts} be a list of tree cuts}
\FOR{each tree cut $\TreeCut{e} \in \AllTreeCuts{T^\ast}{\mathcal{S}(r-1, s)}$}
\STATE{\texttt{TreeCuts.append($\{(T^{e_i}_{k}, s_{k}), (T^{e_i}_{k'}, s_{k'}), e_i\}$)}}
\ENDFOR
\STATE{Choose a tree cut $t = \TreeCut{e}$ distributed $p_\text{cut}( \cdot | T^\ast)$}
\STATE{Discard the edge from $t$ save the two cut trees as the two new region trees}
\end{algorithmic}
\caption{Forest Space Splitting Algorithm}
\label{alg-forest-split}
\end{algorithm}

Algorithm \ref{alg-forest-split} provides the pseudo-code for the
procedure. Notice, we still use Wilson's algorithm to draw an entirely
new spanning tree \(T^\ast\) on the vertex set of \(T_\ell\). The only
significant difference between this and \ref{alg-naive-k-split} is the
use of an arbitrary tree cut selection rule \(p_\text{cut}( \cdot | T)\)
and whereas before we didn't save the tree and instead only cared about
the underlying region, here we stop and keep the specific trees in the
tree cut, discarding only the edge removed. The significance of this
move is since we no longer need to integrate the probability of
splitting a tree over all possible trees that could be drawn on a region
we can use \(p_\text{cut}( \cdot | T)\) instead of the top
\(\mathcal{K}\) method. \(p_\text{cut}( \cdot | T,)\) can be specific so
that whenever there is at least one balanced tree cut it will always be
selected. This dramatically boosts the efficiency of splitting over the
graph space forward kernel as we no longer have to ``waste'' draws when
the number of balanced cuts is less than \(\mathcal{K}_r\) and we don't
select a balanced cut. We formally present the splitting procedure and
the derivation of its sampling probability below. We continue to denote
the sampling probability with \(q(\cdot)\).

\begin{definition}[]\protect\hypertarget{def-arb-tree-cut-kernel}{}\label{def-arb-tree-cut-kernel}

Let \(p_\text{cut}\) be a distribution over tree cuts. We will use the
following notation to indicate a different quantity depending on
context. First, \(p_\text{cut}( \TreeCut{e} \mid T_\ell)\) denotes the
probability we selected that specific tree cut given the region tree
\((T_\ell, s_\ell)\) and \(\mathcal{S}\). In contrast,
\(p_{\text{cut}}(T^\ast, \mathcal{S})\) denotes the random object which
takes a spanning tree \(T^\ast\) and a splitting schedule
\(\mathcal{S}\) and returns a tree cut of \(T^\ast\). The relationship
between the two is as follows, \[
\text{Pr}(p_{\text{cut}}(T^\ast, \mathcal{S}(r-1, s_\ell)) = \TreeCut{e} \mid \Wilson{\IGraph{T_\ell}} = T^\ast) =  p_\text{cut}( \TreeCut{e} \mid T_\ell).
\]

\end{definition}

\begin{definition}[Forest Space
Kernel]\protect\hypertarget{def-new-split-kernel}{}\label{def-new-split-kernel}

Given a multidistrict tree \((T_\ell, s_\ell)\) and a distribution over
tree cuts \(p_\text{cut}( \cdot \mid \cdot)\), the forest space
splitting procedure outlined in Algorithm \ref{alg-forest-split} can be
formally written as the composition of the Wilson function
(Definition~\ref{def-WilsonFunction}) on the subgraph induced by the
vertex set of \(T_\ell\), the tree cut distribution
\(p_\text{cut}( \cdot \mid \cdot)\), and the induced region tree
function (Definition~\ref{def-ITree-cut-func}). Altogether, it can be
written as \[
\ITree{\ArbKernel{\Wilson{\IGraph{T_{\ell}}}, \mathcal{S}(r-1, s_\ell)}} = \set{(T_k, s_k), (T_{k'}, s_{k'})}
\]

Note that all the randomness comes from the \(\Wilson{\cdot}\) and
\(p_\text{cut}( \cdot)\) terms. Further note that since we apply
Wilson's algorithm to the subgraph induced by the vertex set of
\(T_{\ell}\) the spanning tree it returns is still sampled with
probability \(\frac{1}{\tau(\IGraph{T_\ell})}\).

\end{definition}

\begin{definition}[Effective Region Tree Boundary
Length]\protect\hypertarget{def-eff-boundary-len}{}\label{def-eff-boundary-len}

Given a distribution over tree cuts \(p_\text{cut}( \cdot \mid \cdot)\)
and two adjacent region trees \((T_k, s_k), (T_{k'}, s_{k'})\), we
define \(\EffB{T_k, T_{k'}}\), the effective region tree boundary
length, as the following sum, \[
\EffB{T_k, T_{k'}} = \sum_{e \in \mathcal{C}(T_{k}, T_{k'})} p_\text{cut}(\TreeCut{e} | T_{k} \cup \set{e} \cup T_{k'}).
\]

\end{definition}

\begin{proposition}[Forest Splitting
probability]\protect\hypertarget{prp-split-prob-forest}{}\label{prp-split-prob-forest}

Let \((T_{\ell}, s_{\ell})\) be a multidistrict tree and let
\((T_{k}, s_k)\) and \((T_{k'}, s_{k'})\) be the two newly split
balanced regions resulting from Algorithm \ref{alg-forest-split}. Given
our tree cut distribution \(p_\text{cut}( \cdot \mid \cdot)\) the
probability of splitting the two new region trees given the old one is
\[
     q\left(T_{k}, T_{k'} \mid T_{\ell}\right)= \frac{1}{\tau(\IGraph{T_{\ell}})} \cdot \EffB{T_k, T_{k'}}.
\] where
\(\sum_{e \in \mathcal{C}(T_{k}, T_{k'})} p_\text{cut}(\TreeCut{e} \mid T_{k} \cup \set{e} \cup T_{k'})\)
represents the probability the tree cut \(\TreeCut{e}\) formed by
combining \((T_{k}, s_k)\) and \((T_{k'}, s_{k'})\) with \(e\) would be
selected.

\begin{proof}
The proof of this is very similar to the derivation of the splitting probability in \ref{prp-split-prob-graph}. The only difference is instead of the top $\mathcal{K}$ kernel we have an arbitrary $p_\text{cut}$ and we do not apply the induced region function at the end as we keep the sampled trees. 

Once again we begin by writing the event we sample the new region trees given the old in terms of our special functions 
\begin{align*}
    q\left(T_{k}, T_{k'} \mid T_\ell\right) &= q \left( \ITree{\ArbKernel{\Wilson{\IGraph{T_{\ell}}}, \mathcal{S}(r-1, s_\ell)}} = \set{(T_k, s_k), (T_{k'}, s_{k'})} \mid T_\ell\right).
\end{align*}
The details are below but the high level derivation is 
\begin{align*}
q\left(T_{k}, T_{k'} \mid H_\ell\right) &= q \left( \ITree{\ArbKernel{\Wilson{\IGraph{T_{\ell}}}, \mathcal{S}(r-1, s_\ell) }} = \set{(T_k, s_k), (T_{k'}, s_{k'})} \mid T_\ell\right)\\
&= \E{q \left( \ITree{\ArbKernel{\Wilson{\IGraph{T_{\ell}}}, \mathcal{S}(r-1, s_\ell)}} = \set{(T_k, s_k), (T_{k'}, s_{k'})} \mid \Wilson{\IGraph{T_\ell}} = T^\ast, T_\ell \right) \mid T_\ell } \\
&= \E{ \sum_{e \in \mathcal{C}(T_k, T_{k'})} p_{\text{cut}}(\TreeCut{e}\mid T^\ast)  \I{T^\ast = T_{k} \cup \set{e} \cup T_{k'}}\mid T_\ell} \\
&=  \sum_{e \in \mathcal{C}(T_k, T_{k'})} p_{\text{cut}}(\TreeCut{e}\mid T_{k} \cup \set{e} \cup T_{k'})  \E{\I{T^\ast = T_{k} \cup \set{e} \cup T_{k'}}\mid T_\ell}\\
&=  \sum_{e \in \mathcal{C}(T_k, T_{k'})} p_{\text{cut}}(\TreeCut{e}\mid T_{k} \cup \set{e} \cup T_{k'})  q(T^\ast = T_{k} \cup \set{e} \cup T_{k'}\mid T_\ell) \\
&=  \sum_{e \in \mathcal{C}(T_k, T_{k'})} p_{\text{cut}}(\TreeCut{e}|T_{k} \cup \set{e} \cup T_{k'})  \frac{1}{\tau(\IGraph{T_\ell})} \\
&=  \frac{1}{\tau(\IGraph{T_\ell})} \sum_{e \in \mathcal{C}(T_k, T_{k'})} p_{\text{cut}}(\TreeCut{e} \mid T_{k} \cup \set{e} \cup T_{k'})  
\end{align*}

The more detailed derivation is as follows: for a given $T^\ast \in \mathcal{T}(\IGraph{T_\ell})$ consider 
\begin{align*}
&q \left( \ITree{\ArbKernel{T^\ast, \mathcal{S}(r-1, s_\ell) }} = \set{(T_k, s_k), (T_{k'}, s_{k'})} \mid T_\ell, \Wilson{\IGraph{T_\ell}}=T^\ast \right) \\
&= q\left( \ArbKernel{T^\ast, \mathcal{S}(r-1, s_\ell) } \in \ITreeInv{\set{(T_k, s_k), (T_{k'}, s_{k'})}} \mid T_\ell, \Wilson{\IGraph{T_\ell}}=T^\ast \right). \\
\end{align*}
Now, by Lemma \ref{lem-TreeFuncPreImg} we have
\begin{align*}
    \ITreeInv{\set{(T_k, s_k), (T_{k'},s_{k'})}} = \bigcup_{e \in \mathcal{C}(T_k, T_{k'})} \TreeCut{e}
\end{align*}
Thus, we see 

\begin{align*}
&q \left( \ITree{\ArbKernel{T^\ast, \mathcal{S}(r-1, s_\ell) }} = \set{(T_k, s_k), (T_{k'}, s_{k'})} \mid T_\ell, \Wilson{\IGraph{T_\ell}}=T^\ast \right) \\
&= q\left( \ArbKernel{T^\ast, \mathcal{S}(r-1, s_\ell) } \in \bigcup_{e \in \mathcal{C}(T_k, T_{k'})} \TreeCut{e} \mid T_\ell, \Wilson{\IGraph{T_\ell}}=T^\ast \right)  \\
&= \sum_{e \in \mathcal{C}(T_k, T_{k'})} q\left( p_{\text{cut}}(T^\ast = \TreeCut{e} \mid T_\ell, \Wilson{\IGraph{T_\ell}}=T^\ast \right) \I{T^\ast = T_{k} \cup \set{e} \cup T_{k'}} \\
&= \sum_{e \in \mathcal{C}(T_k, T_{k'})} p_{\text{cut}}(\TreeCut{e}|T^\ast) \I{T^\ast = T_{k} \cup \set{e} \cup T_{k'}}
\end{align*}

Now when we take the expectation of the expression above with respect to $\Wilson{\IGraph{T_\ell}}$
Since we sample trees uniformly at random, the expectation of the indicator becomes $\frac{1}{\tau(\IGraph{T_\ell})}$. 
\end{proof}

\end{proposition}

\begin{corollary}[Forest Space Forward
Kernel]\protect\hypertarget{cor-forest-forward-kernel}{}\label{cor-forest-forward-kernel}

Let \(\varphi(\cdot|\xi(F_r))\) be a distribution over multidistrict
trees in \(F_{r-1}\) that only depends on the induced plan \(\xi(F_r)\).
Let \(F_{r-1}\) be a forest plan. If \(F_r\) is a forest plan such that
there exists some \(T_{\ell} \in \xi_{r-1}\),
\(T_{k}, T_{k'} \in \xi_r\) where
\(\IGraph{T_{\ell}} = \IGraph{T_{k}} \cup \IGraph{T_{k'}}\) then the
forward kernel probability is \begin{align*}
    M_r(F_r \mid F_{r-1}) = \varphi(G_{\ell} \mid \xi_{r-1}) \cdot \frac{1}{\tau(G_\ell)} \cdot \EffB{T_k, T_{k'}}
\end{align*} Where \((G_{\ell}, s_\ell) = \IRegion{T_{\ell}, s_\ell}\)
and \(\xi_{r-1} = \xi(F_{r-1})\)

\begin{proof} 
We apply the same proof using the law of total probability from \ref{cor-graph-forward-kernel}, we just use Proposition \ref{prp-split-prob-forest} instead. For notational convenience we replace $\tau(\IGraph{T_\ell})$ with $\tau(G_\ell)$ and write $\varphi$ as $\varphi(G_{\ell} \mid \xi(F_{r-1}))$. 
\end{proof}

\end{corollary}

We can also prove an analogous result to
Corollary~\ref{cor-graph-plan-splittability} about splittability when
the only hard constraints are connectedness and population balance.

\begin{corollary}[Splittability of Forest Space Forward
Kernel]\protect\hypertarget{cor-forest-plan-splittability}{}\label{cor-forest-plan-splittability}

If the only hard constraints are imposed through the score function
\(J\) are population balance and connectedness then under the forward
kernel in Corollary~\ref{cor-forest-forward-kernel}, all forest plans
\(F_r\) in the support of \(\tilde{\pi}_r\) are splittable for
\(r=2,\dots, D\).

\begin{proof} The proof is analogous to Corollary \ref{cor-graph-plan-splittability}. We note that for any $F_r$ where $\tilde{\pi}_r(F_r) > 0$ we can find a $F_{r-1}$ such that $\tilde{\pi}_r(F_{r-1}) M_r(F_{r}|F_{r-1}) > 0$ by simply taking any two adjacent region trees $T_k, T_{k'} \in F_r$ and merging them together to create $F_{r-1}$. 
\end{proof}

\end{corollary}

The optimal weights for forest plans look very similar to the weights in
Equation~\ref{eq-graph-optimal-weights} as well. The main difference in
their derivation is, unlike for graph space plans where there was a
one-to-one correspondence between adjacent regions and previous plans
that could have been split to form the current plan, for forest plans
the function is many-to-one meaning for each pair of adjacent region
trees \(T_k \sim T_{k'}\) there are \(\tau(\IGraph{T_k \cup T_{k'}})\)
1-region ancestors that share all trees but \(T_k, T_{k'}\). We will now
make this more rigorous.

\begin{definition}[Forest Ancestry
Definitions]\protect\hypertarget{def-forest-ancestors}{}\label{def-forest-ancestors}

For all the \(L\)-region ancestor (Definition~\ref{def-LRegionAncestor},
Definition~\ref{def-SetOfLDistrictAncestors}) and new, common, and old
region definitions (Definition~\ref{def-new-old-common-regions}) for
graph space plans we define the analogous forest space definitions to be
the versions replacing regions with region trees.

\begin{refremark}[Region Tree Ancestor Decomposition]
Notice by definition for \(F_{r-1} \in \mathcal{A}_1(F_r)\) we can still
decompose \(F_r\) and \(F_{r-1}\) analogously to
Remark~\ref{rem-RegionAncestorDecomp} as follows \begin{align*}
    F_r = \mathrm{NR}(F_{r-1}, F_r) \sqcup \mathrm{CR}(F_{r-1}, F_r) && F_{r-1} = \mathrm{OR}(F_{r-1}, F_r) \sqcup \mathrm{CR}(F_{r-1}, F_r)
\end{align*}

\label{rem-forest-decomp}

\end{refremark}

We now prove similar results to Lemma~\ref{lem-OldRegionChar} and
Proposition~\ref{prp-AncestorSetCharacterization}.

\begin{lemma}[Old Region Tree
Characterization]\protect\hypertarget{lem-forest-OldRegionChar}{}\label{lem-forest-OldRegionChar}

Let \(r > 1\) and let \(F_r\) be an \(r\)-region forest plan. Let
\(F_{r-1} \in \mathcal{A}_1(F_r)\) be a 1-region tree ancestor. Further,
let \(T_{\ell}\) denote the old region tree and \(T_{k}, T_{k'}\) denote
the new region trees, i.e.,: \begin{align*}
    \abs{\mathrm{OR}(F_{r-1}, F_r)} = \set{T_{\ell}} && \abs{\mathrm{NR}(F_{r-1}, F_r)} = \set{T_k, T_{k'}}
\end{align*} Then the subgraph induced by the old region tree
\(T_{\ell}\) is equal to the subgraph induced by the union of the
vertices in the two new region trees \(T_{k}, T_{k'}\). In other words:
\begin{align*}
    \IGraph{V_{\ell}} = \mathrm{G}(V_{k} \sqcup V_{k'})
\end{align*}

\begin{proof} Recall that we can decompose $F_{r-1}$ and $F_r$ as follows:
\begin{align*}
    F_r = \mathrm{NR}(F_{r-1}, F_r) \sqcup \mathrm{CR}(F_{r-1}, F_r) && F_{r-1} = \mathrm{OR}(F_{r-1}, F_r) \sqcup \mathrm{CR}(F_{r-1}, F_r)
    \end{align*}
Since we know $\mathrm{V}(F_r) = \mathrm{V}(F_{r-1})$ this implies that the set of vertices in $T_{k}$ and $T_{k'}$ is equal to the vertices in $T_{\ell}$. To formally show this consider 
\begin{align*}
    \mathrm{V}(F_r) &= \mathrm{V}(F_{r-1}) \\
    \mathrm{V}(\mathrm{NR}(F_{r-1}, F_r)) \sqcup \mathrm{V}(\mathrm{CR}(F{r-1}, F_r)) &= \mathrm{V}(\mathrm{OR}(F_{r-1}, F_r)) \sqcup \mathrm{V}(\mathrm{CR}(F_{r-1}, F_r)) \implies \\
    \mathrm{V}(\mathrm{NR}(F_{r-1}, F_r)) &= \mathrm{V}(\mathrm{OR}(F_{r-1}, F_r)) \\
    V_{k} \sqcup V_{k'} &= V_{\ell} 
\end{align*}

Since we know that $\IGraph{T_{\ell}} = \mathrm{G}(V_{\ell})$ by definition then that tells us that $H_{\ell} = \mathrm{G}(V_{k} \sqcup V_{k'})$

\end{proof}

\end{lemma}

\begin{proposition}[Forest Ancestor Set
Equivalency]\protect\hypertarget{prp-forest-AncestorSetCharacterization}{}\label{prp-forest-AncestorSetCharacterization}

Fix \(r > 1\) and let \(F_r\) be a \(r\)-region plan. Then the set of
\(1\)-region ancestors \(\mathcal{A}_1(F_r)\) is in bijective
correspondence with the set of region trees formed by the set of all
trees that can be drawn on the vertex set of all pairs of merged
adjacent trees in \(F_r\) so \[
\mathcal{A}_1(F_r) = \bigcup_{T_k \sim T_{k'} \in F_r} \bigcup_{T_\ell \in \mathcal{T}(\IGraph{T_k \cup T_{k'}})} (T_\ell, s_k + s_{k'}).
\]

Where \(T_k \sim T_{k'} \in F_r\) denotes the union of all pairs of
adjacent region trees in \(F_r\).

The bijection is the function that maps a 1-region tree ancestors
\(F_{r-1} \in \mathcal{A}_1(F_r)\) to the old region
\(\mathrm{OR}(F_{r-1}, F_r)\).

\begin{proof}
Define the following functions 
\begin{align*}
f: \mathcal{A}_1(F_r) \longrightarrow \bigcup_{T_k \sim T_{k'} \in F_r} \bigcup_{T_\ell \in \mathcal{T}(\IGraph{T_k \cup T_{k'}})} (T_\ell, s_k + s_{k'}) \\
f(F_{r-1}) = \mathrm{OR}(F_{r-1},F_r)
\end{align*}
\item Part 1. $f$ is injective \\
Let $F_{r-1}, \tilde{F}_{r-1} \in \mathcal{A}_1(F_r)$ such that $f(F_{r-1}) = f(\tilde{F}_{r-1}) = (T_{\ell}, s_\ell)$  \\
Consider that since $F_{r-1}, \tilde{F}_{r-1} \in \mathcal{A}_1(F_r)$ we know by \ref{rem-forest-decomp} that 
\begin{align*}
    F_r &= \mathrm{NR}(F_{r-1}, F_r) \sqcup \mathrm{CR}(F_{r-1}, F_r) = \mathrm{NR}(\tilde{F}_{r-1}, F_r) \sqcup \mathrm{CR}(\tilde{F}_{r-1}, F_r) \\
    F_{r-1} &= \mathrm{OR}(F_{r-1}, F_r) \sqcup \mathrm{CR}(F_{r-1}, F_r) \\
    \tilde{F}_{r-1} &= \mathrm{OR}(\tilde{F}_{r-1}, F_r) \sqcup \mathrm{CR}(\tilde{F}_{r-1}, F_r)
\end{align*}
By definition of $f$ we have $f(F_{r-1}) = \mathrm{OR}(F_{r-1}, F_r)$ and $f(\tilde{F}_{r-1}) = \mathrm{OR}(\tilde{F}_{r-1}, F_r)$ so we see $f(F_{r-1}) = f(\tilde{F}_{r-1})$. That implies 
\begin{align*}
    \tilde{F}_{r-1} = \mathrm{OR}(\tilde{F}_{r-1}, F_r) \sqcup \mathrm{CR}(\tilde{F}_{r-1}, F_r) = f(F_{r-1}) \sqcup \mathrm{CR}(\tilde{F}_{r-1}, F_r)
\end{align*}
Now recall that since both $\tilde{F}_{r-1}, F_{r-1} \in \mathcal{A}_1(\xi_r)$ that means they both must share $r-2$ region trees in common with $F_r$. Since we know they both have the same old region tree we know by \ref{lem-forest-OldRegionChar} that the two new region trees in $F_r$ must be made from the vertex set of the shared old region. That necessarily implies that the $r-2$ other region trees in $F_r$ must be the common trees for both $\tilde{F}_{r-1}, F_{r-1} $ and thus we have 
\begin{align*}
 \mathrm{CR}(\tilde{F}_{r-1}, F_r) =  \mathrm{CR}(F_{r-1}, F_r)
\end{align*}
That in turn implies 
\begin{align*}
    \tilde{F}_{r-1} &= f(F_{r-1}) \sqcup \mathrm{CR}(\tilde{F}_{r-1}, F_r) \\
    &= f(F_{r-1}) \sqcup \mathrm{CR}(F_{r-1}, F_r) \\
    &= F_{r-1}
\end{align*}

\item Part 2. $f$ is surjective \\
Let $(T_\ell, s_\ell)$ be in the image of $f$. Define an $r-1$-region forest plan $F_{r-1}$ as the $r-2$ region trees in $F_r$ that do not correspond to the vertex set $\mathrm{V}(T_\ell)$ and the region $(T_\ell, s_\ell)$. This is an $r-1$ region forest plan because by construction $(T_\ell, s_\ell)$ is formed such that it is disjoint from the $r-2$ other regions in $\xi_r$. Moreover since all but 1 of $F_{r-1}$'s regions are shared with $F_r$ so we see it is actually a 1-region tree ancestor. So by construction we have found $F_{r-1} \in \mathcal{A}_1(F_r)$ such that $f(F_{r-1}) = (T_\ell, s_\ell)$ and thus $f$ is surjective.
\end{proof}

\end{proposition}

\begin{lemma}[]\protect\hypertarget{lem-forest-forward-equivalency}{}\label{lem-forest-forward-equivalency}

Let \(F_{r-1}, \tilde{F}_{r-1}, F_r\) be forest plans such that
\(F_{r-1}, \tilde{F}_{r-1} \in \mathcal{A}_1(F_r)\) and
\(F_{r-1}, \tilde{F}_{r-1}\) have the same common regions with \(F_r\),
in other words

\[
\mathrm{CR}(\tilde{F}_{r-1}, F_r) = \mathrm{CR}(F_{r-1}, F_r)
\] Then their forward kernel probabilities
\(M_r(\tilde{F}_{r-1}| F_r)\), \(M_r(F_{r-1}| F_r)\) are equal.

\begin{proof}
Define $\tilde{T}_\ell = \mathrm{OR}(\tilde{F}_{r-1}, F_r)$ and $T_\ell = \mathrm{OR}(F_{r-1}, F_r)$ respectively. 
First note that $\mathrm{CR}(\tilde{F}_{r-1}, F_r) = \mathrm{CR}(F_{r-1}, F_r)$ necessarily implies that $\mathrm{V}(\tilde{T}_\ell) = \mathrm{V}(T_\ell)$ and thus $\IRegion{\tilde{T}_\ell} = \IRegion{T_\ell} = G_\ell$. 

Now lets consider the forward kernel expressions
\begin{align*}
    M_r(F_r \mid F_{r-1}) &= \varphi(G_{\ell} \mid \xi(F_{r-1})) \cdot \frac{1}{\tau(\IGraph{T_{\ell}})} \cdot \EffB{T_k, T_{k'}} \\
    M_r(F_r \mid \tilde{F}_{r-1}) &= \varphi(G_{\ell} \mid \xi(\tilde{F}_{r-1})) \cdot \frac{1}{\tau(\IGraph{\tilde{T}_{\ell}})} \cdot \EffB{T_k, T_{k'}}
\end{align*}
Notice that since $\mathrm{CR}(\tilde{F}_{r-1}, F_r) = \mathrm{CR}(F_{r-1}, F_r)$ and $\mathrm{V}(\tilde{T}_\ell) = \mathrm{V}(T_\ell)$ then we know that $\xi(F_{r-1}) = \xi(\tilde{F}_{r-1})$. We further know that since $\IRegion{\tilde{T}_\ell} = \IRegion{T_\ell} = G_\ell$ then $\tau(\IGraph{\tilde{T}_{\ell}}) = \tau(\IGraph{\tilde{T}_{\ell}})$. Thus we have 

\begin{align*}
    M_r(F_r \mid \tilde{F}_{r-1}) &= \varphi(G_{\ell} \mid \xi(\tilde{F}_{r-1})) \cdot \frac{1}{\tau(\IGraph{\tilde{T}_{\ell}})} \cdot \EffB{T_k, T_{k'}} \\
    &= \varphi(G_{\ell} \mid \xi(F_{r-1})) \cdot \frac{1}{\tau(\IGraph{T_{\ell}})} \cdot \EffB{T_k, T_{k'}} \\
    &=  M_r(F_r \mid \tilde{F}_{r-1})
\end{align*}
\end{proof}

\end{lemma}

\begin{theorem}[Marginal Forest Proposal
Density]\protect\hypertarget{thm-marg-forest-dens}{}\label{thm-marg-forest-dens}

Given a forest space forward kernel \(M_r\) and a target distribution
\(\tilde{\pi}_r\), the marginal proposal density is given by, \[
f(F_r) = \frac{\gamma_{r}(\xi_r) }{ Z_{r-1}}  \sum_{T_k \sim T_{k'} \in F_r}  \tau(\mathrm{CR}(\xi_r, \tilde{\xi}_{r-1}))^{-1} \frac{\exp{-J(\tilde{\xi}_{r-1})}}{\exp{-J(\xi_r)}}\frac{ \tau(H_\ell)^{\rho-1}}{\tau(G_k)^{\rho} \tau(G_{k'})^{\rho}} \cdot \varphi(G_{\ell} \mid \tilde{\xi}_{r-1}) \cdot \EffB{T_k, T_{k'}} ,
\] where \(\xi_r = \xi(F_r)\), \((G_{k}, s_k) = \IRegion{T_k, s_k}\),
\((G_{k'}, s_{k'}) = \IRegion{T_{k'}, s_{k'}}\) and
\(\tilde{\xi}_{r-1}\) is the plan formed by replacing the adjacent
regions \(G_{k}\) and \(G_{k'}\) in \(\xi_r\) with the merged region
\(H_{\ell} = G_k \cup G_{k'}\).

\begin{proof} 
This proof is very similar to the proof of Theorem \ref{thm-marg-dens}, we just need to adjust for the difference in ancestor sizes. Again, consider the integral needed to derive a closed form of the marginal proposal density 
$$
f_r(F_r) = \int \tilde{\pi}_{r-1}(F_{r-1}) M_r(F_r \mid F_{r-1}) \; dF_{r-1}.
$$

Similar to \ref{thm-marg-dens}, the general idea of the proof is that we know from \ref{cor-forest-forward-kernel} that $M_r(F_r \mid F_{r-1})$ is only non-zero when $F_{r-1}$ is a 1-region tree ancestor of $F_r$ so we see that the integral actually reduces to a sum over all $F_{r-1} \in \mathcal{A}_1(F_r)$ so

$$
f_r(F_r) = \sum_{F_{r-1} \in \mathcal{A}_1(F_r)} \tilde{\pi}_{r-1}(F_{r-1}) M_r(F_r \mid F_{r-1}) 
$$

Now by Proposition \ref{prp-forest-AncestorSetCharacterization} we know this sum is equivalent to summing over all trees that can be drawn on the merged adjacent region tree pairs so 

$$
f_r(F_r) = \sum_{T_k \sim T_{k'} \in F_r} \sum_{T_\ell \in \mathcal{T}(\IGraph{T_k \cup T_{k'}})} \tilde{\pi}_{r-1}(\tilde{F}_{r-1}) M_r(\tilde{F}_r \mid F_{r-1}) 
$$

Where $\tilde{F}_{r-1}$ represents the plan formed by replacing $(T_k, s_k), (T_{k'}, s_{k'}) \in F_r$ with $(T_\ell, s_k + s_{k'})$. Now note the following useful facts. First, we know that $\tilde{\pi}_{r-1}(\tilde{F}_{r-1})$ is equal for all plans $F_{r-1}, \tilde{F}_{r-1}$ where the induced graph plans are equal, i.e., $\xi(F_{r-1}) = \xi(\tilde{F}_{r-1})$. Since all $\tilde{F}_{r-1}$ in the inner sum share all but one of their region trees and the vertex sets of the region tree that differs is all the same we know they all correspond to the same induced graph plan. Thus we know the $\tilde{\pi}_{r-1}$ is constant with respect to the inner sum. Further, by Lemma \ref{lem-forest-forward-equivalency} we also know that the $M_r(F_r \mid \tilde{F}_{r-1})$ is constant with respect to the different $\tilde{F}_{r-1}$. Since $\abs{\mathcal{T}(\IGraph{T_k \cup T_{k'}})} = \tau(\IGraph{T_k \cup T_{k'}})$ we get a simplification to   

$$
f_r(F_r) = \sum_{T_k \sim T_{k'} \in F_r} \tilde{\pi}_{r-1}(\tilde{F}_{r-1}) M_r(\tilde{F}_r \mid F_{r-1}) \tau(\IGraph{T_k \cup T_{k'}})
$$

Where $\tilde{F}_{r-1}$ can be any plan with $\tilde{F}_{r-1} \in \mathcal{A}_1(F_r)$ and $\mathrm{V}(\mathrm{OR}(\tilde{F}_{r-1}, F_r)) = \mathrm{V}(T_k \cup T_{k'})$. We can now further simplify to 

$$
\begin{aligned}
f_r(F_r) &= \sum_{T_k \sim T_{k'} \in F_r} \frac{1}{ Z_{r-1}} \tilde{\gamma}_{r-1}(\tilde{F}_{r-1}) \varphi(G_{\ell} \mid \xi(F_{r-1})) \cdot \frac{1}{\tau(\IGraph{T_k \cup T_{k'}})} \cdot \EffB{T_k, T_{k'}} \tau(\IGraph{T_k \cup T_{k'}}) \\
&= \frac{1}{Z_{r-1}} \sum_{T_k \sim T_{k'} \in F_r}  \tilde{\gamma}_{r-1}(\tilde{F}_{r-1}) \varphi(G_{\ell} \mid \xi(F_{r-1})) \cdot \EffB{T_k, T_{k'}}  \\
&= \frac{1}{ Z_{r-1}} \sum_{T_k \sim T_{k'} \in F_r}  \left( \prod_{T_n \in \tilde{F}_{r-1}} \tau(\IGraph{T_n}) \right)^{-1} \gamma_{r-1}(\xi(\tilde{F}_{r-1})) \varphi(G_{\ell} \mid \xi(F_{r-1})) \cdot \EffB{T_k, T_{k'}} 
\end{aligned}
$$

Now to simplify even further let $\xi_r = \xi(F_r)$, $(G_{k}, s_k) = \IRegion{T_k, s_k}$, $(G_{k'}, s_{k'}) = \IRegion{T_{k'}, s_{k'}}$ and $\tilde{\xi}_{r-1}$ is the plan formed by replacing the adjacent regions $G_{k}$ and $G_{k'}$ in $\xi_r$ with the merged region $H_{\ell} = G_k \cup G_{k'}$ (implying $\tilde{\xi}_{r-1} \in \mathcal{A}_1(\xi_r)$ ) so we have

$$
\begin{aligned}
f_r(F_r) &= \frac{1}{ Z_{r-1}} \sum_{T_k \sim T_{k'} \in F_r} \left( \prod_{G_n\in \tilde{\xi}_{r-1}} \tau(G_n) \right)^{-1} \gamma_{r-1}(\tilde{\xi}_{r-1}) \varphi(G_{\ell} \mid \tilde{\xi}_{r-1}) \cdot \EffB{T_k, T_{k'}} 
\end{aligned}
$$
Now recall from the proof of \ref{thm-marg-dens} we know since $\tilde{\xi}_{r-1} \in \mathcal{A}_1(\xi_r)$ that we have 
$$
\begin{aligned}
\gamma_{r-1}(\tilde{\xi}_{r-1})  &= \frac{\exp{-J(\tilde{\xi}_{r-1})}}{\exp{-J(\xi_r)}}\frac{ \tau(H_\ell)^\rho}{\tau(G_k)^{\rho} \tau(G_{k'})^{\rho}} \cdot \gamma_{r}(\xi_r)  
\end{aligned}
$$
and since $\gamma_{r}(\xi_r)$ is constant with respect to the sum we have 

$$
\begin{aligned}
f_r(F_r) &= \frac{1}{ Z_{r-1}} \sum_{T_k \sim T_{k'} \in F_r} \left( \prod_{G_n\in \tilde{\xi}_{r-1}} \tau(G_n) \right)^{-1} \frac{\exp{-J(\tilde{\xi}_{r-1})}}{\exp{-J(\xi_r)}}\frac{ \tau(H_\ell)^\rho}{\tau(G_k)^{\rho} \tau(G_{k'})^{\rho}} \cdot \gamma_{r}(\xi_r) \varphi(G_{\ell} \mid \tilde{\xi}_{r-1}) \cdot \EffB{T_k, T_{k'}} \\
\end{aligned}
$$

Now consider for the product of $\tau$ terms $\left( \prod_{G_n\in \tilde{\xi}_{r-1}} \tau(G_n) \right)^{-1}$ one of these terms if $\tau(H_\ell)$ and the others are over the regions common to $\xi_r, \tilde{\xi}_{r-1}$ i.e., $\mathrm{CR}(\xi_r, \tilde{\xi}_{r-1})$ so we can simplify further to
\begin{align*}
f_r(F_r) &= \frac{1}{ Z_{r-1}} \sum_{T_k \sim T_{k'} \in F_r} \left( \prod_{G_n\in \tilde{\xi}_{r-1}} \tau(G_n) \right)^{-1} \frac{\exp{-J(\tilde{\xi}_{r-1})}}{\exp{-J(\xi_r)}}\frac{ \tau(H_\ell)^\rho}{\tau(G_k)^{\rho} \tau(G_{k'})^{\rho}} \cdot \gamma_{r}(\xi_r) \varphi(G_{\ell} \mid \tilde{\xi}_{r-1}) \cdot \EffB{T_k, T_{k'}} \\
&= \frac{\gamma_{r}(\xi_r) }{ Z_{r-1}} \sum_{T_k \sim T_{k'} \in F_r} \left( \prod_{G_n\in \mathrm{CR}(\xi_r, \tilde{\xi}_{r-1})} \tau(G_n) \right)^{-1} \frac{\exp{-J(\tilde{\xi}_{r-1})}}{\exp{-J(\xi_r)}}\frac{ \tau(H_\ell)^{\rho-1}}{\tau(G_k)^{\rho} \tau(G_{k'})^{\rho}} \cdot \varphi(G_{\ell} \mid \tilde{\xi}_{r-1}) \cdot \EffB{T_k, T_{k'}} \\
&= \frac{\gamma_{r}(\xi_r) }{ Z_{r-1}} \sum_{T_k \sim T_{k'} \in F_r}  \tau(\mathrm{CR}(\xi_r, \tilde{\xi}_{r-1}))^{-1} \frac{\exp{-J(\tilde{\xi}_{r-1})}}{\exp{-J(\xi_r)}}\frac{ \tau(H_\ell)^{\rho-1}}{\tau(G_k)^{\rho} \tau(G_{k'})^{\rho}} \cdot \varphi(G_{\ell} \mid \tilde{\xi}_{r-1}) \cdot \EffB{T_k, T_{k'}}.
\end{align*}
\end{proof}

\end{theorem}

\begin{proposition}[Optimal Forest
Weights]\protect\hypertarget{prp-forest-opt-wt}{}\label{prp-forest-opt-wt}

Given a forest space forward kernel \(M_r\) and a target distribution
\(\tilde{\pi}_r\), the optimal minimal variance incremental weights are
given by
\begin{equation}\phantomsection\label{eq-forest-optimal-weights}{
w_r(F_{r-1}, F_r)  = \left(\sum_{\substack{ T_{k} \sim T_{k'} \in F_r}} \varphi(G_{k} \cup G_{k'}  \mid \tilde{\xi}_{r-1})  \frac{\exp{-J(\tilde{\xi}_{r-1})}}{\exp{-J(\xi_{r})}} \frac{\tau(H_\ell)^{\rho-1}}{\tau(G_k)^{\rho-1} \tau(G_{k'})^{\rho-1}} \EffB{T_k, T_{k'}}\right)^{-1},
}\end{equation} where \(\xi_r = \xi(F_r)\),
\((G_{k}, s_k) = \IRegion{T_k, s_k}\),
\((G_{k'}, s_{k'}) = \IRegion{T_{k'}, s_{k'}}\) and
\(\tilde{\xi}_{r-1}\) is the plan formed by replacing the adjacent
regions \(G_{k}\) and \(G_{k'}\) in \(\xi_r\) with the merged region
\(H_{\ell} = G_k \cup G_{k'}\).

\begin{proof}
Recall from \ref{prp-opt-wt} we can simplify our optimal weights to 
\begin{align*}
     w_r(F_{r-1}, F_r) &= \frac{1}{Z_{r-1} } \frac{\tilde{\gamma}_r(\xi_r)}{f_r(\xi_r)} 
\end{align*}

Now using by Theorem \ref{thm-marg-forest-dens} we can simplify to 

\begin{align*}
     &w_r(\xi_{r-1}, \xi_r) \\
     &=  \frac{1}{Z_{r-1} } \frac{\tilde{\gamma}_r(\xi_r)}{f_r(\xi_r)} \\
&= \frac{1}{Z_{r-1}} \frac{\frac{\gamma_r(\xi_r)}{\prod_{n=1}^r \tau(G_n)}}{\frac{\gamma_{r}(\xi_r) }{ Z_{r-1}} \sum_{T_k \sim T_{k'} \in F_r}  \tau(\mathrm{CR}(\xi_r, \tilde{\xi}_{r-1}))^{-1} \frac{\exp{-J(\tilde{\xi}_{r-1})}}{\exp{-J(\xi_r)}}\frac{ \tau(H_\ell)^{\rho-1}}{\tau(G_k)^{\rho} \tau(G_{k'})^{\rho}} \cdot \varphi(G_{\ell} \mid \tilde{\xi}_{r-1}) \cdot \EffB{T_k, T_{k'}}} \\
&= \frac{\frac{1}{\prod_{n=1}^r \tau(G_n)}}{ \sum_{T_k \sim T_{k'} \in F_r}  \tau(\mathrm{CR}(\xi_r, \tilde{\xi}_{r-1}))^{-1} \frac{\exp{-J(\tilde{\xi}_{r-1})}}{\exp{-J(\xi_r)}}\frac{ \tau(H_\ell)^{\rho-1}}{\tau(G_k)^{\rho} \tau(G_{k'})^{\rho}} \cdot \varphi(G_{\ell} \mid \tilde{\xi}_{r-1}) \cdot \EffB{T_k, T_{k'}}} \\
&= \frac{1}{\prod_{n=1}^r \tau(G_n)} \left(\sum_{T_k \sim T_{k'} \in F_r}  \tau(\mathrm{CR}(\xi_r, \tilde{\xi}_{r-1}))^{-1} \frac{\exp{-J(\tilde{\xi}_{r-1})}}{\exp{-J(\xi_r)}}\frac{ \tau(H_\ell)^{\rho-1}}{\tau(G_k)^{\rho} \tau(G_{k'})^{\rho}} \cdot \varphi(G_{\ell} \mid \tilde{\xi}_{r-1}) \cdot \EffB{T_k, T_{k'}} \right)^{-1} \\
&= \left(\prod_{n=1}^r \tau(G_n)\right)^{-1}  \left(\sum_{T_k \sim T_{k'} \in F_r}  \tau(\mathrm{CR}(\xi_r, \tilde{\xi}_{r-1}))^{-1} \frac{\exp{-J(\tilde{\xi}_{r-1})}}{\exp{-J(\xi_r)}}\frac{ \tau(H_\ell)^{\rho-1}}{\tau(G_k)^{\rho} \tau(G_{k'})^{\rho}} \cdot \varphi(G_{\ell} \mid \tilde{\xi}_{r-1}) \cdot \EffB{T_k, T_{k'}} \right)^{-1} \\
&=  \left(\left(\prod_{n=1}^r \tau(G_n)\right)\sum_{T_k \sim T_{k'} \in F_r}  \tau(\mathrm{CR}(\xi_r, \tilde{\xi}_{r-1}))^{-1} \frac{\exp{-J(\tilde{\xi}_{r-1})}}{\exp{-J(\xi_r)}}\frac{ \tau(H_\ell)^{\rho-1}}{\tau(G_k)^{\rho} \tau(G_{k'})^{\rho}} \cdot \varphi(G_{\ell} \mid \tilde{\xi}_{r-1}) \cdot \EffB{T_k, T_{k'}} \right)^{-1} \\
&=  \left( \sum_{T_k \sim T_{k'} \in F_r} \left(\prod_{n=1}^r \tau(G_n)\right) \tau(\mathrm{CR}(\xi_r, \tilde{\xi}_{r-1}))^{-1} \frac{\exp{-J(\tilde{\xi}_{r-1})}}{\exp{-J(\xi_r)}}\frac{ \tau(H_\ell)^{\rho-1}}{\tau(G_k)^{\rho} \tau(G_{k'})^{\rho}} \cdot \varphi(G_{\ell} \mid \tilde{\xi}_{r-1}) \cdot \EffB{T_k, T_{k'}} \right)^{-1}
\end{align*}
Now lets consider this term $\left(\prod_{n=1}^r \tau(G_n)\right) \tau(\mathrm{CR}(\xi_r, \tilde{\xi}_{r-1}))^{-1}$. We know that $\prod_{n=1}^r \tau(G_n)$ is a product over the regions in $\xi_r$ and we can decompose $\xi_r$ into $\mathrm{NR}(\xi_r, \tilde{\xi}_r)$ and $\mathrm{CR}(\xi_r, \tilde{\xi}_r)$ so this simplifies into 

$$
\begin{aligned}
\left(\prod_{n=1}^r \tau(G_n)\right) \tau(\mathrm{CR}(\xi_r, \tilde{\xi}_{r-1}))^{-1} &=  \tau(\mathrm{NR}(\xi_r, \tilde{\xi}_{r-1})) \tau(\mathrm{CR}(\xi_r, \tilde{\xi}_{r-1}))^{-1}\\
&= \tau(G_k) \tau(G_{k'})
\end{aligned}
$$

So our expression above simplifies to 

\begin{align*}
w_r(\xi_{r-1}, \xi_r) &=  \left( \sum_{T_k \sim T_{k'} \in F_r} \tau(G_k) \tau(G_{k'}) \frac{\exp{-J(\tilde{\xi}_{r-1})}}{\exp{-J(\xi_r)}}\frac{ \tau(H_\ell)^{\rho-1}}{\tau(G_k)^{\rho} \tau(G_{k'})^{\rho}} \cdot \varphi(G_{\ell} \mid \tilde{\xi}_{r-1}) \cdot \EffB{T_k, T_{k'}} \right)^{-1} \\
&=  \left( \sum_{T_k \sim T_{k'} \in F_r} \frac{\exp{-J(\tilde{\xi}_{r-1})}}{\exp{-J(\xi_r)}}\frac{ \tau(H_\ell)^{\rho-1}}{\tau(G_k)^{\rho-1} \tau(G_{k'})^{\rho-1}} \cdot \varphi(G_{\ell} \mid \tilde{\xi}_{r-1}) \cdot \EffB{T_k, T_{k'}} \right)^{-1} \\
\end{align*}

\end{proof}

\end{proposition}

Notice the optimal weights are almost the same as in
Equation~\ref{eq-graph-optimal-weights} the only major difference is
instead of multiplying the weights by the graph theoretic boundary
length between the two regions it is replaced by the effective region
tree boundary length (Definition~\ref{def-eff-boundary-len}), a sum over
each edge on the boundary of the probability that edge would have been
removed to create the two region trees. The main effect of this change
is it increases the computational complexity of the weights. Before for
graph space sampling computing the weights could be done in \(O(V)\)
time, requiring only a single pass through of the graph (ignoring the
cost to compute any \(J\) terms). Now the complexity is increased and it
is hard to precisely characterize. For each pair of adjacent trees
(which is on the order of \(O(r)\)) we must iterate over the boundary
edges between them and compute the probability that tree cut would have
been chosen from the region tree made by taking
\(T_k \cup e \cup T_{k'}\). This requires a pass through the entire
merged tree \(T_k \cup e \cup T_{k'}\) making it an
\(O(V(T_k \cup T_{k'}))\) operation. For a pair of adjacent region trees
the entire cost is
\(O(V(T_k \cup T_{k'}) \cdot \mathcal{C}(T_{k}, T_{k'}))\). However, for
certain types of tree cut selection rules
Section~\ref{sec-fast-forest-weights} discusses a procedure that makes
the cost \(O(V(T_k \cup T_{k'}) + \abs{\mathcal{C}(T_{k}, T_{k'})})\)
instead. While characterizing the exact computational complexity for a
plan \(F_r\) is difficult, calculating the weights is clearly more
costly than the \(O(V)\) computation required in the previous cases. The
empirical experiments presented in Appendix
Section~\ref{sec-comprehensive-sims} further show that, among the three
sampling spaces considered, computing the weights for spanning forest
space takes the greatest amount of total time.

Overall it is hard to precisely characterize the performance benefit of
sampling on this new space. Empirically we have found that the
acceptance rate for forest space sampling is essentially always higher
than graph space sampling on the same map, making the splitting step
much faster as there are fewer calls to Wilson's algorithm.

\subsubsection{Linking Edge Space}\label{sec-linking-edge}

The algorithm sampling space can be lifted even further to the space of
spanning forests with a linking (or marked) edge. We will refer to this
space as linking edge space. For linking edge space we now keep track of
linked forests \(L_r = (F_r, E_r)\) consisting of a spanning forest
\(F_r\) and a set of \(r-1\) edges \(E_r \subset E\) such that \(E_r\)
is a spanning tree on the plan multigraph \(G/\xi_r\). The plan
multigraph \(G/\xi_r\) is the multigraph where each region is a vertex
and each edge is an edge between the two regions in \(G\). Note this
necessarily implies \(F_r \cup E_r\) forms a spanning tree on \(G\). We
essentially obtain linking edge plan by stopping the splitting process
even earlier. Whereas before for forest plans we saved the trees after
the split we now stop even earlier, saving the edge in the tree cut as
well. Now for each forest plan \(F_r\) we have many linking edge plans
\(L_r\) associated with them. In fact, the precise number of linked
forests is exactly the number of spanning trees which can be drawn on
the plan multigraph \(G/\xi(F_r)\). We denote this number as
\(\tau(G/\xi(L_r))\).

As with forest space, we define a modified target distribution
\(\pi^\ast_r\) to sample from which has a pushforward target of
\(\pi_r\) on graph plan space. The computational cost of splitting is
exactly the same however the cost of computing the optimal weights
changes. The trade-off is instead of computing \(\EffB{T_k, T_{k'}}\),
we only compute the selection probability of a specific tree cut and the
number of linking edges \(\tau(G/\xi(L_r)), \tau(G/\xi(L_{r-1}))\) that
can be drawn on each plan. This makes the complexity of the weights on
the order of \(O(V + r^4)\) allowing for different trade-offs compared
to the forest space weights. A more detailed discussion is presented
after the proof of the optimal weights.

To split a linking edge plan the requirements are the same as with
forest plans. All that is required is a tree cut selection rule
\(p_\text{cut}(\cdot|\cdot)\). We then use the procedure in Algorithm
\ref{alg-link-edge-split}. Notice this is the same as Algorithm
\ref{alg-forest-split} we just save the edge from the tree cut as well.
We now formally present the modified target distribution and derive the
splitting probability and optimal weights.

\begin{algorithm}[H]
\begin{algorithmic}[1]
\REQUIRE{Splitting schedule $\mathcal{S}(\cdot, \cdot)$, region tree $(T_{\ell}, s_\ell) \in F_{r-1}$, and a tree cut selection rule $p_\text{cut}( \cdot | \cdot)$}
\STATE{Draw spanning tree $T^\ast$ on $\IGraph{T_{\ell}}$ using Wilson's algorithm}
\STATE{Let \texttt{TreeCuts} be a list of tree cuts}
\FOR{each tree cut $\TreeCut{e} \in \AllTreeCuts{T^\ast}{\mathcal{S}(r-1, s)}$}
\STATE{\texttt{TreeCuts.append($\{(T^{e_i}_{k}, s_{k}), (T^{e_i}_{k'}, s_{k'}), e_i\}$)}}
\ENDFOR
\STATE{Choose a tree cut $t = \TreeCut{e}$ distributed $p_\text{cut}( \cdot | T^\ast)$}
\STATE{add the cut edge $e$ so $E_r = E_{r-1} \cup \set{e}$ and save the two cut trees as the two new region trees}
\end{algorithmic}
\caption{Linking Edge Space Splitting Algorithm}
\label{alg-link-edge-split}
\end{algorithm}

\begin{definition}[Linking Edge Plan
Definition]\protect\hypertarget{def-linking-plan}{}\label{def-linking-plan}

Given a districting scheme \((G, D, S, [d^-, d^+])\) and \(r \leq D\),
we define a \(r\)-region linking edge plan \(L_r\) as the tuple
\[L_r = (F_r, E_r),\] where \(F_r\) is a \(r\)-region forest plan and
\(E_r \subset E\) is a linking edge set.

We say that \(E_r\) is a linking edge set if \(F_r \cup E_r\) is a
spanning tree on \(G\) (where
\(F_r \cup E_r = E_r \cup \bigcup_{T_k \in F_r} T_k\)).

\end{definition}

We now also define a new special function, the induced forest function.

\begin{definition}[Induced Forest
Function]\protect\hypertarget{def-induced-forest-func}{}\label{def-induced-forest-func}

We define the induced forest function \(F(\cdot)\) as the function that
takes a linking edge plan \(L_r = (F_r, E_r)\) and maps it to the
associated forest plan \(F_r\). In other words, \[
F(L_r) = F_r
\]

\end{definition}

\begin{definition}[Linking Edge Plan Space
Definition]\protect\hypertarget{def-LinkingPlanSpace}{}\label{def-LinkingPlanSpace}

Given a districting scheme \((G, D, S, [d^-, d^+])\) and \(r \leq D\),
we define \(\mathcal{L}_r(G, D, S, [d^-, d^+])\) to be the set of all
\(r\)-region linking edge plans under that districting scheme. For
notational convenience we will write it simply as \(\mathcal{L}_r\).

Note that \(\mathcal{L}_r\) can be equivalently expressed as the
preimage of the space of forest plans, \(\mathcal{F}_r\), under the
induced forest function. In other words, \[
\mathcal{L}_r = F^{-1}(\mathcal{F}_r).
\] This also implies it can be expressed as the preimage of the space of
graph plans, \(\mathcal{P}_r\), under the composition of the induced
forest function and plan function. \[
\mathcal{L}_r = F^{-1}(\xi^{-1}(\mathcal{P}_r)).
\]

\end{definition}

\begin{definition}[Plan Quotient
Multigraph]\protect\hypertarget{def-plan-quotient-graph}{}\label{def-plan-quotient-graph}

Given a plan \(\xi_r\) we define the plan quotient multigraph
\(G/\xi_r\) to be the multigraph where each vertex is a region in
\(\xi_r\) and we say the edge \(e = (u,v) \in E\) is between the regions
\(k\) and \(k'\) if \(v \in G_{k}\) and \(u \in G_{k'}\). In other
words, \(G/\xi_r\) is a multigraph where the regions are vertices and
the number of edges between two regions in the multigraph is equal to
the number of edges between the two regions in \(G\).

Thus, \(\tau(G/\xi_r)\) is equal to the number of spanning trees that
can be drawn on this multigraph.

\end{definition}

\begin{lemma}[Induced Forest Function
Preimage]\protect\hypertarget{lem-induced-forest-func-preim}{}\label{lem-induced-forest-func-preim}

Let \(F_r\) be a forest plan. The preimage of \(F_r\) under the induced
forest plan \(F(\cdot)\) is \[
F^{-1}(F_r) = \bigcup_{T \in \mathcal{T}(G/ \xi(F_r))} (F_r, T),
\] where \(T\) is a spanning tree on \(G/ \xi_r\) and \((F_r, T)\) is
the linking edge plan formed by \(F_r\) and the edges in \(T\)
associated with the underlying edges across regions in \(G\). This
implies

\[
\abs{F^{-1}(F_r)} = \tau(G/ \xi(F_r))
\]

\begin{proof} Recall that we define a linking edge plan as a forest plan $F_r$ and an edge set $E_r$ such that $E_r$ together with the trees in $F_r$ forms a spanning tree on $G$. Since $F_r$ is a disjoint forest on the vertex set of $G$ that tells us that $E_r$ is a valid linking edge set if, and only if, the edges in it form a spanning tree on the plan quotient graph $G/\xi_r$. 

\end{proof}

\end{lemma}

\begin{definition}[Linking Edge Space Target
Distributions]\protect\hypertarget{def-linking-target}{}\label{def-linking-target}

Given a sequence of target distributions \(\pi_1,\dots,\pi_D\) we define
our associated unnormalized linking edge space target distributions
\(\gamma_1^\ast, \dots, \gamma_D^\ast\) as

\[
\gamma^\ast_r(L_r) = \tilde{\gamma}_r(F(L_r))
\frac{1}{\tau(G/\xi(L_r))} = \gamma_r(\xi(L_r)) \frac{1}{\tau(G/\xi(L_r)) \tau(\xi(L_r))}
\] So the associated normalized target densities
\(\pi^\ast_1, \dots, \pi^\ast_D\) are

\begin{equation}\phantomsection\label{eq-linking-target}{
\pi^\ast_r(L_r) = \frac{\gamma^\ast_r(L_r)}{Z^\ast_r}
}\end{equation} Where \(Z^\ast_r\) is normalizing constant. We will soon
prove that, just like for forest space sampling, this normalizing
constant \(Z^\ast_r\) is actually the same as the graph space
normalizing constant \(Z_r\).

\end{definition}

Equation~\ref{eq-linking-target} is proportional to
Equation~\ref{eq-forest-target} with an extra factor of
\(\frac{1}{\tau(G/\xi(L_r))}\), which represents the number of linking
edges associated with a forest space plan. This is analogous to the
extra factor in Definition~\ref{def-linking-target}. Since
\(\pi^\ast_r\) is equal for all linking edge plans that have the same
induced forest plan, we have an additional factor of
\(\tau(G/\xi(L_r))\) in the pushforward measure to forest space. Adding
the corrective term ensures that the pushforward target on forest space
is \(\tilde{\pi}_r\) since we already know the pushforward measure on
graph space is \(\pi_r\).

\begin{lemma}[Pushforward Linking Edge Space Target
Distributions]\protect\hypertarget{lem-linking-pushforward-target}{}\label{lem-linking-pushforward-target}

If we have linking edge plans \(\set{L_r}\) distributed according to
\(\pi^\ast_r\) then the pushforward measure of the associated forest
space plans under \(F(\cdot)\) is distributed according to
\(\tilde{\pi}_r\) from Definition~\ref{def-linking-target}. Furthermore,
the normalizing constant \(Z^{\ast}_r\) of \(\pi^{\ast}_r\) is equal to
the graph space normalizing constant \(Z_r\) of \(\pi_r\).

\begin{proof} 

We begin by proving that $Z^\ast_r = Z_r$. The proof has the same structure as the proof for the forest space normalizing constant. Recall that 

$$
Z_r = \sum_{\xi_r \in \mathcal{P}_r} \gamma_r(\xi_r)
$$
and 
$$
Z^\ast_r = \sum_{L_r \in \mathcal{L}_r} \gamma^\ast_r(L_r)
$$

Now recall that by definition we know that $\mathcal{L}_r = F^{-1}(\mathcal{F}_r)$ so the sum can be equivalently expressed as a double sum over first, each forest plan $F_r \in \mathcal{F}_r$, and second, over each linking edge plan $L_r \in F^{-1}(F_r)$, so we can express the sum as 
$$
\begin{aligned}
Z^\ast_r
&= \sum_{F_r \in \mathcal{F}_r} \sum_{L_r \in F^{-1}(F_r)} \gamma^\ast_r(L_r) 
\end{aligned}
$$

Now, leveraging the fact that if $L_r \in F^{-1}(F_r)$ then $F(L_r) = F_r$ and $\tilde{\gamma}_r(F_r)$ can be expressed solely as a function of $F(L_r)$ we have
$$
\begin{aligned}
Z^\ast_r
&= \sum_{F_r \in \mathcal{F}_r} \sum_{L_r \in F^{-1}(F_r)} \gamma^\ast_r(L_r) \\
&= \sum_{F_r \in \mathcal{F}_r} \sum_{L_r \in F^{-1}(F_r)} \tilde{\gamma}_r(F(L_r))
\frac{1}{\tau(G/\xi(L_r))} \\
&= \sum_{F_r \in \mathcal{F}_r} \sum_{L_r \in F^{-1}(F_r)} \tilde{\gamma}_r(F_r)  
\frac{1}{\tau(G/\xi(L_r))}  &&\text{ because } F_r \in \xi^{-1}(\xi_r) \\
&= \sum_{F_r \in \mathcal{F}_r}  \tilde{\gamma}_r(F_r)  
\frac{1}{\tau(G/\xi(L_r))} \sum_{L_r \in F^{-1}(F_r)}  1  
\end{aligned}
$$
Now recall from Lemma \ref{lem-induced-forest-func-preim} that $\abs{F^{-1}(F_r)} = \tau(G/\xi(L_r))$ so we see $\frac{1}{\tau(G/\xi(L_r))}$ term cancels and we are left with 

$$
\begin{aligned}
Z^\ast_r
&= \sum_{F_r \in \mathcal{F}_r}  \tilde{\gamma}_r(F_r)  
\frac{1}{\tau(G/\xi(L_r))} \sum_{L_r \in F^{-1}(F_r)}  1     \\
&= \sum_{F_r \in \mathcal{F}_r} \tilde{\gamma}_r(F_r)  
\frac{1}{\tau(G/\xi(L_r))} \tau(G/\xi(L_r)) \\
&= \sum_{F_r \in \mathcal{F}_r}  \tilde{\gamma}_r(F_r) \\
&= Z_r && \text{by Lemma \ref{lem-forest-pushforward-target}}
\end{aligned}
$$

Now we will prove the pushforward is $\tilde{\pi}_r$. Assume $L_r \sim \pi^\ast_r$ and consider 
$$
\begin{aligned}
\mathrm{Pr}_{\pi^\ast_r }(F(L_r)= F_r) &= \sum_{\tilde{L}_r \in F^{-1}(F_r)} \pi^\ast_r(\tilde{L}_r) \\
&= \sum_{T \in \mathcal{T}(G/ \xi(L_r))} \pi^\ast_r(F_r, T)\; \; \text{by \ref{lem-induced-forest-func-preim}} 
\end{aligned}
$$

Just like in \ref{lem-forest-pushforward-target} we note that $\pi^\ast_r$ only depends on $L_r$ through the forest plan it induces, meaning the $\pi^\ast_r$ terms in the sum are constant. Since we know there are $\tau(G/\xi(L_r))$ terms in the sum it reduces to  
$$
\begin{aligned}
\mathrm{Pr}_{\pi^\ast_r }(F(L_r)= F_r) &= \sum_{T \in \mathcal{T}(G/ \xi(L_r))} \pi^\ast_r(F_r, T)\; \; \text{by \ref{lem-induced-forest-func-preim}} \\
 &= \tau(G/\xi(L_r)) \pi^\ast_r(L_r) \\
  &= \tau(G/\xi(L_r)) \tilde{\pi}(F(L_r))
\frac{1}{\tau(G/\xi(L_r))} \\
&= \tilde{\pi}(F(L_r))
\end{aligned}
$$

\end{proof}

\end{lemma}

\begin{definition}[Linking Edge Space
Kernel]\protect\hypertarget{def-new-link-split-kernel}{}\label{def-new-link-split-kernel}

Given a multidistrict tree \((T_\ell, s_\ell)\) and a distribution over
tree cuts \(p_\text{cut}( \cdot | \cdot)\), the linking edge space
splitting procedure outlined in \ref{alg-link-edge-split} can be
formally written as the composition of the Wilson function
(Definition~\ref{def-WilsonFunction}) on the subgraph induced by the
vertex set of \(T_\ell\) and the tree cut distribution
\(p_\text{cut}( \cdot | \cdot)\). Altogether it can be written as \[
\ArbKernel{\Wilson{\IGraph{T_{\ell}}}, \mathcal{S}(r-1, s_\ell)} = \TreeCut{e}
\]

Notice this is exactly the same as
Definition~\ref{def-new-split-kernel}, we just don't apply the induced
region tree function at the end.

\end{definition}

\begin{proposition}[Linking Edge Splitting
probability]\protect\hypertarget{prp-split-prob-linking}{}\label{prp-split-prob-linking}

Let \((T_{\ell}, s_{\ell})\) be a multidistrict tree and let
\((T_{k}, s_k), (T_{k'}, s_{k'}) , e\) be the two newly split region
trees and edge resulting from Algorithm \ref{alg-link-edge-split}. Given
our tree cut distribution \(p_\text{cut}( \cdot | \cdot)\) the
probability of this split is \[
     q\left(T_{k}, T_{k'}, e \mid T_{\ell}\right)= \frac{1}{\tau(\IGraph{T_{\ell}})} \cdot p_\text{cut}(\TreeCut{e} | T_{k} \cup \set{e} \cup T_{k'}).
\]

\begin{proof}
The proof of this is the same as \ref{prp-split-prob-forest} we just skip the step taking the preimage of the induced region tree function. As such the details will be skipped. The derivation is as follows:

Once again we begin by writing the event we sample the new region trees given the old in terms of our special functions 
\begin{align*}
    q\left(T_{k}, T_{k'}, e | T_\ell\right) &= q \left( \ArbKernel{\Wilson{\IGraph{T_{\ell}}}, \mathcal{S}(r-1, s_\ell)} = \set{(T_k, s_k), (T_{k'}, s_{k'}), e} \mid T_\ell\right)\\
&= \E{q \left(\ArbKernel{\Wilson{\IGraph{T_{\ell}}}, \mathcal{S}(r-1, s_\ell)} = \set{(T_k, s_k), (T_{k'}, s_{k'}), e} \mid \Wilson{\IGraph{T_\ell}} = T^\ast, T_\ell \right) \mid T_\ell } \\
&= \E{p_{\text{cut}}(\TreeCut{e}|T^\ast)  \I{T^\ast = T_{k} \cup \set{e} \cup T_{k'}}|T_\ell} \\
&= p_{\text{cut}}(\TreeCut{e}|T_{k} \cup \set{e} \cup T_{k'})  \E{\I{T^\ast = T_{k} \cup \set{e} \cup T_{k'}}|T_\ell}\\
&=  \frac{1}{\tau(\IGraph{T_\ell})} p_{\text{cut}}(\TreeCut{e}|T_{k} \cup \set{e} \cup T_{k'})  
\end{align*}

\end{proof}

\end{proposition}

\begin{corollary}[Linking Edge Space Forward
Kernel]\protect\hypertarget{cor-linking-forward-kernel}{}\label{cor-linking-forward-kernel}

Let \(\varphi(\cdot|\xi(F_r))\) be a distribution over multidistrict
trees in \(F_{r-1}\) that only depends on the induced plan \(\xi(F_r)\).
Let \(L_{r-1}\) be a linking edge. If \(L_r\) is a linking edge plan
such that there exists some \(T_{\ell} \in L_{r-1}\),
\(T_{k}, T_{k'} \in L_r\) where there is a linking edge connecting
\(T_k, T_{k'}\)
\(\IGraph{T_{\ell}} = \IGraph{T_{k}} \cup \IGraph{T_{k'}}\) then the
forward kernel probability is \begin{align*}
    M_r(L_r \mid L_{r-1}) = \varphi(G_{\ell} \mid \xi_{r-1}) \cdot \frac{1}{\tau(G_\ell)} \cdot p_{\text{cut}}(\TreeCut{e}|T_{k} \cup \set{e} \cup T_{k'}) 
\end{align*} Where \((G_{\ell}, s_\ell) = \IRegion{T_{\ell}, s_\ell}\)
and \(\xi_{r-1} = \xi(F_{r-1})\)

\begin{proof} 
We apply the same proof using the law of total probability from \ref{cor-forest-forward-kernel}, we just use Proposition \ref{prp-split-prob-linking} instead. For notational convenience we replace $\tau(\IGraph{T_\ell})$ with $\tau(G_\ell)$ and write $\varphi$ as $\varphi(G_{\ell} \mid \xi(F_{r-1}))$. 
\end{proof}

\end{corollary}

We can also prove an analogous result to
Corollary~\ref{cor-forest-plan-splittability} about splittability when
the only hard constraints are connectedness and population balance.

\begin{corollary}[Splittability of Linking Edge Space Forward
Kernel]\protect\hypertarget{cor-link-plan-splittability}{}\label{cor-link-plan-splittability}

If the only hard constraints are imposed through the score function
\(J\) are population balance and connectedness then under the forward
kernel in Corollary~\ref{cor-linking-forward-kernel}, all linking edge
plans \(L_r\) in the support of \(\pi^\ast_r\) are splittable for
\(r=2,\dots, D\).

\begin{proof} The proof is analogous to Corollary \ref{cor-forest-plan-splittability}. Given $L_r$ where $\pi^\ast_r(L_r) > 0$ we just merge any two region trees with a linking edge between them to produce the desired $L_{r-1}$ where $\pi^\ast_r(L_{r-1}) M_r(L_{r}|L_{r-1}) > 0$ and so on.
\end{proof}

\end{corollary}

\begin{theorem}[Marginal Linking Edge Proposal
Density]\protect\hypertarget{thm-marg-linking-dens}{}\label{thm-marg-linking-dens}

Given a linking edge space forward kernel \(M_r\) and a target
distribution \(\pi^\ast_r\), the marginal proposal density is given by,
\[
f(L_r) = \frac{\gamma_{r}(\xi_r) }{ Z_{r-1}}  \sum_{T_k \stackrel{e}{\sim} T_{k'}\in L_r}  \frac{1}{\tau(G/\tilde{\xi}_{r-1})\tau(\mathrm{CR}(\xi_r, \tilde{\xi}_{r-1}))} \frac{\exp{-J(\tilde{\xi}_{r-1})}}{\exp{-J(\xi_r)}}\frac{ \tau(H_\ell)^{\rho-1}}{\tau(G_k)^{\rho} \tau(G_{k'})^{\rho}} \cdot \varphi(G_{\ell} \mid \tilde{\xi}_{r-1}) \cdot p^\ast(T_k, T_{k'},e ),
\] where \(T_k \stackrel{e}{\sim} T_{k'}\in L_r\) denotes adjacent
region trees in \(L_r\) connected by a linking edge,
\(\xi_r = \xi(F_r)\), \((G_{k}, s_k) = \IRegion{T_k, s_k}\),
\((G_{k'}, s_{k'}) = \IRegion{T_{k'}, s_{k'}}\), \(\tilde{\xi}_{r-1}\)
is the plan formed by replacing the adjacent regions \(G_{k}\) and
\(G_{k'}\) in \(\xi_r\) with the merged region
\(H_{\ell} = G_k \cup G_{k'}\), and
\(p^\ast(T_k, T_{k'},e ) = p_\text{cut}(T_{k}, T_{k'} | T_{k} \cup \set{e} \cup T_{k'})\).

\begin{proof} 
The proof is analogous to \ref{thm-marg-forest-dens} and thus only a sketch is provided. Defining the definition of 1-region ancestor analogously for linking edge plans (i.e., they share $r-2$ region trees and $r-2$ linking edges) we first get that the marginal proposal density reduces to a sum over $\tilde{L}_r \in \mathcal{A}_1(L_r)$ so
$$
f(L_r) = \sum_{\tilde{L}_r \in \mathcal{A}_1(L_r)} \pi_{r-1}^\ast(\tilde{L}_r) M_r(L_r|\tilde{L}_r)
$$
Next by the same reasoning as \ref{prp-forest-AncestorSetCharacterization} we can rewrite this sum as

$$
f(L_r) = \sum_{T_k \stackrel{e}{\sim} T_{k'}\in L_r} \sum_{T \in \mathcal{T}(G/\xi(\tilde{L_r}))} \pi_{r-1}^\ast(\tilde{L}_r) M_r(L_r|\tilde{L}_r)
$$
And again we note that $\pi_r^\ast(\tilde{L}_r) M_r(L_r|\tilde{L}_r)$ depends on  $\tilde{L}_r$ only through its induced plan $\xi(\tilde{L}_r)$ meaning it is constant for each term in the sum. Now instead of there being $\tau(\IGraph{T_k \cup T_{k'}})$ terms there are $\tau(\IGraph{T_k \cup T_{k'}}) \cdot \tau(G/\xi(\tilde{L_{r-1} }))$ so the density becomes 
$$
f(L_r) = \sum_{T_k \stackrel{e}{\sim} T_{k'}\in L_r} \tau(\IGraph{T_k \cup T_{k'}}) \cdot \tau(G/\xi(\tilde{L}_{r-1})) \pi_{r-1}^\ast(\tilde{L}_r) M_r(L_r|\tilde{L}_r)
$$

Then everything simplifies analogously to \ref{thm-marg-forest-dens}.

\end{proof}

\end{theorem}

\begin{proposition}[Linking Edge Optimal
weights]\protect\hypertarget{prp-opt-wt-link}{}\label{prp-opt-wt-link}

Given a forward kernel \(M_r\) and target distribution \(\pi_r\), the
optimal minimal variance incremental weights are
\begin{equation}\phantomsection\label{eq-linking-edge-optimal-weights}{
w_r(L_{r-1}, L_r) = \left( \sum_{T_k \stackrel{e}{\sim} T_{k'}\in L_r} \varphi(G_{\ell} |\xi_{r-1}) \frac{\exp{-J(\tilde{\xi}_{r-1})}}{\exp{-J(\xi_r)}} \frac{ \tau(H_\ell)^{\rho-1}}{\tau(G_k)^{\rho-1} \tau(G_{k'})^{\rho-1}} \frac{\tau(G/\xi_{r})}{\tau(G/\xi_{r-1})} p_\text{cut}(T_{k}, T_{k'} | T_{k} \cup \set{e} \cup T_{k'}) \right)^{-1}
}\end{equation}

where \(T_k \stackrel{e}{\sim} T_{k'}\in L_r\) denotes adjacent region
trees in \(L_r\) connected by a linking edge, \(\xi_r = \xi(F_r)\),
\((G_{k}, s_k) = \IRegion{T_k, s_k}\),
\((G_{k'}, s_{k'}) = \IRegion{T_{k'}, s_{k'}}\), \(\tilde{\xi}_{r-1}\)
is the plan formed by replacing the adjacent regions \(G_{k}\) and
\(G_{k'}\) in \(\xi_r\) with the merged region
\(H_{\ell} = G_k \cup G_{k'}\), and
\(p^\ast(T_k, T_{k'},e ) = p_\text{cut}(T_{k}, T_{k'} | T_{k} \cup \set{e} \cup T_{k'})\).

\begin{proof}
The proof is analogous to \ref{prp-forest-opt-wt} (the only difference is an extra term of $\frac{1}{\tau(G/\xi_r)}$ in $\gamma^\ast_r$) and thus omitted. 
\end{proof}

\end{proposition}

Notice for these weights there are two differences from the forest space
weights in Equation~\ref{eq-forest-optimal-weights}. First, instead of
summing over all pairs of adjacent trees in \(F_r\), we only sum over
the \(r-1\) pairs of trees linked by the linking edges in \(E_r\).
Second, we only compute the tree cut selection probability
\(p_\text{cut}(T_{k}, T_{k'} | T_{k} \cup \set{e} \cup T_{k'})\) once
for each linking edge and instead now compute the ratio
\(\frac{\tau(G/\xi_{r})}{\tau(G/\xi_{r-1})}\) as well. This changes the
computational complexity as the selection probability is now merely an
\(O(V(T_k \cup T_{k'}))\) operation but in return we must compute
\(\frac{\tau(G/\xi_{r})}{\tau(G/\xi_{r-1})}\). Thanks to Kirchoff's
Matrix Tree Theorem, computing the number of spanning trees on a plan
multigraph requires computing the determinant of a submatrix of the
multigraph laplacian. This is conservatively on the order of the number
of vertices in the graph minus 1, cubed. That means computing
\(\tau(G/\xi_{r})\) is an \(O((r-1)^3)\) operation and computing
\(\tau(G/\xi_{r-1})\) is an \(O((r-2)^3)\) operation. Since there are
\(r-1\) linking edges we perform the \(O((r-2)^3)\) operation \((r-1)\)
times and the \(O((r-1)^3)\) operation once making the overall
complexity roughly \(O(r^4)\). Building the graph laplacian can be done
with a single pass through of the graph making the total complexity of
the weights \(O(V+r^4)\). In practice we can speed up this computation a
bit by leveraging the fact that the laplacian submatrix is a sparse,
positive definite matrix and using computation methods designed for
faster performance on it. This means, in contrast with forest space, the
weights become more expensive to compute as the number of regions
increases.

\subsection{Convergence Properties of gSMC
Algorithm}\label{sec-smcs-convergence}

The gSMC algorithm can be viewed as a type of sequential particle filter
designed to keep the system alive. At a high level the algorithm works
similar to the standard particle filter algorithm described in Algorithm
\ref{alg-standard-particle-filter}, where particles are sampled from a
markov kernel \(Q_k\) conditional on the previous step and then
reweighted by a selection function \(g_k\) to produce the target
distribution before repeating the process again. The only difference is
now we assume its possible (and indeed highly probable) to sample
particles under \(Q_k\) that have probability zero under the target. To
deal with that we just continue to sample particles until we have \(H\)
alive ones (assuming that is possible, which is sometimes not the case).
Conceptually this is very similar to the particle filter in
\citet{legland2005sequential}, where particles are sampled until the sum
of selection function values normalized by the level \(H \in \bbN\) is
greater than or equal to the supnorm of \(g_k\). However, it differs in
that we propose sampling until there are \(H\) non-zero weight, or
alive, particles and impose weaker assumptions on our overall particle
system. These modifications require some separate proofs, namely the
convergence of empirical measures, although they largely proceed in a
similar manner to \citet{legland2005sequential}. Further, since we can
assume the state space \(E\) is finite that lets us prove much stronger
convergence results (i.e., almost sure convergence in total variation
distance) with minimal additional assumptions.

We first begin by proving convergence results about a more general
sequential particle system algorithm described in Algorithm
\ref{alg-alive-particle-filter} and then in
Section~\ref{sec-gsmc-specific-convergence-results}, we show that the
gSMC algorithm is a special case of that more general algorithm, thus
implying all the convergence results immediately follow.

\subsubsection{General Alive Sequential Particle System
Results}\label{general-alive-sequential-particle-system-results}

We first begin by defining the setup and notation in terms of standard
particle filter/Feynman-Kac flow notation used in papers like
\citet{legland2005sequential} and \citet{AliveSMC}. We define
\((E, \mathcal{E})\) as our common measurable state space, which we will
assume is finite for all the results here. We fix \(n\) and let
\(\set{X_k}_{k=0}^n\) be a (potentiality inhomogenous) markov chain on
\(E\) with an initial state \(X_0 \sim \eta_0\) and transition kernels
\(Q_k\). We also have a sequence of non-negative, bounded functions
\(\set{g_k}_{k=0}^n\) which we refer to as selection or fitness
functions. The goal of the algorithm is to sample particles according to
a sequence of distributions \(\{\eta_k\}_{k=1}^n\) and
\(\{\mu_k\}_{k=0}^n\) which are iteratively defined via mutations by the
Markov kernels \(Q_k\) and ``reweighting'' by the selection functions
\(g_k\). At a high level, the idea of this sequential particle algorithm
is we start with particles distributed according to some initial
distribution \(\eta_0\), ``reweight'' our distribution by the selection
function \(g_0\) to get our post-selection distribution \(\mu_0\), and
then repeat the process over again \(n\) times. At step \(k\) we have
particles distributed roughly according to \(\mu_{k-1}\), we apply our
transition kernel \(Q_k\) to evolve them to the distribution
\(\mu_{k-1} Q_k = \eta_k\) and then ``reweight'' that sample via the
selection function \(g_k\) to obtain \(\mu_k\). This can be represented
diagrammatically as

\[
\mu_{k-1} \longrightarrow \mu_{k-1} Q_k = \eta_k \longrightarrow \mu_k = g_k \cdot \eta_k
\] The precise definition of \(\mu_{k-1} Q_k\) and \(g_k \cdot \eta_k\)
are given in Definition~\ref{def-markov-pushforward} and
Definition~\ref{def-proj-product} respectively. In a standard particle
filter algorithm this is accomplished by replacing \(\eta_k\) and
\(\mu_k\) with empirical approximations \(\eta^N_k\) and \(\mu^N_k\)
obtained using Algorithm \ref{alg-standard-particle-filter}.

The wrinkle in implementing the standard particle filter algorithm for
selection functions where \(g_k\) can take the zero value with
potentially high probability under \(\eta_k\) is if we sampled a fixed
number of particles \(N\) from \(\eta_k\) at each step we might end up
with most, if not all of the sampled particles having a value zero under
the selection function \(g_k\). We refer to these particles where
\(g_k = 0\) as dead particles and particles where \(g_k > 0\) as alive
particles. When all sampled particles are dead we refer to this as
particle system extinction. Under the alive mechanism in Algorithm
\ref{alg-alive-particle-filter} we try to get around this problem by
sampling a random number of particles from our approximation of
\(\eta_k\) until we have \(N\) alive particles. This procedure is based
on the alive method proposed in \citet{legland2005sequential} (and is in
fact exactly the same when \(g_k\) is an indicator function). We adopt
the terminology of referring to the number of alive particles desired as
the level, which we will denote as \(H\). The only assumption we require
is that for each \(k\) we require the support of \(g_k\) to be a subset
of the support of \(\eta_k\). The number of samples until we have \(N\)
alive particles is a stopping time however, with only our very mild
assumption that the support of \(g_k\) is contained with the support of
\(\eta_k\), it is possible for this stopping time to be infinite with
positive probability for some values of \(H\). We refer to this as
halting. However, we can prove that as \(H \to \infty\) the probability
of halting tends to 0, in fact, with probability 1 there exists some
(potentially random) \(H_0\) such that when \(H \geq H_0\) the
probability of halting is 0, and versions of the standard convergence
results for sequential particle system algorithm follow as well our
alive version. We will now define some notation needed for the following
convergence results.

\begin{definition}[Bracket
Notation]\protect\hypertarget{def-bracket-notation}{}\label{def-bracket-notation}

Given a measure \(\lambda\) and a function \(\varphi\) we define the
integral of \(\varphi\) with respect to \(\lambda\) as
\(\inner{\lambda}{\varphi}\) where

\[\inner{\lambda}{\varphi} = \int \varphi(x) \lambda(dx)\]

Since \(E\) is finite this will always be well defined.

\end{definition}

\begin{definition}[Markov Pushforward
Measure]\protect\hypertarget{def-markov-pushforward}{}\label{def-markov-pushforward}

Given a measure \(\lambda\) and a transition kernel \(K(x, dy)\) on
\((E, \mathcal{E})\) we denote the push-forward of \(\lambda\) by the
transition kernel K as \(\lambda K\) where for any \(A \in \mathcal{E}\)
then

\[\lambda K(A) = \int K(x, A) \lambda(dx) = \int K(A|x) \lambda(dx).\]

When \(\lambda\) is a probability measure and \(K\) is a Markov kernel
then this is equivalent to the marginal distribution of \(Y\) if
\(X \sim \lambda\) and \(Y|X \sim K(\cdot|X)\).

\end{definition}

\begin{definition}[Projective
Product]\protect\hypertarget{def-proj-product}{}\label{def-proj-product}

Given a finite measure \(\lambda\) and a non-negative, bounded function
\(g\) we denote \(g \cdot \lambda\) as the projective product, i.e.,
normalized probability measure given by \begin{align*}
(g \cdot \lambda) (A) = \frac{\int_A g(x) \lambda(dx)}{\int g(x) \lambda(dx)} = \frac{\inner{\lambda}{ \Iset{A} g}}{\inner{\lambda}{g}}.
\end{align*} Since we assume our state is finite then that means the
density of \(g \cdot \lambda\) is given by \[
(g \cdot \lambda) (x) = \frac{g(x) \lambda(x)}{\sum_{x' \in E} g(x) \lambda(x)}
\]

\end{definition}

\begin{definition}[Empirical
Distribution]\protect\hypertarget{def-common-empd}{}\label{def-common-empd}

Given a probability distribution \(\mu\) and some \(N \in \bbN\) we let
\(S^N(\mu)\) denote the empirical probability distribution associated
with \(N\) iid samples from \(\mu\). In other words \[
S^N(\mu) = \frac{1}{N} \sum_{i=1}^N \delta_{\xi_i}(\cdot)
\]

Where \(\xi^1, \dots, \xi^N \iid \mu\).

\end{definition}

We will now formally define the definitions of components of Algorithm
\ref{alg-standard-particle-filter}.

\begin{definition}[Algorithm
Distributions]\protect\hypertarget{def-particle-system-distributions}{}\label{def-particle-system-distributions}

The distributions in Algorithm \ref{alg-standard-particle-filter} are
completely determined by the initial distribution \(\eta_0\), our
selection functions \(\{g_k\}_{k=0}^n\), and the Markov kernels
\(Q_k(y, dx)\) for \(k = 1, \dots, n\).

We define our normalized post-selection laws recursively
\(\{\mu_k\}_{k=0}^n\) as \[
\mu_k = g_k \cdot \eta_k
\] and we define our proposal law \(\{\eta_k\}_{k=0}^n\) in a similar
manner. \[
\eta_0  \text{ is given} \; \; \text{ and } \; \; \;\eta_k = \mu_{k-1} Q_k
\] We note that \[
\eta_k(A) = \sum_{y \in \mathrm{Supp}(\mu_{k-1})} \mu_{k-1}(y) Q_k(y, A) 
\] \[
\eta_k^H(A) = \frac{1}{N_k^H} \sum_{i=1}^{N_k^H} \delta_{\xi_k^i}(A)
\]

\[
\mu_k^H(A) = \sum_{i=1}^H W_k^i \delta_{\zeta_k^i}(A)
\]

\end{definition}

\begin{definition}[Unnormalized Post-Selection
Law]\protect\hypertarget{def-unnormalized-flow}{}\label{def-unnormalized-flow}

For Algorithm \ref{alg-standard-particle-filter} we can also define the
unnormalized post-selection law or flows \(\{\gamma_k\}_{k=0}^n\) as

\[
\gamma_0 = g_0 \eta_0 \text{ and } \gamma_k = g_k ( \gamma_{k-1} Q_k) \text{ for } k =1,\dots,n
\]

Note that \(\gamma_k\) is a measure and \(\mu_k\) is the associated
(normalized) probability measure. In other words,

\[
\mu_k(A) = \frac{\gamma_k(A)}{\gamma_k(E)}
\] We can equivalently express \(\gamma_k\) as

\[
\gamma_k = \gamma_{k-1} R_k \text{ where } R_k(y, dx) = Q_k(y, dx) g(x)
\]

\end{definition}

\begin{definition}[Sampled
Particles]\protect\hypertarget{def-sampled-particles}{}\label{def-sampled-particles}

For steps \(k = 0,\dots, n\) we use \(\xi_{k}^i \in E\) to denote the
samples of our algorithm at each step. We refer to these sampled
\(\xi_{k}^i \in E\) as particles. We define \(\mathcal{H}^H_k\) to be
the sigma algebra generated by the algorithm running at level \(H\) up
to and including step \(k-1\).

For \(k=0\) our samples are distributed as iid from \(\eta_0\) \[
\xi_0^1, \xi_0^2, \dots \sim \iid \eta_0
\] and for \(k = 1,\dots,n\) we have our samples as conditionally iid
from \(\mu_{k-1}^H Q_k\) given \(\mathcal{H}^H_k\) \[
\xi_k^1, \xi_k^2, \dots |\mathcal{H}^H_k \sim \iid \mu_{k-1} Q_k
\]

\end{definition}

\begin{definition}[Alive Function and
Sets]\protect\hypertarget{def-alive-funcs}{}\label{def-alive-funcs}

Given selection functions \(\set{g_k}_{k=0}^n\) we define the alive
functions \(\set{a_k}_{k=0}^n\) as the indicator functions \[ 
a_k(x) = \I{g_k(x) > 0}
\] For a particle \(\xi \in E\) we say that it is \emph{alive} if
\(a_k(\xi) = 1\) and \emph{dead} if \(a_k(\xi) = 0\).\\
We define the alive sets \(\set{A_k}_{k=0}^n\) as \[
A_k = \set{x \in E| g_k(x) > 0}
\] Recall, our only requirement for the selection functions is that for
each \(k\), the alive set \(A_k\) is a subset of the support of
\(\eta_k\). Note that this is equivalent to saying \(A_0\) is within the
support of \(\eta_0\) and for each \(k = 1,\dots,n\) for every
\(x \in A_k\) there exists some \(y \in A_{k-1}\) such that
\(Q_k(y, x) > 0\). This follows immediately from the fact that \[
\eta(x) = \sum_{y \in E} \mu_{k-1} (y) Q_k(y, x)
\]

\end{definition}

\paragraph{Algorithm Details}\label{algorithm-details}

\begin{algorithm}[t] 
\caption{Standard particle filter algorithm}
\label{alg-standard-particle-filter}

\begin{enumerate}
\def\labelenumi{\arabic{enumi}.}
\item
  Sample \(\xi_0^1,\dots, \xi_0^{N} \iid \eta_0\).
\item
  Set \[
  \eta_0^N
  =
  S^N(\eta_0)
  =
  \frac{1}{N} \sum_{i=1}^N \delta_{\xi_0^i}(\cdot)
  \] and \[
  \mu_0^N = g_0 \cdot \eta_0^N.
  \]

  Recall \(\mu_0^N\) is the projective product (defined in
  Definition~\ref{def-proj-product}), so for \(A \in \mathcal{E}\) it
  assigns probability \[
  \mu_0^N(A)
  =
  \frac{
      \sum_{i=1}^N g_0(\xi_0^i)\delta_{\xi_0^i}(A)
  }{
      \sum_{i=1}^N g_0(\xi_0^i)
  }.
  \]

  Note that we can equivalently write \(\mu_0^N\) as a weighted sum \[
  \mu_0^N
  =
  \sum_{i=1}^N W_0^i\delta_{\xi_0^i}(\cdot),
  \] where \[
  W_0^i =
   \frac{
      g_0(\xi_0^i)
  }{
      \sum_{i=1}^N g_0(\xi_0^i)
  }.
  \]
\item
  For step \(k\), sample \[
  \xi_k^1,\dots,\xi_k^N \iid \mu^N_{k-1}Q_k
  \] and set \[
  \eta_k^N
  =
  S^N(\mu_{k-1}Q_k) \text{ and }
  \] and \[
  \mu_k^N
  =
  g_k \cdot \eta_k^N.
  \]

  Note that \(\xi_k^1,\dots,\xi_k^N\) are sampled iid conditional on
  \(\mathcal{H}_{k-1}^N\), the \(\sigma\)-algebra generated by the
  particle system up to the previous \(k-1\) step, inclusive.

  Sampling from \(\mu_{k-1}^NQ_k\) is done by sampling one of
  \(\xi_{k-1}^i\) with probability proportional to \(W_{k-1}^i\), and
  then mutating it with the Markov kernel \(Q_k\). We are then left with
  \[
  \eta_k^N
  =
  \frac{1}{N}\sum_{i=1}^N\delta_{\xi_k^i}(\cdot)
  \text{ 
  and
    }
  \mu_k^N
  =
  \sum_{i=1}^N W_k^i\delta_{\xi_k^i}(\cdot),
  \] where \[
  W_k^i
  =
  \frac{
      g_k(\xi_k^i)
  }{
      \sum_{i=1}^N g_k(\xi_k^i)
  }.
  \]
\item
  We repeat this until we have \(\eta_n^N\) and \(\mu_n^N\), which
  approximate \(\eta_n\) and \(\mu_n\) better and better as
  \(N \to \infty\).
\end{enumerate}

\end{algorithm}

\begin{definition}[Alive Particle Sampling
Count]\protect\hypertarget{def-alive-level}{}\label{def-alive-level}

Given a level \(H \in \mathbb{N}^+\) we define \(N_k^H\) to be the
number of draws we need to make from \(\mu^{H}_{k-1} Q_k\) until there
are \(H\) alive samples. In other words, \[
N_k^H = \inf \set{N \geq 1 \big\vert \sum_{i=1}^N a_k(\xi_{k}^i) \geq H}.
\] Note that since this is a function of \(\xi_k^1,\dots\) it is a
stopping time with respect to the filtration generated by the particles.
When \(N_k^H = \infty\) we say that the particle system has halted.

\end{definition}

x This means the number of particles sampled at each step is now random
and we have potentially a lot of sampled particles with non-zero weight
at each step. To help simplify and clarify what is going on at some
steps we now also introduce some new notation \(\zeta\), to represent
the alive particles at each step.

\begin{definition}[Alive Particle
Notation]\protect\hypertarget{def-alive-particles}{}\label{def-alive-particles}

Given a level \(H \in \mathbb{N}^+\) and a (potentially conditional) iid
sample \(\xi_k^i,\dots, \xi_k^{N_{k-1}^H} \iid \mu_{k-1}^{H} Q_k\) we
define the random variables \(\zeta_k^1, \dots, \zeta_k^H\) as the \(H\)
alive (i.e., non-zero weight) sampled particles. In other words \[
\zeta_k^i = \xi_k^j \text{ where $\xi_k^j$ is the $j$-th particle sampled where } g_k(\xi_k^j) > 0
\] This is largely done for convenience of notation.

\end{definition}

\begin{algorithm}[t] 
\caption{Alive particle filter algorithm}
\label{alg-alive-particle-filter}

Given a desired number of non-zero weight particles \(H \in \bbN\) we
obtain empirical approximations \(\eta_k^H\) and \(\mu_k^H\) in the
following manner

\begin{enumerate}
\def\labelenumi{\arabic{enumi}.}
\tightlist
\item
  Sample \(\xi_0^1, \xi_0^2, \dots, \xi_0^{N_{0}^H}, \iid \eta_0\)
\end{enumerate}

This means we have \(H\) alive particles \(\zeta_0^1,\dots, \zeta_0^H\)

\begin{enumerate}
\def\labelenumi{\arabic{enumi}.}
\setcounter{enumi}{1}
\tightlist
\item
  Set \[
  \eta_0^{H}
  =
  S^{N^H_0}(\eta_0)
  =
  \frac{1}{N^H_0} \sum_{i=1}^{N_0^H} \delta_{\xi_0^i}(\cdot)
  \text{ and }
  \mu_0^{H} = g_0 \cdot \eta_0^{H}.
  \]
\end{enumerate}

Note we can equivalently write \(\mu_0^{H}\) as a weighted sum of just
the alive particles \[
   \mu_0^{H}
   =
   \sum_{i=1}^{H} W_0^i\delta_{\zeta_0^i}(\cdot),
   \text{
    where 
   }
   W_0^i =
 \frac{
       g_0(\zeta_0^i)
   }{
       \sum_{i=1}^{H} g_0(\zeta_0^i)
   }.
   \]

\begin{enumerate}
\def\labelenumi{\arabic{enumi}.}
\setcounter{enumi}{2}
\item
  For step \(k\), sample \[
  \xi_k^1,\dots,\xi_k^{N^H_{k-1}} \iid \mu^{H}_{k-1}Q_k
  \] and set \[
  \eta_k^{H}
  =
  S^{N^H_k}(\mu_{k-1}^{H}Q_k) \text{ and }
  \mu_k^{H}
  =
  g_k \cdot \eta_k^{N^H_{k}}.
  \]

  Note that \(\xi_k^1,\dots,\xi_k^{N^H_{k}}\) are sampled iid
  conditional on \(\mathcal{H}_{k-1}^H\), the \(\sigma\)-algebra
  generated by the particle system up to the previous \(k-1\) step,
  inclusive.

  Sampling from \(\mu_{k-1}^{H}Q_k\) is done by sampling one of the
  \(H\) alive particles \(\zeta_{k-1}^i\) with probability proportional
  to \(W_{k-1}^i\), and then mutating it with the Markov kernel \(Q_k\).
  We are then left with \[
  \eta_k^{H}
  =
  \frac{1}{{N^H_{k}}}\sum_{i=1}^{N^H_{k}}\delta_{\xi_k^i}(\cdot)
  \text{ 
  and
    }
  \mu_k^{H}
  =
  \sum_{i=1}^{H} W_k^i\delta_{\zeta_k^i}(\cdot),
  \] where \[
  W_k^i
  =
  \frac{
      g_k(\zeta_k^i)
  }{
      \sum_{i=1}^{H} g_k(\zeta_k^i)
  }.
  \]
\item
  We repeat this until we have \(\eta_n^{H}\) and \(\mu_n^{H}\), which
  we will prove still approximate \(\eta_n\) and \(\mu_n\) better and
  better as \(H \to \infty\).
\end{enumerate}

\end{algorithm}

We will now prove that so long as each alive set \(A_k\) is contained
within the support of \(\eta_k\), then as \(H \to \infty\) the
probability of halting goes to zero, the empirical approximations
\(\set{\eta_k^H}_{k=0}^n\) and \(\set{\mu_k^H}_{k=0}^n\) will converge
almost surely in total variation distance, and any functions
\(\varphi: E \to \bbR\) the quantity
\(\sqrt{H}(\inner{\mu_k^H}{\varphi} - \inner{\mu_k}{\varphi})\) is
asymptotically normal.

We first begin with some useful lemmas.

\begin{lemma}[Continuity of Markov Kernels in Total Variation
Metric]\protect\hypertarget{lem-markov-kernel-continuity}{}\label{lem-markov-kernel-continuity}

Let \(\mathcal{P}(E)\) denote the set of probability measures of
\((E, \mathcal{E})\) and let \(Q\) be a markov kernel on
\((E ,\mathcal{E})\). The map \(f:\mathcal{P}(E) \to \mathcal{P}(E)\)
defined by \(f(\mu) = \mu Q\) is continuous with respect to total
variation distance as a metric.

\begin{proof}
This follows immediately from Proposition 1.1 of \citep{delmoral2003contraction} where $\Phi(x,y) = \abs{x-y}$.
\end{proof}

\end{lemma}

\begin{lemma}[Continuity of Projective Product in Total Variation
Metric]\protect\hypertarget{lem-proj-product-continuity}{}\label{lem-proj-product-continuity}

Let \(\mathcal{P}(E)\) denote the set of probability measures of
\((E, \mathcal{E})\) and let \(g\) be some bounded function defined on
\(E\). The map \(f:\mathcal{P}(E) \to \mathcal{P}(E)\) defined by
\(f(\mu) = g \cdot \mu\) is continuous with respect to total variation
distance as a metric.

\begin{proof}
Lemma 4.3.1 from \citep{delmoral2004feynman} with the projective product represented as a Boltzmann-Gibbs transformation (Definition 2.3.3 in \citep{delmoral2004feynman}) with $g$ as the potential function 
\end{proof}

\end{lemma}

\begin{theorem}[Almost Sure Convergence of Empirical
Measures]\protect\hypertarget{thm-alive-particle-as-convergence}{}\label{thm-alive-particle-as-convergence}

Assume the alive sets \(A_0,\dots, A_n\) are contained within the
supports of \(\eta_0, \dots, \eta_n\) respectively. Let \(\mu_k^{H}\)
and \(\eta_k^{H}\) be the empirical measures from the alive particle
system algorithm in Algorithm \ref{alg-alive-particle-filter} and
\(N_k^H\) be the alive particle sampling count from
\ref{def-alive-level}, then for all \(k=0,\dots, n\), almost surely
there exists \(H_0\) such that for all \(H \geq H_0\) \(N_k^H\) is
finite and as \(H \to \infty\) the empirical measures converge almost
surely in total variation distance to the true \(\mu_k\) and \(\eta_k\)
respectively, so \[
    \mu_k^{H} \toAS \mu_k \; \; \text{as $H \to \infty$   for } k = 0,\dots,n 
\] and \[
    \eta_k^{H} \toAS \eta_k \; \; \text{as $H \to \infty$   for } k = 0,\dots,n 
\] and \[
    P(\exists H_0 \text{ s.t. } \forall H \geq H_0, \; N_k^H \text{ is finite}) = 1 
\] This implies the probability of the particle system halting, i.e.,
\(P(N_k^H = \infty)\), tends to 0 as well.

\begin{proof}
We will prove this inductively.

\medskip
\noindent\textbf{Base case.}

We begin with the base case $k=0$.

\smallskip
\noindent\textit{$N_0^H$ is finite almost surely and $N_0^H \to \infty$ as $H \to \infty$.}

Recall that $N_0^H = \inf\set{N \geq 1 | \sum_{i=1}^N a_0(\xi^i_0) \geq H}$ so if the sum $\sum_{i=1}^\infty a_0(\xi^i_0)$ is infinite almost surely then $N_0^H$ is finite almost surely. Consider since $\xi_0^i \iid \eta_0$ then $a_0(\xi_0^i) \iid \bern{\eta_0(A_0)}$ and since we assume $\eta_0(A_0) > 0$, then we know $\sum_{i=1}^\infty P(a_0(\xi_0^i) = 1) = \infty$. This in turn implies, by the second Borel-Cantelli lemma, that $P(\sum_{i=1}^\infty a_0(\xi^i_0) = \infty) = 1$ which immediately also implies $P(\{N_0^H = \infty\}) = 0$ for all $H$. Further, consider that $N_0^H \geq H$ so we know $N_0^H \to \infty$ as $H \to \infty$.

\smallskip
\noindent\textit{Convergence of $\eta^H_0 \toAS \eta_0$}

Now let $\varphi$ be some function on $E$ (recall since $E$ is finite $\inner{\eta_0}{\varphi} < \infty$). Define 
$$b_N = \frac{1}{N} \sum_{i=1}^N \varphi(\xi_0^i).$$ 
Since $N_0^H$ is finite almost surely then this quantity is well defined. 
It follows immediately from the strong law of large numbers that $b_N \toAS b = \inner{\eta_0}{\varphi}$ as $N \to \infty$.  
Now define the sub-sequence $\{b_{N^H_0}\}_{H=1}^\infty$ where $b_{N^H_0} = \frac{1}{N^H_0} \sum_{i=1}^{N^H_0} \varphi(\xi_0^i)$. We proved earlier that $N^H_0 \to \infty$ as $H \to \infty$. Since $b_N \toAS b$ and $N^H_0 \to \infty$ as $H \to \infty$ then we know with probability 1 the sub-sequence $b_{N^H_0} \toAS b$ as $H \to \infty$. Now note that
$$
b_{N^H_0} = \frac{1}{N^H_0} \sum_{i=1}^{N^H_0} \varphi(\xi_0^i) = \inner{\eta_0^{H}}{\varphi}.
$$
So we have proven $\inner{\eta_0^{H}}{\varphi} \toAS \inner{\eta_0}{\varphi}$. Now recall for countable spaces the total variation distance is given by the formula. 
$$
\dTV{\eta_0^{H}}{\eta_0} = \frac{1}{2} \sum_{x \in E} \abs{\eta_0^{H}(\set{x}) - \eta_0(\set{x})}
$$

So taking our set of functions to be $\varphi_x = \I{x}$ for $x \in E$ so $\inner{\eta_0}{\varphi_x} = \eta_0(\{x\})$, we can write the total variation distance as 

$$
\dTV{\eta_0^{H}}{\eta_0} = \frac{1}{2} \sum_{x \in E} \abs{\inner{\eta_0^{H}}{\varphi_x}  - \inner{\eta_0}{\varphi_x}  }
$$

Since we know $\inner{\eta_0^{H}}{\varphi_x}  \toAS \inner{\eta_0}{\varphi_x}$ for each $\varphi_x$ and $E$ is finite then there exists a set of probability 1 where all $\inner{\eta_0^{H}}{\varphi_x}$ converge almost surely. Since this collection of functions is finite this in turn implies $\frac{1}{2} \sum_{x \in E} \abs{\inner{\eta_0^{H}}{\varphi_x}  - \inner{\eta_0}{\varphi_x}  } \toAS 0$. Thus we have proved $\eta_0^{H} \toAS \eta_0$. 

Now for $\mu_0$, consider that $g_0 \cdot \eta_0^{H} = \mu_0^{H}$ and $g_0 \cdot \eta_0 = \mu_0$. Since we know by \ref{lem-proj-product-continuity} that the projective product is a continuous mapping with respect to total variation distance, then by the continuous mapping theorem we know that $\mu_0^{H} \toAS \mu_0$ as well. 

\medskip
\noindent\textbf{Inductive Step.} Assume for $k-1$ that $P(\limsup_{H > 0} \{N_{k-1}^H = \infty\}) = 0$, $\dTV{\eta_{k-1}^{H}}{\eta_{k-1}} \toAS 0$ and $\dTV{\mu_{k-1}^{H}}{\mu_{k-1}} \toAS 0$.

\smallskip
\noindent\textit{With probability 1, there exists some $H_0$ such that $N_k^H$ is finite for all $H \geq H_0$ (i.e., $P(\limsup_{H > 0} \{N_{k}^H = \infty\}) = 0$).}

We first define the splittable set $S_k$ as the set of elements in the support of $\mu_{k-1}$ that can be "split" (i.e., be mutated by the Markov kernel $Q_k$) into an element in the support of $\mu_{k}$. Formally, it can be written as 

$$
S_{k} = \set{y \in \mathrm{Supp}(\mu_{k-1}) | Q_k(y, A_k) > 0}
$$
By assumption we have $A_k \subset \mathrm{supp}(\eta_k)$. Since for all $x \in A_k$ we have 
$$
\eta_k(x) = \sum_{y \in A_{k-1}} \mu_{k-1}(y) Q_k(y, x)
$$
then there must exist at least one $y \in S_k$ and thus $S_k$ is non-empty. Since $S_k \subseteq \mathrm{Supp}(\mu_{k-1})$ we have
$$
\mu_{k-1}(S_k) > 0.
$$

Now note that if $\mu_{k-1}^H(S_k) > 0$, then
$$
(\mu_{k-1}^H Q_k)(A_k)
=
\sum_{y \in E} \mu_{k-1}^H(y)Q_k(y,A_k)
> 0.
$$
Thus conditional on $\mathcal{H}_{k-1}^H$, every draw at step $k$ from $\nu_k^H$ has some fixed probability
$$
p_k^H = (\mu_{k-1}^H Q_k)(A_k) 
$$
of being alive. Define the event $R_{k-1}^H$ as the event the particle system has not halted up to this point, so
$$
R_{k-1}^H = \set{N_0^H < \infty,\dots, N_{k-1}^H < \infty}.
$$

Therefore, when both $R_{k-1}^H$ and $p_k^H > 0$ are true the number of draws $N_k^H$ needed to obtain $H$ alive particles is finite almost surely, (as in that case $N_k^H$ is conditionally distributed as negative binomial with size $H$ and probability $p_k^H$), meaning $P(\{N_k^H = \infty\} \cap \{\mu_{k-1}^H(S_k) > 0\} \cap R_{k-1}^H) = 0.$ 
Now, by our induction hypothesis,
$$
\dTV{\mu_{k-1}^H}{\mu_{k-1}} \toAS 0,
$$
and therefore
$$
\mu_{k-1}^H(S_k) \toAS \mu_{k-1}(S_k) > 0.
$$
Define the event 
$$
G_H = \set{\mu_{k-1}^H(S_k) > 0}.
$$
The convergence of $\mu_{k-1}^H(S_k)$ implies that $P(\liminf_{H \to \infty} G_H) = 1$ and $P(\limsup_{H \to \infty} G_H^C) = 0$. Further, our assumption that for $j=0,\dots, k-1$ we have $P(\limsup_{H \to \infty} \{N_j^H = \infty\}) = 0$ implies that $P(\limsup_{H \to \infty} (R_{k-1}^H)^C) = 0$

Now consider 
$$
\{N_k^H = \infty\} \subset (R_{k-1}^H)^C \cup G_H^C \cup \left( \{N_k^H = \infty\} \cap G_H \cap R_{k-1}^H \right)
$$
Thus 
\begin{align*}
P\left(\limsup_{H \to \infty} \{N_k^H = \infty\} \right) &\leq P\left(\limsup_{H \to \infty} (R_{k-1}^H)^C\right) \\
&+  P\left(\limsup_{H \to \infty} G_H^C\right) \\
&+ P\left( \limsup_{H \to \infty} \left(\{N_k^H = \infty\} \cap G_H \cap R_{k-1}^H \right)\right)
\end{align*}
We already know the first two terms are zero and we know for $\{N_k^H = \infty\} \cap G_H \cap R_{k-1}^H$ that $R_{k-1}^H$ and $G_H$ being true implies that for large enough $N_k^H$ we will sample $H$ alive particles so we have

$$
P\left( \limsup_{H \to \infty} \left(\{N_k^H = \infty\} \cap G_H \cap R_{k-1}^H \right)\right) = 0.
$$
Thus proving
$$
P\left( \limsup_{H \to \infty} \{N_k^H = \infty\} \right) = 0
$$

\smallskip
\noindent\textit{$\eta_k^{H}$ converges almost surely in total variation distance to $\eta_k$}

First define the measure 
$$\nu_k^H = \mu_{k-1}^H Q_k.$$ 
We know by Lemma \ref{lem-markov-kernel-continuity} that $\mu_{k-1}^H Q_k \toAS \mu_{k-1} Q_k = \eta_k$ so we know $\nu_k^H \toAS \eta_k$. Since we just proved $P\left(\limsup_{H \to \infty} \{N_k^H = \infty\}\right) = 0$ we know there exists some subset $\Omega_0$ of measure 1 where $N_k^H$ is finite for sufficiently large $H$ and thus $\eta_k^H$ is well defined. When $N_k^H = \infty$ then let $\eta_k^H$ be an arbitrary measure.

Now fix $\epsilon > 0$ and define the event 
$$
B^H_{x, \epsilon} = \set{\abs{\eta_k^H(\{x\}) - \nu_k^H(\{x\})} > \epsilon}
$$
Now define 
$$
D^H_{x, \epsilon} = B^H_{x, \epsilon} \cap \{N_k^H < \infty\}
$$

Since $N_k^H \geq H$ is always true consider that 

$$
D^H_{x, \epsilon} = \bigcup_{m=H}^\infty (B^H_{x, \epsilon} \cap \{N_{k}^H = m\})
$$

When $\{N_{k}^H = m\}$ we know that 
$$
\eta_k^H(\{x\}) = \frac{1}{m} \sum_{i=1}^{m} \I{\xi_k^{i} = x}
$$
so
$$
B^H_{x, \epsilon} \cap \{N_{k}^H = m\} \subset \set{\abs{\frac{1}{m} \sum_{i=1}^{m} \I{\xi_k^{i} = x} - \nu_k^H(\{x\})}> \epsilon}
$$
We will now union bound the probability of $D^H_{x, \epsilon}$.

\begin{align*}
P(D^H_{x, \epsilon}) &= P \left(\bigcup_{m=H}^\infty \{B^H_{x, \epsilon} \cap \{N_k^H = m\}\} \right) \\
&\leq \sum_{m=H}^\infty P \left(B^H_{x, \epsilon} \cap \{N_k^H = m\} \right) \\
&\leq \sum_{m=H}^\infty P \left(\abs{\frac{1}{m} \sum_{i=1}^{m} \I{\xi_k^{i} = x} - \nu_k^H(\{x\})}> \epsilon \right)
\end{align*}

Now recall that conditional on $\mathcal{H}_{k-1}^H$ we know $\xi_k^{i}$ are iid from the distribution $\nu_k^H$ so we know that 
$$
P(\xi_{k}^i = x | \mathcal{H}_{k-1}^H) = \nu_k^H(\{x\})
$$
So since $\I{\xi_k^{i} = x}|\mathcal{H}_{k-1}^H \iid \bern{\nu_k^H(\{x\})}$ by the conditional version of Hoeffding's inequality we have the conditional bound 
$$
P \left(\set{\abs{\frac{1}{m} \sum_{i=1}^{m} \I{\xi_k^{i} = x} - \nu_k^H(\{x\})}> \epsilon} \big\vert \mathcal{H}_{k-1}^H \right) \leq 2 \exp{-2m\epsilon^2}
$$
So stringing it all together we have

\begin{align*}
P(D^H_{x, \epsilon}) &\leq \sum_{m=H}^\infty P \left(\abs{\frac{1}{m} \sum_{i=1}^{m} \I{\xi_k^{i} = x} - \nu_k^H(\{x\})}> \epsilon \right) \\
&= \sum_{m=H}^\infty \E{P \left(\set{\abs{\frac{1}{m} \sum_{i=1}^{m} \I{\xi_k^{i} = x} - \nu_k^H(\{x\})}> \epsilon} \big\vert \mathcal{H}_{k-1}^H \right) } \\
&\leq \sum_{m=H}^\infty \E{2 \exp{-2m\epsilon^2} } \\
&= \sum_{m=H}^\infty 2 \exp{-2m\epsilon^2} 
\end{align*}

Now fix $\epsilon > 0$ and consider 
\begin{align*}
P(\dTV{\eta_k^H}{\nu_k^H} > \epsilon  \cap \{ N_k^H < \infty\} ) &= P\left(\frac{1}{2} \sum_{x \in E} \abs{\eta_k^H(\{x\}) - \nu_k^H(\{x\})} > \epsilon \cap \{ N_k^H < \infty\}\right) \\
&= P\left(\sum_{x \in E} \abs{\eta_k^H(\{x\}) - \nu_k^H(\{x\})} > 2\epsilon  \cap \{ N_k^H < \infty\} \right) \\
\end{align*}

Now note that for nonnegative random variables $X_1,\dots X_n$ we know via union bounding that 
$$
P\left( \sum_{i=1}^n X_i > \epsilon \right) \leq \sum_{i=1}^n P\left(X_i > \frac{\epsilon}{n}\right)
$$
so 
\begin{align*}
P(\dTV{\eta_k^H}{\nu_k^H} > \epsilon  \cap \{ N_k^H < \infty\} ) &\leq P\left(\sum_{x \in E} \abs{\eta_k^H(\{x\}) - \nu_k^H(\{x\})} > 2\epsilon  \cap \{ N_k^H < \infty\}  \right) \\
&\leq \sum_{x \in E} P\left( \abs{\eta_k^H(\{x\}) - \nu_k^H(\{x\})} > \frac{2\epsilon}{\abs{E}}  \cap \{ N_k^H < \infty\}  \right)\\
&= \sum_{x \in E} P\left( D_{x,\epsilon'}^H \right) \; \;  \text{ where } \epsilon' = \frac{2\epsilon}{\abs{E}} \\
&\leq \sum_{m=H}^\infty 2 \abs{E} \exp{-2m(\epsilon')^2} 
\end{align*}
Now define $r = \exp{-2(\epsilon')^2}$ so we have 
$$
P(\dTV{\eta_k^H}{\nu_k^H} > \epsilon  \cap \{ N_k^H < \infty\} ) \leq 2  \abs{E}  \sum_{m=H}^\infty r^m
$$

For $\epsilon > 0$ we know $r < 1$ and thus $\sum_{m=H}^\infty r^m = \frac{r^H}{1-r}$. Now consider
\begin{align*}
\sum_{H=1}^\infty P(\dTV{\eta_k^H}{\nu_k^H} > \epsilon  \cap \{ N_k^H < \infty\} ) &\leq \sum_{H=1}^\infty 2  \abs{E}  \sum_{m=H}^\infty r^m \\
&= \frac{2 \abs{E}}{1-r} \sum_{H=1}^\infty r^H \\
&= \frac{2 \abs{E}}{1-r} \frac{r}{1-r}
\end{align*}
Therefore since $\sum_{H=1}^\infty P(\set{\dTV{\eta_k^H}{\nu_k^H} > \epsilon}  \cap \{ N_k^H < \infty\} ) < \infty$ by the first Borel-Cantelli lemma we know that 
$$
P\left(\limsup_{H \to \infty} \left( \set{\dTV{\eta_k^H}{\nu_k^H} > \epsilon} \cap \{ N_k^H < \infty\} \right) \right) = 0
$$
Now we can prove $\dTV{\eta_k^H}{\nu_k^H} \toAS 0$. Consider that 
$$
\set{\dTV{\eta_k^H}{\nu_k^H} > \epsilon} = \left( \set{\dTV{\eta_k^H}{\nu_k^H} > \epsilon} \cap \{ N_k^H < \infty\}  \right) \cup \left( \set{\dTV{\eta_k^H}{\nu_k^H} > \epsilon} \cap \{ N_k^H = \infty\}  \right)
$$
Now recall we already proved that 
$$P(\limsup_{H \to \infty} \{ N_k^H = \infty\}) = 0$$ 
and 
$$P\left(\limsup_{H \to \infty} \left( \set{\dTV{\eta_k^H}{\nu_k^H} > \epsilon} \cap \{ N_k^H < \infty\}  \right) \right) = 0$$
so since 
$$
\limsup_{H \to \infty} \left(\set{\dTV{\eta_k^H}{\nu_k^H} > \epsilon} \cap \{ N_k^H = \infty\}\right) \subset \limsup_{H \to \infty} \{ N_k^H = \infty\}
$$
we know thus know that 
$$
P\left(\limsup_{H \to \infty} \set{\dTV{\eta_k^H}{\nu_k^H} > \epsilon}\right) = 0 
$$
Thus since the probability of $\dTV{\eta_k^H}{\nu_k^H} > \epsilon$ happening infinitely often is zero for arbitrary $\epsilon$, for each $j \in \mathbb{N}^+$, taking $\epsilon = 1/j$ gives
$$
P\left(
\limsup_{H \to \infty}
\left\{\dTV{\eta_k^H}{\nu_k^H} > \frac{1}{j}\right\}
\right)
= 0.
$$
Taking the countable intersection over $j \in \mathbb{N}^+$, there exists a set of probability $1$ on which for every $j$,
$$
\dTV{\eta_k^H}{\nu_k^H} \leq \frac{1}{j}
$$
for all sufficiently large $H$. Therefore,
$$
\dTV{\eta_k^H}{\nu_k^H} \toAS 0.
$$
Now since total variation is a metric by the triangle inequality we have 

$$
\dTV{\eta_k^H}{\eta_k} \leq \dTV{\eta_k^H}{\nu_k^H} + \dTV{\nu_k^H}{\eta_k}
$$
and since we've proved $\dTV{\eta_k^H}{\nu_k^H}, \dTV{\nu_k^H}{\eta_k} \toAS 0$ then that proves $\dTV{\eta_k^H}{\eta_k} \toAS 0$ as well. 

\medskip
\noindent\textit{$\mu_k^{H}$ converges almost surely in total variation distance to $\mu_k$}

Recall by \ref{lem-proj-product-continuity} that the projective product is continuous with respect to the total variation distance metric so since $\mu_k^{H} = g_k \cdot \eta_k^{H}$ then by the continuous mapping theorem $\mu_k^{H} \toAS \mu_k$.
\end{proof}

\end{theorem}

Since \(E\) is finite this result immediately implies strong consistency
of \(\inner{\eta_k^H}{\varphi}\) and \(\inner{\mu_k^H}{\varphi}\) for
all \(k\) and functions \(\varphi\). We now prove that the estimators
\(\inner{\eta_k^H}{\varphi}\) and \(\inner{\mu_k^H}{\varphi}\) are
asymptotically normal as well.

Note that \(\frac{H}{N_k^H} = \inner{\eta_k^H}{a_k}\) so we have \[
\inner{\eta_k^H}{a_k} \toAS \inner{\eta_k}{a_k} = \eta_k(A_k)
\]

\begin{theorem}[Central Limit Theorem for the Alive Particle
Filter]\protect\hypertarget{thm-alive-clt}{}\label{thm-alive-clt}

Assume the alive sets \(A_0,\dots, A_n\) are contained within the
supports of \(\eta_0, \dots, \eta_n\) respectively. Let \(\mu_k^{H}\)
and \(\eta_k^{H}\) be the empirical measures from the alive particle
system algorithm in Algorithm \ref{alg-alive-particle-filter} and
\(N_k^H\) be the alive particle sampling count from
\ref{def-alive-level}, then for all \(k=0,\dots, n\)

\[
\sqrt{H} \left(\inner{\mu_k^H}{\phi} - \inner{\mu_k}{\phi}\right) \toD \gauss{0}{v_k(\phi)}
\] where the asymptotic variance \(v_k(\phi)\) is given by the equation
\[
v_k(\phi) = \sum_{j=0}^k p_j \frac{\mathrm{var}(g_j R_{j+1:k}(\phi - \inner{\mu_k}{\phi}), \eta_j)}{\inner{\eta_j}{g_j R_{j+1:k}1}^2}
\] where \(p_j\) is the probability of the alive set \(A_j\) under
\(\eta_j\) so \(p_j = \eta_j(A_j)\) and \[
R_k(y, dx) = Q_k(y, dx) g_k(x) \text{ and } R_{j+1:k} \phi = R_{j+1}\cdot\cdot\cdot R_k \phi
\] With the convention \(R_{k+1:k} \phi = \phi\)

\begin{proof} 
The proof is almost exactly the same as theorem 4 in \citep{legland2005sequential}. The only difference is \citep{legland2005sequential} uses the following definition for their stopping time 
$$
\tilde{N}_k^H = \inf \set{N \geq 1 \big\vert \sum_{i=1}^N g_k(\xi_{k}^i) \geq H \cdot \sup_{x \in E}g(x)}.
$$
As such, in their alive algorithm 
$$
\frac{\tilde{N}_k^H}{H} \toP \rho_k = \frac{\sup_{x \in E}g(x)}{\inner{\eta_k}{g_k}}
$$
However, for the alive algorithm in Algorithm \ref{alg-alive-particle-filter} we instead have $\frac{H}{N_k^H} = \inner{\eta_k^H}{a_k}$ so by Theorem \ref{thm-alive-particle-as-convergence} 
$$
\inner{\eta_k^H}{a_k} \toAS \inner{\eta_k}{a_k} = \eta_k(A_k) = p_k
$$
and thus $\frac{N_k^H}{H} \toAS \frac{1}{p_k}$.

Everything else in the proof is the same. For our version we know for sufficiently large $H$ that the probability of halting is zero and $\inner{\mu_{k-1}^H Q_k}{g_k} > 0$. Like in \citep{legland2005sequential}, we begin with $\xi_0^1,\dots,$ as iid from $\eta_0$ and for $k=1,\dots,n$ we have $\xi_k^1,\dots,$ as conditionally iid from $\mu_{k-1}^H Q_k$. The only difference from \citep{legland2005sequential} is the stopping rule. 
\end{proof}

\end{theorem}

We will now consider a variant of Algorithm
\ref{alg-alive-particle-filter} that involves applying a
\(\mu_k\)-invariant markov kernel \(K_k\) \(m_H\) times after the
reweighting step. It turns out we can apply the \(K_k\) \(m\) times for
any \(m\) to our alive particles \(\zeta_k^i\) without changing the
weights and the convergence results still hold.

\begin{theorem}[Convergence of the Moved Empirical
Measure]\protect\hypertarget{thm-moved-measure-conv}{}\label{thm-moved-measure-conv}

For any \(k = 0, \dots, n\), let \(\mu_k^H\) be the empirical measure
generated by Algorithm \ref{alg-alive-particle-filter}. Let \(K_k\) be a
\(\mu_k\)-invariant Markov kernel and fix \(m \in \bbN\). When
\(N_k^H < \infty\), define the moved empirical measure

\[
\mu_{k,m}^H = \sum_{i=1}^H W_k^i \delta_{\tilde{\zeta}_i^H}(\cdot)
\] Where \[
\tilde{\zeta}_k^i \sim K_k^m(\zeta_k^i, \cdot) \; \; \text{ independently }
\] Then the moved empirical measure \(\mu_{k,m}^H\) still converges
almost surely in total variation distance to \(\mu_k\)

\begin{proof}

\smallskip
\noindent\textit{The moved empirical measure $\mu_{k,m}^H$ converges almost surely in TV to $\mu_k^H K_{k}^m$ as $H \to \infty$.}

First consider for an arbitrary $x \in E$ that 
$$
\mu_{k,m}^H (\{x\}) = \sum_{i=1}^H W_k^i \I{\tilde{\zeta}_k^i = x}
$$
and
$$
(\mu_k^H K_{k}^m)(\{x\}) = \sum_{i=1}^H W_k^i K^m_k(\zeta_k^i, \{x\})
$$
So 
$$
\mu_{k,m}^H (\{x\}) - (\mu_k^H K_{m}^H)(\{x\}) = \sum_{i=1}^H W_k^i \left( \I{\tilde{\zeta}_k^i = x} -  K^m_k(\zeta_k^i, x)\right)
$$

Now note that condition

Define $\mathcal{F}_k^H = \sigma(\mathcal{H}_{k-1}^H, \set{\zeta_k^i}, \set{W_k^i})$. We know conditioned on $\mathcal{F}_k^H$ that $\tilde{\zeta}_k^i \sim K_k^m(\cdot| \zeta_k^i)$ independently so 
$$
\E{ \I{\tilde{\zeta}_k^i = x} | \mathcal{F}_k^H } = K_k^m(x |\zeta_k^i) = K_k^m(\zeta_k^i, x)
$$
Thus we see that each summand term $W_k^i \left( \I{\tilde{\zeta}_k^i = x} -  K^m_k(\zeta_k^i, x) \right)$ is conditionally independent, mean zero, and bounded by $[-W_k^i, W_k^i]$ given $\mathcal{F}_k^H$. So by conditional Hoeffding's inequality we have 
$$
P\left(\abs{\mu_{k,m}^H (\{x\}) - (\mu_k^H K_{k}^m)(\{x\}) > \epsilon}| \mathcal{F}_k^H \right) \leq 2 \exp{\frac{-\epsilon^2 }{2 \sum_{i=1}^H (W_k^i)^2 }}
$$
Now recall that each $W_k^i$ is positive and bounded in $(0,1)$ so we have the bound 
$$
\sum_{i=1}^H (W_k^i)^2 \leq \left(\max_{1\leq j \leq H} W_j \right) \sum_{i=1}^H W_k^i = \max_{1\leq j \leq H} W_j
$$
and since $\exp{\frac{-\epsilon^2}{2 x}}$ is increasing in $x$ we have the bound 
$$
P\left(\abs{\mu_{k,m}^H (\{x\}) - (\mu_k^H K_{m}^H)(\{x\}) > \epsilon}| \mathcal{F}_k^H \right) \leq 2 \exp{\frac{-\epsilon^2 }{2 \max_{1\leq j \leq H} W_j }}
$$
Since our space $E$ is finite we know that 

$$
0 < \min_{x \in A_k} g(x) \leq g(x) \leq \max_{x \in A_k} g(x)
$$

and since we know for any $i$ that we have 
$$
W_k^i = \frac{g(\zeta_k^i) }{\sum_{j=1}^H g(\zeta_k^j)} \leq \frac{\max_{x \in A_k} g(x)}{H \cdot \min_{x \in A_k} g(x)}
$$
Let $C = \frac{\max_{x \in A_k} g(x)}{\min_{x \in A_k} g(x)}$ Then we have the bound 
$$
\max_{1\leq j \leq H} W_j  \leq \frac{C}{H}
$$
and thus 
$$
P\left(\abs{\mu_{k,m}^H (\{x\}) - (\mu_k^H K_{m}^H)(\{x\}) > \epsilon}| \mathcal{F}_k^H \right) \leq 2 \exp{\frac{-\epsilon^2 H}{2 C}}.
$$
Since the right hand side is constant this gives the unconditional bound 
$$
P\left(\abs{\mu_{k,m}^H (\{x\}) - (\mu_k^H K_{m}^H)(\{x\}) > \epsilon} \right) = \E{P\left(\abs{\mu_{k,m}^H (\{x\}) - (\mu_k^H K_{m}^H)(\{x\}) > \epsilon}| \mathcal{F}_k^H \right)} \leq 2 \exp{\frac{-\epsilon^2 H}{2 C}}
$$

Since the series $\sum_{H=1}^\infty \exp{\frac{-\epsilon^2 H}{2 C}}$ is finite by Borel-Cantelli we have that 
$$
P\left(\limsup_{H \to \infty} \left(\abs{\mu_{k,m}^H (\{x\}) - (\mu_k^H K_{m}^H)(\{x\}) > \epsilon}\right)  \right) = 0.
$$
Since $E$ is finite and the finite intersection of measure one sets is finite then that means there exists a set of measure one where 
$$
P\left(\limsup_{H \to \infty} \left(\sum_{x \in E} \abs{\mu_{k,m}^H (\{x\}) - (\mu_k^H K_{m}^H)(\{x\}) > \epsilon} \right) \right) = 0
$$
Thus implying that $\dTV{\mu_{k,m}^H}{\mu_k^H K_{k}^m} \toAS 0$.

\smallskip
\noindent\textit{The moved empirical measure $\mu_{k,m}^H$ converges almost surely in TV to $\mu_k$ as $H \to \infty$.}

Recall by Lemma \ref{lem-markov-kernel-continuity} that since $\mu_k^H \toAS \mu_k $ in total variation then we know that $\mu_k^H K_k^m \toAS \mu_k K_k^m$ as well. Now we note that since $K^m_k$ is $\mu_k$ invariant then $\mu_k K_k^m = \mu_k$. Now applying the triangle inequality we have 

$$
\dTV{\mu_{k,m}^H}{\mu_k} \leq \dTV{\mu_{k,m}^H}{\mu_k^H K_{k}^m} + \dTV{\mu_k^H K_{k}^m}{\mu_k} \toAS 0
$$
Thus proving the moved empirical measure $\mu_{k,m}^H$ converges to $\mu_k$ almost surely in total variation. 
\end{proof}

\end{theorem}

\begin{proposition}[CLT for Alive Moved Empirical
Measure]\protect\hypertarget{prp-moved-measure-clt}{}\label{prp-moved-measure-clt}

Assume the alive sets \(A_0,\dots, A_n\) are contained within the
supports of \(\eta_0, \dots, \eta_n\) respectively. Let \(\mu_k^{H}\)
and \(\eta_k^{H}\) be the empirical measures from the alive particle
system algorithm in Algorithm \ref{alg-alive-particle-filter}, \(N_k^H\)
be the alive particle sampling count from \ref{def-alive-level}, and
\(K_k\) be a \(\mu_k\)-invariant Markov kernel. Define
\(\hat{\mu}_{k,m}^H\) to be the moved empirical measure for some fixed
\(m \in \bbN\). Then for all \(k=0,\dots, n\)

\[
\sqrt{H} \left(\inner{\hat{\mu}_{k,m}^H}{\phi} - \inner{\mu_k}{\phi}\right) \toD \gauss{0}{\tilde{v}_k(\phi)}
\] for some asymptotic variance \(\tilde{v}_k(\phi)\)

Where \[
\tilde{v}_k(\phi) = v_k(K_k^m \phi) + p_k \frac{\inner{\eta_k}{g_k^2 \mathcal{V}_k^m \phi}}{\inner{\eta_k}{g_k}^2}
\] with \(K_k^m \phi(x)\) being

\[
K_k^m \phi(x) = \int \phi(y) K_k^m(y|x) dy.
\]

\begin{proof} Only a sketch of the proof is provided and the final asymptotic variance term is not presented. Generally, it will not be the same as in Theorem \ref{thm-alive-clt}. We begin by noting that 

$$
\begin{aligned}
\sqrt{H} \left(\inner{\hat{\mu}_{k,m}^H}{\phi} - \inner{\mu_k}{\phi}\right) &= \sqrt{H} \left(\inner{\hat{\mu}_{k,m}^H}{\phi} - \inner{\hat{\mu}_{k}^H}{K_k^m \phi} + \inner{\hat{\mu}_{k}^H}{K_k^m \phi} - \inner{\mu_k}{\phi}\right) \\
&= \sqrt{H} \left(\inner{\hat{\mu}_{k,m}^H}{\phi} - \inner{\hat{\mu}_{k}^H}{K_k^m \phi}\right) +  \sqrt{H}\left( \inner{\hat{\mu}_{k}^H}{K_k^m \phi} - \inner{\mu_k}{\phi}\right)
\end{aligned}
$$
Now define 
$$
A_H(\phi) = \sqrt{H} \left(\inner{\hat{\mu}_{k,m}^H}{\phi} - \inner{\hat{\mu}_{k}^H}{K_k^m \phi}\right)
$$
and
$$
B_H(\phi) = \sqrt{H}\left( \inner{\hat{\mu}_{k}^H}{K_k^m \phi} - \inner{\mu_k}{\phi}\right)
$$

Recall the $K_k^m \phi(x)$ expression can be thought of as $K_k^m \phi(x)= \E{\phi(Y)|X=x}$ where $Y|X \sim K_k^m(\cdot|X)$. So we see we have

$$
\sqrt{H} \left(\inner{\hat{\mu}_{k,m}^H}{\phi} - \inner{\mu_k}{\phi}\right) = A_H(\phi) + B_H(\phi)
$$

\medskip
\noindent\textbf{$A_H(\phi)$ is asymptotically normal}

To prove the Gaussian convergence we will use Theorem A.3 of \citep{SMCCLT}. To do that we must first define some new terms. 

\begin{itemize}
\item Define $\mathcal{F}_k^H = \sigma(\mathcal{H}_{k-1}^H, \set{\zeta_k^i}, \set{W_k^i})$ as the $sigma$-algebra generated by the entire particle system up to the pre-Markov move particles.
\item Define $\mathcal{F}_{k, i}^H$ to be the $\sigma$-algebra
$$
\mathcal{F}_{k, i}^H = \mathcal{F}_{k}^H \lor \sigma(\tilde{\zeta}_k^1,\dots, \tilde{\zeta}_k^i) = \sigma(\mathcal{F}_{k}^H \cup \sigma(\tilde{\zeta}_k^1,\dots, \tilde{\zeta}_k^i))
$$
Where $\mathcal{F}_{k, 0}^H = \mathcal{F}_{k}^H$
\item Define $U_{H,i}$ as  
$$
U_{H,i} = \sqrt{H} W_k^i \phi(\tilde{\zeta}_k^i)
$$
Notice that $U_{H,i}$ is $\mathcal{F}_{k, i}^H$ measurable. 
\item Define $\mathcal{V}_k^m \phi(x)$ as 
$$
\mathcal{V}_{k}^m \phi(x) = K_k^m \phi^2(x) - (K_k^m \phi(x))^2 
$$
We will later show that $\mathcal{V}_k^m \phi(x)$ is the conditional variance of $\tilde{\zeta}_k^i$ given $\mathcal{F}_{k,i-1}^H$, so
$$
\mathcal{V}_k^m \phi(\zeta_k^i) = \mathrm{Var}\left(\phi(\tilde{\zeta}_k^i) | \mathcal{F}_{k,i-1}^H\right) = K_k^m \phi^2(\zeta_k^i) - (K_k^m \phi(\zeta_k^i))^2.
$$
\end{itemize}

We will also now prove that 
$$
\E{U_{H,i} |\mathcal{F}_{k, i-1}^H} = \sqrt{H} W_k^i K_k^m\phi(\zeta_k^i)
$$
and 
$$
\E{U^2_{H,i} |\mathcal{F}_{k, i-1}^H} = H (W_k^i)^2 K_k^m\phi^2(\zeta_k^i)
$$
and

$$
\begin{aligned}
\E{U^2_{H,i} |\mathcal{F}_{k, i-1}^H} - \left(\E{U_{H,i} |\mathcal{F}_{k, i-1}^H}\right)^2 &= H (W_k^i)^2 \mathcal{V}_k^m \phi(\zeta_k^i) 
\end{aligned}
$$

To see this consider first that since $W_k^i \in \mathcal{F}_{k}^H \subset \mathcal{F}_{k, i-1}^H$ then we have 
$$
\E{U_{H,i} |\mathcal{F}_{k, i-1}^H} = \sqrt{H} W_k^i \E{\phi(\tilde{\zeta}_k^i)|\mathcal{F}_{k, i-1}^H}
$$
and
$$
\E{U^2_{H,i} |\mathcal{F}_{k, i-1}^H} = H (W_k^i)^2 \E{\phi(\tilde{\zeta}_k^i)^2|\mathcal{F}_{k, i-1}^H}
$$

Now consider that since $\mathcal{F}_{k, i}^H = \mathcal{F}_{k}^H \lor \sigma(\tilde{\zeta}_k^1,\dots, \tilde{\zeta}_k^i))$ and each $\tilde{\zeta}_k^i$ is conditionally independent given $\mathcal{F}_k^H$, then we know the conditional expectation of any function of $\tilde{\zeta}_k^i$ given $\mathcal{F}_{k, i-1}^H$ is the as given $\mathcal{F}_{k}^H$ so


$$
\E{\phi(\tilde{\zeta}_k^i)|\mathcal{F}_{k, i-1}^H} = \E{\phi(\tilde{\zeta}_k^i)|\mathcal{F}_{k}^H}
$$
and 

$$
\E{\phi(\tilde{\zeta}_k^i)^2|\mathcal{F}_{k, i-1}^H} = \E{\phi(\tilde{\zeta}_k^i)^2|\mathcal{F}_{k}^H}.
$$

Now consider, like in Theorem \ref{thm-moved-measure-conv}, we know conditioned on $\mathcal{F}_k^H$ that $\tilde{\zeta}_k^i \sim K_k^m(\zeta_k^i, \cdot)$ independently so we see the conditional mean of $\tilde{\zeta}_k^i$ is given by

$$
\E{\phi(\tilde{\zeta}_k^i) | \mathcal{F}_k^H} = \int \phi(x) K_k^m(x|\zeta_k^i) dx = K_k^m \phi(\zeta_k^i),
$$
and the second moment is 
$$
\E{\phi(\tilde{\zeta}_k^i)^2 | \mathcal{F}_k^H} = \int \phi(x)^2 K_k^m(x|\zeta_k^i) dx = K_k^m\phi^2(\zeta_k^i).
$$

Then we thus have 

$$
\E{\phi(\tilde{\zeta}_k^i)|\mathcal{F}_{k, i-1}^H} = \E{\phi(\tilde{\zeta}_k^i)|\mathcal{F}_{k}^H} = K_k^m\phi(\zeta_k^i)
$$
and 

$$
\E{\phi(\tilde{\zeta}_k^i)^2|\mathcal{F}_{k, i-1}^H} = \E{\phi(\tilde{\zeta}_k^i)^2|\mathcal{F}_{k}^H} = K_k^m\phi^2(\zeta_k^i),
$$

implying 

$$
\begin{aligned}
\E{U^2_{H,i} |\mathcal{F}_{k, i-1}^H} - \left(\E{U_{H,i} |\mathcal{F}_{k, i-1}^H}\right)^2 &= H (W_k^i)^2 K_k^m\phi^2(\zeta_k^i) - H (W_k^i)^2 (K_k^m\phi(\zeta_k^i))^2 \\
&= H (W_k^i)^2 \left( K_k^m\phi^2(\zeta_k^i) - (K_k^m\phi(\zeta_k^i))^2 \right) \\
&= H (W_k^i)^2 \mathcal{V}_k^m \phi(\zeta_k^i) 
\end{aligned}
$$

Now, to apply Theorem A.3 of \citep{SMCCLT} the following conditions must hold  

\begin{enumerate}
\item For every fixed $H$, $\set{\mathcal{F}_{k, i}^H}_{i=1}^\infty$ is a filtration. 
\item For each $i = 1,\dots,H$ then $U_{H,i}$ is $\mathcal{F}_{k, i}^H$ measurable.
\item $\E{U^2_{H,i} |\mathcal{F}_{k, i-1}^H} < \infty$
\item There exists some $\sigma^2 > 0$ such that 
$$
\sum_{i=1}^H \left(\E{U^2_{H,i} |\mathcal{F}_{k, i-1}^H} - \left(\E{U_{H,i} |\mathcal{F}_{k, i-1}^H}\right)^2 \right) \toP \sigma^2
$$
\item For any $\epsilon > 0$ we have
$$
\sum_{i=1}^H \E{U_{H,i}^2 \I{\abs{U_{H,i}} \geq \epsilon } | \mathcal{F}_{k, i-1}^H } \toP 0
$$
\end{enumerate}

If these conditions hold then by Theorem A.3 of \citep{SMCCLT} for any $u \in \bbR$ we have

$$
\E{\exp{iu \sum_{i=1}^H \left(  U_{H,i}  - \E{U_{H,i} |\mathcal{F}_{k, i-1}^H} \right) } | \mathcal{F}_{k}^H } \toP \exp{-\frac{u^2}{2} \sigma_k^2}
$$
(because $\mathcal{F}_{k, 0}^H = \mathcal{F}_{k}^H$).

\smallskip
\noindent\textit{Condition 1. is satisfied }

For condition 1. this is obvious from the definition of $\mathcal{F}_{k, i}^H$.

\noindent\textit{Condition 2. is satisfied }
This is true because $ W_k^i $ is $\mathcal{F}_k^H$ measurable and $\phi(\tilde{\zeta}_k^i)$ is $\sigma(\tilde{\zeta}_k^i)$ measurable. 

\noindent\textit{Condition 3. is satisfied}
Recall we already proved 
$$
\E{U^2_{H,i} |\mathcal{F}_{k, i-1}^H} = H (W_k^i)^2 K_k^m\phi^2(\zeta_k^i) < \infty.
$$

\noindent\textit{Condition 4. is satisfied }

Recall we already proved that 
$$
\begin{aligned}
\E{U^2_{H,i} |\mathcal{F}_{k, i-1}^H} - \left(\E{U_{H,i} |\mathcal{F}_{k, i-1}^H}\right)^2 &= H (W_k^i)^2 \mathcal{V}_k^m \phi(\zeta_k^i) 
\end{aligned}
$$
So we have 
$$
\begin{aligned}
\sum_{i=1}^H \left(\E{U^2_{H,i} |\mathcal{F}_{k, i-1}^H} - \left(\E{U_{H,i} |\mathcal{F}_{k, i-1}^H}\right)^2 \right) &= \sum_{i=1}^H H (W_k^i)^2 \mathcal{V}_k^m \phi(\zeta_k^i) \\
 &= H\sum_{i=1}^H (W_k^i)^2 \mathcal{V}_k^m \phi(\zeta_k^i) 
\end{aligned}
$$

Now  consider the following 
$$
\begin{aligned}
H \sum_{i=1}^H (W_k^i)^2 \mathcal{V}_k^m \phi(\zeta_k^i) &= H \sum_{i=1}^H \left(\frac{g(\zeta_k^i)}{\sum_{j=1}^H g(\zeta_k^j)}\right)^2 \mathcal{V}_k^m \phi(\zeta_k^i) \\
&= H \frac{1}{\left(\sum_{j=1}^H g(\zeta_k^j)\right)^2}\sum_{i=1}^H g(\zeta_k^i)^2 \mathcal{V}_k^m \phi(\zeta_k^i) \\
&= \frac{(N_k^H)^2}{(N_k^H)^2} H \frac{1}{\left(\sum_{j=1}^H g(\zeta_k^j)\right)^2}\sum_{i=1}^H g(\zeta_k^i)^2 \mathcal{V}_k^m \phi(\zeta_k^i) \\
&= \frac{1}{N_k^H} H \left(\frac{N_k^H}{\sum_{j=1}^H g(\zeta_k^j)}\right)^2\sum_{i=1}^H \frac{g(\zeta_k^i)^2 \mathcal{V}_k^m \phi(\zeta_k^i)}{N_k^H} \\
&= \frac{H}{N_k^H} \left(\frac{N_k^H}{\sum_{j=1}^H g(\zeta_k^j)}\right)^2 \left( \frac{1}{N_k^H} \sum_{i=1}^H  g(\zeta_k^i)^2 \mathcal{V}_k^m \phi(\zeta_k^i)\right) \\
&= \frac{H}{N_k^H} \left(\frac{1}{\inner{\eta_k^H}{g_k}}\right)^2 \left( \frac{1}{N_k^H} \sum_{i=1}^H  g(\zeta_k^i)^2 \mathcal{V}_k^m \phi(\zeta_k^i)\right)
\end{aligned}
$$

Now recall that for a function $\varphi$ we define $\inner{\eta_k^H}{\varphi}$
$$
\inner{\eta_k^H}{\varphi} = \frac{1}{N_k^H} \sum_{i=1}^{N_k^H} \varphi(\xi_k^i)
$$
Notice when $\varphi$ is a function that involves a product with $g_k$ as a term then it reduces to a sum over just the $H$ alive particles $\zeta_k^1,\dots,\zeta_k^H$ so we see that 
$$
\frac{1}{N_k^H} \sum_{i=1}^H  g(\zeta_k^i)^2 \mathcal{V}_k^m \phi(\zeta_k^i) = \frac{1}{N_k^H} \sum_{j=1}^{N_k^H}  g(\xi_k^j)^2 \mathcal{V}_k^m \phi(\xi_k^j) = \inner{\eta_k^H}{g_k^2 \mathcal{V}_k^m \phi}
$$
thus we have
$$
\begin{aligned}
H \sum_{i=1}^H (W_k^i)^2 \mathcal{V}_k^m \phi(\zeta_k^i) &= \frac{H}{N_k^H} \left(\frac{1}{\inner{\eta_k^H}{g_k}}\right)^2 \left( \frac{1}{N_k^H} \sum_{i=1}^{H}  g(\zeta_k^i)^2 \mathcal{V}_k^m \phi(\zeta_k^i)\right) \\
&= \frac{H}{N_k^H} \left(\frac{1}{\inner{\eta_k^H}{g_k}}\right)^2 \inner{\eta_k^H}{g_k^2 \mathcal{V}_k^m \phi} \\
&= \frac{H}{N_k^H} \frac{\inner{\eta_k^H}{g_k^2 \mathcal{V}_k^m \phi}}{\inner{\eta_k^H}{g_k}^2} 
\end{aligned}
$$

Now we know from the almost sure convergence of $\eta_k^H$ that 
$$
\frac{H}{N_k^H} \toAS p_k, \; \;\; \inner{\eta_k^H}{g_k^2 \mathcal{V}_k^m \phi} \toAS \inner{\eta_k}{g_k^2 \mathcal{V}_k^m \phi}, \; \;\; \inner{\eta_k^H}{g_k}^2 \toAS \inner{\eta_k}{g_k}^2
$$
where $p_k = \eta_k(A_k)$ is the alive probability for step $k$. 

So, we have that 
$$
H \sum_{i=1}^H (W_k^i)^2 \mathcal{V}_k^m \phi(\zeta_k^i) \toAS p_k \frac{\inner{\eta_k}{g_k^2 \mathcal{V}_k^m \phi}}{\inner{\eta_k}{g_k}^2}
$$

which of course, implies 
$$
\sum_{i=1}^H \left(\E{U^2_{H,i} |\mathcal{F}_{k, i-1}^H} - \left(\E{U_{H,i} |\mathcal{F}_{k, i-1}^H}\right)^2 \right)  \toP p_k \frac{\inner{\eta_k}{g_k^2 \mathcal{V}_k^m \phi}}{\inner{\eta_k}{g_k}^2}
$$

Note we assume here $p_k \frac{\inner{\eta_k}{g_k^2 \mathcal{V}_k^m \phi}}{\inner{\eta_k}{g_k}^2} > 0$.

\noindent\textit{Condition 5. is satisfied }

Let $\|\phi\|_\infty = \max_{x\in E}\abs{\phi(x)}$. Since $E$ is finite we know $\|\phi\|_\infty$ is finite. Therefore we know for all $i$ that 

$$
\begin{aligned}
\abs{U_{H,i}} &= \sqrt{H} W_k^i \abs{\phi(\tilde{\zeta}_k^i)} \\
&\leq \sqrt{H} \|\phi\|_\infty \max_{1\leq j \leq H} W_k^j 
\end{aligned}
$$
Now recall from Theorem \ref{thm-moved-measure-conv} we proved that
$$
\max_{1\leq j\leq H} W_k^j \leq \frac{C}{H},
$$
where
$$
C = \frac{\max_{x\in A_k}g_k(x)}{\min_{x\in A_k}g_k(x)} < \infty.
$$
Therefore we have
$$
\begin{aligned}
\abs{U_{H,i}} &\leq \frac{\|\phi\|_\infty}{\sqrt{H}} C \to 0
\end{aligned}
$$
as $H \to \infty$. 

Now fix $\epsilon > 0$. By the convergence above, there exists $\delta_\epsilon$ such that for all $\tilde{H} \geq \delta_{\epsilon}$ we have 
$$
\abs{U_{\tilde{H},i}} < \epsilon
$$
thus implying $\I{\abs{U_{\tilde{H},i}} \geq \epsilon } = 0$ as well. Thus we see when $\tilde{H} \geq \delta_{\epsilon}$ then
$$
\sum_{i=1}^{\tilde{H}} \E{U_{\tilde{H},i}^2 \I{\abs{U_{\tilde{H},i}} \geq \epsilon } | \mathcal{F}_{k, i-1}^{\tilde{H}} } = 0,
$$
thus proving 
$$
\lim\limits_{H \to \infty} \sum_{i=1}^H \E{U_{H,i}^2 \I{\abs{U_{H,i}} \geq \epsilon } | \mathcal{F}_{k, i-1}^H } = 0.
$$

\noindent\textbf{The Conditional CLT for $A_H(\phi)$ }

Thus, having proved all the conditions for Theorem A.3 of \citep{SMCCLT} to hold we thus have that for any $u \in \bbR$ then

$$
\E{\exp{iu \left( \sum_{i=1}^H U_{H,i}  - \E{U_{H,i} |\mathcal{F}_{k, i-1}^H} \right) } | \mathcal{F}_{k, 0}^H } \toP \exp{-\frac{u^2}{2} \sigma_k^2}
$$
where $\mathcal{F}_{k, 0}^H = \mathcal{F}_{k}^H$ and 
$$
\sigma_k^2 = p_k \frac{\inner{\eta_k}{g_k^2 \mathcal{V}_k^m \phi}}{\inner{\eta_k}{g_k}^2}
$$

Now notice that 

$$
\begin{aligned}
A_H(\phi) &= \sqrt{H} \left(\inner{\hat{\mu}_{k,m}^H}{\phi} - \inner{\hat{\mu}_{k}^H}{K_k^m \phi}\right) \\
&= \sqrt{H} \sum_{i=1}^H \left(W^i_k \phi(\tilde{\zeta}_k^i)  - W^i_k K_k^m \phi(\zeta_k^i) \right) \\
&= \sqrt{H} \sum_{i=1}^H W^i_k \left( \phi(\tilde{\zeta}_k^i)  - K_k^m \phi(\zeta_k^i) \right) \\
&=  \sum_{i=1}^H \left( \sqrt{H} W^i_k \phi(\tilde{\zeta}_k^i)  - \sqrt{H} W^i_k K_k^m \phi(\zeta_k^i) \right) \\
&=  \sum_{i=1}^H \left( U_{H,i}  - \E{U_{H,i} |\mathcal{F}_{k, i-1}^H} \right)
\end{aligned}
$$

So we have proved exactly that 

$$
\E{\exp{iu A_H(\phi)} | \mathcal{F}_{k}^H } \toP \exp{-\frac{u^2}{2} \sigma_k^2}
$$

\medskip
\noindent\textbf{$B_H(\phi)$ is asymptotically normal}

We begin by noting that since $K_k^m$ is $\mu_k$-invariant then we have 

$$
\mu_k(y) = \int \mu_k(x) K_k^m(y|x) dx
$$

$$
\begin{aligned}
\inner{\mu_k}{\phi} &= \int \phi(y) \mu_k(y) dy \\
&= \int \phi(y) \left(\int \mu_k(x) K_k^m(y|x) dx\right) dy \\
&= \int \mu_k(x) \int \phi(y)  K_k^m(y|x) dy \; dx \\
&= \int \mu_k(x) K_k^m \phi(x) dx \\
&= \inner{\mu_k}{K_k^m \phi}
\end{aligned}
$$
Now notice then that for $B_H(\phi)$ we have
$$
B_H(\phi) = \sqrt{H}\left( \inner{\hat{\mu}_{k}^H}{K_k^m \phi} - \inner{\mu_k}{K_k^m \phi}\right)
$$
so we know from Theorem \ref{thm-alive-clt} that its asymptotically normal with asymptotic variance $v_k(K_k^m \phi)$, in other words
$$
B_H(\phi) \toD \gauss{0}{v_k(K_k^m \phi)}
$$

\noindent\textbf{Full CLT for $A_H(\phi) + B_H(\phi)$}

Now for the full, unconditional CLT consider the following. Fix $u \in \bbR$ and consider

$$
\begin{aligned}
\E{\exp{iu (A_H(\phi) + B_H(\phi))}} &= \E{\E{\exp{iu (A_H(\phi) + B_H(\phi))}|\mathcal{F}_{k}^H }}  \\
&= \E{\E{\exp{iu A_H(\phi) + iuB_H(\phi)}|\mathcal{F}_{k}^H }}  \\
&= \E{\E{\exp{iu A_H(\phi)} \exp{iuB_H(\phi)}|\mathcal{F}_{k}^H }}
\end{aligned}
$$

Now recall that
$$
B_H(\phi) = \sqrt{H}\left( \inner{\mu_{k}^H}{K_k^m \phi} - \inner{\mu_k}{K_k^m \phi}\right)
$$
and thus it is $\mathcal{F}_{k}^H$ measurable so
$$
\begin{aligned}
\E{\exp{iu (A_H(\phi) + B_H(\phi))}} &= \E{\E{\exp{iu A_H(\phi)} \exp{iuB_H(\phi)}|\mathcal{F}_{k}^H } } \\
&= \E{\exp{iuB_H(\phi)} \E{\exp{iu A_H(\phi)} |\mathcal{F}_{k}^H } }
\end{aligned}
$$
Now we know from our results on $A_H(\phi)$ that
$$
\E{\exp{iu A_H(\phi)} |\mathcal{F}_{k}^H } \toP \exp{-\frac{u^2}{2} \sigma_k^2}
$$

To show joint convergence define 
$$
C_H(u) = \E{\exp{iu A_H(\phi)} |\mathcal{F}_{k}^H } 
$$
and
$$
c(u) = \exp{-\frac{u^2}{2} \sigma_k^2}
$$

We know $C_H(u) \toP c(u)$ and since the function $t \mapsto \exp{-it}$ is bounded in absolute value by 1 we know that
$$
\abs{C_H(u) - c(u)} \leq 2.
$$
Thus since $C_H(u) $ is uniformly bounded its convergence in probability implies convergence in $L^1$ as well so 

$$
\E{\abs{C_H(u) - c(u)}} \to 0
$$

$$
\begin{aligned}
\abs{\E{\exp{iu (A_H(\phi) + B_H(\phi))}} - c(u) \E{\exp{iu B_H(\phi)}} } &= \abs{ \E{\exp{iuB_H(\phi)} C_H(u)}  - c(u) \E{\exp{iu B_H(\phi)}} } \\
&= \abs{ \E{\exp{iuB_H(\phi)} ( C_H(u)   - c(u)) }}
\end{aligned}
$$

Now consider by the fact that $t \mapsto \exp{-it}$ is bounded in absolute value by 1

$$
\begin{aligned}
\abs{ \E{\exp{iuB_H(\phi)} ( C_H(u)   - c(u)) }} &\leq  \E{\abs{\exp{iuB_H(\phi)} ( C_H(u)   - c(u)) }} \\
&= \E{\abs{\exp{iuB_H(\phi)}} \abs{C_H(u)   - c(u)) }} \\
&\leq \E{\abs{C_H(u)  - c(u)) }}
\end{aligned}
$$
and since $\abs{C_H(u)}  \toLnum{1} c(u)$ we know that $\E{\abs{C_H(u)  - c(u)) }} \to 0$ so we have proved that 

$$
\E{\exp{iu (A_H(\phi) + B_H(\phi))}} - c(u) \E{\exp{iu B_H(\phi)}} \to 0
$$

Now recall for $B_H(\phi)$ we proved that
$$
B_H(\phi) \toD \gauss{0}{v_k(K_k^m \phi)}
$$

implying
$$
\begin{aligned}
\E{\exp{iu  B_H(\phi)}} \to  \exp{-\frac{u^2}{2} v_k(K_k^m \phi) }
\end{aligned}
$$

so
$$
c(u) \E{\exp{iu B_H(\phi)}} \to \exp{-\frac{u^2}{2} (v_k(K_k^m \phi) + \sigma_k^2) }.
$$
Thus we have proved for arbitrary $u$ that 

$$
\begin{aligned}
\E{\exp{iu (A_H(\phi) + B_H(\phi))}} \to  \exp{-\frac{u^2}{2} (v_k(K_k^m \phi) + \sigma_k^2)}
\end{aligned}
$$

Proving finally that 

$$
\sqrt{H} \left(\inner{\hat{\mu}_{k,m}^H}{\phi} - \inner{\mu_k}{\phi}\right) = A_H(\phi) + B_H(\phi) \toD \gauss{0}{v_k(K_k^m \phi) + p_k \frac{\inner{\eta_k}{g_k^2 \mathcal{V}_k^m \phi}}{\inner{\eta_k}{g_k}^2}}
$$

\end{proof}

\end{proposition}

\newpage

\subsubsection{Alive SMC Sampler
Results}\label{sec-alive-smcs-convergence-results}

Given what we have proved generally about our alive particle system we
will now demonstrate that Algorithm \ref{alg-alive-smcs}, referred to as
the alive SMC sampler algorithm, (of which the gSMC algorithm in
Algorithm \ref{alg-gsmcs} is a special case of) has all the desired
convergence properties.

\begin{algorithm}[H]
\begin{algorithmic}[1]
\REQUIRE{target distributions $\set{\pi_r}_{r=1}^D$, initial proposal distribution $q_1$, forward kernels $\{M_r\}^D_{r=2}$, backward kernels $\{L_r\}^{D-1}_{r=1}$ }
\STATE{set  $T_{1} = 0$}
\FOR{$i=1$ to $N$}
\STATE{set $p=0$}
\WHILE{$p = 0$}
\STATE{sample $\xi_1^\ast \sim q_1$ }
\STATE{set $p=\pi_{1}(\xi^\ast_{1})$}
\STATE{set $T_{1} = T_{1} + 1$}
\ENDWHILE
\STATE{set $\xi^{(i)}_{1}=\xi^\ast_{1}$}
\STATE{compute the initial importance weight as
\begin{equation*}
w_{1}^{(i)}(\xi^{(i)}_{1})
= \frac{\gamma_1(\xi^{(i)}_{1})}{q_1(\xi^{(i)}_{1})}
\end{equation*}
}
\ENDFOR

\FOR{$r=1$ to $D-1$}
\STATE{Set $T_{r+1} = 0$}
\FOR{$i=1$ to $N$}
\STATE{set $p=0$}
\WHILE{$p = 0$}
\STATE{sample parent index $i^\prime \in \set{1,\dots,N}$ with $\Pr(i^\prime = j) \propto w_r^{(j)}$ for each $j=1,\ldots,N$}
\STATE{sample $\xi_{r+1}^\ast$ from the forward kernel $M_{r+1}(\cdot \mid \xi^{(i^\prime)}_{r})$}
\STATE{set $p=\pi_{r+1}(\xi^\ast_{r+1})$}
\STATE{Set $T_{r+1} = T_{r+1} + 1$}
\ENDWHILE
\STATE{set $\xi^{(i)}_{r+1}=\xi^\ast_{r+1}$}
\STATE{compute the incremental weight as
\begin{equation*}
w_{r+1}^{(i)}(\xi^{(i')}_{r}, \xi_{r+1}^{(i)})
= \frac{\gamma_{r+1}(\xi^{(i)}_{r+1}) L_{r}( \xi^{(i^\prime)}_{r} \mid \xi_{r+1}^{(i)}) }{ \gamma_{r}(\xi^{(i^\prime)}_{r}) M_{r+1}( \xi_{r+1}^{(i)}\mid \xi^{(i^\prime)}_{r})}
\end{equation*}
}
\ENDFOR
\ENDFOR
\end{algorithmic}
\caption{The Alive SMC Sampler algorithm}
\label{alg-alive-smcs}
\end{algorithm}

We will now show that as long as a forward reachability requirement
(i.e., all particles in the support of \(\pi_r\) can be obtained from
applying the forward kernels) and a backwards support compatibility
requirements (i.e., any time a forward move from the support of
\(\pi_{r-1}\) to \(\pi_r\) has positive probability then the backwards
move also does) are satisfied then strong convergence results hold for
Algorithm \ref{alg-gsmcs} (at least one finite sample spaces).

\begin{definition}[Reachability]\protect\hypertarget{def-reachability}{}\label{def-reachability}

Suppose we have a sequence of target distributions
\(\{\pi_r\}_{r=1}^D\), an initial proposal \(q_1\), and forward and
backward kernels \(\{M_r\}_{r=2}^D\) and \(\{L_r\}_{r=1}^{D-1}\) on a
finite space \(\tilde{E}\). We say a particle \(\xi_1\) in the support
of \(\pi_1\) (so \(\xi_s \in \mathrm{Supp}(\pi_1)\)) is reachable if it
has positive probability under \(q_1\), i.e., \(q(\xi_1) > 0\). For
\(r>1\) we say a particle \(\xi_r \in \mathrm{Supp}(\pi_r)\) is
reachable if there exists a reachable
\(\xi_{r-1} \in \mathrm{Supp}(\pi_{r-1})\) such that
\((\xi_r, \xi_{r-1})\) have positive forward probability, i.e.,

\[M_r(\xi_r|\xi_{r-1}) >0 \]

We say that a target distribution \(\pi_r\) is reachable is all
particles \(\xi_r\) in its support are reachable. We say that the entire
sequence \(\{\pi_r\}_{r=1}^D\) is reachable if each \(\pi_r\) is
reachable.

\end{definition}

The reachability requirement essentially says for all \(r\), for any
particle \(\xi_r\) in the support of \(\pi_r\) we can find a sequence of
particles \(\xi_1,\dots,\xi_{r-1}\) such that the forward kernel
probabilities are all positive.

\begin{definition}[Support-Compatible Backward
Kernels]\protect\hypertarget{def-support-compatible-backward}{}\label{def-support-compatible-backward}

Suppose we have a sequence of target distributions
\(\{\pi_r\}_{r=1}^D\), an initial proposal \(q_1\), and forward and
backward kernels \(\{M_r\}_{r=2}^D\) and \(\{L_r\}_{r=1}^{D-1}\) on a
finite space \(\tilde{E}\).

For \(r = 2,\dots, D\) we say that the pair \(M_r, L_{r-1}\) are
support-compatible if for every \(\xi_r, \xi_{r-1}\) the probability of
going forward between the pair
(\(\gamma_{r-1}(\xi_{r-1}) M_{r}(\xi_{r}|\xi_{r-1})\)) is non-zero if,
and only if, the probability of going backwards between them
(\(\gamma_r(\xi_r) L_{r-1}(\xi_{r-1}|\xi_r)\)) is also non-zero. In
other words

\[
\gamma_r(\xi_r) L_{r-1}(\xi_{r-1}|\xi_r) > 0 \iff \gamma_{r-1}(\xi_{r-1}) M_{r}(\xi_{r}|\xi_{r-1}) > 0.
\]

This can be equivalently expressed as stating, for all
\(\xi_r \in \mathrm{Supp}(\pi_r)\) that the support of
\(L_{r-1}(\cdot|\xi_r)\) is equal to the set of
\(\xi_{r-1} \in \mathrm{Supp}(\pi_{r-1})\) where
\(M_r(\xi_r|\xi_{r-1})\) is positive, so \[
\mathrm{Supp}(L_{r-1}(\cdot|\xi_r)) = \set{\xi_{r-1} \in \mathrm{Supp}(\pi_{r-1}) \big\vert M_r (\xi_r|\xi_{r-1}) > 0}.
\]

If all pairs are support compatible then we say the forward and backward
kernels \(\{M_r\}_{r=2}^D\) and \(\{L_r\}_{r=1}^{D-1}\) are
support-compatible.

\end{definition}

\begin{corollary}[Reachability and Support-Compatible
Implications]\protect\hypertarget{cor-reachability-implications}{}\label{cor-reachability-implications}

Suppose we have a sequence of target distributions
\(\{\pi_r\}_{r=1}^D\), an initial proposal \(q_1\), and forward and
backward kernels \(\{M_r\}_{r=2}^D\) and \(\{L_r\}_{r=1}^{D-1}\) that
are both reachable and support-compatible. Then for all \(r=2,\dots, D\)
we have that the incremental weight function \(w_r(\xi_{r-1}, \xi_r)\)
is non-zero if, and only if, \(\gamma_r(\xi_r) > 0\) and
\(\gamma_{r-1}(\xi_{r-1}) M_{r}(\xi_r|\xi_{r-1}) > 0\). In other words,

\[
w_r(\xi_{r-1}, \xi_r) > 0 \iff \gamma_r(\xi_r), \;\gamma_{r-1}(\xi_{r-1}) M_{r}(\xi_r|\xi_{r-1}) > 0
\]

\begin{proof} 
The forward direction is trivial. The backwards direction follows immediately from Definition \ref{def-support-compatible-backward}. 
\end{proof}

\end{corollary}

We will now demonstrate Algorithm \ref{alg-alive-smcs} can be framed as
a special case of Algorithm \ref{alg-alive-particle-filter}, allowing
all our previous results to immediately hold for the Alive SMC Sampler
as well. Note that while these results are all presented in the context
of redistricting, they generalize to any such sampling procedure on a
finite state space. Unfortunately, there is a clash of notation involved
between Algorithm \ref{alg-alive-smcs} and
\ref{alg-alive-particle-filter}, so we will temporarily change the
notation for the plans we actually want from Algorithm
\ref{alg-alive-smcs} from \(\xi_r\) to \(z_r\) for these results. We
will also introduce a dummy ``empty'' plan \(z_0\). There is only 1 of
these.

\begin{definition}[Selection Function and Transition Kernel for the
Alive SMC Sampler
Algorithm]\protect\hypertarget{def-gsmc-to-particle-filter}{}\label{def-gsmc-to-particle-filter}

Suppose we actually want to sample the distributions
\(\{\pi_r\}_{r=1}^D\) on the space \(\tilde{E}\) using Algorithm
\ref{alg-alive-smcs} with an initial proposal \(q_1\) and forward and
backward kernels \(\{M_r\}_{r=2}^D\) and \(\{L_r\}_{r=1}^{D-1}\)
respectively. We will now define a new state space, selection functions,
and transition kernels such that performing Algorithm
\ref{alg-alive-smcs} on \(\tilde{E}\) is equivalent to performing
Algorithm \ref{alg-alive-particle-filter} on \(E\).

Define our new space \(E = \tilde{E} \times \tilde{E}\), so for
\(x \in E\) it is actually the tuple \(x = (z, \tilde{z})\) where
\(z , \tilde{z} \in \tilde{E}\). We set \(n = D-1\) and define our
selection functions, initial distribution, and Markov kernels as
follows.

For \(k=0\), we define our selection function \(g_0\) as the importance
weight function

\[
g_0(x) = g_0((z_0,z_1)) = \frac{\gamma_1(z_1)}{q_1(z_1)}
\] where \(q_1(z_1) > 0\) and 0 otherwise. We define our initial
distribution \(\eta_0\) as

\[
\eta_0 (z_0, z_1) \sim q_1(z_1).
\]

Recall that \(z_0\) is a dummy placeholder (the ``empty'' particle). The
Dirac delta point mass on \(z_0\) is omitted.

\noindent For \(k = 1,\dots, n\), we define our selection function
\(g_k\) as the incremental weight function:

\[
g_k(x) = g_k((z_k, z_{k+1})) = w_{k+1}(z_k, z_{k+1}) = \frac{\gamma_{k+1}(z_{k+1}) L_k(z_k|z_{k+1})}{\gamma_{k}(z_{k}) M_{k+1}(z_{k+1}|z_{k})}.
\] if the denominator is non-zero. If the denominator is zero define it
to be 0.

For the Markov kernels \(\{Q_k\}_{k=1}^{n}\), given elements
\(x_{k-1} = (z_{k-1}, \tilde{z}_k)\) and \(x_{k} = (z_{k}, z_{k+1})\),
we define our mutation kernel \(Q_k(x_k|x_{k-1})\) as the function which
takes the child plan \(\tilde{z}_k \in x_{k-1}\) and applies the SMC
Sampler forward kernel to sample
\(z_{k+1} \sim M_{k+1}(\cdot|\tilde{z}_k)\), returning
\(x_{k} = (\tilde{z}_{k}, z_{k+1})\). We can write this formally as

\[
    Q_k(x_k|x_{k-1}) = Q_k((z_{k}, z_{k+1})|(z_{k-1}, \tilde{z}_k)) = M_{k+1}(z_{k+1}|z_k) I(\tilde{z}_k = z_k).
\]

Note that because of the indicator function that
\(Q_k((z_{k}, z_{k+1})|(z_{k-1}, \tilde{z}_k))\) will always be zero
unless \(\tilde{z}_k = z_k\).

\end{definition}

We will now prove that defining the problem like in
Definition~\ref{def-gsmc-to-particle-filter} gives us plans which are
marginally distribution as \(\pi_r\) at each step.

\begin{proposition}[Alive Algorithm Distributions Yields Desired
Results]\protect\hypertarget{prp-alive-smcs-distribs}{}\label{prp-alive-smcs-distribs}

Suppose we wish to run Algorithm \ref{alg-alive-smcs} and we have a
sequence of target distributions \(\{\pi_r\}_{r=1}^D\), an initial
proposal \(q_1\), and forward and backward kernels \(\{M_r\}_{r=2}^D\)
and \(\{L_r\}_{r=1}^{D-1}\) that are both reachable and
support-compatible.

\noindent Then if we define \(E\), \(\eta_0\), \(\set{g_k}_{k=0}^n\),
\(\set{Q_k}_{k=1}^n\) as in Definition~\ref{def-gsmc-to-particle-filter}
and run Algorithm \ref{alg-alive-particle-filter}, then \(\mu_0(x)\) is
given by the equation

\[
\mu_0(x) = \mu_0((z_0, z_{1})) = \pi_{1}(z_{1}).
\]

For \(k = 1,\dots, n\), the resulting distributions \(\eta_k\) and
\(\mu_k\) are \[
\eta_k(x) = \eta_k((z_k, z_{k+1})) = \pi_{k}(z_{k}) M_{k+1}(z_{k+1}|z_{k})
\] and \[
\mu_k(x) = \mu_k((z_k, z_{k+1})) = \pi_{k+1}(z_{k+1}) L_k(z_k|z_{k+1}).
\]

\begin{proof}

We will prove this sequentially, beginning with $k=0$ (which reduces to essentially importance sampling).

\medskip
\noindent\textbf{Initial Step $k=0$}  

Recall that we assume our initial distribution is 

$$\eta_0(x) = \eta_0((z_0,z_1)) \sim q(z_1)$$
with $z_0$ just being a placeholder. Since our measure is discrete fix an arbitrary $\tilde{x}_0 = (\tilde{z}_0, \tilde{z}_1)$ and consider 
$$
\begin{aligned}
    P_{\mu_0}(X_0 = \tilde{x}_0) = \frac{g_0(\tilde{x}_0) \eta_0(\tilde{x}_0))}{\int g_0(x_0) \eta_0(x_0) dx_0} = \frac{g_0((\tilde{z}_0, \tilde{z}_1)) \eta_0((\tilde{z}_0, \tilde{z}_1)}{\int \int g_0((z_0, z_1)) \eta_0((z_0, z_1)) dz_0 \; dz_1}
\end{aligned}
$$
Now recall that we define 
$$
g_0(x_0) = g_0((z_0,z_1)) = \frac{\gamma_1(z_1)}{q_1(z_1)}
$$
so the integral simplifies to 
$$
\begin{aligned}
\frac{g_0((\tilde{z}_0, \tilde{z}_1)) \eta_0((\tilde{z}_0, \tilde{z}_1)}{\int g_0((z_0, z_1)) \eta_0((z_0, z_1)) dz_0 \; dz_1} &= \frac{\frac{\gamma_1(\tilde{z}_1)}{q_1(\tilde{z}_1)} q_1(\tilde{z}_1)}{\int\int \frac{\gamma_1(z_1)}{q_1(z_1)} q_1(z_1) dz_0 \; dz_1} \\
&= \frac{\frac{\gamma_1(\tilde{z}_1)}{q_1(\tilde{z}_1)} q_1(\tilde{z}_1)}{\int \frac{\gamma_1(z_1)}{q_1(z_1)} q_1(z_1)  dz_1} 
\end{aligned}
$$
Now by the reachability assumption this guarantees that $\frac{\gamma_1(\tilde{z}_1)}{q_1(\tilde{z}_1)}$ is always well defined on the support of $\pi_1$ and thus 
$$
\int \frac{\gamma_1(z_1)}{q_1(z_1)} q_1(z_1)  dz_1 = \int \gamma_1(z_1)  dz_1 = Z_1
$$
so our expression simplifies to 
$$
\begin{aligned}
\frac{g_0((\tilde{z}_0, \tilde{z}_1)) \eta_0((\tilde{z}_0, \tilde{z}_1)}{\int g_0((z_0, z_1)) \eta_0((z_0, z_1)) dz_0 \; dz_1} &= \frac{\frac{\gamma_1(\tilde{z}_1)}{q_1(\tilde{z}_1)} q_1(\tilde{z}_1)}{\int \frac{\gamma_1(z_1)}{q_1(z_1)} q_1(z_1)  dz_1} \\
&= \frac{\gamma_1(\tilde{z}_1)}{\int \gamma_1(z_1) dz_1} \\
&= \frac{\gamma_1(\tilde{z}_1)}{Z_1} \\
&= \pi_1(\tilde{z}_1)
\end{aligned}
$$

and thus we've proved that $\mu_0((z_0, z_1)) = \pi_1(z_1)$.

\medskip
\noindent\textbf{Proving for $\eta_1((z_1, z_2)) = M_{2}(z_{2}|z_1)   \pi_1(z_1) $ } 

Now for $k=1$ consider that we define $\eta_1 = \mu_0 Q_1$. We thus have 

$$ 
\begin{aligned}
P_{\eta_1}(X_1 = \tilde{x}_1) &= \int Q_1(\tilde{x}_1|x_0) \mu_0(x_0) dx_0 \\
&= \int \int Q_1((\tilde{z}_1, \tilde{z}_2)|(z_0,z_1)) \mu_0((z_0,z_1)) dz_0 dz_1 \\
&= \int  \int M_{2}(\tilde{z}_{2}|\tilde{z}_1) I(\tilde{z}_1 = z_1)  \pi_1(z_1) dz_0 dz_1 \\
&= \int M_{2}(\tilde{z}_{2}|\tilde{z}_1) I(\tilde{z}_1 = z_1)  \pi_1(z_1) dz_1 
\end{aligned} 
$$

Now consider that for $\int M_{2}(\tilde{z}_{2}|\tilde{z}_1) I(\tilde{z}_1 = z_1)  \pi_1(z_1) dz_1$ the integrand is only potentially non-zero when $I(\tilde{z}_1 = z_1) = 1$ so it actually just simplifies to the value at $\tilde{z}_1$ yielding 
$$ 
\begin{aligned}
P_{\eta_1}(X_1 = \tilde{x}_1) &= \int M_{2}(\tilde{z}_{2}|\tilde{z}_1) I(\tilde{z}_1 = z_1)  \pi_1(z_1) dz_1 \\
&= M_{2}(\tilde{z}_{2}|\tilde{z}_1)   \pi_1(\tilde{z}_1) 
\end{aligned} 
$$
proving that for $k=1$ then $\eta_1((z_1, z_2)) = \pi_1(z_1) M_2(z_2|z_1)$. 

\medskip
\noindent\textbf{Proving for $k=1,\dots,n$ that $\mu_k((z_k, z_{k+1})) = \pi_{k+1}(z_{k+1}) L_{k}(z_{k}|z_{k+1}) $ }  

Given that 
$$
\eta_k((z_k, z_{k+1})) = \pi_k(z_k) M_{k+1}(z_{k+1}|z_k),
$$ 

consider the term $g(x_k)\eta_k(x_k) = g_k((z_k, z_k)) \eta_k((z_k, z_{k+1})$ can be written as 
$$
\begin{aligned}
g_k(x_k) \eta_k(x_k) &= \frac{\gamma_{k+1}(z_{2}) L_k(z_k|z_{k+1})}{\gamma_{k}(z_{k}) M_{k+1}(z_{k+1}|z_{k})}  \pi_k(z_k) M_{k+1}(z_{k+1}|z_k) \\
&= \frac{\gamma_{k+1}(z_{k+1}) L_k(z_k|z_{k+1})}{\gamma_{k}(z_{k}) M_{k+1}(z_{k+1}|z_{1})}  \frac{\gamma_k(z_k)}{Z_k} M_{k+1}(z_{k+1}|z_k) 
\end{aligned}
$$

Given the support-compatibility assumption we know that whenever the numerator $\gamma_{k+1}(z_{k+1}) L_k(z_k|z_{k+1}) > 0$ then we also have $\gamma_{k}(z_{k}) M_{k+1}(z_{k+1}|z_{k}) > 0$ and vice versa, so the entire ratio $\frac{\gamma_{k+1}(z_{k+1}) L_k(z_k|z_{k+1})}{\gamma_{k}(z_{k}) M_{k+1}(z_{k+1}|z_{k})}$ is well defined on the support of $\gamma_{k}(z_{k}) M_{k+1}(z_{k+1}|z_{k})$. Thus we see it simplifies to
$$
\begin{aligned}
g_k(x_k) \eta_k(x_k) &= \gamma_{k+1}(z_{k+1}) L_k(z_k|z_{k+1}) \frac{1}{Z_k}
\end{aligned}
$$

so when we consider $P_{\mu_k}(X_k = \tilde{x}_k)$ we see that we have 

$$ 
\begin{aligned}
P_{\mu_k}(X_k = \tilde{x}_k) &= \frac{g_k((\tilde{z}_k, \tilde{z}_{k+1})) \eta_k((\tilde{z}_k, \tilde{z}_{k+1}))}{\int \int g_k((z_k, z_{k+1})) \eta_k((z_k, z_{k+1})) dz_k \; dz_{k+1}} \\
&= \frac{\frac{\gamma_{k+1}(\tilde{z}_{k+1}) L_k(\tilde{z}_k|\tilde{z}_{k+1})}{Z_k}}{\int \int \frac{\gamma_{k+1}(z_{k+1}) L_k(z_k|z_{k+1})}{Z_k} dz_k \; dz_{k+1}} \\
&=  \frac{\gamma_{k+1}(\tilde{z}_{k+1}) L_k(\tilde{z}_k|\tilde{z}_{k+1})}{\int \gamma_{k+1}(z_{k+1})\left(\int  L_k(z_k|z_{k+1}) \; dz_k\right) \; dz_{k+1}}
\end{aligned} 
$$

and recall that $L_k$ is a Markov kernel so $\int  L_k(z_k|z_{k+1}) \; dz_k = 1$ and thus the integral simplifies to 
$$ 
\begin{aligned}
P_{\mu_k}(X_k = \tilde{x}_k) &=  \frac{\gamma_{k+1}(\tilde{z}_{k+1}) L_k(\tilde{z}_k|\tilde{z}_{k+1})}{\int \gamma_{k+1}(z_{k+1})\left(\int  L_k(z_k|z_{k+1}) \; dz_k\right) \; dz_{k+1}} \\
 &=  \frac{\gamma_{k+1}(\tilde{z}_{k+1}) L_k(\tilde{z}_k|\tilde{z}_{k+1})}{\int \gamma_{k+1}(z_{k+1}) dz_{k+1}} \\
 &=  \frac{\gamma_{k+1}(\tilde{z}_{k+1}) L_k(\tilde{z}_k|\tilde{z}_{k+1})}{Z_{k+1}}\\
 &=  \pi_{k+1}(\tilde{z}_{k+1}) L_k(\tilde{z}_k|\tilde{z}_{k+1})
\end{aligned} 
$$

\medskip
\noindent\textbf{Proving for $k=2,\dots,n$ that  $\eta_k((z_k, z_{k+1})) = M_{k+1}(z_{k+1}|z_k)   \pi_k(z_k) $ }  

Given that $\mu_{k-1}((z_{k-1}, z_{k})) = \pi_{k}(z_{k}) L_{k-1}(z_{k-1}|z_{k}) $ and $\eta_k = \mu_{k-1} Q_k$ consider

$$ 
\begin{aligned}
P_{\eta_k}(X_k= \tilde{x}_k) &= \int Q_k(\tilde{x}_k|x_{k-1}) \mu_{k-1}(x_{k-1}) dx_{k-1} \\
&= \int \int Q_k((\tilde{z}_{k+1}, \tilde{z}_k)|(z_{k-1},z_k)) \mu_{k-1}((z_{k-1},z_k)) dz_{k-1} dz_k \\
&= \int  \int M_{k+1}(\tilde{z}_{k+1}|\tilde{z}_k) I(\tilde{z}_k = z_k)  \pi_k(z_k) L_{k-1}(z_{k-1}|z_{k}) dz_{k-1} dz_k \\
&= \int   M_{k+1}(\tilde{z}_{k+1}|\tilde{z}_k) I(\tilde{z}_k = z_k)  \pi_k(z_k) \left(\int  L_{k-1}(z_{k-1}|z_{k}) dz_{k-1} \right) dz_k \\
&= \int   M_{k+1}(\tilde{z}_{k+1}|\tilde{z}_k) I(\tilde{z}_k = z_k)  \pi_k(z_k)  dz_k \\
&= M_{k+1}(\tilde{z}_{k+1}|\tilde{z}_k)   \pi_k(\tilde{z}_k) 
\end{aligned} 
$$

\end{proof}

\end{proposition}

Now that we have proved running Algorithm
\ref{alg-alive-particle-filter} with the definition
Definition~\ref{def-gsmc-to-particle-filter} yields plans that are
marginally distribution \(\pi_r\) after each step, we can tie it all
together to prove that running Algorithm \ref{alg-alive-smcs} yields the
desired strong convergence results.

\begin{proposition}[Convergence of the Alive SMC
Sampler]\protect\hypertarget{prp-alive-smcs-convergence}{}\label{prp-alive-smcs-convergence}

Given a sequence of target distributions \(\set{\pi_r}_{r=1}^D\), an
initial proposal distribution \(q_1\), forward kernels
\(\{M_r\}^D_{r=2}\) and backward kernels \(\{L_r\}^{D-1}_{r=1}\), let
\(\set{(\xi_{r}^{(i)}, W^{(i)}_r)}\) be the weighted sample formed from
running Algorithm \ref{alg-alive-smcs}. If they are reachable
(Definition~\ref{def-reachability}) and support-compatible
(Definition~\ref{def-support-compatible-backward}) then the following
results hold for all \(r=1,\dots, D\):

\begin{enumerate}
\def\labelenumi{\arabic{enumi}.}
\tightlist
\item
  Let \(\TV{\cdot}\) represent the total variation distance between
  probability measures. Then, as \(N \to\infty\), the weighted empirical
  measure
  \(\hat{\pi}_r^N(\cdot) = \sum_{i=1}^N W_r^{(i)} \delta_{\xi_r^{(i)}}(\cdot)\)
  converges almost surely in total variation distance to \(\pi_r\),
  i.e., \[
  \TV{\hat{\pi}^N_r - \pi_r} \toAS 0
  \]
\item
  For any function \(h: \tilde{E} \to \bbR\) we have, \[
   \E[\hat{\pi}^N_r]{h(\xi)} \toAS \E[\pi_r]{h(\xi)}
  \] and \[
   \sqrt{N} (\E[\hat{\pi}^N_r]{h(\xi)} - \E[\pi_r]{h(\xi)})
   \toD \mathcal{N}(0, V_\text{SMC}(h))
  \] for some asymptotic variance \(V_\text{SMC}(h)\).
\end{enumerate}

\begin{proof} 
To prove this we must first prove that running Algorithm \ref{alg-alive-smcs} is equivalent to running 
Algorithm \ref{alg-alive-particle-filter} with the space and functions defined in \ref{def-gsmc-to-particle-filter}. Once that is proved, then we must also prove that the reachability and support-compatibility conditions imply that under our definitions in Definition \ref{def-gsmc-to-particle-filter}, the support of the alive sets $A_k$ are contained in the proposal distributions $\eta_k$. 

We first begin with proving that running Algorithm \ref{alg-alive-smcs} is equivalent to running 
Algorithm \ref{alg-alive-particle-filter} with the space and functions defined in Definition \ref{def-gsmc-to-particle-filter}. Recall, to avoid a clash of notation we let $n = D-1$, write $z_{k+1}^{(i)}$ for the $i$-th particle yielded by Algorithm \ref{alg-alive-smcs}, and write $\zeta_k^{i}$ for the $i$-th alive particle yielded by Algorithm \ref{alg-alive-particle-filter}. 

There are two crucial insight for this equivalency. First, due to the reachability and support-compatibility assumptions, the contrived alive function $g_k((z_k, z_{k+1}))$ is non-zero if, and only if, the associated $z_{k+1}$ has $\gamma_{k+1}(z_{k+1}) > 0$, guaranteeing the conditions for accepting a sample in the two algorithms are the same. Second, sampling from $\mu_{k-1}^N Q_k$ is equivalent to the sampling procedure for $z_{k+1}^\ast$ because the weights in $\mu_{k-1}^N$ are exactly the normalized incremental weights. We also note that for a particle $\xi_k = (z_k, z_{k+1})$ in our contrived space $E$, after sampling and normalizing all subsequent operations only depend on $z_{k+1}$, so we are justified in discarding it in Algorithm \ref{alg-alive-smcs} after computing the incremental weight. 

To show this formally, starting with $k=0$, we see that sampling $(z_0^\ast, z_1^\ast)$ until $\gamma_1(z_1) > 0$ is equivalent to sampling $\xi_0^\ast = (z_0^\ast, z_1^\ast)$ until $g_0(\xi_0^\ast) > 0$ as, due to the reachability assumption, we know 

$$
\gamma_1(z_1) > 0 \iff g_0(\xi_0^\ast) = \frac{\gamma_1(z_1)}{q_1(z_1)} > 0
$$

For $k=1,\dots,n$ consider that under Algorithm \ref{alg-alive-smcs} to sample a proposed $z_{k+1}^\ast$ we first sample $i^\prime$ where $\mathrm{Pr}(i^\prime = j) \propto w_k^{(j)}$ then sample $z_{k+1}^\ast \sim M_{k+1}(\cdot|z_k^{(i^\prime)})$. We then accept if $\gamma_{k+1}(z_{k+1}^\ast) > 0$. For Algorithm \ref{alg-alive-particle-filter} we sample $\xi_k^\ast \sim \mu_{k-1}^N Q_k$ where
$$
\mu_{k-1}^N = \sum_{i=1}^N W_{k-1}^i \delta_{\zeta_{k-1}^i} \; \; \text{ and } W_{k-1}^i = \frac{g_{k-1}(\zeta_{k-1}^i)}{\sum_{j=1}^N g_{k-1}(\zeta_k^i)}
$$
where $\zeta_{k-1}^{i} = (z_{k-1}^{(i)}, z_{k}^{(i)})$. Notice that due to our choice $g_{k-1}$ we have that $W_{k-1}^i$ is simply the normalized incremental weights $w_k^{(i)}$ from Algorithm \ref{alg-alive-smcs}. So since the mutation kernel $Q_k(\cdot|\zeta_{k-1}^{i}) = Q_k(\cdot|(z_{k-1}^{(i)}, z_{k}^{(i)}))$ only depends on $z_{k}^{(i)}$, we see that sampling $\xi_k^\ast \sim \mu_{k-1}^N Q_k$ is exactly equivalent to sampling $\mathrm{Pr}(i^\prime = j) \propto w_k^{(j)}$ then $z_{k+1}^\ast \sim M_{k+1}(\cdot|z_k^{(i^\prime)})$. Now all that is left is proving that the stopping conditions are the same. In other words, accepting $\xi_k^\ast \sim \mu_{k-1}^N Q_k$ when $g_k(\xi_k^\ast) > 0$ is equivalent to accepting $z_{k+1}^{\ast} \sim M_{k+1}(\cdot|z_k^{(i^\prime)})$ when $\gamma_{k+1}(z_{k+1}^{\ast}) > 0$. Thanks to our reachability and support-compatibility assumptions this equivalency immediately follows from Corollary \ref{cor-reachability-implications}.

Now we need to prove that the reachability and support-compatibility conditions imply what we want about the alive sets.  Consider from Definition \ref{def-alive-funcs} we know the alive sets are defined as. 
$$
A_k = \set{x \in E| g_k(x) > 0}
$$
and we want to prove that $A_k \subset \mathrm{Supp}(\eta_k)$ for all $k = 0,\dots,n$. 

Consider for $k = 0$ case that $A_0$ can be expressed as
$$
\begin{aligned}
A_0 &= \set{(z_{0}, z_{1}) \in \tilde{E} \times \tilde{E} | g_0((z_{k}, z_{k+1})) > 0} \\
&= \set{ z \in \tilde{E} | \frac{\gamma_1(z)}{q_1(z)} > 0}
\end{aligned}
$$
and so we see that since $\eta_0$ is essentially just $q_1$ then since reachability directly implies $\mathrm{Supp}(\pi_1) \subset \mathrm{Supp}(q_1)$, it implies $A_0 \subset \mathrm{Supp}(\eta_0)$.

\noindent
Now for $k = 1,\dots, n$, consider that we can write the alive set $A_k$ as the set of plans $(z_{k}, z_{k+1})$ where the incremental weight function $w_r(z_k, z_{k+1})$ is non-zero, so

$$
\begin{aligned}
A_k &= \set{(z_{k}, z_{k+1}) \in \tilde{E} \times \tilde{E} | g_k((z_{k}, z_{k+1})) > 0} \\
&= \set{(z_{k}, z_{k+1}) \in \tilde{E} \times \tilde{E} \big\vert \; \frac{\gamma_{k+1}(z_{k+1}) L_k(z_k|z_{k+1})}{\gamma_{k}(z_{k}) M_{k+1}(z_{k+1}|z_{k})} > 0}.
\end{aligned}
$$

From our reachability and support-compatibility assumptions it follows immediately from Corollary \ref{cor-reachability-implications} that 

$$
\begin{aligned}
A_k &=  \set{(z_{k}, z_{k+1}) \in \tilde{E} \times \tilde{E} \big\vert \; \frac{\gamma_{k+1}(z_{k+1}) L_k(z_k|z_{k+1})}{\gamma_{k}(z_{k}) M_{k+1}(z_{k+1}|z_{k})} > 0} \\
&=  \set{(z_{k}, z_{k+1}) \in \tilde{E} \times \tilde{E} \big\vert \; \gamma_{k+1}(z_{k+1}), \; \gamma_{k}(z_{k}) M_{k+1}(z_{k+1}|z_{k}) > 0}.
\end{aligned}
$$
So we see immediately if $(z_k, z_{k+1}) \in A_k$ then we know $\gamma_{k}(z_{k}) M_{k+1}(z_{k+1}|z_{k}) > 0$ or equivalently
$$
(z_k, z_{k+1}) \in A_k \implies \pi_{k}(z_{k}) M_{k+1}(z_{k+1}|z_{k}) >0
$$

Now recall we know from Proposition \ref{prp-alive-smcs-distribs} that 

$$
\eta_k(x) = \eta_k((z_k, z_{k+1})) = \pi_{k}(z_{k}) M_{k+1}(z_{k+1}|z_{k})
$$

so we see that immediately implies if $(z_k, z_{k+1}) \in A_k$ then $(z_k, z_{k+1}) \in \mathrm{Supp}(\eta_k)$. 

Therefore since we have proved $A_k \subset \mathrm{Supp}(\eta_k)$ , we know that the results of Theorem \ref{thm-alive-particle-as-convergence} and \ref{thm-alive-clt} hold. Technically with these results we've actually proved something stronger, namely the convergence of the weighted empirical measure of the pairs $(z_k, z_{k+1})$ jointly to the joint distribution $\pi_{k+1}(z_{k+1}) L_{k}(z_k|z_{k+1})$. Since joint distributions completely determine the marginals, this result automatically implies for the projected measure $(z_k, z_{k+1}) \mapsto z_{k+1}$ all the same convergence properties hold for the marginal distribution $z_{k+1} \sim \pi_{k+1}$ and thus we are done. 
\end{proof}

\end{proposition}

We can also prove the strong consistency of the normalizing constant
estimator.

\begin{proposition}[Alive SMCS Consistent Normalizing Constant
Estimator]\protect\hypertarget{prp-alive-smcs-norm-estimate}{}\label{prp-alive-smcs-norm-estimate}

Suppose the three conditions from
Proposition~\ref{prp-alive-smcs-convergence} hold. For each
\(r=1,\dots,D\) define the estimator

\[
\hat{Z}_r = \prod_{s=1}^r \left( \frac{1}{T_s} \sum_{i=1}^N w_s^{(i)} \right)
\] where \(\set{w_s^{(i)}}_{i=1}^N\) are the unnormalized incremental
weights from Algorithm \ref{alg-alive-smcs}, then
\(\hat{Z}_r \toAS Z_r\)

\begin{proof} We begin by first noting for $s=1$ that we can write $\frac{1}{T_s} \sum_{i=1}^N w_s^{(i)}$ as 
$$
\frac{1}{T_s} \sum_{i=1}^N w_s^{(i)} = \frac{1}{T_s} \sum_{i=1}^{T_s} w_s^{(i)} = \inner{\eta_{s-1}^N}{g_{s-1}},
$$
where $\eta_s$ and $g_s$ are as defined in Definition \ref{def-alive-funcs}. Now recall from Theorem \ref{thm-alive-particle-as-convergence} we know that $\inner{\eta_{s-1}^N}{g_{s-1}} \toAS  \inner{\eta_{s-1}}{g_{s-1}}$. Recall that 
$$
\inner{\eta_{s-1}}{g_{s-1}} = \int g_{s-1}(x) \eta_{s-1}(x) dx
$$
and we proved in Proposition @prp-alive-smcs-convergence that for $s=1$ we have
$$
\inner{\eta_0}{g_0} = \int g_0(x) \eta_0(x) dx = Z_1
$$
and for $s = 1, \dots D$ we have 
$$
\begin{aligned}
g_{s-1}(x_{s-1}) \eta_{s-1}(x_{s-1}) &= \gamma_{s}(z_{s}) L_{s-1}(z_{s-1}|z_{s}) \frac{1}{Z_{s-1}}
\end{aligned}
$$
and thus for $\inner{\eta_{s-1}}{g_{s-1}}$ we have

$$
\begin{aligned}
\inner{\eta_{s-1}}{g_{s-1}} &= \int g_{s-1}(x_{s-1}) \eta_{s-1}(x_{s-1}) dx \\
&= \int \int \gamma_{s}(z_{s}) L_k(z_{s-1}|z_{s}) \frac{1}{Z_{s-1}} \; dz_{s-1} \; dz_s \\
&= \frac{1}{Z_{s-1}}  \int \gamma_{s}(z_{s}) \left(\int  L_k(z_{s-1}|z_{s})  \; dz_{s-1} \right) \; dz_s \\
&= \frac{1}{Z_{s-1}} \int \gamma_{s}(z_{s})  \; dz_s \\
&= \frac{Z_s}{Z_{s-1}}.
\end{aligned}
$$

So putting it altogether, since 
$$
\begin{aligned}
\hat{Z}_r &= \prod_{s=1}^r \left( \frac{1}{T_s} \sum_{i=1}^N w_s^{(i)} \right) \\
&= \prod_{s=1}^r \inner{\eta_{s-1}^N}{g_{s-1}}
\end{aligned}
$$
then by repeated application of Slutsky's theorem we have that 
$$
\begin{aligned}
\hat{Z}_r \toAS Z_1 \prod_{s=2}^r \frac{Z_s}{Z_{s-1}} = Z_r
\end{aligned}
$$

\end{proof}

\end{proposition}

\begin{refremark}[Unbiased Normalizing Constant Estimator]
While this estimator in Proposition~\ref{prp-alive-smcs-norm-estimate}
is strongly consistent, it is generally not unbiased. An unbiased
estimator of \(Z_r\) can be obtained by modifying Algorithm
\ref{alg-alive-smcs} as follows: after each sampling for loop continue
sampling and until an \(N+1\) particle is obtained with non-zero target
probability. Discard that particle but save \(T_{r,\text{extra}}\), the
number of sampling attempts that had to be made (analogous to how
\(T_r\) counts the number of attempts for the saved particles). Use that
to construct the following estimator \[
Z^\ast_r = \prod_{s=1}^r \left( \frac{\sum_{i=1}^N w_s^{(i)}}{T_s + T_{s,\text{extra}} - 1}  \right)
\] This estimator is now an unbiased estimator of \(Z_r\). No proof is
given but the argument is the same as that of the unbiased estimator
constructed in \citet{UnbiasedBootstrap}.

\label{rem-unbiased-norm-estimator}

\end{refremark}

\begin{definition}[Moved Empirical Alive SMC Sampler
Measure]\protect\hypertarget{def-moved-alive-smcs-measure}{}\label{def-moved-alive-smcs-measure}

For Algorithm \ref{alg-alive-smcs} suppose we have a \(\pi_r\) invariant
Markov kernel \(\tilde{M}_r\) for any \(r = 1,\dots,D\). For a fixed
\(m\), define the moved empirical measure \(\hat{\pi}_{r,m}^N\) as \[
\hat{\pi}_{r,m}^N(\cdot) = \sum_{i=1}^N W_r^{(i)} \delta_{\xi_{r, m}^{(i)}}(\cdot)
\] where \(\xi_{r, m}^{(i)}\) is the particle obtained by independently
applying the Markov kernel \(\tilde{M}_r\) to each particle \(m\) times.
In other words, sampling
\(\xi_{r, 1}^{(i)} \sim \tilde{M}_r(\cdot| \xi_{r}^{(i)})\),
\(\xi_{r, 2}^{(i)} \sim \tilde{M}_r(\cdot| \xi_{r,1}^{(i)})\), \dots,
\(\xi_{r, m}^{(i)} \sim \tilde{M}_r(\cdot| \xi_{r,m-1}^{(i)})\). Or
equivalently,
\(\xi_{r, m}^{(i)} \sim \tilde{M}_r^m(\cdot| \xi_{r}^{(i)})\) where
\(\tilde{M}_r^m\) represents \(m\) iterations of the Markov kernel.

\end{definition}

We will now prove that just like for the Alive particle filter algorithm
in Algorithm \ref{alg-alive-particle-filter}, the same strong
convergence results still hold for \(\hat{\pi}_{r,m}^N(\cdot)\). First,
we begin by again translating
Definition~\ref{def-moved-alive-smcs-measure} to an Algorithm
\ref{alg-alive-particle-filter} framework.

\begin{definition}[Markov Kernel for the Alive SMC Sampler Algorithm in
Alive Particle Filter
Framework]\protect\hypertarget{def-smcs-mcmc-to-particle-filter}{}\label{def-smcs-mcmc-to-particle-filter}

Suppose we already have the scenario in
Definition~\ref{def-gsmc-to-particle-filter} but we now also have a
Markov kernel \(\tilde{M}_r\) on \(\tilde{E}\) as well. Then, define the
following Markov kernel \(K_k\) on \(E\) as \[
K_k(\tilde{x}_k|x_k) = K_k((\tilde{z}_k, \tilde{z}_{k+1})|(z_k, z_{k+1})) = \tilde{M}_{k+1}(\tilde{z}_{k+1} | z_{k+1}) L_k(\tilde{z}_k\mid\tilde{z}_{k+1})
\] In other words, given a pair \((z_k, z_{k+1})\) then \(K_k\) samples
\(\tilde{z}_{k+1} \sim \tilde{M}_r(\cdot|z_{k+1})\) and then
\(\tilde{z}_k \sim L_k(\cdot|\tilde{z}_{k+1})\).

\end{definition}

We will now prove that if \(\tilde{M}_r\) is \(\pi_r\) invariant then
\(K_k\) is \(\mu_k\) invariant as well.

\begin{corollary}[Markov
Invariance]\protect\hypertarget{cor-smcs-mcmc-invar}{}\label{cor-smcs-mcmc-invar}

Suppose the conditions in Proposition~\ref{prp-alive-smcs-distribs} hold
and for the Markov kernel \(\tilde{M}_r\) we define \(K_k\) as in
Definition~\ref{def-smcs-mcmc-to-particle-filter}. If \(\tilde{M}_r\) is
\(\pi_r\) invariant then \(K_k\) is \(\mu_k\) invariant.

\begin{proof}
Recall from Proposition \ref{prp-alive-smcs-distribs} we know that for $k=1,\dots, n$ that 
$$
\mu_k(x) = \mu_k((z_k, z_{k+1})) = \pi_{k+1}(z_{k+1}) L_k(z_k|z_{k+1}).
$$

Now consider 
$$
P_{\mu_k K_k}(X_k = \tilde{x}_k)
$$

$$ 
\begin{aligned}
P_{\mu_k K_k}(X_k = \tilde{x}_k) &= \int K_k(\tilde{x}_k|x_{k}) \mu_{k}(x_{k}) dx_{k} \\
&= \int \int \tilde{M}_{k+1}(\tilde{z}_{k+1} | z_{k+1}) L_k(\tilde{z}_{k}|\tilde{z}_{k+1}) \pi_{k+1}(z_{k+1}) L_k(z_k|z_{k+1}) dz_{k} dz_{k+1} \\
&= \int \tilde{M}_{k+1}(\tilde{z}_{k+1} | z_{k+1}) L_k(\tilde{z}_{k}|\tilde{z}_{k+1})  \pi_{k+1}(z_{k+1}) \left( \int   L_k(z_k|z_{k+1}) dz_{k} \right) dz_{k+1} \\
&= \int \tilde{M}_{k+1}(\tilde{z}_{k+1} | z_{k+1}) \pi_{k+1}(z_{k+1})  L_k(\tilde{z}_{k}|\tilde{z}_{k+1})  dz_{k+1} \\
&=  L_k(\tilde{z}_{k}|\tilde{z}_{k+1})  \int \tilde{M}_{k+1}(\tilde{z}_{k+1} | z_{k+1}) \pi_{k+1}(z_{k+1})  dz_{k+1} 
\end{aligned} 
$$
Now since we have that $\tilde{M}_{k+1}$ is $\pi_{k+1}$ invariant then we have 
$$ 
\begin{aligned}
P_{\mu_k K_k}(X_k = \tilde{x}_k) &=  L_k(\tilde{z}_{k}|\tilde{z}_{k+1})  \int \tilde{M}_{k+1}(\tilde{z}_{k+1} | z_{k+1}) \pi_{k+1}(z_{k+1})  dz_{k+1} \\
&=  L_k(\tilde{z}_{k}|\tilde{z}_{k+1}) \pi_{k+1}(\tilde{z}_{k+1} ) \\
&= P_{\mu_k}(X_k = \tilde{x}_k)
\end{aligned} 
$$
and thus we've proved $K_k$ is $\mu_k$-invariant. 
\end{proof}

\end{corollary}

\begin{proposition}[Convergence of the Alive SMC Sampler with MCMC
Steps]\protect\hypertarget{prp-alive-smcs-mcmc-convergence}{}\label{prp-alive-smcs-mcmc-convergence}

Suppose we have the conditions in
Proposition~\ref{prp-alive-smcs-convergence}. For any \(r=1,\dots,D\) if
\(\tilde{M}_r\) is a \(\pi_r\)-invariant Markov kernel then for any
\(m \in \bbN\) for the moved empirical measure \(\hat{\pi}^N_{r,m}\) the
following results hold:

\begin{enumerate}
\def\labelenumi{\arabic{enumi}.}
\tightlist
\item
  Let \(\TV{\cdot}\) represent the total variation distance between
  probability measures. Then, as \(N \to\infty\), the weighted empirical
  measure \(\hat{\pi}^N_{r,m}\) converges almost surely in total
  variation distance to \(\pi_r\), i.e., \[
  \TV{\hat{\pi}^N_{r,m} - \pi_r} \toAS 0
  \]
\item
  For any function \(h: \tilde{E} \to \bbR\) we have, \[
   \E[\hat{\pi}^N_{r,m}]{h(\xi)} \toAS \E[\pi_r]{h(\xi)}
  \] and \[
   \sqrt{N} (\E[\hat{\pi}^N_{r,m}]{h(\xi)} - \E[\pi_r]{h(\xi)})
   \toD \mathcal{N}(0, \tilde{V}_\text{SMC}(h))
  \] for some asymptotic variance \(\tilde{V}_\text{SMC}(h)\).
\end{enumerate}

\begin{proof} By Corollary \ref{cor-smcs-mcmc-invar} we have that $K_k$ is $\mu_k$-invariant and thus by Theorem \ref{thm-moved-measure-conv} and Proposition \ref{prp-moved-measure-clt} we have almost sure convergence of the empirical measure and a CLT for the distribution $\mu_k$ on pairs. We note since sampling $(\tilde{z}_k,\tilde{z}_{k+1}) \sim Q_k(\cdot|(z_{k-1},z_{k}))$ in the Alive particle filter framework only depends on $z_k$, we do not actually need to sample from $L_k$ in $K_k$ nor store the old particle so we can continue to just use Algorithm \ref{alg-alive-smcs} without modification and the convergence results apply to the marginal distribution of $z_{k+1} = \xi_r$ as before. 
\end{proof}

\end{proposition}

\subsubsection{gSMC
Convergence}\label{sec-gsmc-specific-convergence-results}

With these more general results for Algorithm \ref{alg-alive-smcs}, we
can now prove the strong convergence of Algorithm \ref{alg-gsmcs} with
and without MCMC moves. Notice that Algorithm \ref{alg-gsmcs} is
actually a special case of this algorithm where the initial importance
sampling step isn't needed, meaning we don't need to specify a \(q_1\)
and instead of sampling \(N\) plans \(\xi_1^{(i)}\), we just skip that
step and set all the initial weights to \(w_1^{(i)} = \gamma_1(\xi_1)\).
We can do that in Algorithm \ref{alg-gsmcs} because our forward kernel
in all three sampling spaces only depends on the multidistrict it splits
through its vertex set and for a plan with 1 region the only possible
vertex set it could have is \(V\), all the vertices in the entire map.
For graph space sampling there actually only exists one 1-region plan.
For forest and linking edge space there are \(\tau(G)\) such plans in
\(\mathcal{F}_1\) and \(\mathcal{L}_1\) respectively but that does not
matter as \(\tilde{\gamma}_1\) and \(\gamma^\ast_1\) only depend on the
associated induced graph plan and are thus constant. However, if we were
using either a forward kernel that depended on more than the vertex set
for the first split (like an SMC version of \citep{cyclewalk}) or a
target distribution that depended on more than the induced graph plan
then the initial importance sampling step would matter again.

\begin{theorem}[Strong Convergence of the gSMC
Algorithm]\protect\hypertarget{thm-gsmc-convergence}{}\label{thm-gsmc-convergence}

Given a sequence of target distributions \(\set{\pi_r}_{r=1}^D\) of the
form in Equation~\ref{eq-target} with a splittable score function (see
Definition~\ref{def-splittable-score-func}), if the forward kernel and
optimal weights described in Section~\ref{sec-graph-space-splitting},
Section~\ref{sec-spanning-forest}, or Section~\ref{sec-linking-edge} are
used in Algorithm \ref{alg-alive-smcs} then for each \(r = 2,\dots,D\),
the following results hold

\begin{enumerate}
\def\labelenumi{\arabic{enumi}.}
\tightlist
\item
  Let \(\TV{\cdot}\) represent the total variation distance between
  probability measures. Then, as \(N \to\infty\), the weighted empirical
  measure
  \(\hat{\pi}^N_r(\cdot) = \sum_{i=1}^N W_r^{(i)} \delta_{\xi_r^{(i)}}(\cdot)\)
  converges almost surely in total variation distance to the true
  \(\pi_r\), i.e., \[
  \TV{\hat{\pi}^N_r - \pi_r} \toAS 0
  \]
\item
  For any real valued function \(h\) on unlabeled plans we have, \[
   \E[\hat{\pi_r}^N]{h(\xi)} \toAS \E[\pi_r]{h(\xi)}
  \] and \[
   \sqrt{N} (\E[\hat{\pi}^N_r]{h(\xi)} - \E[\pi_r]{h(\xi)})
   \toD \mathcal{N}(0, V_\text{SMC}(h))
  \] for some asymptotic variance \(V_\text{SMC}(h)\).
\end{enumerate}

\begin{proof} 
First recall that Algorithm \ref{alg-gsmcs} is just a special case of Algorithm \ref{alg-alive-smcs} where we don't even need to sample from an initial $q_1$. Thus for Proposition \ref{prp-alive-smcs-convergence} to apply, we just need to prove reachability and backwards support-compatibility. For reachability, recall we know from Corollaries \ref{cor-graph-plan-splittability}, \ref{cor-forest-plan-splittability}, and \ref{cor-link-plan-splittability} that if the only hard constraints imposed are population balance and connectedness then all plans on each space are splittable. The definition of splittability is exactly the same as reachability just without the initial assumption on $\xi_1$, which is not needed here. If we assume our score functions are splittable then reachability continues to hold. 

Next we note that, by definition, for the optimal weights reachability implies backwards support-compatibility. Recall that the backwards kernel associated with the optimal weights is given by 

$$
L^\text{opt}_{r-1}(\xi_{r-1}\mid \xi_r) = \frac{\pi_{r-1}(\xi_{r-1})M_r( \xi_r \mid \xi_{r-1})}{f_r(\xi_r)}.
$$
where $f_r$ denotes the marginal proposal density obtained by integrating out the previous partial plan $\xi_{r-1}$ according to its target distribution,
$$
f_r(\xi_r) = \int \pi_{r-1}(\xi_{r-1}) M_r(\xi_r \mid \xi_{r-1}) \; d\xi_{r-1}.
$$

Thus, we see that for any $\xi_{r}, \xi_{r-1}$ if we have $\pi_{r-1}(\xi_{r-1}) M_r(\xi_r \mid \xi_{r-1}) > 0$ then that implies $f_r(\xi_r) > 0$ and thus $L^\text{opt}_{r-1}(\xi_{r-1}\mid \xi_r) > 0$. Conversely, if we have $L^\text{opt}_{r-1}(\xi_{r-1}\mid \xi_r) > 0$ then that implies $\pi_{r-1}(\xi_{r-1}) M_r(\xi_r \mid \xi_{r-1}) > 0$ and thus by Definition \ref{def-support-compatible-backward} support-compatibility is implied. 

Thus we know the conditions from Proposition \ref{prp-alive-smcs-convergence} all hold and so the convergence results from there apply as well. 
\end{proof}

\end{theorem}

We can also succinctly prove how to estimate the normalizing constant

\begin{corollary}[gSMC Consistent Normalizing Constant
Estimator]\protect\hypertarget{cor-gsmc-norm-estimate}{}\label{cor-gsmc-norm-estimate}

Suppose the conditions from Theorem~\ref{thm-gsmc-convergence} hold. On
any of the three sampling spaces, for each \(r=1,\dots,D\) define the
estimator

\[
\hat{Z}_r = \gamma_1(\xi_1) \prod_{s=2}^r \left( \frac{1}{T_s} \sum_{i=1}^N w_s^{(i)} \right)
\] where \(\set{w_s^{(i)}}_{i=1}^N\) are the unnormalized incremental
weights from Algorithm \ref{alg-gsmcs}, then \(\hat{Z}_r \toAS Z_r\)

\begin{proof} We begin by noting, as proved in Lemmas \ref{lem-forest-pushforward-target} and \ref{lem-linking-pushforward-target}, that for the forest and linking edge space targets $\tilde{\pi}_r$ and $\pi^\ast_r$ respectively, the normalizing constants are the exact same as $\pi_r$. Therefore, by Proposition \ref{prp-alive-smcs-norm-estimate} we know for each of the three spaces the estimator 
$$
\tilde{Z}_r = \prod_{s=1}^r \left( \frac{1}{T_s} \sum_{i=1}^N w_s^{(i)} \right)
$$
is a strongly consistent estimator of $Z_r$. However, we can simplify this even further. Recall that for any map there is only a single 1-region plan $\xi_1$ (the 1-region plan consisting of the entire map). So we actually know that $Z_1 = \gamma_1(\xi_1)$ and thus we can replace the estimator of $Z_1$ with just $Z_1$ itself ($\gamma_1(\xi_1)$). So we see that 
$$
\hat{Z}_r = \gamma_1(\xi_1) \prod_{s=2}^r \left( \frac{1}{T_s} \sum_{i=1}^N w_s^{(i)} \right)
$$
is a strongly consistent estimator of $Z_r$ as well. 
\end{proof}

\end{corollary}

\begin{refremark}[Unbiased gSMC Normalizing Constant Estimator]
The suggested unbiased estimator in
Remark~\ref{rem-unbiased-gsmc-norm-estimator} applied to the gSMC
algorithm is of the form \[
Z^\ast_r = \gamma_1(\xi_1) \prod_{s=2}^r \left( \frac{\sum_{i=1}^N w_s^{(i)}}{T_s + T_{s,\text{extra}} - 1}  \right)
\] Again, no proof is given.

\label{rem-unbiased-gsmc-norm-estimator}

\end{refremark}

We now prove the strong convergence of the gSMC-MCMC algorithm as well.

\begin{theorem}[Convergence Results for the gSMC-MCMC
Algorithm]\protect\hypertarget{thm-gsmc-mcmc-convergence}{}\label{thm-gsmc-mcmc-convergence}

Suppose we have a sequence of target distributions
\(\set{\pi_r}_{r=1}^D\) of the form in Equation~\ref{eq-target} with a
splittable score function (see
Definition~\ref{def-splittable-score-func}) and the forward kernel and
optimal weights described in Section~\ref{sec-graph-space-splitting},
Section~\ref{sec-spanning-forest}, or Section~\ref{sec-linking-edge} are
used in Algorithm \ref{alg-alive-smcs}. For any \(r = 2,\dots,D\) if
\(\tilde{M_r}\) is a \(\pi_r\)-invariant Markov kernel (or
\(\tilde{\pi}_r\)/\(\pi^\ast_r\) invariant for spanning forest and
linking edge space respectively) then for the moved empirical measure

\[
\hat{\pi}^N_{r,m}(\cdot) = \sum_{i=1}^N W_r^{(i)} \delta_{\xi_{r, m}^{(i)}}(\cdot)
\]

where \(\xi_{r, m}^{(i)}\) is the particle obtained by independently
applying the Markov kernel \(\tilde{M}_r\) to each plan \(m\) times. In
other words, sampling
\(\xi_{r, 1}^{(i)} \sim \tilde{M}_r(\cdot| \xi_{r}^{(i)})\),
\(\xi_{r, 2}^{(i)} \sim \tilde{M}_r(\cdot| \xi_{r,1}^{(i)})\), \dots,
\(\xi_{r, m}^{(i)} \sim \tilde{M}_r(\cdot| \xi_{r,m-1}^{(i)})\).

The following results hold:

\begin{enumerate}
\def\labelenumi{\arabic{enumi}.}
\tightlist
\item
  Let \(\TV{\cdot}\) represent the total variation distance between
  probability measures. Then, as \(N \to\infty\), the weighted empirical
  measure \(\hat{\pi}^N_{r,m}\) converges almost surely in total
  variation distance to \(\pi_r\), i.e., \[
  \TV{\hat{\pi}^N_{r,m} - \pi_r} \toAS 0
  \]
\item
  For any function \(h: \tilde{E} \to \bbR\) we have, \[
   \E[\hat{\pi}^N_{r,m}]{h(\xi)} \toAS \E[\pi_r]{h(\xi)}
  \] and \[
   \sqrt{N} (\E[\hat{\pi}^N_{r,m}]{h(\xi)} - \E[\pi_r]{h(\xi)})
   \toD \mathcal{N}(0, \tilde{V}_\text{SMC}(h))
  \] for some asymptotic variance \(\tilde{V}_\text{SMC}(h)\).
\end{enumerate}

\begin{proof} This follows immediately from Theorem \ref{thm-gsmc-convergence}'s result that the choice of forward kernels and a splittable score function imply reachability and backward support-compatibility and from Proposition \ref{prp-alive-smcs-mcmc-convergence}.
\end{proof}

\end{theorem}

The preceding convergence results also apply to the hierarchical sampler
provided that its intermediate target sequence is reachable under the
chosen splitting schedule.

\subsection{Mergesplit Kernels}\label{sec-mcmc-appendix}

We now present the formulas for the proposal distributions of the
mergesplit kernels used in the various sample spaces. They are nearly
identical to the analogous forward kernels shown in the main text except
that \(\varphi(\cdot\mid\cdot)\) is now a distribution over pairs of
adjacent regions in a plan rather than a distribution over
multidistricts in a plan. We assume whenever the proposed plans
\(\xi_r^\ast\) has probability zero under its target distribution then
the acceptance probability is zero as well.

\subsubsection{Graph Space MCMC Kernel}\label{graph-space-mcmc-kernel}

The graph space MCMC kernel is based on the kernel developed in
\citep{recom}.

\begin{proposition}[Graph Space Mergesplit Proposal
Density]\protect\hypertarget{prp-graph-ms-prob}{}\label{prp-graph-ms-prob}

Let \(\xi_r, \xi_r^\ast\) be balanced plans such that there exists
\(G_{k}, G_{k'} \in \xi_r\), \(G^\ast_{k}, G^\ast_{k'} \in \xi^\ast_r\)
where \(G_{k} \cup G_{k'} = G^\ast_{k}\cup G^\ast_{k'}\) and all other
regions are the same. If we choose \(\mathcal{K}\) such that \[
    \mathcal{K} \geq \max_{T \in \mathcal{T}(G_{k} \cup G_{k'} )} \abs{\mathrm{ok}(T, \mathcal{S}_{r}(s_k + s_{k'} ))},
\] then the mergesplit probability \(M_r(\xi_r^\ast|\xi_r)\) is given by
\[
M_r(\xi_r^\ast|\xi_r) = \varphi(G_{k}, G_{k'}\mid \xi_{r})\frac{1}{\mathcal{K}_r} \frac{\tau(G^\ast_{k})\tau(G^\ast_{k'}) }{\tau(G^\ast_{k} \cup G^\ast_{k'})} \abs{\mathcal{C}(G^\ast_{k}, G^\ast_{k'})} 
\]

\begin{proof}
This follows immediately from \ref{prp-split-prob-graph}, replacing $H_\ell$ with $G_k \cup G_{k'}$ and noting that $\tau(G^\ast_{k} \cup G^\ast_{k'}) = \tau(G_{k} \cup G_{k'})$.
\end{proof}

\end{proposition}

\begin{corollary}[Graph Space Mergesplit
MH-Ratio]\protect\hypertarget{cor-graph-ms-ratio}{}\label{cor-graph-ms-ratio}

Given \(\xi_r, \xi_r^\ast\) and \(\mathcal{K}\) as in
Proposition~\ref{prp-graph-ms-prob} the associated acceptance
probability of going from \(\xi_r \to \xi_r^\ast\) is

\[
A(\xi_r^\ast, \xi_r) = \mathrm{min}\left(1, \frac{\varphi(G_{k}^\ast, G_{k'}^\ast \mid \xi_{r}^\ast)}{\varphi(G_{k}, G_{k'}\mid \xi_{r})} \frac{\exp{-J(\xi_r^\ast)}}{\exp{-J(\xi_r)}} \cdot \left(\frac{\tau(G^\ast_{k})\tau(G^\ast_{k'})}{\tau(G_{k})\tau(G_{k'})}\right)^{\rho-1} \cdot \frac{\abs{\mathcal{C}(G_{k}, G_{k'})}}{\abs{\mathcal{C}(G^\ast_{k}, G^\ast_{k'})}} \right).
\]

\end{corollary}

\subsubsection{Forest Space MCMC Kernel}\label{forest-space-mcmc-kernel}

The forest space MCMC kernel is based on the kernel developed in
\citep{McmcForests}.

\begin{proposition}[Forest Space Mergesplit Proposal
Density]\protect\hypertarget{prp-forest-ms-prob}{}\label{prp-forest-ms-prob}

~

Let \(F_r, F_r^\ast\) be balanced forest plans such that there exists
\(T_{k}, T_{k'} \in F_r\), \(T^\ast_{k}, T^\ast_{k'} \in F^\ast_r\)
where
\(\mathrm{V}(T_{k} \cup T_{k'}) = \mathrm{V}(T^\ast_{k}\cup T^\ast_{k'})\)
and all other region trees are the same. Given a tree cut selection
kernel \(\ArbKernel{\cdot|\cdot}\) then the mergesplit probability
\(M_r(F_r^\ast|F_r)\) is given by \[
M_r(F_r^\ast|F_r) = \varphi(T_{k}, T_{k'}\mid F_{r}) \frac{1}{\tau(G^\ast_{k} \cup G^\ast_{k'})} \cdot \EffB{T_k^\ast, T_{k'}^\ast},
\] where \[
\EffB{T_k, T_{k'}} = \sum_{e \in \mathcal{C}(T_{k}, T_{k'})} p_\text{cut}(\TreeCut{e} | T_{k} \cup \set{e} \cup T_{k'}).
\]

\begin{proof}
This follows immediately from \ref{cor-forest-forward-kernel} replacing $T_\ell$ with $T_k \cup T_{k'}$ and noting that $\tau(G^\ast_{k} \cup G^\ast_{k'}) = \tau(G_{k} \cup G_{k'})$.
\end{proof}

\end{proposition}

\begin{corollary}[Forest Space Mergesplit
MH-Ratio]\protect\hypertarget{cor-forest-ms-ratio}{}\label{cor-forest-ms-ratio}

Given \(F_r, F_r^\ast\) and \(\ArbKernel{\cdot|\cdot}\) as in
Proposition~\ref{prp-forest-ms-prob} the associated acceptance
probability of going from \(F_r \to F_r^\ast\) is

\[
A(F_r^\ast, F_r) = \mathrm{min}\left(1, \frac{\varphi(T_{k}^\ast, T_{k'}^\ast \mid F_{r}^\ast)}{\varphi(T_{k}, T_{k'}\mid F_{r})} \frac{\exp{-J(\xi_r^\ast)}}{\exp{-J(\xi_r)}} \cdot \left(\frac{\tau(G^\ast_{k})\tau(G^\ast_{k'})}{\tau(G_{k})\tau(G_{k'})}\right)^{\rho-1} \cdot \frac{\EffB{T_k, T_{k'}}}{\EffB{T_k^\ast, T_{k'}^\ast}} \right),
\] where \(\xi^\ast_r = \xi(F_r^\ast)\), \(\xi_r = \xi(F_r)\)

\end{corollary}

\subsubsection{Linking Edge Space MCMC
Kernel}\label{linking-edge-space-mcmc-kernel}

The linking edge space MCMC kernel is based on the kernel developed in
\citep{McmcLinkingEdge}.

\begin{proposition}[Linking Edge Space Mergesplit Proposal
Density]\protect\hypertarget{prp-linking-ms-prob}{}\label{prp-linking-ms-prob}

Let \(L_r, L_r^\ast\) be balanced linking edge plans such that there
exists \((T_{k}, T_{k'}, e) \in L_r\),
\((T^\ast_{k}, T^\ast_{k'}, e^\ast) \in L^\ast_r\) where
\(\mathrm{V}(T_{k} \cup T_{k'}) = \mathrm{V}(T^\ast_{k}\cup T^\ast_{k'})\),
and all other region trees and linking edges are the same. Given a tree
cut selection kernel \(\ArbKernel{\cdot|\cdot}\) then the mergesplit
probability \(M_r(L_r^\ast|L_r)\) is given by \[
M_r(L_r^\ast|L_r) = \varphi(T_{k}, T_{k'}, e\mid F_{r}) \frac{1}{\tau(G^\ast_{k} \cup G^\ast_{k'})} \cdot p_\text{cut}( (T^\ast_k, s_k), (T^\ast_{k'}, s_{k'}), e^\ast | T_{k}^\ast \cup \set{e^\ast} \cup T_{k'}^\ast).
\]

\begin{proof}
This follows immediately from \ref{cor-linking-forward-kernel} replacing $T_\ell$ with $T_k \cup T_{k'}$ and noting that $\tau(G^\ast_{k} \cup G^\ast_{k'}) = \tau(G_{k} \cup G_{k'})$.
\end{proof}

\end{proposition}

\begin{corollary}[Linking Edge Space Mergesplit
MH-Ratio]\protect\hypertarget{cor-linking-ms-ratio}{}\label{cor-linking-ms-ratio}

Given \(L_r, L_r^\ast\) and \(\ArbKernel{\cdot|\cdot}\) as in
Proposition~\ref{prp-linking-ms-prob} the associated acceptance
probability of going from \(L_r \to L_r^\ast\) is

\[
A(L_r^\ast, L_r) = \mathrm{min}\left(1, \frac{\varphi(T_{k}^\ast, T_{k'}^\ast, e^\ast \mid L_{r}^\ast)}{\varphi(T_{k}, T_{k'}, e\mid L_{r})} \frac{\exp{-J(\xi_r^\ast)}}{\exp{-J(\xi_r)}} \cdot \frac{\tau(G/\xi_r)}{\tau(G/\xi_r^\ast)} \cdot \left(\frac{\tau(G^\ast_{k})\tau(G^\ast_{k'})}{\tau(G_{k})\tau(G_{k'})}\right)^{\rho-1} \cdot \frac{p(T_k, T_{k'}, e)}{p(T^\ast_k, T^\ast_{k'}, e^\ast)} \right),
\] where \(\xi^\ast_r = \xi(L_r^\ast)\), \(\xi_r = \xi(L_r)\) and

\[
p(T^\ast_k, T^\ast_{k'}, e^\ast) = p_\text{cut}( (T^\ast_k, s_k), (T^\ast_{k'}, s_{k'}), e^\ast | T_{k}^\ast \cup \set{e^\ast} \cup T_{k'}^\ast)
\]

\end{corollary}

\subsection{Hierarchical
Sampling}\label{sec-hierarchical-sampling-proofs}

As mentioned earlier, gSMC can be modified to perform the same kind of
hierarchical sampling as described in \citet{mccartan2023}. This
modification also changes the target distribution and space, forward
kernel, optimal weights, and MCMC kernel. We will begin by assuming only
one level of administrative hierarchy, although as
Section~\ref{sec-gen-hier} explains, this can be generalized to
arbitrary levels of nested administrative hierarchies. At a high level,
this change is all driven by the use of a modified, hierarchical version
of Wilson's algorithm which draws trees hierarchically with respect to
the administrative units.

\subsubsection{Modified Target Space}\label{modified-target-space}

The hierarchical nature of the modified splitting procedure changes the
space of plans being sampled from. We define the plans produced by the
procedure as hierarchical plans. For hierarchical plans their most
useful properties are sampled plans are now guaranteed to have no more
than \(D-1\) administrative unit splits and any two regions will overlap
in at most one administrative unit but the differences go even further
than that. We now formally define hierarchical plans, the modified
target distribution, and several crucial related definitions. Throughout
we assume that for all administrative units their induced subgraphs are
connected.

\begin{definition}[Administrative
Units]\protect\hypertarget{def-admin-mapping}{}\label{def-admin-mapping}

We formalize the notion of administrative boundaries by defining a set
of administrative units denoted by \(A\) and an associated map
\(\eta: V \to A\) which maps vertices in \(G\) to their associated
administrative unit. This function induces an equivalence relation
\(\sim_\eta\) on vertices where \(v\sim_\eta u\) for nodes \(v\) and
\(u\) if and only if \(\eta(v)=\eta(u)\). We often use a simplifying
notation \(\eta^{-1}(a)\) to denote the subgraph induced by the vertices
\(v \in V\) where \(\eta(v) = a\).

\end{definition}

\begin{definition}[Administrative
Splits]\protect\hypertarget{def-admin-splits}{}\label{def-admin-splits}

Given a map \(G\) with a set of administrative units \((\eta, A)\) and a
\(r\)-region plan \(\xi_r\) we define the number of administrative
splits of the plan as the sum of the number of connected components of
each region intersect administrative unit minus the number of units:
\begin{align*}
    \Split{\xi_r} = \left(\sum_{a \in A} \sum_{k = 1}^r C(\eta^{-1}(a) \cap G_{k}) \right) - \abs{A}.
\end{align*} Where \(C(\cdot)\) counts the number of connected
components in the subgraph \(\eta^{-1}(a) \cap G_k\).

\end{definition}

\begin{definition}[Hierarchically
Connected]\protect\hypertarget{def-hier-connected}{}\label{def-hier-connected}

Given a set of administrative units \((\eta, A)\) we say a region
\((G_k, s_k)\) is hierarchically connected if for every administrative
unit \(a \in A\) that unit intersected with the region has at most one
connected component. In other words: \begin{align*}
    \forall a \in A \; \; \implies \mathrm{C}(G_k \cap \eta^{-1}(a)) \leq 1
\end{align*} We say a plan \(\xi_r\) is hierarchically connected if
every region is hierarchically connected.

\end{definition}

\begin{definition}[Administrative Quotient
Graph]\protect\hypertarget{def-admin-quo-graph}{}\label{def-admin-quo-graph}

Suppose we have a map \(G\) with a set of administrative units
\((\eta, A)\). Let \(H \subset G\) be a subgraph of \(G\). We define the
administrative level quotient multigraph \(H / \sim_{\eta}\) as the
multigraph produced by quotienting by \(\eta\). In other words, this is
the multigraph where each vertex is a unit \(a \in A\) such that
\(H \cap \eta^{-1}(a) \neq \emptyset\) and for \(a,b \in A\) each edge
in the multigraph corresponds to an edge between two vertices in \(H\)
across \(a,b\).

\end{definition}

\begin{definition}[Hierarchical
Tree]\protect\hypertarget{def-hier-tree}{}\label{def-hier-tree}

~

Given a map \(G\) and an administrative mapping \(\eta : V \to A\) we
say a spanning tree \(T_\eta\) on some subgraph \(H \subset G\) is a
valid \(\eta\)-hierarchical tree if

\begin{enumerate}
    \item $T_\eta$ is a spanning tree on $H$
    \item For all units $a \in A$ then $T_\eta$ restricted to $a$ is still a spanning tree. Formally 
\begin{align*}
    \forall a \in A \implies  T_\eta \cap \eta^{-1}(a) \text{ is still a spanning tree}
\end{align*}
Ignoring the case where $T_\eta \cap \eta^{-1}(a) = \emptyset$
\item The administrative level quotient multigraph $T_\eta / \sim_{\eta}$ is still a spanning tree on $H / \sim_{\eta}$
\end{enumerate}

\end{definition}

\begin{definition}[Hierarchical Spanning Tree
Count]\protect\hypertarget{def-hier-tree-count}{}\label{def-hier-tree-count}

Given a map \(G\) with a set of administrative units \((\eta, A)\) for a
connected subgraph \(H \subset G\) we define the number of hierarchical
spanning trees with respect to \(\eta\) as \begin{align*}
    \tau_{\eta}(H) = \tau(H / \sim_{\eta} ) \cdot \prod_{a \in A} \tau(H \cap \eta^{-1}(a))
\end{align*} where \(\tau(\emptyset) = 1\) by convention.

Instead of just the number of spanning trees that can be drawn on \(H\),
the hierarchical tree count is the product of

\begin{itemize}
\tightlist
\item
  The number spanning trees that can be drawn on the quotient multigraph
\item
  For each unit \(a\), the number of spanning trees that can be drawn on
  \(H\) intersect the vertices associated with \(a\)
\end{itemize}

\end{definition}

\begin{definition}[Hierarchical Plan
Tree]\protect\hypertarget{def-hier-plan-tree}{}\label{def-hier-plan-tree}

~

Given a plan \(\xi_r\) we say that a spanning tree \(T^\eta\) on \(G\)
is a \((\xi_r, \eta)\)-hierarchical plan tree (or just a hierarchical
plan tree for short) if it satisfies the following two properties

\begin{enumerate}
    \item $T_\eta$ is an $\eta$-hierarchical tree on $G$
    \item For each region $G_{k} \in \xi_r$ the restriction $T^\eta \cap G_{k}$ is an $\eta$-hierarchical tree on $G_{k}$.
\end{enumerate}

Essentially a hierarchical plan tree is a hierarchical tree on the
entire map such that the restriction to any region is also a
hierarchical tree on that region. This type of tree is useful as we will
prove later that the existence of a hierarchical plan tree is equivalent
to proving a plan is hierarchically splittable.

\end{definition}

\begin{definition}[Hierarchical
Plan]\protect\hypertarget{def-hier-plan}{}\label{def-hier-plan}

We say a plan \(\xi_r\) is an \(\eta\)-hierarchical plan (or simply a
hierarchical plan) if it is possible to draw a plan hierarchical tree on
it.

\end{definition}

We can now define the modified hierarchical target distribution which is
designed to sample hierarchical plans on \(G, (A, \eta)\) where
compactness is now parameterized by \(\tau_\eta(\cdot)\), not
\(\tau(\cdot)\).

\begin{definition}[Hierarchical Target
Distribution]\protect\hypertarget{def-hier-target}{}\label{def-hier-target}

The modified hierarchical version of gSMC is designed to sample plans
from the following class of target distributions:

\begin{equation}\phantomsection\label{eq-hier-target}{
    \pi(\xi) \propto {\bm 1}\{\xi \text{ is hierarchical}\} \exp{-J(\xi)} \prod_{k=1}^D \tau_\eta(G_k)^\rho,
}\end{equation}

Likewise the modified intermediate target distributions now become

\begin{equation}\phantomsection\label{eq-hier-intermediate-target}{
    \pi_r(\xi_r) \propto {\bm 1}\{\xi_r \text{ is hierarchical}\} \exp{-J(\xi_r)} \prod_{k=1}^r \tau_\eta(G_k)^\rho,
}\end{equation}

\end{definition}

We now present an equivalent characterization of hierarchical plans in
Proposition~\ref{prp-hier-plan-char}. But first we must define two new
concepts.

\begin{definition}[Administratively
Adjacent]\protect\hypertarget{def-admin-adj}{}\label{def-admin-adj}

We say two regions \(G_{k}, G_{k'}\) are administratively adjacent if
there exists at least one administrative unit \(z \in A\) such that
\(G_{k}, G_{k'}\) are adjacent within \(z\). In other words if
\begin{align*}
    \exists z \in A \; \; \mathcal{C}(G_{k} \cap \eta^{-1}(z), G_{k'} \cap \eta^{-1}(z)) \neq \emptyset
\end{align*}

\end{definition}

\begin{definition}[Administratively Adjacent Quotient
Graph]\protect\hypertarget{def-admin-adj-quo-graph}{}\label{def-admin-adj-quo-graph}

For a plan \(\xi_r\), we define the administrative region quotient graph
\(G/ (\sim_\eta \xi_r)\) to be a subgraph of \(G/\xi_r\) where each
vertex is a region and we only count an edge \((u,v) = e\) if \(u\) and
\(v\) are in the same administrative unit (\(\eta(u) = \eta(v)\)) but
different regions.

\end{definition}

This is essentially the plan quotient multigraph but instead of counting
all edges between regions we only count edges between regions that are
in the same administrative unit. In other words,
\(G/ (\sim_\eta \xi_r)\) is the administrative adjacency graph between
regions in \(\xi_r\).

We will show that a plan \(\xi_r\) is hierarchical if and only if it is
both administratively connected and for each connected component of
\(G/ (\sim_\eta \xi_r)\) the number of splits is equal to the number of
regions in the component minus 1. We present this formally in
Proposition~\ref{prp-hier-plan-char} but we first introduce some
supporting results.

\begin{lemma}[Joining Disjoint Trees at a Vertex]\label{lemma:Join1VertexOverlapTree}
Let $\tilde{G}$ be an undirected multigraph and let $T_1, T_2$ be spanning trees on the subgraphs $H_1, H_2 \subset \tilde{G}$ where $H_1, H_2$ have $n_1,n_2$ vertices respectively . Suppose that $T_1 \cap T_2 = (\set{v}, \emptyset)$. Then $T = T_1 \cup T_2$ is also a spanning tree on $H_1 \cup H_2$.
\begin{proof} To show $T$ is a tree we just need to show its connected and has $\abs{\IVertex{T}} - 1$ edges
\item Part 1. $T$ is connected \\
To show $T$ is connected pick an arbitrary $u \in T$ and WLOG suppose $u \in T_1$. Since $T_1$ is connected starting from $u$ we can visit every vertex in $T_1$. This includes $v$. Since we can visit $v$ and $v \in T_2$ then since $T_2$ is connected we can also visit every vertex in $T_2$, therefore $T$ is connected.
\item Part 2. $T$ has $\abs{\IVertex{T}} - 1$ edges \\
Note that since $T_1 \cap T_2 = (\set{v}, \emptyset)$ then we know $\IEdge{T} = \IEdge{T_1} \sqcup \IEdge{T_2}$ and since $T_1$ and $T_2$ are trees we know $\abs{\IEdge{T}} = n_1-1+n_2-1$. \\
Now consider that since $T_1, T_2$ only share one vertex $v$ then we know the vertex set of $T$ are equal to the vertices in $T_1$ and $T_2$ without $v$ and $v$ so
\begin{align*}
    \IVertex{T} &= (\IVertex{T_1} \setminus \set{v}) \cup \set{v} \cup (\IVertex{T_2} \setminus \set{v}) \implies \\
    \abs{\IVertex{T}} &= \abs{(\IVertex{T_1} \setminus \set{v})} +\abs{\set{v}} + \abs{(\IVertex{T_2} \setminus \set{v})} \\
    \abs{\IVertex{T}} &= n_1 - 1 + 1 + n_2-1 \\
    \abs{\IVertex{T}} &= n_1  + n_2-1
\end{align*}
Thus we've shown that $T$ has $n_1+n_2 - 2$ edges and $n_1+n_2-1$ vertices. Therefore by definition of spanning trees we know $T$ is a spanning tree on $H_1 \cup H_2$.
\end{proof}
\end{lemma}

\begin{lemma}[Joining Spanning Trees]\label{lemma:MergedTree}
Let $G_a, G_b$ be disjoint regions of $G$. Now let $T_1$ be a spanning tree on $H_1$ and $T_2$ be a spanning tree on $H_2$. Then for any $e \in \mathcal{C}(T_1, T_2)$ the subgraph $T = T_1 \cup \set{e} \cup T_2$ is a spanning tree on the merged region $G_L = G_a \cup G_b$.
\begin{proof} Let $n_a = \abs{\IVertex{G_a}}$ and $n_b = \abs{\IVertex{G_b}}$. Since $T_1, T_2$ are spanning trees on those regions we know $\IVertex{G_a} = \IVertex{T_1}$ and $\IVertex{G_b} = \IVertex{T_2}$ and thus by definition of being a spanning tree we know that $T_1$ has $n_a-1$ edges and $T_2$ has $n_b -1$ edges. \\
Now let $e \in \mathcal{C}(T_1, T_2)$ and consider $T = T_1 \cup \set{e} \cup T_2$. Since $T_1$ and $T_2$ are connected then we see that $T$ must also be connected as if we pick an arbitrary $v \in T_1$ since $T_1$ is connected we can visit every vertex there. Further, since $e \in \mathcal{C}(T_1, T_2)$ we can use $e$ to traverse from $T_1$ to $T_2$ and then since $T_2$ is connected we can traverse the rest of $T_2$.\\
Now consider since $T_1$ and $T_2$ have disjoint vertex sets then we know $e \notin T_1, T_2$. Therefore we know 
\begin{align*}
     \abs{\IEdge{T}} &= \abs{\IEdge{T_1}} + \abs{\set{e}} + \abs{\IEdge{T_2}} \\
     &= n_a - 1 +1 + n_b - 1  \\
    &= n_a + n_b - 1
\end{align*}
Thus since $T$ is connected and has $n_a + n_b - 1$ vertices then by definition of spanning trees $T$ is a spanning tree on $G_a \cup G_b$.
\end{proof}
\end{lemma}

\begin{theorem}[Joining Hierarchical Trees]\label{thrm:HierTreeJoining}
Let $T_\eta^a, T^b_\eta$ be hierarchical trees on the adjacent, hierarchically connected regions $(G_a, s_a), (G_b, s_b)$. \\
Let $(u,v) = e \in \mathcal{C}(G_a, G_b)$ with $\eta(u) = x, \eta(v) = y$ and define $T = T_\eta^a \cup \set{e} \cup T^b_\eta$. \\
$T$ \textbf{is a hierarchical tree if either}
\begin{enumerate}
    \item $x=y$ and $\IVertex{G_a / \sim_{\eta}} \cap \IVertex{G_b / \sim_{\eta}} = \set{x}$ - meaning $u,v$ are both in the same administrative unit and the only administrative unit the regions overlap in is $x$
    \item $x \neq y$ and $\IVertex{G_a / \sim_{\eta}} \cap \IVertex{G_b / \sim_{\eta}} = \emptyset$ - meaning $u,v$ are in different counties and the two regions don't overlap in any administrative units.
\end{enumerate}
$T$ \textbf{is not a hierarchical tree if either}
\begin{enumerate}
    \item $\abs{\IVertex{G_a / \sim_{\eta}} \cap \IVertex{G_b / \sim_{\eta}}} > 1$ - Meaning that the two regions overlap in more than 1 administrative unit 
    \item $x \neq y$ and $\abs{\IVertex{G_a / \sim_{\eta}} \cap \IVertex{G_b / \sim_{\eta}}} > 0$ - Meaning that the edge crosses an administrative boundary and the regions overlap in at least one administrative unit
\end{enumerate}

\begin{proof} Before we proceed define the following for convenience 
\begin{itemize}
    \item  $(u,v) = e \in \mathcal{C}(G_a, G_b)$ with $\eta(u) = x, \eta(v) = y$ 
    \item Let $C_a$ be the administrative units in $G_a$ (i.e., $z \in A$ such that $G_a \cap \eta^{-1}(z) \neq \emptyset$)
    \item Let $C_b$ be the administrative units in $G_b$
    \item Let $G_L = G_a \cup G_b$ so $G_L$ is the region formed by combining $G_a$ and $G_b$. Note then that $T$ is a spanning tree of $G_L$ (although it may not always be a hierarchical tree)
\end{itemize}
We will prove each of the statements above. 
\item Part 1. If $x=y$ (equivalently $\eta(u) = \eta(v)$) and $C_a \cap C_b = \set{x}$ then $T$ is a hierarchical tree \\
We already know that $T$ is a spanning tree so we just need to show its a hierarchical tree 
\begin{enumerate}
    \item For each $z \in A$, $T \cap \eta^{-1}(z)$ is a spanning tree on $G_L \cap \eta^{-1}(z)$ \\
First lets consider $z \neq x$. For both $T^a_\eta \cap \eta^{-1}(z)$, $T^b_\eta \cap \eta^{-1}(z)$ those restricted trees are completely unchanged since we only added an edge in $\eta^{-1}(x)$ so that is fine.\\
For $x$ we know that $T^a_\eta \cap \eta^{-1}(x)$, $T^b_\eta \cap \eta^{-1}(x)$ are spanning trees. Since $G_a$ and $G_b$ are disjoint regions then we know any trees drawn on any subset of those regions must also be disjoint. That means that $T^a_\eta \cap \eta^{-1}(x)$, $T^b_\eta \cap \eta^{-1}(x)$ are disjoint spanning trees on a subset of $\eta^{-1}(x)$ and there is an underlying edge $(u,v) = e \in \mathcal{C}(T^a_\eta, T^b_\eta)$ where $u,v \in \eta^{-1}(x)$. Thus by \ref{lemma:MergedTree} we know that \\
$(T^a_\eta \cap \eta^{-1}(x)) \cup \set{e} \cup (T^b_\eta \cap \eta^{-1}(x))$ is a spanning tree on $(G_a \cup G_b) \cap \eta^{-1}(x)$. This is equivalent to $T \cap \eta^{-1}(x)$ is a spanning tree on $G_L \cap \eta^{-1}(x)$.
    \item $T / \sim_{\eta}$ is a spanning tree on $G_L / \sim_{\eta}$ \\
Note that since $x = y$ and $C_a \cap C_b = \set{x}$ then we know that the only administrative unit that is in both regions is $x$. Thus we know that $(T_\eta^a / \sim_{\eta}) \setminus \set{x}$ and $(T_\eta^b / \sim_{\eta}) \setminus \set{x}$ are disjoint trees on the administrative quotient multigraph that only share the vertex $x$. Thus by \ref{lemma:Join1VertexOverlapTree} we know that \\
$T_\eta^a / \sim_{\eta} \cup T_\eta^b / \sim_{\eta} = T / \sim_{\eta}$ is a tree on the administrative quotient multigraph 
\end{enumerate}
\item Part 2. If $x \neq y$ and $C_a \cap C_b = \emptyset$ then $T$ is a hierarchical tree \\
Again we need to show $T$ is a hierarchical tree.
\begin{enumerate}
    \item For each $z \in A$, $T \cap \eta^{-1}(z)$ is a spanning tree on $G_L \cap \eta^{-1}(z)$ \\
First recall that since $G_a, G_b$ are disjoint regions and they don't overlap in any administrative units then we know for any administrative unit $z \in A$ that is contained in at least one of the regions it must actually be in only 1 region. WLOG assume that region is $G_a$ meaning that $T_\eta^a \cap \eta^{-1}(z)$ is a spanning tree and $T_\eta^a \cap \eta^{-1}(z) = \emptyset$. \\
Now since we know that $e$ is an edge between counties and thus cannot be fully contained in $\eta^{-1}(z)$ then we know that $T$ restricted to $\eta^{-1}(z)$ is just equal to  $T_\eta^a \cap \eta^{-1}(z)$ which we already know is a spanning tree. Thus we've shown $T \cap \eta^{-1}(z)$ is a spanning tree 
    \item $T / \sim_{\eta}$ is a spanning tree on $G_L / \sim_{\eta}$ \\
Note that since $C_a \cap C_b = \emptyset$ then that means $T^a_\eta / \sim_\eta$ and $T^b_\eta / \sim_\eta$ are disjoint spanning trees on the administrative quotient multigraph. Furthermore since $e$ is an edge between units $x$ and $y$ then we know it is an edge on the underlying administrative quotient multigraph between the nodes $x$ and $y$. Therefore by \ref{lemma:MergedTree} we know that $T^a_\eta / \sim_\eta \cup \set{e} \cup T^b_\eta / \sim_\eta$ is a spanning tree on the administrative quotient multigraph. Thus since we know $T / \sim_\eta = T^a_\eta / \sim_\eta \cup \set{e} \cup T^b_\eta / \sim_\eta$ then we have shown $T / \sim_\eta$ is a tree on $G_L / \sim_\eta$.
\end{enumerate}
\item Part 3. If $\abs{C_a \cap C_b} > 1$ then $T$ is not a hierarchical tree \\
Since $\abs{C_a \cap C_b} > 1$  we know there are at least two administrative units which are in both regions. Lets denote them $z, w \in C_a \cap C_b$. Note for $T$ to be a hierarchical spanning tree we need both $T \cap \eta^{-1}(z)$ and $T \cap \eta^{-1}(w)$ to be spanning trees. \\
We also know that since the two regions are disjoint then that means we know $T_\eta^a, T_\eta^b$ are disjoint which also means those two trees restricted to $w$ and $z$ are disjoint. The only way to make both $(T_\eta^a \cap \eta^{-1}(z)) \cap  (T_\eta^b \cap \eta^{-1}(z))$ and $(T_\eta^a \cap \eta^{-1}(w)) \cap  (T_\eta^b \cap \eta^{-1}(w))$ trees is to add an edge two both restricted trees. However we know that $T = T^a_\eta \cup \set{e} \cup T^b_\eta$ so no matter what $e$ is an edge from it is impossible to make both of the restricted trees connected and thus at least one of $T \cap \eta^{-1}(z)$, $T \cap \eta^{-1}(w)$ must not be a spanning tree meaning $T$ cannot be a hierarchical tree. 

\item Part 4. If $x \neq y$ and $\abs{C_a \cap C_b} > 0$ then $T$ is not a hierarchical tree \\
Let $z \in C_a \cap C_b$ be one of the administrative units contained in both regions. Now consider since $\eta(v) \neq \eta(u)$ then we know at least one of $u$ or $v$ is not contained in $\eta^{-1}(z)$. WLOG suppose $v \notin \eta^{-1}(z)$. Now consider like before since the two regions are disjoint we know $T_\eta^a, T_\eta^b$ are disjoint which also means those two trees restricted to $z$ are disjoint. Now recall that $T = T^a_\eta \cup \set{e} \cup T^b_\eta$ and since $e$ is not an edge within $\eta^{-1}(z)$ then $T \cap \eta^{-1}(z) = T_\eta^a \cap \eta^{-1}(z) \sqcup T_\eta^b \cap \eta^{-1}(z)$. In other words $T$ restricted to $\eta^{-1}(z)$ is unchanged and still the union of two disjoint spanning trees. That is not a spanning tree so since $T \cap \eta^{-1}(z)$ is not a spanning tree is cannot be a hierarchical tree. 
\end{proof}
    
\end{theorem}

\begin{lemma}[Drawing A Hierarchical Tree on Administratively Adjacent
Connected
Component]\protect\hypertarget{lem-AdjComponentTree}{}\label{lem-AdjComponentTree}

~

Let \(\xi_r\) be a hierarchically connected plan. Now, let
\(\mathscr{C}_i\) be a connected component of the administratively
adjacent quotient graph \(G/ (\sim_\eta \xi_r)\). If the number of
splits in \(\mathscr{C}_i\) is equal to the number of regions in the
component minus one, i.e.,
\(\Split{\mathscr{C}_i} = \abs{\IVertex{\mathscr{C}_i}} - 1\), then it
is possible to draw a tree \(T^\eta_i\) such that

\begin{itemize}
    \item $T^\eta_i$ restricted to any region $G_k \in \mathscr{C}_i$ is a $\eta$-hierarchical tree
    \item $T^\eta_i$ is a hierarchical tree on the region made merging all the regions in $\mathscr{C}_i$ i.e., $T^\eta_i$ is a hierarchical tree on $\bigcup_{G_k \in \mathscr{C}_i} G_k$ 
\end{itemize}

\begin{proof}
For the rest of the proof lets assume that $\Split{\mathscr{C}_i} = s$ and $s > 0$. We know the $s=0$ case is trivial because that means the component has no splits which means its simply a single hierarchically connected region and its always possible to draw a hierarchically connected tree on a hierarchically connected region. We will proceed in parts now 
\item Part 1. The entire $\mathscr{C}_i$ does not split any administrative units \\
This part is very straightforward. We just need to prove for any administrative unit $z \in A$ then it is either completely contained in $\mathscr{C}_i$ or not in it at all. This is obvious by definition of $G/ (\sim_\eta \xi_r)$. Recall for this quotient graph we only count edges between two regions that are wholly within an administrative unit. So if we have a connected component of $G/ (\sim_\eta \xi_r)$ then it must be the case that an administrative unit is either wholly contained in it or not in it at all. Its not possible to only contain part of an administrative unit because if it did that would imply one of the regions was administratively adjacent to a region in the component but by definition the component should contain that already. \\
The reason we care about this is because it means we can decompose $\Split{\xi_r}$ into a sum of splits of the connected components in $G/ (\sim_\eta \xi_r)$.
\item Part 2. For any two $G_k, G_{r,k'} \in \mathscr{C}_i$ they overlap in at most one administrative unit \\
Since $\mathscr{C}_i$ is a connected and has $s+1$ vertices we know there exists a path $P = \set{e_1,\dots,e_s}$ that visits every region once. In other words we know $e_j = (u_{j}, v_j)$ where $\eta(v_j) = \eta(u_j)$ and for $j \neq j'$ we know at least one of the regions in $e_j, e_{j'}$ is not in the other. So that is to say each edge is associated with a distinct pair of regions in $\mathscr{C}_i$. \\
Now consider that we know the number of splits in $\mathscr{C}_i$ must be at least equal to the number of edges in $P$ as consider the following. Since each $e_j$ is wholly contained within an administrative unit consider an arbitrary $z \in \mathscr{C}_i$ that contains $\ell$ edges, i.e., $\eta(e_1),\dots,\eta(e_\ell) \in \eta(z)^{-1}$. Since each edge corresponds to a distinct pair of regions that means there must be at least $\ell + 1$ regions in $\eta^{-1}(z)$ which means $\Components{z, \xi_r} \geq \ell + 1$. That tells us that the number of splits in $\mathscr{C}_i$ must be at least $\ell$. Since this is true for arbitrary $z \in \eta(\mathscr{C}_i)$ we thus see it must be true that if we sum over all units there must be at least $s+1-1$ splits in $\mathscr{C}_i$ because of the edges in $P$. \\
Now lets suppose that two regions $G_k, G_{r,k'} \in \mathscr{C}_i$ overlapped in at least two administrative units. That means there is another edge $\tilde{e}$ between the two regions that is contained within a different administrative unit from the edge associated with the pair of regions in $P$. That implies that there is at least one extra split in $\mathscr{C}_i$ that we didn't count earlier by summing over the edges in $P$. That in turn implies that $\mathscr{C}_i$ has at least $s+1$ splits however that contradicts our assumption that $\mathscr{C}_i$ has the number of regions minus 1, so $s$ splits. Therefore this cannot be the case and it must be true that for any two $G_k, G_{r,k'} \in \mathscr{C}_i$ they overlap in at most one administrative unit 
\item Part 3. Constructing the tree \\
To construct the promised tree $T_i^\eta$ we proceed as follows. Let $G_{r,k_1},\dots,G_{r,k_{s+1}}$ be the regions in $\mathscr{C}_i$. On each region draw an $\eta$-hierarchical tree $T_{k_j}^\eta$ (which we can do since each region is hierarchically connected). Now we are going to connect them iteratively using our path $P$ from part 2. \\
Start with $e_1 = (v_1, u_1)$. These vertices have associated regions $G_{r,\xi_r(v_1)}, G_{r,\xi_r(u_1)}$ and we know from part 2 that they don't overlap in more than 1 administrative unit. Since we know they are administratively adjacent that means $G_{r,\xi_r(v_1)}, G_{r,\xi_r(u_1)}$ are adjacent in exactly one administrative unit. So if we take the hierarchical trees we've drawn on those regions and join them with $e_1$ to create $T^\eta_{\xi_r(v_1)} \cup e_1 \cup T^\eta_{\xi_r(u_1)}$ then by theorem \ref{thrm:HierTreeJoining} we know that the joined tree is a hierarchical tree on the merged region $G_{r,\xi_r(v_1)} \cup G_{r,\xi_r(u_1)}$. We can continue this process for $e_2$, leveraging the fact that the two regions are administratively adjacent for exactly 1 county and $e_2$ is an edge within that unit, to see that if we join $T^\eta_{\xi_r(v_1)} \cup e_1 \cup T^\eta_{\xi_r(u_1)}$ with $e_2$ and the hierarchical tree on the other region then we know have a hierarchical tree on the region formed by merging the three regions in $e_1,e_2$. We can continue doing this for every edge $e \in P$ and we see since $P$ is a path visiting all regions in the entire component $\mathscr{C}_i$ then the merged tree is a valid hierarchical tree on the region formed by merging all the regions in $\mathscr{C}_i$. 
\end{proof}

\end{lemma}

\begin{proposition}[Equivalent Characterization of Hierarchical
Plan]\protect\hypertarget{prp-hier-plan-char}{}\label{prp-hier-plan-char}

~

Given a plan \(\xi_r\) it is possible to draw at least one
\((\xi_r,\eta)\)-hierarchical plan tree on \(\xi_r\) (and thus \(\xi_r\)
is a hierarchical plan) if, and only if, the following are true

\begin{itemize}
    \item $\xi_r$ is hierarchically connected 
    \item For each connected component $\mathscr{C}_i$ of $G/ (\sim_\eta \xi_r)$ the number of splits of the component is equal to the number of regions in the component minus 1, in other words 
\begin{align*}
    \Split{\mathscr{C}_i} = \abs{\IVertex{\mathscr{C}_i}}-1
\end{align*}
\end{itemize}

\begin{proof}
We begin by proving the forward direction. 
The high level idea of the proof is to apply \ref{lem-AdjComponentTree} to each connected component of $G/(\sim_\eta \xi_r)$ and then connect the components in a tree. \\
\item Part 1. Drawing a tree on each $\mathscr{C}_i$ \\
Suppose $G/ (\sim_\eta \xi_r)$ has $\ell$ connected components $\mathscr{C}_i$, $i = 1,\dots,\ell$. By \ref{lem-AdjComponentTree} we know we can draw trees $T^\eta_i$ for $i = 1,\dots,\ell$ such that for each $i$ then we know $T^\eta_i$ is a hierarchical tree on each region in $\mathscr{C}_i$ and a hierarchical tree on the region made by merging every region in $\mathscr{C}_i$. 
\item Part 2. Combining all the $T^\eta_i$ into a plan hierarchical tree \\
First note that since $G/\xi_r$ is a connected multigraph and $G/(\sim_\eta \xi_r)$ is a subgraph of that we know there exists a path $P = \set{e_1,\dots,e_{\ell-1}}$ connecting all the connected components of $G/(\sim_\eta \xi_r)$. Further since these are edges across connected components of administratively adjacent regions we know that each one of the $e_j \in P$ is an edge that crosses administrative boundaries and that for the two regions associated with each $e_j$ they do not overlap in any administrative units (or else they would be in the same connected component). \\
Now consider $e_1 \in P$. Let $T^\eta_i, T_k^\eta$ are the two trees associated with the two respective components $\mathscr{C}_i, \mathscr{C}_k$. Further define $\tilde{G}_i, \tilde{G}_k$ as the regions made by merging all the regions in $\mathscr{C}_i, \mathscr{C}_k$ so
\begin{align*}
    \tilde{G}_i = \bigcup_{G_k \in \mathscr{C}_i} G_k && \tilde{G}_k = \bigcup_{G_{r,k'} \in \mathscr{C}_k} G_{r,k'}
\end{align*}
Now we know by \ref{lem-AdjComponentTree} that $T^\eta_i, T_k^\eta$ are hierarchical trees on $\tilde{G}_i, \tilde{G}_k$. Now consider that by definition of being a connected component in $G/(\sim_\eta \xi_r)$ we know that for any region in $\mathscr{C}_i$ is it not administratively adjacent to any region in $\mathscr{C}_k$ and vice versa. Therefore, we know that $\tilde{G}_i, \tilde{G}_k$ do not overlap in any administrative units. Thus we see that since $e_1$ is an edge between $\tilde{G}_i, \tilde{G}_k$ across administrative boundaries then we know by \ref{thrm:HierTreeJoining} that $T^\eta_i \cup \set{e_1} \cup T_k^\eta$ is a hierarchical tree on $\tilde{G}_i \cup \tilde{G}_k$. \\
Now take $e_2 \in P$. By the same argument as before we can take the merged tree we created using $e_1$ and merge it with the new component tree using $e_2$ and have the resulting tree be a hierarchical tree on the region made by merging the three connected components. This is because again since the trees are on administratively connected components we know that $e_2$ is an edge across administrative units linking regions that do not overlap in any administrative units. \\
So we see we can continue this process for every edge $e_j \in P$ until we have a hierarchical tree on the region made by merging all the regions in $\xi_r$ which is simply $G$. Thus we now have a tree that is a hierarchical tree on every region and on the entire map. This is exactly a plan hierarchical tree and we are thus done.

The reverse direction is trivial. If a hierarchical plan tree exists that necessarily implies $\xi_r$ is hierarchically connected and the connected component requirement is satisfied. 
\end{proof}

\end{proposition}

\subsubsection{Modification to Forward Kernels and SMC
Weights}\label{modification-to-forward-kernels-and-smc-weights}

Given administrative units \((A, \eta)\) and a multidistrict \(G_k\) the
hierarchical version of Wilson's algorithm works by first drawing a
spanning tree on the portion of each administrative unit contained in
\(G_k\), and then drawing a spanning tree on the quotient multigraph
\(G_k/\sim_\eta\). Taken together this forms a hierarchical tree on
\(G_k\) and the removal of any edge creates two new hierarchical trees.
Thus we see this hierarchical version of Wilson's algorithm is designed
to sample hierarchical plans.

The relevance of this change shows up in several different aspects of
the forward kernels and optimal weights for the three different sampling
spaces. At a high level, for all three it changes any \(\tau\) terms to
\(\tau_eta\) to reflect the fact that trees are now sampled
hierarchically. For graph and forest space sampling it changes both the
number of pairs of adjacent regions we sum over in the weights as well
as the boundary length terms to reflect the fact that sometimes certain
merges and edges are hierarchically invalid. Specifically we sum over
hierarchically adjacent pairs (Definition~\ref{def-hier-adj-region}) and
edges in the administrative boundary
set(Definition~\ref{def-admin-boundary}). For linking edge space it also
changes the calculation of the linking edge correction term
\(\tau(G/\xi)\) to reflect the hierarchical nature of linking edges now.
We do not present proofs of the modified forward kernel and weights as
they are essentially the same as the standard version but we do present
the modified final results. We now present everything in full detail.

\begin{definition}[Administrative Boundary
Set]\protect\hypertarget{def-admin-boundary}{}\label{def-admin-boundary}

~

Suppose we have a map \(G\) with a set of administrative units
\((\eta, A)\). Let \(G_{a}\), \(G_{b}\) be two disjoint subgraphs of
\(G\). We define the administrative boundary set
\(\mathcal{C}_\eta(G_a, G_b)\) as \begin{align*}
    \mathcal{C}_\eta(G_a, G_b) = \begin{cases}
        \mathcal{C}(G_a, G_b) & \text{If  } \eta(G_a) \cap \eta(G_b) = \emptyset \\
        \mathcal{C}(G_a \cap \eta^{-1}(x), G_b \cap \eta^{-1}(x)) & \text{If  } \eta(G_a) \cap \eta(G_b) = \set{x} \\
        \emptyset & \text{if }  \abs{\eta(G_a) \cap \eta(G_b)} \geq 2
    \end{cases}
\end{align*} In other words this set is

\begin{enumerate}
    \item All normal boundary edges between $G_a$ and $G_b$ if they don't share any administrative units
    \item If $G_a$ and $G_b$ only overlap in the county $\eta(G_a) = \eta(G_b) = \set{x}$ then its just the edges between $G_a$ and $G_b$ in $\eta^{-1}(x)$
    \item If $G_a$ and $G_b$ overlap in more than 1 county then the set is empty. 
\end{enumerate}

\end{definition}

\begin{definition}[Hierarchically Adjacent
Regions]\protect\hypertarget{def-hier-adj-region}{}\label{def-hier-adj-region}

Suppose we have a map \(G\) with a set of administrative units
\((\eta, A)\). Let \(\xi_r\) be a hierarchical \(r\)-region plan. We say
that two adjacent regions \(G_k, G_{k'}\) are hierarchically adjacent if
the plan \(\xi_{r-1}\) formed by merging \(G_k\) and \(G_{k'}\) is still
a hierarchical plan. We denote this property as
\(G_k \sim_\eta G_{k'}\).

\end{definition}

\begin{proposition}[Characterization of Hierarchically Adjacent
Regions]\protect\hypertarget{prp-hier-adj-cond}{}\label{prp-hier-adj-cond}

Suppose we have a map \(G\) with a set of administrative units
\((\eta, A)\). Let \(\xi_r\) be a hierarchical \(r\)-region plan with
\(G_k, G_{k'}\) adjacent in \(\xi_r\). If one of the following holds

\begin{enumerate}
\def\labelenumi{\arabic{enumi}.}
\tightlist
\item
  \(G_{k}\) and \(G_{k'}\) are administratively adjacent (i.e., adjacent
  in the same administrative unit)
\item
  \(G_{k}\) and \(G_{k'}\) are not administratively adjacent and they
  are not in the same connected component of \(G/(\sim_\eta \xi_r)\)
\end{enumerate}

Then \(G_k\) and \(G_{k'}\) are hierarchically adjacent.

If \(G_{k}\) and \(G_{k'}\) are not administratively adjacent but they
are in the same connected component of \(G/(\sim_\eta \xi_r)\) then they
are not hierarchically adjacent.

\begin{proof}
This follows immediately from \ref{thrm:HierTreeJoining}.

\end{proof}

\end{proposition}

For graph space sampling the forward kernel and optimal weights change
as follows.

\begin{proposition}[Hierarchical Graph Space Forward
Kernel]\protect\hypertarget{prp-hier-graph-forward-kernel}{}\label{prp-hier-graph-forward-kernel}

Let \(\xi_{r-1}\) be a balanced hierarchical plan. If \(\xi_r\) is a
hierarchical balanced plan such that there exists some
\(H_{\ell} \in \xi_{r-1}\), \(G_{k}, G_{k'} \in \xi_r\) where
\(H_{\ell} = G_{k} \cup G_{k'}\) and if we choose \(\mathcal{K}\) such
that \[
    \mathcal{K} \geq \max_{T \in \mathcal{T}_\eta(H_{\ell})} \abs{\mathrm{ok}(T_\ell, \mathcal{S}_{r-1}(s_\ell))},
\] then under the hierarchical sampling procedure the forward kernel
probability is \[
    M_r(\xi_r \mid \xi_{r-1}) = \varphi(H_{\ell}\mid \xi_{r-1}) \cdot \frac{1}{\mathcal{K}_r} \frac{\tau_\eta(G_{k})\tau_\eta(G_{k'}) }{\tau_\eta(H_{\ell})} \abs{\mathcal{C}_\eta(G_{k}, G_{k'})} 
\]

\end{proposition}

\begin{proposition}[Hierarchical Graph Space Optimal
weights]\protect\hypertarget{prp-hier-opt-wt}{}\label{prp-hier-opt-wt}

Given a hierarchical forward kernel \(M_r\) and target distribution
\(\pi_r\), the optimal minimal variance incremental weights are \[
w_r(\xi_{r-1}, \xi_r) = \mathcal{K}_r \cdot \left( \sum_{\substack{ G_{k} \sim_\eta G_{k'} \in \xi_r}} \varphi(G_{k} \cup G_{k'} \mid \tilde{\xi}_{r-1})  \frac{\exp{-J(\tilde{\xi}_{r-1})}}{\exp{-J(\xi_{r})}} \left(\frac{\tau_\eta(G_{k} \cup G_{k'})}{\tau_\eta(G_k) \tau_\eta(G_{k'})}\right)^{\rho-1} \abs{\mathcal{C}_\eta(G_{k}, G_{k'})}  \right)^{-1},
\] where \(G_{k} \sim_\eta G_{k'}\) denotes hierarchically adjacent
regions (Definition~\ref{def-hier-adj-region}) in \(\xi_r\),
\(\tilde{\xi}_{r-1}\) is the plan formed by merging \(G_{k}\) and
\(G_{k'}\), and \(\mathcal{C}_\eta(G_{k}, G_{k'})\) is the
administrative boundary set (Definition~\ref{def-admin-boundary})

\end{proposition}

For forest space sampling we define a hierarchical forest plan \(F_r\)
as a forest plan such that the induced plan \(\xi(F_r)\) is a
hierarchical plan. We define the hierarchical effective boundary length
as follows:

\begin{definition}[Hierarchical Effective Region Tree Boundary
Length]\protect\hypertarget{def-hier-eff-boundary-len}{}\label{def-hier-eff-boundary-len}

Given a distribution over tree cuts \(p_\text{cut}( \cdot | \cdot)\) and
two adjacent region trees \((T_k, s_k), (T_{k'}, s_{k'})\), we define
\(\EffB[\eta]{T_k, T_{k'}}\), the effective region tree boundary length,
as the following sum

\[
\EffB[\eta]{T_k, T_{k'}} = \sum_{e \in \mathcal{C}_\eta(T_{k}, T_{k'})} p_\text{cut}(\TreeCut{e} | T_{k} \cup \set{e} \cup T_{k'})
\]

\end{definition}

The forest space forward kernel and optimal weights change as follows.

\begin{proposition}[Hierarchical Forest Space Forward
Kernel]\protect\hypertarget{prp-hier-forest-forward-kernel}{}\label{prp-hier-forest-forward-kernel}

Let \(\varphi(\cdot|\xi(F_r))\) be a distribution over multidistrict
trees in \(F_{r-1}\) that only depends on the induced plan \(\xi(F_r)\).
Let \(F_{r-1}\) be a hierarchical forest plan. If \(F_r\) is a
hierarchical forest plan such that there exists some
\(T_{\ell} \in \xi_{r-1}\), \(T_{k}, T_{k'} \in \xi_r\) where
\(\IGraph{T_{\ell}} = \IGraph{T_{k}} \cup \IGraph{T_{k'}}\) then the
forward kernel probability is \begin{align*}
    M_r(F_r \mid F_{r-1}) = \varphi(G_{\ell} \mid \xi_{r-1}) \cdot \frac{1}{\tau_\eta(G_\ell)} \cdot \EffB[\eta]{T_k, T_{k'}}
\end{align*} Where \((G_{\ell}, s_\ell) = \IRegion{T_{\ell}, s_\ell}\)
and \(\xi_{r-1} = \xi(F_{r-1})\)

\end{proposition}

\begin{proposition}[Hierarchical Optimal Forest
Weights]\protect\hypertarget{prp-hier-forest-opt-wt}{}\label{prp-hier-forest-opt-wt}

For hierarchical forest plans \(F_r, F_{r-1}\), given a hierarchical
forest space forward kernel \(M_r\) and target distribution
\(\tilde{\pi}_r\), the optimal minimal variance incremental weights are
given by
\begin{equation}\phantomsection\label{eq-hier-forest-optimal-weights}{
w_r(F_{r-1}, F_r)  = \left(\sum_{\substack{ T_{k} \sim_\eta T_{k'} \in F_r}} \varphi(G_{k} \cup G_{k'}  \mid \tilde{\xi}_{r-1})  \frac{\exp{-J(\tilde{\xi}_{r-1})}}{\exp{-J(\xi_{r})}} \frac{\tau_\eta(H_\ell)^{\rho-1}}{\tau_\eta(G_k)^{\rho-1} \tau_\eta(G_k')^{\rho-1}} \EffB[\eta]{T_k, T_{k'}}\right)^{-1},
}\end{equation} where \(\xi_r = \xi(F_r)\),
\((G_{k}, s_k) = \IRegion{T_k, s_k}\),
\((G_{k'}, s_{k'}) = \IRegion{T_{k'}, s_{k'}}\) and
\(\tilde{\xi}_{r-1}\) is the plan formed by replacing the adjacent
regions \(G_{k}\) and \(G_{k'}\) in \(\xi_r\) with the merged region
\(H_{\ell} = G_k \cup G_{k'}\).

\end{proposition}

For linking edge space we must first define the following concepts.

\begin{definition}[Hierarchical Linking Edge Plan
Definition]\protect\hypertarget{def-hier-linking-plan}{}\label{def-hier-linking-plan}

Given administrative units \((A, \eta)\) we define a hierarchical
\(r\)-region linking edge plan \(L_r\) as the tuple \(L_r = (F_r, E_r)\)
where \(F_r\) is a hierarchical \(r\)-region forest plan and
\(E_r \subset E\) is a linking edge set such that \(F_r \cup E_r\) is a
hierarchical plan tree on \(\xi(F_r)\) (where
\(F_r \cup E_r = E_r \cup \bigcup_{T_k \in F_r} T_k\)).

\end{definition}

\begin{definition}[]\protect\hypertarget{def-hier-link-edge-count}{}\label{def-hier-link-edge-count}

For a hierarchical plan \(\xi_r\) we define the hierarchical linking
edge count \(\tau_{\eta}(G/\xi_r)\) to be the number of spanning trees
that can be drawn on the \(G/ (\sim_\eta \xi_r)\), administratively
adjacent quotient graph, times the number of spanning trees that can be
drawn on each of the connected components of \(G/ (\sim_\eta \xi_r)\)
(i.e., connected subgraphs of \(G/ (\sim_\eta \xi_r)\)). If there are
\(n\) such components we can write this value as

\[
\tau_{\eta}(G/\xi_r) = \tau(G/ (\sim_\eta \xi_r)) \prod_{i=l}^n \tau(\mathscr{C}_i).
\]

Notice that this computation is similar to the decomposition of
\(\tau_\eta(G_k)\) into a product over the spanning trees within each
region intersect administrative unit and then the count of trees across
on the quotient multigraph.

\end{definition}

\begin{proposition}[Hierarchical Linking Edge Space Forward
Kernel]\protect\hypertarget{prp-hier-linking-forward-kernel}{}\label{prp-hier-linking-forward-kernel}

Let \(\varphi(\cdot|\xi(F_r))\) be a distribution over multidistrict
trees in \(F_{r-1}\) that only depends on the induced plan \(\xi(F_r)\).
Let \(L_{r-1}\) be a hierarchical linking edge plan. If \(L_r\) is a
hierarchical linking edge plan such that there exists some
\(T_{\ell} \in L_{r-1}\), \(T_{k}, T_{k'} \in L_r\), with
\(T_{k}, T_{k'}\) connected by a linking edge, where
\(\IGraph{T_{\ell}} = \IGraph{T_{k}} \cup \IGraph{T_{k'}}\) then the
forward kernel probability is \begin{align*}
    M_r(L_r \mid L_{r-1}) = \varphi(G_{\ell} \mid \xi_{r-1}) \cdot \frac{1}{\tau_\eta(G_\ell)} \cdot p_{\text{cut}}(\TreeCut{e}|T_{k} \cup \set{e} \cup T_{k'}) 
\end{align*} Where \((G_{\ell}, s_\ell) = \IRegion{T_{\ell}, s_\ell}\)
and \(\xi_{r-1} = \xi(F_{r-1})\)

\end{proposition}

\begin{proposition}[Linking Edge Optimal
weights]\protect\hypertarget{prp-hier-opt-wt-link}{}\label{prp-hier-opt-wt-link}

Given a hierarchical forward kernel \(M_r\) and target distribution, the
optimal minimal variance incremental weights are \[
w_r(L_{r-1}, L_r) = \left( \sum_{T_k \stackrel{e}{\sim} T_{k'}\in L_r} \varphi(G_{\ell} |\xi_{r-1}) \frac{\exp{-J(\tilde{\xi}_{r-1})}}{\exp{-J(\xi_r)}} \frac{ \tau_\eta(H_\ell)^{\rho-1}}{\tau(G_k)_\eta^{\rho-1} \tau_\eta(G_{k'})^{\rho-1}} \frac{\tau_\eta(G/\xi_{r})}{\tau_\eta(G/\xi_{r-1})} p_\text{cut}(T_{k}, T_{k'} | T_{k} \cup \set{e} \cup T_{k'}) \right)^{-1}
\]

where \(T_k \stackrel{e}{\sim} T_{k'}\in L_r\) denotes adjacent region
trees in \(L_r\) connected by a linking edge, \(\xi_r = \xi(F_r)\),
\((G_{k}, s_k) = \IRegion{T_k, s_k}\),
\((G_{k'}, s_{k'}) = \IRegion{T_{k'}, s_{k'}}\), \(\tilde{\xi}_{r-1}\)
is the plan formed by replacing the adjacent regions \(G_{k}\) and
\(G_{k'}\) in \(\xi_r\) with the merged region
\(H_{\ell} = G_k \cup G_{k'}\), and
\(p^\ast(T_k, T_{k'},e ) = p_\text{cut}(T_{k}, T_{k'} | T_{k} \cup \set{e} \cup T_{k'})\).

\end{proposition}

\subsubsection{Modification to MCMC}\label{modification-to-mcmc}

All MCMC kernels are modified in the same manner as the forward kernel
and weights above meaning all relevant \(\tau(\cdot)\),
\(\mathcal{C}(\cdot,\cdot)\), and \(\EffB{\cdot,\cdot}\) terms are
replaced by their hierarchical versions \(\tau_\eta(\cdot)\),
\(\mathcal{C}_\eta(\cdot,\cdot)\), and \(\EffB[\eta]{\cdot,\cdot}\). In
addition, the rules for selecting which pairs of adjacent regions to
merge changes slightly. All hierarchically adjacent regions can be
safely merged but merges can also be attempted for any two adjacent
regions in the same hierarchically connected component however the
proposed plan must be rejected if it is a non-hierarchical plan.

\subsubsection{Generalizing Further}\label{sec-gen-hier}

The hierarchical sampling modification can be extended to an arbitrary
number of layers of nested administrative boundaries. For example, given
Census tracts that nest perfectly within municipalities which in turn
nest perfectly within counties the number of splits for each level can
be limited to \(D-1\) each. We would just modify Wilson's algorithm
again to draw spanning trees on each administrative unit subgraph and
across administrative unit multigraphs. The forward kernels and weights
would follow a similar modification where we must take into account both
which pairs of adjacent regions can be merged while still leaving the
results merged plan a valid hierarchical plan and what boundary edges
could have been split.

\section{Validation Examples}\label{sec-validation}

In this section we assess the empirical performance of the gSMC
algorithm on a single-member and multi-member map example. We only
display results for graph space sampling with and without MCMC steps.
Similar results for all other sampling space/splitting schedule
combinations were omitted.

\subsection{Single-Member Map}\label{single-member-map}

\subsubsection{Setup}\label{setup-3}

For the single-member districting map, we adopt a validation setting
similar to the one used in \citet{mccartan2023}. The map is a 7-by-7
grid, where each grid square has the same population. There are a total
of 158,753,814 exactly balanced 7-district plans on this map. The
enumerations were carried out using the Julia code from
\citet{schutzman2019enumerator}.

We use a target distribution with \(\rho=1\) and impose no additional
constraints through the \(J\) term other than exact population balance
and contiguity. As noted in Section~\ref{sec-target}, values of \(\rho\)
other than one may not be computationally feasible for much larger
real-world redistricting problems.

We evaluate both the gSMC algorithm with and without additional MCMC
steps (Section~\ref{sec-addl-mcmc}) across logarithmically spaced sample
sizes ranging from \(N=10\) to \(N=10,000\). For each MCMC round the
expected number of successful steps targeted was 7. The smaller sample
sizes are not intended for practical use; rather, they help illustrate
the behavior of the method when the sample size is too small. At each
value of \(N\), we run the algorithm independently 30 times so that
bias, standard errors, and \(\hat R\) can be estimated with reasonable
precision.

To assess sampling accuracy, as in \citet{mccartan2023}, we evaluate the
algorithms performance with respect to
\(\mathrm{rem}(\xi)\cdot|\IEdge{G}|\), the number of edges that have to
be removed to create a plan \(\xi\). As discussed in
\citet{mccartan2023}, this value is strongly correlated with the target
distribution when \(\rho = 1\) and no additional constraints are imposed
through \(J\).

\subsubsection{Results}\label{results-2}

\begin{figure}

\centering{

\pandocbounded{\includegraphics[keepaspectratio]{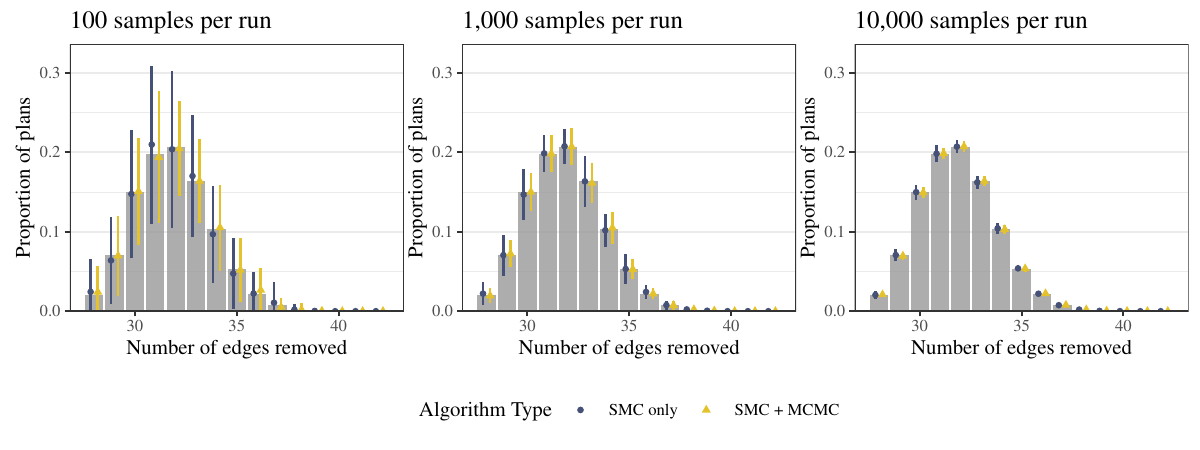}}

}

\caption{\label{fig-7x7}The panels show the histogram for the number of
removed edges based on the enumerated true distribution (grey bars) and
the corresponding empirical estimates from the gSMC algorithm with and
without MCMC steps. Each plot shows the results for different sample
sizes of \(N\). The dots and vertical lines denote the mean estimates
and their 90\% confidence intervals that are estimated using the 30
independent runs.}

\end{figure}%

Figure~\ref{fig-7x7} plots the exact ground truth distribution for the
number of edges removed \(\mathrm{rem}(\xi)\cdot|\IEdge{G}|\) (grey
histogram) and superimposes the corresponding estimates from the
simulated samples, both with (yellow) and without (black) the added MCMC
updates, under graph space sampling with any-valid splits. The plotted
intervals summarize uncertainty across the 30 independent runs. As the
number of samples grows, the variability of the estimates from the
simulated plans shrinks and the estimates approximate the true
distribution values well, providing strong empirical support for the
correctness of the gSMC algorithm.

We next examine whether the diagnostics from
Section~\ref{sec-diagnostics} track true sampling quality in this
validation setting. To make the comparison concrete, we focus on a
single estimand: the probability that a plan removes exactly 32 edges,
which is the median of the enumerated distribution and has a ground
truth probability of .207. For each value of \(N\), we compute the
\(\hat R\) and standard error of the estimand from the simulated plans.
We then compare these observable quantities (they do not require access
to the true value to compute) with the actual bias and root mean squared
error, which can be calculated here only because the ground truth is
known. The hope is that the readily available \(\hat R\) and standard
error diagnostics provide a reliable indication of the otherwise
unavailable bias and RMSE.

\begin{figure}

\centering{

\pandocbounded{\includegraphics[keepaspectratio]{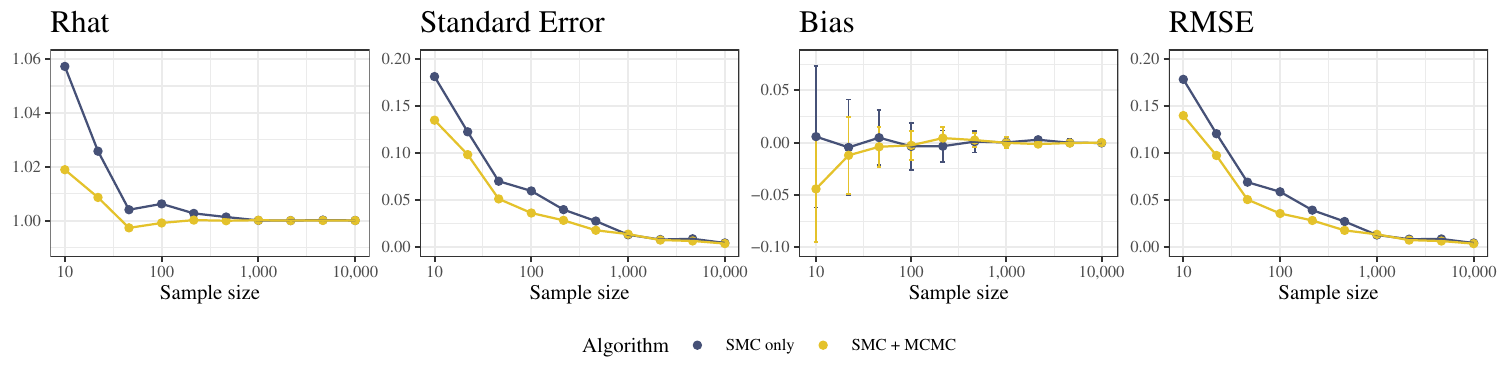}}

}

\caption{\label{fig-diagn-7x7}The \(\hat R\) statistic, standard errors,
bias, and RMSE for the median edges removed statistic, calculated across
30 independent runs of the gSMC algorithms for the graph partition
sampling spaces using any-valid splits with and without MCMC steps added
in. Values are plotted versus sample size \(N\) (on a log scale) per
run.}

\end{figure}%

Figure~\ref{fig-diagn-7x7} summarizes the results. As the number of
samples increases, all of the expected indicators move in the right
direction: the \(\hat R\) statistics fall, standard errors decline, and
RMSE decreases. For both variants of gSMC, \(\hat R\) drops below the
1.05 threshold by \(S=22\), suggesting that the independent runs are
producing consistent answers and the remaining error is being driven
largely by sampling variability, which decreases as the sample size
grows. When the sample size is small, the gSMC algorithm with MCMC steps
tends to perform better than the original SMC though their performance
becomes indistinguishable as the sample size increases. The results
suggest that, as in \citet{mccartan2023}, the standard errors and
\(\hat R\) statistics continue to perform well as proxies for RMSE and
convergence (and thus unbiasedness).

\subsection{Multi-Member Plans}\label{sec-mmd-validation}

We now perform another validation for multi-member district plans. We
consider a 5-by-7 grid map shown in Figure~\ref{fig-map-5x7} with 3
districts, 7 seats, and districts of size 2 or 3. Each vertex has an
equal population and there are a total of 420,993 balanced plans. The
enumerated plans were generated using \citet{schutzman2019enumerator}.
We present the analogous plots to those shown in
Section~\ref{sec-validation} except that we perform district-only splits
rather than any-valid splits. Figure~\ref{fig-5x7} shows the true
distribution of the edge removed statistic as a histogram. As with the
7-by-7 grid, by \(N=100\) samples the SMC estimates are all close to the
true values. Figure~\ref{fig-diagn-5x7} demonstrates the diagnostics
continue to perform well.

\begin{figure}

\centering{

\includegraphics[width=\linewidth,height=2.125in,keepaspectratio]{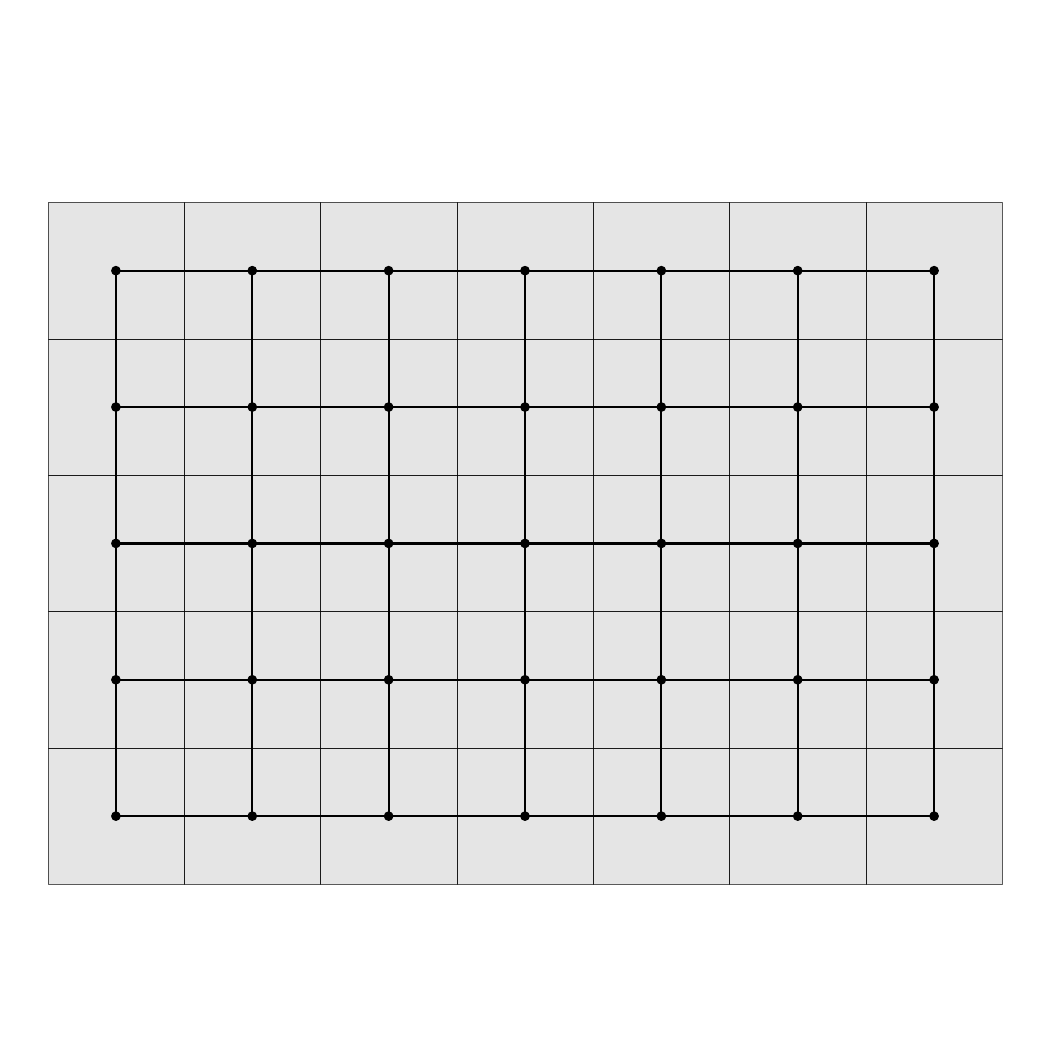}

}

\caption{\label{fig-map-5x7}The 5-by-7 map used in multi-member
validation.}

\end{figure}%

\begin{figure}

\centering{

\pandocbounded{\includegraphics[keepaspectratio]{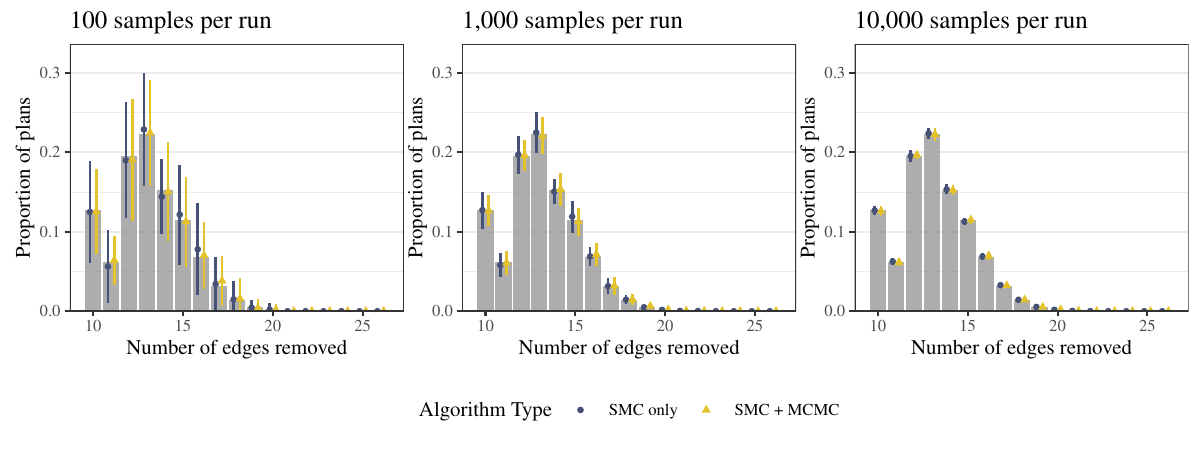}}

}

\caption{\label{fig-5x7}The panels show the proportion of removed edges
for the enumerated true distribution (grey bars) and the empirical
distribution from the gSMC algorithm with and without MCMC steps. Each
plot shows the results for different sample sizes of \(N\). The dots and
vertical lines denote the mean estimates and their 90\% confidence
intervals that are estimated using the 30 independent runs.}

\end{figure}%

\begin{figure}

\centering{

\pandocbounded{\includegraphics[keepaspectratio]{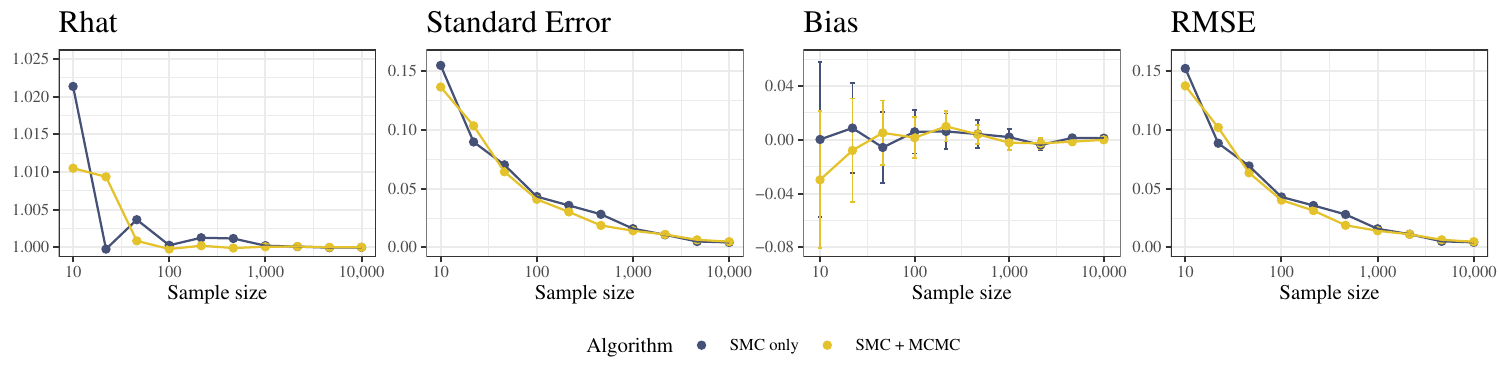}}

}

\caption{\label{fig-diagn-5x7}\(\hat R\), standard errors, bias, and
RMSE for the median edges removed statistic, calculated across 30
independent runs of the gSMC algorithms for the graph partition sampling
spaces using any district-only splits with and without MCMC steps added
in. Values are plotted versus sample size \(N\) (on a log scale) per
run.}

\end{figure}%

\section{Algorithm Implementation Details}\label{sec-appendixC}

\subsection{Choosing a multidistrict selection
probability}\label{sec-Choosing-varphi}

The selection probability \(\varphi\) is used during the sampling stage
to select a multidistrict to split and to compute the retroactive
selection probability of a pair of merged regions when computing
weights. Any choice that is a function of the region only is
theoretically valid. We have found it computationally convenient to
select a region \((H_\ell, s_\ell)\) with probability proportional to
its size (i.e., \((s_\ell)^\alpha\)) for some \(\alpha \in [0,1]\).
Empirical experiments in Section~\ref{sec-varphi-tests} suggest that
convergence and runtime is largely unaffected by \(\alpha\) in this
range.

\subsection{Choosing a Splitting
Schedule}\label{sec-Choosing-Split-Schedule}

We have implemented two types of splitting schedules: district-only and
any-valid (this is only implemented for single-member districting
schemes). District-only splits is the same method as the one used in
\citet{mccartan2023}, where one of the split regions is always a
district. In contrast, any-valid splits means that for a multidistrict,
we allow it to be split into two regions of any size greater than or
equal to \(d^-\). In theory, the splitting schedule could be chosen to
allow for other types of splits. For example, one could split a map with
\(12\) single-member districts into two regions of size 6, then each
into regions of size 3, and so on.

However, implementing these schedules is computationally challenging in
practice. This is because the choice of splitting schedule changes the
intermediate target distribution spaces and the weights for intermediate
steps (see Section~\ref{sec-custom-split-sched} for details).
Empirically, we find that district-only splits yield weights with higher
variance at the intermediate steps than any valid splits. However, both
appear to yield final weights with similar variances.

The choice of splitting schedule also affects the overall computational
cost. For a region \((H_\ell, s_\ell)\), Wilson's algorithm has a
complexity on the order of \(O(V(H_\ell)^2)\). It is costly to split
multidistricts of greater size because they generally have more
vertices. This is especially true for maps with more districts or lower
population tolerances where more calls to Wilson's algorithm must be
made. Overall, using any-valid splits typically results in a runtime
reduction of 5\% to 25\% when compared to district-only splits with
gSMC.\footnote{For large maps with more than 30 districts or so, this
  trend can sometimes disappear as runs will frequently become ``stuck''
  (i.e., none of the partial plans at some step \(r\) can be split
  further) or almost stuck (the acceptance rate craters precipitously
  low). The Any-valid schedule seems more prone to these issues.
  However, using a sufficient number of MCMC steps appears to prevent
  this from occurring.} When MCMC steps are added as discussed in
Section~\ref{sec-addl-mcmc}, the reduction can be as great as 50\%. See
Appendix Section~\ref{sec-comprehensive-sims} for more discussion.

\subsection{\texorpdfstring{Choosing \(\mathcal{K}_r\) for Graph
Space}{Choosing \textbackslash mathcal\{K\}\_r for Graph Space}}\label{sec-split-k-estimation}

For graph space sampling the theoretical validity of the gSMC algorithm
rests on the appropriate choice of \(\mathcal{K}_r\) at each stage. In
practice, however, it is computationally infeasible to exactly calculate
this parameter for a given region. Computing a global maximum over all
possible regions in plans in the sample of size \(r\) is also
intractable, although there are practical workarounds. In our open
source implementation \texttt{redist}, we follow the procedure suggested
by \citet{mccartan2023} and estimate the value of \(\mathcal{K}_r\)
before each step. We have found this estimation procedure works well and
passes the validation checks described in Appendix
Section~\ref{sec-validation}. However, we do not have any theoretical
results about the effect on convergence if \(\mathcal{K}_r\) is
estimated incorrectly. Using forest or linking edge space sidesteps this
problem altogether.

\subsection{Forward Kernels for Forest and Linking Edge
Space}\label{forward-kernels-for-forest-and-linking-edge-space}

For forest space and linking edge any choice of tree cut distribution
\(\ArbKernel{\cdot|\cdot}\) may be specified. We have implemented two
different ones in practice in our \texttt{redist} package. The first one
is simply uniform over the number of balanced tree cuts so \[
p_\text{cut}(\TreeCut{e}| T) \propto \abs{\mathrm{ok}(T_\ell, \mathcal{S}_{r-1})}
\] The second is \[
p_\text{cut}(\TreeCut{e}| T) \propto \exp{-\alpha \cdot \mathrm{MaxAbsDev}(\TreeCut{e})}
\]

for some value \(\alpha \in \bbR\). Empirically we have observed almost
no difference in the performance of these two different distributions
and as such recommend uniform over balanced tree cuts as a default.

\subsection{Custom Splitting Schedules}\label{sec-custom-split-sched}

As discussed earlier, there is a subtle interaction between the
splitting schedule and the intermediate target distributions. While the
final target distribution \(\pi = \pi_D\) is not changed by choice of
target distribution, the intermediate distributions are. This is
relevant when computing the weights as terms in the sum where merging
two regions would create a plan impossible to generate under the
splitting schedule must be set to zero. For district-only and any-valid
splits we can check if the plan associated with merging two regions is
valid under the respective splitting schedule by simply checking if the
merged plan conforms with the splitting schedule for the previous step.
However, this is not true of arbitrary splitting schedules in general
where it is possible to have merged plans where it is not clear how to
figure out if it is valid under the schedule.

To illustrate that consider the following example. We take our map to be
a \(6 \times 6\) grid where each vertex has the same population and we
wish to draw six single-member districts. Suppose we wanted to use a
custom splitting schedule where we split \(\xi_1\) into two regions of
size 2 and 4, then we split the region of size 4 into 2 and 2, and then
we split each region of size 2 into districts. We can represent this
symbolically in terms of the allowable region sizes at each step as \[
(6) \to (2,4) \to (2,2,2) \to (1,1,2,2) \to (1,1,1,1,2) \to (1,1,1,1,1,1).
\]

Now suppose we are calculating the graph space weights for the plan in
Figure \ref{fig-custom-split-example}. On first glance it would seem
permissible to merge the districts to create the 5-region plan as this
merged plan appears to conform with the splitting schedule requirement
that the sizes be \((1,1,1,1,2)\) when \(r=5\). However, upon further
consideration we see that this plan is not splittable under the custom
schedule as it is impossible to create a \(3\)-region plan with sizes
\((2,2,2)\) from merging regions. To see this consider the two possible
\(4\)-region plans in Figure \ref{fig-custom-split-example} with sizes
\((1,1,2,2)\) that can be made by merging adjacent regions in the
\(5\)-region plan. We see once we perform these merges it is impossible
to create a plan with sizes \((2,2,2)\) as the two remaining districts
are not adjacent. Thus we see when computing the weights for the
original \(r=6\) plan that initial merge of the two districts should
have probability zero in the sum.

\begin{figure}[t]
\centering
\begin{tikzpicture}[
    x=1cm, y=1cm,
    >=stealth,
    arr/.style={->, line width=0.9pt}
]

\node (r6) at (0,0)
    {\includegraphics[width=3.8cm]{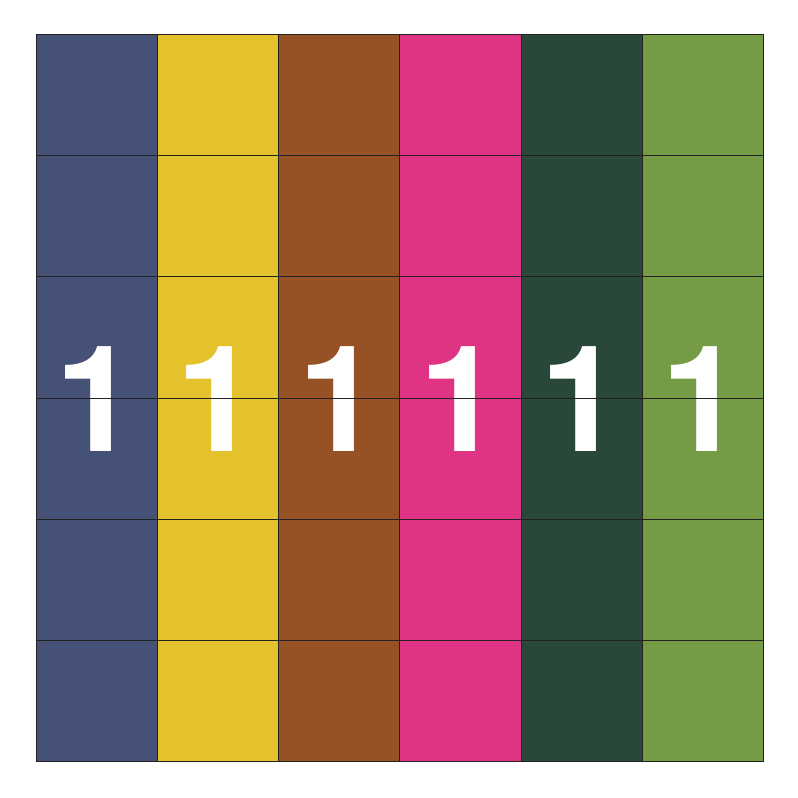}};

\node (r5) at (5.5,0)
    {\includegraphics[width=3.8cm]{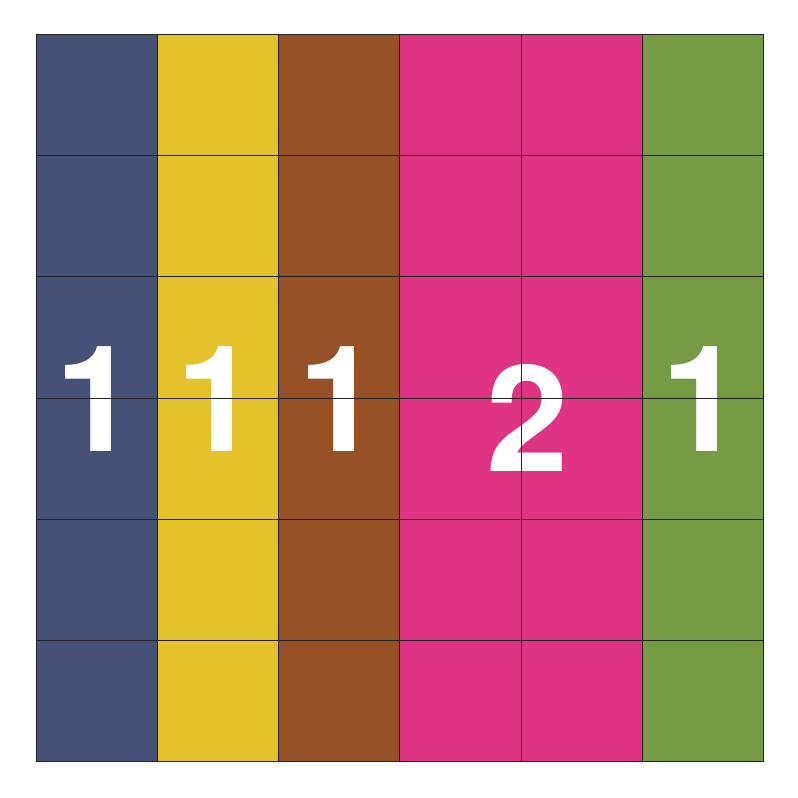}};

\node (r41) at (10.8,2.2)
    {\includegraphics[width=3.8cm]{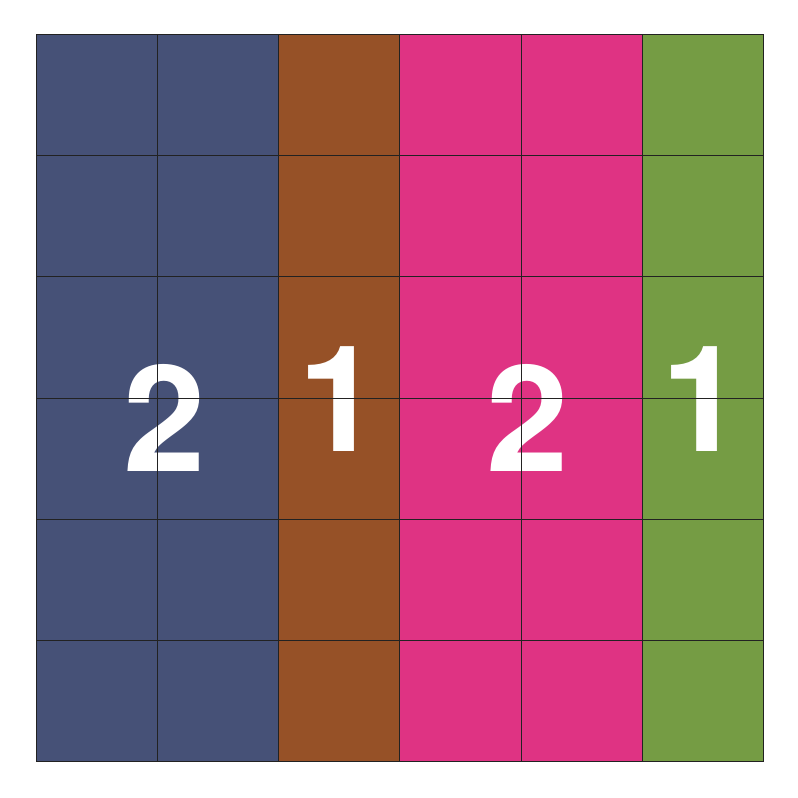}};

\node (r42) at (10.8,-2.2)
    {\includegraphics[width=3.8cm]{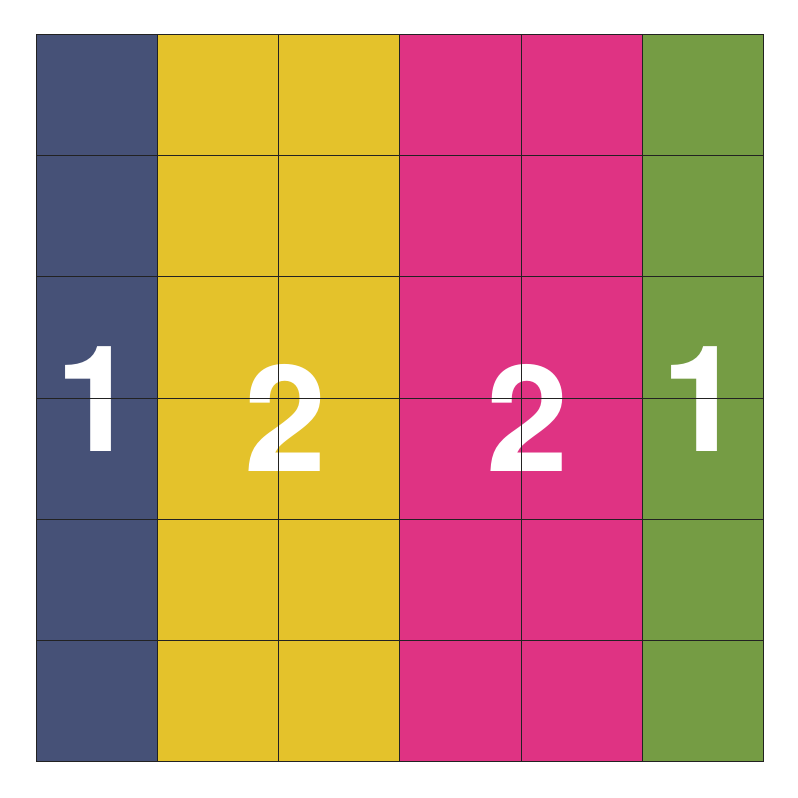}};

\draw[arr] (r6.east) -- (r5.west);
\draw[arr] (r5.east) .. controls +(1.6,1.0) and +(-1.6,-0.8) .. (r41.west);
\draw[arr] (r5.east) .. controls +(1.6,-1.0) and +(-1.6,0.8) .. (r42.west);

\end{tikzpicture}

\caption{Example of merges which are not actually possible under the custom splitting schedule. Regions in each plan are labeled by size.}

\label{fig-custom-split-example}
\end{figure}

This is just one example of why extreme care is needed for custom
splitting schedules. \newpage

\subsection{Choosing the Number of MCMC Moves}\label{sec-choose-mcmc}

In our open source implementation \texttt{redist}, we parameterize the
number of Markov steps to be performed with the variable
\verb|mh_accept_per_smc|, which represents the expected number of
proposals successfully accepted after each SMC step. The specific number
of kernel moves will thus depend on the map, districting scheme, and
constraints imposed. We set the number of steps globally using
\verb|ceiling(mh_accept_per_smc * (1/p))|, where \(p\) is the average
MCMC acceptance rate among all plans from the previous step. While we
currently do not have a theoretical justification for this choice, in
practice, we have found it helpful to set \verb|mh_accept_per_smc| to
about one third of the number of districts in the map. If necessary, we
iteratively increase the \verb|mh_accept_per_smc| by increments of 5 to
15 and then check if that leads to convergence. If this does not lead to
substantial progress, we increase the sample size. We leave a
theoretical investigation of this issue to future research.

\subsection{Local Sensitivity Analysis}\label{sec-local-sens}

Altering the target distribution by modifying the \(J\) function, the
value of \(\rho\), or other parameters can potentially affect downstream
analyses. This issue is especially relevant when the law imposes
qualitative requirements on districting plans, such as compactness or
the avoidance of splits in administrative units, without specifying how
these requirements should be operationalized or what thresholds should
be used. For example, Article II, § 16 of the Pennsylvania Constitution
requires districts to be ``compact and contiguous'' and states that
``unless absolutely necessary no county, city, incorporated town,
borough, township or ward shall be divided in forming either a
senatorial or representative district.'' Researchers must therefore
translate such qualitative legal requirements into quantitative choices
when specifying the simulation procedure. In the Pennsylvania case,
hierarchical sampling and constraints imposed through the \(J\) term can
be used to minimize county and municipality splits, but the appropriate
choice of specific parameter values may nevertheless remain unclear.

We describe two possible approaches to local sensitivity analysis for
this problem. First, researchers can vary the parameter values over a
plausible range and examine the sensitivity of their substantive
conclusions to these changes. This approach directly addresses the issue
but can be computationally intensive. Moreover, in applications such as
the Pennsylvania analysis, moving parameter values beyond a certain
range may substantially reduce sampling efficiency, making this approach
impractical. In such cases, an alternative is to examine the
relationship between the sensitivity parameter of interest and key
statistics from the downstream analysis. A strong relationship indicates
that even a small change in the parameter could meaningfully affect
empirical conclusions. The rationale for this approach parallels that of
the Bayesian robustness literature, where derivatives of posterior
expectations with respect to hyperparameters can be expressed in terms
of posterior covariances; see, e.g., \citet{giordano2018covariances} and
\citet{Gustafson2000}.

In the case of the Pennsylvania analysis, this local sensitivity
analysis approximates the expected number of Democratic seats under a
perturbation of \(\rho\) as \[
\E[\rho_1]{\text{dem seats}} \approx \E[\rho_0]{\text{dem seats}} + (\rho_1 - \rho_0) \mathrm{Cov}(\text{edges kept}, \text{dem seats}),
\] where \(\rho_0 = 1\) and \(\rho_1 > 1\). Thus, the weak correlation
between the number of edges kept and the number of Democratic seats
observed in Figure~\ref{fig-cor-plot} suggests that a small perturbation
of \(\rho\) around \(\rho=1\) would not substantially alter the partisan
conclusions drawn from the simulations.

\begin{figure}

\centering{

\pandocbounded{\includegraphics[keepaspectratio]{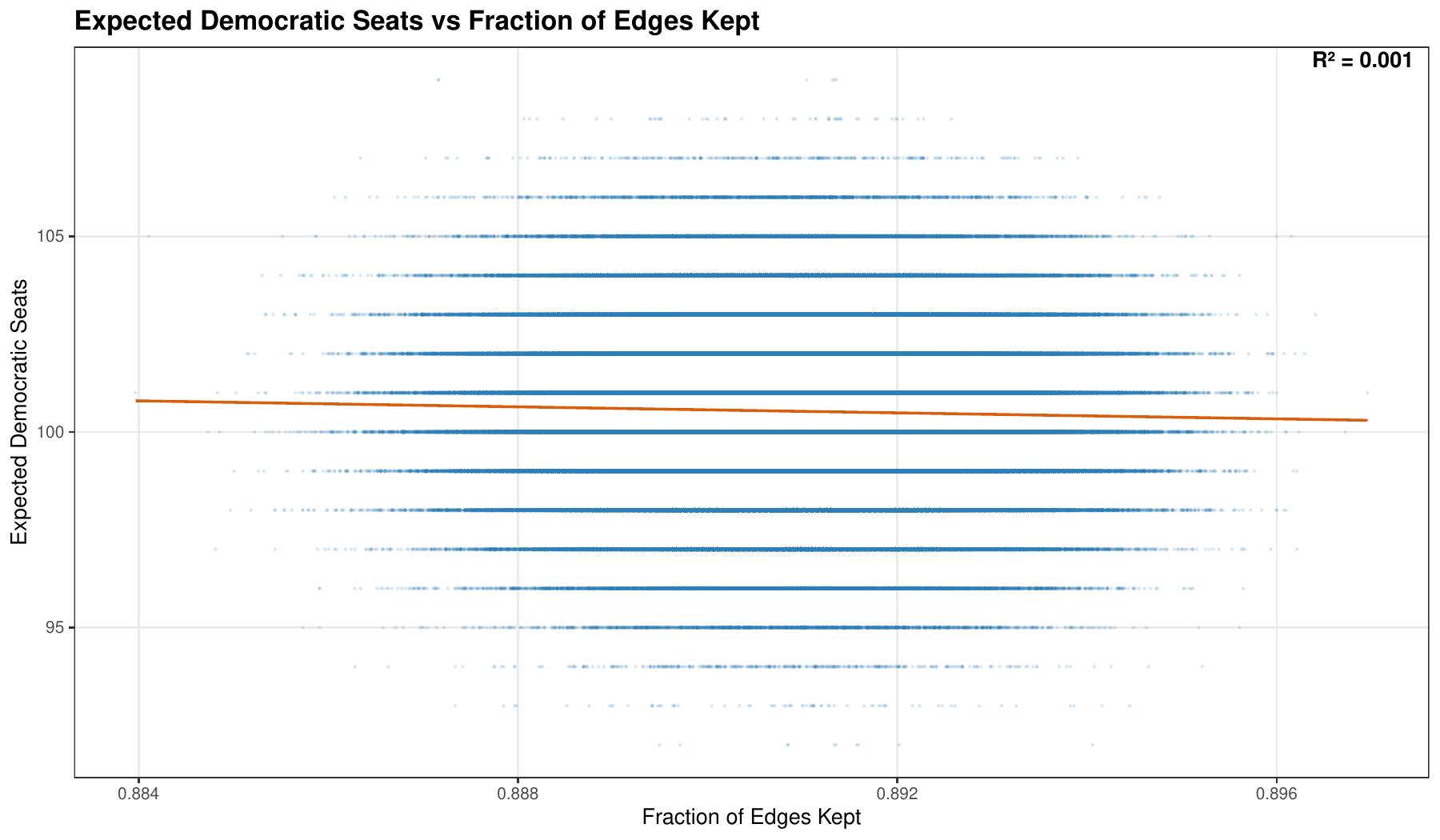}}

}

\caption{\label{fig-cor-plot}Sample Correlation Between Fraction of
Edges Kept and Expected Democratic Seats}

\end{figure}%

\subsection{Faster Forest Weights}\label{sec-fast-forest-weights}

\subsubsection{Procedure}\label{procedure}

For a generic tree cut distribution \(p_{\text{cut}}\) and two adjacent
region trees \((T_A, s_A)\) and \((T_B, s_b)\) in the forest plan
\(F_r\), computing the effective region tree boundary length
\(\EffB{T_A, T_B}\) is an
\(O((\abs{T_A} + \abs{T_B}) \cdot \mathcal{C}(T_A, T_B))\) operation, as
for each boundary edge \(e \in \mathcal{C}(T_A, T_B)\) we must traverse
all of \(T_A\) and \(T_B\) to compute the probability of choosing
\(\{(T_{A}, s_{A}), (T_{B}, s_{B}), e\}\) under \(p_{\text{cut}}\).
However, if the selection probability is solely a function of the
vertices in \(T_A\) and \(T_B\) and the sizes \(s_A\) and \(s_B\), i.e.,
\[p_{\text{cut}}((T_{A'}, s_{A'}), (T_{B'}, s_{B'}), e |T_A \cup \set{e} \cup T_B , s_A+s_B) \propto q((\IVertex{T_{A'}}, s_{A'}), (\IVertex{T_{B'}}, s_{B'})),\]
then through some specific computation it is actually possible to
compute the selection probability in
\(O(\abs{T_A} + \abs{T_B} + \mathcal{C}(T_A, T_B))\) time. The procedure
goes as follows

\begin{enumerate}
\def\labelenumi{\arabic{enumi}.}
\tightlist
\item
  Choose an arbitrary vertex \(r_A \in T_A\) and ``root'' the tree from
  that edge (i.e., treat \(T_A\) like a directed graph with all edges
  flowing away from the root). Given a non-root vertex \(v\) we let
  \(\text{parent}(v)\) denote the parent edge in this rooted tree.
\item
  For every \(v \in T_A\) define \(S_v\) as the subtree of \(T_A\)
  rooted at \(v\) and compute the following values \begin{align*}
  W_v^{\text{out}} &= \Phi(\IVertex{T_A \setminus S_v} \cup \IVertex{T_B}, \IVertex{S_v}) \\
  W_v^{\text{in}} &= \Phi(\IVertex{T_A \setminus S_v} , \IVertex{S_v} \cup \IVertex{T_B}) \\
  \triangle_v &= W_v^{\text{in}} - W_v^{\text{out}}
  \end{align*}
\end{enumerate}

Where we define \(\Phi(V_1, V_2)\) as the sum of all unnormalized
selection probabilities associated with the vertex sets \(V_1\) and
\(V_2\) over all possible split size configurations under the splitting
schedule \(\mathcal{S}(r, s_A+s_B, \xi(F_r))\), i.e.,

\[
\Phi(V_1, V_2) = \sum_{(s_1, s_2) \in \mathcal{S}(r, s_A+s_B, \xi(F_r))} q((V_1, s_{1}), (V_2, s_2))
\]

Normally for a specific edge \(e\) there is at most only one possible
balanced tree cut over all the splitting size configurations, however,
theoretically the algorithm could be run with such loose population
bounds that for an edge \(e\) and size \(s\) there are multiple possible
split size configurations that are still balanced. The \(\Phi\) function
accounts for that, although in practice it will almost always be the
case that

\[
\Phi(V_1, V_2) = q((V_1, s_{A}), (V_2, s_B))
\] 3. For the root set
\(w(r_A) = \sum_{v \in \IVertex{T_A}, v \neq r_a} W_v^{\text{out}}\).
Then, beginning with \(r_A\)'s children we successively compute
\[ w(v) = w(\text{parent}(v)) + \triangle_v\] 4. Repeat this rooting and
computation process for \(T_B\). 5. Using the computed \(w(\cdot)\)
values from the two trees we can compute the effective region tree
boundary \(\EffB{T_A, T_B}\) as \[
\EffB{T_A, T_B} = \sum_{(a,b) \in \mathcal{C}(T_A,T_B)} \frac{q((\IVertex{T_{A}}, s_{A}), (\IVertex{T_{B}}, s_{B}))}{w(a) + w(b) + \Phi(\IVertex{T_A}, \IVertex{T_B})}
\]

This entire operation is
\(O(\abs{T_A} + \abs{T_B} + \abs{\mathcal{C}(T_A, T_B)})\) as computing
the \(w(\cdot)\) values requires a constant time number of walks through
the tree (the \(\triangle_v\) and \(W_{v}^{\text{out}}\) can be computed
in a single pass through and the \(w(\cdot)\) values also only require a
single walk through the tree), leading to work on the order of
\(O(\abs{T_A})\) and \(O(\abs{T_B})\), and summing the values over the
boundary edges is on the order of \(\abs{\mathcal{C}(T_A, T_B)}\).

\subsubsection{Proof Sketch}\label{proof-sketch}

A detailed proof is omitted but the idea is provided. To see why the
procedure above works, we first begin by assuming we have chosen a root
vertex \(r_A\). We now note that since \(T_A\) is a tree we know there
is a unique path between any two edges in the tree. Thus, we know there
is a unique directed path from the root, \(r_A\), to any non-root vertex
\(v\). That means for every non-root vertex \(v\) we can associate it
with the unique edge \(e_v \in T_A\) given by \((\text{parent}(v), v)\)
where \(\text{parent}(v)\) is the vertex on the path from \(r_A\) to
\(v\) right before \(v\) and \(e_v\) is the last edge in the directed
path.

Now recall we define \begin{align*}
W_v^{\text{out}} &= \Phi(\IVertex{T_A \setminus S_v} \cup \IVertex{T_B}, \IVertex{S_v}) \\
W_v^{\text{in}} &= \Phi(\IVertex{T_A \setminus S_v} , \IVertex{S_v} \cup \IVertex{T_B})
\end{align*}

Now let \(a \in \IVertex{T_A}\) be a vertex in \(T_A\) on the boundary
between \(T_A\) and \(T_B\) in \(G\), (i.e., there exists at least one
edge \((a,b) \in \mathcal{C}(T_A,T_B)\)). There are two cases to
consider when we consider attaching \(T_B\) to our rooted tree.

\begin{enumerate}
\def\labelenumi{\arabic{enumi}.}
\tightlist
\item
  Case 1. \(a\) is ``above'' \(v\), i.e., \(a\) is ``outside'' the
  subtree rooted at \(v\) (\(a \notin \IVertex{S_v}\))
\end{enumerate}

In this case removing the edge \(e_v = (\text{parent}(v), v)\) creates
two vertex sets: the vertex set
\(\IVertex{T_A \setminus S_v} \cup \IVertex{T_B}\) associated with
\(T_A\) rooted at \(r_A\) with \(T_B\) joined to it and the subtree
\(S_v\) removed, and the subtree \(S_v\). So we see when considering the
possible unnormalized tree cut selection probability weight contributed
by the edge \(e_v\) when the edge \((a,b)\) joins \(T_A, T_B\), we see
it is equal to
\(\Phi(\IVertex{T_A \setminus S_v} \cup \IVertex{T_B}, \IVertex{S_v}) = W_v^{\text{out}}\).

\begin{enumerate}
\def\labelenumi{\arabic{enumi}.}
\setcounter{enumi}{1}
\tightlist
\item
  Case 2. \(a\) is ``below'' \(v\), i.e., \(a\) is ``inside'' the
  subtree rooted at \(v\) (\(a \in \IVertex{S_v}\))
\end{enumerate}

In this case removing the edge \(e_v = (\text{parent}(v), v)\) creates
two vertex sets: the vertex set \(\IVertex{T_A \setminus S_v}\)
associated with \(T_A\) rooted at \(r_A\) (without \(T_B\) joined to it)
and the subtree \(S_v\) removed, and the subtree
\(S_v \cup \IVertex{T_B}\) where \(T_B\) is attached to it. So we see
when considering the possible unnormalized tree cut selection
probability weight contributed by the edge \(e_v\) when the edge
\((a,b)\) joins \(T_A, T_B\), we see it is equal to
\(\Phi(\IVertex{T_A \setminus S_v}, \IVertex{S_v} \cup \IVertex{T_B}) = W_v^{\text{in}}\).

Now define \(W_v(a)\) as \[
W_v(a) = \begin{cases}
        W_v^{\text{in}} & \text{when } a \in S_A  \\
        W_v^{\text{out}} & \text{when } a \notin S_A  \\
\end{cases} 
\] Notice that this is exactly the total unnormalized probability weight
contribution from the edge \(e_v\) to the normalizing constant for the
tree cut selection probability for \(T_A \cup (a,b) \cup T_B\).
\textbackslash{}

Also we can express this as

\[
W_v(a) = W_v^{\text{out}} + {\bm 1}\{a \in S_v\}(W_v^{\text{in}} - W_v^{\text{out}} )
\]

Now when we consider the normalizing constant for
\(T_A \cup (a,b) \cup T_B\) we see that it can be decomposed into a sum
over all the internal edges in \(T_A\) and \(T_B\) and finally
\((a,b)\). The idea is that if we define \(F(v)\) to be the total
unnormalized weight from \(T_A\) if we joined \(T_B\) to \(v \in T_A\)
then we have this recurrence:

\begin{enumerate}
\def\labelenumi{\arabic{enumi}.}
\tightlist
\item
  \(F(r_A) = \sum_{v \neq r_A} W_v^{\text{out}}(a)\)
\item
  \(F(v) = F(\text{parent}(v)) + \triangle_v\) WHERE the joining vertex
  \(a\) is considered to still be part of the rooted tree
\end{enumerate}

This relies on the fact that the total contribution is equal to the sum
\(\sum_{v \in T_A} W_v(a)\). Now consider the following recurrence. We
know that since \(r_A\) is the root of the tree then for any vertex
\(a\) it will be attached ``inside'' of it so we know that

Define

\begin{align*}
F(a) &= \sum_{v \neq r_A} W_v(a) \\
&= \sum_{v \neq r_A} W_v^{\text{out}} + {\bm 1}\{a \in S_v\}(W_v^{\text{in}} - W_v^{\text{out}} ) \\
&= \sum_{v \neq r_A} W_v^{\text{out}} + \sum_{v \neq r_A} {\bm 1}\{a \in S_v\}(W_v^{\text{in}} - W_v^{\text{out}}) \\
&= \sum_{v \neq r_A} W_v^{\text{out}}(a) + \sum_{v \neq r_A} {\bm 1}\{a \in S_v\}(\triangle_v) \\
&= \sum_{v \neq r_A} W_v^{\text{out}}(a) + \sum_{v \neq r_A , a \in S_v} \triangle_v
\end{align*} So we use the fact
\(\sum_{v \neq r_A} W_v^{\text{out}}(a) = F(r_A)\) now to handle
\(\sum_{v \neq r_A , a \in S_v} \triangle_v\) we note that since its a
tree there is only one unique path from \(r_A\) to \(v\) so it becomes

\begin{align*}
F(a) &=  \sum_{v \neq r_A} W_v^{\text{out}}(a) + \sum_{v \neq r_A , a \in S_v} \triangle_v \\
&=  F(r_A) + \sum_{v \neq r_A , v \in \text{Path}(r,a) \setminus \set{r}} \triangle_v \\ 
&=  F(r_A) + \sum_{v \in \text{Path}(r,a) \setminus \set{r}} \triangle_v 
\end{align*}

\[
F(a) = \sum_{v \neq r_A} W_v(a)
\]

\(F_{r_A}(a) = \sum_{v \in T_A, v \neq r_A} W_v^{\text{out}}(a)\)
because if we join at \(a\) it is ``outside'' of the subtrees of every
other vertex in the tree.

Now consider for \(v\) whose parent is \(r_A\). if \(a\) is attached to
\(v\) then
\(F_v(a) =  W_{r_A}^{\text{out}}(a) + \sum_{u \in S_v, v \neq u}  W_v^{\text{out}}(a)\)

\newpage

\section{Additional Figures}\label{sec-addl-figs}

\subsection{Ireland}\label{ireland}

\FloatBarrier

\begin{figure}

\begin{minipage}{0.50\linewidth}

\centering{

\pandocbounded{\includegraphics[keepaspectratio]{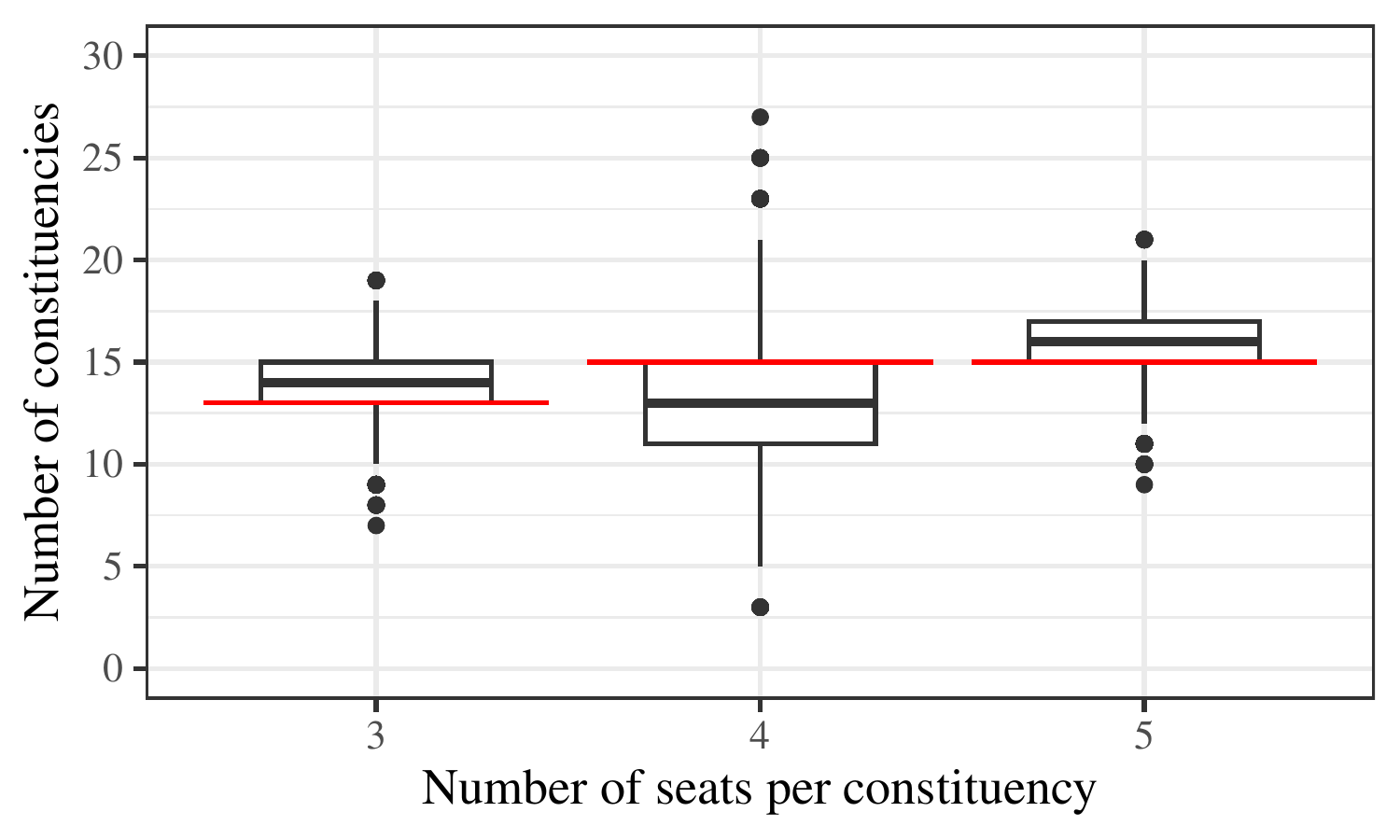}}

}

\subcaption{\label{fig-seats-current}Constituencies with three, four, or
five seats}

\end{minipage}%
\begin{minipage}{0.50\linewidth}

\centering{

\pandocbounded{\includegraphics[keepaspectratio]{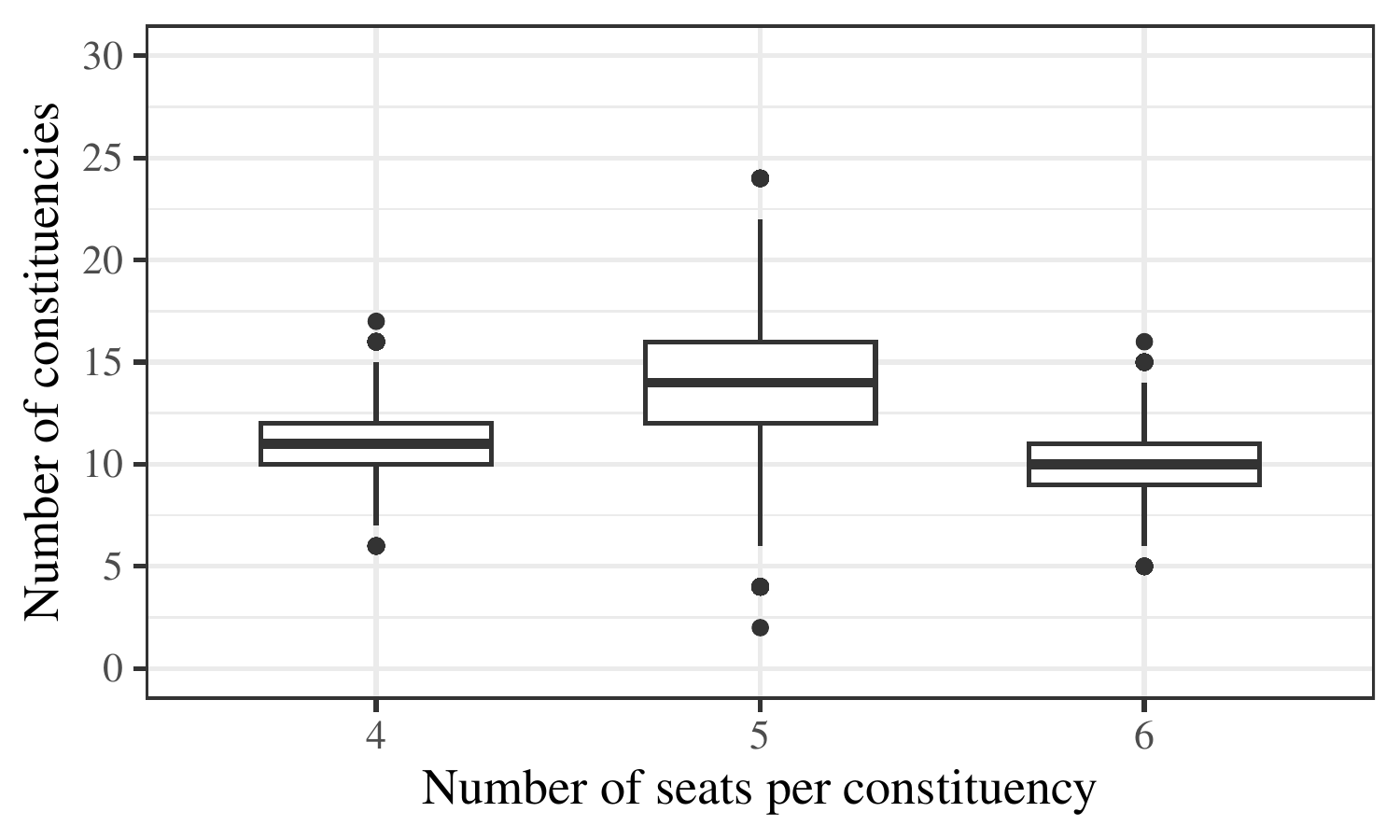}}

}

\subcaption{\label{fig-seats-sims}Constituencies with four, five, or six
seats}

\end{minipage}%

\caption{\label{fig-ireland-props}Constituency-seat counts under the
current (panel (a); 3--5 seats) and alternative (panel (b); 4--6 seats)
redistricting schemes. Box plots show the distribution of the number of
simulated constituencies for each size while the red horizontal line
shows the number of constituencies for each size under the enacted
plan.}

\end{figure}%

\newpage

\subsection{Pennsylvania State House}\label{sec-Penn-House}

\FloatBarrier

\begin{figure}

\centering{

\pandocbounded{\includegraphics[keepaspectratio]{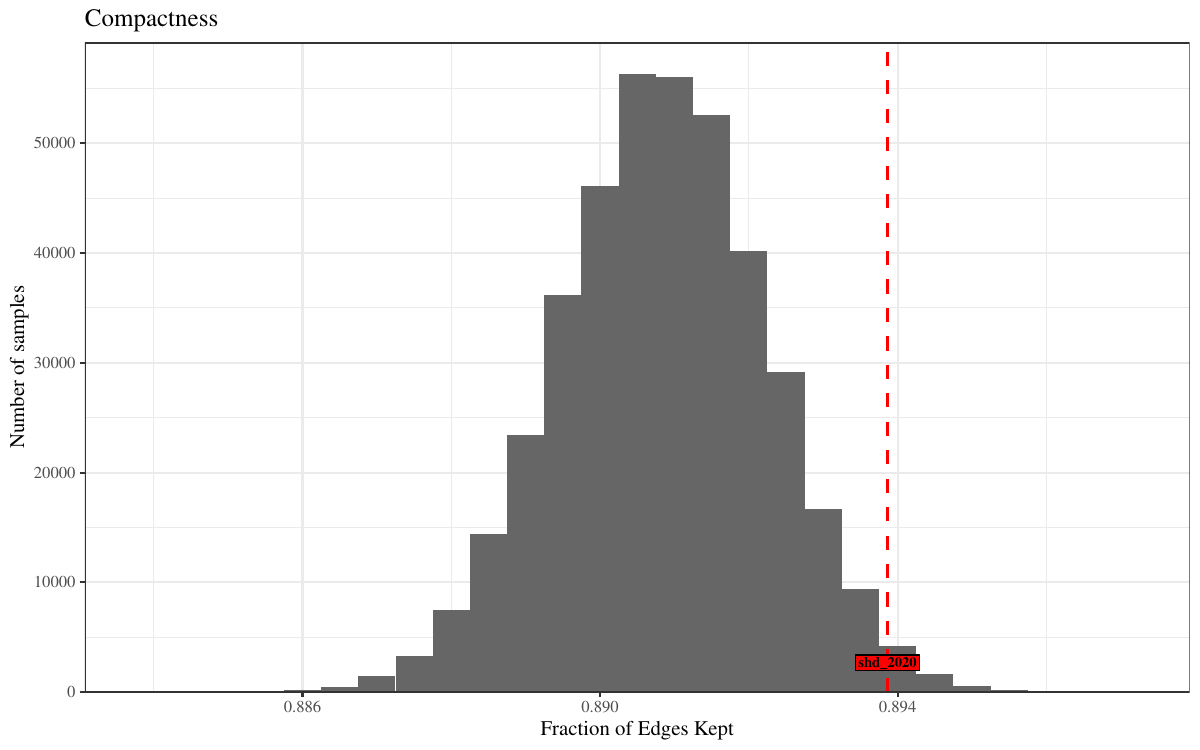}}

}

\caption{\label{fig-PA-House-Sims-Compactness}Compactness (larger means
more compact) statistics for the sampled and enacted plan.}

\end{figure}%

\begin{figure}

\centering{

\pandocbounded{\includegraphics[keepaspectratio]{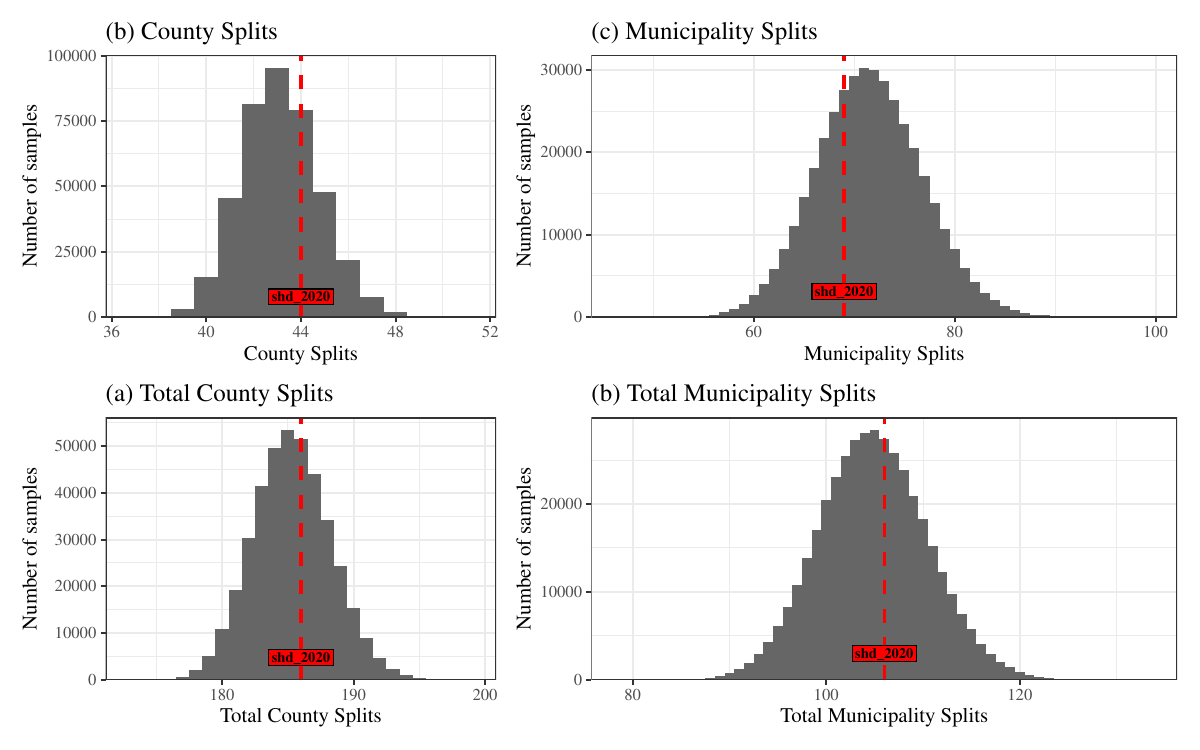}}

}

\caption{\label{fig-PA_split_figs}Partisan statistics for the sampled
and enacted plan: County and Municipality split figures}

\end{figure}%

\begin{figure}

\centering{

\pandocbounded{\includegraphics[keepaspectratio]{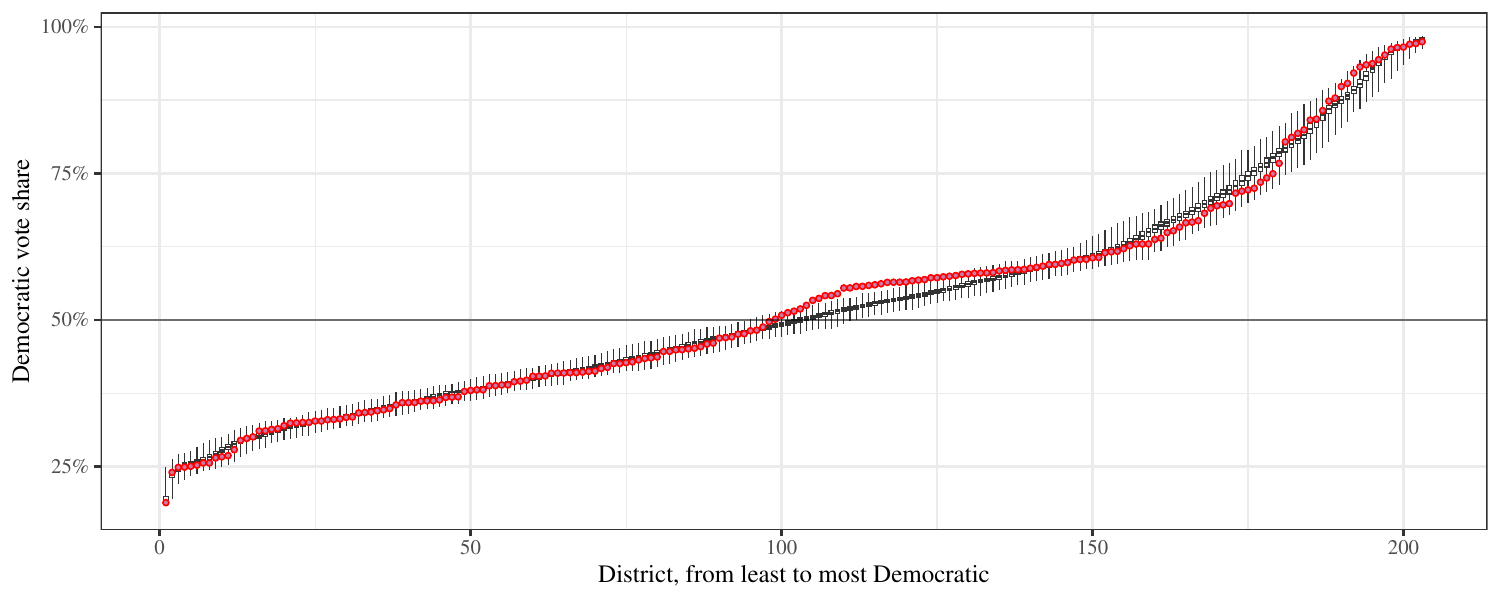}}

}

\caption{\label{fig-full-dem-share}Order statistics of Democratic
two-party vote share (i.e., within each plan districts are ordered by
Democratic vote share). Whiskers are drawn such that they span the full
observed range.}

\end{figure}%

\section{Additional Empirical Examples}\label{sec-addl-empirical}

\counterwithin*{figure}{subsubsection}
\renewcommand{\thefigure}{\thesubsubsection.\arabic{figure}}

We present some additional empirical results that explore the
performance of the proposed algorithms. In
Section~\ref{sec-comprehensive-sims}, we use ten different maps
encompassing a variety of sizes, population tolerances, and number of
districts. We examine the performance of the algorithms across a range
of sample sizes and all three sampling spaces for both any-valid and
district-only splitting schedules. We also consider a range of MCMC
steps (including no MCMC at all).

In Section~\ref{sec-extreme-balance-sims}, we present simulations that
assess the performance of various simulation algorithms under target
distributions with tight constraints. Specifically, we consider
extremely tight population bounds as often required in real-world US
Congressional redistricting. We present the results based on four maps;
two of the maps have a maximum deviation of two people per district,
another has a maximum of one, and the remaining one has perfect balance.
We demonstrate that gSMC and gSMC-MCMC are able to sample convergent
plans even in these extremely constrained maps when compared to MCMC
algorithms.

In Section~\ref{sec-rho-tests}, we present simulations for five
different maps exploring the performance of the new algorithms over a
range of different \(\rho\) values. These results suggest that there is
a neighborhood around \(\rho\) where both gSMC and gSMC-MCMC can sample
with reasonable efficiency and the addition of MCMC steps can help boost
performance. However, the size of the neighborhood around \(\rho\) is
somewhat problem dependent.

Finally, in Section~\ref{sec-varphi-tests} we present simulations of the
2020 Wisconsin congressional map over a range of different \(\varphi\)
choices. These results suggest performance and runtime are relatively
unaffected by the choice of \(\varphi\).

Unless stated otherwise, all simulations were done using our open source
implementation \texttt{redist} on a research computing cluster. Since
SMC is parallelizable, 112 cores were used for each SMC-based run unless
stated otherwise.\footnote{Cluster reports for compute heavy jobs (i.e.,
  jobs where most time is spent simulating plans as opposed to saving
  them) indicate CPU core utilization rates above 98\%.} The memory
usage required for SMC sampling tends to be roughly on the order of
\(3 \times V \times N\), where \(V\) is the number of vertices in the
map and \(N\) is the number of samples. Our open source implementation,
\texttt{redist}, stores spanning forests in bitpacked boolean edge list
arrays so the additional memory footprint for forest and linking edge
space sampling relative to graph space is extremely minimal.

All maps used 2020 election cycle data from the \texttt{alarmdata}
package. Convergence is measured in terms of the performance of the
\(\hat{R}\) values for all summary statistics for a map. The summary
statistics are computed from the data associated with the
\texttt{alarmdata} maps and generally the number of statistics depends
on the map and the number of districts. The two common quantities of
interest used to measure convergence are the maximum \(\hat{R}\) among
all summary statistics and the 99th quantile \(\hat{R}\) value. Unless
stated otherwise, 30 independent runs per configuration were used.

\subsection{Comprehensive Simulations}\label{sec-comprehensive-sims}

Here, we comprehensively examine the performance of the algorithm across
all six possible sampling space/splitting schedule combinations for a
variety of MCMC steps (including none at all) over a range of sample
sizes. Table~\ref{tbl-sys-empirical-examples} the ten of the maps used
for simulations. Because most jurisdictions in the US mandate their
districts preserve counties and other administrative units, all sampling
is done hierarchically with respect to their counties or
pseudo-counties. Two of the maps, Washington Congressional and Tennessee
State House, have additional constraints imposed through the \(J\)
function. As discussed in Section~\ref{sec-choose-mcmc}, we parameterize
the number of Markov steps to be performed after each SMC round with an
expected number of successful steps value (\texttt{mh\_accept\_per\_smc}
in \texttt{redist}). We use this same value to determine the number of
MCMC moves to apply after each SMC step, with the idea being if the
number of expected successful MCMC steps if \(k\), then we will run
\(m\) steps where we expect, on average, \(k\) of them to be successful
per plan. We leave the question of whether it is more optimal to vary
the expected number of successful steps throughout the process (e.g.,
should the number be higher at the beginning, end, etc.) to future
research.

\begin{longtable}[]{@{}
  >{\raggedright\arraybackslash}p{(\linewidth - 10\tabcolsep) * \real{0.2745}}
  >{\raggedright\arraybackslash}p{(\linewidth - 10\tabcolsep) * \real{0.0980}}
  >{\raggedright\arraybackslash}p{(\linewidth - 10\tabcolsep) * \real{0.3137}}
  >{\raggedright\arraybackslash}p{(\linewidth - 10\tabcolsep) * \real{0.0784}}
  >{\raggedright\arraybackslash}p{(\linewidth - 10\tabcolsep) * \real{0.0588}}
  >{\raggedleft\arraybackslash}p{(\linewidth - 10\tabcolsep) * \real{0.1765}}@{}}
\caption{Empirical Redistricting Examples used in the comprehensive
simulation study. Population bounds are rounded to the nearest whole
number.}\label{tbl-sys-empirical-examples}\tabularnewline
\toprule\noalign{}
\begin{minipage}[b]{\linewidth}\raggedright
Name
\end{minipage} & \begin{minipage}[b]{\linewidth}\raggedright
Districts
\end{minipage} & \begin{minipage}[b]{\linewidth}\raggedright
Population Tolerance
\end{minipage} & \begin{minipage}[b]{\linewidth}\raggedright
Vertices
\end{minipage} & \begin{minipage}[b]{\linewidth}\raggedright
Edges
\end{minipage} & \begin{minipage}[b]{\linewidth}\raggedleft
\(J\) Constraints
\end{minipage} \\
\midrule\noalign{}
\endfirsthead
\toprule\noalign{}
\begin{minipage}[b]{\linewidth}\raggedright
Name
\end{minipage} & \begin{minipage}[b]{\linewidth}\raggedright
Districts
\end{minipage} & \begin{minipage}[b]{\linewidth}\raggedright
Population Tolerance
\end{minipage} & \begin{minipage}[b]{\linewidth}\raggedright
Vertices
\end{minipage} & \begin{minipage}[b]{\linewidth}\raggedright
Edges
\end{minipage} & \begin{minipage}[b]{\linewidth}\raggedleft
\(J\) Constraints
\end{minipage} \\
\midrule\noalign{}
\endhead
\bottomrule\noalign{}
\endlastfoot
Arkansas Congressional & 4 & .1\% (\(752,881 \pm 752.9\)) & 2,747 &
7,509 & No \\
Wisconsin Congressional & 8 & .1\% (\(736,715 \pm 737\)) & 7,059 &
19,281 & No \\
Tennessee Congressional & 9 & .5\% (\(767,871 \pm 3,839\)) & 1,965 &
5,604 & No \\
Washington Congressional & 10 & .1\% (\(770,528 \pm 771\)) & 7.434 &
19,517 & Yes \\
Pennsylvania Congressional & 17 & .1\% (\(764,861 \pm 765\)) & 9,178 &
25,519 & No \\
Illinois Congressional & 17 & .1\% (\(753,677 \pm 754\)) & 10,084 &
28,219 & No \\
New York Congressional & 26 & .1\% (\(776,971 \pm 777\)) & 14,191 &
39,089 & No \\
California Congressional & 52 & .1\% (\(760,350 \pm 776\)) & 9,129 &
24,977 & No \\
Washington State Senate & 49 & 5\% (\(157,251 \pm 7,863\)) & 7,434 &
19,517 & No \\
Tennessee State House & 99 & 5\% (\(69,806 \pm 3,490\)) & 1,965 & 5,604
& Yes \\
\end{longtable}

We also compare the performance of the new algorithms against a modified
version of the SMC method in \citet{mccartan2023}, dubbed ``Original
SMC.'' The only significant difference from \citet{mccartan2023} is
instead of performing a labeling correction in the weights, we use a
``simple'' backwards kernel which is uniform on the number of 1-district
ancestors for a plan. Formally, if we have a plan \(\xi_r\) where
\(r-1\) of the regions are districts, then the backwards kernel for
\(\xi_{r-1}\) is given by \[
L_{r-1}(\xi_r| \xi_{r-1}) = \frac{1}{\abs{\tilde{\mathcal{A}}_1(\xi_r)}} \I{\xi_{r-1} \in \mathcal{A}_1(\xi_r)}(\xi_{r-1})
\] where \(\tilde{\mathcal{A}}_1(\xi_r)\) is the set of 1-district
ancestors. This is analogous to Definition~\ref{def-LRegionAncestor},
but instead of \(\xi_r\) and \(\xi_{r-1}\) sharing all but 1 region,
they must share all but 1 district. This modified Original SMC is
implemented using the same \texttt{redist} code for everything but the
weights.

For each map the specific details on the sample size, range of MCMC
steps, administrative units, and any additional constraints used are
contained in their respective subsection. For all ten maps we create the
following figures:

\begin{itemize}
\tightlist
\item
  Figures \ref{fig-ar-cd-gsmc-rhats}, \ref{fig-wi-cd-gsmc-rhats},
  \ref{fig-tn-cd-gsmc-rhats}, \ref{fig-wa-cd-gsmc-rhats},
  \ref{fig-pa-cd-gsmc-rhats}, \ref{fig-il-cd-gsmc-rhats},
  \ref{fig-ny-cd-gsmc-rhats}, \ref{fig-ca-cd-gsmc-rhats},
  \ref{fig-wa-ssd-gsmc-rhats}, and \ref{fig-tn-shd-gsmc-rhats} plot
  boxplots of \(\hat{R}\) for each sample size and expected number of
  successful mcmc steps configuration faceted by sampling space and
  splitting schedule. These provide a quick comparison of how all the
  algorithm variants performed for a given map and the influence of
  increasing the number of MCMC steps throughout.
\item
  Figures \ref{fig-ar-cd-gsmc-time}, \ref{fig-wi-cd-gsmc-time},
  \ref{fig-tn-cd-gsmc-time}, \ref{fig-wa-cd-gsmc-time},
  \ref{fig-pa-cd-gsmc-time}, \ref{fig-il-cd-gsmc-time},
  \ref{fig-ny-cd-gsmc-time}, \ref{fig-ca-cd-gsmc-time},
  \ref{fig-wa-ssd-gsmc-time}, and \ref{fig-tn-shd-gsmc-time} plot wall
  time breakdowns for gSMC only runs for a specific sample size. The
  time is categorized into time spent performing the splitting step,
  computing weights, estimating parameters (this is only relevant for
  graph space sampling), and other. Other includes time spent allocating
  memory and returning objects to R.\footnote{The timing does not
    include some time spent in R before the c++ code is called or
    afterwards when the \texttt{plans} object is returned in R, but this
    extra time is negligible.}
\item
  Figures \ref{fig-ar-cd-gsmc-mcmc-time},
  \ref{fig-wi-cd-gsmc-mcmc-time}, \ref{fig-tn-cd-gsmc-mcmc-time},
  \ref{fig-wa-cd-gsmc-mcmc-time}, \ref{fig-pa-cd-gsmc-mcmc-time},
  \ref{fig-il-cd-gsmc-mcmc-time}, \ref{fig-ny-cd-gsmc-mcmc-time},
  \ref{fig-ca-cd-gsmc-mcmc-time}, \ref{fig-wa-ssd-gsmc-mcmc-time}, and
  \ref{fig-tn-shd-gsmc-mcmc-time} plot wall time breakdowns for
  gSMC-MCMC runs for a specific sample size. This time is categorized
  into time spent calling Wilson's algorithm (this is shared among SMC
  and MCMC code), computing SMC weights, computing the metropolis
  hastings ratio, and other.
\item
  Figures \ref{fig-ar-cd-weight-time}, \ref{fig-wi-cd-weight-time},
  \ref{fig-tn-cd-weight-time}, \ref{fig-wa-cd-weight-time},
  \ref{fig-pa-cd-weight-time}, \ref{fig-il-cd-weight-time},
  \ref{fig-ny-cd-weight-time}, \ref{fig-ca-cd-weight-time},
  \ref{fig-wa-ssd-weight-time}, and \ref{fig-tn-shd-weight-time} plot
  the average median weight computation time per plan by SMC step for a
  specific sample size.
\item
  Figures \ref{fig-ar-cd-pareto-frontier},
  \ref{fig-wi-cd-pareto-frontier}, \ref{fig-tn-cd-pareto-frontier},
  \ref{fig-wa-cd-pareto-frontier}, \ref{fig-pa-cd-pareto-frontier},
  \ref{fig-il-cd-pareto-frontier}, \ref{fig-ny-cd-pareto-frontier},
  \ref{fig-ca-cd-pareto-frontier}, \ref{fig-wa-ssd-pareto-frontier}, and
  \ref{fig-tn-shd-pareto-frontier} plot the Pareto frontier of mean wall
  time and maximum/99th quantile \(\hat{R}\) value for Original SMC and
  all gSMC and gSMC-MCMC configurations run. Each point is a pair of
  mean wall time and maximum or 99th quantile \(\hat{R}\) value which is
  not dominated by any other pair. This plot provides a quick summary
  of, on a pure wall time basis, the performance of gSMC and gSMC-MCMC
  versus Original SMC.
\item
  Figures \ref{fig-ar-cd-gsmc-div}, \ref{fig-wi-cd-gsmc-div},
  \ref{fig-tn-cd-gsmc-div}, \ref{fig-wa-cd-gsmc-div},
  \ref{fig-pa-cd-gsmc-div}, \ref{fig-il-cd-gsmc-div},
  \ref{fig-ny-cd-gsmc-div}, \ref{fig-ca-cd-gsmc-div},
  \ref{fig-wa-ssd-gsmc-div}, and \ref{fig-tn-shd-gsmc-div} plot the mean
  percentage of unique districts per run (i.e., the total number of
  unique districts in a given run divided by the total number of
  districts sampled in that same run) for each sampling space over all
  sample sizes tested faceted by splitting schedule. Original SMC is
  also displayed as a separate black line in the district-only faceted
  plot.
\item
  Figures \ref{fig-ar-cd-gsmc-mcmc-div}, \ref{fig-wi-cd-gsmc-mcmc-div},
  \ref{fig-tn-cd-gsmc-mcmc-div}, \ref{fig-wa-cd-gsmc-mcmc-div},
  \ref{fig-pa-cd-gsmc-mcmc-div}, \ref{fig-il-cd-gsmc-mcmc-div},
  \ref{fig-ny-cd-gsmc-mcmc-div}, \ref{fig-ca-cd-gsmc-mcmc-div},
  \ref{fig-wa-ssd-gsmc-mcmc-div}, and \ref{fig-tn-shd-gsmc-mcmc-div}
  plot the mean percentage of unique districts per run for each sampling
  space over all expected successful number of MCMC step multiples
  faceted by splitting schedule and sample size.
\end{itemize}

Tables \ref{tab:rhat-q99-gsmc} and \ref{tab:rhat-max-gsmc} provide a
succinct summary of the fastest gSMC configurations for each map to have
its 99th quantile \(\hat{R}\) value and maximum \(\hat{R}\) value
respectively for all summary statistics below the 1.05 threshold. Empty
values for a specific map indicate that no convergent gSMC
configurations run were convergent. Tables \ref{tab:rhat-q99-gsmc-mcmc}
and \ref{tab:rhat-max-gsmc-mcmc} displays the same information but only
for all gSMC-MCMC configurations run for a map. Finally, Tables
\ref{tab:rhat-q99-combined} and \ref{tab:rhat-max-combined} displays the
results among all gSMC and gSMC-MCMC configurations run for each map.
Figure~\ref{fig-split-schedule-reduction} provides a summary of the
runtime reduction percentage for using any-valid splitting as opposed to
district-only splitting grouped by the number of successful MCMC steps
and averaged across all sample sizes.

We summarize the findings from our analyses below:

\begin{itemize}
\tightlist
\item
  As the contrast between Tables \ref{tab:rhat-q99-gsmc} and
  \ref{tab:rhat-max-gsmc} versus Tables \ref{tab:rhat-q99-gsmc-mcmc} and
  \ref{tab:rhat-max-gsmc-mcmc} makes clear, gSMC alone is not able to
  achieve convergence on any of the maps with more districts than New
  York congressional (26 districts), whereas gSMC-MCMC is able to handle
  all of the maps.
\item
  Tables \ref{tab:rhat-q99-combined} and \ref{tab:rhat-max-combined}
  along with Figures \ref{fig-ar-cd-pareto-frontier},
  \ref{fig-wi-cd-pareto-frontier}, \ref{fig-tn-cd-pareto-frontier},
  \ref{fig-wa-cd-pareto-frontier}, \ref{fig-pa-cd-pareto-frontier},
  \ref{fig-il-cd-pareto-frontier}, suggest that beginning with maps with
  a moderate number of districts (around 10 or so), it is more
  computationally efficient to add MCMC steps as opposed to increasing
  the sample size for gSMC. Future research is needed to determine how
  to optimally decide between increasing the number of MCMC steps run as
  opposed to sample size.
\item
  Tables \ref{tab:rhat-q99-combined} and \ref{tab:rhat-max-combined}
  suggest that the two new sampling spaces (spanning forest and linking
  edge) strictly outperform graph space sampling on a wall time basis.
  Given that linking edge space is the fastest to converge in a majority
  of maps, we recommend practitioners use linking edge space as a
  default first choice. Figures \ref{fig-pa-cd-gsmc-rhats},
  \ref{fig-il-cd-gsmc-rhats}, \ref{fig-ny-cd-gsmc-rhats}, and
  \ref{fig-ca-cd-gsmc-rhats} also suggest that linking edge appears to
  perform better relative to the other sampling spaces for maps with
  tighter population bounds, in contrast to Figures
  \ref{fig-wa-ssd-gsmc-rhats} and \ref{fig-tn-shd-gsmc-rhats} where all
  three sampling spaces with any-valid splits perform well.
\item
  Figure~\ref{fig-split-schedule-reduction} empirically demonstrates
  that the any-valid splitting schedule provides a reduction in wall
  time over district-only splitting where the reduction percentage grows
  as the size of the map, the number of districts, and the number of
  MCMC steps run increases. Further, Tables \ref{tab:rhat-q99-combined}
  and \ref{tab:rhat-max-combined} suggest that this wall time reduction
  does not appear to cause any reduction in performance. As such, we
  recommend practitioners use the any-valid splitting schedule as a
  default first choice.
\item
  The mean percentage of unique district plots suggest that for both
  gSMC and gSMC-MCMC the any-valid splitting schedule tends to produce a
  higher percentage of unique districts across sampling spaces. Both
  splitting schedules appear to do strictly better than Original SMC in
  terms of percentage of unique districts.
\item
  Finally, the Pareto frontier plots confirm that the new algorithms are
  a significant improvement over the \citet{mccartan2023} method for
  larger maps. For Arkansas, Wisconsin, and Washington congressional,
  the SMC still converges, albeit a little slower. However, beginning
  with Pennsylvania Congressional, it is no longer able to achieve fully
  convergent samples even though gSMC alone can converge up until New
  York Congressional.
\end{itemize}

Overall, these results demonstrate that the combination of optimal
weights, new sampling spaces, and the addition of MCMC steps
dramatically extend the range of problems the new algorithms can handle.

\begin{figure}[H]

\centering{

\pandocbounded{\includegraphics[keepaspectratio]{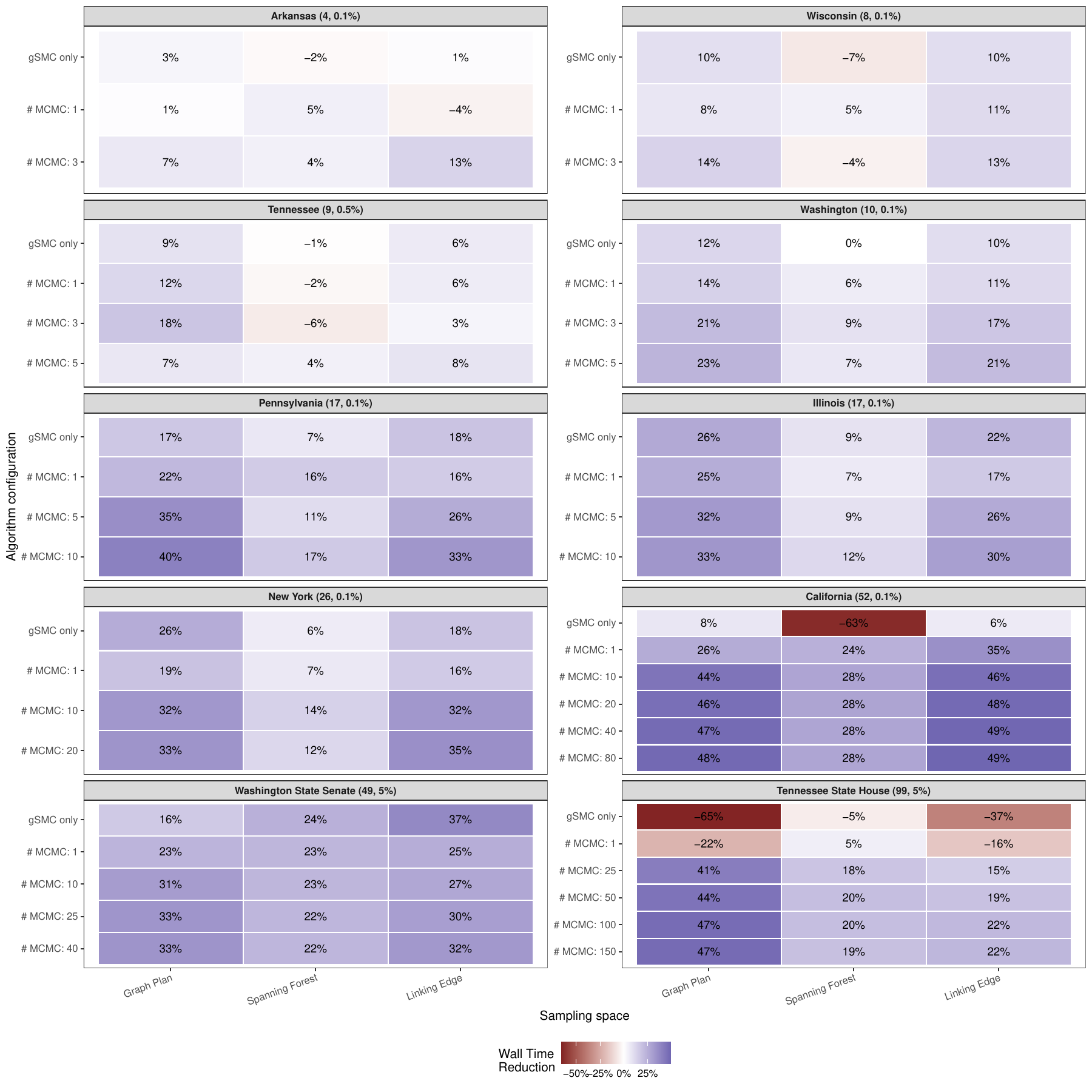}}

}

\caption{\label{fig-split-schedule-reduction}Breakdown of time spent
during runtime across the three gSMC sampling spaces and two splitting
schedules. A reduction of 20\% means any-valid splitting took 80\% of
the time compared to district-only splitting.}

\end{figure}%

\begin{table}[!htbp]
\centering
\small
\renewcommand{\arraystretch}{1.1}
\setlength{\tabcolsep}{4pt}
\begin{tabular}{@{}l l l r r@{}}
\hline
\textbf{Name} & \textbf{Sampling space} & \textbf{Splitting schedule} & \textbf{Sample size} & \textbf{Mean wall time} \\
\hline
Arkansas (4, 0.1\%) & Spanning Forest & District-only & 1,000 & 0.3 s \\
Balanced Arkansas (4, 0) & Linking Edge & Any-valid & 1,000 & 48.4 s \\
Wisconsin (8, 0.1\%) & Spanning Forest & Any-valid & 1,000 & 1.3 s \\
Tennessee (9, 0.1\%) & Linking Edge & Any-valid & 1,000 & 0.4 s \\
Washington (10, 0.1\%) & Spanning Forest & District-only & 10,000 & 7.7 s \\
Pennsylvania (17, 0.1\%) & Spanning Forest & Any-valid & 25,000 & 30.7 s \\
Illinois (17, 0.1\%) & Graph Plan & District-only & 10,000 & 30.0 s \\
New York (26, 0.1\%) & Linking Edge & Any-valid & 50,000 & 124.2 s \\
California (52, 0.1\%) & — & — & — & — \\
Washington State (49, 5\%) & — & — & — & — \\
Tennessee State House (99, 5\%) & — & — & — & — \\
\hline
\end{tabular}
\caption{Fastest gSMC only configurations with 99th quantile R-hat at or below 1.05.}
\label{tab:rhat-q99-gsmc}
\end{table}

\begin{table}[!htbp]
\centering
\small
\renewcommand{\arraystretch}{1.1}
\setlength{\tabcolsep}{4pt}
\begin{tabular}{@{}l l l r r r@{}}
\hline
\textbf{Name} & \textbf{Sampling space} & \textbf{Splitting schedule} & \textbf{N. MCMC} & \textbf{Sample size} & \textbf{Mean wall time} \\
\hline
Arkansas (4, 0.1\%) & Spanning Forest & Any-valid & 1 & 1,000 & 0.5 s \\
Balanced Arkansas (4, 0) & Linking Edge & Any-valid & 1 & 1,000 & 74.3 s \\
Wisconsin (8, 0.1\%) & Linking Edge & Any-valid & 1 & 1,000 & 2.0 s \\
Tennessee (9, 0.1\%) & Linking Edge & Any-valid & 1 & 1,000 & 0.6 s \\
Washington (10, 0.1\%) & Linking Edge & District-only & 1 & 2,500 & 4.9 s \\
Pennsylvania (17, 0.1\%) & Linking Edge & Any-valid & 10 & 1,000 & 12.7 s \\
Illinois (17, 0.1\%) & Linking Edge & Any-valid & 10 & 1,000 & 13.7 s \\
New York (26, 0.1\%) & Linking Edge & District-only & 10 & 1,000 & 39.9 s \\
California (52, 0.1\%) & Linking Edge & Any-valid & 40 & 5,000 & 382.7 s \\
Washington State (49, 5\%) & Linking Edge & Any-valid & 10 & 5,000 & 70.4 s \\
Tennessee State House (99, 5\%) & Linking Edge & Any-valid & 100 & 2,500 & 189.5 s \\
\hline
\end{tabular}
\caption{Fastest gSMC-MCMC configurations with 99th quantile R-hat at or below 1.05. N. MCMC stands for number of expected successful MCMC steps.}
\label{tab:rhat-q99-gsmc-mcmc}
\end{table}

\begin{table}[!htbp]
\centering
\small
\renewcommand{\arraystretch}{1.1}
\setlength{\tabcolsep}{3.5pt}
\resizebox{\linewidth}{!}{%
\begin{tabular}{@{}l l l l r r r@{}}
\hline
\textbf{Name} & \textbf{Algorithm variant} & \textbf{Sampling space} & \textbf{Splitting schedule} & \textbf{N. MCMC} & \textbf{Sample size} & \textbf{Mean wall time} \\
\hline
Arkansas (4, 0.1\%) & gSMC & Spanning Forest & District-only & — & 1,000 & 0.3 s \\
Balanced Arkansas (4, 0) & gSMC & Linking Edge & Any-valid & — & 1,000 & 48.4 s \\
Wisconsin (8, 0.1\%) & gSMC & Spanning Forest & Any-valid & — & 1,000 & 1.3 s \\
Tennessee (9, 0.1\%) & gSMC & Linking Edge & Any-valid & — & 1,000 & 0.4 s \\
Washington (10, 0.1\%) & gSMC-MCMC & Linking Edge & District-only & 1 & 2,500 & 4.9 s \\
Pennsylvania (17, 0.1\%) & gSMC-MCMC & Linking Edge & Any-valid & 10 & 1,000 & 12.7 s \\
Illinois (17, 0.1\%) & gSMC-MCMC & Linking Edge & Any-valid & 10 & 1,000 & 13.7 s \\
New York (26, 0.1\%) & gSMC-MCMC & Linking Edge & District-only & 10 & 1,000 & 39.9 s \\
California (52, 0.1\%) & gSMC-MCMC & Linking Edge & Any-valid & 40 & 5,000 & 382.7 s \\
Washington State (49, 5\%) & gSMC-MCMC & Linking Edge & Any-valid & 10 & 5,000 & 70.4 s \\
Tennessee State House (99, 5\%) & gSMC-MCMC & Linking Edge & Any-valid & 100 & 2,500 & 189.5 s \\
\hline
\end{tabular}
}
\caption{Fastest gSMC or gSMC-MCMC configurations with 99th quantile R-hat at or below 1.05. N. MCMC stands for number of expected successful MCMC steps.}
\label{tab:rhat-q99-combined}
\end{table}

\begin{table}[!htbp]
\centering
\small
\renewcommand{\arraystretch}{1.1}
\setlength{\tabcolsep}{4pt}
\begin{tabular}{@{}l l l r r@{}}
\hline
\textbf{Name} & \textbf{Sampling space} & \textbf{Splitting schedule} & \textbf{Sample size} & \textbf{Mean wall time} \\
\hline
Arkansas (4, 0.1\%) & Spanning Forest & District-only & 1,000 & 0.3 s \\
Balanced Arkansas (4, 0) & Linking Edge & Any-valid & 1,000 & 48.4 s \\
Wisconsin (8, 0.1\%) & Linking Edge & Any-valid & 2,500 & 1.9 s \\
Tennessee (9, 0.1\%) & Linking Edge & Any-valid & 1,000 & 0.4 s \\
Washington (10, 0.1\%) & Spanning Forest & District-only & 10,000 & 7.7 s \\
Pennsylvania (17, 0.1\%) & Spanning Forest & District-only & 25,000 & 33.4 s \\
Illinois (17, 0.1\%) & Spanning Forest & District-only & 25,000 & 35.0 s \\
New York (26, 0.1\%) & Spanning Forest & District-only & 50,000 & 159.0 s \\
California (52, 0.1\%) & — & — & — & — \\
Washington State (49, 5\%) & — & — & — & — \\
Tennessee State House (99, 5\%) & — & — & — & — \\
\hline
\end{tabular}
\caption{Fastest gSMC only configurations with maximum R-hat at or below 1.05.}
\label{tab:rhat-max-gsmc}
\end{table}

\begin{table}[!htbp]
\centering
\small
\renewcommand{\arraystretch}{1.1}
\setlength{\tabcolsep}{4pt}
\begin{tabular}{@{}l l l r r r@{}}
\hline
\textbf{Name} & \textbf{Sampling space} & \textbf{Splitting schedule} & \textbf{N. MCMC} & \textbf{Sample size} & \textbf{Mean wall time} \\
\hline
Arkansas (4, 0.1\%) & Spanning Forest & Any-valid & 1 & 1,000 & 0.5 s \\
Balanced Arkansas (4, 0) & Linking Edge & Any-valid & 1 & 1,000 & 74.3 s \\
Wisconsin (8, 0.1\%) & Linking Edge & Any-valid & 1 & 1,000 & 2.0 s \\
Tennessee (9, 0.1\%) & Linking Edge & Any-valid & 1 & 1,000 & 0.6 s \\
Washington (10, 0.1\%) & Linking Edge & Any-valid & 3 & 2,500 & 6.3 s \\
Pennsylvania (17, 0.1\%) & Linking Edge & Any-valid & 10 & 1,000 & 12.7 s \\
Illinois (17, 0.1\%) & Linking Edge & Any-valid & 10 & 1,000 & 13.7 s \\
New York (26, 0.1\%) & Linking Edge & Any-valid & 10 & 5,000 & 100.6 s \\
California (52, 0.1\%) & Linking Edge & Any-valid & 80 & 5,000 & 742.9 s \\
Washington State (49, 5\%) & Linking Edge & Any-valid & 40 & 2,500 & 119.3 s \\
Tennessee State House (99, 5\%) & Linking Edge & Any-valid & 150 & 5,000 & 537.1 s \\
\hline
\end{tabular}
\caption{Fastest gSMC-MCMC configurations with maximum R-hat at or below 1.05. N. MCMC stands for number of expected successful MCMC steps.}
\label{tab:rhat-max-gsmc-mcmc}
\end{table}

\begin{table}[!htbp]
\centering
\small
\renewcommand{\arraystretch}{1.1}
\setlength{\tabcolsep}{3.5pt}
\resizebox{\linewidth}{!}{%
\begin{tabular}{@{}l l l l r r r@{}}
\hline
\textbf{Name} & \textbf{Algorithm variant} & \textbf{Sampling space} & \textbf{Splitting schedule} & \textbf{N. MCMC} & \textbf{Sample size} & \textbf{Mean wall time} \\
\hline
Arkansas (4, 0.1\%) & gSMC & Spanning Forest & District-only & — & 1,000 & 0.3 s \\
Balanced Arkansas (4, 0) & gSMC & Linking Edge & Any-valid & — & 1,000 & 48.4 s \\
Wisconsin (8, 0.1\%) & gSMC & Linking Edge & Any-valid & — & 2,500 & 1.9 s \\
Tennessee (9, 0.1\%) & gSMC & Linking Edge & Any-valid & — & 1,000 & 0.4 s \\
Washington (10, 0.1\%) & gSMC-MCMC & Linking Edge & Any-valid & 3 & 2,500 & 6.3 s \\
Pennsylvania (17, 0.1\%) & gSMC-MCMC & Linking Edge & Any-valid & 10 & 1,000 & 12.7 s \\
Illinois (17, 0.1\%) & gSMC-MCMC & Linking Edge & Any-valid & 10 & 1,000 & 13.7 s \\
New York (26, 0.1\%) & gSMC-MCMC & Linking Edge & Any-valid & 10 & 5,000 & 100.6 s \\
California (52, 0.1\%) & gSMC-MCMC & Linking Edge & Any-valid & 80 & 5,000 & 742.9 s \\
Washington State (49, 5\%) & gSMC-MCMC & Linking Edge & Any-valid & 40 & 2,500 & 119.3 s \\
Tennessee State House (99, 5\%) & gSMC-MCMC & Linking Edge & Any-valid & 150 & 5,000 & 537.1 s \\
\hline
\end{tabular}
}
\caption{Fastest gSMC or gSMC-MCMC configurations with maximum R-hat at or below 1.05. N. MCMC stands for number of expected successful MCMC steps.}
\label{tab:rhat-max-combined}
\end{table}

\raggedbottom

\newpage

\subsubsection{Arkansas 2020 Congressional Map}\label{sec-ar-cd-2020}

We use a map with \(V=2,747\) precincts, \(E = 7,509\) edges, and a
median population of 620 per precinct. We use counties as the units for
hierarchical sampling. No additional constraints are imposed through the
\(J\) function. We sample both gSMC and gSMC-MCMC at a sample size of
1,000. The expected number of successful MCMC steps used were: 1 and 3.
All configurations run converged on this map.

\begin{figure}[H]

\centering{

\pandocbounded{\includegraphics[keepaspectratio]{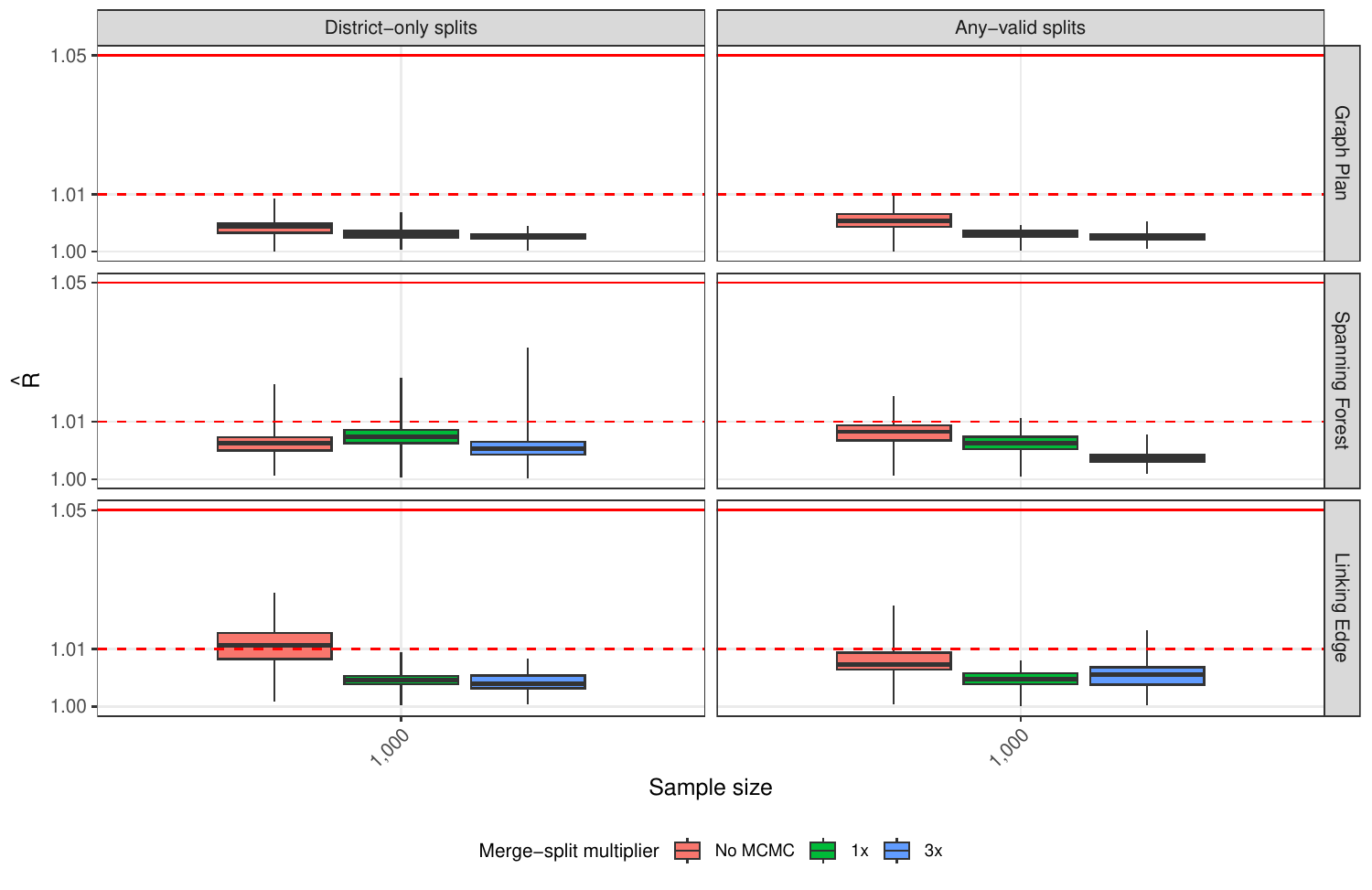}}

}

\caption{\label{fig-ar-cd-gsmc-rhats}Arkansas Congressional:
Distribution of \(\hat{R}\) values for all gSMC and gSMC-MCMC
configurations run, faceted by sampling space and splitting schedule.}

\end{figure}%

\begin{figure}[H]

\begin{minipage}{\linewidth}

\begin{figure}[H]

\centering{

\includegraphics[width=0.9\linewidth,height=\textheight,keepaspectratio]{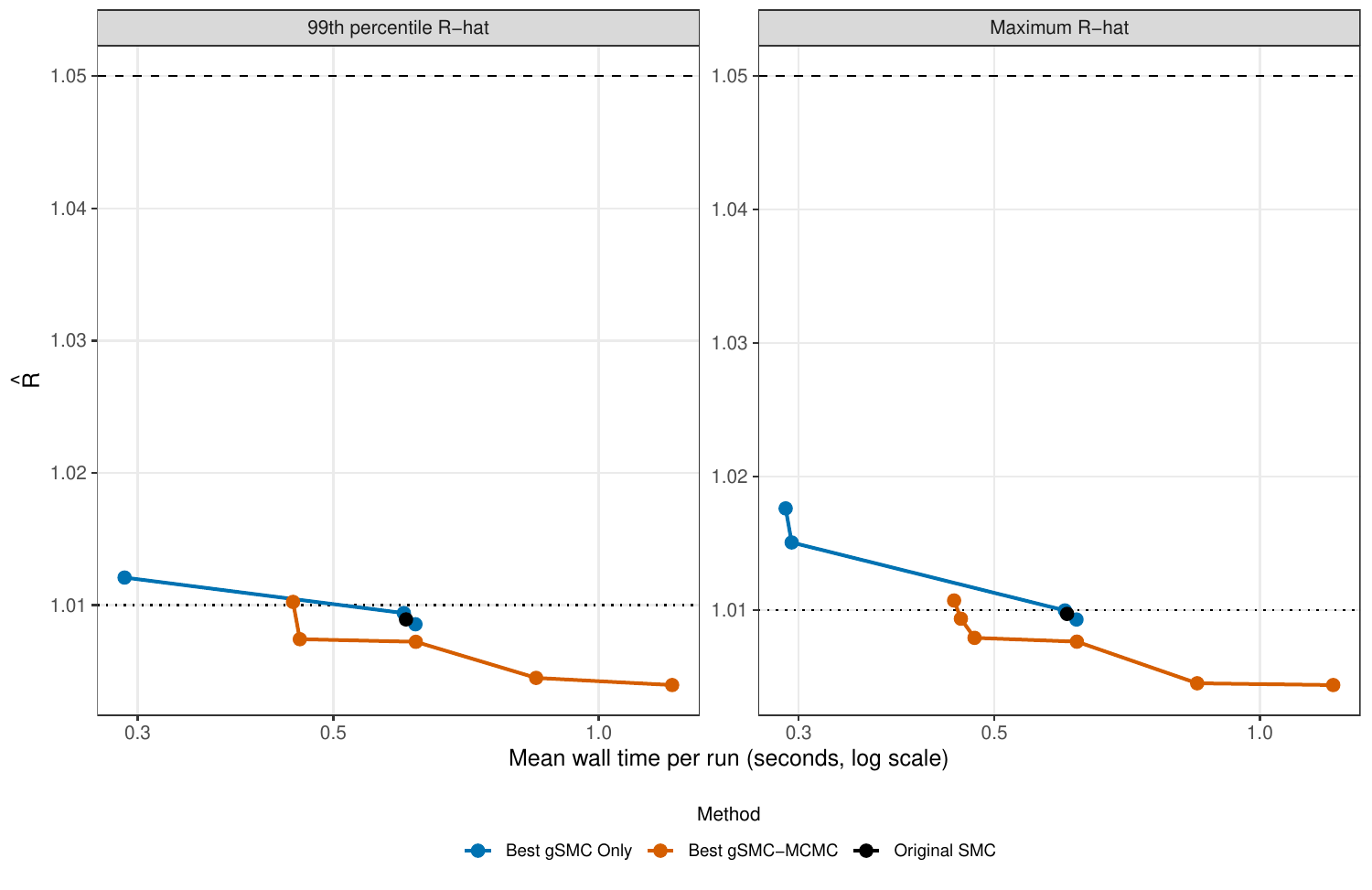}

}

\caption{\label{fig-ar-cd-pareto-frontier}Arkansas Congressional:
Empirical Pareto frontier comparing mean wall time and convergence of
the best-performing gSMC and gSMC-MCMC configurations and Original SMC.
Each point represents an algorithm configuration, with mean wall time
per run on the horizontal axis and either the 99th percentile of the
resulting \(\hat{R}\) values (left) or the maximum \(\hat{R}\) value
(right) on the vertical axis. Moving along a frontier corresponds to
configurations for which improved convergence came at the cost of
greater wall time.}

\end{figure}%

\end{minipage}%
\newline
\begin{minipage}{\linewidth}

\begin{figure}[H]

\centering{

\includegraphics[width=0.9\linewidth,height=\textheight,keepaspectratio]{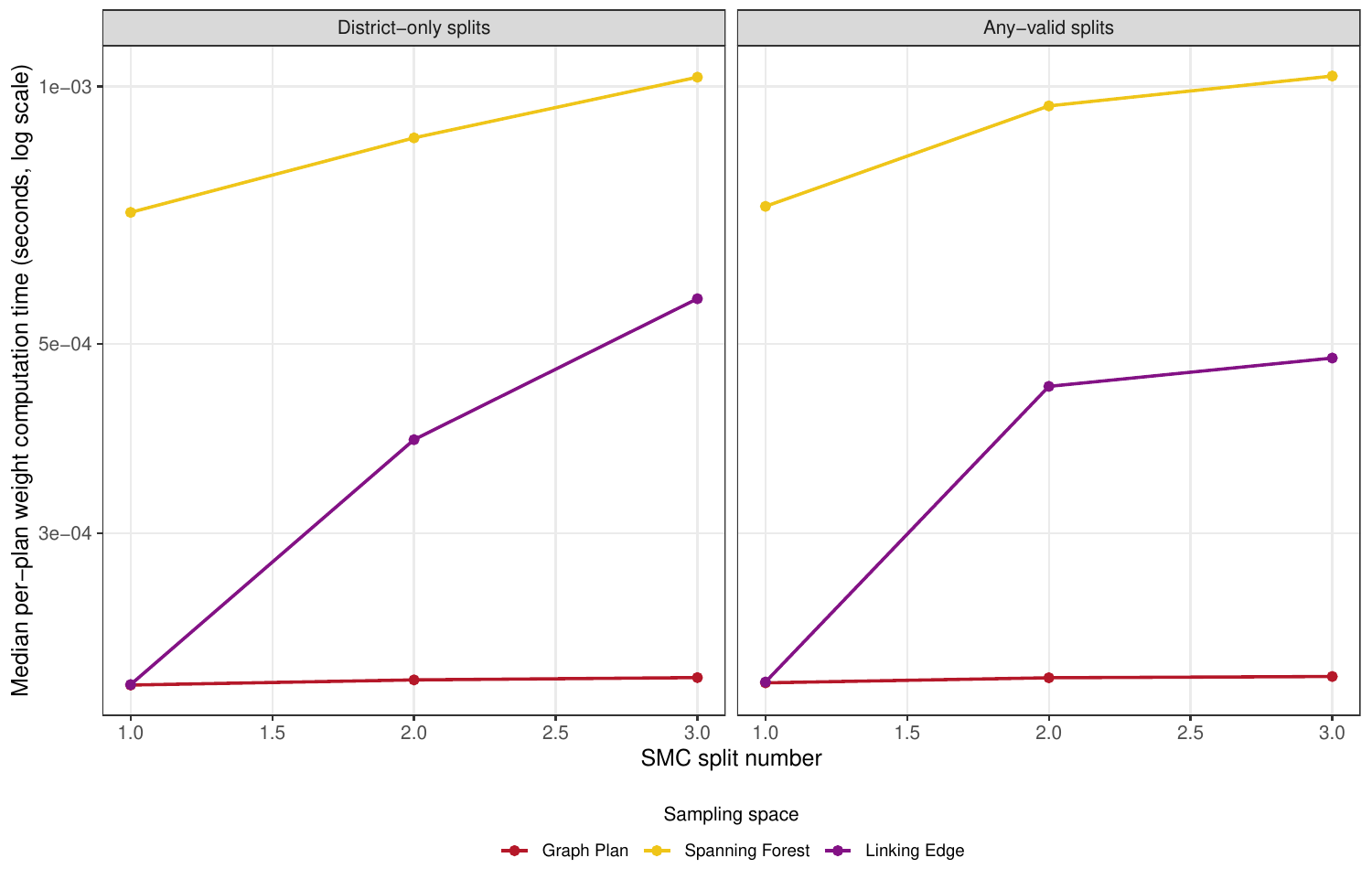}

}

\caption{\label{fig-ar-cd-weight-time}Average median per-plan weight
computation time across sampling spaces for a sample size of 1,000,
faceted by splitting schedule.}

\end{figure}%

\end{minipage}%

\end{figure}%

\begin{figure}[H]

\begin{minipage}{\linewidth}

\centering{

\includegraphics[width=0.8\linewidth,height=\textheight,keepaspectratio]{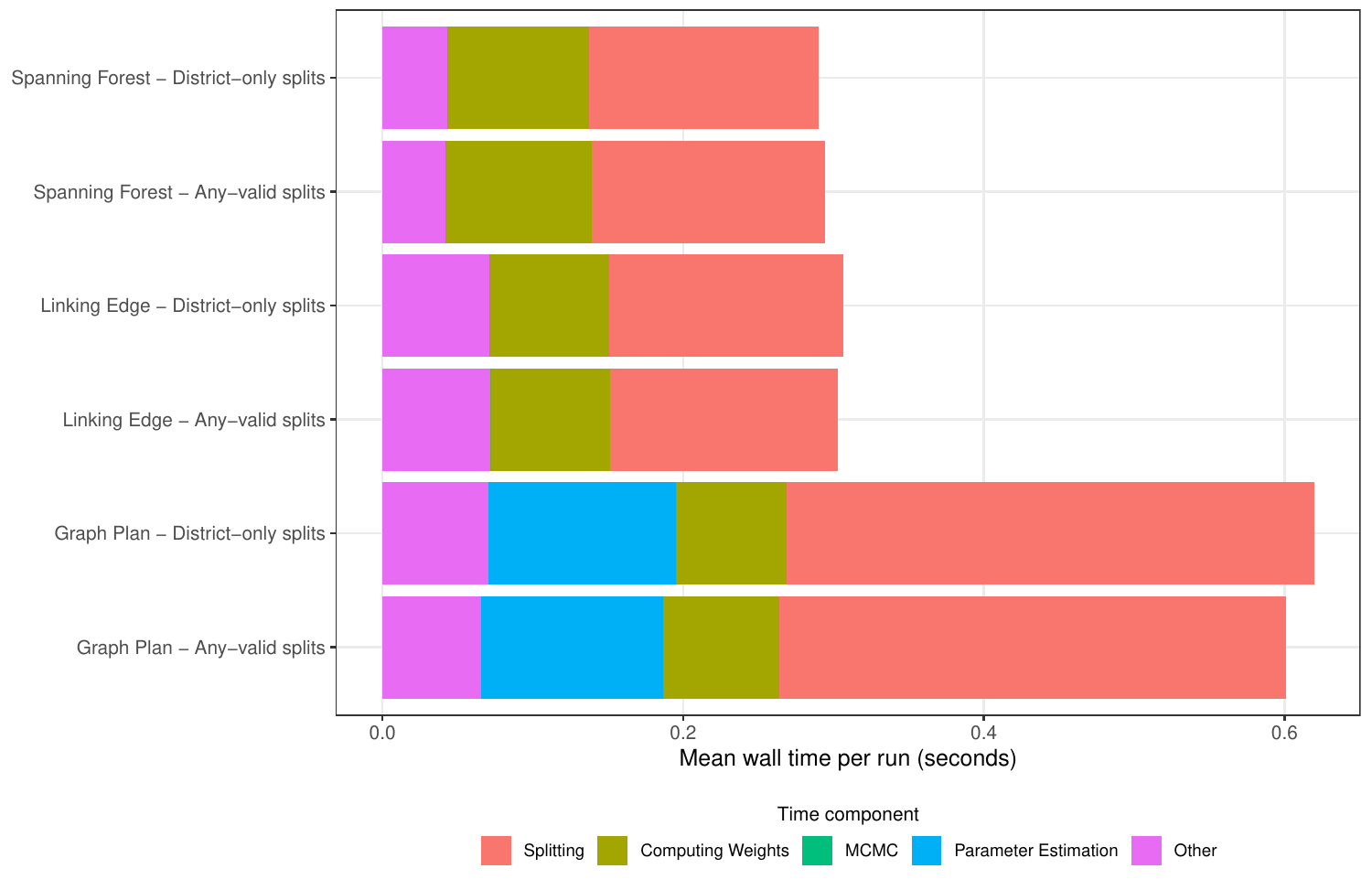}

}

\subcaption{\label{fig-ar-cd-gsmc-time}Breakdown of wall-time categories
across the three gSMC sampling spaces and two splitting schedules for a
sample size of 1,000.}

\end{minipage}%
\newline
\begin{minipage}{\linewidth}

\centering{

\pandocbounded{\includegraphics[keepaspectratio]{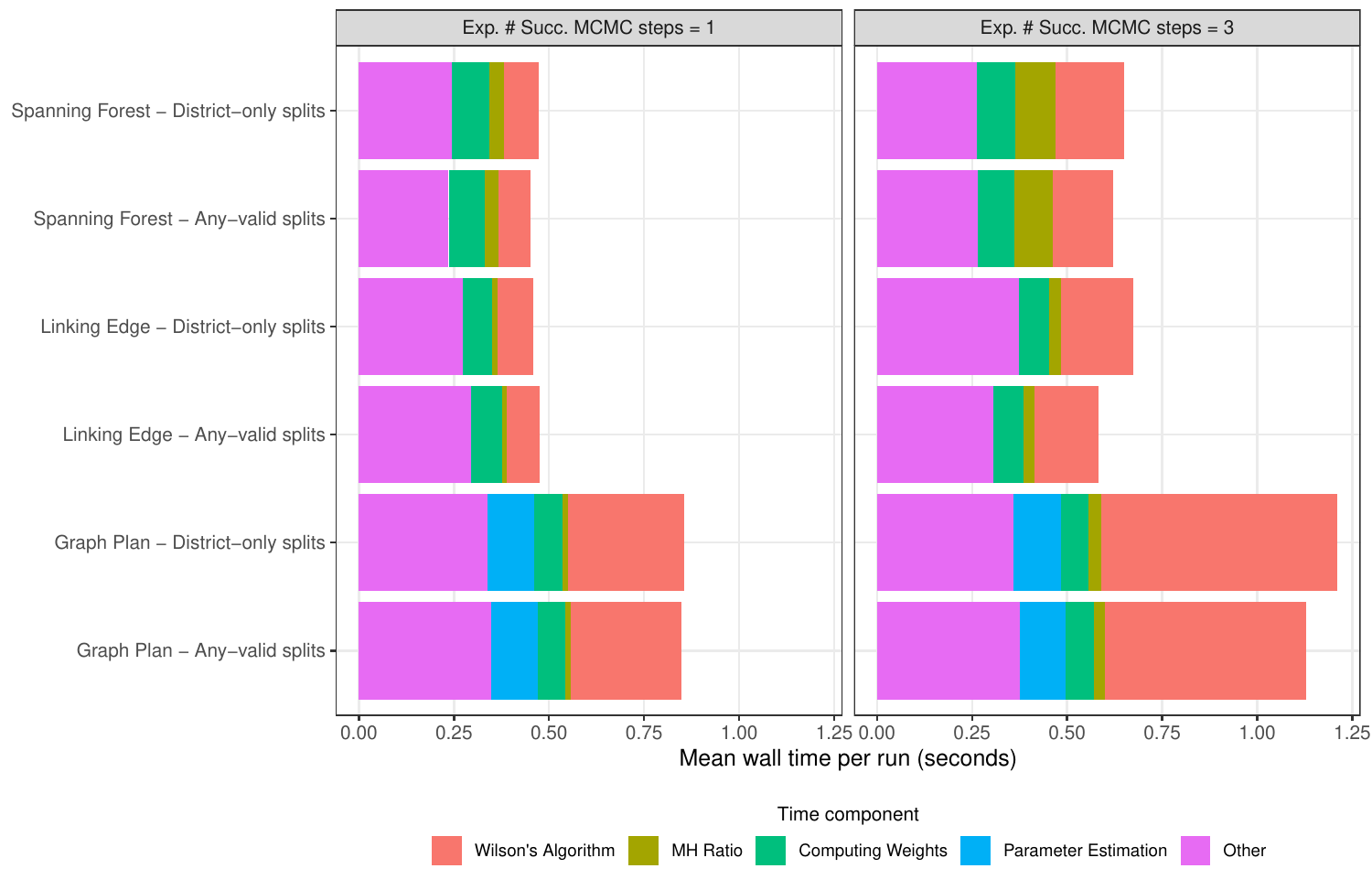}}

}

\subcaption{\label{fig-ar-cd-gsmc-mcmc-time}Breakdown of wall-time
categories across the three gSMC sampling spaces and two splitting
schedules, faceted by expected number of successful MCMC steps, for a
sample size of 1,000.}

\end{minipage}%

\caption{\label{fig-ar-cd-time-breakdown}Arkansas Congressional:
Breakdown of wall time for gSMC and gSMC-MCMC configurations at a sample
size of 1,000.}

\end{figure}%

\begin{figure}[H]

\begin{minipage}{\linewidth}

\centering{

\includegraphics[width=0.95\linewidth,height=\textheight,keepaspectratio]{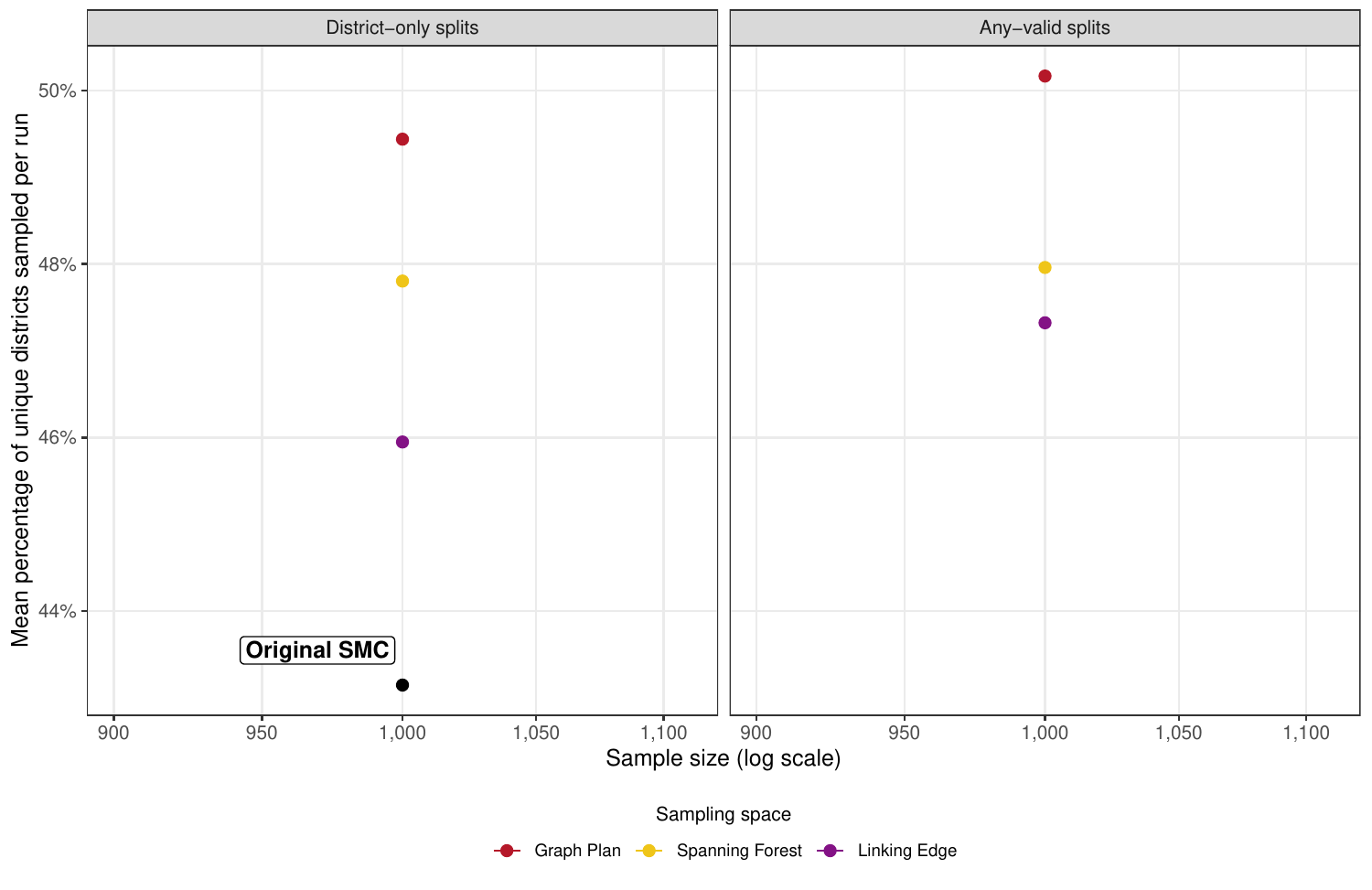}

}

\subcaption{\label{fig-ar-cd-gsmc-div}Arkansas Congressional: Mean
percentage of unique districts per run versus sample size for each of
the three gSMC sampling spaces, faceted by splitting schedule.}

\end{minipage}%
\newline
\begin{minipage}{\linewidth}

\centering{

\includegraphics[width=0.95\linewidth,height=\textheight,keepaspectratio]{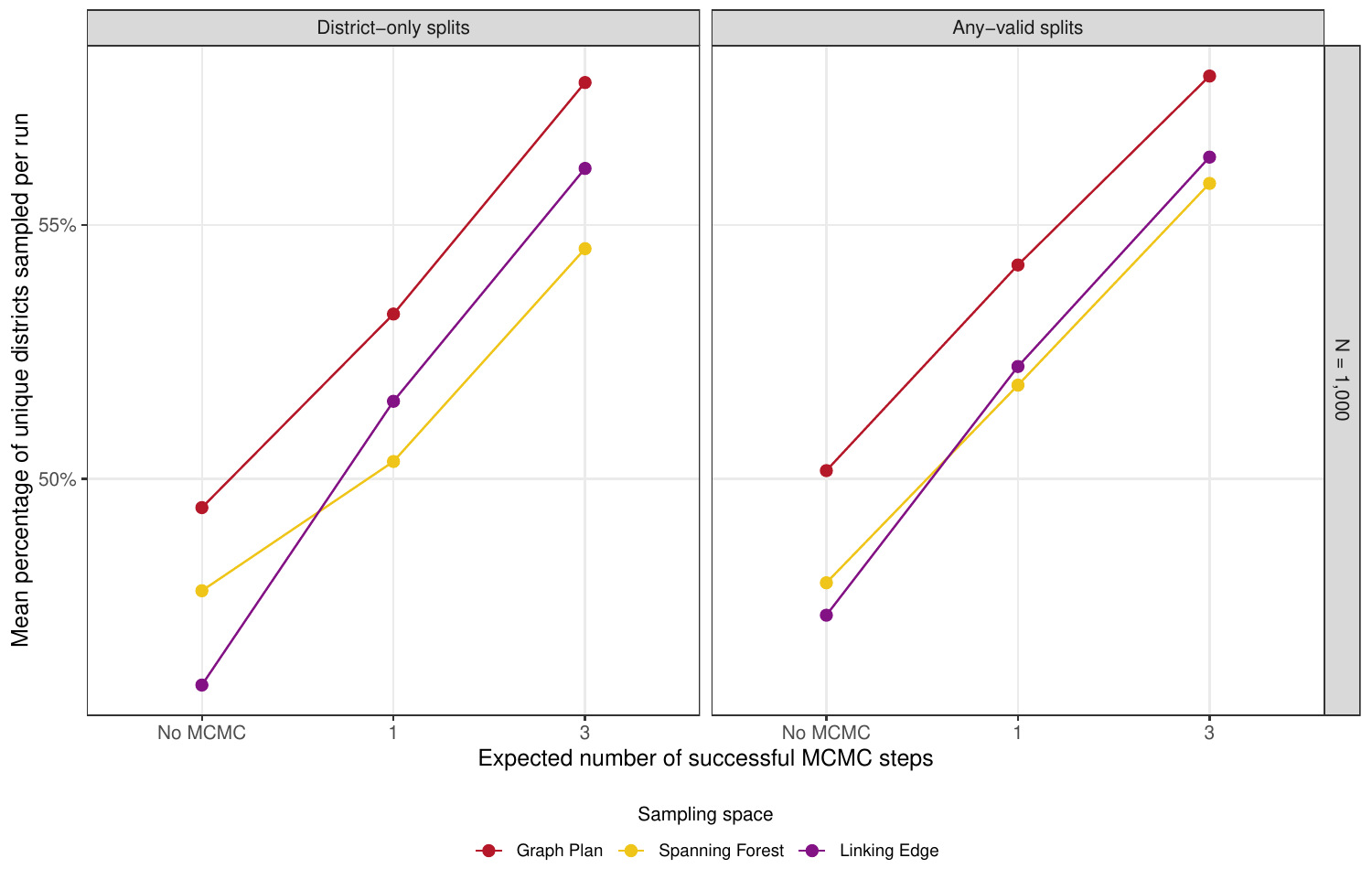}

}

\subcaption{\label{fig-ar-cd-gsmc-mcmc-div}Arkansas Congressional: Mean
percentage of unique districts per run across all expected numbers of
successful MCMC steps for each of the three gSMC sampling spaces,
faceted by splitting schedule and sample size.}

\end{minipage}%

\caption{\label{fig-ar-cd-div}Arkansas Congressional: Mean percentage of
unique districts per run across all gSMC and gSMC-MCMC configurations
run.}

\end{figure}%

\subsubsection{Wisconsin 2020 Congressional Map}\label{sec-wi-cd-2020}

We use a map with \(V=7,059\) precincts, \(E = 19,281\) edges, and a
median population of 697 per precinct. We use the provided
pseudo-counties from the \texttt{alarmdata} map as the units for
hierarchical sampling. No additional constraints are imposed through the
\(J\) function. We sample both gSMC and gSMC-MCMC at sample sizes of
1,000 and 2,500. We also sample gSMC at a sample size of 5,000. The
expected number of successful MCMC steps used were: 1 and 3.

\begin{figure}[H]

\centering{

\pandocbounded{\includegraphics[keepaspectratio]{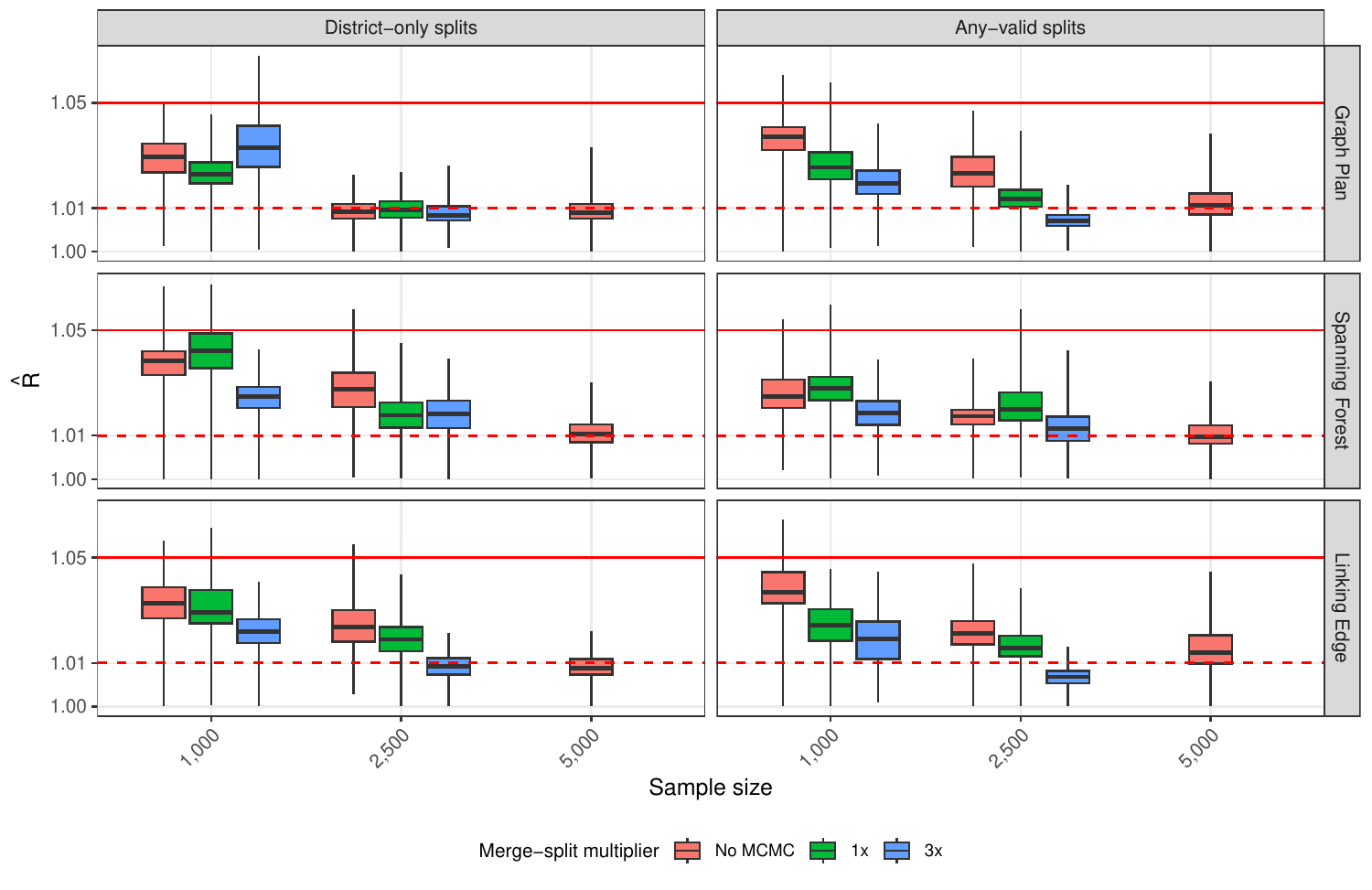}}

}

\caption{\label{fig-wi-cd-gsmc-rhats}Wisconsin Congressional:
Distribution of \(\hat{R}\) values for all gSMC and gSMC-MCMC
configurations run, faceted by sampling space and splitting schedule.}

\end{figure}%

\begin{figure}[H]

\begin{minipage}{\linewidth}

\begin{figure}[H]

\centering{

\includegraphics[width=0.9\linewidth,height=\textheight,keepaspectratio]{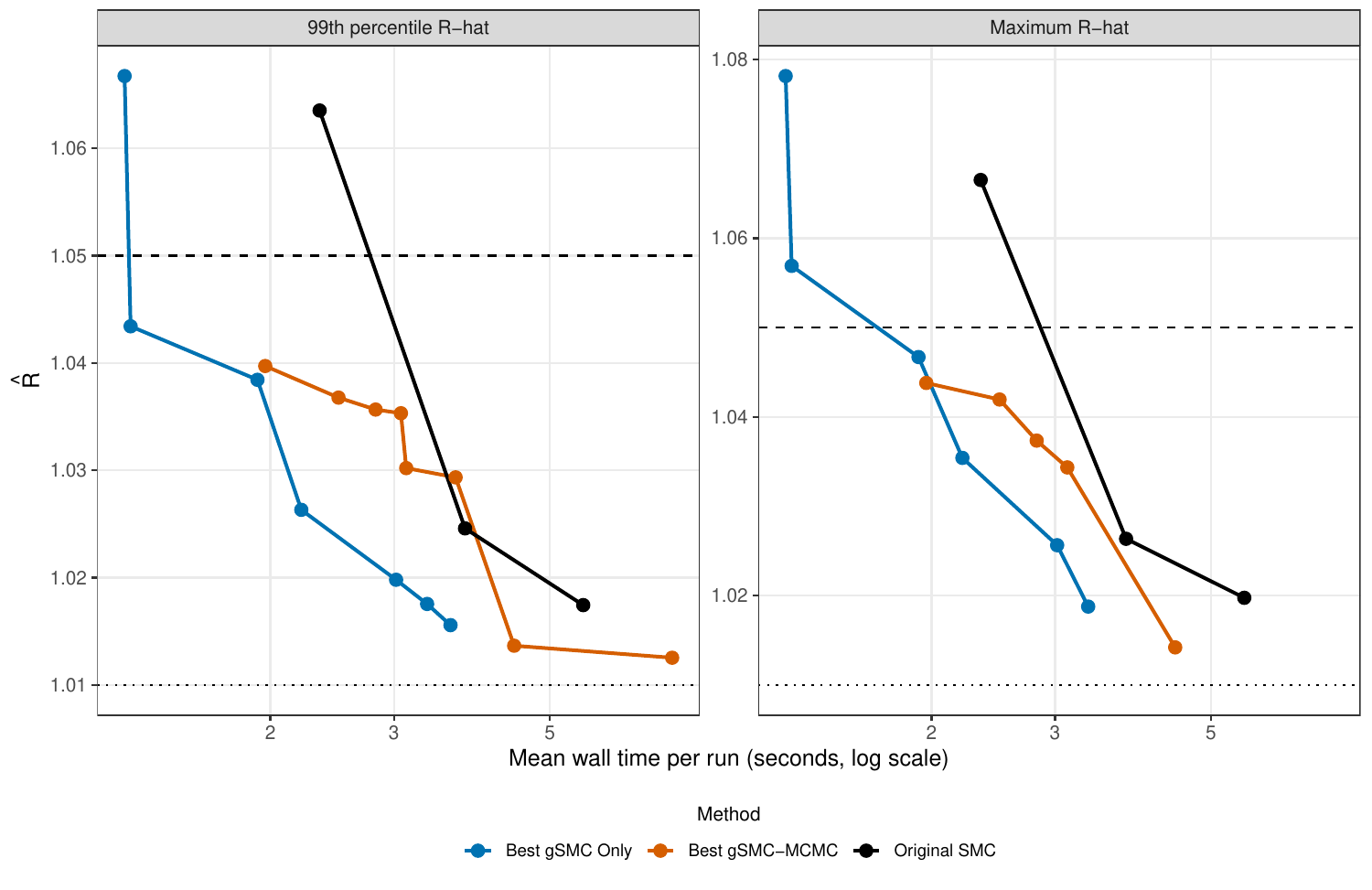}

}

\caption{\label{fig-wi-cd-pareto-frontier}Wisconsin Congressional:
Empirical Pareto frontier comparing mean wall time and convergence of
the best-performing gSMC and gSMC-MCMC configurations and Original SMC.
Each point represents an algorithm configuration, with mean wall time
per run on the horizontal axis and either the 99th percentile of the
resulting \(\hat{R}\) values (left) or the maximum \(\hat{R}\) value
(right) on the vertical axis. Moving along a frontier corresponds to
configurations for which improved convergence came at the cost of
greater wall time.}

\end{figure}%

\end{minipage}%
\newline
\begin{minipage}{\linewidth}

\begin{figure}[H]

\centering{

\includegraphics[width=0.9\linewidth,height=\textheight,keepaspectratio]{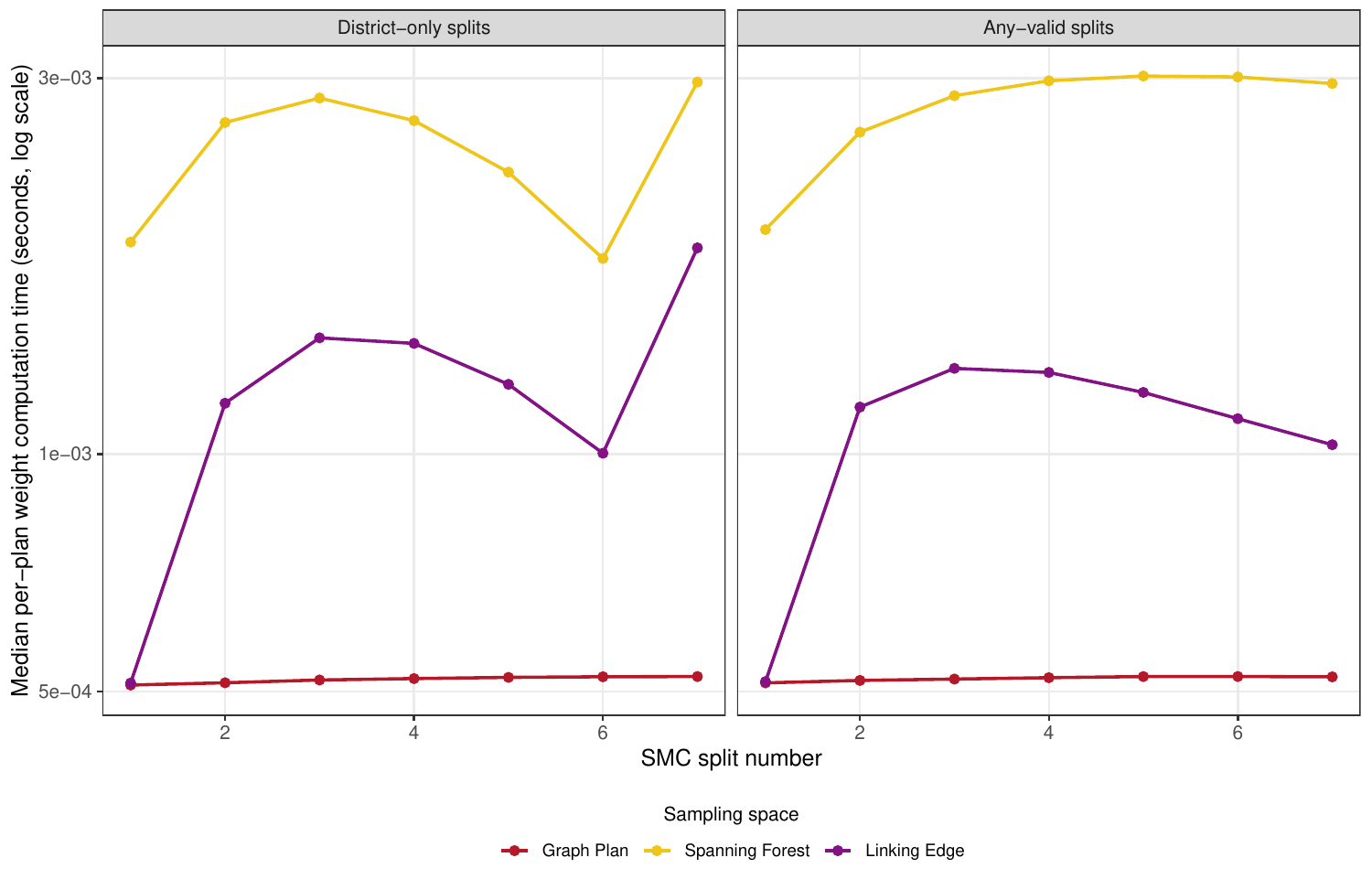}

}

\caption{\label{fig-wi-cd-weight-time}Average median per-plan weight
computation time across sampling spaces for a sample size of 2,500,
faceted by splitting schedule.}

\end{figure}%

\end{minipage}%

\end{figure}%

\begin{figure}[H]

\begin{minipage}{\linewidth}

\centering{

\includegraphics[width=0.8\linewidth,height=\textheight,keepaspectratio]{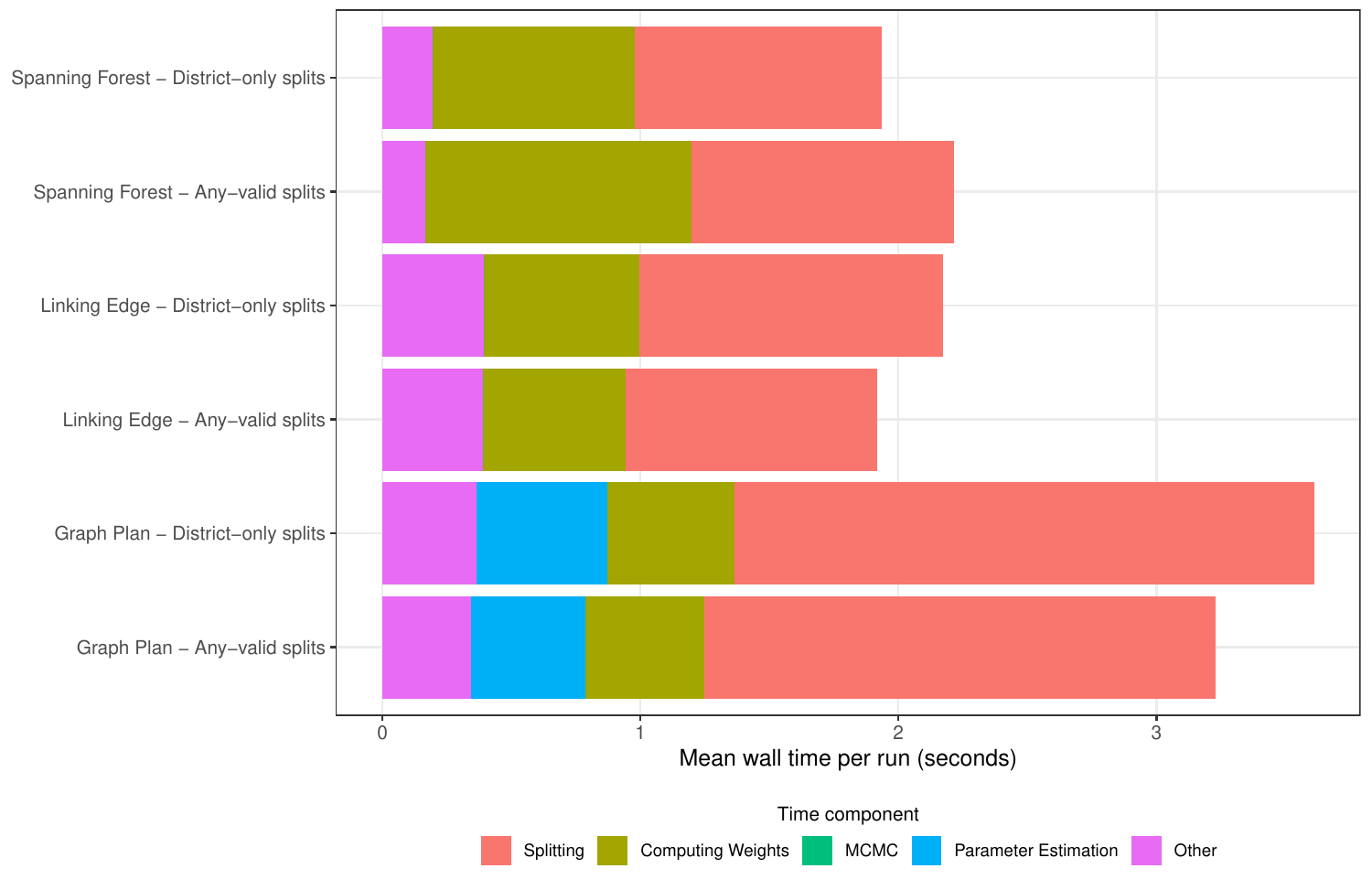}

}

\subcaption{\label{fig-wi-cd-gsmc-time}Breakdown of wall-time categories
across the three gSMC sampling spaces and two splitting schedules for a
sample size of 2,500.}

\end{minipage}%
\newline
\begin{minipage}{\linewidth}

\centering{

\pandocbounded{\includegraphics[keepaspectratio]{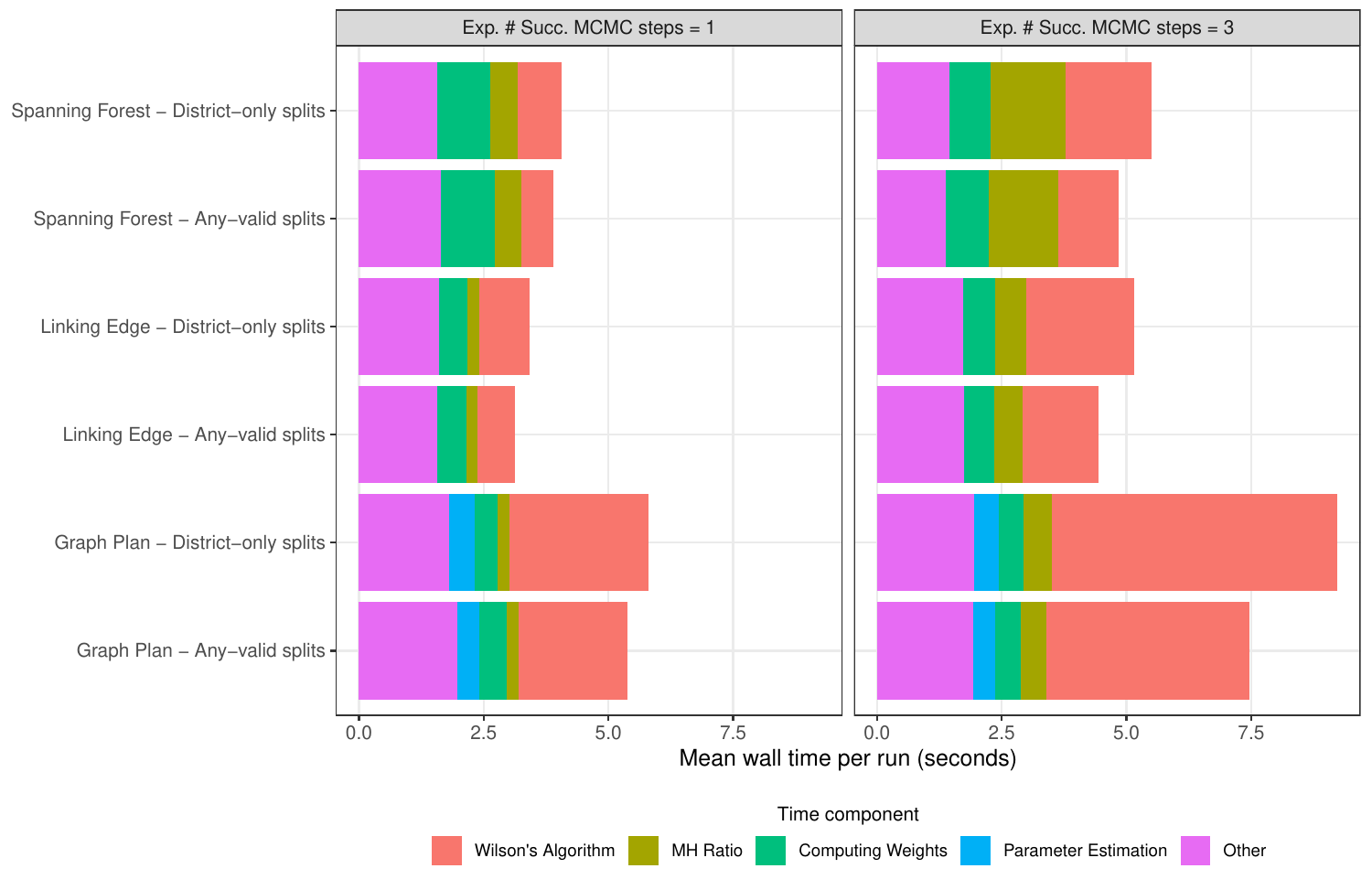}}

}

\subcaption{\label{fig-wi-cd-gsmc-mcmc-time}Breakdown of wall-time
categories across the three gSMC sampling spaces and two splitting
schedules, faceted by expected number of successful MCMC steps, for a
sample size of 2,500.}

\end{minipage}%

\caption{\label{fig-wi-cd-time-breakdown}Wisconsin Congressional:
Breakdown of wall time for gSMC and gSMC-MCMC configurations at a sample
size of 2,500.}

\end{figure}%

\begin{figure}[H]

\begin{minipage}{\linewidth}

\centering{

\includegraphics[width=0.95\linewidth,height=\textheight,keepaspectratio]{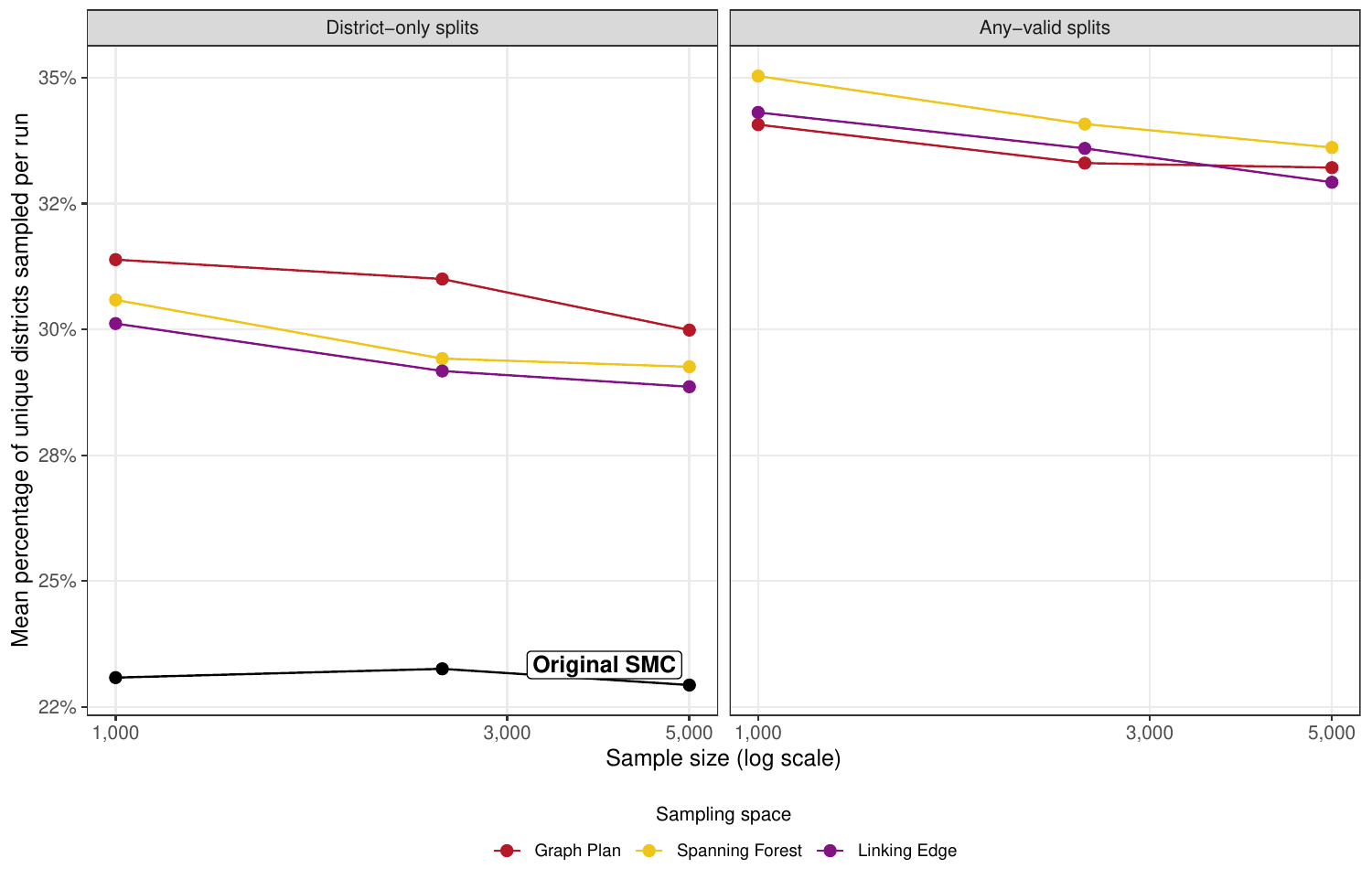}

}

\subcaption{\label{fig-wi-cd-gsmc-div}Wisconsin Congressional: Mean
percentage of unique districts per run versus sample size for each of
the three gSMC sampling spaces, faceted by splitting schedule.}

\end{minipage}%
\newline
\begin{minipage}{\linewidth}

\centering{

\includegraphics[width=0.95\linewidth,height=\textheight,keepaspectratio]{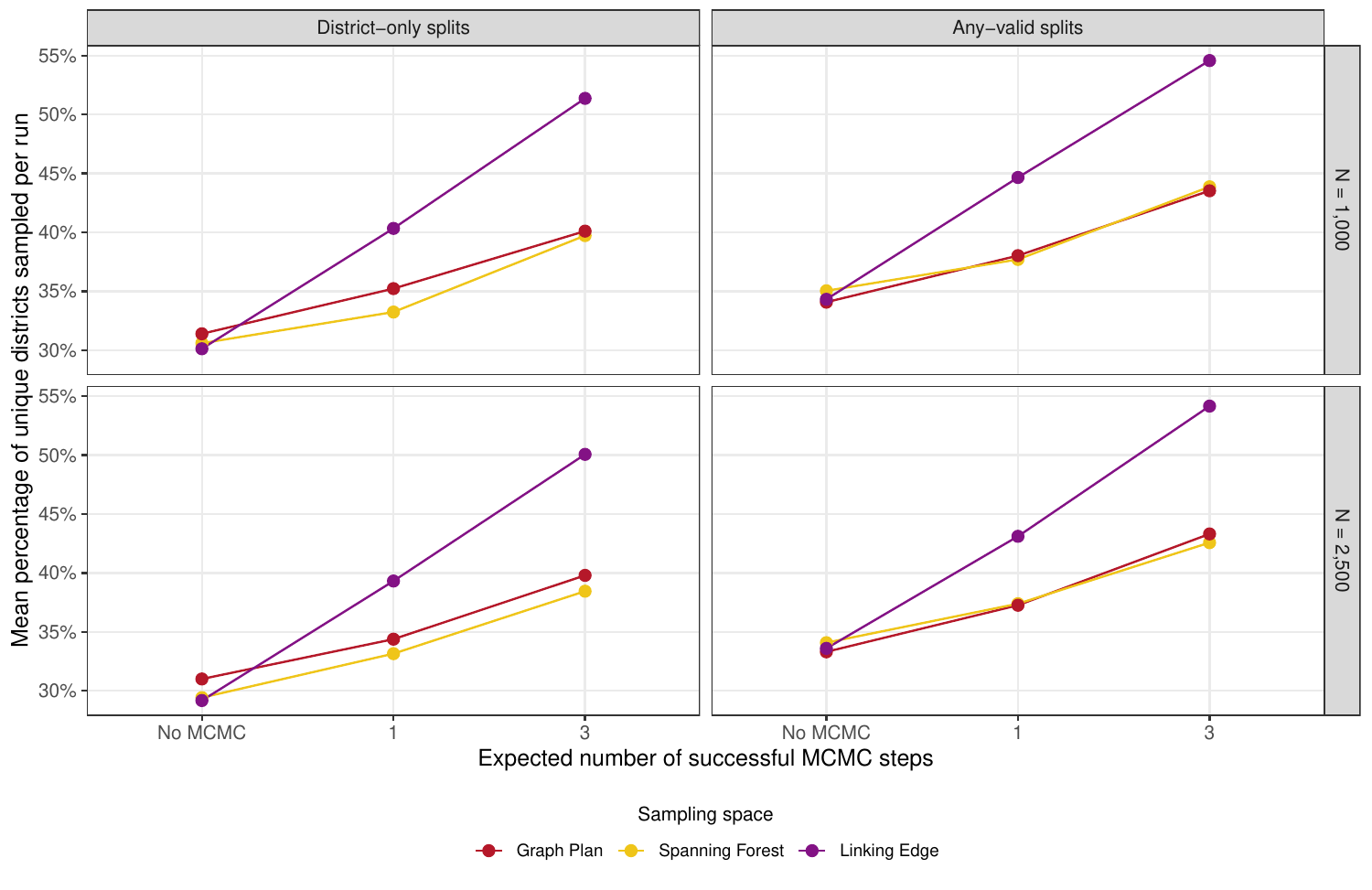}

}

\subcaption{\label{fig-wi-cd-gsmc-mcmc-div}Wisconsin Congressional: Mean
percentage of unique districts per run across all expected numbers of
successful MCMC steps for each of the three gSMC sampling spaces,
faceted by splitting schedule and sample size.}

\end{minipage}%

\caption{\label{fig-wi-cd-div}Wisconsin Congressional: Mean percentage
of unique districts per run across all gSMC and gSMC-MCMC configurations
run.}

\end{figure}%

\subsubsection{Tennessee 2020 Congressional Map}\label{sec-tn-cd-2020}

We use a map with \(V=1,965\) precincts, \(E = 5,604\) edges, and a
median population of 2,857 per precinct. We use the provided
pseudo-counties from the \texttt{alarmdata} map as the units for
hierarchical sampling. No additional constraints are imposed through the
\(J\) function. We sample both gSMC and gSMC-MCMC at sample sizes of
1,000, 2,500, and 5,000. The expected number of successful MCMC steps
used were: 1, 3, and 5. In contrast to Wisconsin, as
Figure~\ref{fig-tn-cd-gsmc-rhats} demonstrates, all configurations with
a sample size of 2,500 or higher converged. In fact, the majority of
configurations with 1,000 samples also converged, suggesting that both
the number of districts and the size of the map graph influence the
difficulty of obtaining convergent samples.

\begin{figure}[H]

\centering{

\pandocbounded{\includegraphics[keepaspectratio]{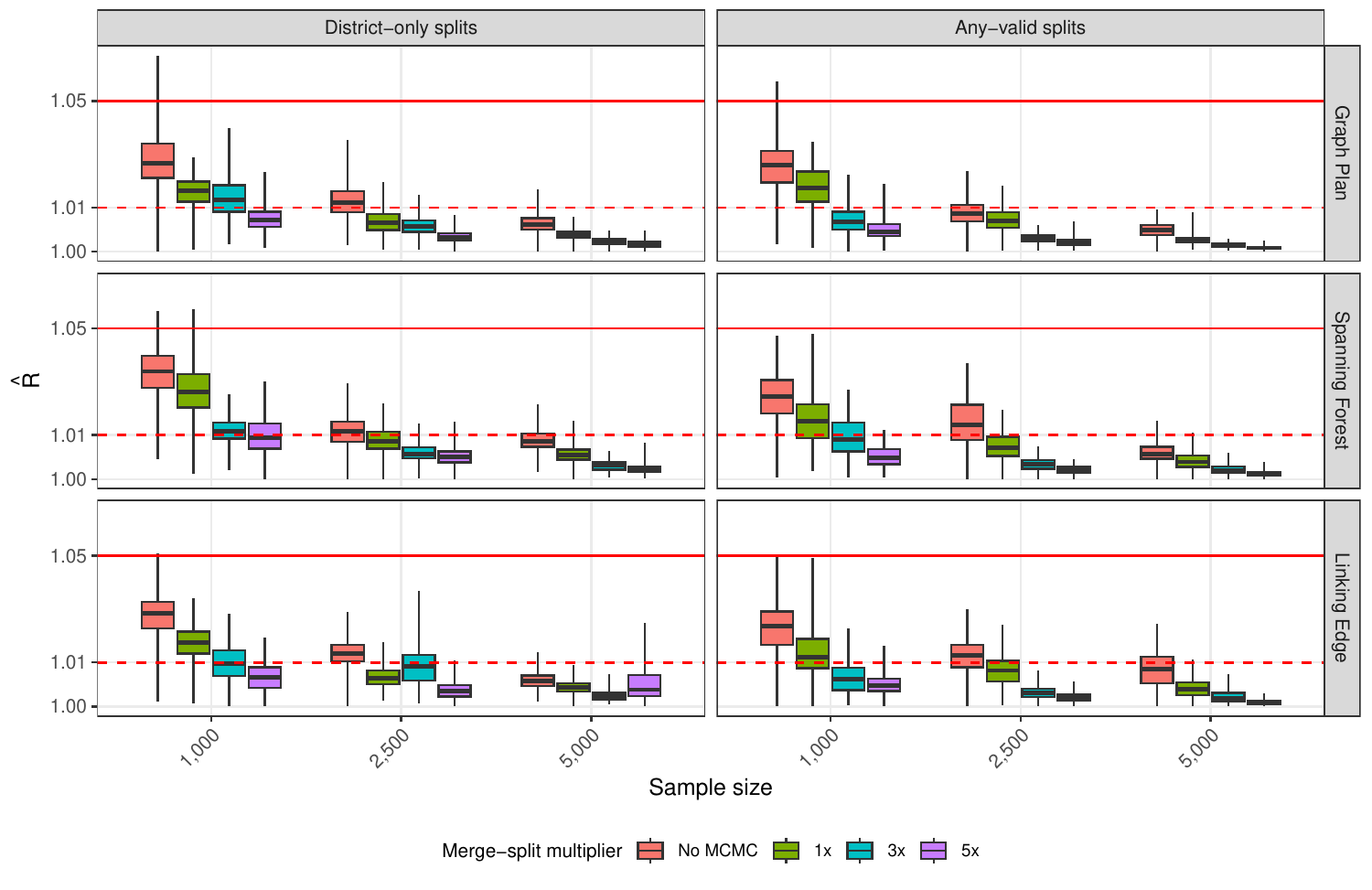}}

}

\caption{\label{fig-tn-cd-gsmc-rhats}Tennessee Congressional:
Distribution of \(\hat{R}\) values for all gSMC and gSMC-MCMC
configurations run, faceted by sampling space and splitting schedule.}

\end{figure}%

\begin{figure}[H]

\begin{minipage}{\linewidth}

\begin{figure}[H]

\centering{

\includegraphics[width=0.9\linewidth,height=\textheight,keepaspectratio]{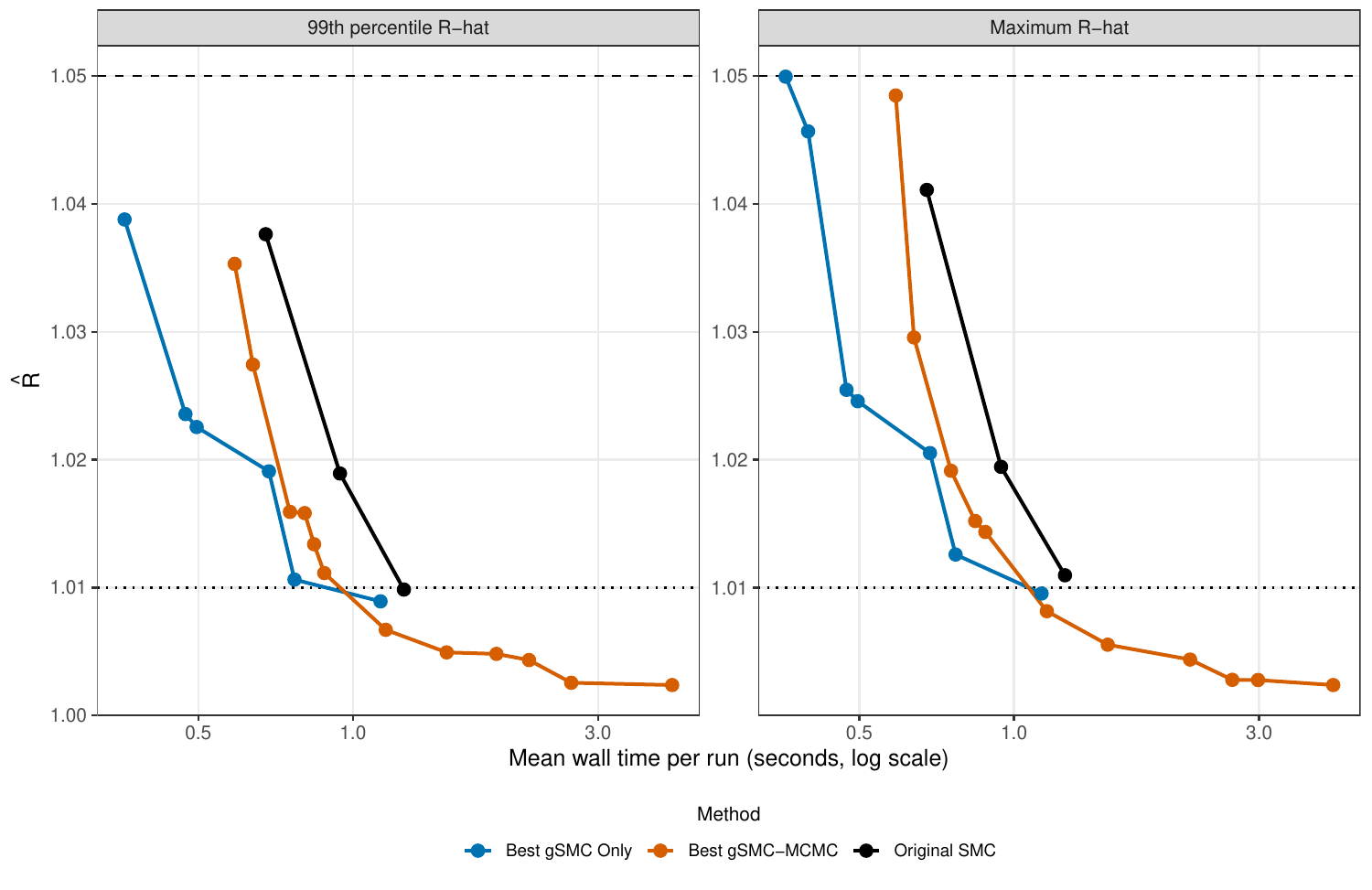}

}

\caption{\label{fig-tn-cd-pareto-frontier}Tennessee Congressional:
Empirical Pareto frontier comparing mean wall time and convergence of
the best-performing gSMC and gSMC-MCMC configurations and Original SMC.
Each point represents an algorithm configuration, with mean wall time
per run on the horizontal axis and either the 99th percentile of the
resulting \(\hat{R}\) values (left) or the maximum \(\hat{R}\) value
(right) on the vertical axis. Moving along a frontier corresponds to
configurations for which improved convergence came at the cost of
greater wall time.}

\end{figure}%

\end{minipage}%
\newline
\begin{minipage}{\linewidth}

\begin{figure}[H]

\centering{

\includegraphics[width=0.9\linewidth,height=\textheight,keepaspectratio]{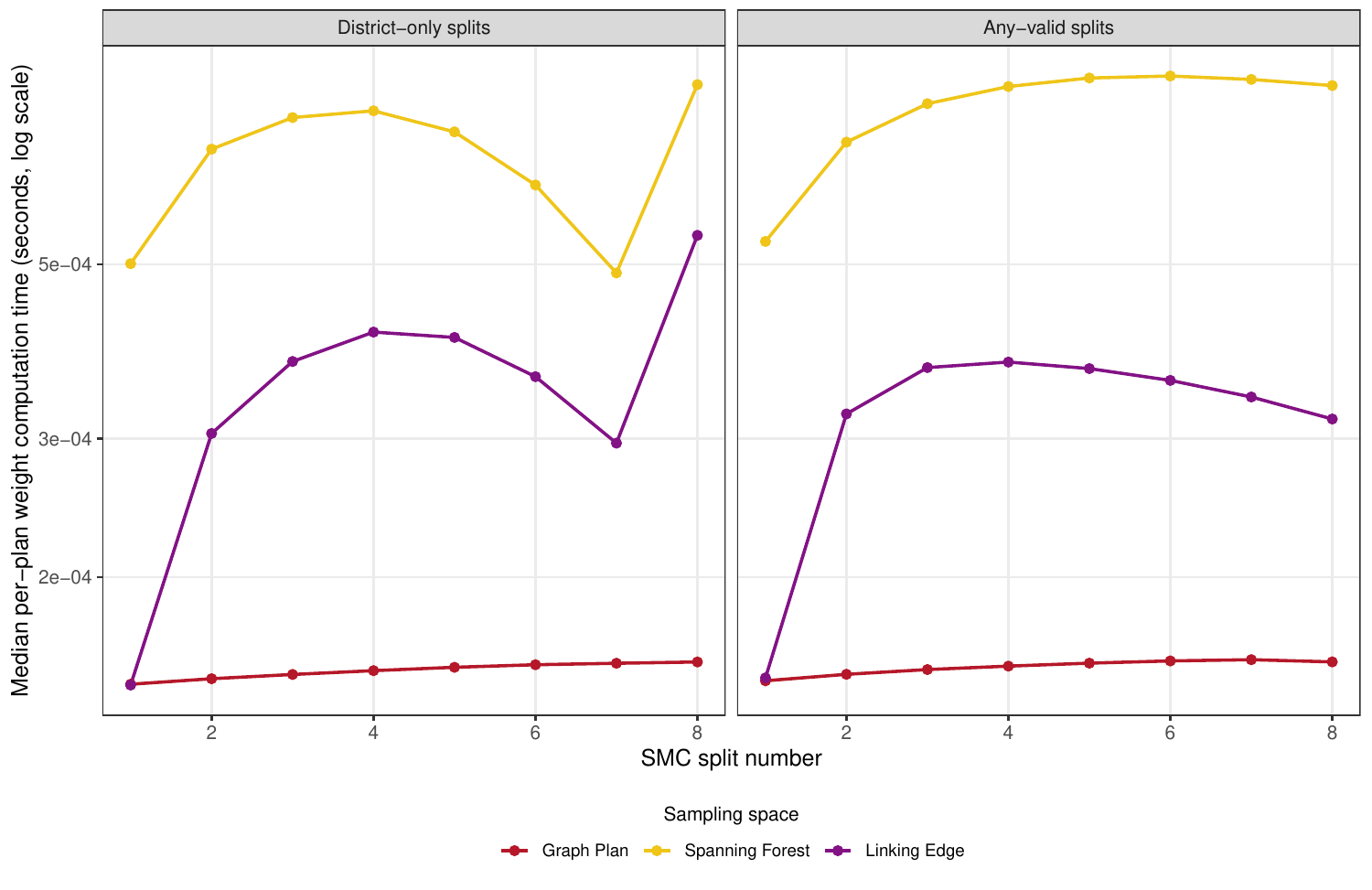}

}

\caption{\label{fig-tn-cd-weight-time}Average median per-plan weight
computation time across sampling spaces for a sample size of 5,000,
faceted by splitting schedule.}

\end{figure}%

\end{minipage}%

\end{figure}%

\begin{figure}[H]

\begin{minipage}{\linewidth}

\centering{

\includegraphics[width=0.8\linewidth,height=\textheight,keepaspectratio]{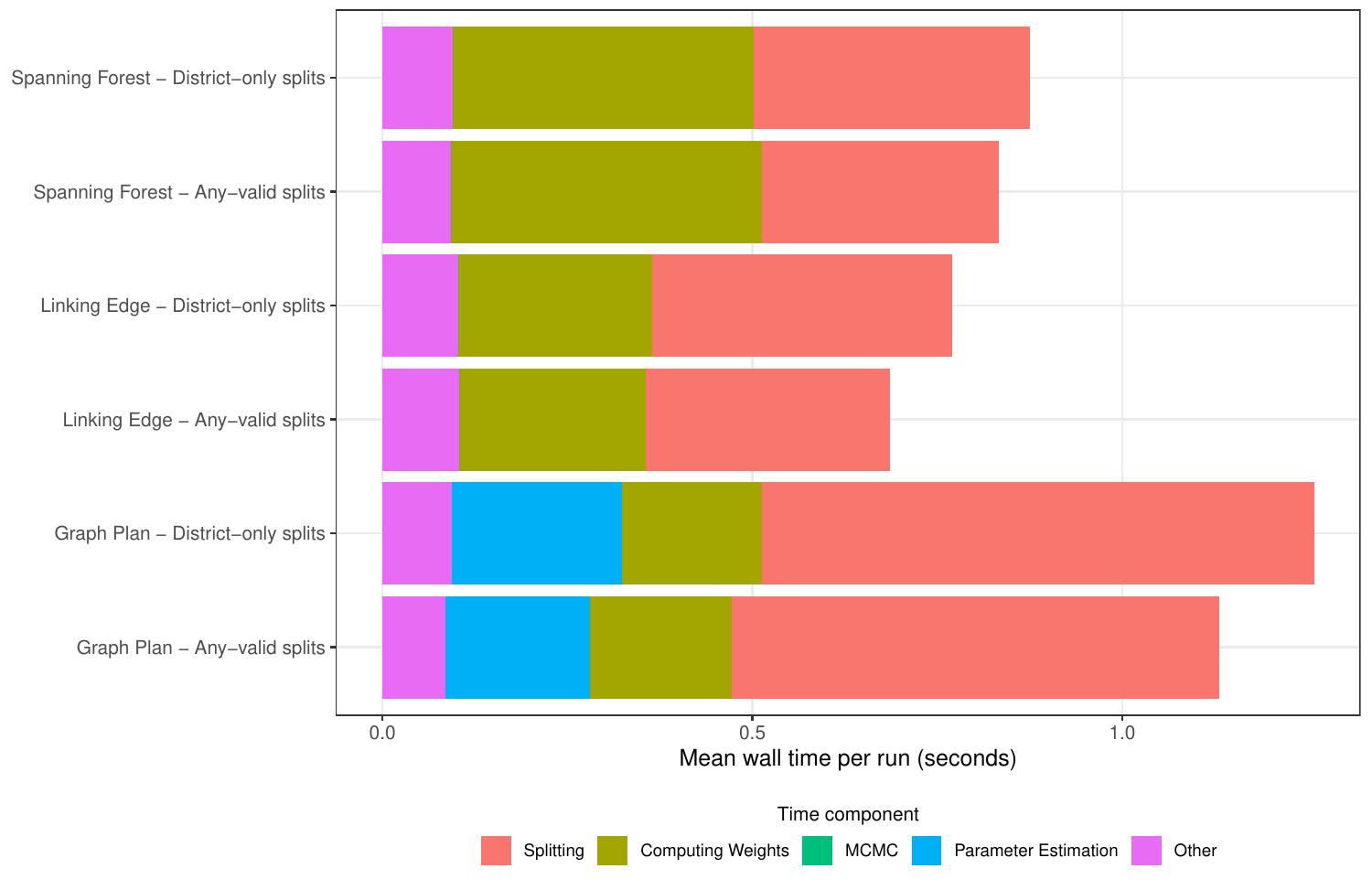}

}

\subcaption{\label{fig-tn-cd-gsmc-time}Breakdown of wall-time categories
across the three gSMC sampling spaces and two splitting schedules for a
sample size of 5,000.}

\end{minipage}%
\newline
\begin{minipage}{\linewidth}

\centering{

\pandocbounded{\includegraphics[keepaspectratio]{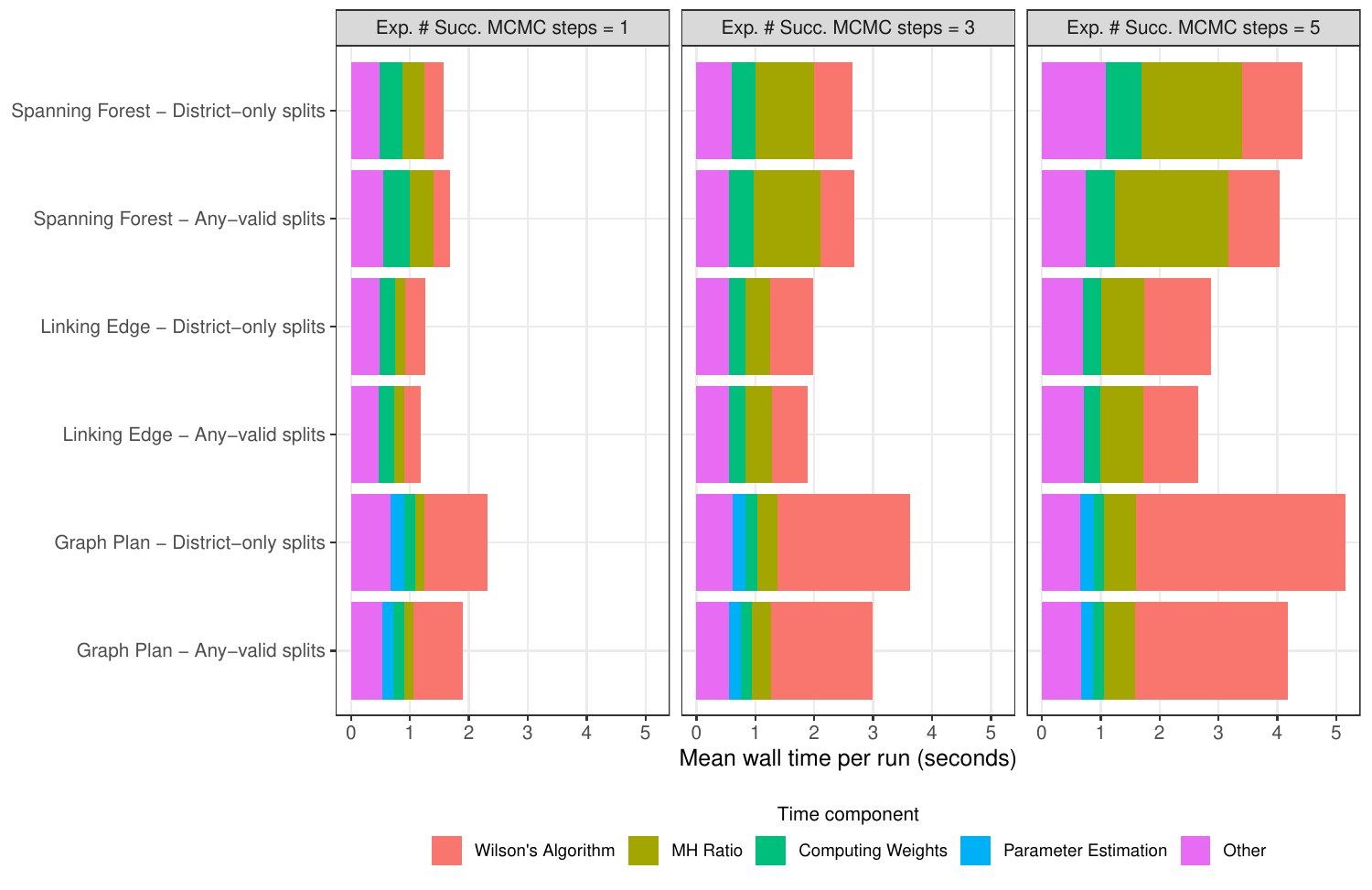}}

}

\subcaption{\label{fig-tn-cd-gsmc-mcmc-time}Breakdown of wall-time
categories across the three gSMC sampling spaces and two splitting
schedules, faceted by expected number of successful MCMC steps, for a
sample size of 5,000.}

\end{minipage}%

\caption{\label{fig-tn-cd-time-breakdown}Tennessee Congressional:
Breakdown of wall time for gSMC and gSMC-MCMC configurations at a sample
size of 5,000.}

\end{figure}%

\begin{figure}[H]

\begin{minipage}{\linewidth}

\centering{

\includegraphics[width=0.95\linewidth,height=\textheight,keepaspectratio]{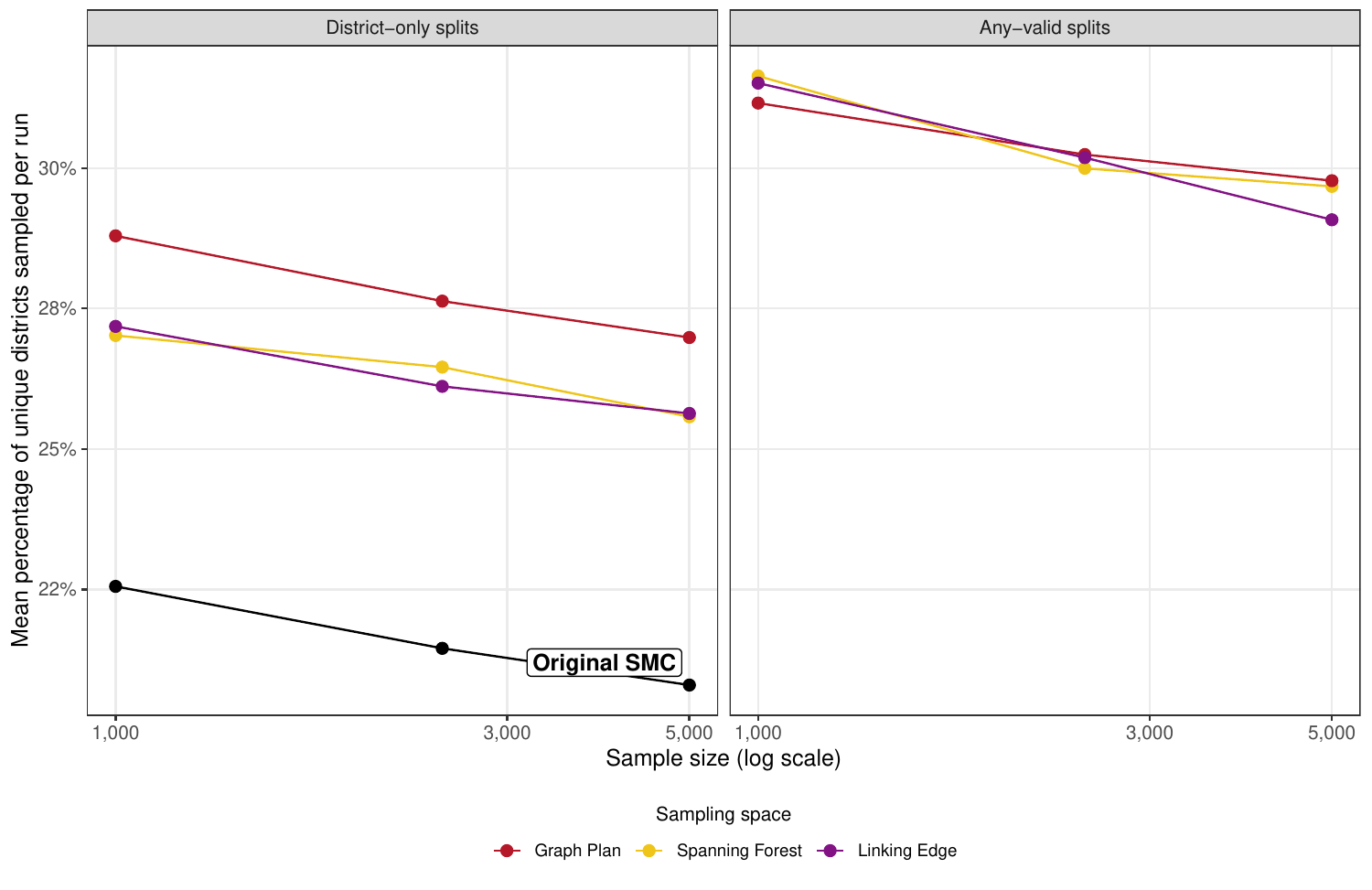}

}

\subcaption{\label{fig-tn-cd-gsmc-div}Tennessee Congressional: Mean
percentage of unique districts per run versus sample size for each of
the three gSMC sampling spaces, faceted by splitting schedule.}

\end{minipage}%
\newline
\begin{minipage}{\linewidth}

\centering{

\includegraphics[width=0.95\linewidth,height=\textheight,keepaspectratio]{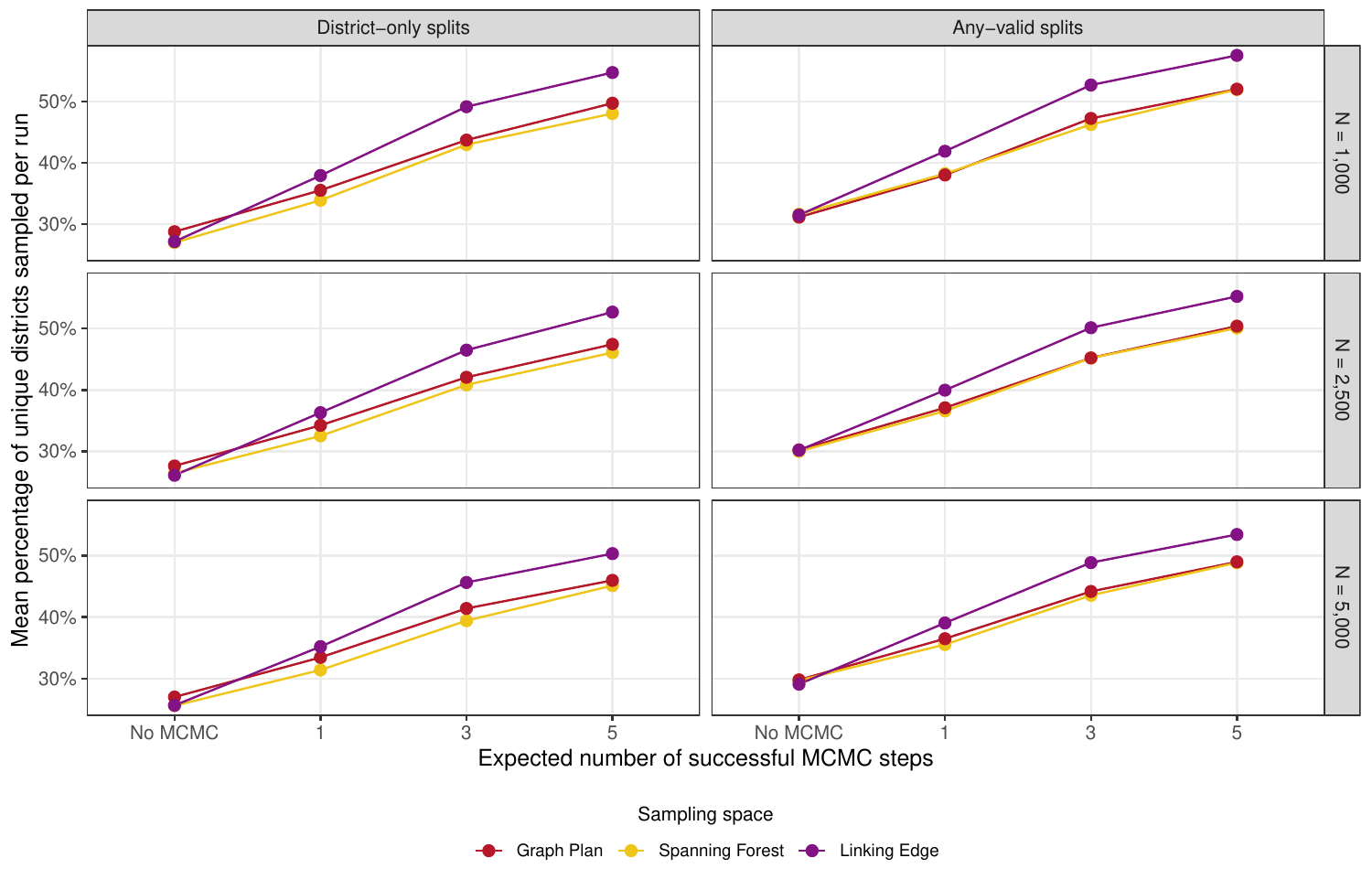}

}

\subcaption{\label{fig-tn-cd-gsmc-mcmc-div}Tennessee Congressional: Mean
percentage of unique districts per run across all expected numbers of
successful MCMC steps for each of the three gSMC sampling spaces,
faceted by splitting schedule and sample size.}

\end{minipage}%

\caption{\label{fig-tn-cd-div}Tennessee Congressional: Mean percentage
of unique districts per run across all gSMC and gSMC-MCMC configurations
run.}

\end{figure}%

\subsubsection{Washington 2020 Congressional Map}\label{sec-wa-cd-2020}

We use a map with \(V=7,434\) precincts, \(E = 19,517\) edges, and a
median population of 940.5 per precinct. We use the provided
pseudo-counties from the \texttt{alarmdata} map as the units for
hierarchical sampling. We add two additional hinge constraints through
the \(J\) function. These hinge constraints score each region in a plan
and given a population, in this case white voting age population, and
target percentages, they penalize districts which are below their target
percentage. We apply the same constraints and use the same strength used
in the \texttt{alarmdata} Washington 2020 congressional plan
simulations. We sample both gSMC and gSMC-MCMC at sample sizes of 1,000,
2,500, and 5,000. We also sample gSMC at sample sizes of 10,000 and
20,000. The expected number of successful MCMC steps used were: 1, 3,
and 5. Unlike the previous Figures \ref{fig-ar-cd-weight-time},
\ref{fig-wi-cd-weight-time}, \ref{fig-tn-cd-weight-time}, the graph
space weight computation time here in Figure~\ref{fig-wa-cd-weight-time}
does increase noticeably as the number of steps increases. That is
because while graph space weights are \(O(V)\) to compute when there are
no additional constraints, here we use constraints whose cost is linear
in the number of regions in the plan so it makes a noticeable impact. In
contrast, since the linking edge and forest space weights are already
more expensive to compute the impact is less noticeable.

\begin{figure}[H]

\centering{

\pandocbounded{\includegraphics[keepaspectratio]{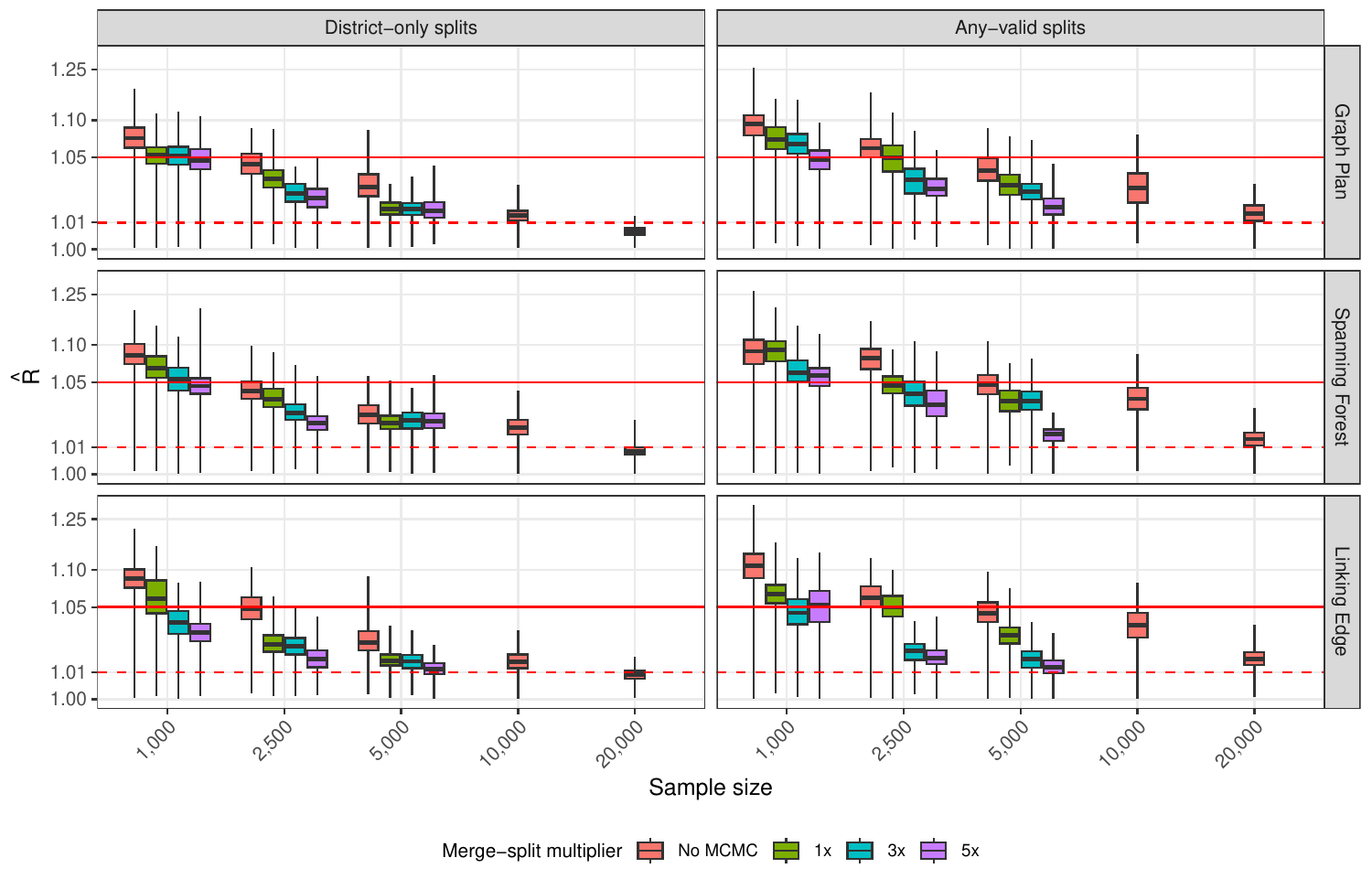}}

}

\caption{\label{fig-wa-cd-gsmc-rhats}Washington Congressional:
Distribution of \(\hat{R}\) values for all gSMC and gSMC-MCMC
configurations run, faceted by sampling space and splitting schedule.}

\end{figure}%

\begin{figure}[H]

\begin{minipage}{\linewidth}

\begin{figure}[H]

\centering{

\includegraphics[width=0.9\linewidth,height=\textheight,keepaspectratio]{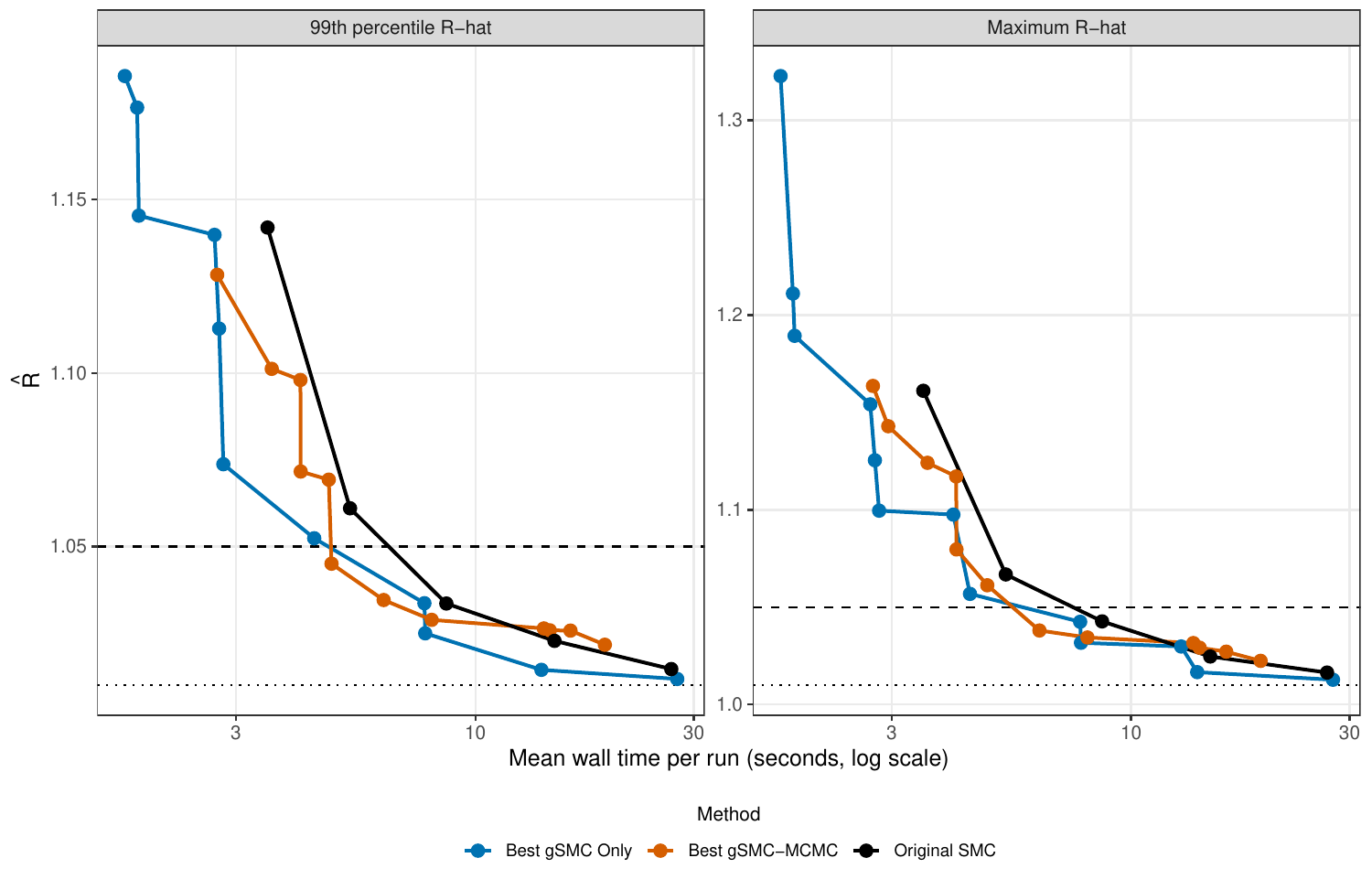}

}

\caption{\label{fig-wa-cd-pareto-frontier}Washington Congressional:
Empirical Pareto frontier comparing mean wall time and convergence of
the best-performing gSMC and gSMC-MCMC configurations and Original SMC.
Each point represents an algorithm configuration, with mean wall time
per run on the horizontal axis and either the 99th percentile of the
resulting \(\hat{R}\) values (left) or the maximum \(\hat{R}\) value
(right) on the vertical axis. Moving along a frontier corresponds to
configurations for which improved convergence came at the cost of
greater wall time.}

\end{figure}%

\end{minipage}%
\newline
\begin{minipage}{\linewidth}

\begin{figure}[H]

\centering{

\includegraphics[width=0.9\linewidth,height=\textheight,keepaspectratio]{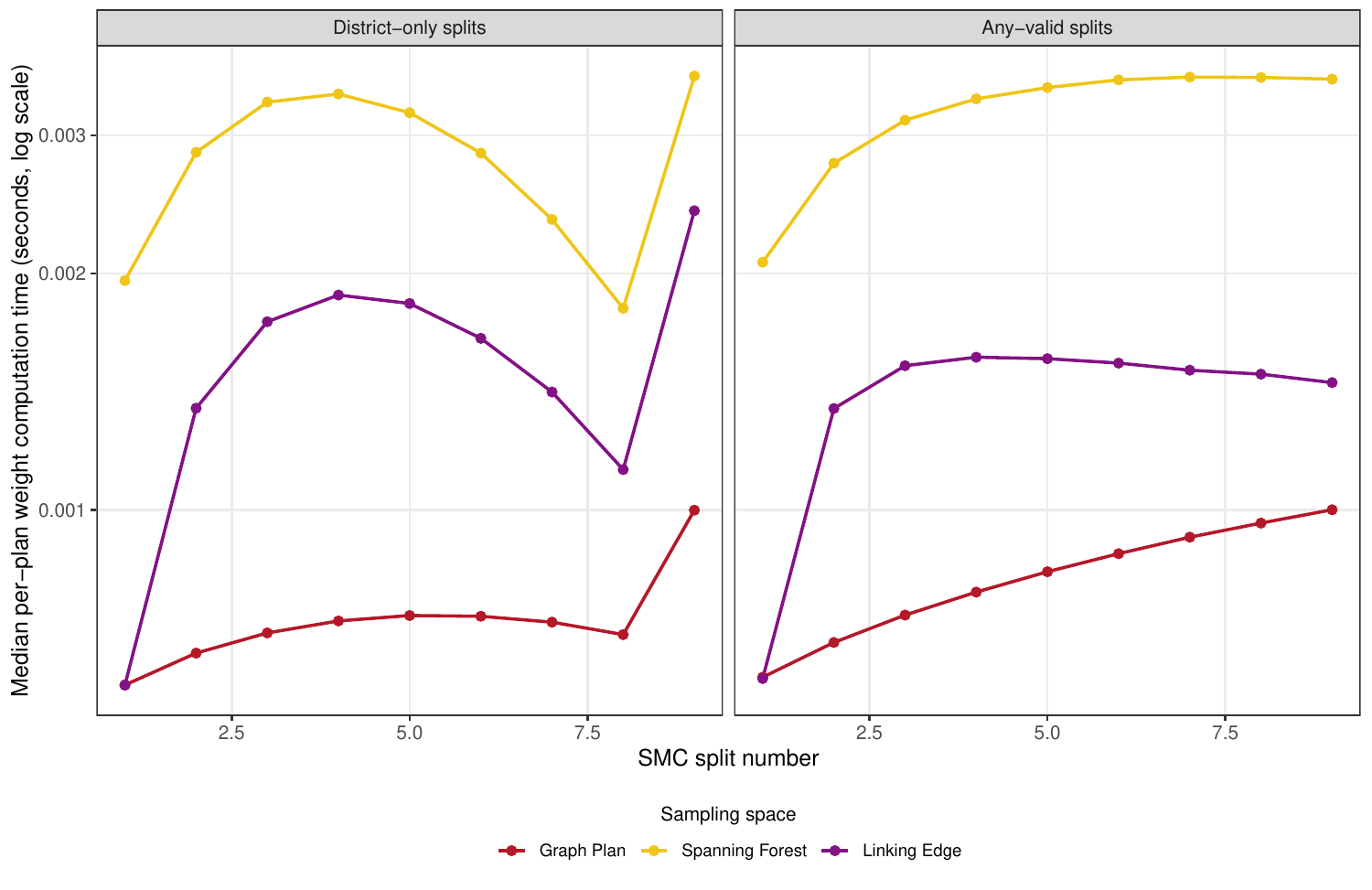}

}

\caption{\label{fig-wa-cd-weight-time}Average median per-plan weight
computation time across sampling spaces for a sample size of 5,000,
faceted by splitting schedule.}

\end{figure}%

\end{minipage}%

\end{figure}%

\begin{figure}[H]

\begin{minipage}{\linewidth}

\centering{

\includegraphics[width=0.8\linewidth,height=\textheight,keepaspectratio]{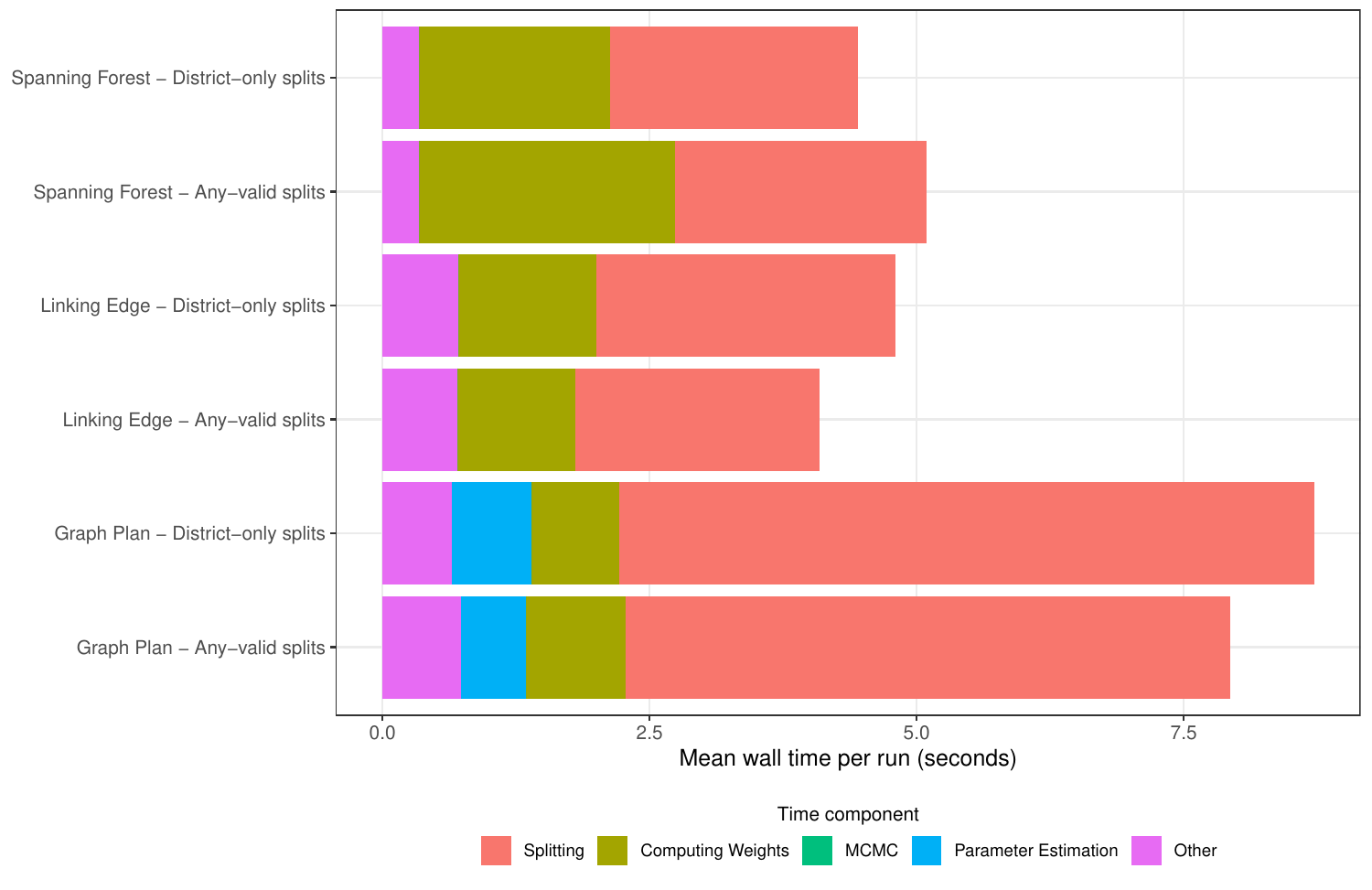}

}

\subcaption{\label{fig-wa-cd-gsmc-time}Breakdown of wall-time categories
across the three gSMC sampling spaces and two splitting schedules for a
sample size of 5,000.}

\end{minipage}%
\newline
\begin{minipage}{\linewidth}

\centering{

\pandocbounded{\includegraphics[keepaspectratio]{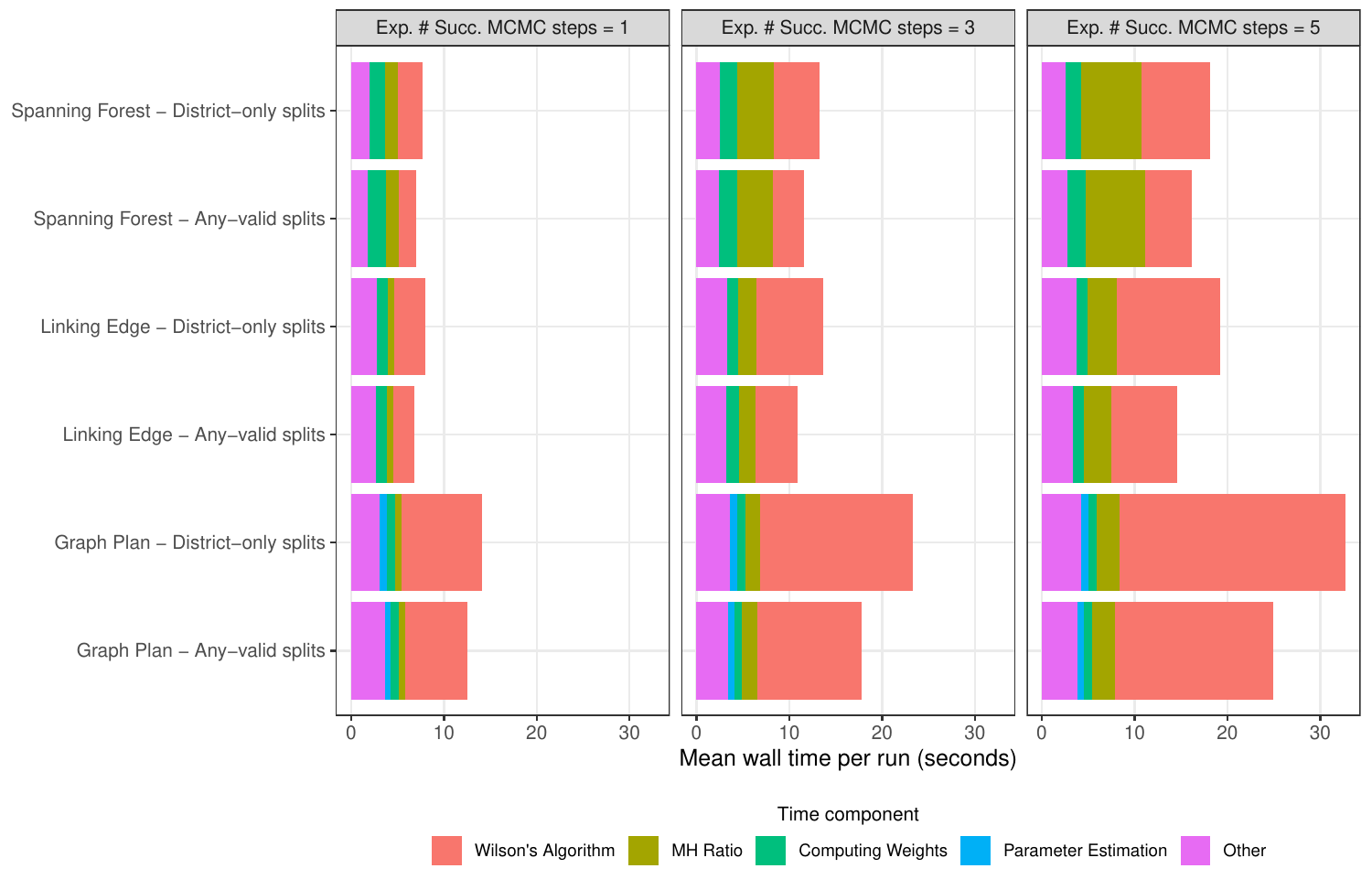}}

}

\subcaption{\label{fig-wa-cd-gsmc-mcmc-time}Breakdown of wall-time
categories across the three gSMC sampling spaces and two splitting
schedules, faceted by expected number of successful MCMC steps, for a
sample size of 5,000.}

\end{minipage}%

\caption{\label{fig-wa-cd-time-breakdown}Washington Congressional:
Breakdown of wall time for gSMC and gSMC-MCMC configurations at a sample
size of 5,000.}

\end{figure}%

\begin{figure}[H]

\begin{minipage}{\linewidth}

\centering{

\includegraphics[width=0.95\linewidth,height=\textheight,keepaspectratio]{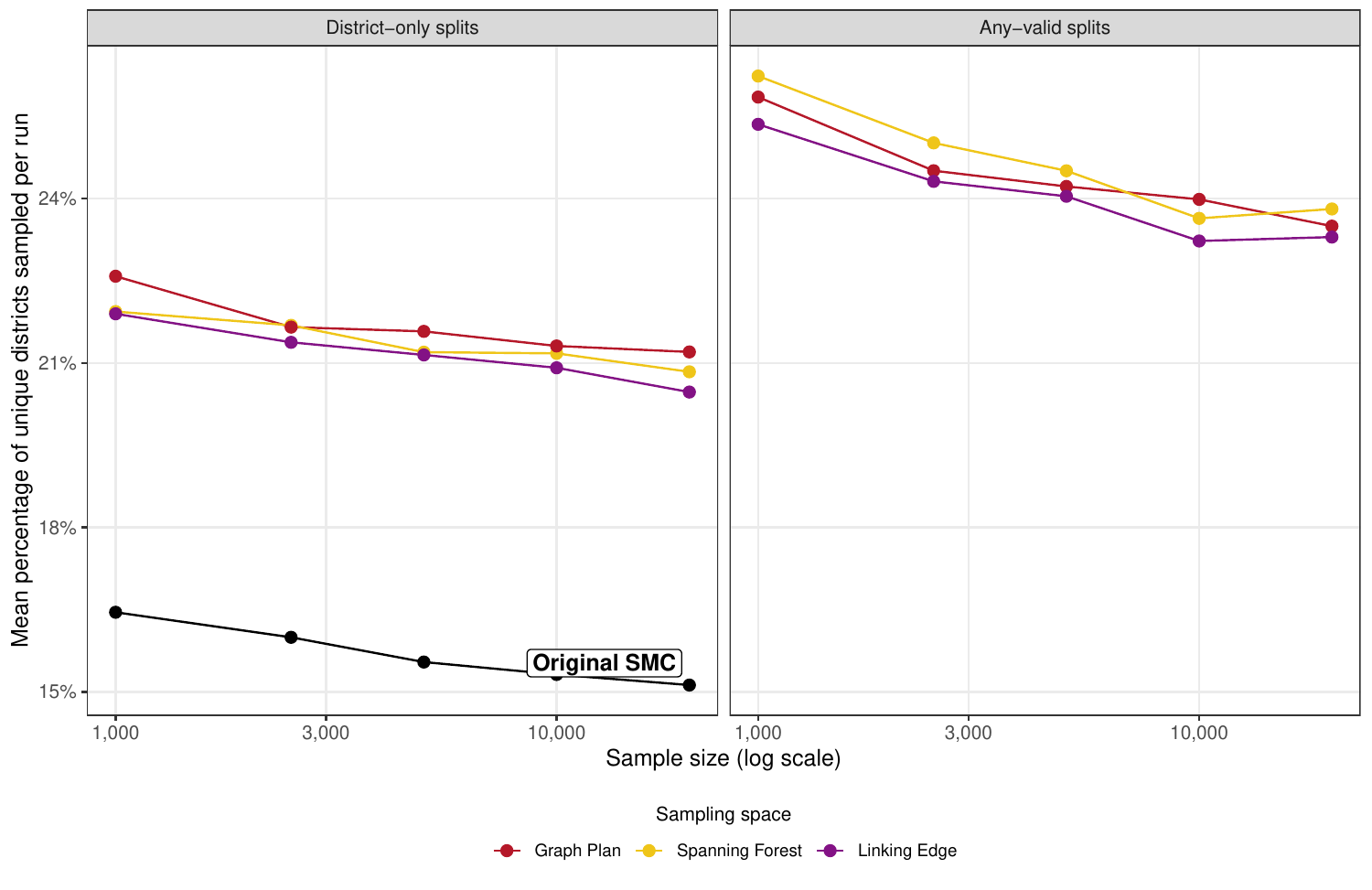}

}

\subcaption{\label{fig-wa-cd-gsmc-div}Washington Congressional: Mean
percentage of unique districts per run versus sample size for each of
the three gSMC sampling spaces, faceted by splitting schedule.}

\end{minipage}%
\newline
\begin{minipage}{\linewidth}

\centering{

\includegraphics[width=0.95\linewidth,height=\textheight,keepaspectratio]{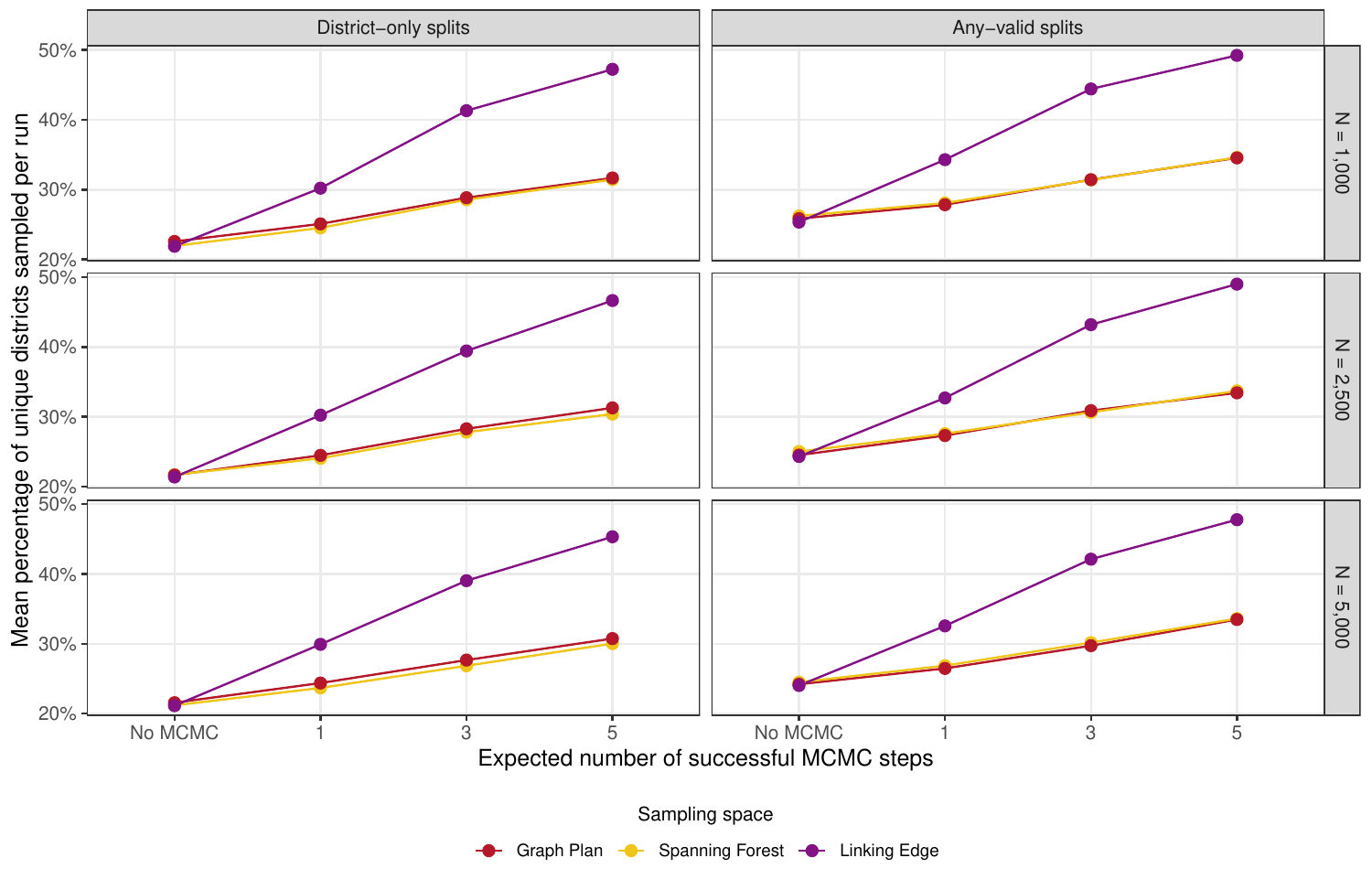}

}

\subcaption{\label{fig-wa-cd-gsmc-mcmc-div}Washington Congressional:
Mean percentage of unique districts per run across all expected numbers
of successful MCMC steps for each of the three gSMC sampling spaces,
faceted by splitting schedule and sample size.}

\end{minipage}%

\caption{\label{fig-wa-cd-div}Washington Congressional: Mean percentage
of unique districts per run across all gSMC and gSMC-MCMC configurations
run.}

\end{figure}%

\subsubsection{Pennsylvania 2020 Congressional
Map}\label{pennsylvania-2020-congressional-map}

We use a map with \(V=9,178\) precincts, \(E = 25,519\) edges, and a
median population of 1,176 per precinct. We use the provided
pseudo-counties from the \texttt{alarmdata} map as the units for
hierarchical sampling. No additional constraints are imposed through the
\(J\) function. We sample both gSMC and gSMC-MCMC at sample sizes of
1,000, 2,500, and 5,000. We also sample gSMC at sample sizes of 10,000
and 25,000. The expected number of successful MCMC steps used were: 1,
5, and 10. As Figure~\ref{fig-pa-cd-gsmc-rhats} demonstrates, linking
edge space sampling with any-valid splits performs exceptionally well on
this map, converging with just 1,000 samples when using 10 expected
successful MCMC steps. In contrast, this is the first map sampled where
Original SMC in unable to fully converge with the given sample sizes.

\begin{figure}[H]

\centering{

\pandocbounded{\includegraphics[keepaspectratio]{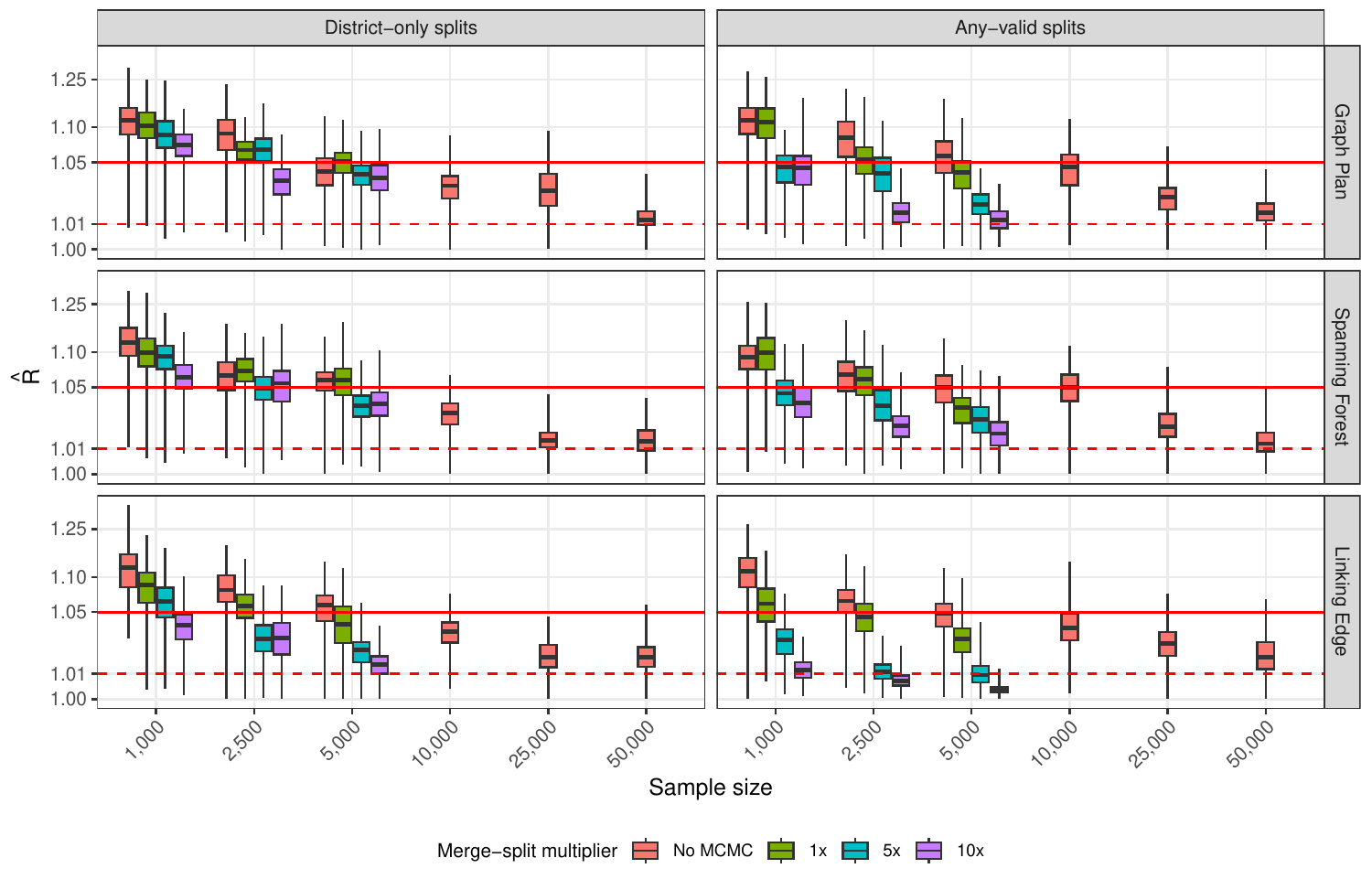}}

}

\caption{\label{fig-pa-cd-gsmc-rhats}Pennsylvania Congressional:
Distribution of \(\hat{R}\) values for all gSMC and gSMC-MCMC
configurations run, faceted by sampling space and splitting schedule.}

\end{figure}%

\begin{figure}[H]

\begin{minipage}{\linewidth}

\begin{figure}[H]

\centering{

\includegraphics[width=0.9\linewidth,height=\textheight,keepaspectratio]{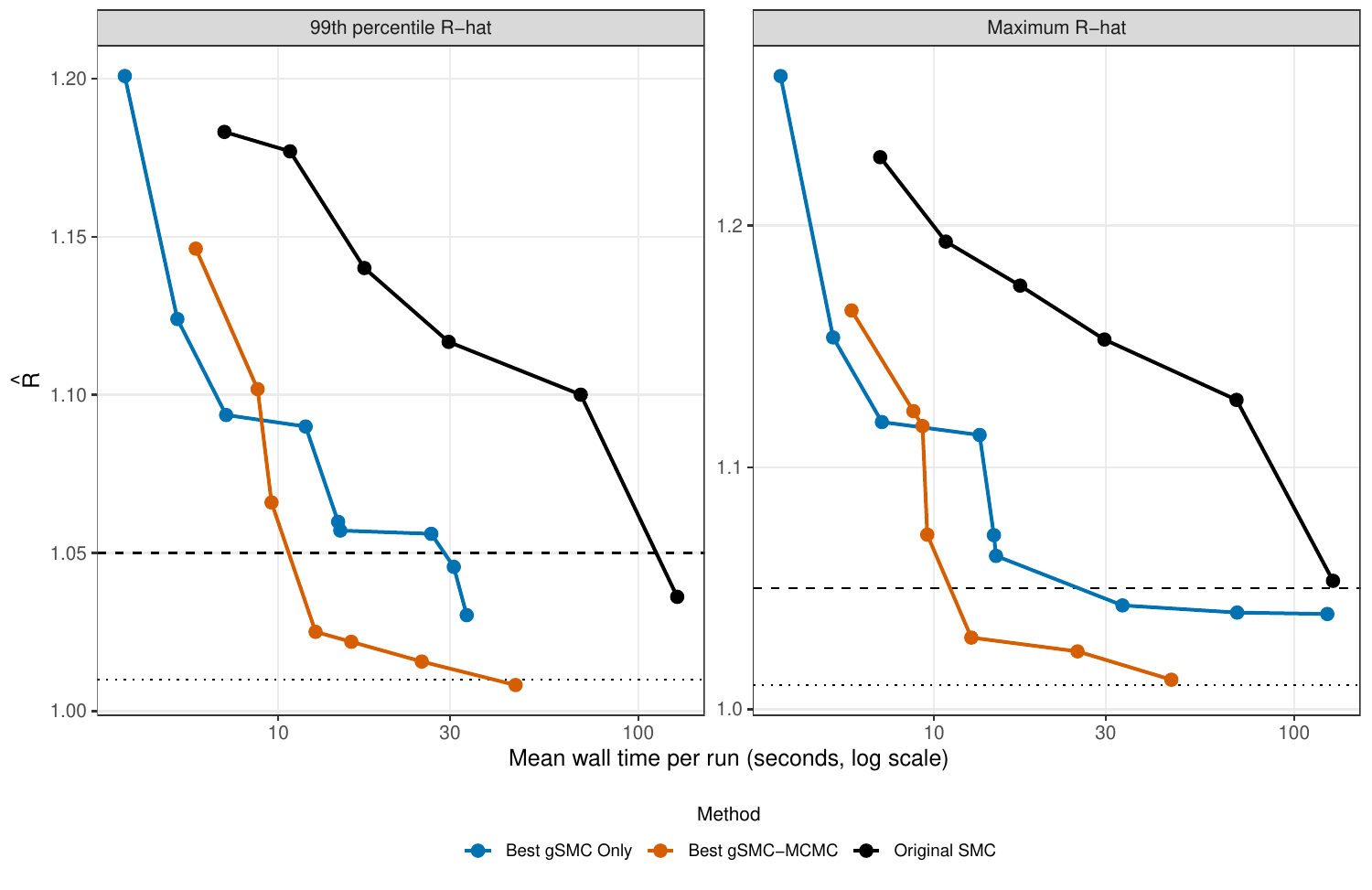}

}

\caption{\label{fig-pa-cd-pareto-frontier}Pennsylvania Congressional:
Empirical Pareto frontier comparing mean wall time and convergence of
the best-performing gSMC and gSMC-MCMC configurations and Original SMC.
Each point represents an algorithm configuration, with mean wall time
per run on the horizontal axis and either the 99th percentile of the
resulting \(\hat{R}\) values (left) or the maximum \(\hat{R}\) value
(right) on the vertical axis. Moving along a frontier corresponds to
configurations for which improved convergence came at the cost of
greater wall time.}

\end{figure}%

\end{minipage}%
\newline
\begin{minipage}{\linewidth}

\begin{figure}[H]

\centering{

\includegraphics[width=0.9\linewidth,height=\textheight,keepaspectratio]{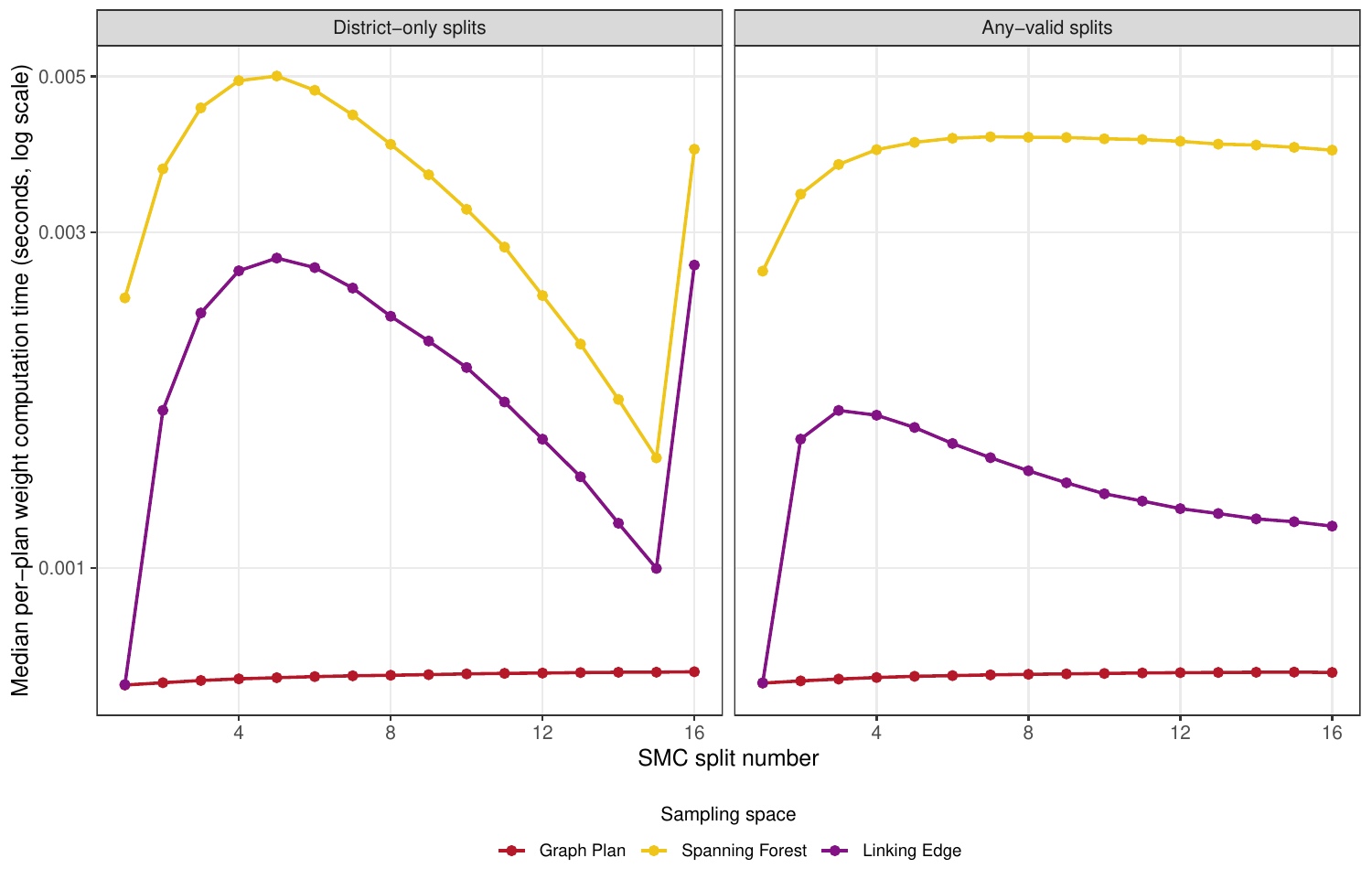}

}

\caption{\label{fig-pa-cd-weight-time}Average median per-plan weight
computation time across sampling spaces for a sample size of 5,000,
faceted by splitting schedule.}

\end{figure}%

\end{minipage}%

\end{figure}%

\begin{figure}[H]

\begin{minipage}{\linewidth}

\centering{

\includegraphics[width=0.8\linewidth,height=\textheight,keepaspectratio]{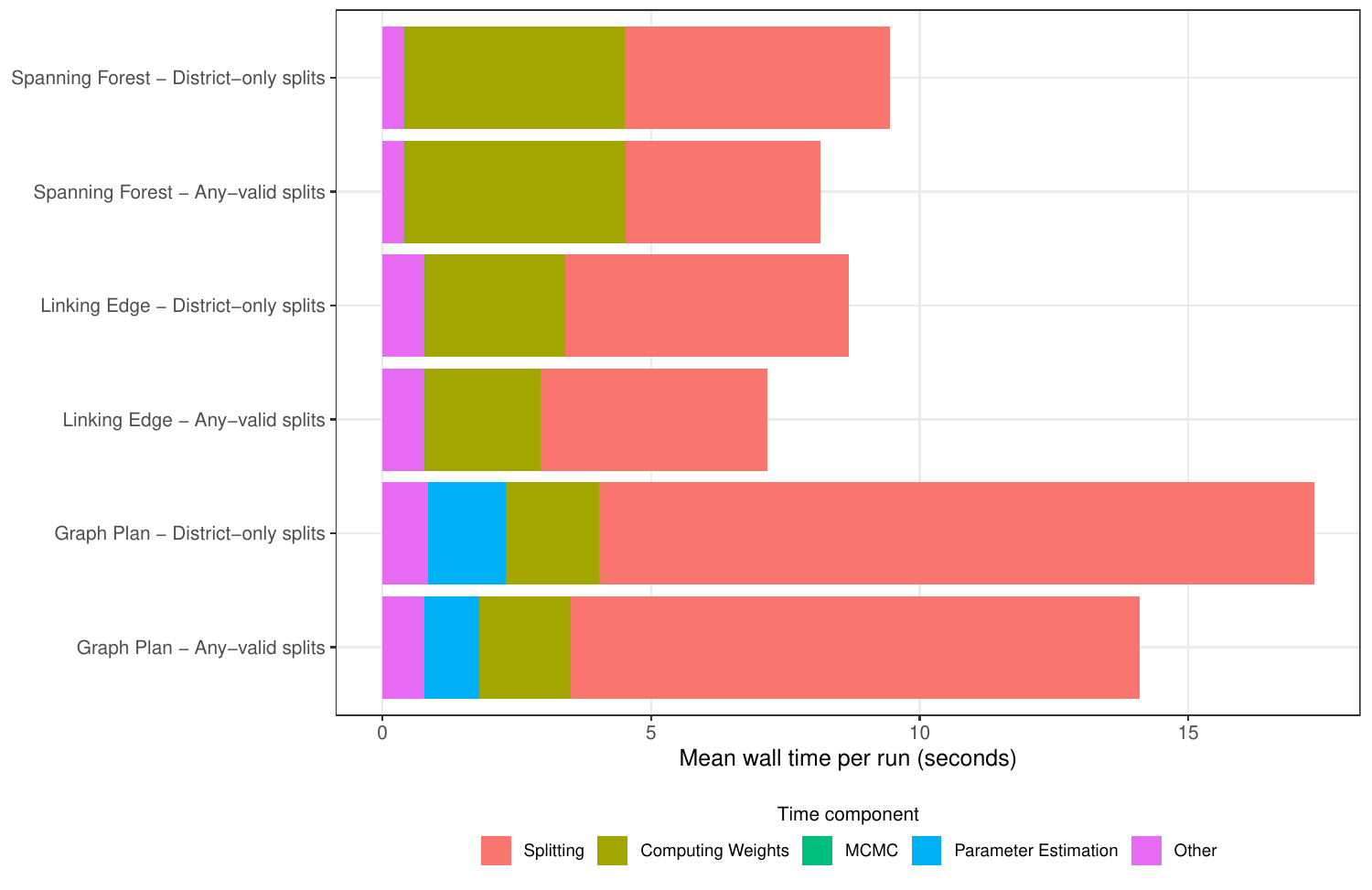}

}

\subcaption{\label{fig-pa-cd-gsmc-time}Breakdown of wall-time categories
across the three gSMC sampling spaces and two splitting schedules for a
sample size of 5,000.}

\end{minipage}%
\newline
\begin{minipage}{\linewidth}

\centering{

\pandocbounded{\includegraphics[keepaspectratio]{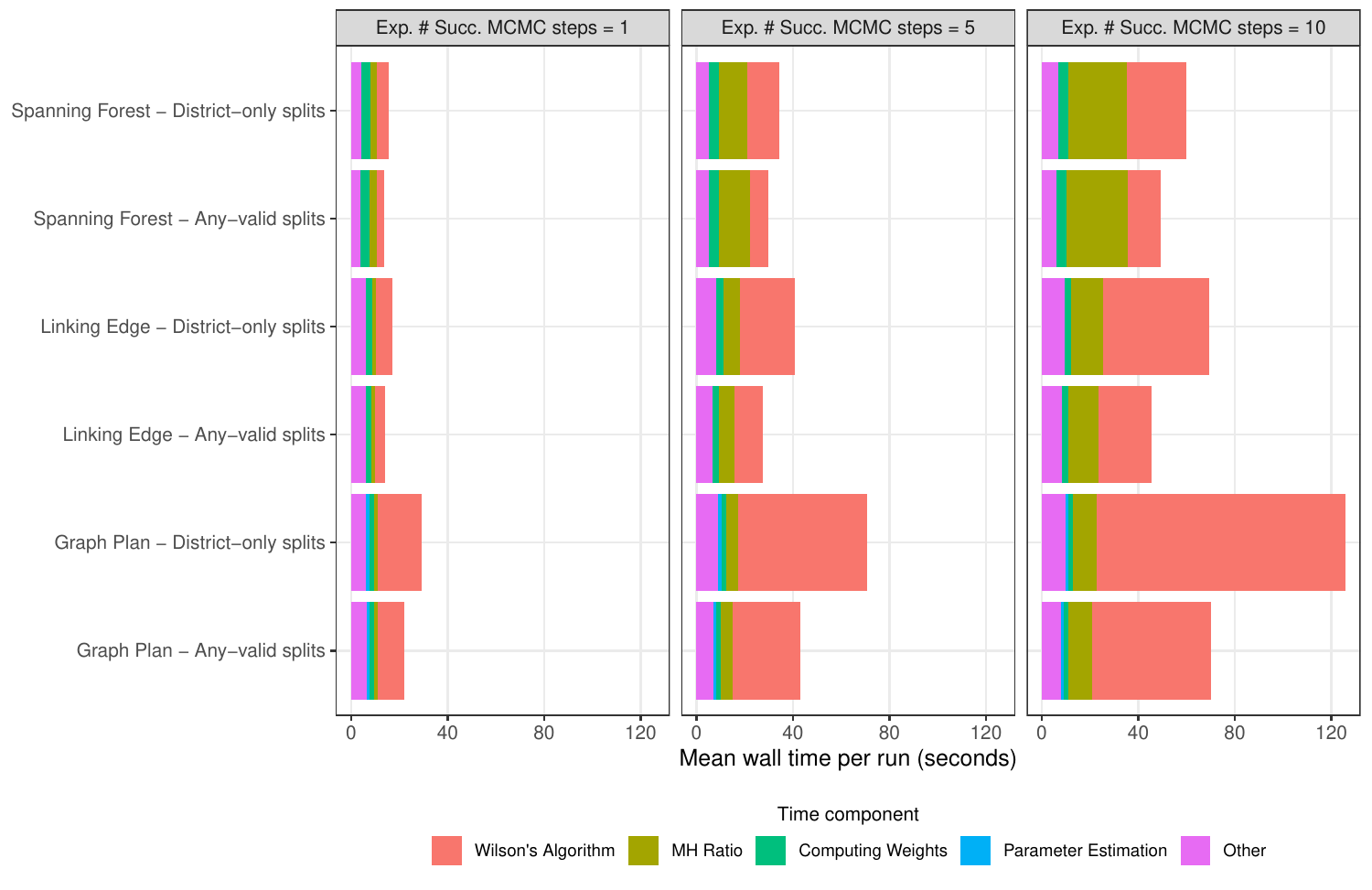}}

}

\subcaption{\label{fig-pa-cd-gsmc-mcmc-time}Breakdown of wall-time
categories across the three gSMC sampling spaces and two splitting
schedules, faceted by expected number of successful MCMC steps, for a
sample size of 5,000.}

\end{minipage}%

\caption{\label{fig-pa-cd-time-breakdown}Pennsylvania Congressional:
Breakdown of wall time for gSMC and gSMC-MCMC configurations at a sample
size of 5,000.}

\end{figure}%

\begin{figure}[H]

\begin{minipage}{\linewidth}

\centering{

\includegraphics[width=0.95\linewidth,height=\textheight,keepaspectratio]{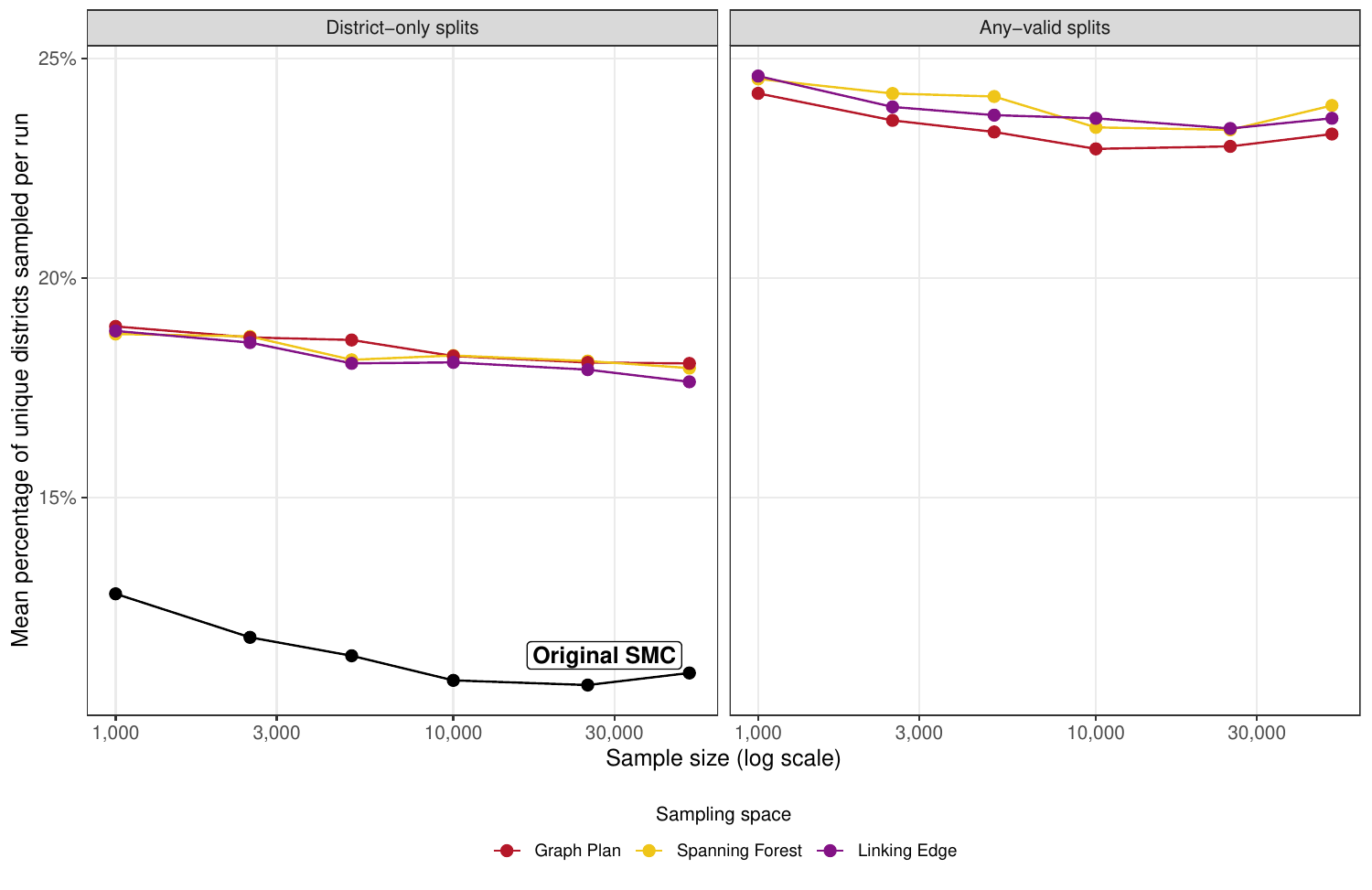}

}

\subcaption{\label{fig-pa-cd-gsmc-div}Pennsylvania Congressional: Mean
percentage of unique districts per run versus sample size for each of
the three gSMC sampling spaces, faceted by splitting schedule.}

\end{minipage}%
\newline
\begin{minipage}{\linewidth}

\centering{

\includegraphics[width=0.95\linewidth,height=\textheight,keepaspectratio]{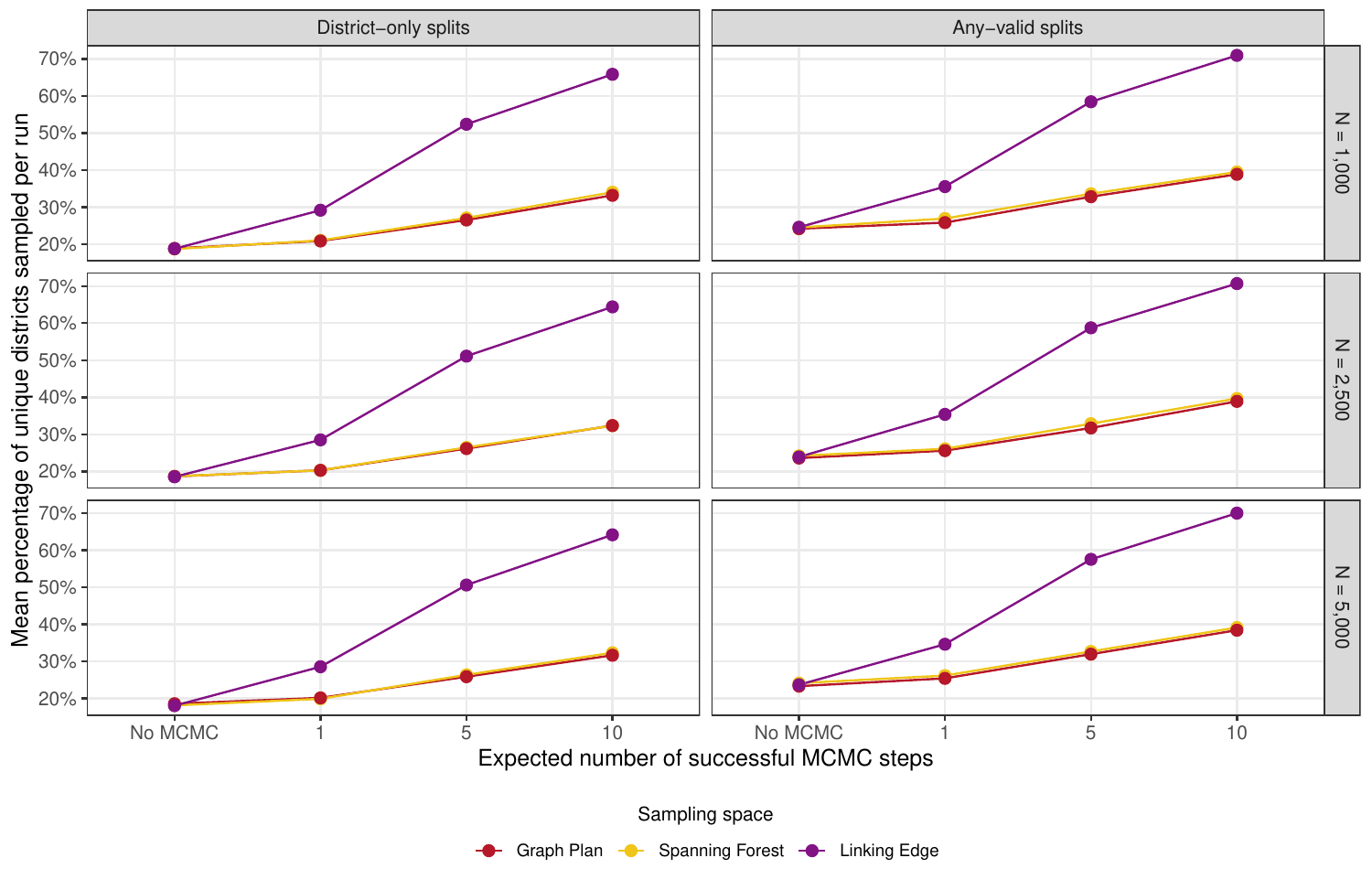}

}

\subcaption{\label{fig-pa-cd-gsmc-mcmc-div}Pennsylvania Congressional:
Mean percentage of unique districts per run across all expected numbers
of successful MCMC steps for each of the three gSMC sampling spaces,
faceted by splitting schedule and sample size.}

\end{minipage}%

\caption{\label{fig-pa-cd-div}Pennsylvania Congressional: Mean
percentage of unique districts per run across all gSMC and gSMC-MCMC
configurations run.}

\end{figure}%

\subsubsection{Illinois 2020 Congressional Map}\label{sec-il-cd-2020}

We use a map with \(V=10,084\) precincts, \(E = 28,219\) edges, and a
median population of 1148.5 per precinct. We use the provided
pseudo-counties from the \texttt{alarmdata} map as the units for
hierarchical sampling. No additional constraints are imposed through the
\(J\) function. We sample both gSMC and gSMC-MCMC at sample sizes of
1,000, 2,500, and 5,000. We also sample gSMC at sample sizes of 10,000,
25,000 and 50,000. The expected number of successful MCMC steps used
were: 1, 5, and 10. As Figure~\ref{fig-il-cd-gsmc-rhats} demonstrates,
linking edge space sampling with any-valid splits continues to perform
exceptionally well on this map, just like with Pennsylvania.

\begin{figure}[H]

\centering{

\pandocbounded{\includegraphics[keepaspectratio]{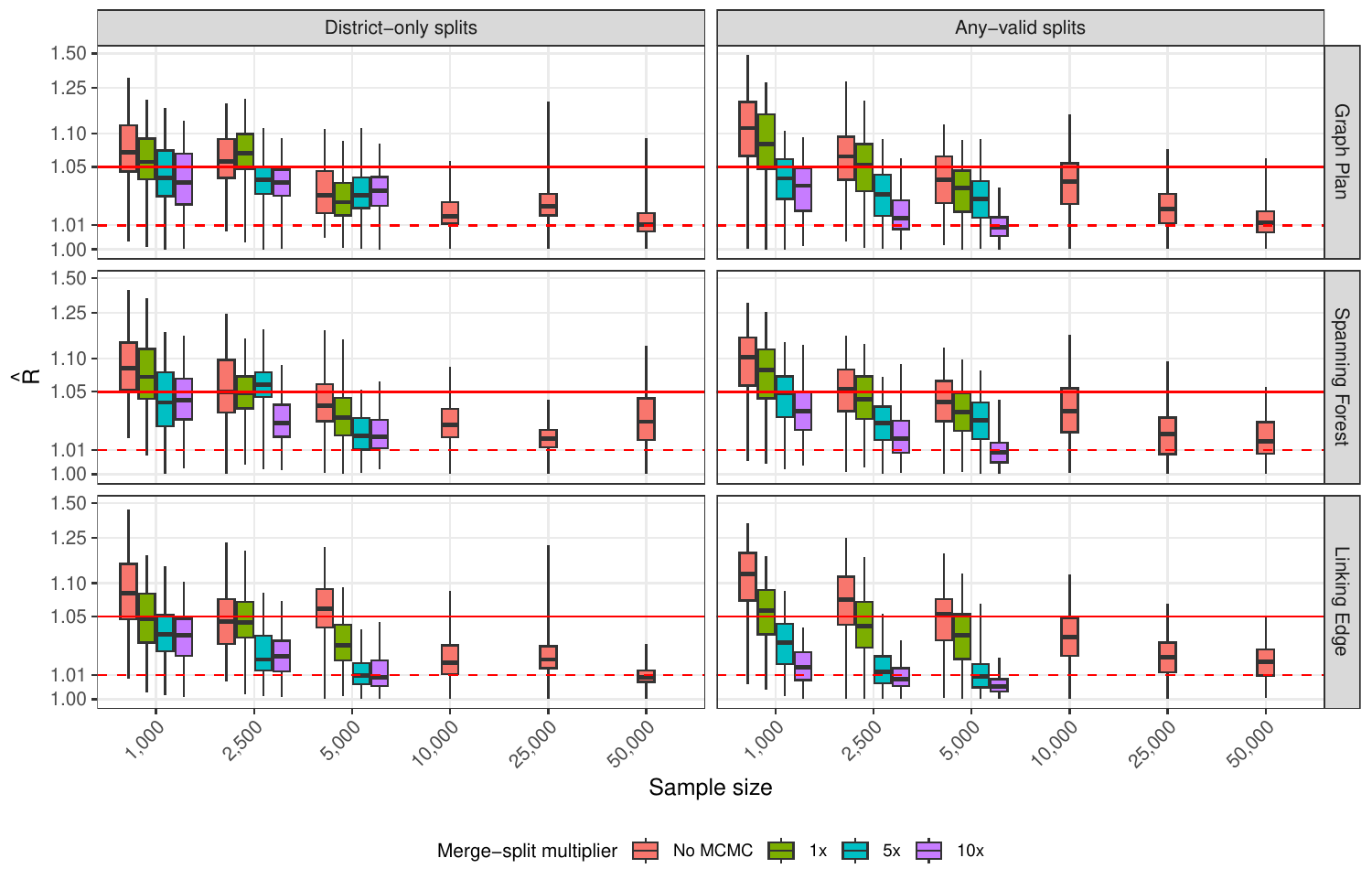}}

}

\caption{\label{fig-il-cd-gsmc-rhats}Illinois Congressional:
Distribution of \(\hat{R}\) values for all gSMC and gSMC-MCMC
configurations run, faceted by sampling space and splitting schedule.}

\end{figure}%

\begin{figure}[H]

\begin{minipage}{\linewidth}

\begin{figure}[H]

\centering{

\includegraphics[width=0.9\linewidth,height=\textheight,keepaspectratio]{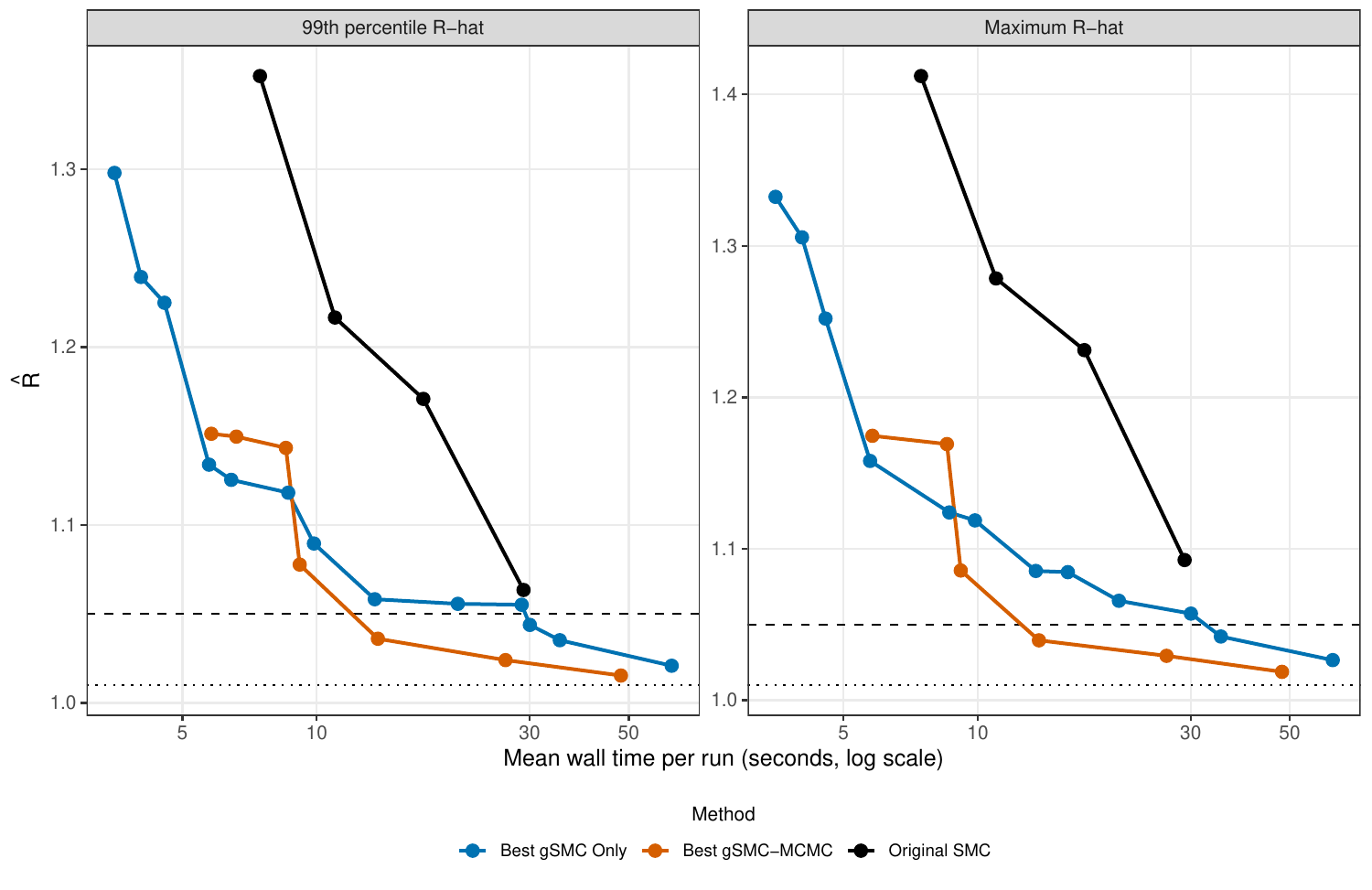}

}

\caption{\label{fig-il-cd-pareto-frontier}Illinois Congressional:
Empirical Pareto frontier comparing mean wall time and convergence of
the best-performing gSMC and gSMC-MCMC configurations and Original SMC.
Each point represents an algorithm configuration, with mean wall time
per run on the horizontal axis and either the 99th percentile of the
resulting \(\hat{R}\) values (left) or the maximum \(\hat{R}\) value
(right) on the vertical axis. Moving along a frontier corresponds to
configurations for which improved convergence came at the cost of
greater wall time.}

\end{figure}%

\end{minipage}%
\newline
\begin{minipage}{\linewidth}

\begin{figure}[H]

\centering{

\includegraphics[width=0.9\linewidth,height=\textheight,keepaspectratio]{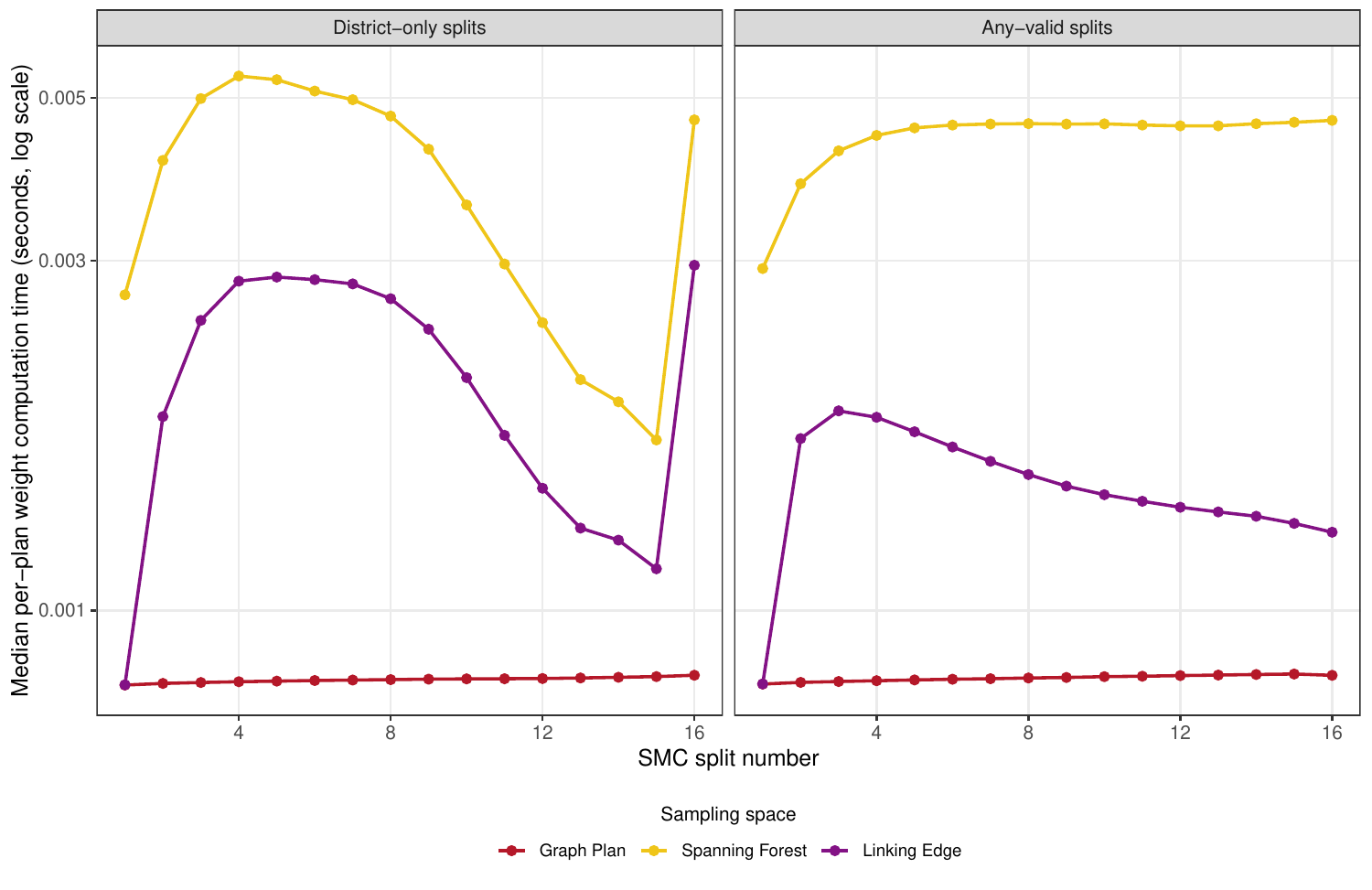}

}

\caption{\label{fig-il-cd-weight-time}Average median per-plan weight
computation time across sampling spaces for a sample size of 5,000,
faceted by splitting schedule.}

\end{figure}%

\end{minipage}%

\end{figure}%

\begin{figure}[H]

\begin{minipage}{\linewidth}

\centering{

\includegraphics[width=0.8\linewidth,height=\textheight,keepaspectratio]{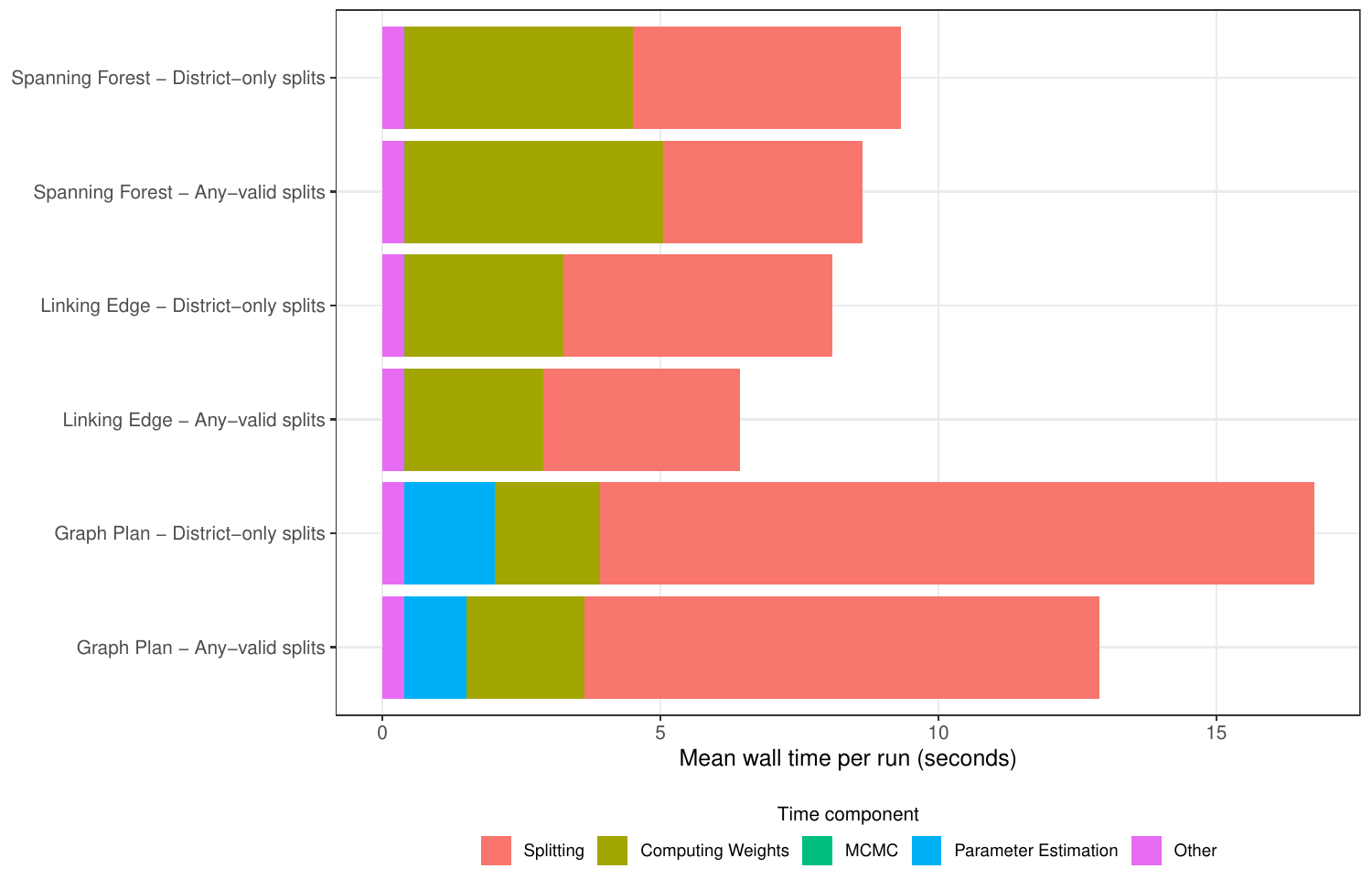}

}

\subcaption{\label{fig-il-cd-gsmc-time}Breakdown of wall-time categories
across the three gSMC sampling spaces and two splitting schedules for a
sample size of 5,000.}

\end{minipage}%
\newline
\begin{minipage}{\linewidth}

\centering{

\pandocbounded{\includegraphics[keepaspectratio]{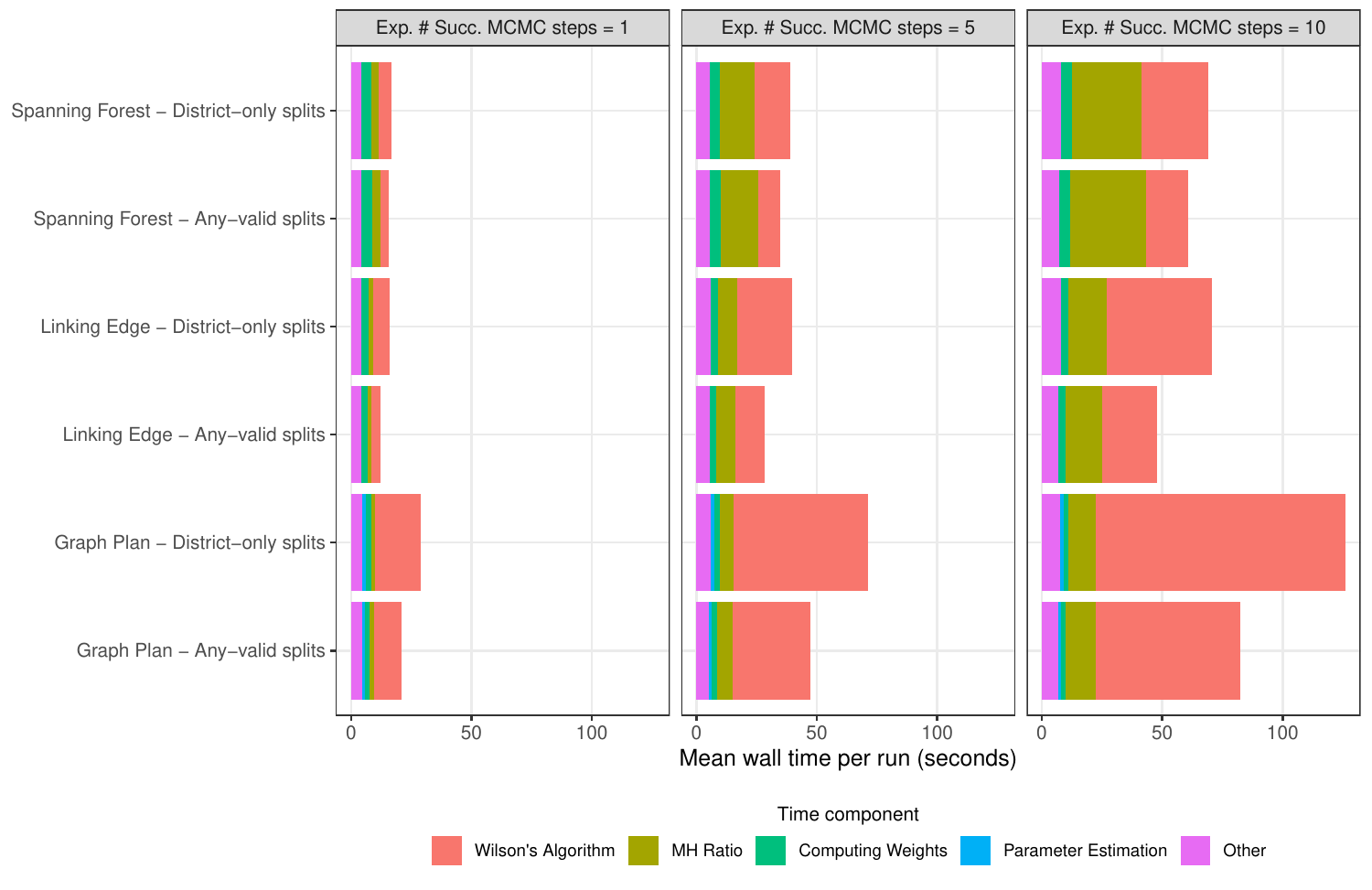}}

}

\subcaption{\label{fig-il-cd-gsmc-mcmc-time}Breakdown of wall-time
categories across the three gSMC sampling spaces and two splitting
schedules, faceted by expected number of successful MCMC steps, for a
sample size of 5,000.}

\end{minipage}%

\caption{\label{fig-il-cd-time-breakdown}Illinois Congressional:
Breakdown of wall time for gSMC and gSMC-MCMC configurations at a sample
size of 5,000.}

\end{figure}%

\begin{figure}[H]

\begin{minipage}{\linewidth}

\centering{

\includegraphics[width=0.95\linewidth,height=\textheight,keepaspectratio]{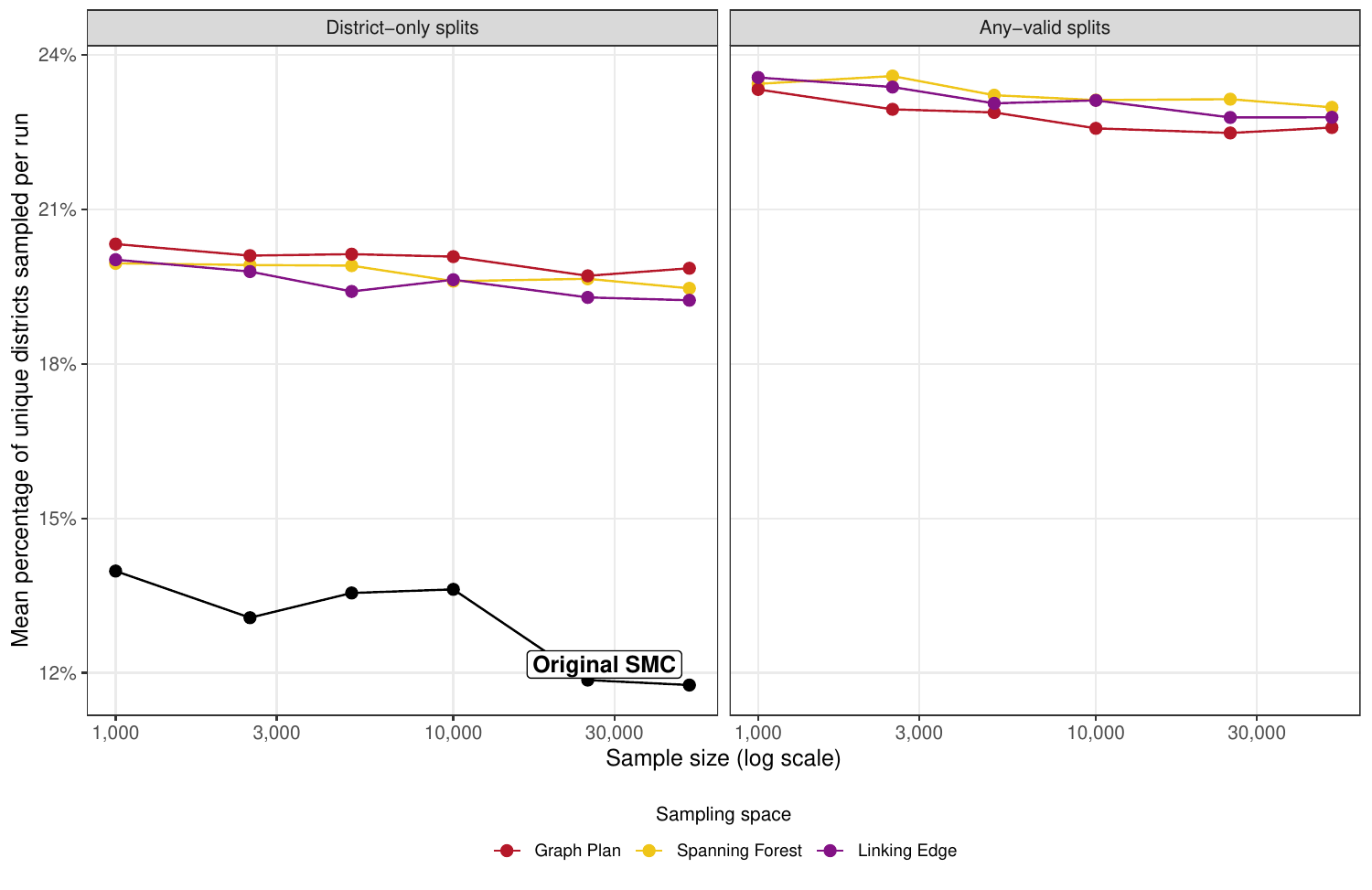}

}

\subcaption{\label{fig-il-cd-gsmc-div}Illinois Congressional: Mean
percentage of unique districts per run versus sample size for each of
the three gSMC sampling spaces, faceted by splitting schedule.}

\end{minipage}%
\newline
\begin{minipage}{\linewidth}

\centering{

\includegraphics[width=0.95\linewidth,height=\textheight,keepaspectratio]{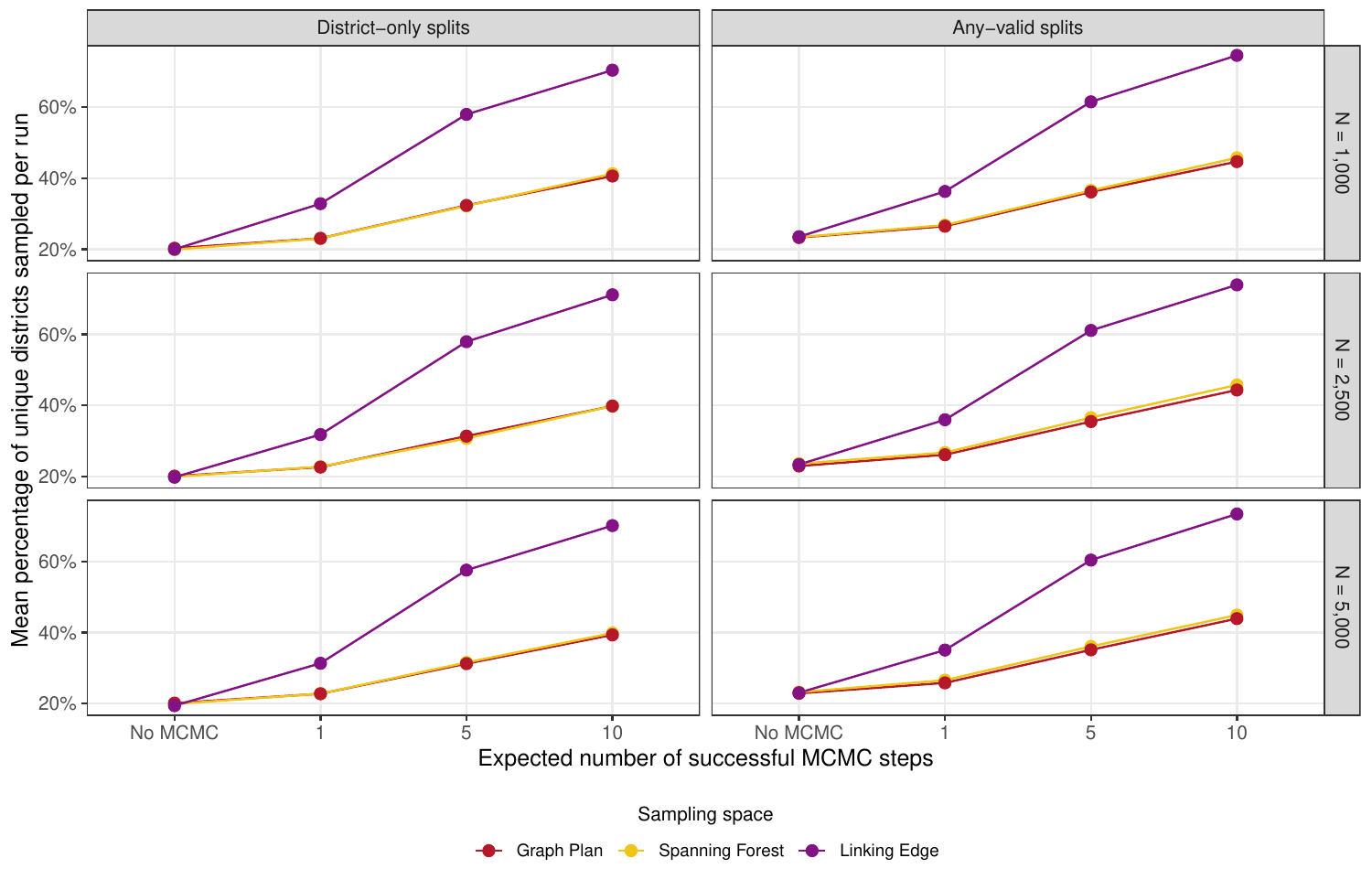}

}

\subcaption{\label{fig-il-cd-gsmc-mcmc-div}Illinois Congressional: Mean
percentage of unique districts per run across all expected numbers of
successful MCMC steps for each of the three gSMC sampling spaces,
faceted by splitting schedule and sample size.}

\end{minipage}%

\caption{\label{fig-il-cd-div}Illinois Congressional: Mean percentage of
unique districts per run across all gSMC and gSMC-MCMC configurations
run.}

\end{figure}%

\subsubsection{New York 2020 Congressional Map}\label{sec-ny-cd-2020}

We use a map with \(V=14,191\) precincts, \(E = 39,089\) edges, and a
median population of 1,301 per precinct. This is the largest map graph
tested. We use the provided pseudo-counties from the \texttt{alarmdata}
map as the units for hierarchical sampling. No additional constraints
are imposed through the \(J\) function. We sample both gSMC and
gSMC-MCMC at sample sizes of 1,000, 5,000, and 10,000. We also sample
gSMC at sample sizes of 20,000, and 50,000. The expected number of
successful MCMC steps used were: 1, 10, and 20. As
Figure~\ref{fig-ny-cd-gsmc-rhats} demonstrates, gSMC alone struggles to
fully converge on this map, even with a sample size of 50,000. Indeed,
this is the last example test where gSMC alone converges.

\begin{figure}[H]

\centering{

\pandocbounded{\includegraphics[keepaspectratio]{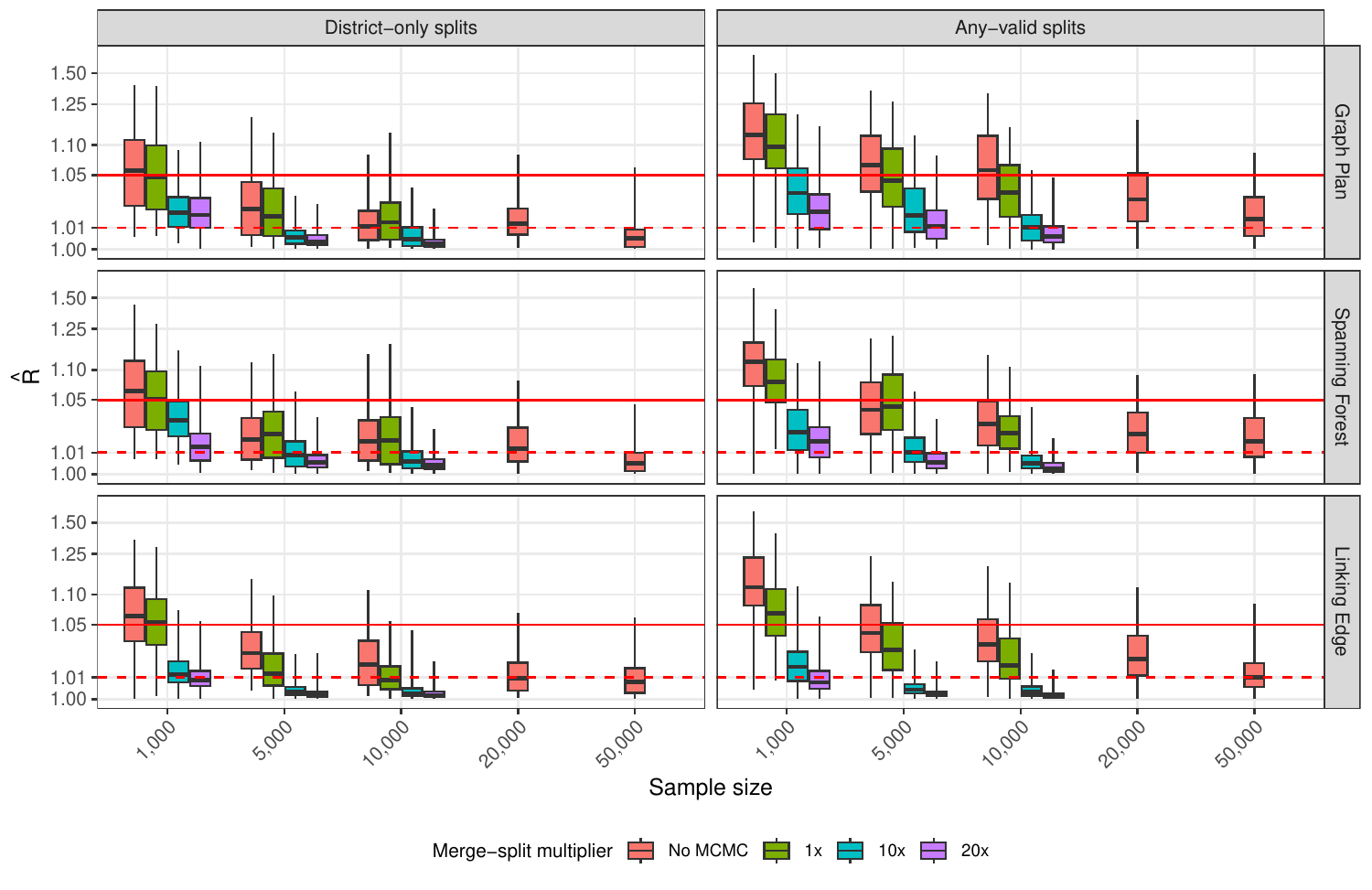}}

}

\caption{\label{fig-ny-cd-gsmc-rhats}New York Congressional:
Distribution of \(\hat{R}\) values for all gSMC and gSMC-MCMC
configurations run, faceted by sampling space and splitting schedule.}

\end{figure}%

\begin{figure}[H]

\begin{minipage}{\linewidth}

\begin{figure}[H]

\centering{

\includegraphics[width=0.9\linewidth,height=\textheight,keepaspectratio]{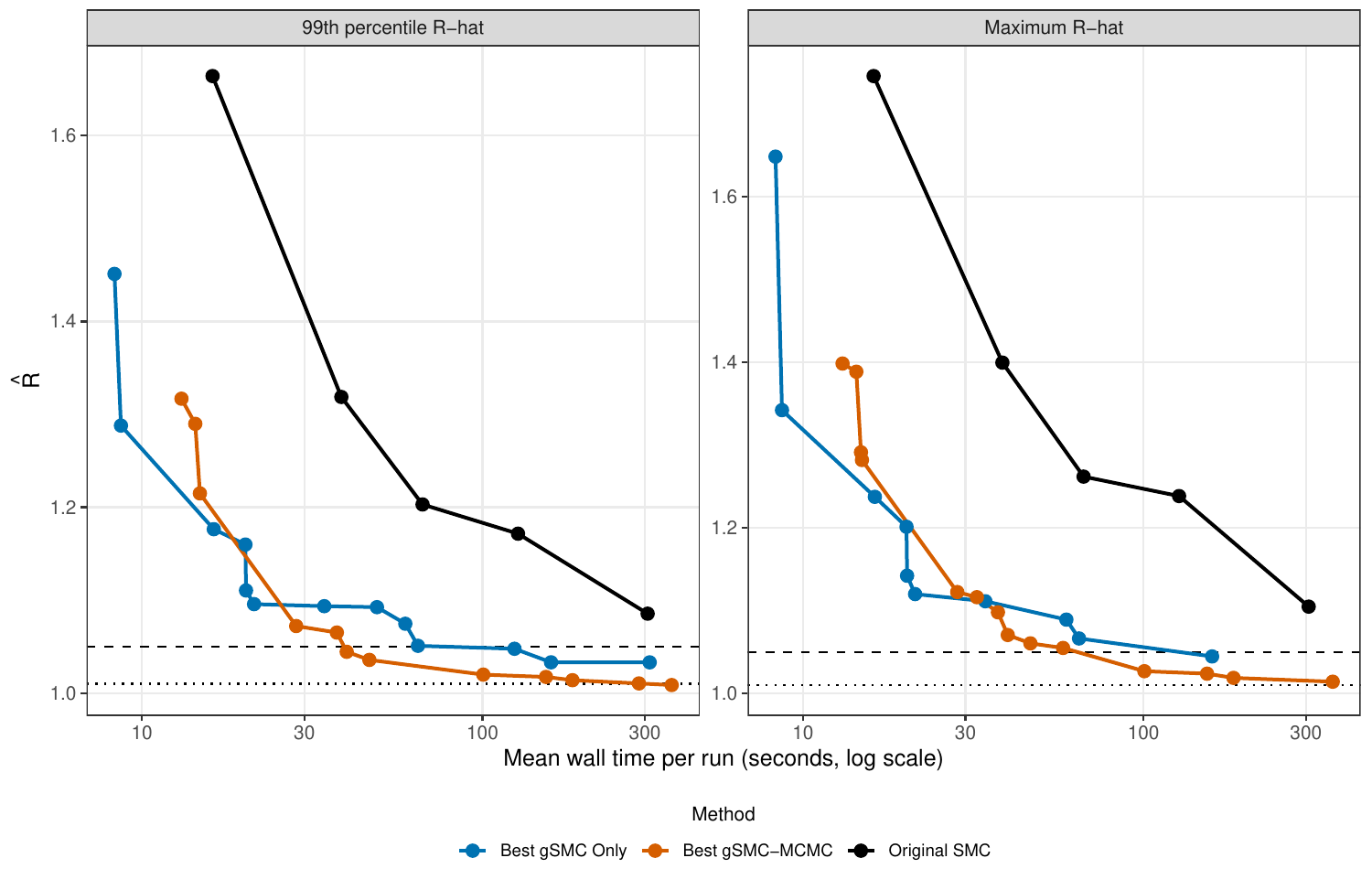}

}

\caption{\label{fig-ny-cd-pareto-frontier}New York Congressional:
Empirical Pareto frontier comparing mean wall time and convergence of
the best-performing gSMC and gSMC-MCMC configurations and Original SMC.
Each point represents an algorithm configuration, with mean wall time
per run on the horizontal axis and either the 99th percentile of the
resulting \(\hat{R}\) values (left) or the maximum \(\hat{R}\) value
(right) on the vertical axis. Moving along a frontier corresponds to
configurations for which improved convergence came at the cost of
greater wall time.}

\end{figure}%

\end{minipage}%
\newline
\begin{minipage}{\linewidth}

\begin{figure}[H]

\centering{

\includegraphics[width=0.9\linewidth,height=\textheight,keepaspectratio]{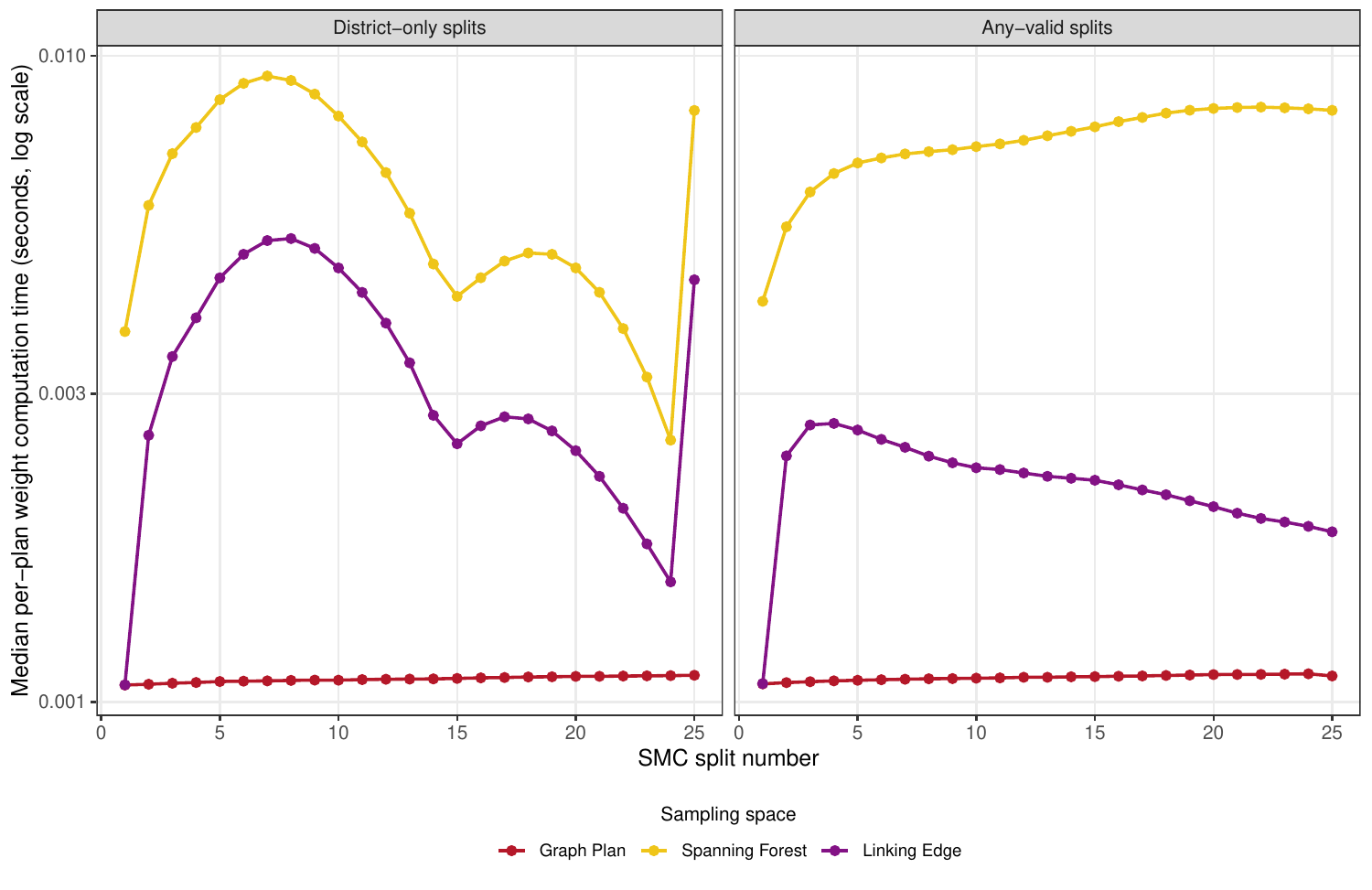}

}

\caption{\label{fig-ny-cd-weight-time}Average median per-plan weight
computation time across sampling spaces for a sample size of 10,000,
faceted by splitting schedule.}

\end{figure}%

\end{minipage}%

\end{figure}%

\begin{figure}[H]

\begin{minipage}{\linewidth}

\centering{

\includegraphics[width=0.8\linewidth,height=\textheight,keepaspectratio]{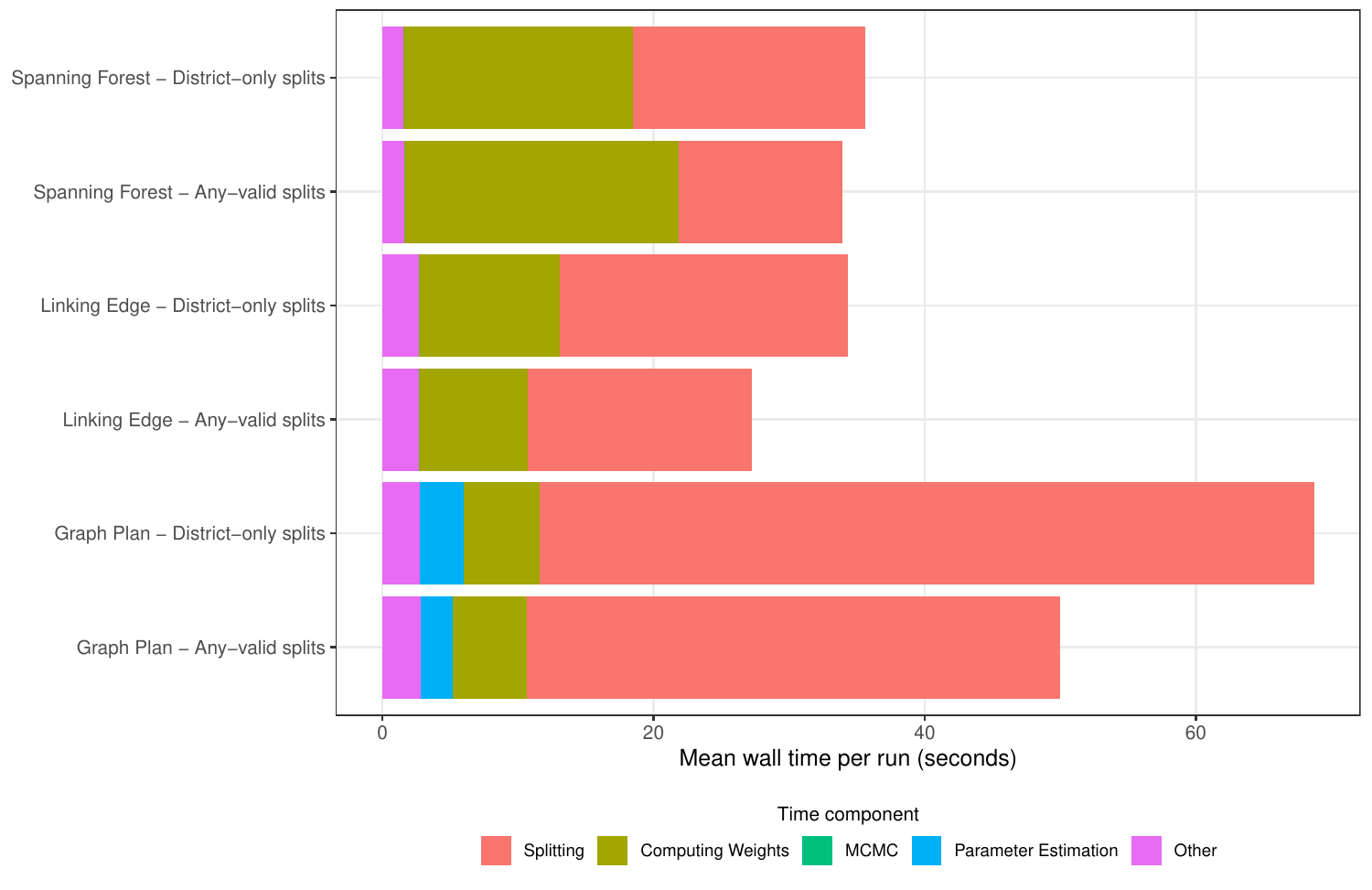}

}

\subcaption{\label{fig-ny-cd-gsmc-time}Breakdown of wall-time categories
across the three gSMC sampling spaces and two splitting schedules for a
sample size of 10,000.}

\end{minipage}%
\newline
\begin{minipage}{\linewidth}

\centering{

\pandocbounded{\includegraphics[keepaspectratio]{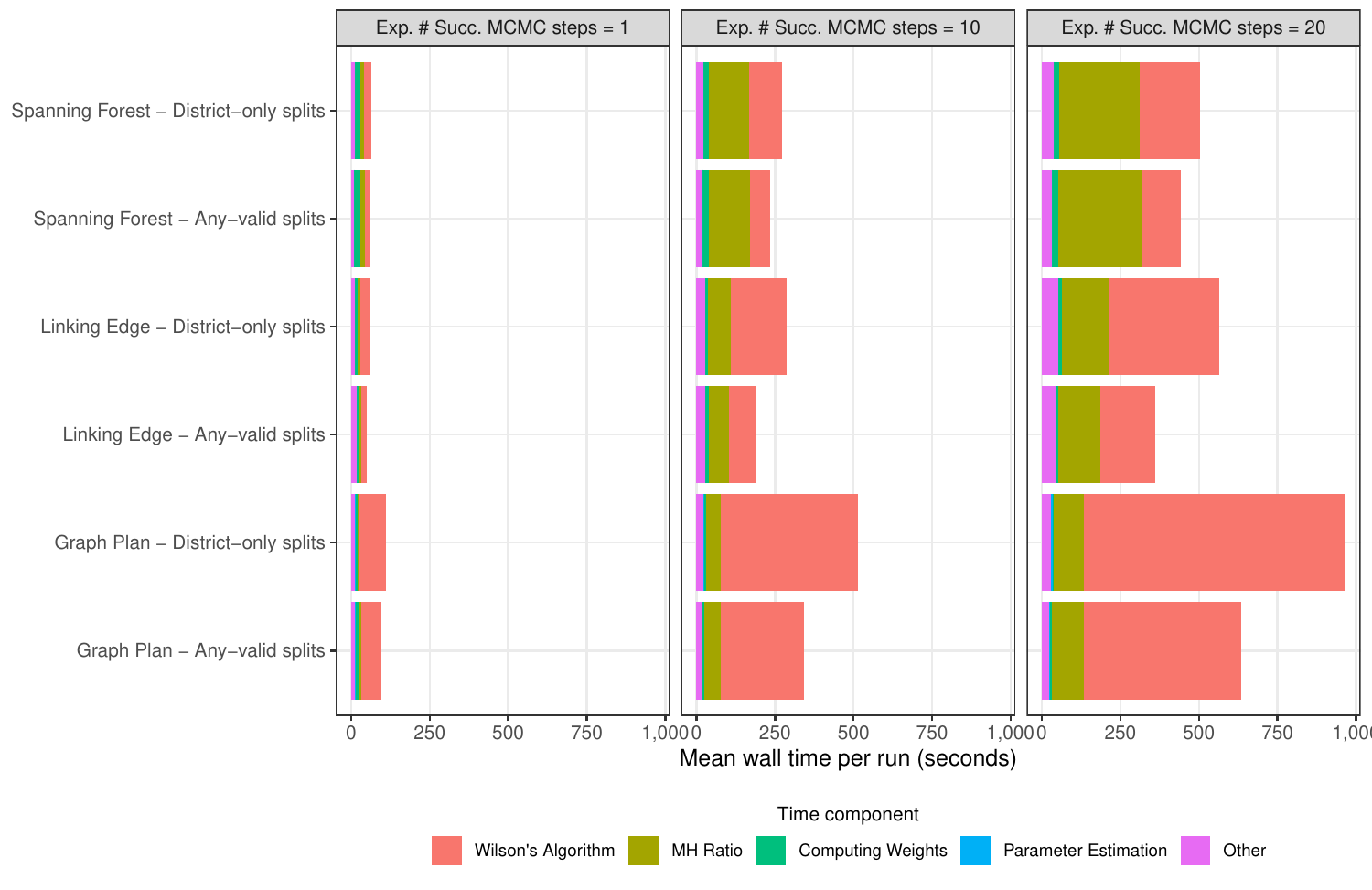}}

}

\subcaption{\label{fig-ny-cd-gsmc-mcmc-time}Breakdown of wall-time
categories across the three gSMC sampling spaces and two splitting
schedules, faceted by expected number of successful MCMC steps, for a
sample size of 10,000.}

\end{minipage}%

\caption{\label{fig-ny-cd-time-breakdown}New York Congressional:
Breakdown of wall time for gSMC and gSMC-MCMC configurations at a sample
size of 10,000.}

\end{figure}%

\begin{figure}[H]

\begin{minipage}{\linewidth}

\centering{

\includegraphics[width=0.95\linewidth,height=\textheight,keepaspectratio]{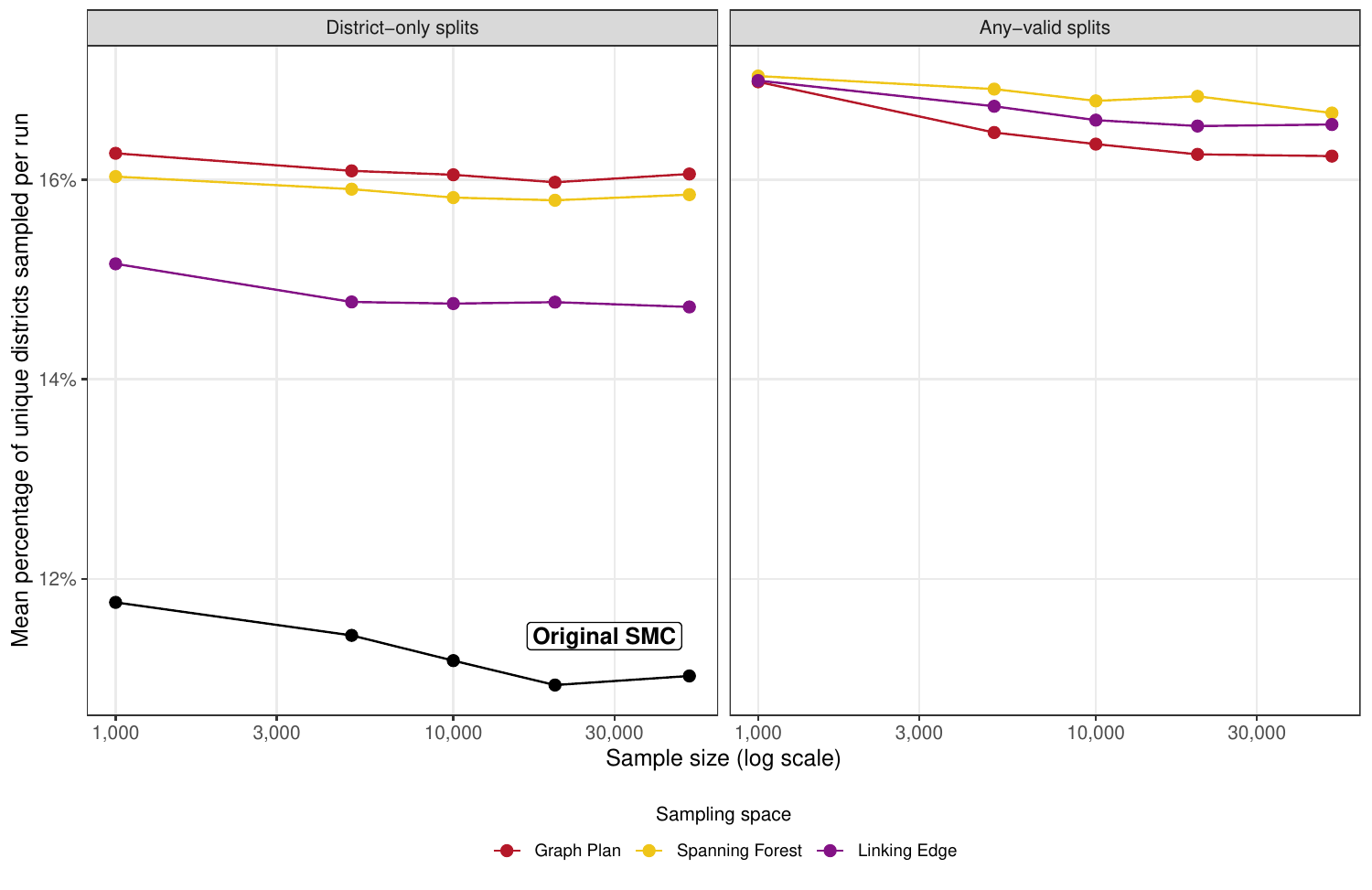}

}

\subcaption{\label{fig-ny-cd-gsmc-div}New York Congressional: Mean
percentage of unique districts per run versus sample size for each of
the three gSMC sampling spaces, faceted by splitting schedule.}

\end{minipage}%
\newline
\begin{minipage}{\linewidth}

\centering{

\includegraphics[width=0.95\linewidth,height=\textheight,keepaspectratio]{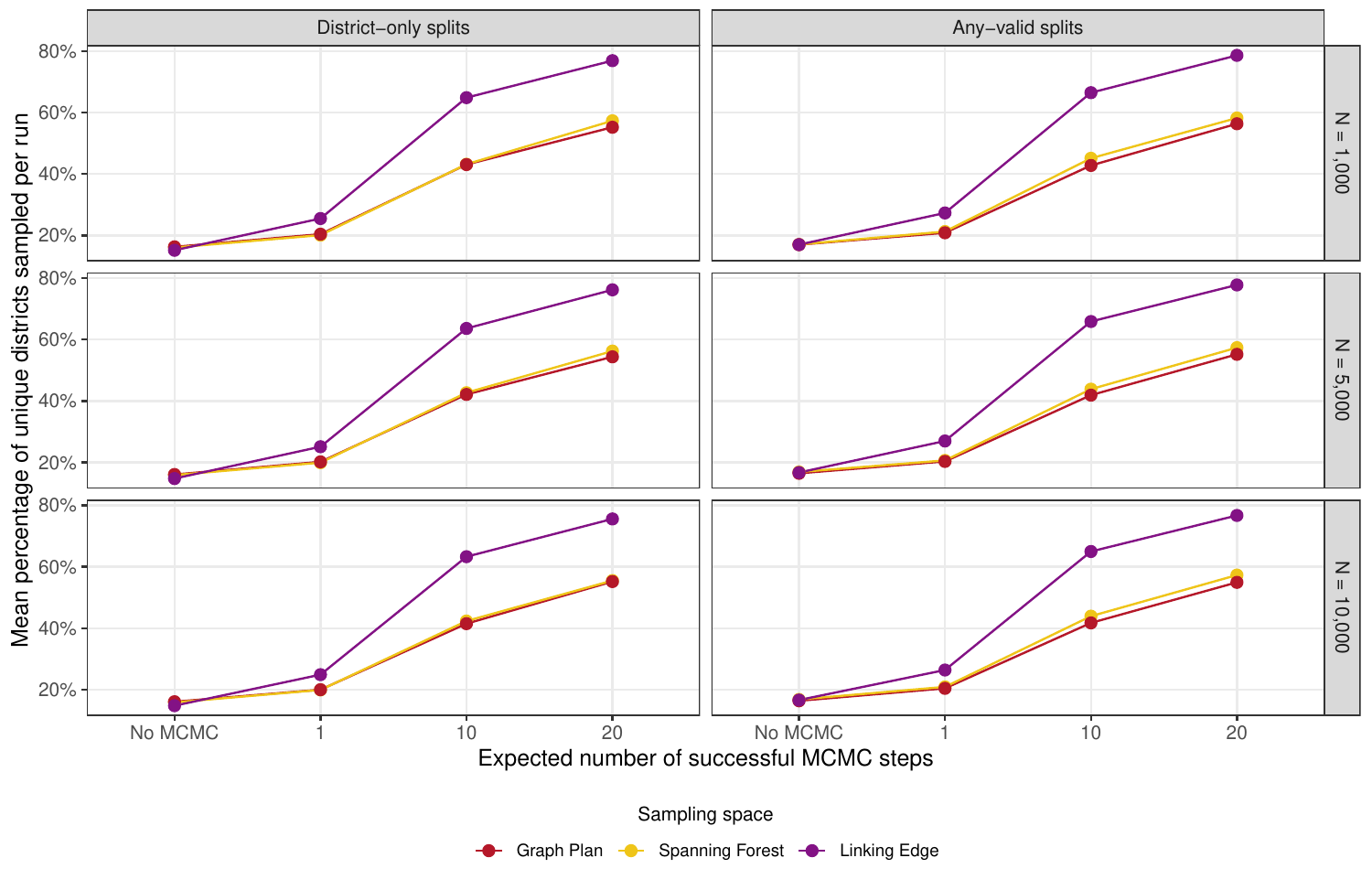}

}

\subcaption{\label{fig-ny-cd-gsmc-mcmc-div}New York Congressional: Mean
percentage of unique districts per run across all expected numbers of
successful MCMC steps for each of the three gSMC sampling spaces,
faceted by splitting schedule and sample size.}

\end{minipage}%

\caption{\label{fig-ny-cd-div}New York Congressional: Mean percentage of
unique districts per run across all gSMC and gSMC-MCMC configurations
run.}

\end{figure}%

\subsubsection{California 2020 Congressional Map}\label{sec-ca-cd-2020}

We use a map with \(V=9,129\) precincts and \(E = 24,977\) edges and a
median population of 4,190 per precinct. We use the provided
pseudo-counties from the \texttt{alarmdata} map as the units for
hierarchical sampling. No additional constraints are imposed through the
\(J\) function. We sample both gSMC and gSMC-MCMC at sample sizes of
2,500, 5,000, 10,000, and 20,000. The expected number of successful MCMC
steps used were: 1, 10, 20, 40, and 80. As
Figure~\ref{fig-ca-cd-gsmc-rhats} demonstrates, among the configurations
run, only gSMC-MCMC using linking edge space with any-valid splits
produced fully convergent samples for this map.

\begin{figure}[H]

\centering{

\pandocbounded{\includegraphics[keepaspectratio]{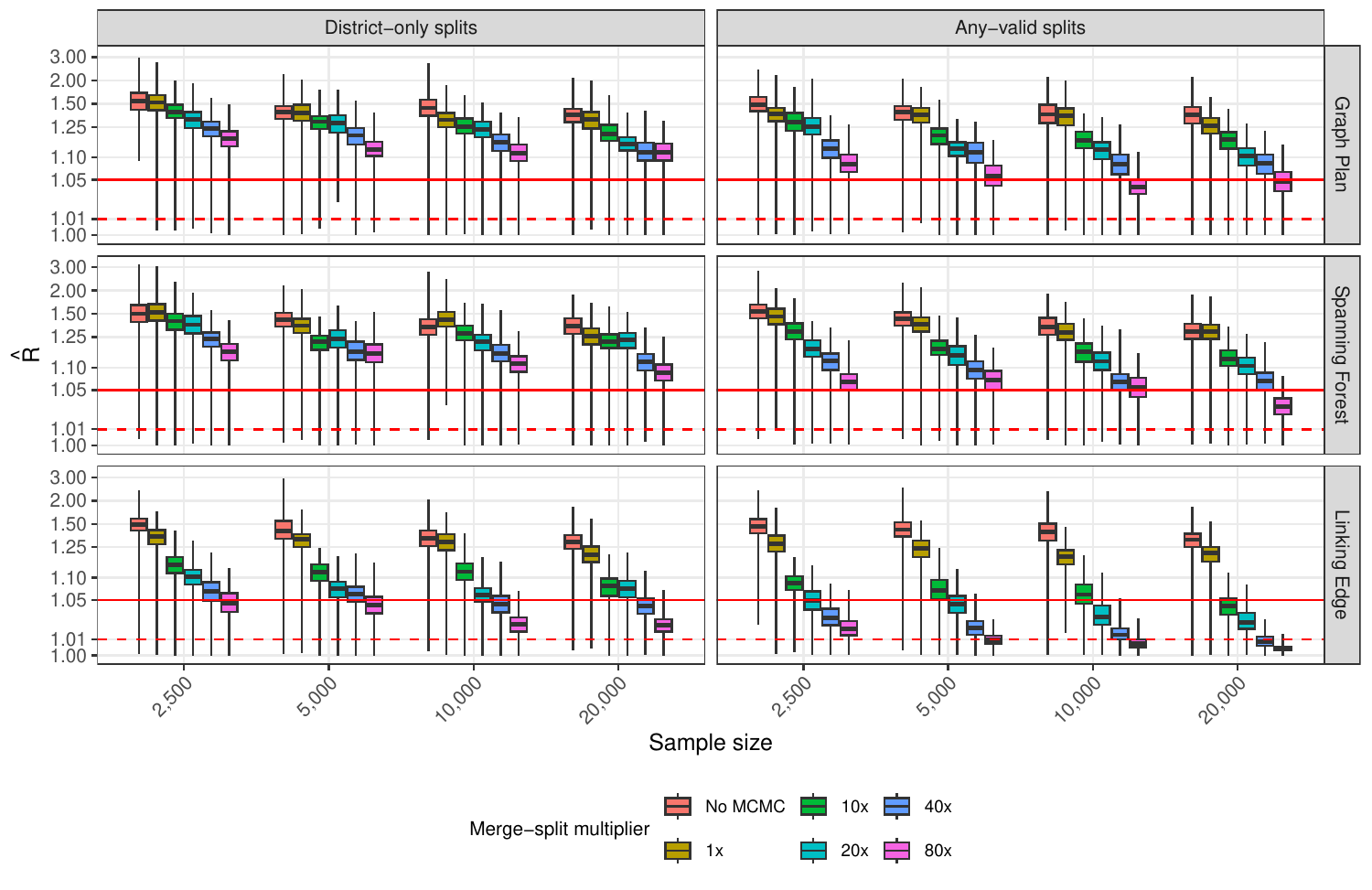}}

}

\caption{\label{fig-ca-cd-gsmc-rhats}California Congressional:
Distribution of \(\hat{R}\) values for all gSMC and gSMC-MCMC
configurations run, faceted by sampling space and splitting schedule.}

\end{figure}%

\begin{figure}[H]

\begin{minipage}{\linewidth}

\begin{figure}[H]

\centering{

\includegraphics[width=0.9\linewidth,height=\textheight,keepaspectratio]{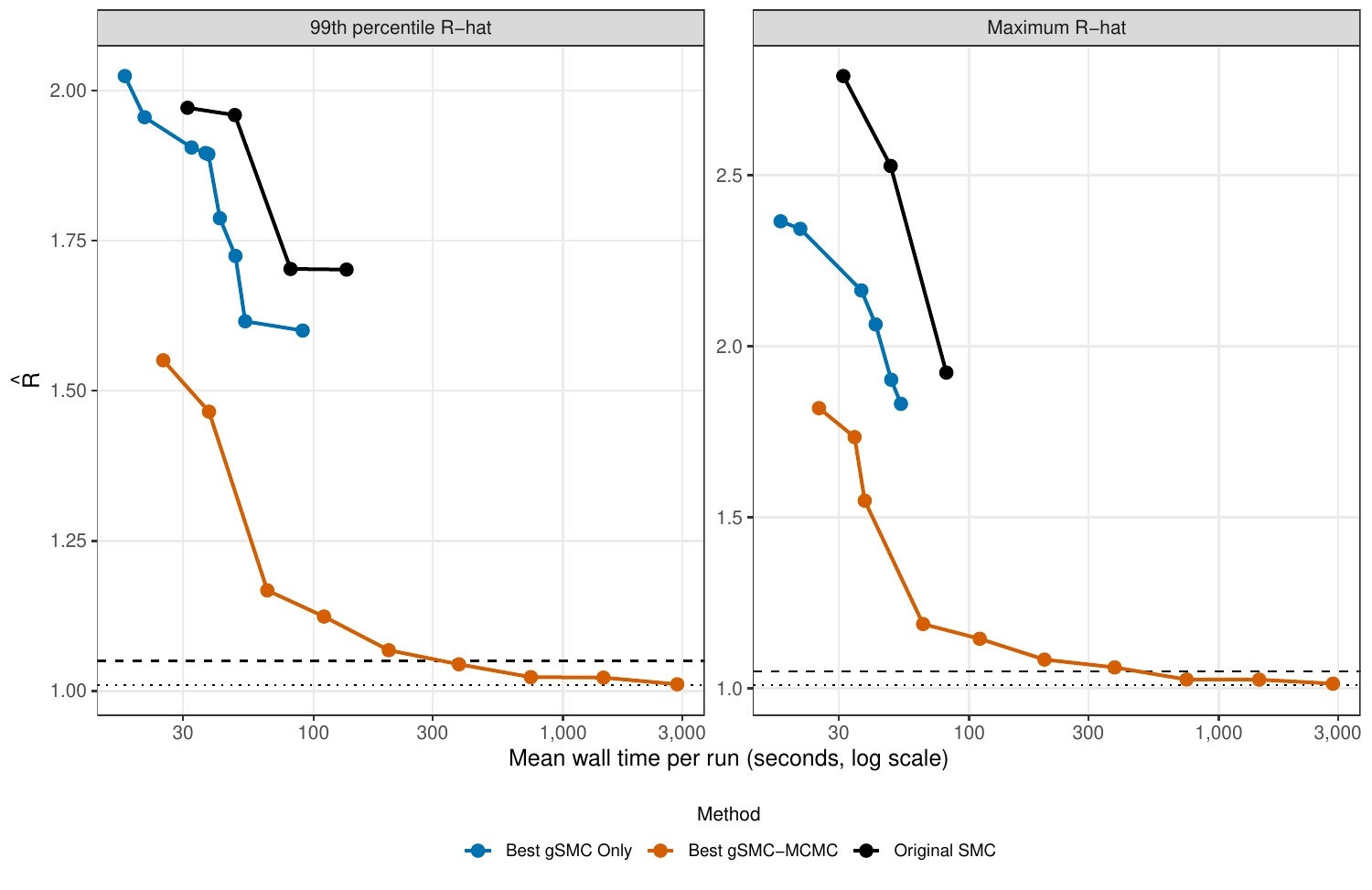}

}

\caption{\label{fig-ca-cd-pareto-frontier}California Congressional:
Empirical Pareto frontier comparing mean wall time and convergence of
the best-performing gSMC and gSMC-MCMC configurations and Original SMC.
Each point represents an algorithm configuration, with mean wall time
per run on the horizontal axis and either the 99th percentile of the
resulting \(\hat{R}\) values (left) or the maximum \(\hat{R}\) value
(right) on the vertical axis. Moving along a frontier corresponds to
configurations for which improved convergence came at the cost of
greater wall time.}

\end{figure}%

\end{minipage}%
\newline
\begin{minipage}{\linewidth}

\begin{figure}[H]

\centering{

\includegraphics[width=0.9\linewidth,height=\textheight,keepaspectratio]{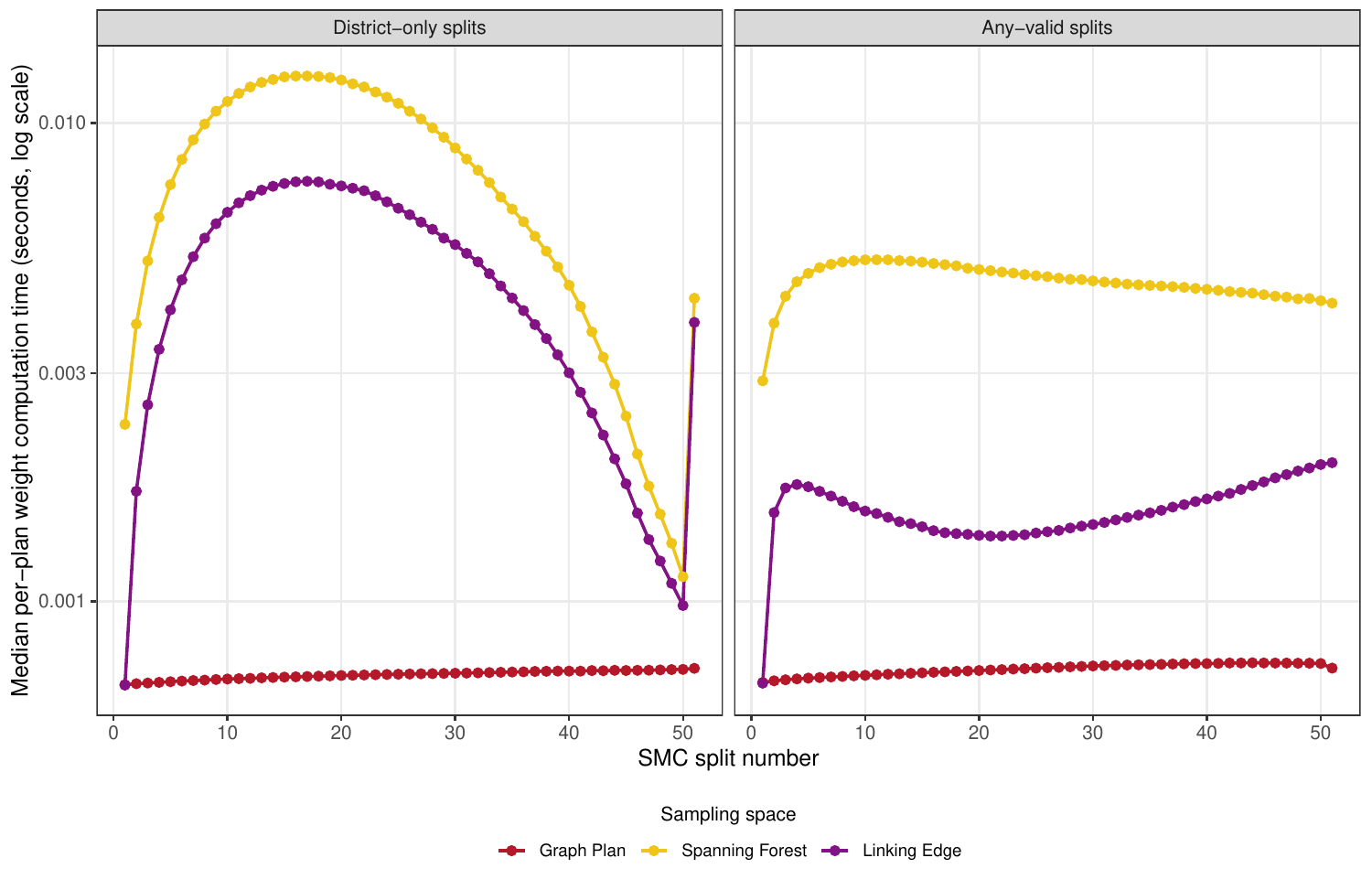}

}

\caption{\label{fig-ca-cd-weight-time}Average median per-plan weight
computation time across sampling spaces for a sample size of 20,000,
faceted by splitting schedule.}

\end{figure}%

\end{minipage}%

\end{figure}%

\begin{figure}[H]

\begin{minipage}{\linewidth}

\centering{

\includegraphics[width=0.8\linewidth,height=\textheight,keepaspectratio]{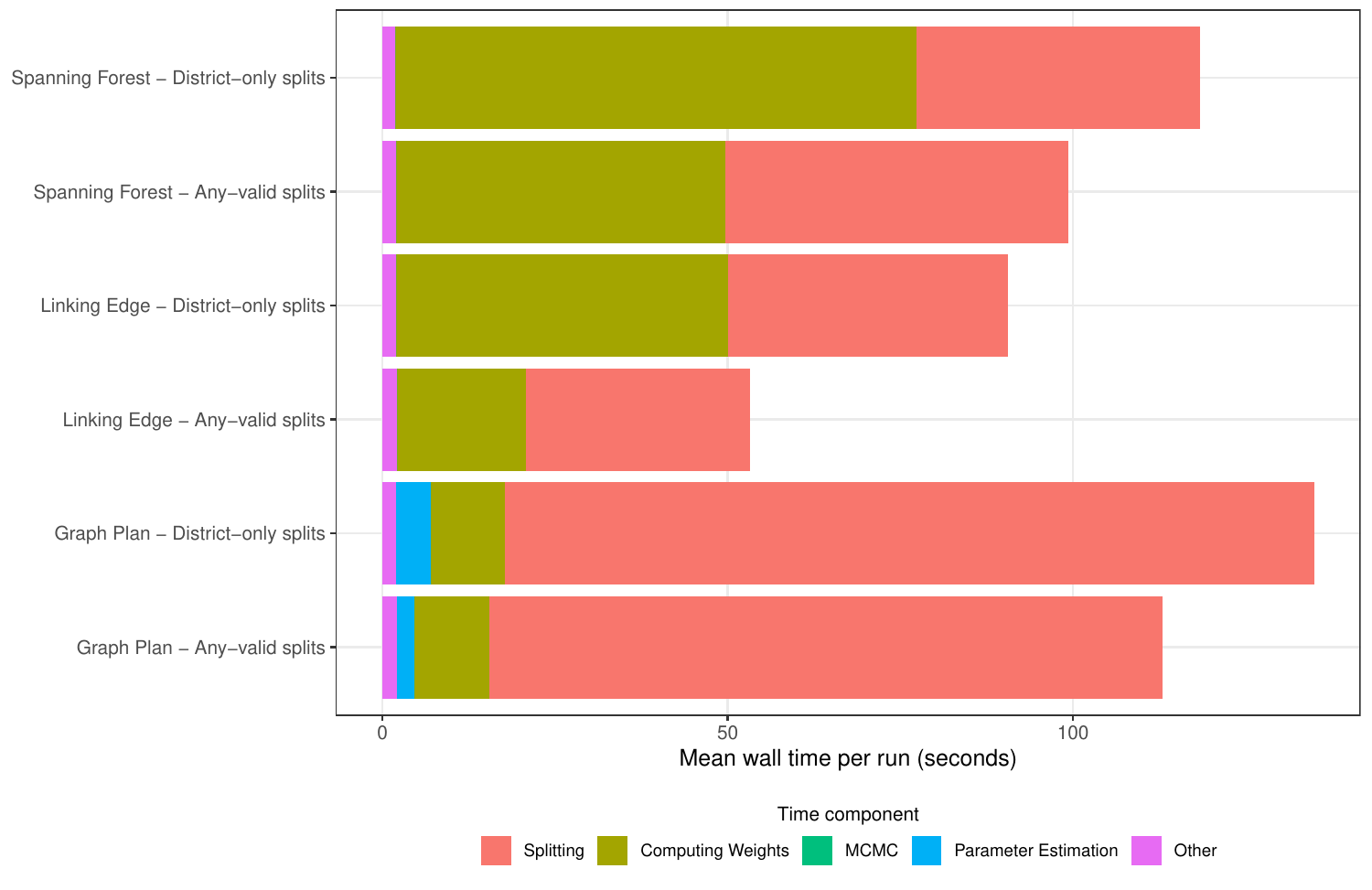}

}

\subcaption{\label{fig-ca-cd-gsmc-time}Breakdown of wall-time categories
across the three gSMC sampling spaces and two splitting schedules for a
sample size of 20,000.}

\end{minipage}%
\newline
\begin{minipage}{\linewidth}

\centering{

\pandocbounded{\includegraphics[keepaspectratio]{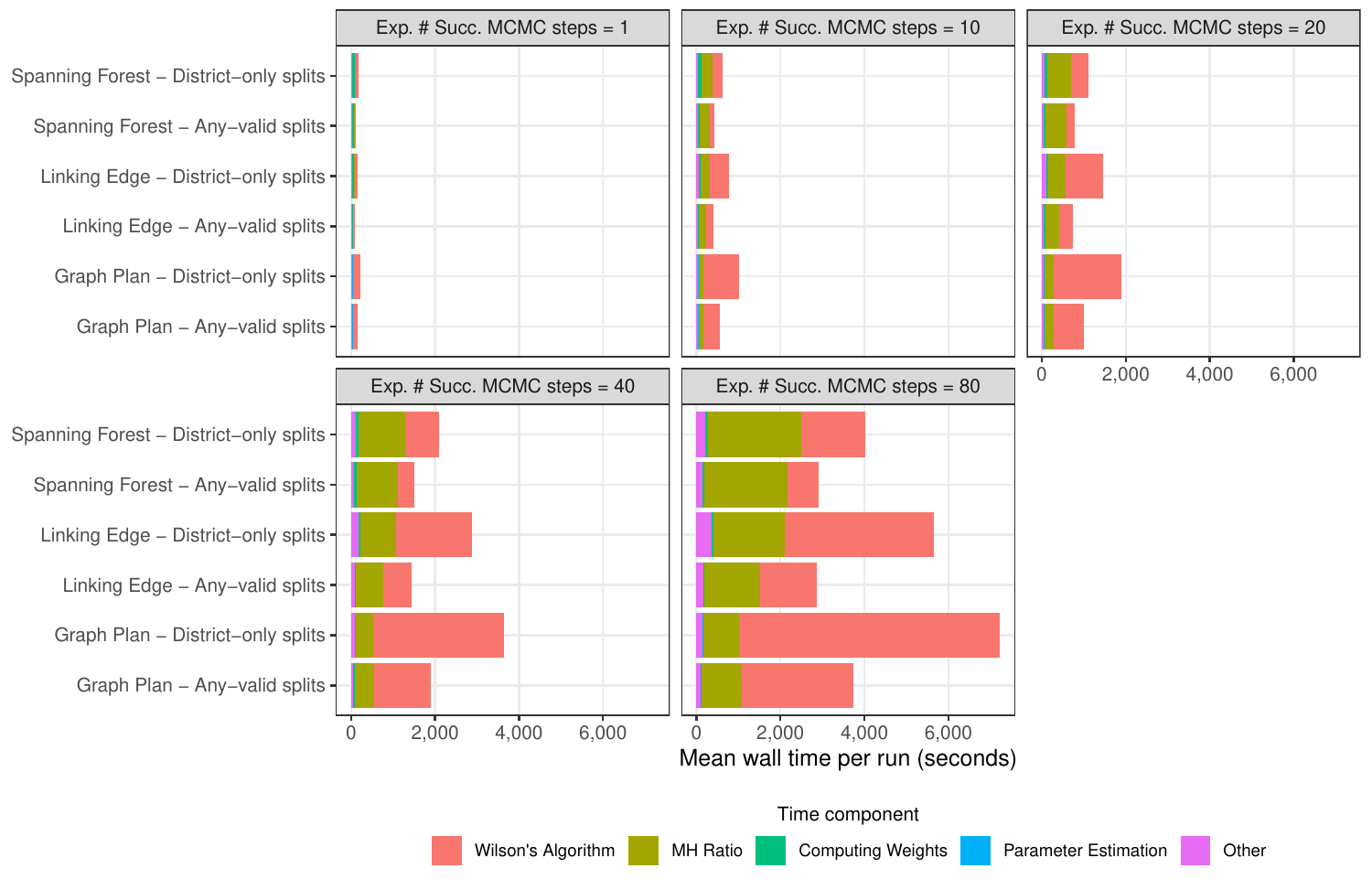}}

}

\subcaption{\label{fig-ca-cd-gsmc-mcmc-time}Breakdown of wall-time
categories across the three gSMC sampling spaces and two splitting
schedules, faceted by expected number of successful MCMC steps, for a
sample size of 20,000.}

\end{minipage}%

\caption{\label{fig-ca-cd-time-breakdown}California Congressional:
Breakdown of wall time for gSMC and gSMC-MCMC configurations at a sample
size of 20,000.}

\end{figure}%

\begin{figure}[H]

\begin{minipage}{\linewidth}

\centering{

\includegraphics[width=0.95\linewidth,height=\textheight,keepaspectratio]{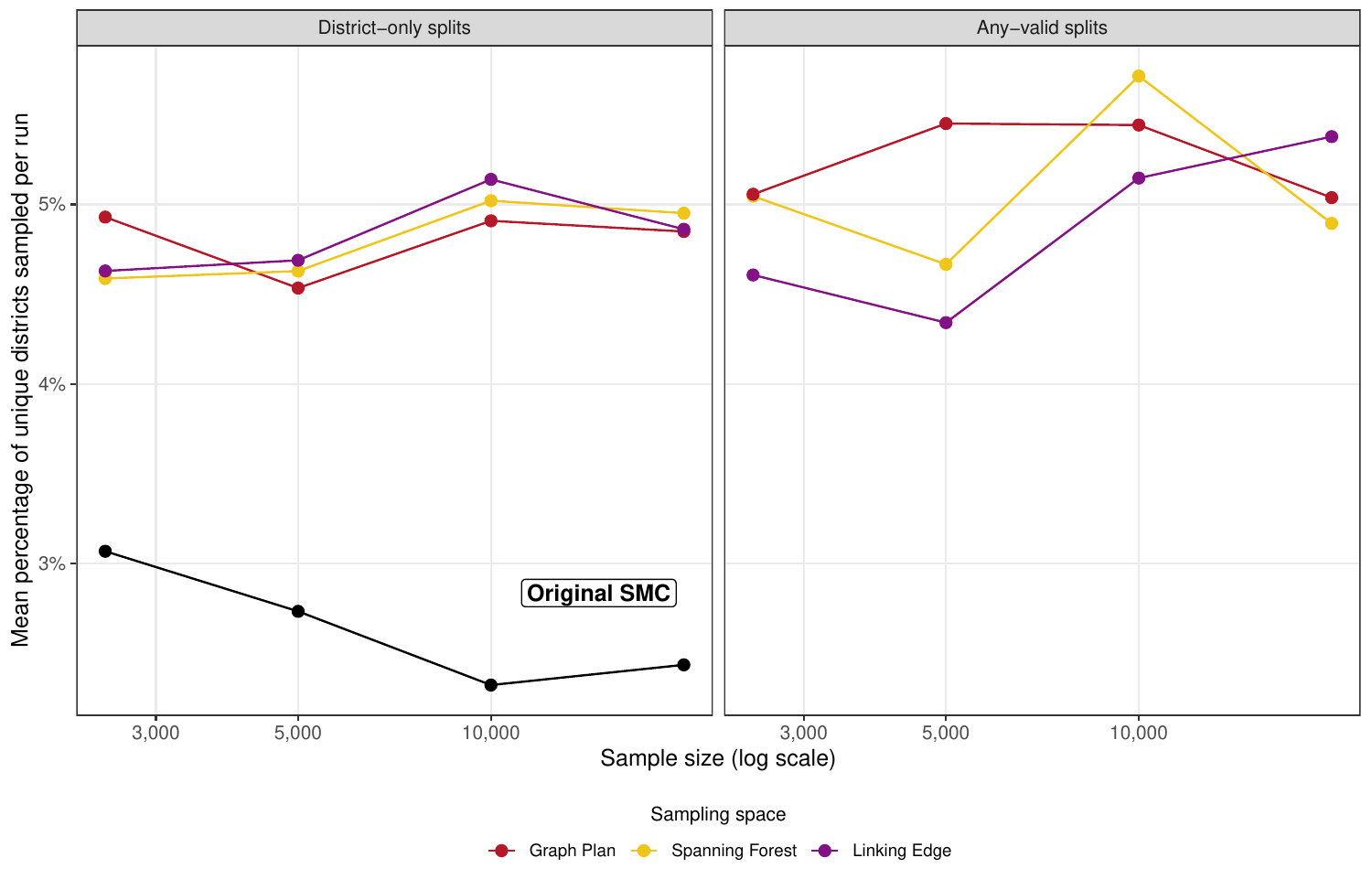}

}

\subcaption{\label{fig-ca-cd-gsmc-div}California Congressional: Mean
percentage of unique districts per run versus sample size for each of
the three gSMC sampling spaces, faceted by splitting schedule.}

\end{minipage}%
\newline
\begin{minipage}{\linewidth}

\centering{

\includegraphics[width=0.95\linewidth,height=\textheight,keepaspectratio]{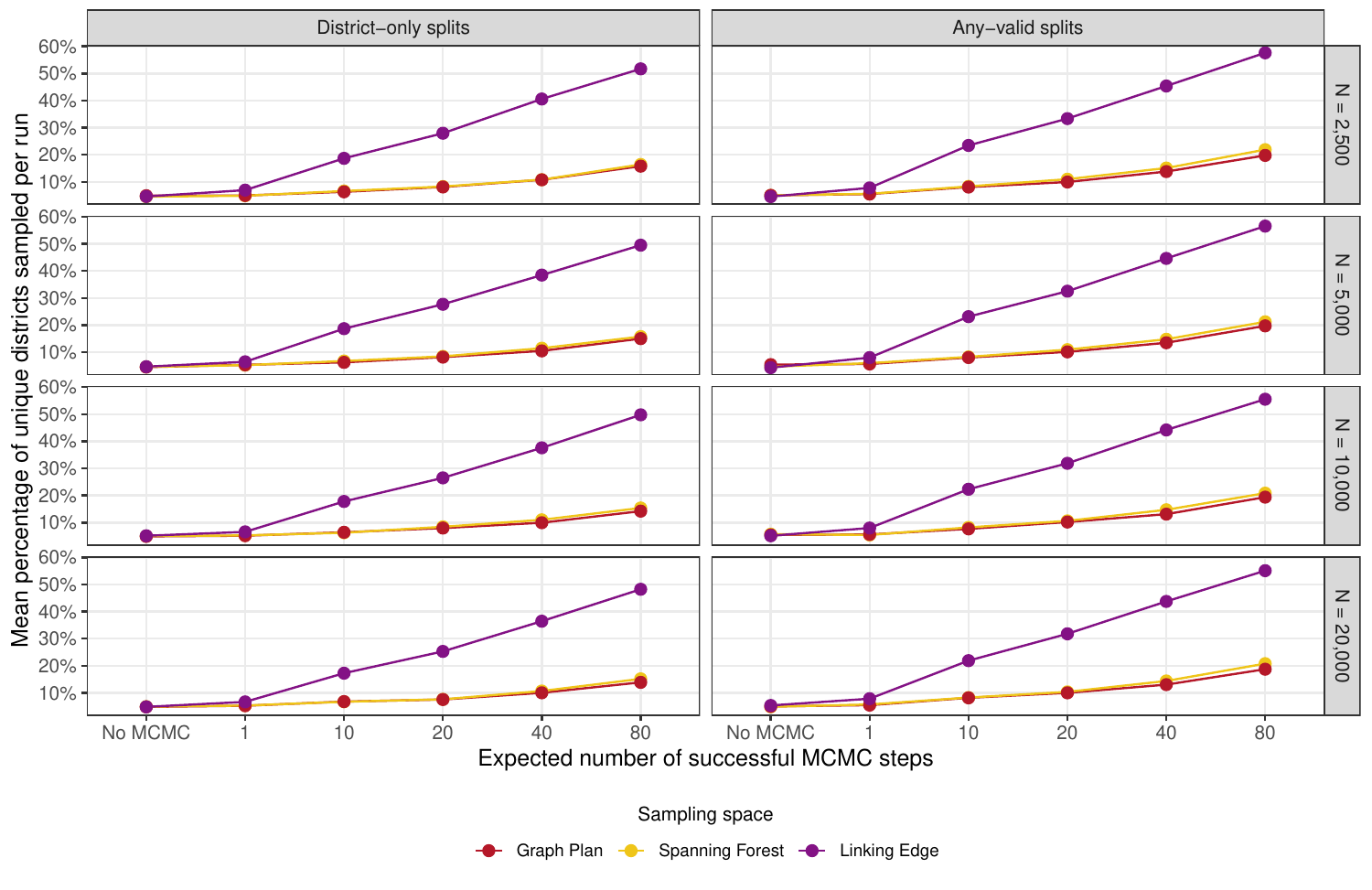}

}

\subcaption{\label{fig-ca-cd-gsmc-mcmc-div}California Congressional:
Mean percentage of unique districts per run across all expected numbers
of successful MCMC steps for each of the three gSMC sampling spaces,
faceted by splitting schedule and sample size.}

\end{minipage}%

\caption{\label{fig-ca-cd-div}California Congressional: Mean percentage
of unique districts per run across all gSMC and gSMC-MCMC configurations
run.}

\end{figure}%

\subsubsection{Washington 2020 State Senate Map}\label{sec-wa-ssd-2020}

We use the same map with \(V=7,434\) precincts, \(E = 19,517\) edges,
and a median population of 940.5 per precinct as in
Section~\ref{sec-wa-cd-2020}. We use the same provided pseudo-counties
as well but impose no additional constraints are imposed through the
\(J\) function. This is the first map with a relatively loose population
tolerance (5\%, as opposed to the other which were all .5\% or .1\%). We
sample both gSMC and gSMC-MCMC at sample sizes of 2,500, 5,000, 10,000.
The expected number of successful MCMC steps used were: 1, 10, 25, and
40. Unlike in Figure~\ref{fig-ca-cd-gsmc-rhats}, as
Figure~\ref{fig-wa-ssd-gsmc-rhats} demonstrates, any-valid splitting on
all three spaces converges once a sufficiently large number of MCMC
steps are applied. Unlike the other states up to this point, as
Figure~\ref{fig-wa-ssd-gsmc-mcmc-div} demonstrates, all three sampling
spaces have roughly the same mean percentage of unique districts when
MCMC steps are added.

\begin{figure}[H]

\centering{

\pandocbounded{\includegraphics[keepaspectratio]{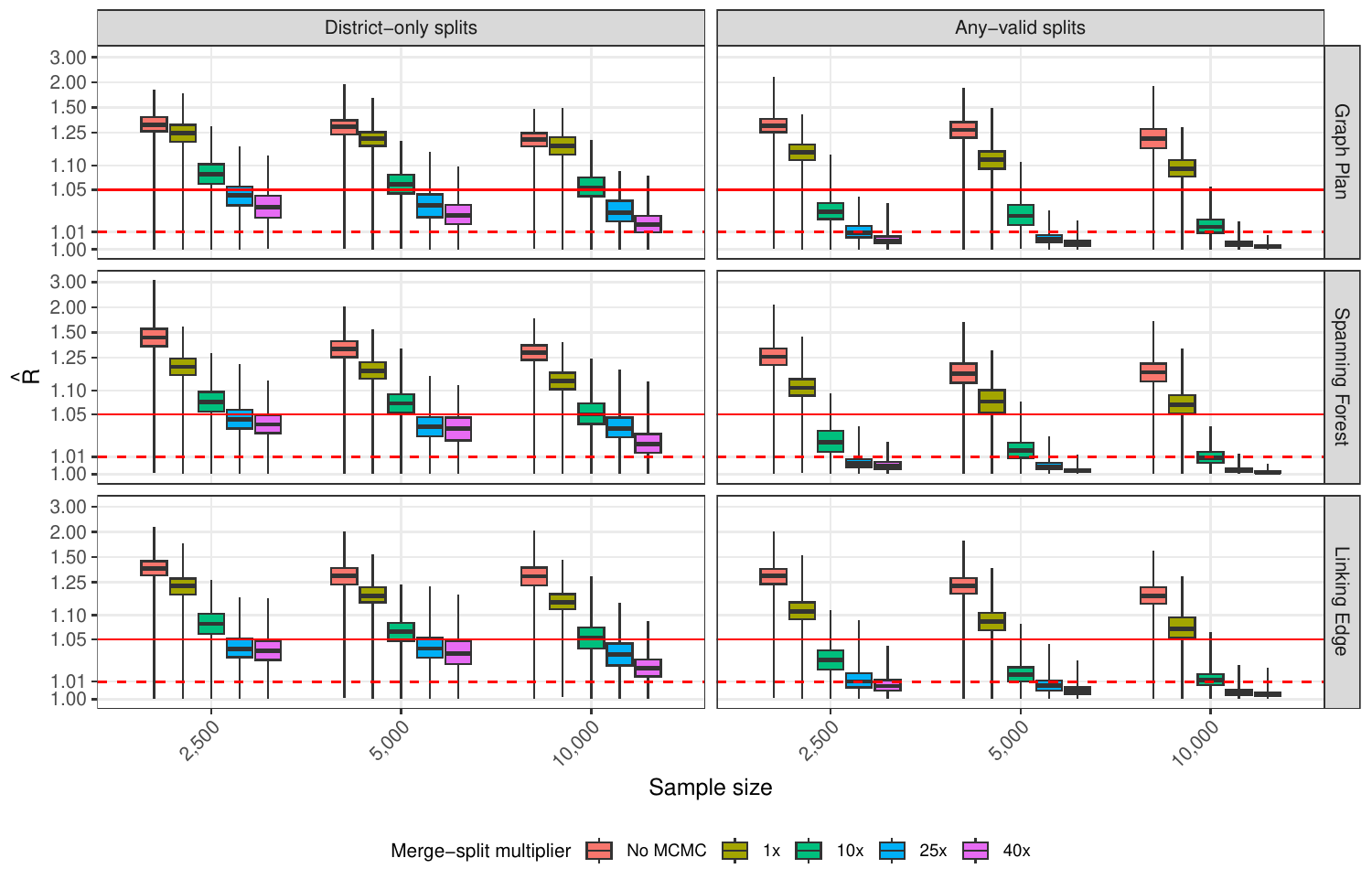}}

}

\caption{\label{fig-wa-ssd-gsmc-rhats}Washington State Senate:
Distribution of \(\hat{R}\) values for all gSMC and gSMC-MCMC
configurations run, faceted by sampling space and splitting schedule.}

\end{figure}%

\begin{figure}[H]

\begin{minipage}{\linewidth}

\begin{figure}[H]

\centering{

\includegraphics[width=0.9\linewidth,height=\textheight,keepaspectratio]{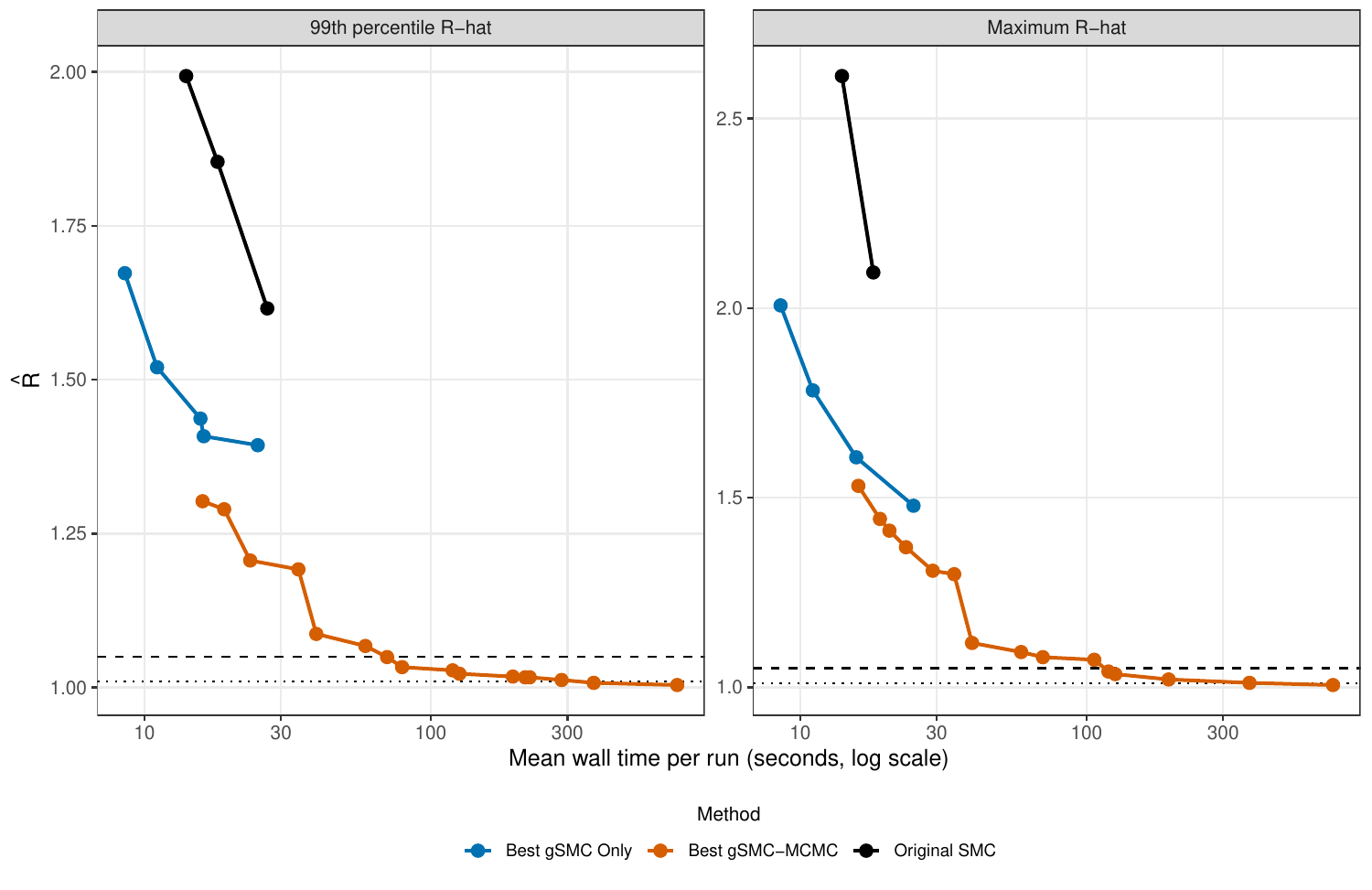}

}

\caption{\label{fig-wa-ssd-pareto-frontier}Washington State Senate:
Empirical Pareto frontier comparing mean wall time and convergence of
the best-performing gSMC and gSMC-MCMC configurations and Original SMC.
Each point represents an algorithm configuration, with mean wall time
per run on the horizontal axis and either the 99th percentile of the
resulting \(\hat{R}\) values (left) or the maximum \(\hat{R}\) value
(right) on the vertical axis. Moving along a frontier corresponds to
configurations for which improved convergence came at the cost of
greater wall time.}

\end{figure}%

\end{minipage}%
\newline
\begin{minipage}{\linewidth}

\begin{figure}[H]

\centering{

\includegraphics[width=0.9\linewidth,height=\textheight,keepaspectratio]{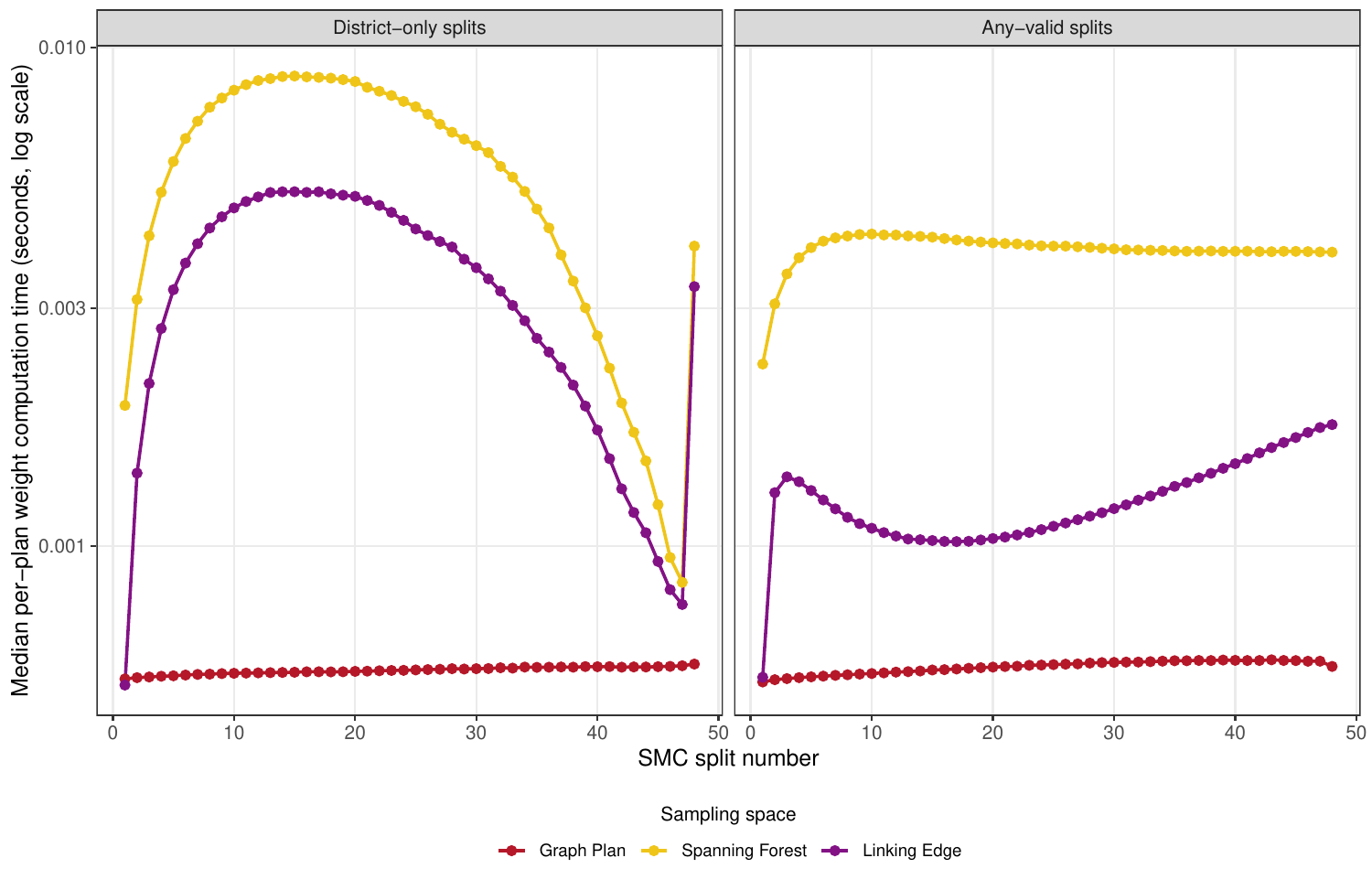}

}

\caption{\label{fig-wa-ssd-weight-time}Average median per-plan weight
computation time across sampling spaces for a sample size of 10,000,
faceted by splitting schedule.}

\end{figure}%

\end{minipage}%

\end{figure}%

\begin{figure}[H]

\begin{minipage}{\linewidth}

\centering{

\includegraphics[width=0.8\linewidth,height=\textheight,keepaspectratio]{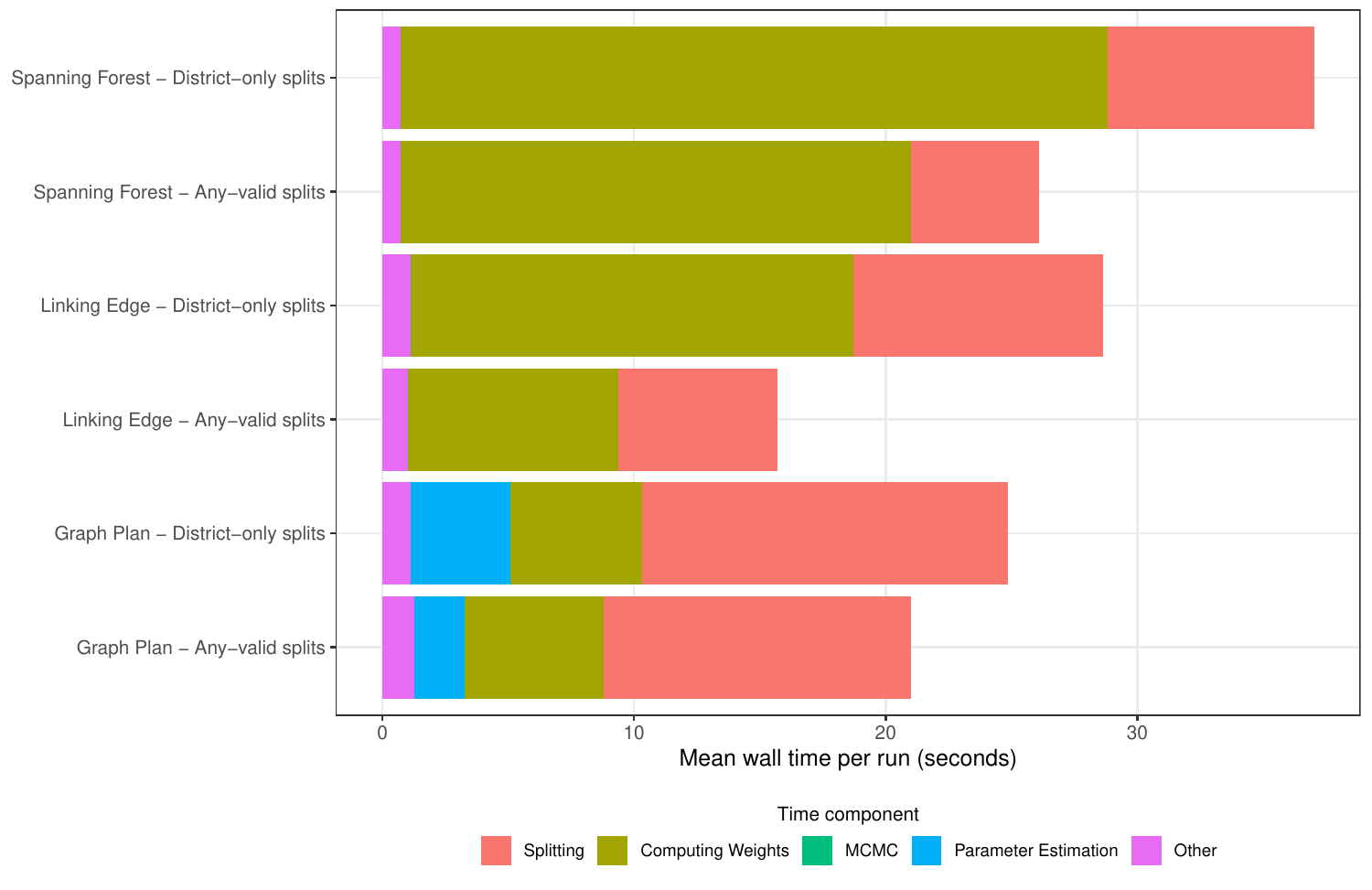}

}

\subcaption{\label{fig-wa-ssd-gsmc-time}Breakdown of wall-time
categories across the three gSMC sampling spaces and two splitting
schedules for a sample size of 10,000.}

\end{minipage}%
\newline
\begin{minipage}{\linewidth}

\centering{

\pandocbounded{\includegraphics[keepaspectratio]{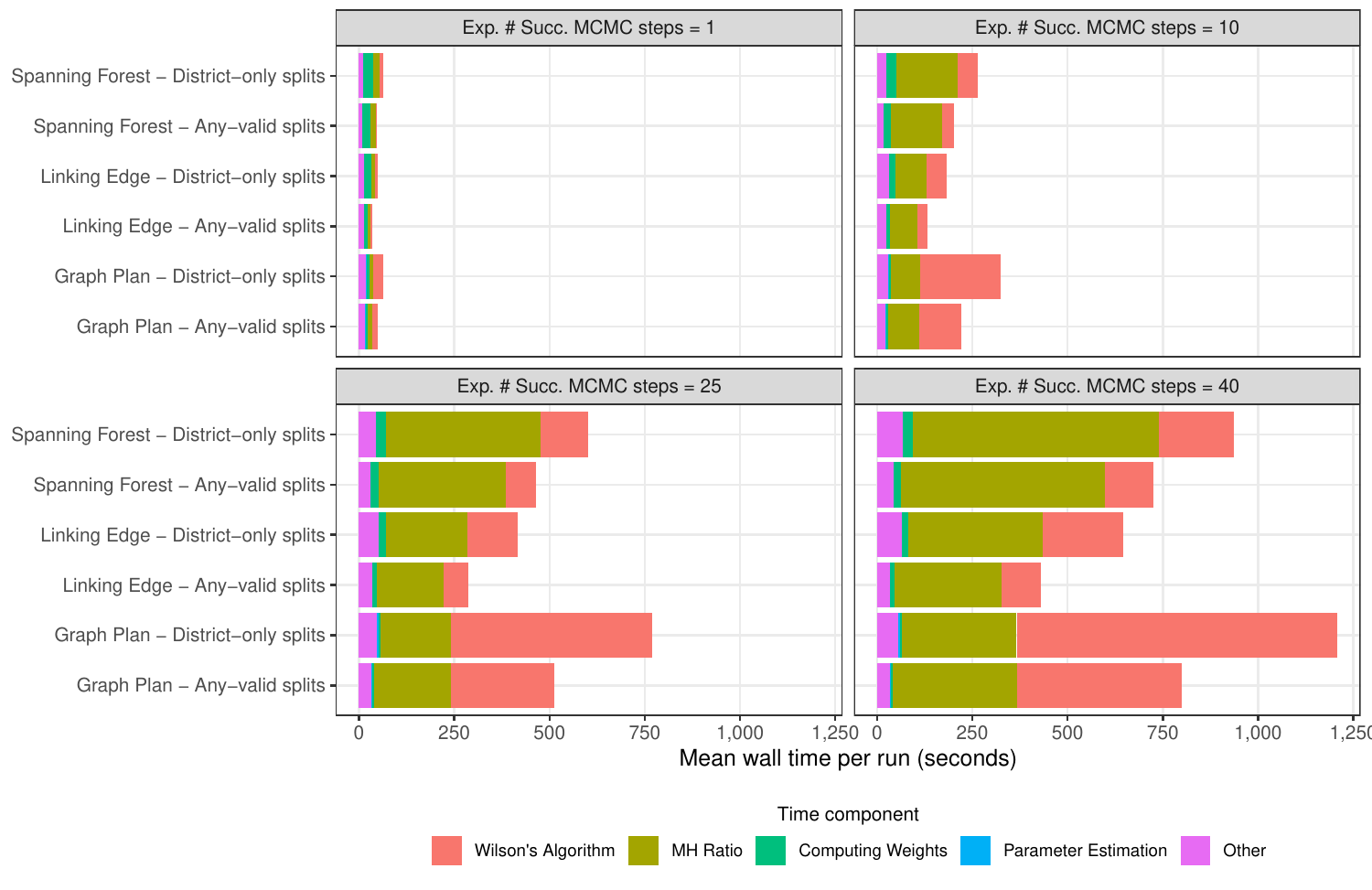}}

}

\subcaption{\label{fig-wa-ssd-gsmc-mcmc-time}Breakdown of wall-time
categories across the three gSMC sampling spaces and two splitting
schedules, faceted by expected number of successful MCMC steps, for a
sample size of 10,000.}

\end{minipage}%

\caption{\label{fig-wa-ssd-time-breakdown}Washington State Senate:
Breakdown of wall time for gSMC and gSMC-MCMC configurations at a sample
size of 10,000.}

\end{figure}%

\begin{figure}[H]

\begin{minipage}{\linewidth}

\centering{

\includegraphics[width=0.95\linewidth,height=\textheight,keepaspectratio]{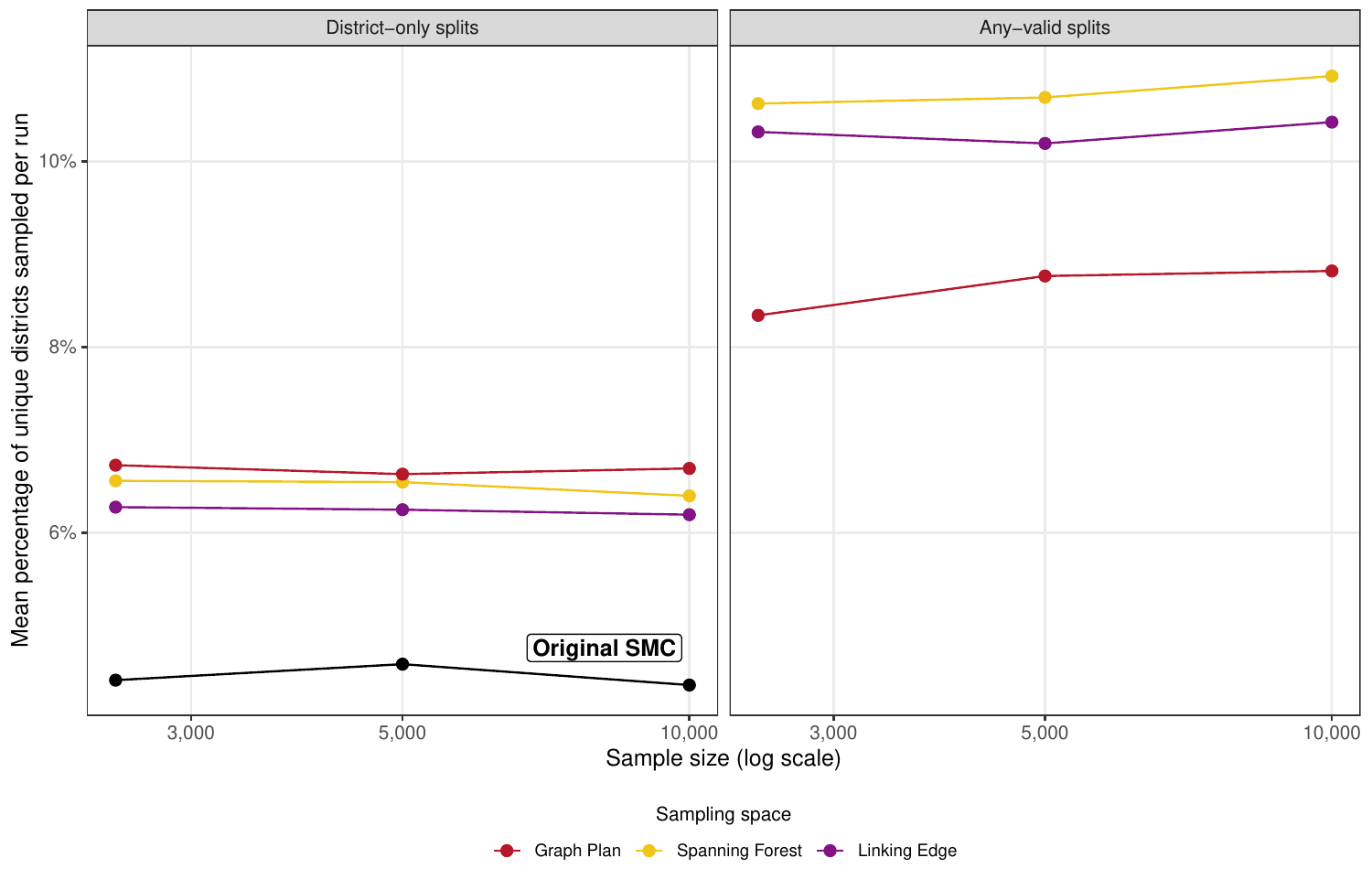}

}

\subcaption{\label{fig-wa-ssd-gsmc-div}Washington State Senate: Mean
percentage of unique districts per run versus sample size for each of
the three gSMC sampling spaces, faceted by splitting schedule.}

\end{minipage}%
\newline
\begin{minipage}{\linewidth}

\centering{

\includegraphics[width=0.95\linewidth,height=\textheight,keepaspectratio]{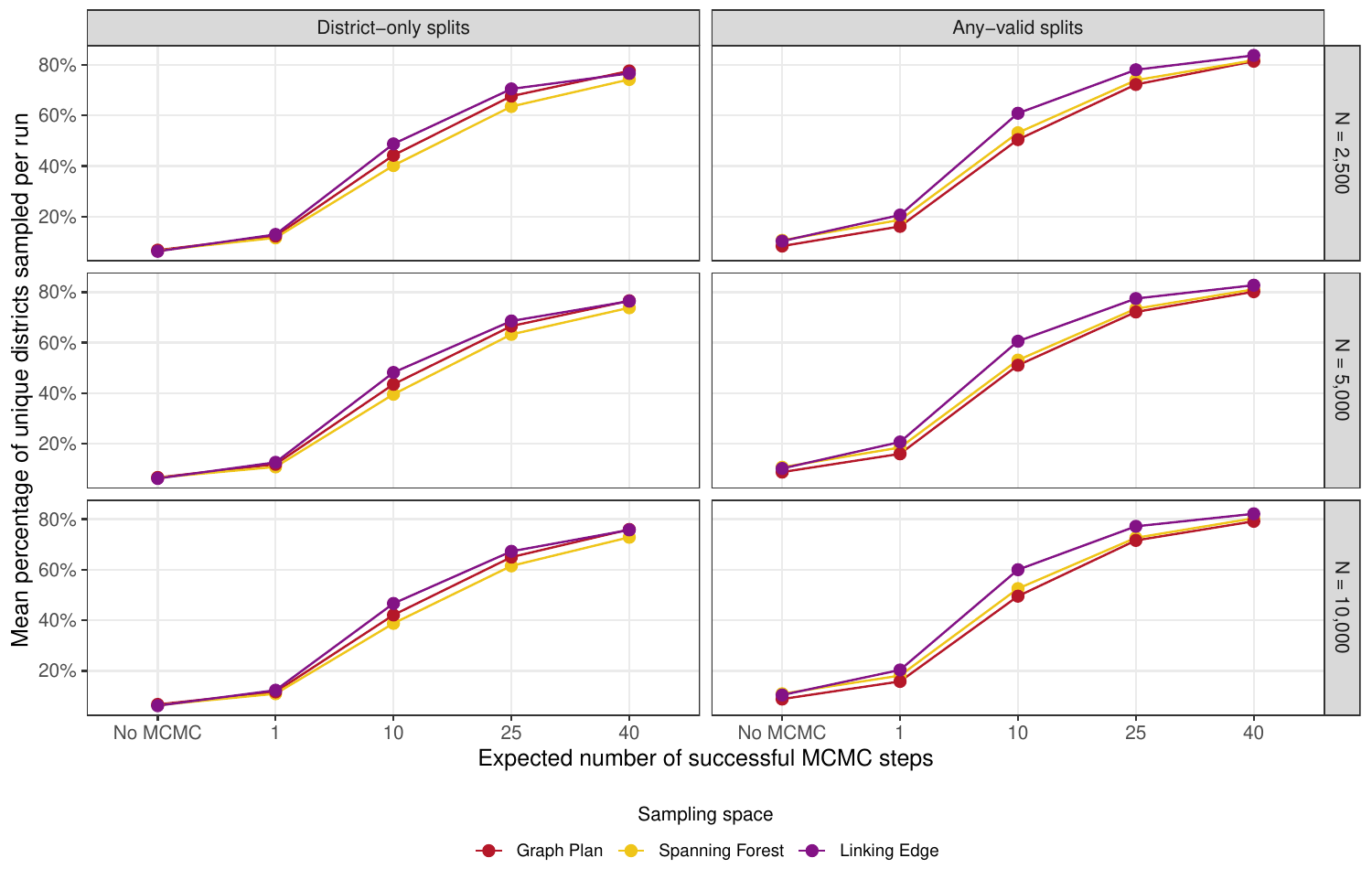}

}

\subcaption{\label{fig-wa-ssd-gsmc-mcmc-div}Washington State Senate:
Mean percentage of unique districts per run across all expected numbers
of successful MCMC steps for each of the three gSMC sampling spaces,
faceted by splitting schedule and sample size.}

\end{minipage}%

\caption{\label{fig-wa-ssd-div}Washington State Senate: Mean percentage
of unique districts per run across all gSMC and gSMC-MCMC configurations
run.}

\end{figure}%

\subsubsection{Tennessee 2020 State House Map}\label{sec-tn-shd-2020}

We use the same map with \(V=1,965\) precincts, \(E = 5,604\) edges, and
a median population of 2,857 per precinct as in
Section~\ref{sec-tn-cd-2020}. We use different pseudo-counties, instead
replacing all counties with a population larger than 837,678 with their
constituent municipalities. We also impose a total county split
constraint, similar to that used in the Pennsylvania state house example
from Section~\ref{sec-pa-house-main}. This map has the largest number of
districts of any of the maps tested. We sample both gSMC and gSMC-MCMC
at sample sizes of 2,500, 5,000, 10,000. The expected number of
successful MCMC steps used were: 1, 25, 50, 100, and 150. Like in
Section~\ref{sec-wa-ssd-2020}, as Figure~\ref{fig-wa-ssd-gsmc-rhats}
demonstrates, any-valid splitting on all three spaces converges once a
sufficiently large number of MCMC steps are applied, although here graph
space's performance lags behind on a sample size/number of MCMC steps
basis. As Figure~\ref{fig-tn-shd-weight-time} demonstrates, this map is
also the first map where, by the final SMC steps, the weight computation
time for linking edge space is longer than forest space. Although, as
Figure~\ref{fig-tn-shd-gsmc-mcmc-time} demonstrates when MCMC steps are
added this does not appear to actually lead to a shorter wall time for
forest space as opposed to linking edge.

\begin{figure}[H]

\centering{

\pandocbounded{\includegraphics[keepaspectratio]{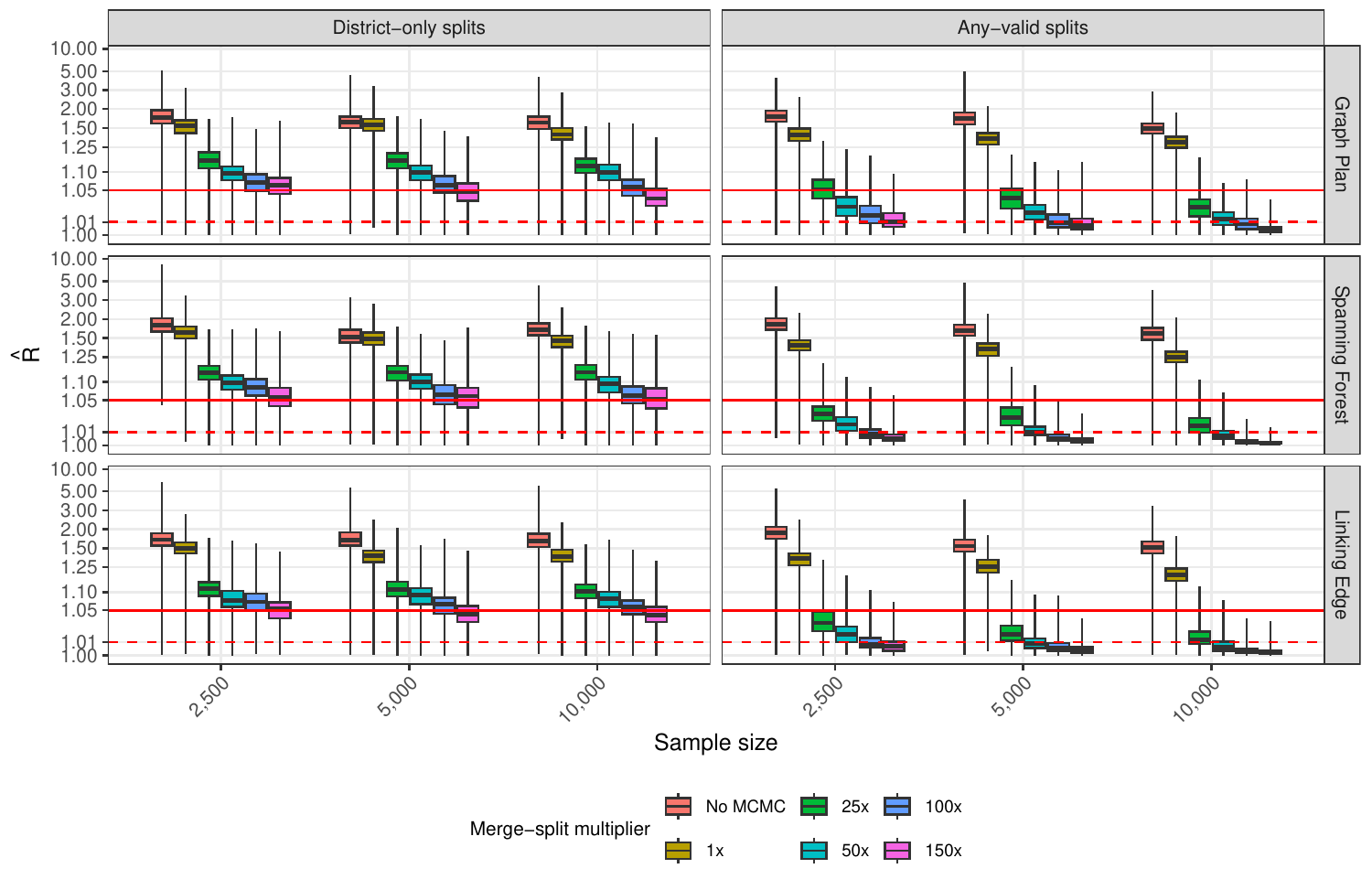}}

}

\caption{\label{fig-tn-shd-gsmc-rhats}Tennessee State House:
Distribution of \(\hat{R}\) values for all gSMC and gSMC-MCMC
configurations run, faceted by sampling space and splitting schedule.}

\end{figure}%

\begin{figure}[H]

\begin{minipage}{\linewidth}

\begin{figure}[H]

\centering{

\includegraphics[width=0.9\linewidth,height=\textheight,keepaspectratio]{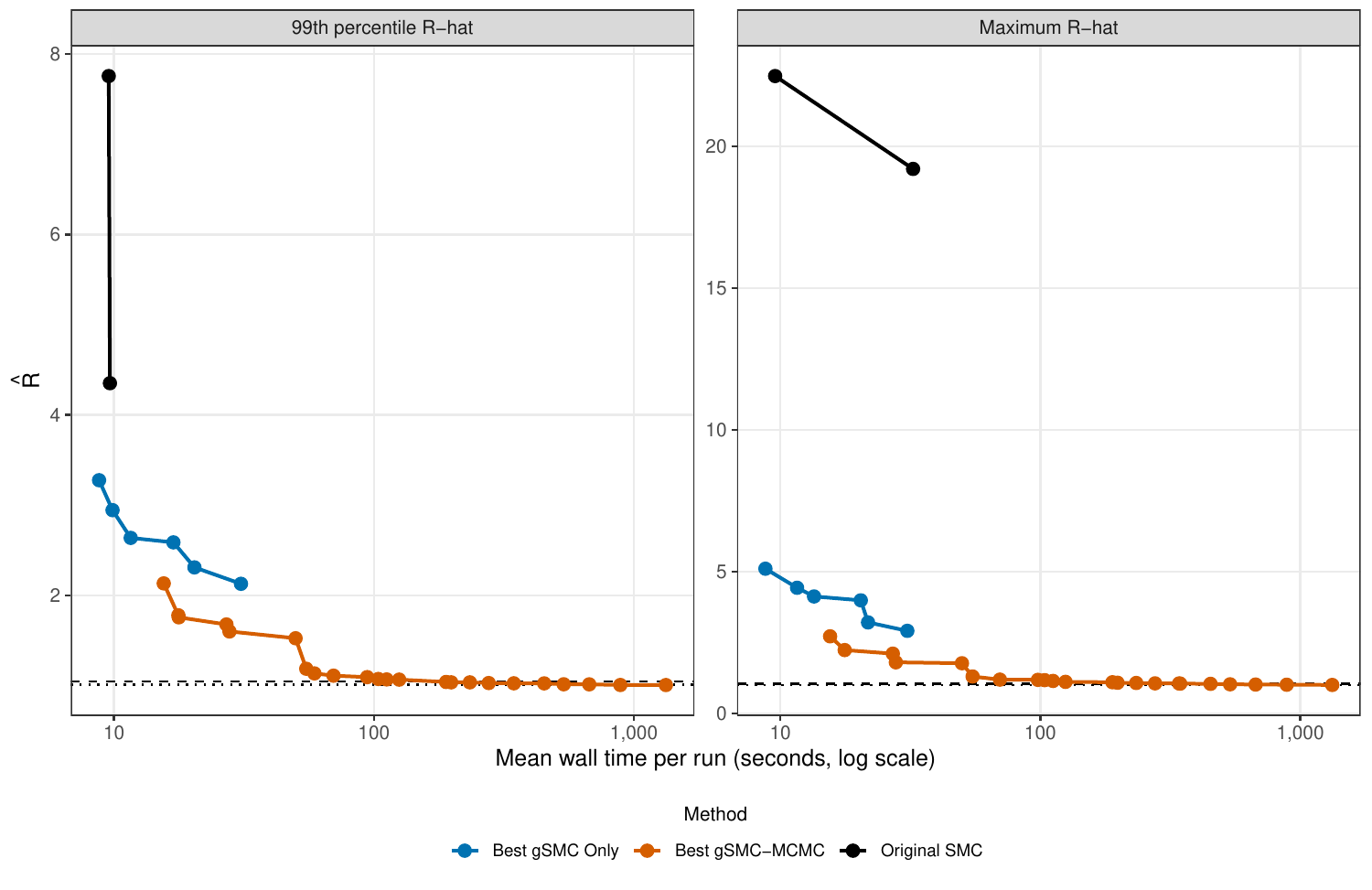}

}

\caption{\label{fig-tn-shd-pareto-frontier}Tennessee State House:
Empirical Pareto frontier comparing mean wall time and convergence of
the best-performing gSMC and gSMC-MCMC configurations and Original SMC.
Each point represents an algorithm configuration, with mean wall time
per run on the horizontal axis and either the 99th percentile of the
resulting \(\hat{R}\) values (left) or the maximum \(\hat{R}\) value
(right) on the vertical axis. Moving along a frontier corresponds to
configurations for which improved convergence came at the cost of
greater wall time.}

\end{figure}%

\end{minipage}%
\newline
\begin{minipage}{\linewidth}

\begin{figure}[H]

\centering{

\includegraphics[width=0.9\linewidth,height=\textheight,keepaspectratio]{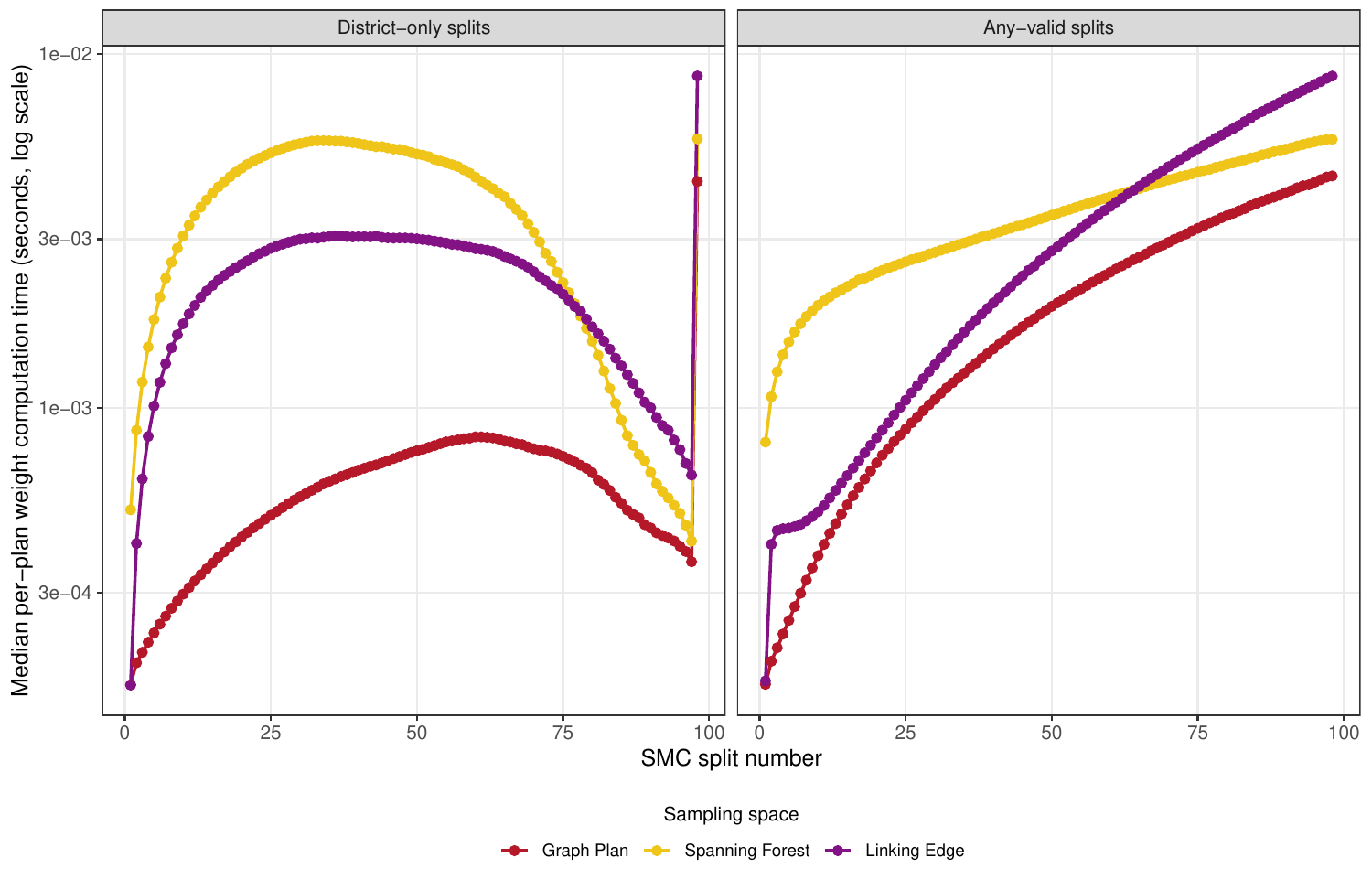}

}

\caption{\label{fig-tn-shd-weight-time}Average median per-plan weight
computation time across sampling spaces for a sample size of 10,000,
faceted by splitting schedule.}

\end{figure}%

\end{minipage}%

\end{figure}%

\begin{figure}[H]

\begin{minipage}{\linewidth}

\centering{

\includegraphics[width=0.8\linewidth,height=\textheight,keepaspectratio]{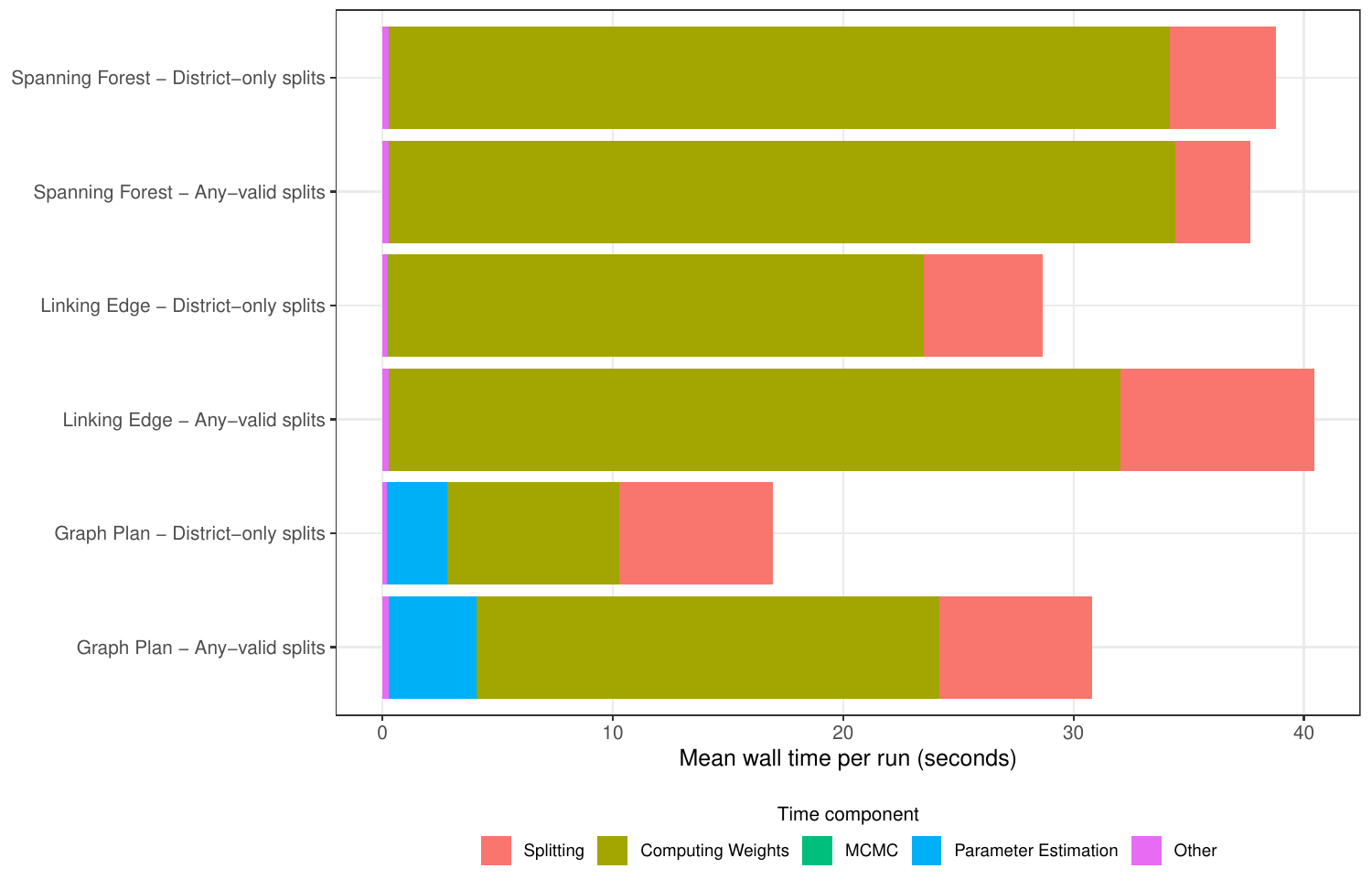}

}

\subcaption{\label{fig-tn-shd-gsmc-time}Breakdown of wall-time
categories across the three gSMC sampling spaces and two splitting
schedules for a sample size of 10,000.}

\end{minipage}%
\newline
\begin{minipage}{\linewidth}

\centering{

\pandocbounded{\includegraphics[keepaspectratio]{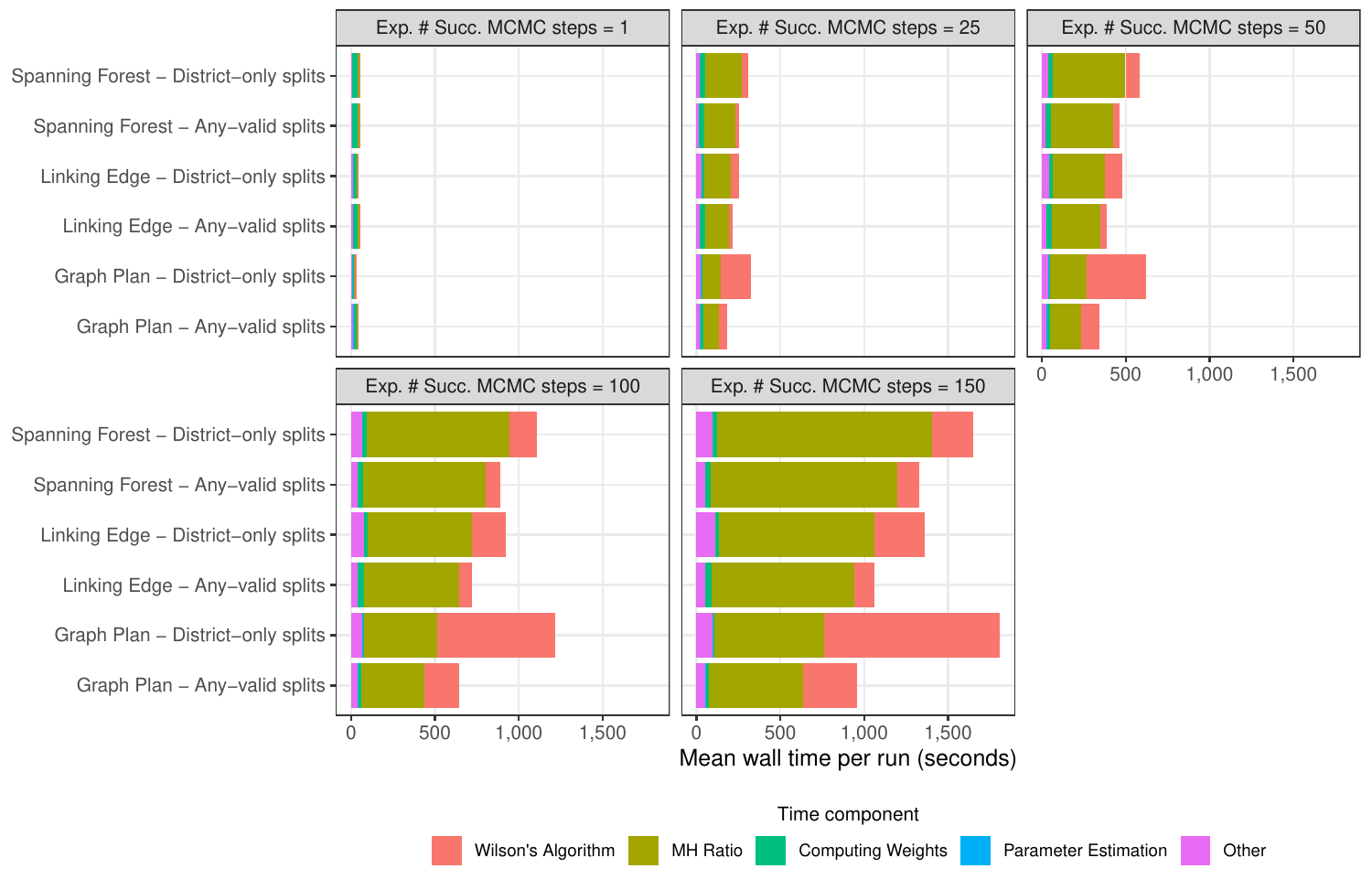}}

}

\subcaption{\label{fig-tn-shd-gsmc-mcmc-time}Breakdown of wall-time
categories across the three gSMC sampling spaces and two splitting
schedules, faceted by expected number of successful MCMC steps, for a
sample size of 10,000.}

\end{minipage}%

\caption{\label{fig-tn-shd-time-breakdown}Tennessee State House:
Breakdown of wall time for gSMC and gSMC-MCMC configurations at a sample
size of 10,000.}

\end{figure}%

\begin{figure}[H]

\begin{minipage}{\linewidth}

\centering{

\includegraphics[width=0.95\linewidth,height=\textheight,keepaspectratio]{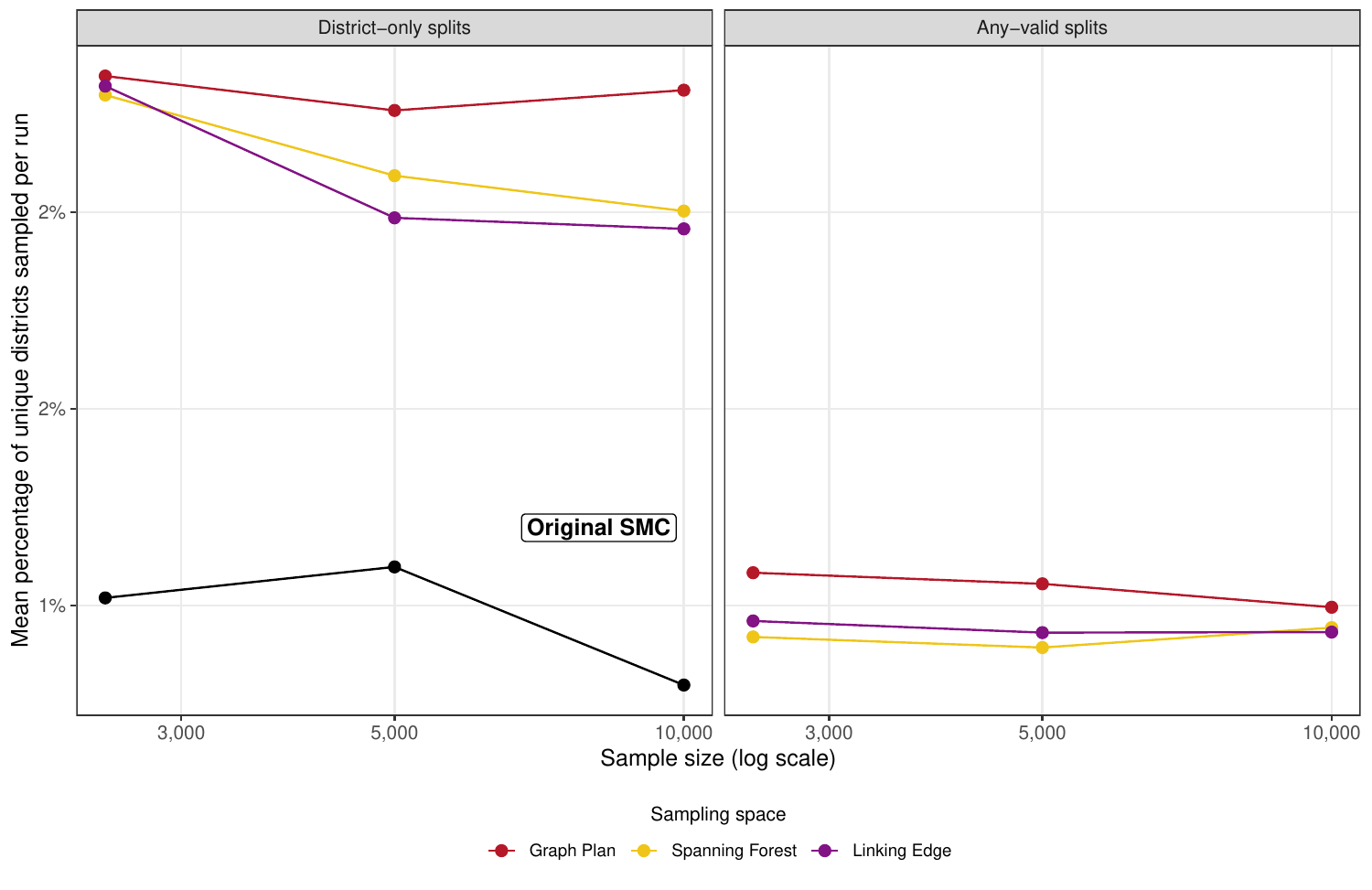}

}

\subcaption{\label{fig-tn-shd-gsmc-div}Tennessee State House: Mean
percentage of unique districts per run versus sample size for each of
the three gSMC sampling spaces, faceted by splitting schedule.}

\end{minipage}%
\newline
\begin{minipage}{\linewidth}

\centering{

\includegraphics[width=0.95\linewidth,height=\textheight,keepaspectratio]{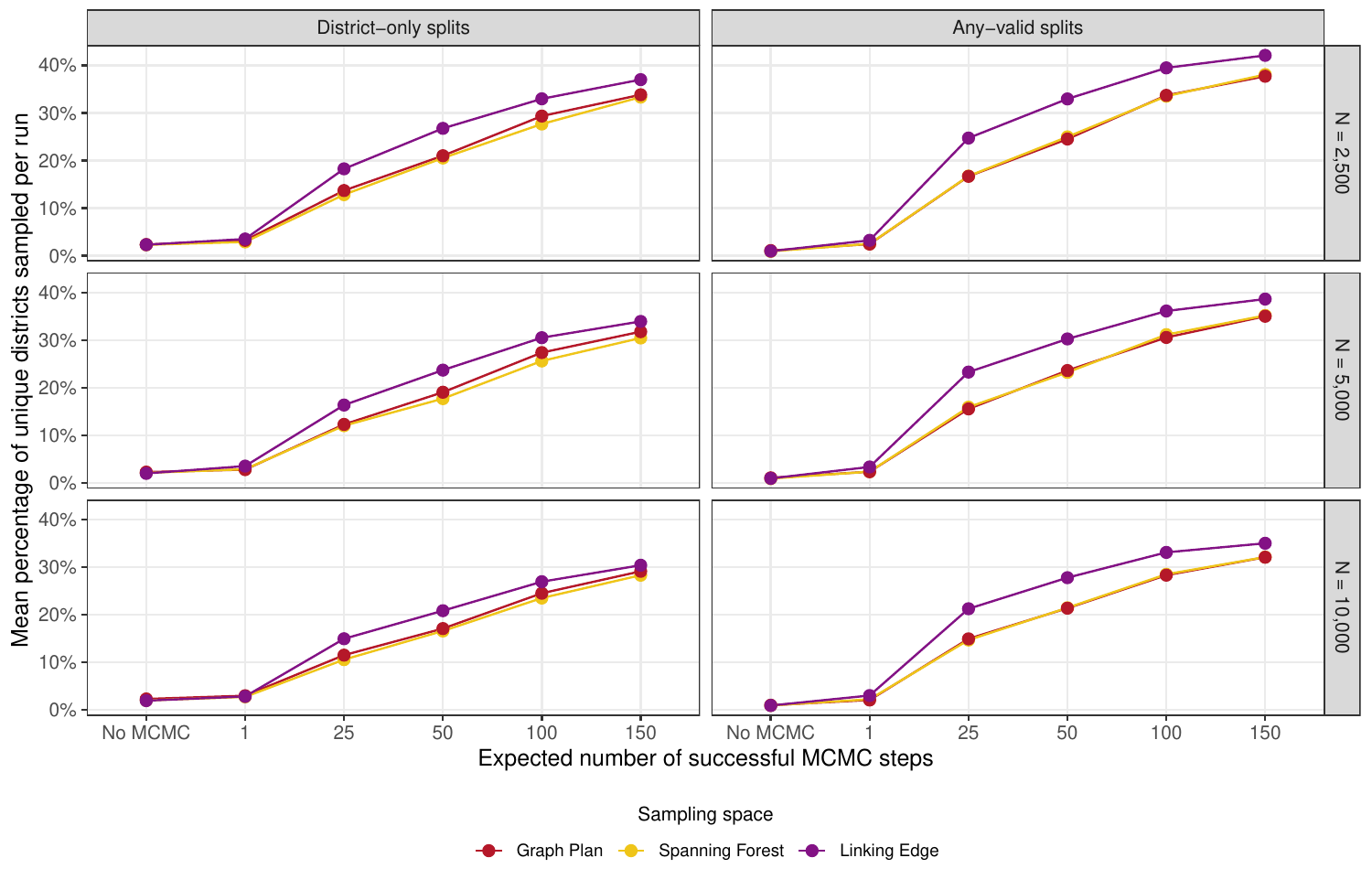}

}

\subcaption{\label{fig-tn-shd-gsmc-mcmc-div}Tennessee State House: Mean
percentage of unique districts per run across all expected numbers of
successful MCMC steps for each of the three gSMC sampling spaces,
faceted by splitting schedule and sample size.}

\end{minipage}%

\caption{\label{fig-tn-shd-div}Tennessee State House: Mean percentage of
unique districts per run across all gSMC and gSMC-MCMC configurations
run.}

\end{figure}%

\subsection{Extreme Balance Simulations}\label{sec-extreme-balance-sims}

We now examine the performance of the new algorithms in the following
maps listed in Table~\ref{tbl-bal-empirical-examples} with extremely
strong population tolerance constraints.

\begin{longtable}[]{@{}llllr@{}}
\caption{Balanced maps used in the simulation
study.}\label{tbl-bal-empirical-examples}\tabularnewline
\toprule\noalign{}
Name & Districts & Population Bounds & Vertices & Edges \\
\midrule\noalign{}
\endfirsthead
\toprule\noalign{}
Name & Districts & Population Bounds & Vertices & Edges \\
\midrule\noalign{}
\endhead
\bottomrule\noalign{}
\endlastfoot
Arkansas Congressional & 4 & \(752,881 \pm 0\) & 2,747 & 7,509 \\
Wisconsin Congressional & 8 & \(736,715 \pm 1\) & 7,059 & 19,281 \\
Tennessee Congressional & 9 & \(767,871 + 1\) & 1,965 & 5,604 \\
Illinois Congressional & 17 & \(753,677 \pm 1\) & 10,084 & 28,219 \\
\end{longtable}

As before, unless stated otherwise, we perform 30 independent runs per
sample size/sampling space/splitting schedule/MCMC step variation. The
specific sample sizes and expected number of successful MCMC steps are
listed in each respective map subsection. Unlike
Section~\ref{sec-comprehensive-sims}, we do not always sample across the
same range of sample sizes and MCMC steps for each sampling space and
splitting schedule variant. We explore the performance of the new
algorithm variants in terms of all \(\hat{R}\) values for all summary
statistics of interest being below our 1.05 threshold and the percentage
of districts that are unique within a given run.

We also compare the new gSMC variants on these two metrics against
equivalent Markov chains (MCMC) with the same target distribution as
their SMC counterparts on the three respective sampling spaces on a wall
time basis. These Markov chains are also implemented in \texttt{redist}
and share the same underlying code. Note that for forest space sampling,
the Markov chain is essentially equivalent to the MCMC method presented
in \citet{McmcForests} where \(\beta = 1\) and \(\gamma = 1 - \tau\).
For linking edge space sampling, the Markov chain is almost identical to
the one developed in \citet{McmcLinkingEdge} except that the linking
edge term is corrected for and thus does not appear in the final
pushforward distribution on plans. Each MCMC chain run was on a single
core of the same research cluster as the one used for SMC. The sample
size for the MCMC runs was set as the sample size of the largest gSMC
samples and the runtime was adjusted by increasing the amount of
thinning. All chains were run for 10,000 warmup steps. A small gSMC
sample of 250 plans was used as starting points for each map.

We caution that precisely comparing SMC versus MCMC based methods is
difficult because of the different computational and memory demands of
the two algorithms. As mentioned earlier, SMC is parallelizable, as both
the splitting and weight computation step can be done independently in
parallel for each plan via multithreading. Even the MCMC in the
gSMC-MCMC algorithm can be run in parallel, as the Markov kernel is
applied independently to each plan. In contrast, MCMC methods can
typically only use a single core per independent run, although some
methods like \citet{cannon2025} allow for multithreading to overcome
very low acceptance rates. Thus, our wall time comparison should not be
interpreted as a definitive evaluation.

In terms of memory requirement, for SMC it grows linearly in the number
of plans sampled. Because of the interdependent nature of the SMC
samples, the entire \(V \times N\) matrix of plans and the associated
matrix of region sizes must be stored in memory throughout the entire
run. In fact, two copies must be stored for the splitting step as the
previous step's plans must be kept until all new plans have been split.
In contrast, for MCMC since only the current plan is needed memory usage
can be drastically reduced by writing each sampled plan directly to disk
and only storing the current and proposed plan in memory. However, we
note that the Markov chains implemented in \texttt{redist} do not do
this, instead storing the entire plans matrix in memory.

Accepting this important caveat, Tables \ref{tab:gsmc-bal-rhat},
\ref{tab:gsmc-mcmc-bal-rhat}, and \ref{tab:mcmc-bal-rhat}, present, for
each map, the fastest algorithm variant on a mean wall time basis to
have all \(\hat{R}\) values under the 1.05 threshold for gSMC,
gSMC-MCMC, and MCMC respectively. Figures \ref{fig-bal-ar-gsmc-rhats},
\ref{fig-bal-wi-gsmc-rhats}, \ref{fig-bal-tn-gsmc-rhats}, and
\ref{fig-bal-il-gsmc-rhats} display for each map, boxplots of the
\(\hat{R}\) values for all configurations of gSMC and gSMC-MCMC over all
sample sizes ran. Figures \ref{fig-bal-ar-mcmc-gsmc-walltime-rhats},
\ref{fig-bal-wi-mcmc-gsmc-walltime-rhats},
\ref{fig-bal-tn-mcmc-gsmc-walltime-rhats}, and
\ref{fig-bal-il-mcmc-gsmc-walltime-rhats} display for each map, boxplots
of \(\hat{R}\) values for all MCMC chains run versus mean wall time
faceted the three sampling spaces. For each sampling space, the gSMC or
gSMC-MCMC configuration with the lowest mean wall time among all
configurations that had all \(\hat{R}\) values under the threshold (if
any such configurations exist) is also displayed.

Finally, Figures \ref{fig-bal-ar-distr-div}, \ref{fig-bal-wi-distr-div},
\ref{fig-bal-tn-distr-div}, and \ref{fig-bal-il-distr-div} display the
mean percentage of unique districts within a run for each algorithm
configuration faceted by sample size. Note that since the MCMC chains
were only ran with the same sample size as the largest gSMC run,
generally they will not be present on all the plot facets. Each
subsection provides more information on the algorithm variants and
configurations run.

Overall, we find that the gSMC and gSMC-MCMC algorithms substantially
outperforms the MCMC algorithms in terms of \(\hat{R}\) convergence and
percentage of unique districts on a wall time basis. As Tables
\ref{tab:gsmc-bal-rhat}, \ref{tab:gsmc-mcmc-bal-rhat}, and
\ref{tab:mcmc-bal-rhat} succinctly demonstrate, both gSMC and gSMC-MCMC
are able to achieve convergence much faster, if at all, compared to
MCMC. This comports with the results in Theorems
\ref{thm-gsmc-convergence} and \ref{thm-gsmc-mcmc-convergence}, holding
that the gSMC algorithm does not require the state space to be connected
for theoretical convergence guarantees to hold, although the lower
population tolerance makes the problem much harder.

Interestingly, for Illinois and Tennessee, as Figures
\ref{fig-bal-tn-mcmc-gsmc-walltime-rhats} and
\ref{fig-bal-il-mcmc-gsmc-walltime-rhats} demonstrate, gSMC and MCMC
alone were not able to achieve convergence but gSMC-MCMC was, suggesting
that the addition of MCMC steps to gSMC allows for better performance
than either one is individually capable of, at least for extremely
constrained settings. Figures \ref{fig-bal-ar-distr-div},
\ref{fig-bal-wi-distr-div}, \ref{fig-bal-tn-distr-div}, and
\ref{fig-bal-il-distr-div} also demonstrate that gSMC and gSMC-MCMC are
able to achieve, on average, a higher percentage of unique districts
much faster than MCMC alone, suggesting that they are able to explore
the space of plans and avoid getting stuck in local modes more
efficiently than MCMC alone in these maps.

Future work is needed to develop more effective methods of comparing SMC
versus MCMC based methods as well as comparing the performance of other
MCMC methods like \citep{cyclewalk, BUD, cannon2025} in extremely
constrained maps like the ones here.

\begin{table}[H]
\centering
\small
\renewcommand{\arraystretch}{1.1}
\setlength{\tabcolsep}{5pt}
\begin{tabular}{@{}lllrr@{}}
\hline
\textbf{Name} & \textbf{Sampling Space} & \textbf{Splitting Schedule} & \textbf{Sample Size} & \textbf{Mean Wall Time} \\
\hline
Arkansas & Linking Edge & Any-valid & 1,000 & 48.4 s \\
Wisconsin & Linking Edge & District-only & 50,000 & 51.5 min \\
Tennessee & — & — & — & — \\
Illinois & — & — & — & — \\
\hline
\end{tabular}
\caption{Fastest gSMC only configurations with maximum R-hat at or below 1.05. Blank entry indicates no convergent samples.}
\label{tab:gsmc-bal-rhat}
\end{table}

\begin{table}[H]
\centering
\small
\renewcommand{\arraystretch}{1.1}
\setlength{\tabcolsep}{5pt}
\begin{tabular}{@{}lllrrr@{}}
\hline
\textbf{Name} & \textbf{Sampling Space} & \textbf{Splitting Schedule} & \textbf{Exp. Num. MCMC Acc.} & \textbf{Sample Size} & \textbf{Mean Wall Time} \\
\hline
Arkansas & Linking Edge & Any-valid & 1 & 1,000 & 1.2 min \\
Wisconsin & Graph Plan & Any-valid & 20 & 10,000 & 40.5 min \\
Tennessee & Linking Edge & Any-valid & 60 & 20,000 & 2.7 hr \\
Illinois & Linking Edge & Any-valid & 30 & 10,000 & 2.5 hr \\
\hline
\end{tabular}
\caption{Fastest gSMC-MCMC configurations with maximum R-hat at or below 1.05. Blank entry indicates no convergent samples.}
\label{tab:gsmc-mcmc-bal-rhat}
\end{table}

\begin{table}[H]
\centering
\small
\renewcommand{\arraystretch}{1.1}
\setlength{\tabcolsep}{5pt}
\begin{tabular}{@{}llrrrr@{}}
\hline
\textbf{Name} & \textbf{Sampling Space} & \textbf{Warmup} & \textbf{Thin} & \textbf{Total Steps} & \textbf{Mean Wall Time} \\
\hline
Arkansas & Linking Edge & 10,000 & 256,500 & 641,260,000 & 21.3 hr \\
Wisconsin & — & — & — & — & — \\
Tennessee & — & — & — & — & — \\
Illinois & — & — & — & — & — \\
\hline
\end{tabular}
\caption{Fastest MCMC configurations with maximum R-hat at or below 1.05. Blank entry indicates no convergent samples.}
\label{tab:mcmc-bal-rhat}
\end{table}

\newpage

\subsubsection{Balanced Arkansas 2020 Congressional
Map}\label{balanced-arkansas-2020-congressional-map}

We use the same map as in Section~\ref{sec-ar-cd-2020} but impose a
perfect balance requirement (i.e., each of the four congressional
districts must have the same population of 752,881). Just as in
Section~\ref{sec-wi-cd-2020}, we sample with \(\rho=1\), use counties as
the administrative units for hierarchical sampling, and impose no
additional constraints through the \(J\) term. As presented in
Figure~\ref{fig-bal-ar-gsmc-rhats}, for gSMC and gSMC-MCMC we sample
1,000 and 2,500 plans across all gSMC and gSMC-MCMC configurations. All
MCMC chains were run with 2,500 samples. Table
\ref{tab:mcmc-config-balanced-ar-cd-2020} lists all the MCMC chain
configurations ran.

\begin{table}[H]
\centering
\small
\renewcommand{\arraystretch}{1.1}
\setlength{\tabcolsep}{5pt}
\resizebox{\linewidth}{!}{%
\begin{tabular}{@{}l r r r r r@{}}
\hline
\textbf{Sampling Space} & \textbf{Sample Size} & \textbf{Warmup Steps} & \textbf{Thinning} & \textbf{Total Steps} & \textbf{Mean Wall Time} \\
\hline
Graph Plan & 2,500 & 10,000 & 13,500 & 33,760,000 & 1.5 hr \\
Graph Plan & 2,500 & 10,000 & 27,000 & 67,510,000 & 2.9 hr \\
Graph Plan & 2,500 & 10,000 & 67,500 & 168,760,000 & 7.2 hr \\
Graph Plan & 2,500 & 10,000 & 202,500 & 506,260,000 & 21.8 hr \\
Spanning Forest & 2,500 & 10,000 & 8,400 & 21,010,000 & 1.6 hr \\
Spanning Forest & 2,500 & 10,000 & 16,800 & 42,010,000 & 3.0 hr \\
Spanning Forest & 2,500 & 10,000 & 42,000 & 105,010,000 & 7.4 hr \\
Spanning Forest & 2,500 & 10,000 & 126,000 & 315,010,000 & 22.5 hr \\
Linking Edge & 2,500 & 10,000 & 17,100 & 42,760,000 & 1.5 hr \\
Linking Edge & 2,500 & 10,000 & 34,200 & 85,510,000 & 2.9 hr \\
Linking Edge & 2,500 & 10,000 & 85,500 & 213,760,000 & 7.1 hr \\
Linking Edge & 2,500 & 10,000 & 256,500 & 641,260,000 & 21.3 hr \\
\hline
\end{tabular}
}
\caption{MCMC configurations for Arkansas (4 districts, ± 0 people). Total steps equal warmup plus sample size times thinning.}
\label{tab:mcmc-config-balanced-ar-cd-2020}
\end{table}

\begin{figure}[H]

\centering{

\pandocbounded{\includegraphics[keepaspectratio]{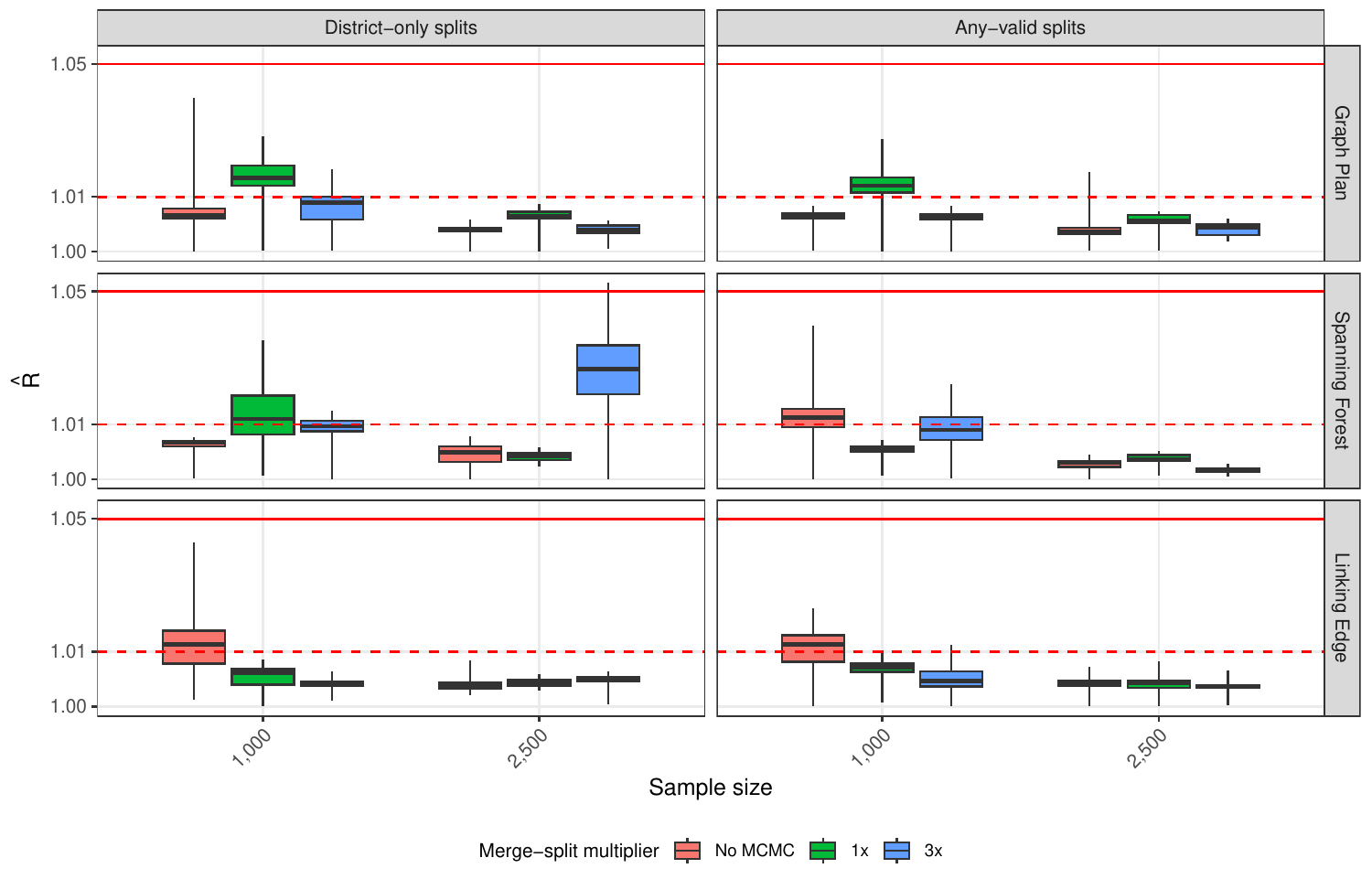}}

}

\caption{\label{fig-bal-ar-gsmc-rhats}Balanced Arkansas Congressional:
Distribution of \(\hat{R}\) values for all gSMC and gSMC-MCMC
configurations run, faceted by sampling space and splitting schedule.}

\end{figure}%

\begin{figure}[H]

\begin{minipage}{\linewidth}

\begin{figure}[H]

\centering{

\includegraphics[width=0.95\linewidth,height=\textheight,keepaspectratio]{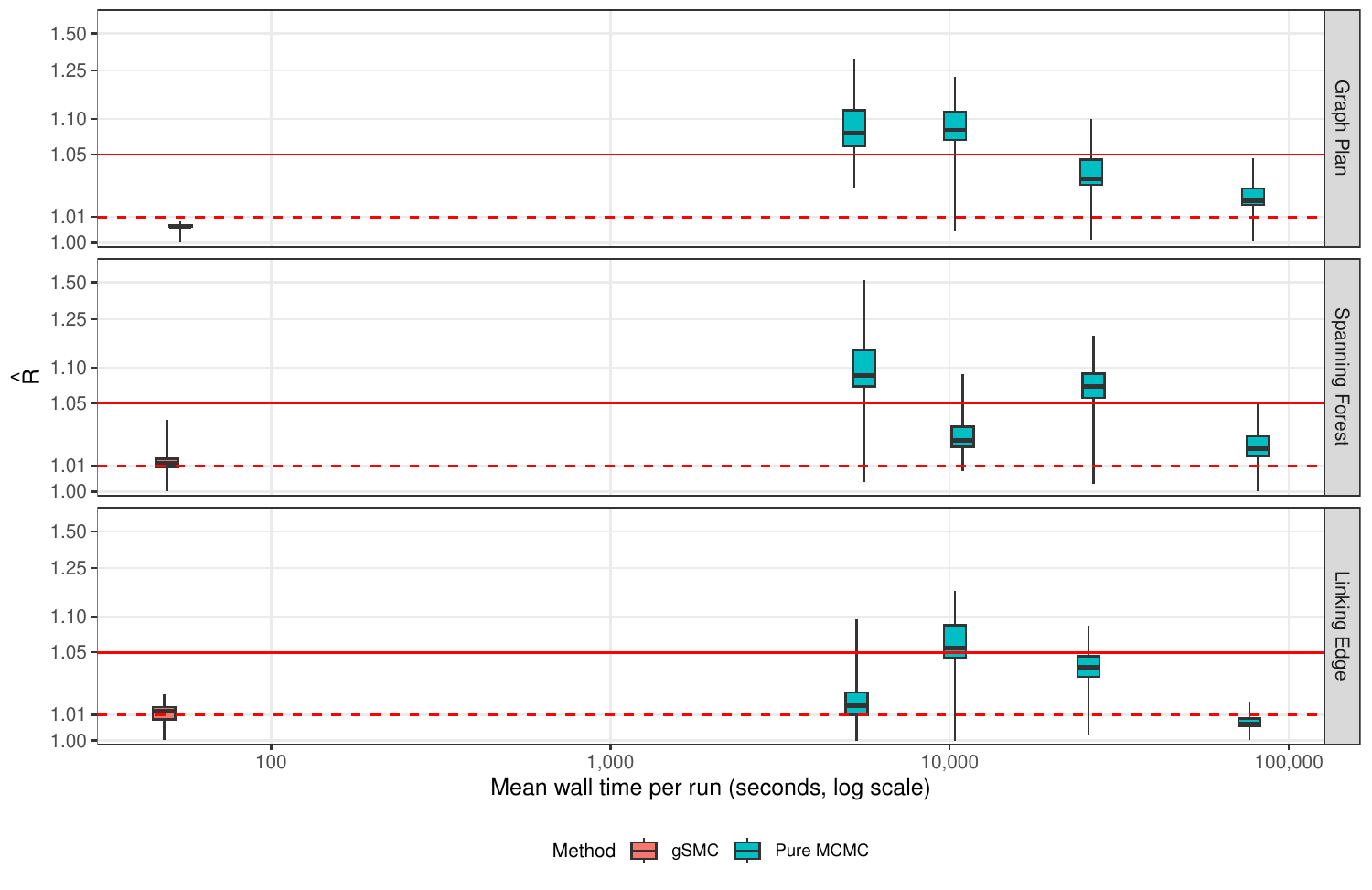}

}

\caption{\label{fig-bal-ar-mcmc-gsmc-walltime-rhats}Balanced Arkansas
Congressional: Distribution of \(\hat{R}\) values versus mean wall time
for all MCMC runs, faceted by sampling space. For each sampling space,
if it exists, the gSMC or gSMC-MCMC configuration with the lowest mean
wall time among all configurations with all \(\hat{R}\) values below the
1.05 threshold is also displayed.}

\end{figure}%

\end{minipage}%
\newline
\begin{minipage}{\linewidth}

\begin{figure}[H]

\centering{

\includegraphics[width=0.93\linewidth,height=\textheight,keepaspectratio]{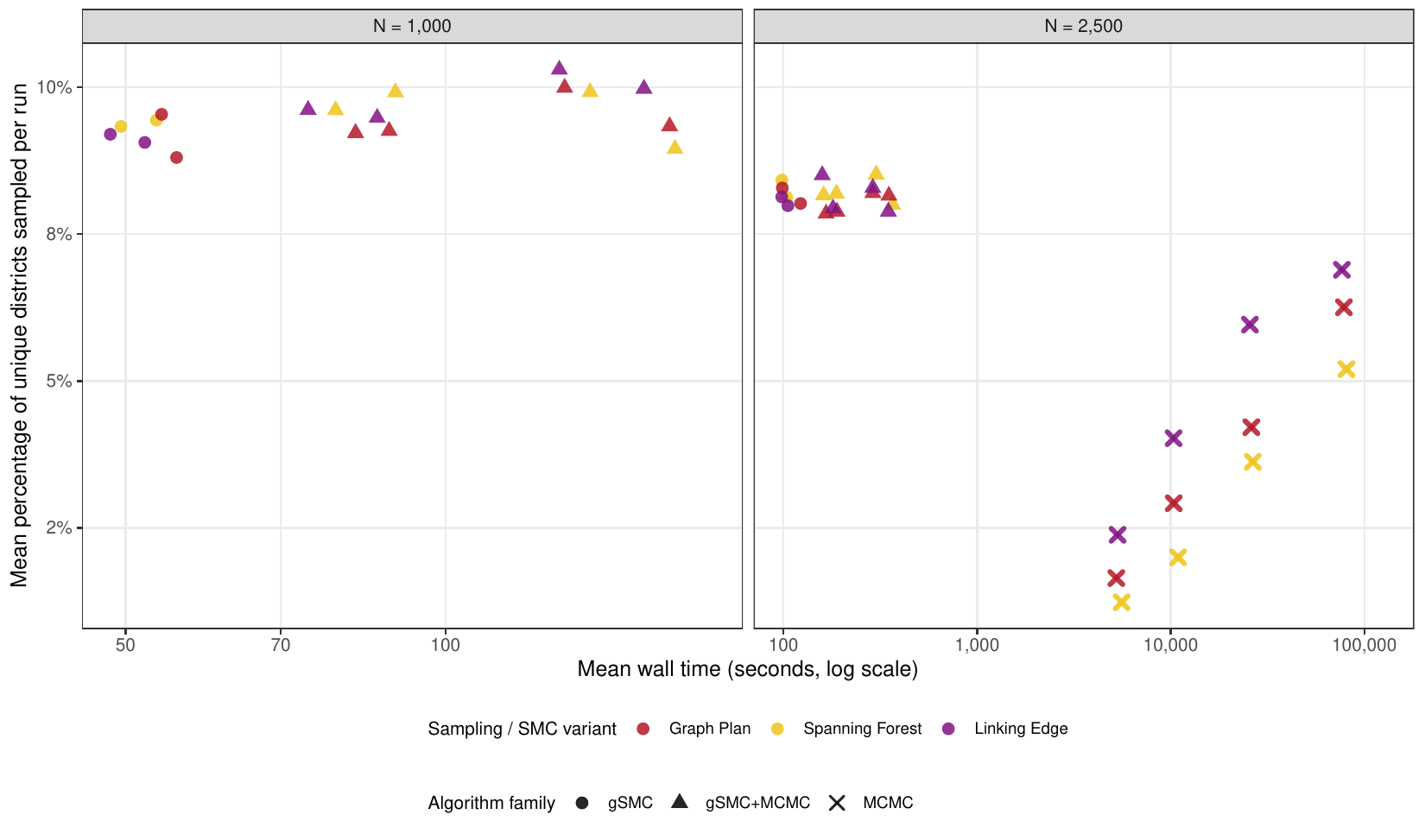}

}

\caption{\label{fig-bal-ar-distr-div}Balanced Arkansas Congressional:
Mean percentage of unique districts per run versus mean wall time for
all gSMC, gSMC-MCMC, and MCMC configurations run.}

\end{figure}%

\end{minipage}%

\end{figure}%

\subsubsection{Balanced Wisconsin 2020 Congressional
Map}\label{balanced-wisconsin-2020-congressional-map}

We use the same map as in Section~\ref{sec-wi-cd-2020} but limit the
maximum population deviation to \(736,715 \pm 1\), so any two districts
differ by at most two people. Just as in Section~\ref{sec-wi-cd-2020},
we sample with \(\rho=1\), use pseudo-counties as the administrative
units for hierarchical sampling, and impose no additional constraints
through the \(J\) term. As presented in
Figure~\ref{fig-bal-wi-gsmc-rhats}, for gSMC and gSMC-MCMC variants we
sample some, but not all possible configurations, for sample sizes of
1,000, 10,000, and 50,000. All MCMC chains were run with 50,000 samples.
Table \ref{tab:mcmc-config-balanced-wi-cd-2020} lists all the MCMC chain
configurations ran.

\begin{table}[H]
\centering
\small
\renewcommand{\arraystretch}{1.1}
\setlength{\tabcolsep}{5pt}
\resizebox{\linewidth}{!}{%
\begin{tabular}{@{}l r r r r r@{}}
\hline
\textbf{Sampling Space} & \textbf{Sample Size} & \textbf{Warmup Steps} & \textbf{Thinning} & \textbf{Total Steps} & \textbf{Mean Wall Time} \\
\hline
Graph Plan & 50,000 & 10,000 & 100 & 5,010,000 & 19.2 min \\
Graph Plan & 50,000 & 10,000 & 16,000 & 800,010,000 & 47.6 hr \\
Spanning Forest & 50,000 & 10,000 & 100 & 5,010,000 & 42.8 min \\
Spanning Forest & 50,000 & 10,000 & 6,000 & 300,010,000 & 37.3 hr \\
Linking Edge & 50,000 & 10,000 & 100 & 5,010,000 & 12.9 min \\
Linking Edge & 50,000 & 10,000 & 27,000 & 1,350,010,000 & 57.5 hr \\
\hline
\end{tabular}
}
\caption{MCMC configurations for Wisconsin (8 districts, ± 1 person). Total steps equal warmup plus sample size times thinning.}
\label{tab:mcmc-config-balanced-wi-cd-2020}
\end{table}

\begin{figure}[H]

\centering{

\pandocbounded{\includegraphics[keepaspectratio]{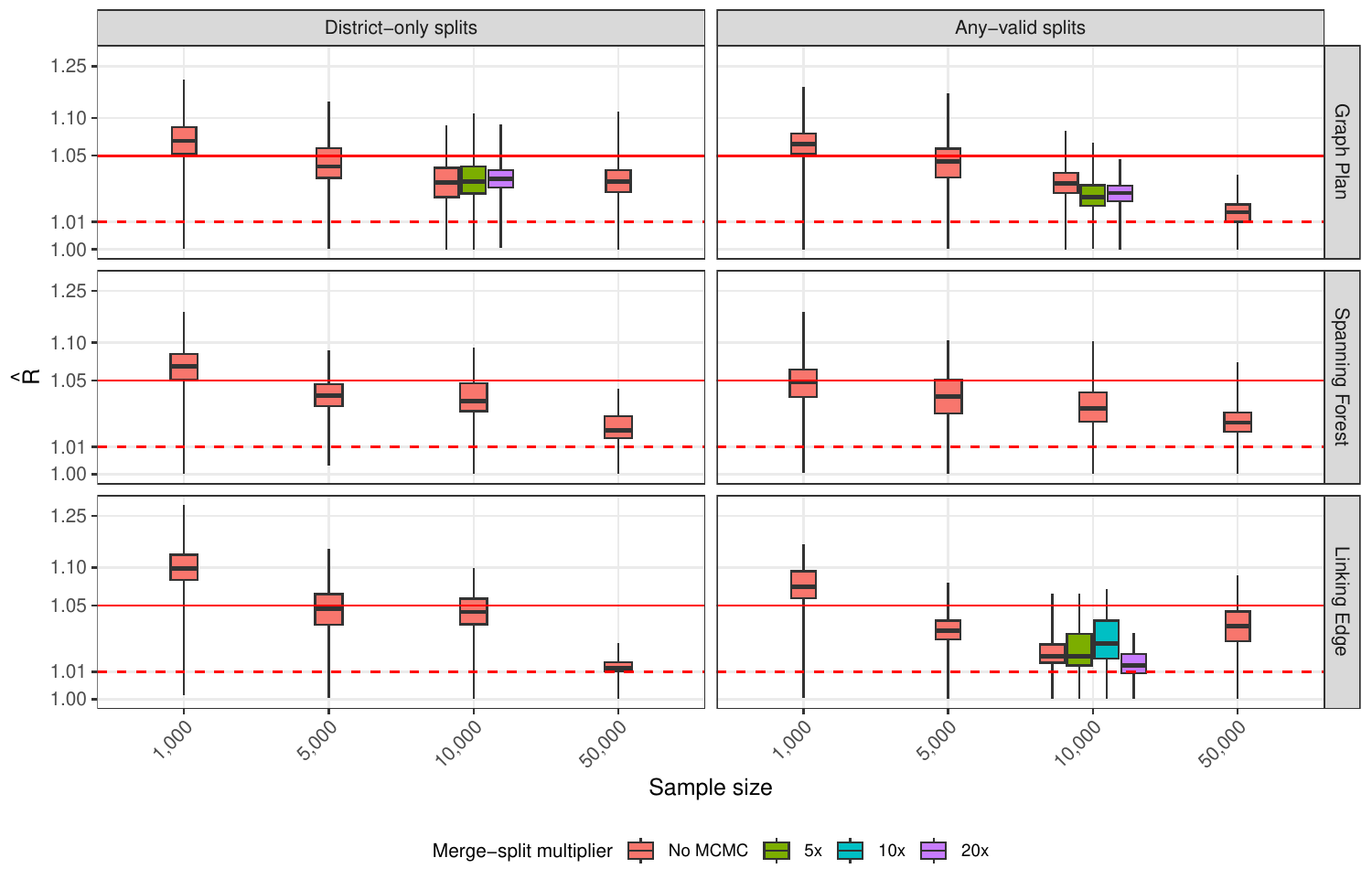}}

}

\caption{\label{fig-bal-wi-gsmc-rhats}Balanced Wisconsin Congressional:
Distribution of \(\hat{R}\) values for all gSMC and gSMC-MCMC
configurations run, faceted by sampling space and splitting schedule.}

\end{figure}%

\begin{figure}[H]

\begin{minipage}{\linewidth}

\begin{figure}[H]

\centering{

\includegraphics[width=0.95\linewidth,height=\textheight,keepaspectratio]{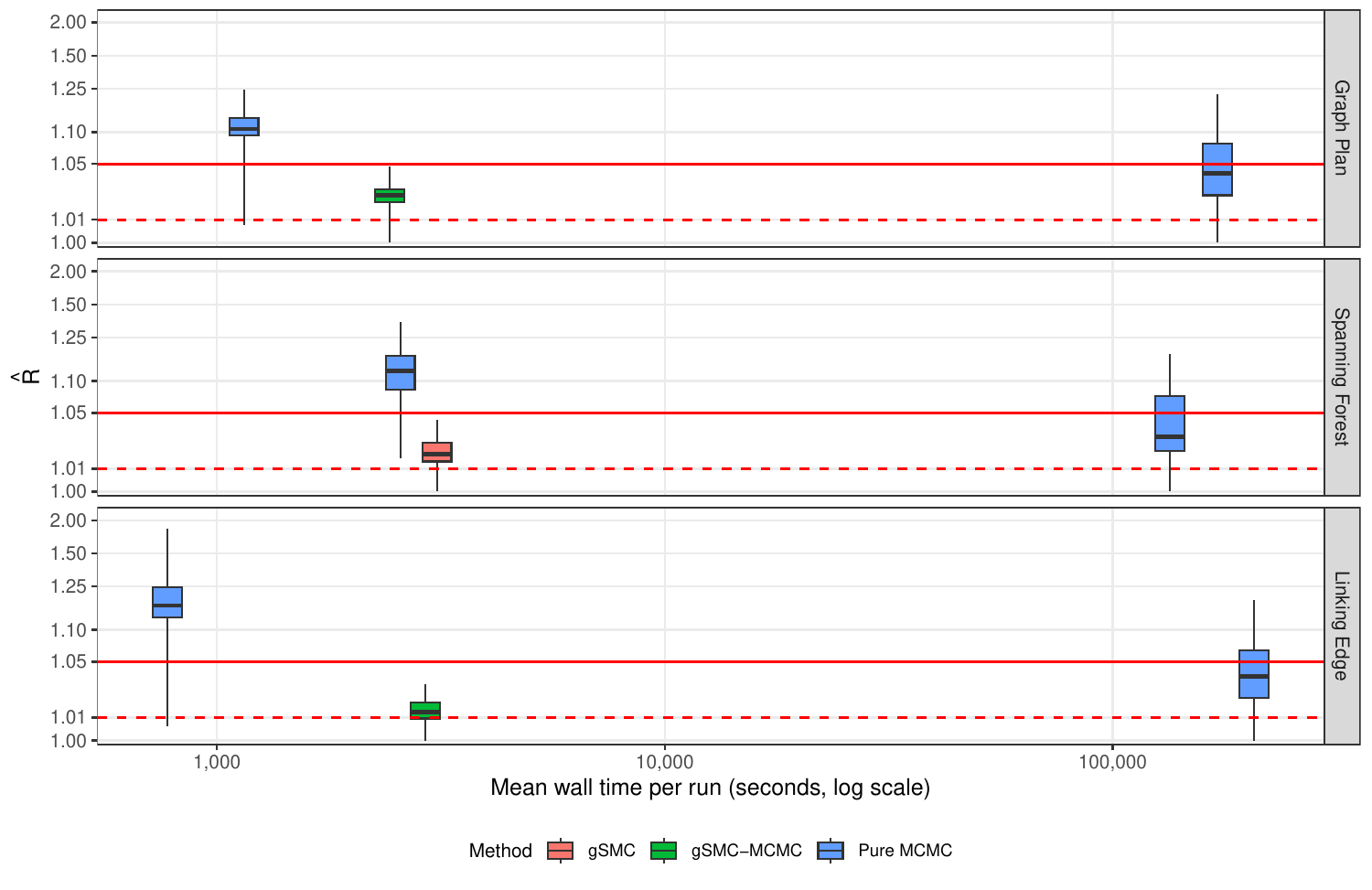}

}

\caption{\label{fig-bal-wi-mcmc-gsmc-walltime-rhats}Balanced Wisconsin
Congressional: Distribution of \(\hat{R}\) values versus mean wall time
for all MCMC runs, faceted by sampling space. For each sampling space,
if it exists, the gSMC or gSMC-MCMC configuration with the lowest mean
wall time among all configurations with all \(\hat{R}\) values below the
1.05 threshold is also displayed.}

\end{figure}%

\end{minipage}%
\newline
\begin{minipage}{\linewidth}

\begin{figure}[H]

\centering{

\includegraphics[width=0.93\linewidth,height=\textheight,keepaspectratio]{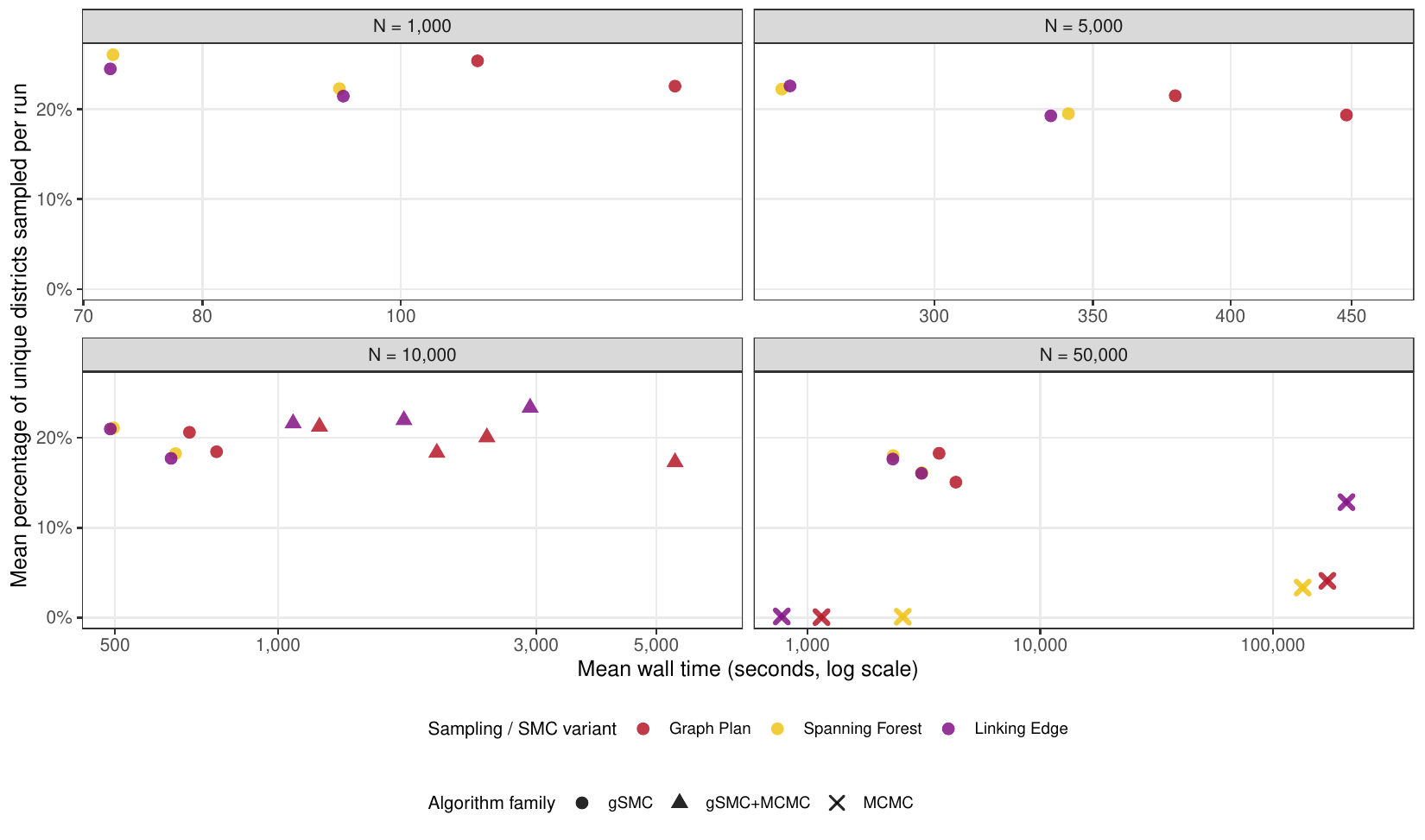}

}

\caption{\label{fig-bal-wi-distr-div}Balanced Wisconsin Congressional:
Mean percentage of unique districts per run versus mean wall time for
all gSMC, gSMC-MCMC, and MCMC configurations run.}

\end{figure}%

\end{minipage}%

\end{figure}%

\subsubsection{Balanced Tennessee 2020 Congressional
Map}\label{balanced-tennessee-2020-congressional-map}

We use the same map as in Section~\ref{sec-tn-cd-2020} but limit the
maximum population deviation to \(767,871 + 1\), so any two districts
differ by at most one person. Just as in Section~\ref{sec-tn-cd-2020},
we sample with \(\rho=1\), use pseudo-counties as the administrative
units for hierarchical sampling, and impose no additional constraints
through the \(J\) term. As presented in
Figure~\ref{fig-bal-tn-gsmc-rhats}, for gSMC and gSMC-MCMC variants we
sample some, but not all possible configurations, for sample sizes of
1,000, 5,000, 10,000, 20,000 and 50,000. All MCMC chains were run with
50,000 samples. Table \ref{tab:mcmc-config-balanced-tn-cd-2020} lists
all the MCMC chain configurations ran.

\begin{table}[H]
\centering
\small
\renewcommand{\arraystretch}{1.1}
\setlength{\tabcolsep}{5pt}
\resizebox{\linewidth}{!}{%
\begin{tabular}{@{}l r r r r r@{}}
\hline
\textbf{Sampling Space} & \textbf{Sample Size} & \textbf{Warmup Steps} & \textbf{Thinning} & \textbf{Total Steps} & \textbf{Mean Wall Time} \\
\hline
Graph Plan & 50,000 & 10,000 & 185 & 9,260,000 & 9.9 min \\
Graph Plan & 50,000 & 10,000 & 12,210 & 610,510,000 & 10.8 hr \\
Graph Plan & 50,000 & 10,000 & 24,420 & 1,221,010,000 & 21.6 hr \\
Spanning Forest & 50,000 & 10,000 & 75 & 3,760,000 & 8.1 min \\
Spanning Forest & 50,000 & 10,000 & 5,625 & 281,260,000 & 9.9 hr \\
Spanning Forest & 50,000 & 10,000 & 11,250 & 562,510,000 & 19.6 hr \\
Linking Edge & 50,000 & 10,000 & 175 & 8,760,000 & 6.9 min \\
Linking Edge & 50,000 & 10,000 & 16,450 & 822,510,000 & 10.5 hr \\
Linking Edge & 50,000 & 10,000 & 32,900 & 1,645,010,000 & 21.0 hr \\
\hline
\end{tabular}
}
\caption{MCMC configurations for Tennessee (9 districts, + 1 person). Total steps equal warmup plus sample size times thinning.}
\label{tab:mcmc-config-balanced-tn-cd-2020}
\end{table}

\begin{figure}[H]

\centering{

\pandocbounded{\includegraphics[keepaspectratio]{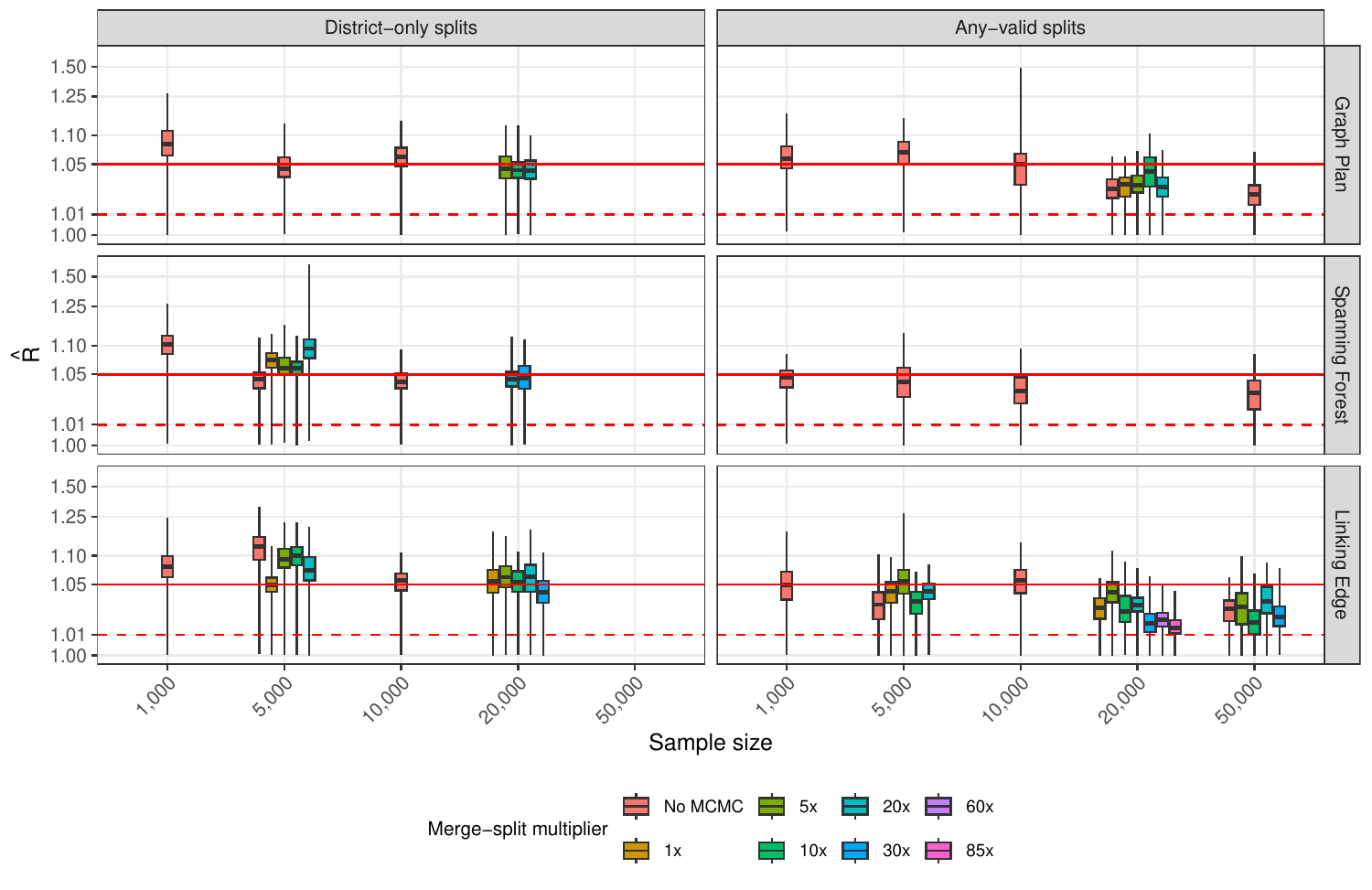}}

}

\caption{\label{fig-bal-tn-gsmc-rhats}Balanced Tennessee Congressional:
Distribution of \(\hat{R}\) values for all gSMC and gSMC-MCMC
configurations run, faceted by sampling space and splitting schedule.}

\end{figure}%

\begin{figure}[H]

\begin{minipage}{\linewidth}

\begin{figure}[H]

\centering{

\includegraphics[width=0.95\linewidth,height=\textheight,keepaspectratio]{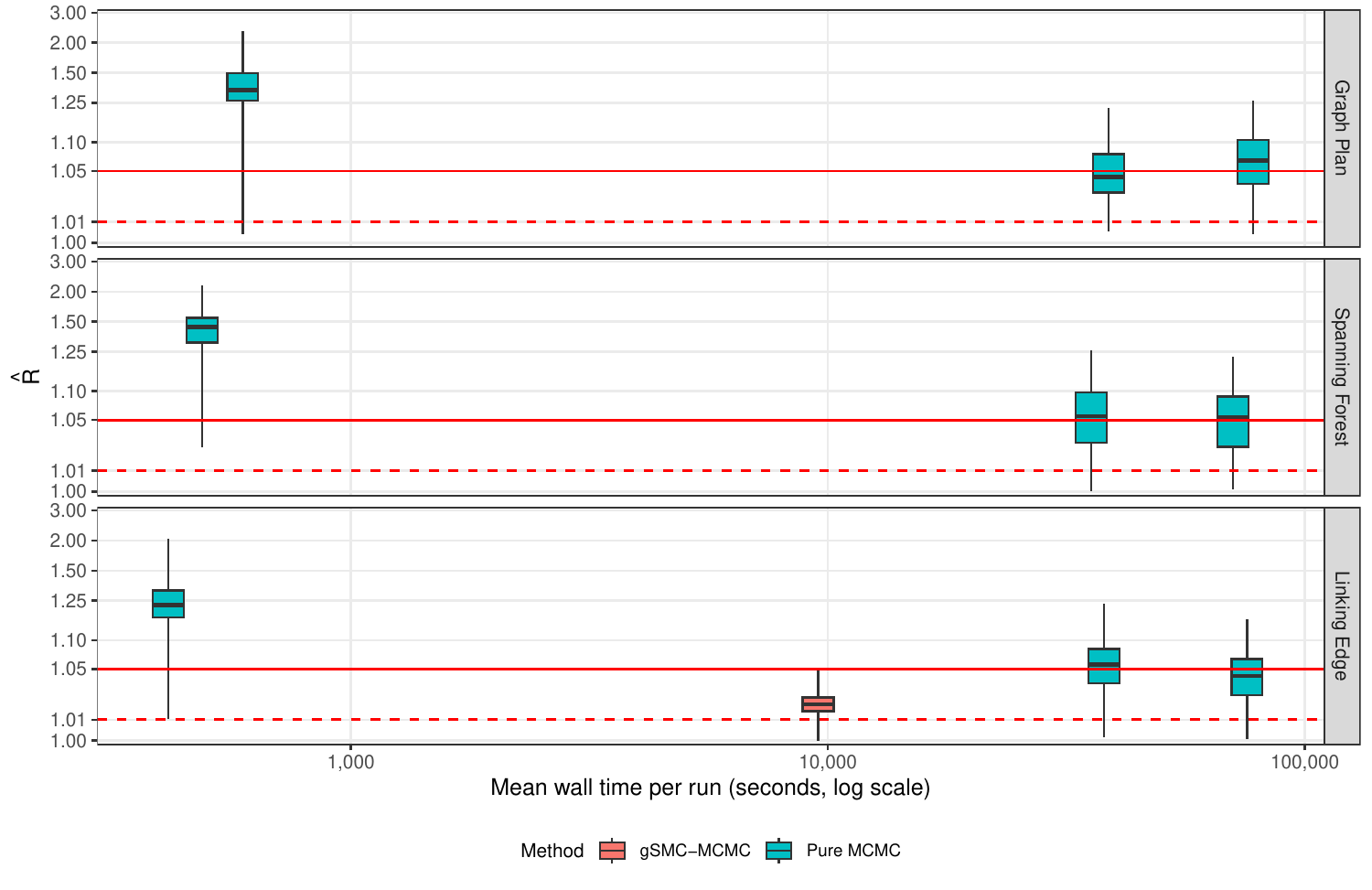}

}

\caption{\label{fig-bal-tn-mcmc-gsmc-walltime-rhats}Balanced Tennessee
Congressional: Distribution of \(\hat{R}\) values versus mean wall time
for all MCMC runs, faceted by sampling space. For each sampling space,
if it exists, the gSMC or gSMC-MCMC configuration with the lowest mean
wall time among all configurations with all \(\hat{R}\) values below the
1.05 threshold is also displayed.}

\end{figure}%

\end{minipage}%
\newline
\begin{minipage}{\linewidth}

\begin{figure}[H]

\centering{

\includegraphics[width=0.93\linewidth,height=\textheight,keepaspectratio]{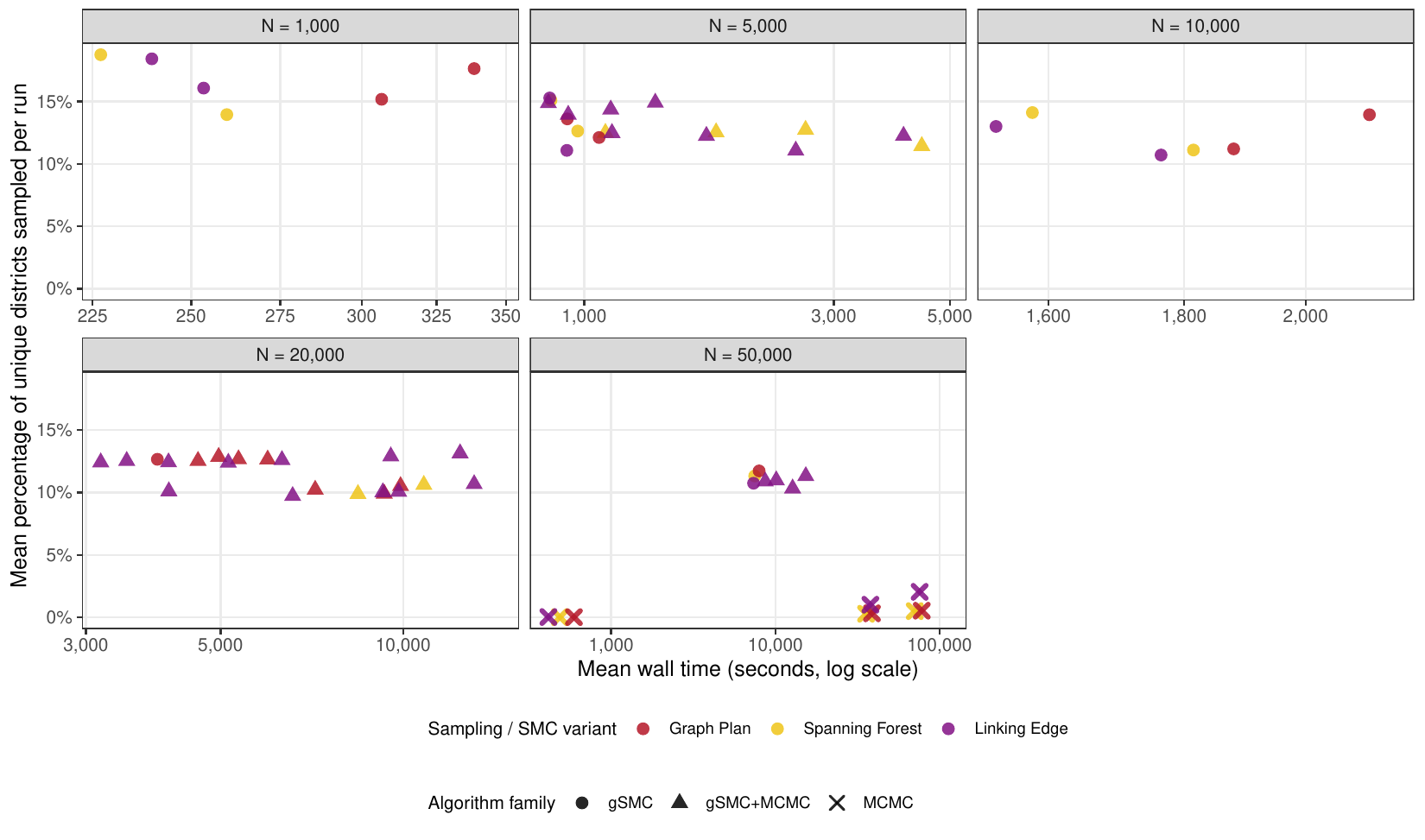}

}

\caption{\label{fig-bal-tn-distr-div}Balanced Tennessee Congressional:
Mean percentage of unique districts per run versus mean wall time for
all gSMC, gSMC-MCMC, and MCMC configurations run.}

\end{figure}%

\end{minipage}%

\end{figure}%

\subsubsection{Balanced Illinois 2020 Congressional
Map}\label{balanced-illinois-2020-congressional-map}

We use the same map as in Section~\ref{sec-il-cd-2020} but limit the
maximum population deviation to \(753,677 \pm 1\), so any two districts
differ by at most two people. Just as in Section~\ref{sec-il-cd-2020},
we sample with \(\rho=1\), use pseudo-counties as the administrative
units for hierarchical sampling, and impose no additional constraints
through the \(J\) term. As presented in
Figure~\ref{fig-bal-il-gsmc-rhats}, for gSMC and gSMC-MCMC variants we
sample some, but not all possible configurations, for sample sizes of
1,000, 5,000, 10,000, and 50,000. All MCMC chains were run with 50,000
samples. Table \ref{tab:mcmc-config-il-cd-2020} lists all the MCMC chain
configurations ran.

\begin{table}[H]
\centering
\small
\renewcommand{\arraystretch}{1.1}
\setlength{\tabcolsep}{5pt}
\resizebox{\linewidth}{!}{%
\begin{tabular}{@{}l r r r r r@{}}
\hline
\textbf{Sampling Space} & \textbf{Sample Size} & \textbf{Warmup Steps} & \textbf{Thinning} & \textbf{Total Steps} & \textbf{Mean Wall Time} \\
\hline
Graph Plan & 50,000 & 10,000 & 10 & 510,000 & 2.1 min \\
Graph Plan & 50,000 & 10,000 & 18,000 & 900,010,000 & 60.0 hr \\
Spanning Forest & 50,000 & 10,000 & 30 & 1,510,000 & 9.2 min \\
Spanning Forest & 50,000 & 10,000 & 12,750 & 637,510,000 & 72.2 hr \\
Linking Edge & 50,000 & 10,000 & 100 & 5,010,000 & 12.3 min \\
Linking Edge & 50,000 & 10,000 & 27,500 & 1,375,010,000 & 54.7 hr \\
\hline
\end{tabular}
}
\caption{MCMC configurations for Illinois (17 districts, ± 1 person). Total steps equal warmup plus sample size times thinning.}
\label{tab:mcmc-config-il-cd-2020}
\end{table}

\begin{figure}[H]

\centering{

\pandocbounded{\includegraphics[keepaspectratio]{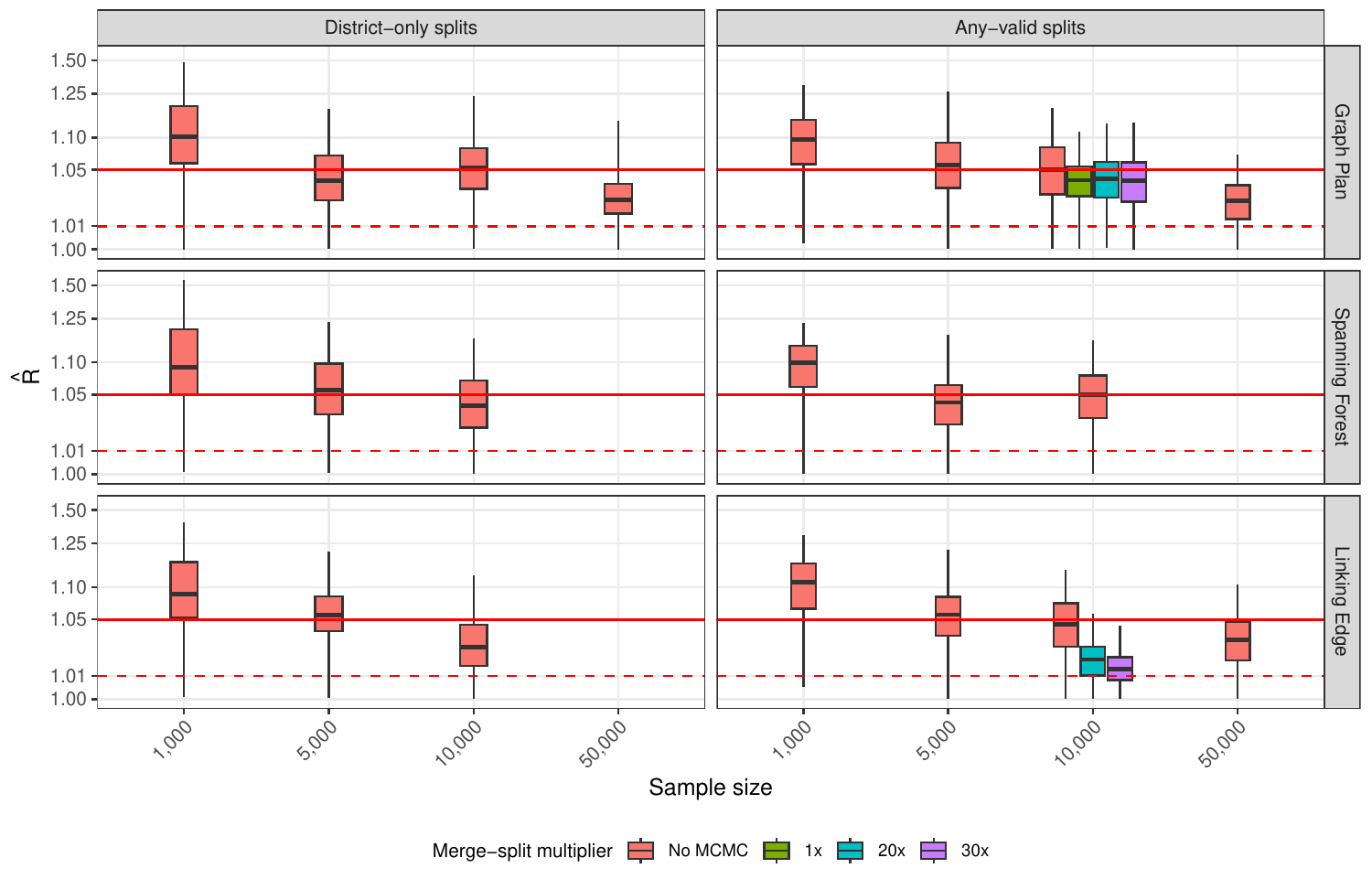}}

}

\caption{\label{fig-bal-il-gsmc-rhats}Balanced Illinois Congressional:
Distribution of \(\hat{R}\) values for all gSMC and gSMC-MCMC
configurations run, faceted by sampling space and splitting schedule.}

\end{figure}%

\begin{figure}[H]

\begin{minipage}{\linewidth}

\begin{figure}[H]

\centering{

\includegraphics[width=0.95\linewidth,height=\textheight,keepaspectratio]{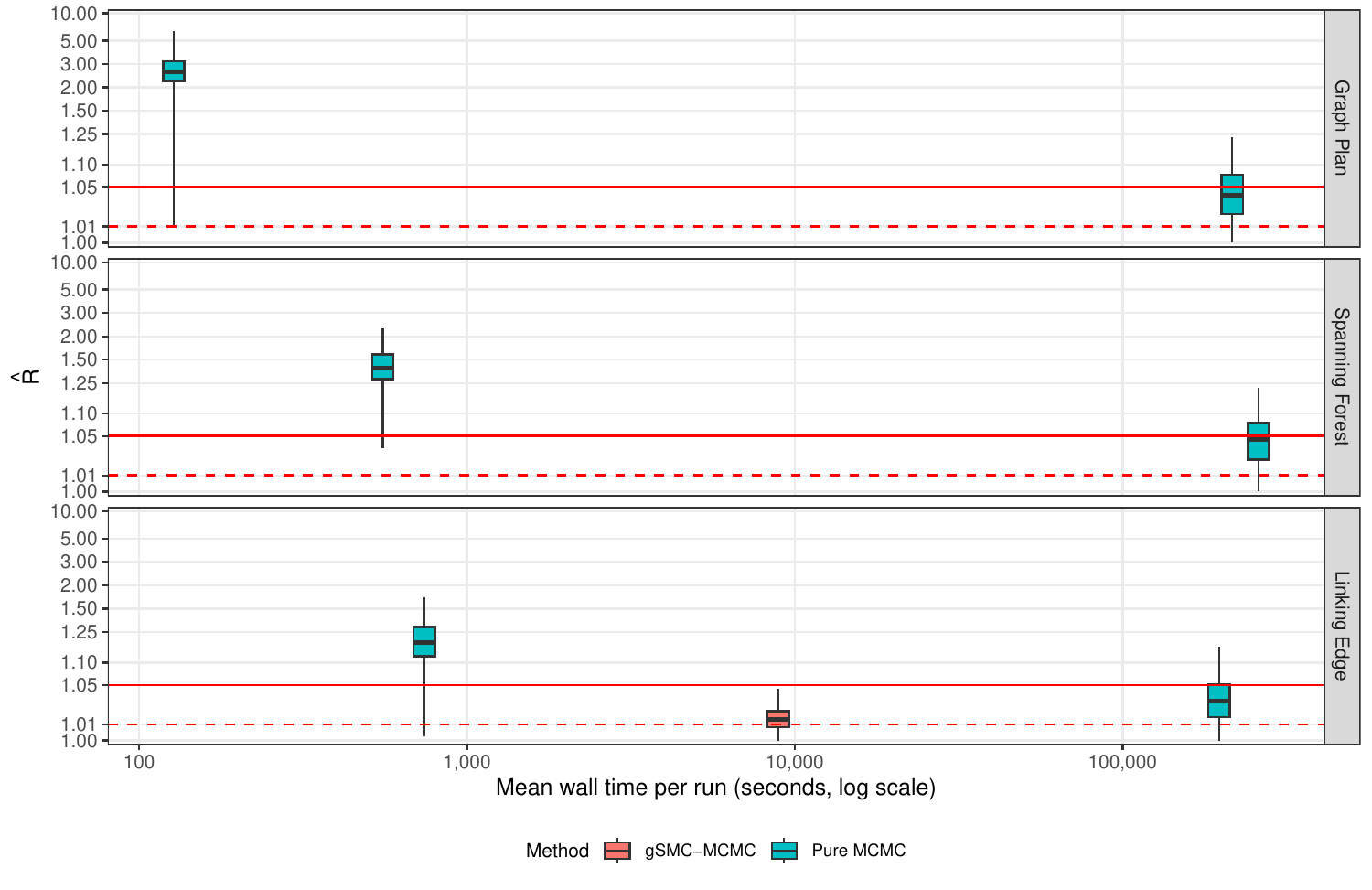}

}

\caption{\label{fig-bal-il-mcmc-gsmc-walltime-rhats}Balanced Illinois
Congressional: Distribution of \(\hat{R}\) values versus mean wall time
for all MCMC runs, faceted by sampling space. For each sampling space,
if it exists, the gSMC or gSMC-MCMC configuration with the lowest mean
wall time among all configurations with all \(\hat{R}\) values below the
1.05 threshold is also displayed.}

\end{figure}%

\end{minipage}%
\newline
\begin{minipage}{\linewidth}

\begin{figure}[H]

\centering{

\includegraphics[width=0.93\linewidth,height=\textheight,keepaspectratio]{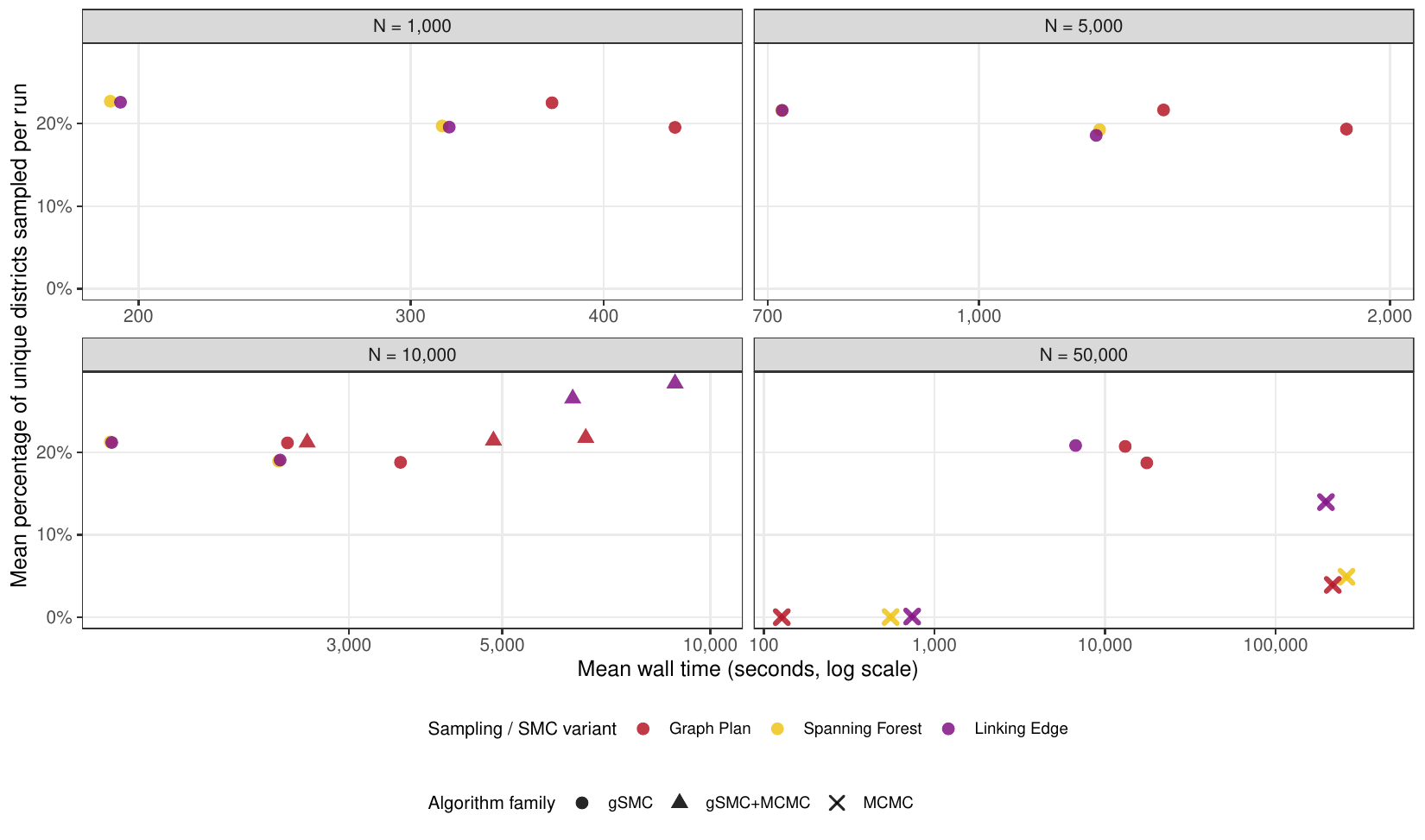}

}

\caption{\label{fig-bal-il-distr-div}Balanced Illinois Congressional:
Mean percentage of unique districts per run versus mean wall time for
all gSMC, gSMC-MCMC, and MCMC configurations run.}

\end{figure}%

\end{minipage}%

\end{figure}%

\counterwithout*{figure}{subsubsection}
\counterwithin*{figure}{subsection}
\renewcommand{\thefigure}{\thesubsection.\arabic{figure}}

\newpage

\subsection{Compactness Sensitivity}\label{sec-rho-tests}

As mentioned in Section~\ref{sec-target}, the performance of algorithms
tends to degrade the further the value of \(\rho\) gets from 1. In this
section, we empirically investigate this on the five maps summarized in
Table \ref{tbl-rho-sensitivity-settings}. We examine performance for a
fixed sample size over a range of \(\rho\) values. For all maps but
Pennsylvania Congressional, we set the sample size such that gSMC with
no MCMC steps converges for \(\rho=1\).

\begin{center}
\begin{minipage}{\textwidth}

\captionof{table}{Sensitivity analyses across states and values of $\rho$. Population bounds, $J$ constraints, and administrative units are all the same as in Table \ref{tbl-sys-empirical-examples}. For Iowa the population bounds were $\approx 797,592 \pm 80$ with no additional $J$ constraints.}
\label{tbl-rho-sensitivity-settings}

\centering
\resizebox{\textwidth}{!}{%
\begin{tabular}{lrrrrrrr}
\toprule
Name &
Districts &
$\rho$ Range Tested &
Vertices &
Edges &
Sample Size &
\multicolumn{2}{c}{Convergent $\rho$ Range} \\
\cmidrule(lr){7-8}
&
&
&
&
&
&
gSMC only &
gSMC--MCMC \\
\midrule
Iowa Congressional
    & 4  & $[0.5, 5]$    & 99    & 222    & 5,000  & $[.75, 5]$ & $[.75, 5]$ \\
Arkansas Congressional
    & 4  & $[0, 5]$      & 2,747 & 7,509  & 5,000  & $[.85, 1.25]$ & $[.85, 1.5]$ \\
Wisconsin Congressional
    & 8  & $[0.5, 2]$    & 7,059 & 19,281 & 5,000  & $[.9, 1.1]$ & $[.9, 1.5]$ \\
Tennessee Congressional
    & 9  & $[0.5, 2]$    & 1,965 & 5,604  & 5,000  & $[.85, 1.25]$ & $[.75, 2]$ \\
Pennsylvania Congressional
    & 17 & $[0.75, 2]$   & 9,178 & 25,519 & 10,000 & $[.95, .95]$ & $[.95, 1.15]$ \\
\bottomrule
\end{tabular}%
}

\end{minipage}
\end{center}

In Table \ref{tbl-rho-sensitivity-settings} we also present the
``Convergent \(\rho\) Range'' for each map tested. This range is the
widest interval of \(\rho\) values where at least one gSMC/gSMC-MCMC
configuration had all \(\hat{R}\) values below the 1.05 threshold. In
Figures \ref{fig-ia-rho-rhats}, \ref{fig-ar-rho-rhats},
\ref{fig-wi-rho-rhats}, \ref{fig-tn-rho-rhats}, and
\ref{fig-pa-rho-rhats}, we plot the distribution of \(\hat{R}\) values
across \(\rho\) values faceted by sample space and splitting schedule
for gSMC and several different gSMC-MCMC runs for each state.

As the table demonstrates, generally the more districts and vertices in
the map, the smaller the convergent \(\rho\) range is and the bigger the
boost from adding MCMC steps. However, the precise relationship between
the two is hard to characterize. For example, Tennessee has more
districts but fewer vertices and edges than Arkansas and while its
convergent \(\rho\) range for gSMC alone is the same, the range with
gSMC-MCMC is wider.

There also appears to be an asymmetry in the algorithms performance in
neighborhoods around \(\rho=1\). All of the \(\hat{R}\) distribution
plots suggest demonstrate there is a general ``U''-like shape centered
at \(\rho=1\) to the performance but it is not symmetric. For the same
displacement from \(\rho=1\), the algorithm seems to do better if its
positive as opposed to negative, especially for gSMC-MCMC. For all the
gSMC-MCMC \(\rho\) convergence ranges, none of them are symmetric about
\(\rho=1\), instead converging for more \(\rho\) values greater than 1
as opposed to less than it. Figure~\ref{fig-ia-rho-rhats} is the most
striking example of this where by the time \(\rho = 5\) the distribution
is an almost degenerate point mass, with most sample just producing
5,000 copies of the same plan across runs. Across all the states, there
does not appear to be a strong trend in terms of which sampling space or
splitting schedule performs best.

Overall these results suggest that adding more MCMC steps can
dramatically improve performance as \(\rho\) gets further away from 1
but there is still likely a point where the number of additional MCMC
steps required for convergent becomes computationally infeasible. Future
research is needed to examine the trade-off between increasing the
sample size versus the number of MCMC steps when it comes to boosting
performance for non-unit \(\rho\) values.

Finally, in Figure~\ref{fig-comp-edge-vs-rho}, we show box plots of the
sample means of the edge-cut compactness values combined across all
configurations, for each value of \(\rho\) faceted by map. These plots
empirically demonstrate that adjusting \(\rho\) is a useful way to
adjust the edge-cut compactness of the sampled plans.

\begin{figure}[H]

\centering{

\includegraphics[width=1\linewidth,height=\textheight,keepaspectratio]{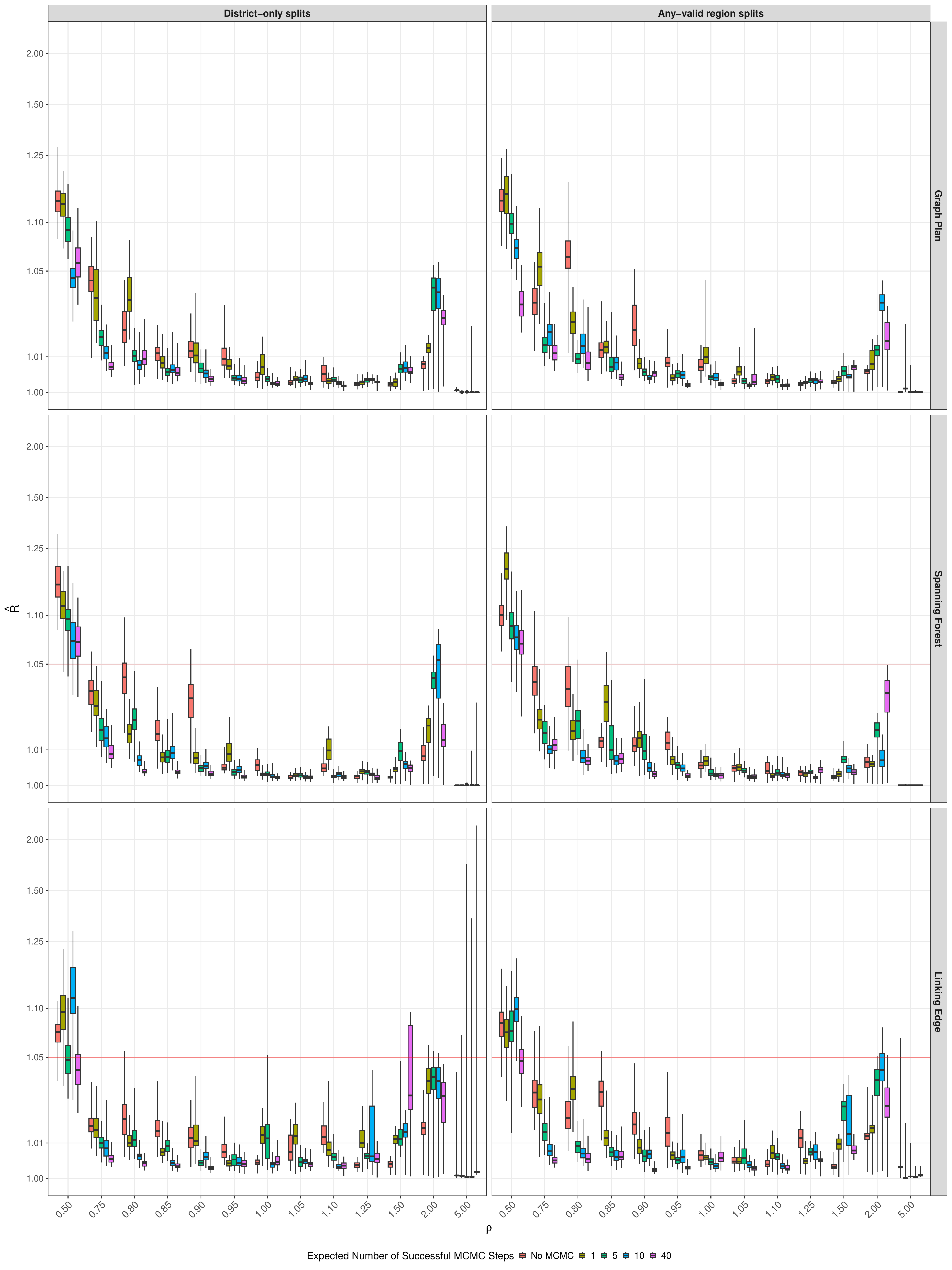}

}

\caption{\label{fig-ia-rho-rhats}Iowa Congressional: Distribution of
Rhat values across sampling space and splitting schedule for a range of
\(\rho\) values.}

\end{figure}%

\begin{figure}[H]

\centering{

\includegraphics[width=1\linewidth,height=\textheight,keepaspectratio]{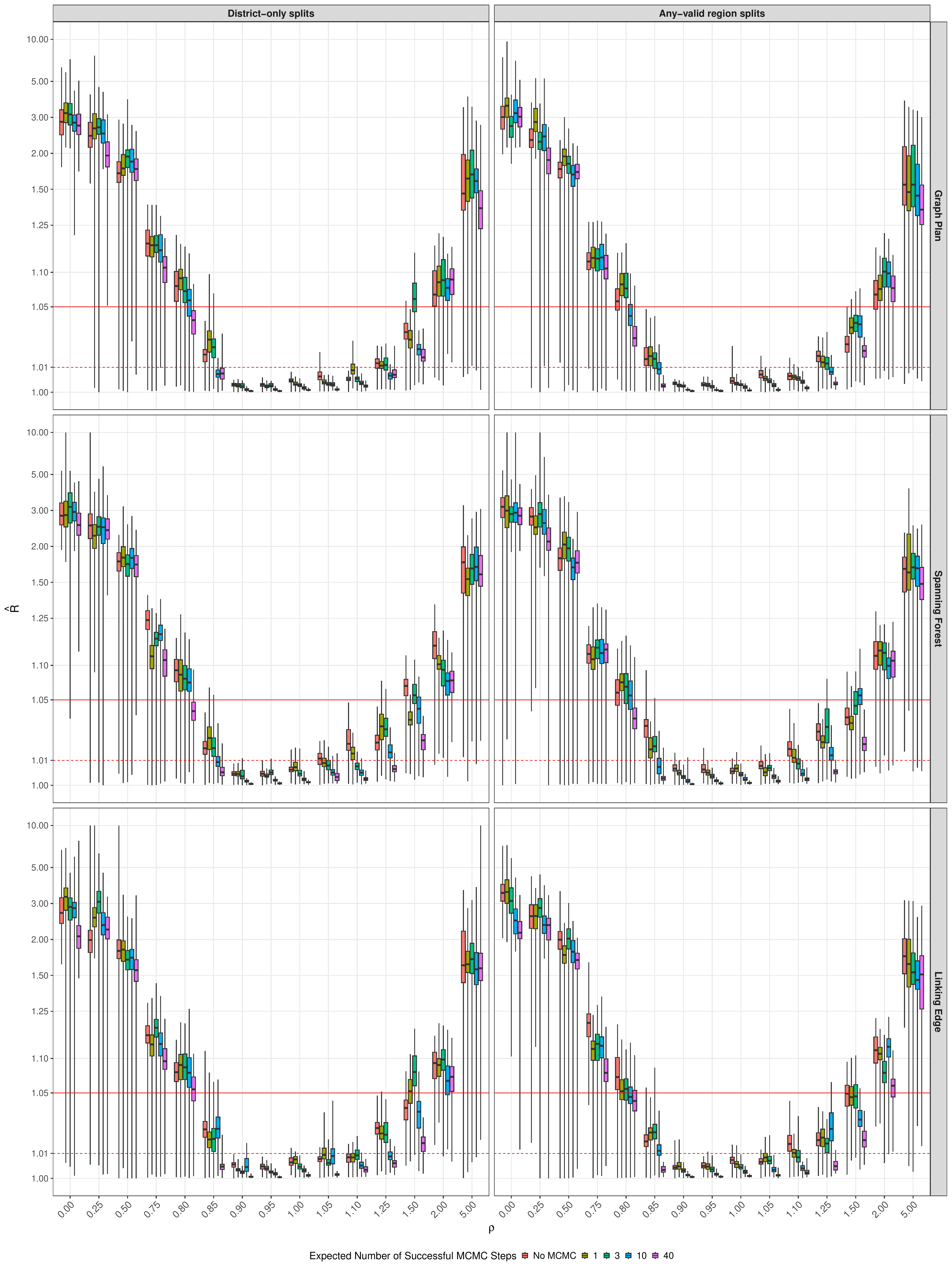}

}

\caption{\label{fig-ar-rho-rhats}Arkansas Congressional: Distribution of
Rhat values across sampling space and splitting schedule for a range of
\(\rho\) values. \(\hat{R}\) values larger than 10 trunacted at 10.}

\end{figure}%

\begin{figure}[H]

\centering{

\includegraphics[width=1\linewidth,height=\textheight,keepaspectratio]{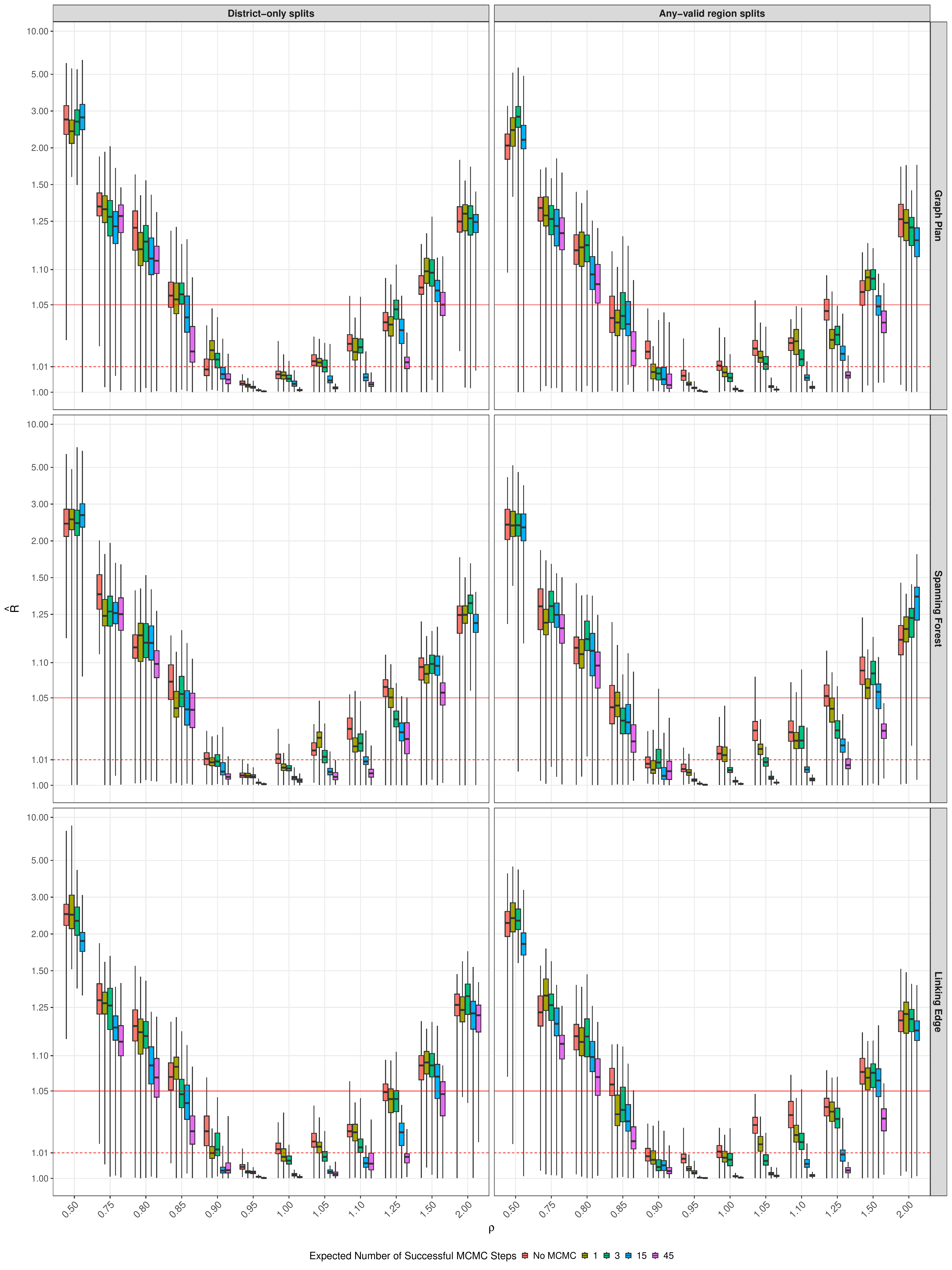}

}

\caption{\label{fig-wi-rho-rhats}Wisconsin Congressional: Distribution
of Rhat values across sampling space and splitting schedule for a range
of \(\rho\) values.}

\end{figure}%

\begin{figure}[H]

\centering{

\includegraphics[width=1\linewidth,height=\textheight,keepaspectratio]{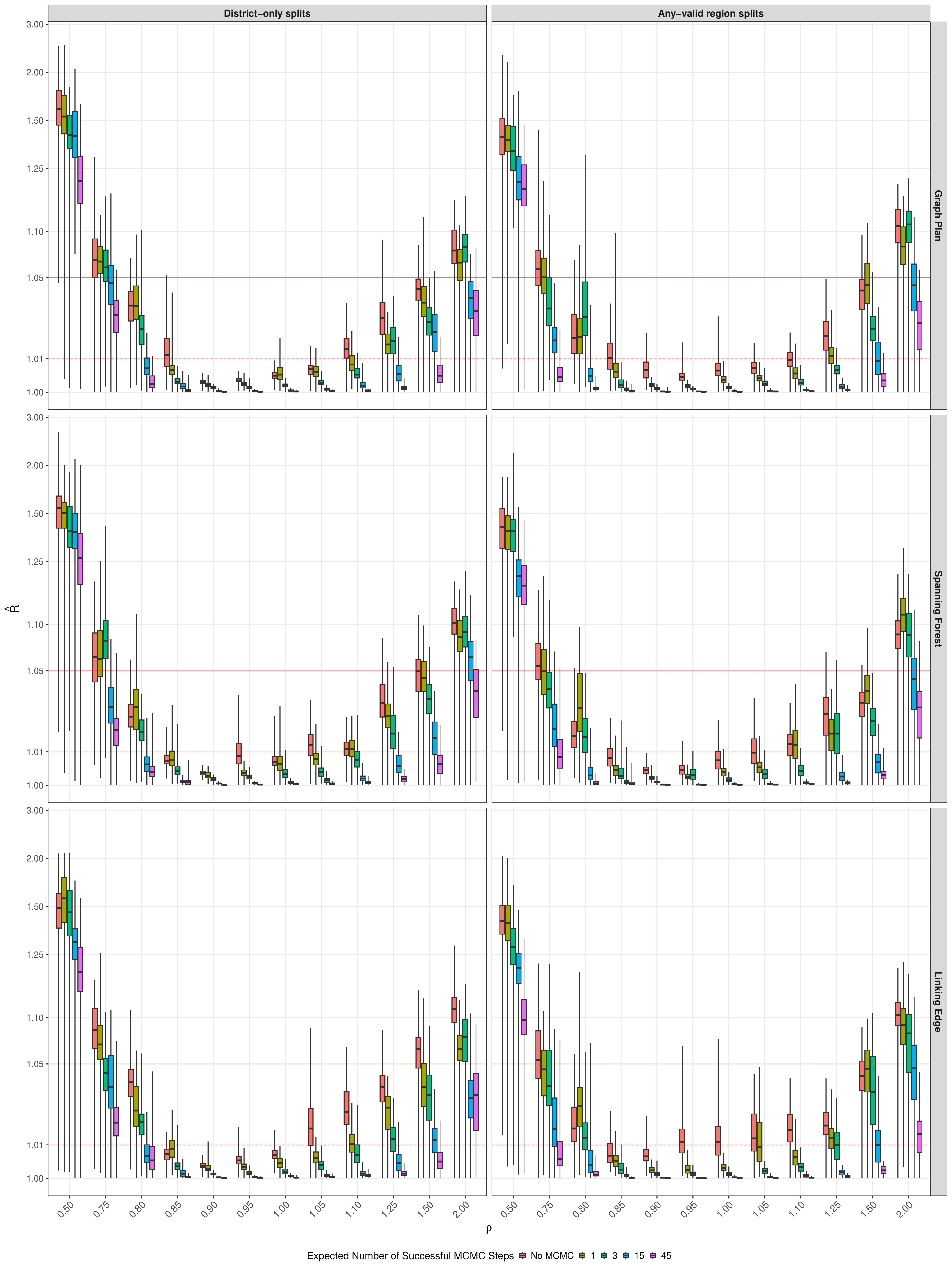}

}

\caption{\label{fig-tn-rho-rhats}Tennessee Congressional: Distribution
of Rhat values across sampling space and splitting schedule for a range
of \(\rho\) values.}

\end{figure}%

\begin{figure}[H]

\centering{

\includegraphics[width=1\linewidth,height=\textheight,keepaspectratio]{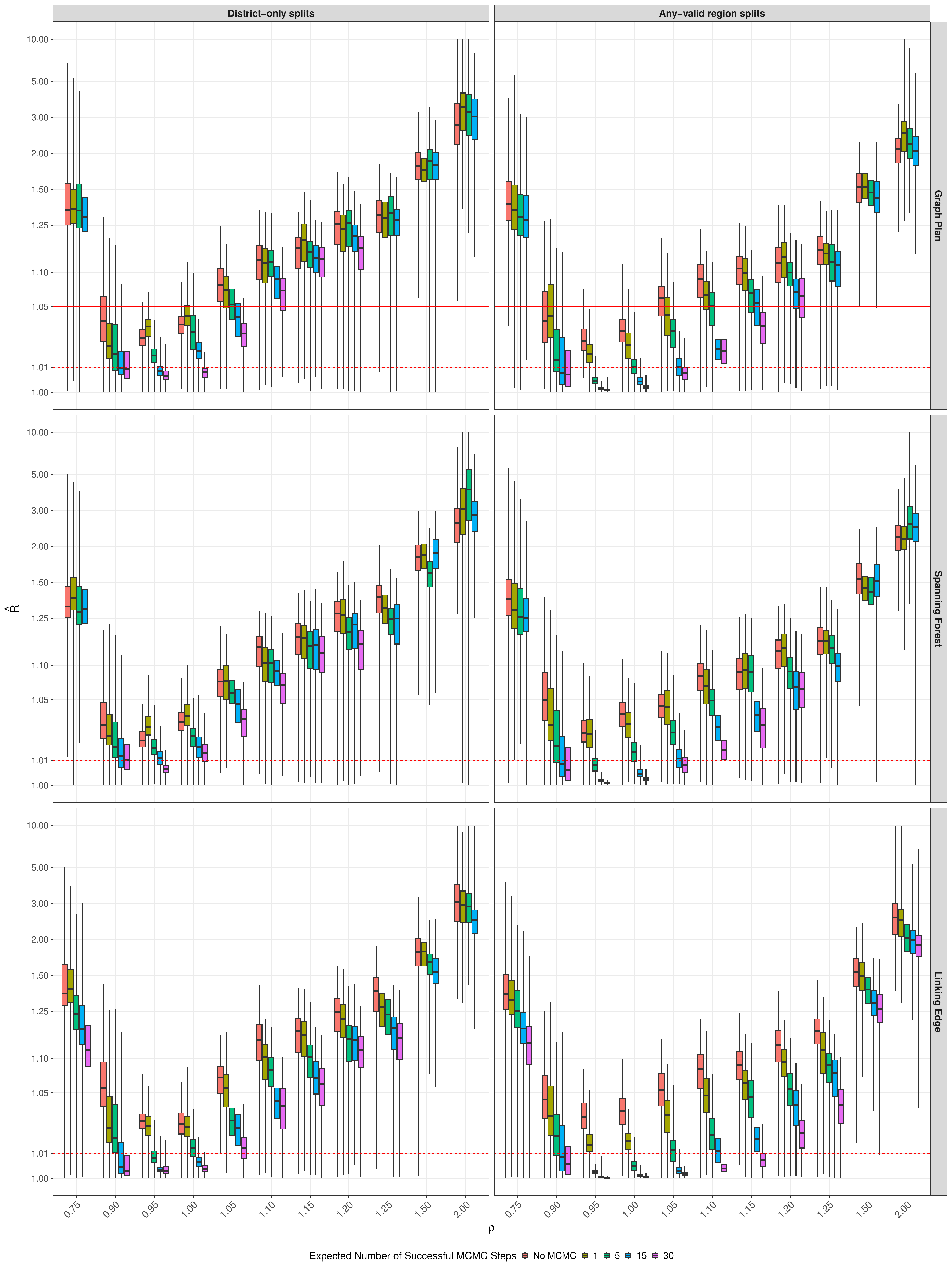}

}

\caption{\label{fig-pa-rho-rhats}Pennsylvania Congressional:
Distribution of Rhat values across sampling space and splitting schedule
for a range of \(\rho\) values. \(\hat{R}\) values larger than 10
trunacted at 10.}

\end{figure}%

\begin{figure}[H]

\centering{

\includegraphics[width=1\linewidth,height=\textheight,keepaspectratio]{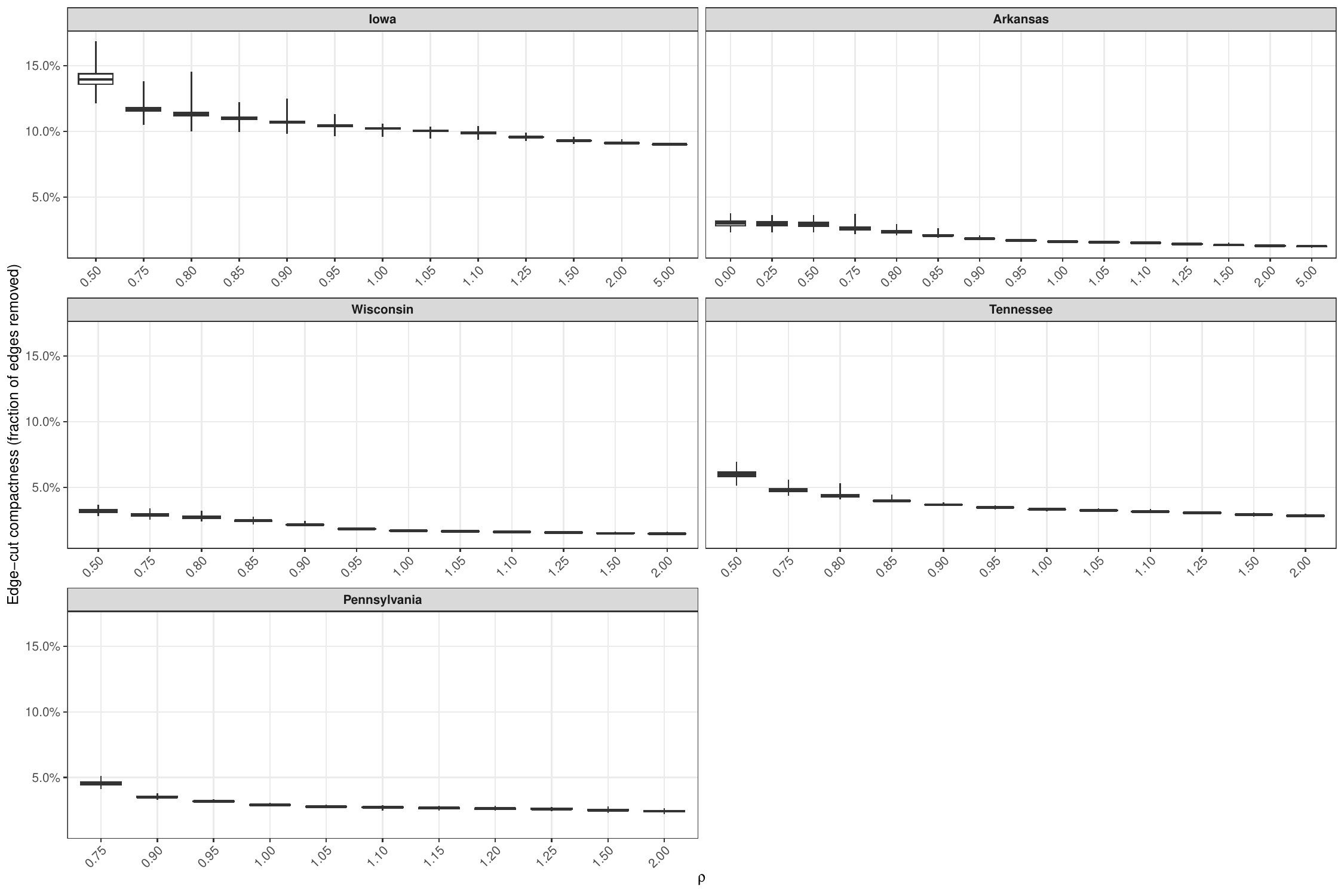}

}

\caption{\label{fig-comp-edge-vs-rho}Distribution of edge-cut sample
means across \(\rho\) values for each map. (Lower edge-cut value means
more compact.)}

\end{figure}%

\newpage

\subsection{Multidistrict Selection Sensitivity}\label{sec-varphi-tests}

In this section, we examine the influence of certain choices of the
\(\varphi\) function used to select a multidistrict to split in the
splitting step of Algorithm \ref{alg-gsmcs}. Specifically, we examine
\(\varphi\) where, for a multidistrict \((H_\ell, s_\ell) \in \xi_r\),
its probability of selection is proportional to its size raised to the
power \(\alpha\) for some \(\alpha \in \bbR\), i.e.,
\(\varphi((H_\ell, s_\ell) | \xi_r) \propto (s_\ell)^\alpha\).

Using the 2020 Wisconsin Congressional map from
Section~\ref{sec-wi-cd-2020}, we examine the performance of the
any-valid splitting schedule across all three sampling spaces for 30
independent runs of 5,000 samples each looking at
\(\alpha \in \{-.5, 0, .25, .5, .75, 1, 5\}\).
Figure~\ref{fig-varphi-rhats} displays the distribution of \(\hat{R}\)
values for each of the different choices of \(\alpha\) faceted by sample
space. Figure~\ref{fig-varphi-runtime} displays the distribution of
runtimes for each of the different choices of \(\alpha\) faceted by
sample space. Based on these two figures, both the runtime and
\(\hat{R}\) convergence metrics appear relatively insensitive to the
choice of \(\alpha\).

\begin{figure}[H]

\centering{

\includegraphics[width=1\linewidth,height=\textheight,keepaspectratio]{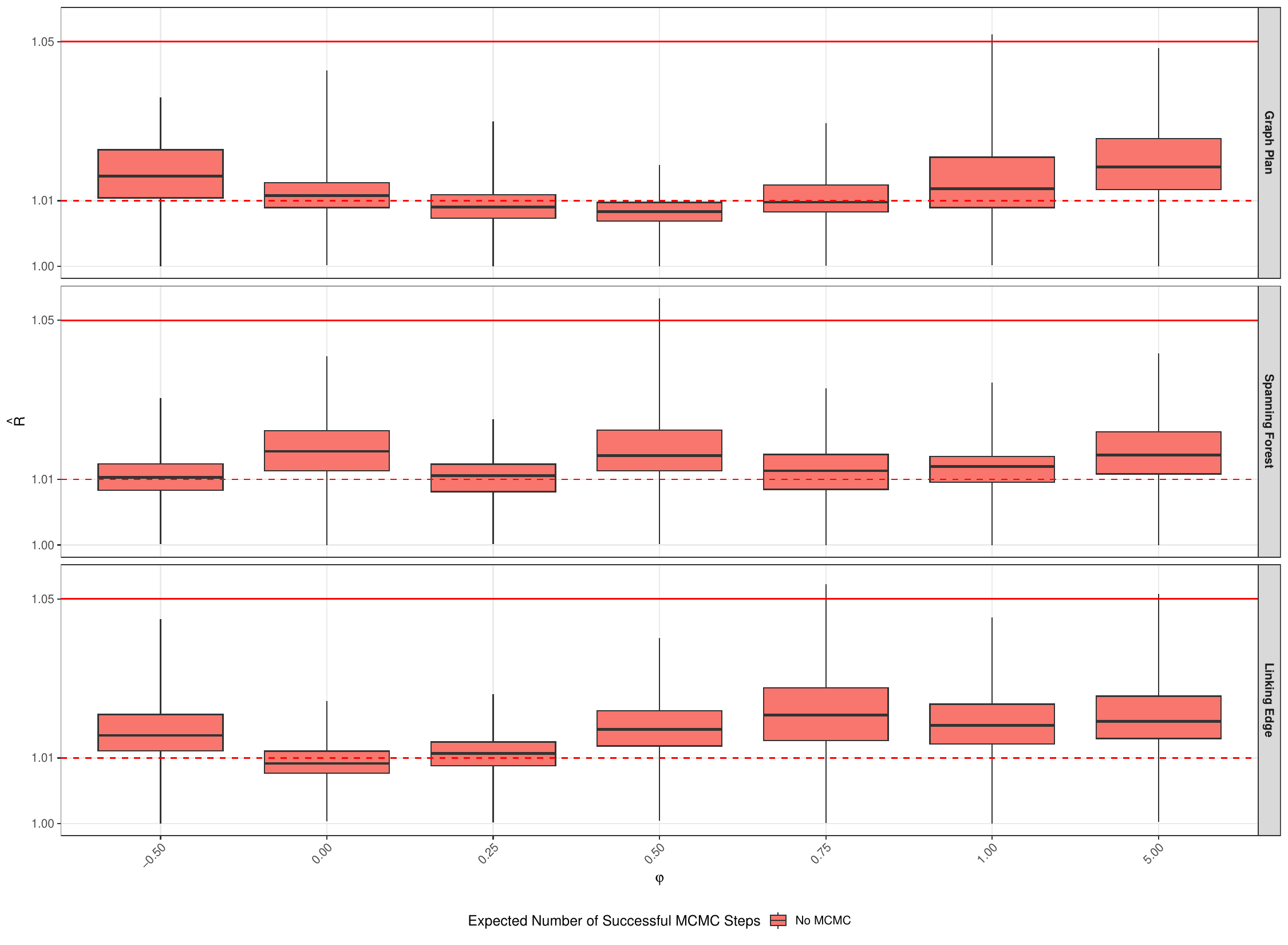}

}

\caption{\label{fig-varphi-rhats}Wisconsin Congressional: Distribution
of Rhat values across sampling space for any-valid splitting schedule
for a range of \(\alpha\) values.}

\end{figure}%

\begin{figure}[H]

\centering{

\includegraphics[width=1\linewidth,height=\textheight,keepaspectratio]{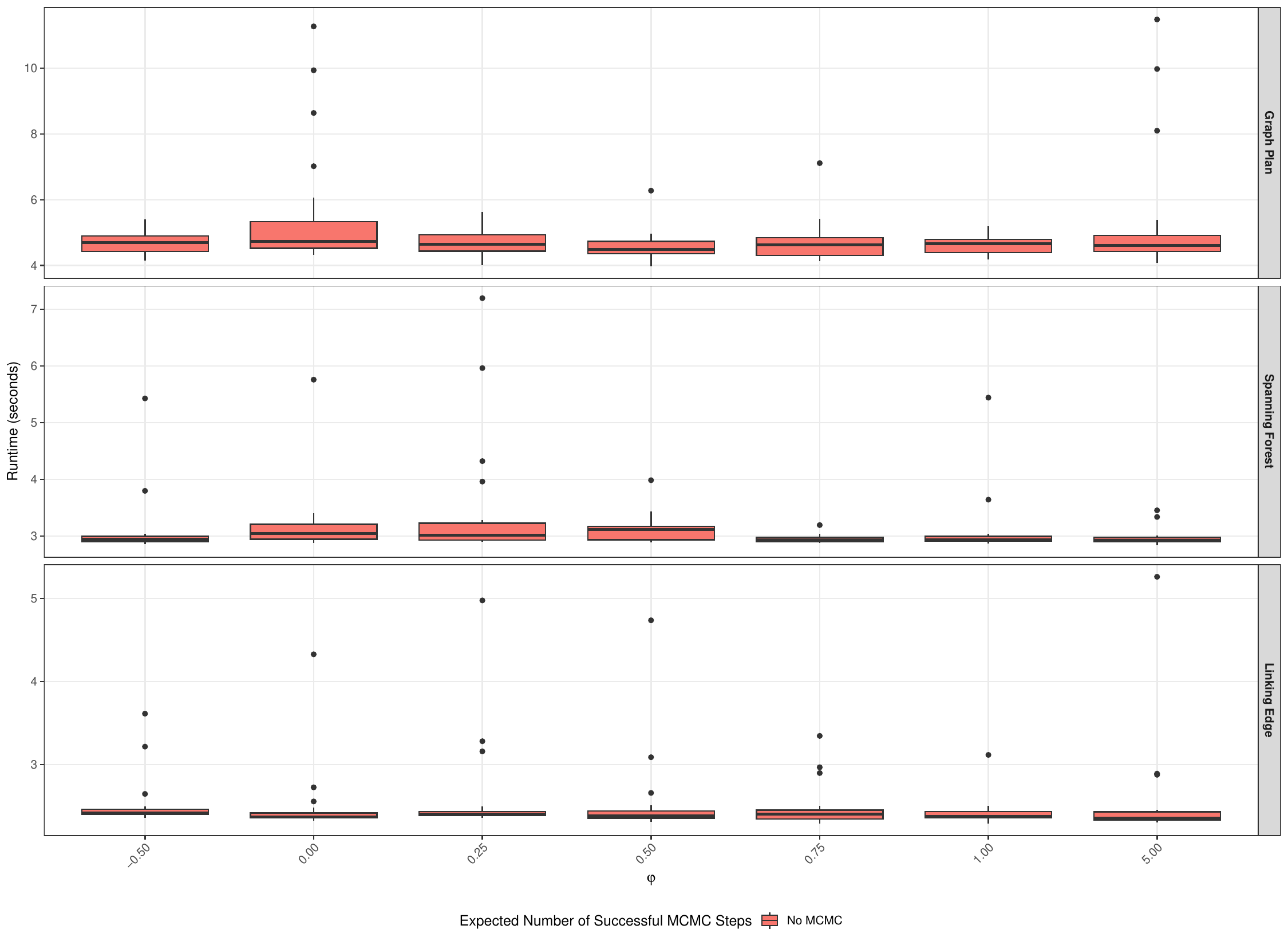}

}

\caption{\label{fig-varphi-runtime}Wisconsin Congressional: Distribution
of run wall times across sampling space for any-valid splitting schedule
for a range of \(\alpha\) values.}

\end{figure}%

\end{definition}

\end{definition}


\end{document}